\documentclass[thmsa]{book}

\usepackage{amssymb,latexsym}  % latexsym added by arXiv admin to give \Join, it
\input{tcilatex}
\begin{document}

\author{Bert Schroer \\
%EndAName
Institut f\"{u}r Theoretische Physik der FU-Berlin, Arnimallee 14, 14195
Berlin Germany.\\
e-mail: schroer@physik.fu-berlin.de}
\title{A Course on:\\
``Modular Localization and Nonperturbative Local Quantum Physics'' \\
CBPF, Rio de Janeiro, March 1998}
\maketitle
\tableofcontents

\section{Introduction}

A new text on an apparently old and established subject as QFT should
justify and measure itself relative to the many existing review articles and
textbooks. The main pragmatic motivation underlying these notes consists in
the desire to unify two presently largely disconnected branches of QFT:

(1) the standard (canonical, functional) approach which is mainly
perturbative in the sense of an infinitesimal ``deformation'' on free fields
and

(2) nonperturbative constructions of low-dimensional models as in the
formfactor-bootstrap approach (which for the time being is limited to
factorizable models in d=1+1 space-time dimensions) as well as the
non-Lagrangian construction of conformal chiral QFT's.

The synthesis requires a significant step beyond the concepts which were
used in order to formulate the two mentioned separate branches of QFT. On
the physical side, the S-matrix regains some of its early prominence;
however unlike in the old proposals of Heisenberg as well as in the later
S-matrix bootstrap approach of Chew et al., it remains subservient to the
locality and causality principles encapsulated in the theory of local
observables which is the heartpiece of algebraic QFT.

In the new context of the TCP- and the Tomita-reflection operators
(explained in detail in these notes), the S-matrix takes the role of a
powerful constructive tool in QFT. In theories with a mass gap, the validity
of scattering theory immediately provides a natural arena for the
nonperturbative description: the Hilbert space of incoming particles which
for Fermions/Bosons is a Fock space, i.e. an orthogonal sum of multiparticle
spaces which have the anti/symmetric tensor product structure in terms of
the Wigner one particle spaces. Whereas the connected Poincar\'{e}
representations of the incoming and interacting fields are identical%
\footnote{%
Note that in general the free incoming hamiltonian is different from the
unperturbed hamiltonian of standard perturbation theory. Only for special on
shell normalization conditions they agree.}, those interacting reflection
operators which contain the time reflections ($T,TP,TCP)$ differ from their
free incoming expressions. According to an old observation of Bisognano and
Wichmann this implies that the modular data for the algebra of the wedge
regions are known: the modular group is the one parametric group of wedge
affiliated Lorentz boosts whereas the modular conjugation differs from its
incoming value by the S-matrix. The fundamental significance of the wedge
region to the basics of QFT will perhaps not be surprising to physicist who
are familiar with Unruh's work on the thermal properties of the horizon of
the Rindler wedge as the simplest illustration of Hawking's and Bekensteins
thermal and entropical properties of black hole horizons. The point in these
notes is that the new concepts of modular localization causes the ``outing''
of this quasiclassical behavior associated to Killing vectors in CST QFT 
\textit{as a generic property of nonperturbative QFT.}

At this place it seems to be natural to make some explanatory remarks about
the Tomita-Takesaki modular theory. Whereas the detailed technical aspects
will be reserved for the mathematical appendix, some intuitive understanding
can be obtained by looking at its fascinating history. Mathematically it is
a vast generalization of the modular factor which accounts for the
difference between right and left Haar measure in the case of non unimodular
groups as e.g. SL(2,R). Tomita and later Takesaki succeeded to convert this
idea into a powerful tool for the investigation of von Neumann algebras. In
fact Alain Connes could not have carried out his path-breaking work on the
classification of von Neumann factor algebras without this Tomita-Takesaki
theory. Modular theory is also behind the subfactor theory of Vaughn Jones.

The physics side is equally impressive. At the time when Tomita presented
his theory, Haag, Hugenholz and Winnink \cite{Haag} published their
fundamental work on (heat bath) thermal aspects of QFT. The ''KMS''
condition (a name which was coined in that paper) was used before by various
physicist (in particular \textbf{K}ubo, \textbf{M}artin and \textbf{S}%
chwinger) as a clever mathematical trick in order to avoid to compute
cumbersome traces in evaluating Gibbs thermal ensembles in relativistic
theories. In the hands of Haag Hugenholz and Winnink this formula became the
key for their formulation of equilibrium quantum statistical mechanics
directly in the thermodynamic limit (in which the Gibbs representation
becomes meaningless as a consequence of volume divergencies in its vacuum
fluctuations). The generator for the modular operator turns out to be a
''thermal Hamiltonian'' with two-sided spectrum and finite fluctuations. It
acts on the original algebra as well as on its commutant which represents a
kind of an non-geometric ``shadow world'' (corresponding in the above
analogy to group theory to the opposite action ). The modular conjugation $J$
turns out to be the flip operation which maps the left into the right
action. Later Bisognano and Wichmann found another very nontrivial field
theoretic illustration of the Tomita-Takesaki theory in their study of the
algebra which is generated by fields restricted to the wedge region (which,
as already mentioned, corresponds to the Rindler region that played an
important role in Unruh's discussion of the Hawking effect), briefly called
the wedge algebra. In their case the modular operator turned out to be the
wedge-affiliated Lorentz boost and the modular conjugation $J$ was (up to a
180 degree rotation around the boost direction) the field theoretical TCP
operator. This time the von Neumann commutant was part of the real world,
namely the algebra of the causally disjoint region behind the wedge horizon.
This work (as well as some special prior observations on free fields \cite
{Oster Leyland}) strongly suggested that there was a deep relation between
modular theory and relativistic causality and localization. In more recent
times Borchers and Wiesbrock as well as Araki and Zsido showed that with two
subalgebras in appropriate modular position, one can build up space-time
symmetries (d=1+1 conformal and, with somewhat stronger assumptions, also
Poincar\'{e} covariance).

This set the stage for asking some fundamental questions about
nonperturbative QFT. The framework of algebraic QFT attributes less
importance to individual fields and prefers nets of algebras in order to
describe the physical content of local quantum physics. On the other hand
the conventional introduction of free fields via Wigner's particle theory of
positive energy representation of the Poincar\'{e} group is known to lead to
many free fields which belong to the same (m,s) Wigner representation and
live in the same Fock space. They are members of one local equivalence
(Borchers-) class of fields and are identical in their physical content;
though only very few posses a free Lagrangian $L_{0}$, i.e. can be
interpreted as resulting from canonical quantization of a classical field
theory. The unicity on the field theoretic level is achieved by a modular
construction (explained in chapter 3) which produces the net of interaction
free theories directly, i.e. without the intervention of point-like fields
and in this way leads to a one to one relation of the (m,s) Wigner
representation with one (m,s) algebraic net. Therefore these more
fundamental concepts leads not only to a de-emphasis of Lagrangians and
quantization ideas, but even delegates the very point-like covariant fields,
the apparent heart-piece of QFT, to the status of ``field coordinates'' for
the description of local ``nets'' of algebras. In this way it re-captures
the unicity of the Wigner representation theory which was lost by the
existence of a multitude of intertwiners relating the unique Wigner
representation to the infinitely many possible covariant representations.
This new point of view, which is consistent with fields (and hence
incorporates QFT), but places them into a much bigger and richer conceptual
arena, will be named Local Quantum Physics \cite{Haag} or LQP. The name QFT
is often linked with quantization, actions, functional integrals etc. which
do not appear in our nonperturbative approach but only in the chapter 4 on
perturbation theory.

This leads to the interesting question of whether such a construction of
field-coordinate independent approach is feasible in the presence of
interactions. It was natural to start again with the wedge situation and the
previously mentioned observation that if one replaces the reference to the
Wigner one-particle space by the incoming Fock space of scattering theory,
then the modular group will still be \textit{the L-boost of the free
incoming net but the }$J$\textit{\ operator will deviate from the free one
by the true scattering matrix.}

In setting up a scenario for a nonperturbative construction of nets in
interacting QFT's, an important building block is the ``modular M\o ller
operator'', a kind of square root out of the S-matrix. In the context of
factorizing d=1+1 theories this operator is closely related to the physical
representation of the Zamolodchikov-Faddeev algebra. There are good reasons
to believe (see chapter6) that such operators exists in general QFT and that
they are always associated with \textit{semi-local fields without vacuum
polarization} (just like the Z-F operators in x-space). They are somewhat
reminiscent of the elusive light cone fields, but different from those they
are not dependent on prerequisites of good short distance behavior for the
validity of the canonical formalism and their conceptual auxiliary position
relative to the local fields and compactly (double cone) localized algebras
is very clear. The absence of pair states after applying them to the vacuum
requires that they cannot be localized in smaller compact subregions inside
the noncompact wedge. The associated physical picture is that the wedge
region with its L-boost invariance group still permits to resolve massive
one particle states (a mass gap is assumed)\footnote{%
However the higher particle scattering states cannot be resolved. The
S-matrix cannot be extracted from the interacting wedge algebra alone, but
one needs the relation to the relatively extremely nonlocally positioned
incoming wedge algebra.}, whereas the regions of double cones are too small
in order to still carry particle properties. The construction of the modular
wedge localization spaces is achieved in terms of Unruh-like KMS states on
these semilocal generators.

Zero mass situations can give rise to infrared problems which lead to the
loss of the reference in-space (the large time limits vanish for example in
QED, and scattering cross sections must be computed directly in terms of
expectation values and probabilities instead of amplitudes); the associated
''infraparticles'' do not have a Fock space structure. In that case a Fock
reference space can only be associated with compact space-time regions of
the interacting theory, e.g. double cones which have no infrared problems.
The double cone algebras have a very big ``folium'' of states which includes
the double cone restriction of e.g. free massless states, but on the other
hand their modular structure is non-geometric (in the massive case) and less
susceptible to physical interpretation.

If the zero mass theories are conformally invariant, the modular objects are
again known and turn out to be geometric; however the deviation of the
modular involution from its free value is generally not relatable to
scattering. This opens the possibility to use standard Fock space methods
instead of introducing ad hoc new algebras (example: W-algebras), which are
not natural in the setting of particle representations\footnote{%
QFT which starts from Wigner's positive energy representation leads to Fock
spaces as the arena for interacting QFT. Theorems on positivity are
therefore not required (apart from gauge theories) because the embedding
Fock space is automatically positive.}, with bring in the risk of loosing
the Hilbert space setting due to indefinite metric. As a result of \textit{%
local unitary equivalence} of any conformal quantum algebra to e.g. zero
mass Fermions, we always have a (positive) Hilbert reference space for any
conformal QFT. In these notes we use the prefix ''quantum'' only for Hilbert
space representations, leaving aside certain indefinite metric
representations (which may have some use in statistical mechanics). Despite
their conceptual complexity, zero mass situations tend to be analytically
simpler, a fact which contributed e.g. to the popularity of chiral conformal
QFT.

We emphasized the gain in constructive insight into nonperturbative QFT in
order to convince the reader that the times have passed when the
contribution of algebraic QFT to problems of high energy physics and the
quasiparticle structure of condensed matter physics was ``smaller than any
preassigned epsilon'' (if this ever was true), as some people who worked on
long forgotten fashionable subjects of the 60$^{ies}$ used to say at that
time.

The gain in understanding of the foundations of QFT is equally impressive.
For example certain well-known ``technical assumptions'' concerning domains
of unbounded (smeared) field operators, which appeared first in Wightman's
axiomatic formulation, become now related with the basic physical properties
of the thermal modular localization subspaces (the domain spaces of the
Tomita operator $S)$. The Wightman domain (which appears in Wightman's
axiomatics \cite{wightman} ) turns out to be the intersection of all these
subspaces and the natural domain for fields restricted to a region is simply
the intersection of all modular subspaces corresponding to the double cones
which fit into that region. This modular characterization suggests that the
basic role of point-like (and hence operator-valued singular functions)
fields, as opposed to nets of bounded operators, is precisely to bring these
modular localization properties into the open, i.e. to make them manifest.
The space obtained by smearing a field vector $A(x)\Omega $ with arbitrary
test functions, i.e. the ``one field space'', is easily seen in chiral
conformal theories to be in one to one correspondence with an irreducible
representation space of the \textit{infinite dimensional group generated by
all modular transformations} of double cones (or rather intervals in the
chiral case)\footnote{%
This conjecture about the relation of individual fields with a kind of
universal modular group is due to Fredenhagen (private communication) who
observed it in chiral conformal QFT \cite{Fre Joe}}. If this correspondence
turns out to be a general property of algebraic field theory, one would have
achieved a group theoretical construction of fields in a theory of local
observables and in this way equipped them with a new fundamental role which
goes far beyond the before mentioned role of being just local
coordinatizations of algebras. Compared with the standard introduction of
local fields via quantization this would be a major conceptual achievement
since the modular properties are the most ''quantum'' or ''noncommutative''
in the sense that they \textit{do not permit} (unlike noncommutative
geometry) \textit{a commutative version at all}. As we will see the modular
algebraic properties insure that we maintain the biggest distance to
quasiclassical or perturbative concepts, which is just what we want.

In this modular role the ``field coordinates'' are more than just generators
of local algebras which carry information about the range of a local
observable algebra applied to the vacuum in their domain property. It is
through them that the somewhat hidden modular properties manifest themselves
through covariance laws and localization properties. The reason why these
modular properties which are so basic to QFT have been discovered so late,
lies in the fact that Lagrangians, euclidean methods and functional
integrals which dominated the scene of QFT, are too close to the
(quasi)classical quantization parallelism. Whereas quantization has played
an important intuitive role in the understanding of QFT from the side of
quasiclassical concepts and the formalism of renormalized perturbation
theory, the totally noncommutative modular structure promises to make
systematic inroads into the largely unknown nonperturbative regime.

It would have been nice to use the gain in nonperturbative insight in order
to classify free d=2+1 anyons and plektons by constructing their correlators
and related nonrelativistic quasiparticles. Unfortunately at the time of
writing the study of free plektonic charge carriers had barely begun
(chapter 7.7), but from the few indications it appears that the
nonrelativistic limit will remain a QFT i.e. there are no fields which
applied to the ground state yield a one (quasi)particle state as there are
for Fermions/Bosons in the Schr\"{o}dinger theory. In order to maintain the
plektonic relation between spin and statistics in the nonrelativistic limit
one needs the presence of virtual particles. This is related to the
paradoxical sounding statement that there are no relativistic free fields
(on shell, without vacuumpolarization clouds) for the description of free
anyons or plektons.

The credo that all the physics of field-coordinate-free QFT is carried by
the net of observables has withstood the passing of time and gained growing
acceptance, but it still transcends standard textbook formalism. I hope in
these notes I will convince some of the remaining sceptics about the
viability of the idea that all of local quantum physics is already contained
in the net of modular localization subspaces by demonstrating how it works
in d=1+1 factorizing theories.

At this point it should be abundantly clear that the central theme of these
lectures is\textit{\ localization of states and locality} (Einstein
causality in the case of observables) \textit{of algebras}. These concepts
are not only of great physical relevance in condensed matter physics
(localized and itinerant states), but also pivotal for the \textit{%
interpretation} of QFT. Attempts to bypass this and to use instead global
concepts from the beginning (as e.g. path integrals, topological field
theories) will occasionally receive some critical attention, but do not
enter the main line of this notes. Due to the proximity of localization with
the notion of temperature, the standard ''heat bath'' Hamiltonian
temperature will be introduced already in chapter 2. The first chapter is
reserved for some introductory remarks on general quantum theory. The Wigner
theory and free fields in chapter 3 is already presented with the modular
hindsight. Subsequently we will learn about the basic distinction between
''heat bath''-and ''localization''-temperature. In that same chapter we also
collect other constructions which stays close to free fields.

The subsequent chapter 5 on the general Wightman and LSZ framework and its
euclidean counterpart should be considered as a preparation for the modular
nonperturbative approach in chapter 6 which contains a detailed application
to the formfactor program of factorizing theories. It also serves as a
preparation for the pure algebraic approach in chapter 7. In order to keep
the notes reasonably self-contained, a brief collection of mathematical
results can be found in an appendix at the end.

It is the proximity of the nonperturbative concepts to TCP symmetry and
other fundamental notions in QFT which generate confidence in our approach.
The change of paradigm, which accompanies this new approach, is that those
structures of QFT which were always thought of as fundamental but somewhat
rigid and almost ``kinematical'' properties as TCP, now move into the center
of the ``dynamical'' stage\footnote{%
Of course such statements should be taken with the full awareness that the
cut between ``kinematics'' and what was hitherto considered as ``dynamics''
was never as rigid and a priori as the textbooks make us believe.}. In fact
in chapter 6 the reader will be led to take notice of the fact that all the
interaction resides in the TCP-reflection and that the S-matrix, in addition
to its well-known role as the large-time asymptotic operator describing
scattering, also plays an important role in modular localization. It is this
second property which e.g. explains its fundamental role in the formfactor
construction of factorizable QFT's. In this way the latter important
concepts used in that construction as crossing symmetry construction becomes
intimately connected with the thermal KMS and pair-creation properties of
the Rindler-Hawking Unruh issue and with physics near more general horizons.

A new approach is also expected to cast some new light on past successes.
For this reason we will attempt a rather systematic presentation which
includes a substantial part of the standard material of QFT (especially
chapter 4), but occasionally somewhat different from the standard textbook
treatments by emphasis and interpretation. Algebraic QFT acknowledges the
fact that it has a common cradle with other versions of relativistic
particle theories which is the renormalized Feynman-Schwinger Tomonaga
perturbation theory. In order to maintain and emphasize this common
heritage, we present in chapter 4 a simple derivation of the most important
effects in QED with a minimum on formalism but emphasis on physical
concepts. In the same chapter we also expose the physics of renormalization
in greater detail. Our subsequent presentation of perturbative gauge theory
uses the linear version of the BRS formalism of cohomological representation
formalism of selfinteracting spin 1 theories. This casts a new light on
perturbative massive spin 1 theories which in the standard gauge approach
were interpreted as spontaneously broken. Our investigations show that the
conceptual problems in gauge theories are still ill-understood and many
popular ideas of a preliminary nature.

In fact QFT, despite its age, is still in its conceptual infancy, especially
from the new nonperturbative viewpoint. This goes contrary to the widespread
(but unfounded) belief that the basic equations of the world below the
Planck scale are already known and that they predict a great ''desert'' and
that if we would be more clever in our computations (bigger computers etc.),
we would have a description of all phenomena below that scale.

The algebraic approach has many aspects in common with the philosophy
underlying Wilson's renormalization group. It also rejects the idea of a
``theory of everything''. Its picture of physical reality is more that of
many shells which are hierarchical ordered with respect to decreasing
distances. In every shell of this infinite ``onion'' there is the
possibility of having a consistent mathematical theory and the principles of
the next shell contain the previous ones as a limiting case. In fact the
mathematical description is very much interwoven with the definition of one
shell and its separation to the next one. Lagrangians, actions etc. are
interpreted as characterizing equivalence classes of theories which are
indistinguishable with respect to certain physical properties in the present
regime, similar to universality classes in condensed matter physics. Its
underlying philosophy is consistent with the use of effective Lagrangians a
la Weinberg. However it is less optimistic concerning what one can learn
from such phenomenological concepts and its notion of equivalence classes is
somewhat different and in some sense sharper. In addition to the well known
short distance universality classes with their unique scale invariant
representatives there are ``long distance'' classes (defined via cluster
limits of the S-matrix) which have the same charge superselection rules and
particle structure but simpler (or no) interactions. In d=1+1 they have
factorizing representatives.

The strategy for classifying and constructing models is in a certain sense
opposite to that proposed on the basis of renormalization group
transformations. Whereas the Wilson formalism approaches the construction of
fixed point theories by starting outside the fixed point theories and
reaches the latter by iterated renormalization group transformations, the
algebraic approach aims directly at an intrinsic construction of those
representatives and proposes to explore the more complicated terrain around
these simpler theories by perturbative ideas in the vein of Zamolodchikov's
picture of massive perturbations on chiral conformal theories.

A nice illustration of the difference in the underlying philosophy is given
by chiral conformal theories. In the algebraic approach the (imaginary time)
theory of critical indices becomes related to the computational easier
classifiable braid group statistics of the associated (real time) chiral QFT
and in this way the spectrum of possible critical indices gets related to
the superselection rules and their fusion and braid group characteristics of
the noncommutative local charges (critical indices modulo 1 = statistical
phases). Hence if one wants to compare the philosophy in these notes to the
earlier contributions on the connection between Statistical Mechanics and
QFT, it would be more appropriate to point towards Kadanoff's use of the
Coulomb gas as an analogue system for fusion of superselected charges.

There is however a quite serious discrepancy in points of view related to
certain aspects of the post Wilson renormalization group philosophy; notably
to the cavalier use of the notion of ``cutoffs'' in real time QFT\footnote{%
A statistical mechanics with a momentum space cut-off cannot be the analytic
continuation of a physically interpretable real time QFT.}. Whereas the
algebraic approach does not negate the usefulness of representing field
theoretic models as the scaling limit of lattice models ( however this only
has been achieved for a few cases of d=1+1 factorizing models), it is
totally incompatible with the direct use of (relativistic) cutoffs as a
method to dump some unknown physics of the next shell in a physically
acceptable way. Whereas the proponents of such ideas usually think in terms
of cutting off some integrals in their euclidean approach, the algebraic
framework is fully aware of the fact that nonlocal ``theories'' do not
permit a self-sustaining physical interpretation (example: the causality
origin of the crossing symmetric cluster properties as one needs them for
the derivation of scattering theory gets lost together with causality and
with it crumbles the whole interpretation). Here it is helpful to remember
that almost as long as QFT exists there were attempts to find a physical way
to regularize or cut-off light cone singularities or to introduce covariant
formfactors in Lagrangians or to add pairs of (conjugate)complex poles into
Feynman rules. Each attempt led to a total loss of physical interpretation.
LQP takes the 60 years of failure to give physical meaning to such notions
as cutoffs elementary length and noncausal interactions very serious and
protects the Einstein causality and localization principles of the first
half of this century against their present sell out\footnote{%
The renormalization group approach only maintains its aura of physical
modesty as long as it is not directed towards a sellout of the Einstein
causality principles by claiming that there is a physically interpretable
QFT in the presence of cutoffs and that it can define physically viable
nonlocal relativistic QFT.}. To put it very bluntly: whereas one can
approximate operators as Hamiltonians, it is not possible to approximate
principles inasmuch as it is not meaningful in everyday life to use the
phrase ``partially pregnant''. The transcendence of a physical principle is
not a ``bit of a non-principle'', but rather a new principle containing the
old one as a limiting case. None of the constructive aspects based on
modular localization makes sense without Einstein causality. Their is no
chance to manipulate a relativistic cutoff into any of the factorizing
models without wrecking irrevocably the entire physical interpretation. From
the algebraic point of view one may however entertain the speculative idea
that there exists a new principle beyond Einstein causality and that it
underlies the still very elusive quantum gravity.

The content of this report rather tries to attract those physicists who, on
the one hand still have not lost their conceptual curiosity, and on the
other hand feel uneasy about the present predominance of formalism over
conceptional insight. Such a reader may enjoy the unique charm and surprise
of finding deep physical concepts in rather mundane physical problems, far
from big deserts and theories of ''everything'' and well below the Planck
length. This should however not be misunderstood as a return to a
mathematical stone age.

The understanding of the material presented in these notes is mathematically
as demanding as the investment into geometry and topology, which the typical
reader probably already made in good faith and in the hope to cover most of
the mathematical structure of quantum physics. However the logical and
conceptual structure of local quantum physics is in many respects different
from geometry. Some of the new mathematical tools in local quantum physics
as the Tomita-Takesaki modular theory and the Vaughn Jones subfactor theory
are sketched in the appendix. Many of the important contributions were also
made by physicist, but apart from perhaps one exception \cite{Haag}, they
have not entered textbooks on QFT. Such a refinement of textbook knowledge
would be particularly necessary if it comes to low dimensional QFT. After
all, the main physical \textit{value of low dimensional models is their use
for a better conceptual and structural understanding of general QFT} (
perhaps also for the purpose of basing universality explanations in
condensed matter physics on firm grounds) and only in second place the
increase of our mathematical culture. For this reason the presentation of
low dimensional models (in particular chiral conformal QFT) may appear
different from the way the reader may have met this material in other
reviews.

Some ideas which apparently are natural from a differential geometric view
and presently enjoy great popularity as e.g. supersymmetry appear somewhat
artificial and unphysical from the LQP viewpoint. Apart from the fact that
no compelling theoretical principle which leads to such structures is known,
supersymmetry reveals itself within the conceptual frame of algebraic QFT as
an ``accidental symmetry'', unlike any other space-time or internal
symmetry. This peculiarity one expects to manifest itself e.g. in its
thermal behavior.. Indeed, very recent investigations of the thermal
behavior of supersymmetry showed \cite{Bu O} that this symmetry is
completely unstable in thermal KMS states\footnote{%
The usual (broken) symmetry picture, which holds for locally generates
internal and external symmetries in a heat bath does not apply here. Rather
supersymmetry suffers a total thermal collapse \cite{Bu O}.} and thus
confirm old suspicions that one is dealing here with an accidental symmetry.
In those low-dimensional supersymmetric models which are soluble (e.g. the
conformal Ising model and various massive factorizing models), the
supersymmetry plays no role in the nonperturbative analysis and its
placement within a family of soluble models. The short distance properties
of its charge-carrying fields are not better than that of its non
supersymmetric neighbors.. Whereas usual inner symmetries are deeply related
to the charge structure of the representation theory of its observables, the
systematic superselection analysis never asks for a supersymmetric encoding.
And whereas usual space-time symmetries are profoundly related to the group
generated by modular groups of local algebras, supersymmetry points toward
nothing but itself. For this reason there is no place for SUSY in these
notes.

Needless to say that our approach is rather conservative in its use of
existing concepts. The revolutionary aspect to the degree that there is one
at all, lies more in precise observations and discoveries than inventions.
In particular we will keep to physical notions as equivalence classes of
real time fields, TCP- and Tomita J-reflections rather than geometrical
concepts, imaginary time formalism, commutative cohomology and S- and
T-mirror-reflections etc.

On the other hand the reader will find a detailed presentation of ``Haag
Duality'' which is a pivotal property of nonperturbative local QFT and is
known to lead to such fundamental issues as (braid group) statistics and an
intrinsic understanding of spontaneous symmetry breaking. In d=1+1 we also
study its relation to the QFT formulation of Kadanoff order-disorder Duality
which is the local version of the global Kramers-Wannier Duality in lattice
systems.

I elaborated this material in the conviction that the best strategy in a
time of stagnation and crisis\footnote{%
Instead of a detailed verbal description, I refer to the picture at the end.}
is to return to the roots of QFT and re-analyze the underlying principles in
the light of new concepts. For this reason there is heavy emphasis on
Wigner's approach to particles which, long before algebraic QFT came into
existence, was the first successful attempt to do relativistic quantum
physics in an intrinsic fashion i.e. without recourse to quantization.

The relation of the present nonperturbative approach of LQP to standard
textbook QFT and its more recent differential-geometric refinements is
somewhat analogous to that of a wave optical treatment of QM relative to
proper QM in its own conceptual setting. In both cases one has to redo and
relearn the conceptual basis of the theory. The main distinction between the
present algebraic method and standard quantization approach (canonical
quantization, functional integrals etc.) is that in the first case the
properties which are assumed at the start of a construction are really
reflected in the result. On the other hand the quantization approach is more
artistic: canonical commutation relations outside free fields are almost
never valid for the renormalized results (exception: certain
super-renormalizable two-dimensional theories) and the same holds for
functional integral (Feynman-Kac) representations. Such structures are
working hypothesis which serve to start a chain of calculational ideas but
which themselves get lost on the way and often lead to other correct
properties. It seems to be a general feature of a quantization method that
it leads invariably to a connection of the existence of a QFT with a
sufficiently good short distance behavior of the underlying basic fields.
This is not borne out in our nonperturbative approach which is based on
modular localization and uses ideas which are very different from
quantization.

Anybody who knows me, understands that my attachment to Brazil (where most
of these notes were written) is not only due to its natural beauty, but
there are also deep scientific roots. With feelings of nostalgia I remember
those wonderful years (1968-1980) of collaboration with J. A. Swieca. These
were times of free-roaming scientific endeavour, long before the
globalization of physics, which nowadays forces young physicists to build
their career around some trendy formal ideas, started to do incalculable
harm to QFT. In these notes I remain faithful to the spirit of that
collaboration in that I shun away from inventions. All the new developments
and ideas the reader finds in this article are discoveries or observations
which follow from the principles of local quantum physics. The style is a
bit more informal than that of a text book, for example there are many
instances of repetitions. With new material originating from a different
conceptual framework than that of standard QFT it may actually be an
advantage for the reader to have repetitions of the same material from
slightly different perspectives.

This presentation should not be misread as a moral judgement against one or
the other approach to QFT. It rather is an attempt to revive some of the
critical conceptually based Bohr-Heisenberg-Wigner spirit in the present
times of: ``everything goes'' (at least as long as it lives up to that
entertaining high caliber scientific journalism which characterizes many
contributions to high energy physics, see the hep-th server\footnote{%
I strongly advice any young person still in search of a permanent position,
not to say such things in public.}). .

CBPF, Rio de Janeiro, March 1998\thinspace \thinspace \thinspace \thinspace
\thinspace \thinspace \thinspace \thinspace \thinspace \thinspace \thinspace
\thinspace \thinspace \thinspace \thinspace \thinspace \thinspace \thinspace
\thinspace \thinspace \thinspace \thinspace \thinspace \thinspace \thinspace
\thinspace \thinspace \thinspace \thinspace \thinspace Bert Schroer

\chapter{The Basics of Quantum Theory}

\section{Multiparticle Wave Functions, Particle Statistics}

It is well-known that quantum mechanical multiparticle systems can be
obtained by the tensor product construction from one-particle spaces. Let: 
\begin{equation}
\psi (x,s_{3})\in \mathcal{H}_{1}=\mathcal{L}^{2}\left\{ \mathcal{R}%
^{3},\left\{ -s,\ldots ,+s\right\} \right\}
\end{equation}
be a wave function of a nonrelativistic particle of spin s where, as usual, s%
$_{3}$ is the component of spin with respect to an arbitrarily chosen
quantization axis. The $L^{2}-$space has the standard inner product 
\begin{equation}
(\psi _{2},\psi _{1}):=\sum_{s_{3}=-s}^{s}\int d^{3}x\psi
_{2}^{*}(x,s_{3)}\psi _{1}(x,s_{3})
\end{equation}
and norm $\left\| \psi \right\| =(\psi ,\psi )^{\frac{1}{2}}$

The two-particle space: 
\begin{equation}
\mathcal{H}_{2}=\mathcal{H}_{1}^{a}\overline{\otimes _{\mathbf{C}}}\mathcal{H%
}_{1}^{b}\qquad \hbox{a b..type of\ particle}\mathcal{\ }
\end{equation}
is simply the closure of the algebraic tensor product over the complex
numbers generated by vectors: $\sum_{ik}\psi _{i}^{a}\otimes _{\mathbf{C}%
}\psi _{k}^{b}$ equipped with the scalar product (in the following we omit
the subscript $\mathbf{C}$) induced by the formula: 
\begin{equation}
(\psi _{i^{\prime }}^{a}\otimes \psi _{k^{\prime }}^{b},\psi _{i}^{a}\otimes
\psi _{k}^{b}):=(\psi _{i^{\prime }}^{a},\psi _{i}^{a})\cdot (\psi
_{k^{\prime }}^{b},\psi _{k}^{b})
\end{equation}
continued by linearity in the right and antilinearity in the left factor.
If, as it is usually the case, the one-particle algebra forms a complete
(irreducible) set of dynamical variables in the one-particle space, the
totality of all observables which are compatible with measurements of
subsytem $a$ (form the commutant of the $a$-algebra) inside the composed
system $a+b$ are just those which live in the tensor factor $b$; $a$ and $b$
are stochastically independent in a very strong sense. The generalization to 
$N$ \-factors: 
\begin{equation}
\mathcal{H}_{N}=\mathcal{H}_{1}\overline{\otimes }\mathcal{H}_{1}.....%
\overline{\otimes }\mathcal{H}_{1}
\end{equation}
is straightforward. The tensorproduct structure is the mathematical
formulation of kinematical ``statistical independence'' in the sense of
quantum theory. The dynamical variables of the subsystems (say particles
with spin) at a given time factorize: 
\begin{eqnarray}
\vec{X}_{i} &=&1\otimes ...\otimes \vec{x}_{i}\otimes 1...\otimes 1\,\,\,\,
\\
&&\,\hbox{and\ similarly\ for }\vec{P}_{i}\hbox{ and }\vec{S}_{i}  \nonumber
\end{eqnarray}
They act on N-particle wave functions $\psi (x_{1}s_{1},...x_{N}s_{N})\,$
with $s_{i}$ (omitting the subscript 3 for brevity) the third component of
the $i^{th}$spin. Global symmetry transformations, as translations and
rotations, act multiplicatively: 
\begin{equation}
\left( U(\vec{a})\psi \right) (x_{1}s_{1},...x_{N}s_{N})=\left( U_{1}(\vec{a}%
)\otimes ...\otimes U_{1}(\vec{a})\psi \right) (x_{1}s_{1},...x_{N}s_{N})
\end{equation}
\begin{equation}
\left( U(\vec{n},\theta )\psi \right) (x_{1}s_{1},...x_{N}s_{N})=\left(
U_{1}(\vec{n},\theta )\otimes ...\otimes U_{1}(\vec{n},\theta )\right) \psi
(x_{1}s_{1},...x_{N}s_{N})\,\,\,\,\,\,\,
\end{equation}
\begin{equation}
U(a)=e^{i\vec{P}\vec{x}},\,\,\,\,\,\,\,\,\,\,\vec{P}=\sum_{i}1\otimes
...\otimes \vec{P}_{i}\otimes ...\otimes 1
\end{equation}
$U(\vec{n},\theta )=e^{i\vec{J}\vec{n}\theta },\,\,\,\,\,\,\,\,\,\vec{J}%
=\sum_{i}1\otimes ...\otimes \vec{J}_{i}\otimes ...\otimes 1\,\,,\,\,\,\,%
\vec{J}_{i}=\vec{x}_{i}\times \vec{p}_{i}+\frac{1}{2}\vec{\sigma}_{i}$

Here $\vec{\sigma}$ are the Pauli-matrices. Operators which implement an
interaction between particles, as the hamiltonian $H$, violate this
one-particle factorization: 
\begin{equation}
U(t)=e^{-iHt},\,\,\,\,\,H=H_{0}+H_{int},\,\,\,\,H_{int}=\sum_{i<k}1\otimes
...V_{ik}(x_{i}-x_{k})...\otimes 1
\end{equation}

In the last expression the identity operator at the $i^{th}$ and $k^{th}$
place has been replaced by a conventional local pair interaction which, in
the exponentiated form looses the pairing property of the infinitesimal
generator. Therefore at first sight it appears, that the localized
pair-interaction leaves no mark at all on $U(t)$. Fortunately this is not
quite correct, its marks are the important ``cluster properties''. In our
context they ar\.{e} : 
\begin{equation}
\begin{array}{c}
H_{N}\rightarrow H_{N_{1}}\oplus H_{N_{2}},\,\,\,\Omega _{N}\rightarrow
\Omega _{N_{1}}\otimes \Omega _{N_{2}},\,\,\,S_{N}\rightarrow
S_{N_{1}}\otimes S_{N_{2}} \\ 
\hbox{on\ clustering\ wave\ functions:}\  \\ 
lim_{a\rightarrow \infty }\psi
(x_{1,....}x_{n_{1}},x_{n_{1}+1}+a,...x_{N_{1}+N_{2}}+a)
\end{array}
\end{equation}
Here $\Omega ^{\pm }=\lim_{t\rightarrow \infty }e^{iH_{0}t}e^{-iHt}$ are the
M\o ller operators and $S$ is the S-matrix $S=\Omega ^{+*}\Omega ^{-}.$
Equivalently one may introduce a partial translation $U_{C_{N_{2}}}(a)$
which translates the particles in the $N_{2}$-cluster $C_{N_{2}}$ infinitely
far away from the rest: 
\begin{equation}
\lim_{a\rightarrow \infty
}U_{C_{N_{2}}}(a)H_{N}U_{C_{N_{2}}}^{*}(a)=H_{N_{1}}\oplus H_{N_{2}},
\end{equation}
and similarly for $e^{iHt},$ $\Omega ^{\pm }$ and $S$ with $\otimes $
instead of $\oplus .$ The cluster property is therefore a kind of asymptotic
factorization, expressing statistical independence for long distances. As we
will see, it follows from more fundamental locality structures in QFT and
the existence of a spatially homogeneous reference state. Although it is
trivially satisfied for short-range quantum mechanical interactions, it
cannot always be taken for granted if the interactions become long-range. In
that case the cluster decomposition requirement is expected to affect the
boundary conditions of scattering theory. For example for the relative
Aharonov-Bohm interaction between ``dyons'' (electrically and magnetically
charged particles in spacetime dimension d=2+1), the cluster decomposition
property is expected to become a \textit{nontrivial imposition on the a
priori unknown scattering boundary conditions} $\neq $ plane wave condition,
which is the short range interaction boundary condition for stationary
scattering on multiparticle scattering states. These quantum theoretical
subtleties are of course irrelevant for the calculation of the Aharonov-Bohm
phase shift, which can be understood in entirely quasiclassical terms. In
the tentative applications to braid group statistics of dyons however, the
boundary condition is apriori not known. In the presence of an A-B long
range interactions that scattering boundary condition is the correct one,
which yields the cluster decomposition property of the scattering amplitude,
thus assuring that the $N$-particle S-matrix passes to the previously
determined $(N-1)$-particle S-matrix upon removal of one particle to
infinity (plane wave scattering conditions for long range interactions
generally destroy this property). In this way the $N$-particle problem
becomes related to the ($N-1)$-particle problem via the cluster property,
even though there is no actual creation of particles. Hence what appeared at
the beginning as a standard quantum mechanical problem, turns out to be a
problem in which particle states for all $N$ enter, i.e. a problem with a
field theoretic aspect. Presently not even a description of nonrelativistic
anyons exists since no solution of the N-particle Aharonov-Bohm scattering
with cluster properties has been elaborated.

In local QFT, cluster properties are a consequence of the general framework
and do not have to be imposed on interactions as in nonrelativistic physics.
In fact it is appropriate to think of the cluster decomposition property as
the relic of LQP (Haag) in the nonrelativistic limit. The statistics of
particles must have this inclusive structure lelating $N-1$ to $N$ and the
fact that the identity of particles is expressed by the braid group $%
B_{\infty }$ which is the inductive limit of $..\subset B_{i}\subset
B_{i+1}\subset ...$ (the same holds for the permutation group $S_{\infty }$
which is a special quotient of the braid group) reflects precisely the
inclusive ``russian matrushka'' structure of the clustering property on
purely algebraic level with the spacetime localization removed. In fact, as
will become clearer in a later chapter, this removal in LQP corresponds to a
theory with full localization ``flesh'' too a combinatorical theory of
intertwiners together with a natural tracial state. That theory of ``bones''
in the geometric approach is called ``topological field theory''.

It is well-known that the dynamical variables obtained as tensor products
form an irreducible system ($\simeq $validity of Schur's lemma) if the
single particle variables have this property in the one-particle Hilbert
space (true in our example of $\vec{p},$ $\vec{x},$ $\vec{s}$). In that case
the requirement of identity of particles permits only one-dimensional
(abelian) representations of the permutation group i.e. fermionic or bosonic
representations. In order to understand this, we have to incorporate the
notion of indistinguishability into our tensor products. Writing
symbolically $\psi _{n}(1,2,...n)$ as a short hand for our $x$- and
spin-dependent wave functions, a permuted wave function (as usual in QM, the
permutations act on the indices as $P^{-1}$ in order to comply with the
group composition, a behaviour well known from the action of the rotation
group on spatial coordinates): 
\begin{equation}
\begin{array}{c}
\left( U(P)\psi \right) _{n}(1,2,...n):=\psi _{n}^{P}(1,2,...,n):= \\ 
=\psi _{n}(P^{-1}(1),P^{-1}(2),...;P^{-1}(n))
\end{array}
\end{equation}
must describe the same physical state\footnote{%
i.e. the same expectation values in permuted wave functions for quantum
mechanical observables. Later we will introduce a more general concept of
``state'' which does not require a Hilbert space but rather leads to one.},
even though the permutation leads to a different vector in Hilbert space.
This distinction between vectors and physical states (in the sense of
defining expectation values on the algebra of observables) is crucial for
the understanding of the (later presented) superselection rules in the later
sections. In terms of observables, this simply means that the $U(P)^{\prime
}s$ commute with the observables. If, as in standard Schr\"{o}dinger theory,
the observables form an irreducible set of operators, the representation $%
U(P)$ of the permutation group $S_{n}$ must be abelian and hence: 
\begin{eqnarray}
\psi _{n}^{P}(1,2,...n) &=&\omega (P)\psi _{n}(1,2,...n)\, \\
\,\,\,\,\omega (P) &=&\left\{ 
\begin{array}{c}
1\,\,\,\,\,\,\,\,\,\,\,\,\,\,\,\,\,\,\,\,Bose \\ 
sign(P)\,\,\,\,\,\,\,Fermi\,\,\,\,
\end{array}
\,\,\right.  \nonumber
\end{eqnarray}
\newline
Mathematically this statement is a tautology and physically there is no
reason to assume irreducibility. In fact the occurrence of nonabelian
representations of $S_{n}$ is equivalent to the appearance of reducible
quantum mechanical observable algebras.

In order to construct such a nonabelian example, a rudimentary knowledge of
the representation theory of the permutation group is helpful. The
equivalence classes of irreducible representations of $S_{N}$ are
characterized by partitions with height $n$: 
\begin{equation}
N_{1}\geq N_{2}\geq ....\geq N_{n}\,,\,\,\,\,\,\sum_{i}N_{i}=N  \label{par}
\end{equation}
They are pictured by so called Young tableaus, an array of $N$ boxes in rows
of nonincreasing size (the admissibility condition for tableaus). Young
tableaux are also in a one to one relation to irreducible representations of 
$SU(N)$.

The representation of $S_{N}$ corresponding to each tableau of depth $d$ is
most conveniently described by decomposing the natural representation on the 
$N$-fold tensor product of a $d$-dimensional complex vector space $%
V^{\otimes N}$ (of dimension $d\cdot N$) into irreducibles. Up to
equivalencies, one obtains a subset of irreducible components by applying $%
S_{N}$ to the reference vector: 
\begin{equation}
\begin{array}{c}
e_{1}^{\otimes \left( N_{1}-N_{2}\right) }\otimes \left( e_{1}\wedge
e_{2}\right) ^{\otimes \left( N_{2}-N_{3}\right) }\otimes ....\otimes \left(
e_{1}\wedge e_{2}...\wedge e_{d}\right) ^{\otimes N_{d}} \\ 
e_{1}....e_{d}\,\,%
\hbox{\thinspace basis\thinspace
\thinspace in\thinspace \thinspace }\,V
\end{array}
\label{cy 1}
\end{equation}
with the tensor product action: 
\begin{equation}
\begin{array}{c}
U(P)\xi _{1}\otimes \xi _{2}\otimes ....\otimes \xi _{N}=\xi _{P(1)}\otimes
\xi _{P(2)}....\otimes \xi _{P(N)}\,,\; \\ 
\;\xi _{i}\in \left\{ e_{1}....e_{d}\right\}
\end{array}
\label{cy 2}
\end{equation}

The cyclically generated subspace is the irreducible representation space
corresponding to (\ref{cy 1}). Here the $\wedge $ designates the wedge
product leading to completely antisymmetric tensors. Clearly the vectors (%
\ref{cy 1} are in a one-to-one correspondence with partitions (\ref{par})
with height $n=d,$ which are also cyclic vectors with respect to irreducible 
$SU(d)$ representations. An ``admissible'' numbering of a given tableaux is
any numbering which is increasing in each row and column. The different
admissible numberings correspond to the multiplicity with which the
irreducible representation occurs in the regular representation (which has a
larger dimension than the above natural representation) of the group algebra 
$\mathbf{C}S_{N}$ (see next section).

Another method to construct the representation theory is the inductive
construction of tableaus according to Schur's rules of adding a small box to
a previously constructed tableaux corresponding to $S_{n-1}$: 
\begin{equation}
\pi _{T}\times box=\oplus _{adm.}\pi _{T+box}
\end{equation}
The admissible ways to add a box are such that the resulting tableaus are
admissible in the aforementioned sense. But now there is no maximal height $%
d $ and instead of the natural representation on tensor products on $d$-dim.
vectorspaces we are inducing the so called regular representation. The
iteration starting from $n=1$ gives a reducible representation of $S_{N}$
with $m(T_{N})$=multiplicity of occurrance of the tableaux $T_{N}$:

\begin{equation}
\oplus _{T_{N}}m(T_{N})\pi _{T_{N}}\,\,\,\,
\end{equation}
The multiplicity is the same as that of $T_{N}$ in the group algebra\textbf{%
\ }$\mathbf{C}S_{N}$ (next section). It is obviously equal to the number of
possibilities to furnish admissible numbering (the box numbering inherited
from the inductive Schur construction). It agrees with the dimensionality of
the representation. Furthermore the regular representation contains all
equivalence classes of irreducible representations whose number, as will be
shown in the next section for general finite groups, equals the number of
conjugacy classes.

Now we are able to sketch a (rather trivial) counterexample against the
irreducibility of the observable algebra. Imagine that we have Bose particle
which carry besides spin a ``hidden'' quantum number (``flavor''or
``color'') i.e. an internal degree of freedom which can take d values.
Assume that the measurability is restricted to flavor neutral operators:

\begin{equation}
\underline{\vec{x}}=\frac{1}{d}\sum_{a=1}^{d}\vec{x}^{\left( a\right)
},\,\,\,\,\,\,\,\,\underline{\vec{p}}=....\,\,\underline{\vec{s}}=.....
\end{equation}

Clearly the flavor averaged multiparticle observables act cyclically on a
smaller ``neutral'' reduced Hilbert space which may be described in the
following way: 
\begin{equation}
\mathcal{H}_{N}^{red.}=\mathcal{H}_{1}\otimes ....\mathcal{H}_{1}\otimes 
\mathbf{C}S_{N}
\end{equation}
Here the one-particle wave function spaces \QTR{cal}{H}$_{1}$have no flavor
degree of freedoms and $\mathbf{C}S_{N}$ is the $N$! dim. representation
space of the regular S$_{N}$ representation. We connect the reduced inner
product with the one in the flavor description: 
\begin{equation}
\begin{array}{c}
\left( \Phi \otimes P^{-1},\Psi \otimes Q^{-1}\right) _{red}= \\ 
\sum_{i_{1}..i_{N}}\frac{1}{d^{N}}\left( \varphi _{1}\otimes
e_{i_{P(1)}}\otimes ..\varphi _{N}\otimes e_{i_{P(N)}},\psi _{1}\otimes
e_{i_{Q(1)}}\otimes ..\otimes \psi _{N}\otimes e_{i_{Q(N)}}\right) _{sym}
\end{array}
\end{equation}

Here the $\varphi ^{\prime }s$ and $\psi ^{\prime }s$ are the spatial wave
functions without flavor and the e's are from the basis in d-dimensional
flavor space $V$. The inner product is the natural scalar product in the
symmetrized tensor space $\left( H\otimes V\right) _{symm.}^{\otimes _{N}}.$
On the left hand side $\Phi $ and $\Psi $ are the tensor products of the $%
\varphi _{i}$ and $\psi _{i}.$ The reduction is implemented through
averaging over the flavor degrees of freedom. The orthonormality relations
of the e's allow to simplify the result to: 
\begin{equation}
\begin{array}{c}
\sum_{i_{1}...i_{N}}d^{-N}\sum_{S}\prod_{j}{}\left( \varphi _{j},\psi
_{S(j)}\right) \left( e_{i_{P(j)}},e_{i_{Q(j)}}\right) = \\ 
=\sum_{S}\prod_{j}\left( \varphi _{j},\psi _{S(j)}\right) \varphi ^{\left(
N\right) }(PSQ^{-1})
\end{array}
\end{equation}
$\varphi ^{\left( N\right) }$denotes a tracial linear functional on CS$_{N}$
which on the basis elements $P$ is given by: 
\begin{equation}
\varphi ^{\left( N\right) }(P)=\frac{1}{d^{N}}TrU_{N}(P)\equiv tr(p)
\end{equation}
where $U_{N}(P)$ is the previously introduced natural representation of $S_{N%
\hbox{ }}$ on $V^{\otimes N}$ and Tr stands for the natural matrix trace.

The memory on the averaged flavor is completely absorbed in the multiplicity
factor. Note that flavored fermions would have given a similar result with
the only difference of $signP$ factors.

The conceptually aware reader will note, that this ``cooked up''
parastatistics illustration is precisely what an experimentalizer confronts,
if in a nonrelativistic atomic problem the electron spin would have no
dynamical manifestation (neglegible spin-orbit coupling) or if for a
neutron-proton system he would not be able to measure electric charge.
Internal symmetry is a very clever theoretical invention which trades the
unpleasant nonabelian Young-tableaux against the more physical (\textit{more
local!}) standard compact internal symmetry-group description. However it is
not universally applicable, see the later discussion of attempts to encode
nonabelian braid group statistics into a ``quantum symmetry'' concept.

It is now easy to see that the normalized trace of the natural
representation has an intrinsic characterization in terms of a tracial state 
$\varphi $ (a positive linear function) on the group algebra $\mathbf{C}%
S_{\infty }.$ Here $S_{\infty }$ is the inductive limit of the $S_{N}$
groups: 
\begin{equation}
S_{2}\subset ........\subset S_{N}\subset S_{N+1}\subset ......\vec{\subset}%
S_{\infty }
\end{equation}
The normalized tracial functional $\varphi ^{\left( N\right) }(P)$ on CS$%
_{N} $ (the extension from S$_{N}$ to the group algebra CS$_{N}$ is by
linearity) has a natural extension $\varphi $ to the inductive limit CS$%
_{\infty }.$ It is characterized by the following three properties: 
\begin{eqnarray}
&&\varphi \left( x\right) \,\hbox{\thinspace is\ tracial\ state\ on\ CS}_{N}
\\
i.e.\,\,\,\,\,\varphi (x^{*}x) &\geq &0,\,\,\,\,\varphi
(1)=1,\,\,\,\,\varphi (xy)=\varphi (yx)  \nonumber
\end{eqnarray}
\begin{equation}
\varphi \hbox{\ fullfills\thinspace the\thinspace Markov-property:\ }\varphi
(P_{1}P_{2})=\varphi (P_{1})\varphi (P_{2})
\end{equation}
\begin{equation}
\varphi (P=\hbox{transposition})=\pm \frac{1}{d},\quad \ 
\hbox{for\
flavoured\ fermions}
\end{equation}
where\thinspace $P_{1}=P_{1}$($\tau _{1}$,$\tau _{2}$....$\tau _{N_{1}-1}$%
)\thinspace is\thinspace \thinspace a\thinspace \textit{\thinspace }%
permutation involving the first $N_{1}-1$ generators{} and

$P_{2}=P_{2}$($\tau _{N_{1}+1}$,....$\tau _{N-1}$)\thinspace is a $\,$%
permutation involving the remaining generators except $\tau _{N_{1}}$.

\thinspace The generators $\tau _{i}$ (transpositions) are most conveniently
pictured as crossings of the $i^{th}$ strand with its neighbor $i+1$,
whereas all the other strands are running parallel (say upward).

These generators are subject to three relations: 
\begin{eqnarray}
\tau _{i}\tau _{j} &=&\tau _{j}\tau _{i},\,\,\,\,if\,\,\,\left| i-j\right|
\geq 2 \\
\tau _{i}\tau _{i+1}\tau _{i} &=&\tau _{i+1}\tau _{i}\tau
_{i+1},\,\,\,\,\,\,\,\,\,\tau _{i}^{2}=1
\end{eqnarray}
The first relation by its own is most appropriately pictured by allowing
over- and under-crossings in the $\tau _{i}$ i.e.by introducing $\tau
_{i}^{\pm }=\left\{ \tau _{i},\tau _{i}^{-1}\right\} $ see Fig 2. If one
adds to this Artin relation the second one $\tau ^{2}=1$, the braid group
(which even for finite number of strands is an infinite group) B$_{\infty }$
passes to the permutation group $S_{\infty }$. Both groups owe their
physical relevance to the fact that they are crucial for the understanding
of particle statistics and normal commutation relations between
charge-carrying fields. Their natural inclusive structure should be seen as
an algebraic counterpart of the physical cluster property. As we already
pointed out, the $S_{\infty }$ representations described by Young tableaus
occur in the centralizer algebras of tensor products of group
representations. Take e.g. the algebra generated by the d-dimensional
defining matrix representation $\pi $ of SU(n): 
\begin{equation}
\pi (g)\otimes \pi (g).......\otimes \pi (g)\,\,\,\,\,%
\hbox{in\thinspace
\thinspace \thinspace }V^{\otimes _{N}}
\end{equation}

In this case the commutant (or centralizer of the group representation) of
these operators does not only contain the algebra $\mathbf{C}S_{N}$ (this is
evidently the case for all tensor representations of groups ) but this
algebra is even identical to the centralizer. The following two questions
are relevant for the statistics classification:

(1)\thinspace \thinspace \thinspace \thinspace \thinspace Can one argue that
the indistinguishability requirement of particles (or other localizable
objects) together with other physical principles leads to representations
which are multiples of the above tracial representations of $S_{\infty }$ ?

(2) Is it natural to interpret the appearance of a tracial state with the
Markov property in terms of \thinspace an inner symmetry, and does one gain
anything by introducing the symmetry multiplicities (Heisenberg's isospin
and its ``flavor'' generalization) manifestly into the formalism?

Both questions have an affirmative answer, i.e.statistics and internal
symmetry are inexorably linked. The relevant theorem is the following:

\begin{theorem}
(following Doplicher, Roberts \cite{DR}) The most general statistics allowed
in d=3+1 dimensions is abelian (Bose-Fermi) \thinspace \thinspace together
with a compact internal symmetry group. The Boson-Fermion alternative is
\thinspace related to that of integer versus halfinteger spin (the
spin-statistics relation). The parastatistics associated with nonabelian
Young-tableaux (and described mathematically by a Markov trace on the
infinite permutation group algebra $\mathbf{C}S_{\infty })$ can be traded
for a compact internal symmetry group $G$ and Bosons/Fermions. The latter
description leads to a more local and less noncommutative description of QFT
which is also more susceptible to a ``quantization interpretation'' with
natural semiclassical limits.
\end{theorem}

Note that in the DR formulation the statistics problem is not only linked
with the spin, but also with the internal group symmetry.

Although the tracial nature and the Markov property can be derived from a
properly adapted (to the indistinguishability principle ) cluster
decomposition property, the natural place for its understanding and proof is
relativistic QFT (as is the case with other structural properties as spin
and statistics). The QFT locality and positive energy requirements naturally
imply the cluster property and the inclusive picture relating N particles
with N+1. We will give a derivation in a later chapters 5 and 7 on QFT in
connection with the theory of superselection sectors. It is this picture,
and not the N-particle quantum mechanics for fixed N (the standard quantum
mechanical ``proof'' for the Fermi-Bose alternative in most books on QM is a
tautology) which is responsible for the results on statistics. This
continues to be true in the case of the new quantizations of statistical
dimensions for braid group statistics which one finds in $d\leq 2+1$
dimensional QFT.

Superselection rules appeared first in the 1952 work of Wick, Wightman and
Wigner\cite{Haag}. These authors pointed out that the unrestricted
superposition principle of quantum mechanics or equivalently, the
unrestricted identification of self adjoint operators with observables (as
formulated by von Neumann) suffers a limitation through the appearance of
superselection rules. Their main example was standard quantum theory which
describes integer as well as halfinteger spin. Its Hilbert space is a sum of 
$\mathcal{H}^{\pm }$ where (-)+ corresponds to (half)integer spin. A linear
combination of vectors from both spaces changes its relative sign under $%
2\pi $ -rotation: 
\begin{equation}
\psi =\alpha \psi _{-}+\beta \psi _{+}\rightarrow \,\psi ^{2\pi }:=U(2\pi
)\psi =-\alpha \psi _{-}+\beta \psi _{+}
\end{equation}
Whereas the projective nature allows state vectors to suffer phase changes
(the quantum mechanical origin of halfinteger spin!), observables and states
(in the sense of expectation values) are unchanged under such a 2$\pi $
rotation and in this sense behave classically. The following calculation
shows that this is only possible iff the observables have vanishing matrix
elements between $\mathcal{H^{-}}$ and $\mathcal{H^{+}}$ 
\begin{eqnarray}
\left( \psi ,A\psi \right) &=&\left( \psi ^{2\pi },A\psi ^{2\pi }\right)
\Longleftrightarrow \left( \psi _{-},A\psi _{+}\right) =0%
\hbox{ \thinspace
\thinspace } \\
&&\,\,\hbox{for\ all\ observables}\ A\in \mathcal{A}  \nonumber
\end{eqnarray}
The proof just follows by inserting the above linear combinations. This
selection rule is called the ``univalence rule''. In contradistinction to
e.g. the $\Delta l=\pm 1$ angular momentum selection rules of atomic physics
which suffer changes in higher order radiative corrections, \textit{super}%
selection rules are universally valid. The vector state $\psi $ above cannot
be distinguished from a density matrix $\rho $: 
\begin{equation}
\left( \psi ,A\psi \right) =tr\rho A\,\,\,\,\,\,with\,\,\,\rho =\left|
\alpha \right| ^{2}\left| \psi _{-}\right\rangle \left\langle \psi
_{-}\right| +\left| \beta \right| ^{2}\left| \psi _{+}\right\rangle
\left\langle \psi _{+}\right|
\end{equation}
The formal generalization for the Hilbert spaces and observables is
obviously: 
\begin{equation}
\mathcal{H}=\oplus _{i}\mathcal{H}_{i}\,,\,\,\,\,A=\oplus
_{i}A_{i}\,,\,\,\,\,\,A_{i}=A\mid _{\mathcal{H}_{i}}
\end{equation}
Such observable algebras in block form \thinspace have a nontrivial center
given by the block projections. In the following we will illustrate this
decomposition theory by a simple but rich mathematical example, the
superselection rules of the group algebra.

\section{\thinspace \thinspace The Superselection Sectors of \textbf{C}G}

As a mathematical illustration of superselection rules we are going to
explain the representation theory of the group algebras.

Let $G$ be a (not necessarily commutative) finite group.We affiliate a
natural $\mathbf{C}^{*}$-algebra, the group-algebra $\mathbf{C}G$ with $G$
in the following way:

\begin{itemize}
\item  (i)\thinspace \thinspace \thinspace \thinspace \thinspace \thinspace
The group elements g$\in G$ including the unit e form the basis of a linear
vectorspace over $\mathbf{C}$: 
\begin{equation}
x\in \mathbf{C}G,\,\,\,\,\,x=\sum_{g}x(g)g\,,\,\,\,\,with\hbox{ }x(g)\in 
\mathbf{C}
\end{equation}

\item  (ii)\thinspace \thinspace \thinspace \thinspace \thinspace This
finite dimensional vector space $\mathbf{C}G$ inherits a natural convolution
product structure from G: 
\begin{equation}
\left( \sum_{g\in G}x(g)g\right) \cdot \left( \sum_{h\in G}y(h)h\right)
=\sum_{g,h\in G}x(g)y(h)g\cdot h=\sum_{k\in G}z(k)k
\end{equation}
\[
with\,\,z(k)=\sum_{h\in G}x(kh^{-1})y(h)=\sum_{g\in G}x(g)y(k^{-1}g) 
\]

\item  (iii)\thinspace \thinspace \thinspace \thinspace A *-structure, i.e.
an antilinear involution: 
\begin{equation}
x\rightarrow x^{*}=\sum_{g\in
G}x(g)^{*}g^{-1}\,\,,\,\,\,\,i.e.x^{*}(g)=x(g^{-1})^{*}
\end{equation}
Since\thinspace \thinspace : 
\begin{equation}
\left( x^{*}x\right) \left( e\right) =\sum_{g\in G}\left| x(g)\right|
^{2}\geq 0,\,\,\,\,(=iff\,\,\,\,\,x=0)
\end{equation}
\thinspace \thinspace \thinspace \thinspace this *- structure is
nondegenerate and defines a positive definite inner product: 
\[
\left( y,x\right) \equiv (y^{*}x)(e) 
\]

\item  (iv)\thinspace \thinspace \thinspace \thinspace The last formula
converts $\mathbf{C}G$ into a Hilbert space and hence, as a result of its
natural action on itself, it also gives a $C^{*}$ norm (as any operator
algebra): 
\begin{equation}
\left| \left| x\right| \right| =\sup_{\left\| y\right\| =1}\left\|
xy\right\| ,\,\,\,\,\,\,\,\,C^{*}-condition:\,\left| \left| x^{*}x\right|
\right| =\left| \left| x^{*}\right| \right| \left| \left| x\right| \right|
\end{equation}
A $\mathbf{C}^{*}-$norm on a *-algebra is necessarily unique (if it exists
at all). It can be introduced through the notion of spectrum ( mathematical
appendix).
\end{itemize}

It is worthwhile to note that (iii) also serves to introduce a tracial state
on $\mathbf{C}G$ i.e.a positive linear functional $\varphi \,$with the trace
property: 
\begin{equation}
\varphi \left( x\right) :=x(e),\,\,\,\,\varphi (x^{*}x)\geq 0,\,\,\,\varphi
(xy)=\varphi (yx)
\end{equation}
This state (again as a result of (iii)) is even faithful, i.e. the scalar
product defined by: 
\begin{equation}
\left( \hat{x},\hat{y}\right) :=\varphi (x^{*}y)
\end{equation}
is nondegenerate. On the left hand side the elements of $\mathbf{C}G$ are
considered as members of a vector space. The nondegeneracy and the
completeness of the algebra with respect to this inner product (a result of
the finite dimensionality of $\mathbf{C}G$) give a natural representation
(the regular representation of $\mathbf{C}G$) on this Hilbert space: 
\begin{equation}
x\hat{y}:=\widehat{xy}
\end{equation}
The norm of these operators is identical to the previous one.

This construction of this ``regular'' representation from the tracial state
on the $\mathbf{C}$*-group-algebra is a special case of the general
Gelfand-Neumark-Segal (GNS-)construction presented in a later section.

Returning to the group theoretical structure, we define the conjugacy
classes $K_{g}\,$and study their composition properties. 
\begin{equation}
K_{g}:=\left\{ hgh^{-1},h\in G\right\}
\end{equation}
In particular we have $K_{e}$=$\left\{ e\right\} $. These sets form disjoint
classes and hence: 
\begin{equation}
G=\cup _{i}K_{i},\,\,\,\left| G\right| =\sum_{i=0}^{r-1}\left| K_{i}\right|
,\,\,\,\,K_{e}=K_{0},\,\,\,K_{1,}....K_{r-1},\,\,\,\,r=\#classes
\end{equation}
We now define central ``charges'': 
\begin{equation}
Q_{i}:=\sum_{g\in K_{i}}g\,\in \mathcal{Z}(\mathbf{C}G):=\left\{ z,\,\left[
z,x\right] =0\,\,\,\,\,\forall x\in CG\right\} \,\,  \label{QB}
\end{equation}
It is easy to see that the center $\mathcal{Z}(\mathbf{C}G)$ consists
precisely of those elements whose coefficient functions $z(g)$ are constant
on conjugacy classes i.e. $z(g)=z(hgh^{-1})$ for all h. The coefficient
functions of $Q_{i}$: 
\begin{equation}
Q_{i}(g)=\left\{ 
\begin{array}{c}
1\,\,\,\,if\,\,\,g\in K_{i} \\ 
0\,\,\,otherwise
\end{array}
\right.
\end{equation}
evidently form a complete set of central functions. The composition of two
such charges is therefore a linear combination of the r independent $%
Q_{i}^{\prime }s$ with positive integer valued coefficients (as a result of
the previous formula (\ref{QB})): 
\begin{equation}
Q_{i}Q_{j}=\sum_{l}N_{ij}^{l}Q_{l}
\end{equation}
The fusion coefficients $N$ can be arranged in terms of $r$ commuting
matrices 
\begin{equation}
\mathbf{N}_{j},\,\,\,with\,\,\left( \mathbf{N}_{j}\right) _{i}^{l}=N_{ij}^{l}
\end{equation}
The associativity of the 3-fold product $QQQ$ is the reason for this
commutativity, whereas the abelianess of the algebra (only valid for abelian
groups!) results in the $i-j$ symmetry of the fusion matrices: 
\[
Q_{i\,\,}\,\,central\,\,\,\curvearrowright
\,\,Q_{i}Q_{j}=Q_{j}Q_{i}\,\,\,\curvearrowright N_{ij}=N_{ji}\,\,\, 
\]

Functions on conjugacy classes also arise naturally from characters $\chi $
of representations $\pi $: 
\begin{equation}
\chi ^{\pi }(g)=Tr\pi (g)\,,\,\,\,\,\,\,\chi ^{\pi }(g)=\chi ^{\pi
}(hgh^{-1})
\end{equation}
This applies in particular to the previously defined left regular
representation $\lambda $ with $\left( \lambda _{g}x\right) \left( h\right)
=x(g^{-1}h).$ Its decomposition in terms of irreducible characters goes hand
in hand with the central decomposition of\textbf{\ }$\mathbf{C}G$: 
\begin{equation}
\mathbf{C}G=\sum_{l}P_{l}\mathbf{C}G,\,\,\,\,\,\,\,\,\,Q_{i}=\sum_{l}Q_{i}^{%
\pi _{l}}P_{l}
\end{equation}
The central projectors $P_{l}$ are obtained from the algebraic spectral
decomposition theory of the $Q_{i}^{\prime }s\,$ by inverting the above
formula. The ``physical'' interpretation of the coefficients is: $Q_{i}^{\pi
_{l}}=\pi _{l}(Q_{i})$ i.e. the value of the $i^{th}$ charge in the $l^{th}$
irreducible representation. The central projectors $P_{l}$ are simply the
projectors on the irreducible components contained in the left regular
representation. Since any representation of $G$ is also a representation of
the group algebra, every irreducible representation must occur in $\lambda (%
\mathbf{C}G)$. One therefore is supplied with a complete set of irreducible
representations, or in more intrinsic terms, with a complete set of $r$
equivalence classes of irreducible representations. As we met the intrinsic
(independent of any basis choices) fusion rules of the charges, we now
encounter the intrinsic fusion laws for equivalence classes of irreducible
representations. 
\begin{equation}
\pi _{k}\otimes \pi _{l}\simeq \sum_{m}\tilde{N}_{kl}^{m}\pi _{m}
\end{equation}
Whereas the matrix indices of the $N^{\prime }s$ label conjugacy classes,
those of $\tilde{N}$ refer to irreducible representation equivalence
classes. The difference of these two fusions is typical for nonabelian
groups and corresponds to the unsymmetry of the character table: although
the number of irreducible representations equals the number of central
charges (=\thinspace \# conjugacy classes), the two indices in $\pi
_{k}(Q_{a})$ have a different meaning. With an appropriate renormalization
this mixed matrix which measures the value of the $i^{th}$ charge in the $%
l^{th}$ representation we obtain the unitary \textit{character matrix} $%
S_{ka}\equiv \sqrt{\frac{\left| K_{a}\right| }{\left| G\right| }}Tr\pi
_{k}(g_{a})$ (Tr is the normalized trace) which diagonalizes the commuting
system of $N^{\prime }s$ as well as $\tilde{N}^{\prime }s$: 
\begin{eqnarray}
\frac{S_{ka}}{S_{0a}}\cdot \frac{S_{la}}{S_{0a}} &=&\sum_{m}\tilde{N}%
_{kl}^{m}\frac{S_{ma}}{S_{0a}},\quad  \label{S} \\
\frac{\sqrt{\left| K_{a}\right| }S_{ka}}{S_{k0}}\cdot \frac{\sqrt{\left|
K_{b}\right| }S_{kb}}{S_{k0}} &=&\sum_{c}N_{ab}^{c}\frac{\sqrt{\left|
K_{c}\right| }S_{kc}}{S_{k0}}\quad  \nonumber
\end{eqnarray}
The surprise is that $S$ shows up in two guises, once as the unitary which
diagonalizes this $N(\tilde{N})$-system, and then also as the system of
eigenvalues which can be arranged in matrix form. We will not elaborate on
this point.

In passing we mention that closely related to the group algebra $\mathbf{C}G$
is the so-called ``double'' of the group (Drinfeld): 
\begin{equation}
D(G)=C(G)\Join _{ad}G
\end{equation}
In this crossed product designated by $\bowtie $ , the group acts on the
functions on the group $\mathbf{C}(G)$ via the adjoint action: 
\begin{equation}
\alpha _{h}(f)(g)=f(h^{-1}gh)
\end{equation}
The dimension of this algebra is $\left| G\right| ^{2}$ as compared to dim$%
\mathbf{C}G=\left| G\right| $. Its irreducible representations are labeled
by pairs $(\left[ \pi _{irr}\right] ,K)$ of irreducible representation and
conjugacy class and therefore their matrices $N$ and $S$ are \textit{selfdual%
}. In this sense group doubles are ``more symmetric'' than groups. In
chapter 7 we will meet selfdual matrices $S$ which cannot be interpreted as
a double of a group and which resemble the $S$ of abalian groups.

Finally we may notice that the equivalence classes of irreducible
representations appear with the natural multiplicity: 
\begin{equation}
mult(\pi _{l}\,\,in\,\,\lambda _{reg})=\,dim\pi _{l}
\end{equation}
The results may easily be generalized to compact groups where they are known
under the name of Peter-Weyl theory.

Since group algebras are very special, some remarks on general finite
dimensional algebras are in order.

Any finite dimensional $C$*-algebra $\mathcal{R}$ may be decomposed into
irreducible components.and any finite dimensional irreducible C*-algebra is
isomorphic to a matrix algebra $Mat_{n}(\mathbf{C})$ . If the irreducible
component $Mat_{n_{i}}(\mathbf{C})$ occurs with the multiplicity $m_{i}$ ,
the algebra $\mathcal{R}$ has the form is isomorphic to the following matrix
algebra: 
\begin{equation}
\mathcal{R}=\bigoplus_{i}Mat_{n_{i}}(\mathbf{C})\otimes
1_{m_{i}}\,\,\,\,\,\,in\,\,\,\mathcal{H}=\oplus \mathcal{H}_{n_{i}}\otimes 
\mathcal{H}_{m_{i}}
\end{equation}
and the multiplicities are unrelated to the dimensionalities of the
components. The commutant of $\mathcal{R}$ in \QTR{cal}{H\,\,}is: 
\begin{equation}
\mathcal{R}^{\prime }=\oplus _{i}1_{n_{i}}\otimes Mat_{m_{i}}(\mathbf{C}%
)\,,\,\,\,\,\,\,\,\,Z:=R\cap R^{\prime }=\oplus _{i}\mathbf{C\cdot }%
1_{n_{i}}\otimes 1_{m_{i}}
\end{equation}
The last formula defines the center.

Let us conclude with some remarks on states over (finite dim.) $C$%
*-algebras. Since finite dimensional $C$*-algebras decompose into
irreducible components (this decomposition agrees with the central
decomposition), it suffices to look at states on irreducibles i.e.on the
matrix algebra $Mat_{n}(\mathbf{C})$. The linear functionals can be
identified with these matrix spaces since one can use the unique normalized
tracial state $\varphi (A)=\frac{1}{n}TrA$ to define a nondegenerate inner
product which does this identification. By restricting these linear
functionals to be positive and normalized one obtains the well known
representation of states in terms of density matrices: 
\begin{equation}
\varphi _{\rho }(A)=tr\rho A,\,\,\,\,\rho \geq 0,\,\,\,tr\rho =1
\end{equation}
States are simply positive normalized linear fuctionals $\varphi $ on (in
the present case finite dimensional) $^{*}$-algebras $\mathcal{A}.$ From the
definition one immediatly abstracts two properties: 
\begin{eqnarray}
\varphi (A^{*}) &=&\overline{\varphi (A)}  \nonumber \\
\left| \varphi (A^{*}B)\right| ^{2} &\leq &\varphi (A^{*}A)\varphi
(B^{*}B),\,\,\forall A,B\in \mathcal{A}  \label{Schwa}
\end{eqnarray}
The reality property follows directly from the positivity $\varphi
((A+cB)^{*}(A+cB))\geq 0$ for arbitrary complex numbers c and the second
(Cauchy-Schwartz) inequality follows from the positivity of the sesquilinear
form $\varphi ((cA+dB)^{*}(cA+dB))$ in c,d. The validity of the above
density matrix representation of $\varphi $ follows from the linearity which
on $Mat_{n}(\mathbf{C})$ which is spanned in terms of matrix units $%
e_{ik},\varphi (A)=\sum a_{ik}\varphi (e_{ik})$ and the positivity which in
terms of $\varphi (e_{ik})$ leads to the positivity of the matrix ($\rho
)_{ik}\equiv \varphi (e_{ik}).$

In the simplest case of $Mat_{2}(C),$\thinspace which corresponds to the
spin algebra generated by the Pauli matrices together with the identity
matrix, the convex space of states (the space of density matrices $\rho $)
is a 3-dim.ball: 
\begin{equation}
\,\,\rho =\frac{1}{2}\left( 1+\vec{r}\vec{\sigma}\right) ,\quad \,\,\vec{r}%
^{2}\leq 1\quad
\end{equation}
\[
\sigma _{1}=\left( 
\begin{array}{cc}
0 & 1 \\ 
1 & 0
\end{array}
\right) ,\sigma _{2}=\left( 
\begin{array}{cc}
0 & -i \\ 
i & 0
\end{array}
\right) ,\sigma _{3}=\left( 
\begin{array}{cc}
1 & 0 \\ 
0 & -1
\end{array}
\right) ,\,\, 
\]
Besides the normalization we used the positivity of $\rho $ (which requires
positive eigenvalues and hence a positive determinant) to derive the
inequality. The pure states correspond to one-dimensional projectors and
therefore cover the surface of the ball. Note that pure states can be
defined intrinsically (i.e.without referring to one-dimensional projectors
or state vectors) by the property of indecomposability of $\varphi :$%
\begin{equation}
\not{\exists}\,\,\,\varphi _{1},\varphi _{2}\,\,s.t.\quad \varphi =\alpha
\varphi _{1}+\beta \varphi _{2},\,\,\alpha ,\beta \geq 0,\,\,\,\,\,\alpha
+\beta =1,\,\varphi _{1}\neq \varphi _{2}\,
\end{equation}
Starting from $Mat_{3}(\mathbf{C}),$ one meets the new phenomenon of a
stratification of the surface into convex subregions called ``faces''.
Density matrices with 3 different eigenvalues correspond to faithfull states
inside the convex state space. If one of the eigenvalues vanishes, one
looses faithfulness and although these states are ``purer'', they are not
pure in the sense of indecomposability. They form a ``face'' which looks
precisely like the previous ball. In higher dimensions one finds lower
dimensional faces inside higher dimensional ones. Once one is on a face, any
further purification takes place inside this face. The facial structure is
important for recognizing that a normalized convex space is the state space
of a $C^{*}$-algebra. Returning to the situation of a general finite
dimensional $\mathbf{C}$*algebra, we now see that a state is described by a
collection of positive traces or $\rho ^{\prime }s$ (one for each central
component) with normalized total trace.

In order to appreciate the structural differences to classical observables
and states, one should remember that the classical observable algebra
consists simply of (continuous) functions on phase space and pure states are
represented by Dirac $\delta $-functions whereas the Liouville measure and
the Radon-Nikodym derivatives with respect to it represent the mixtures and
may be viewed as continuously smeared out $\delta -$functions. Since the
notion of coherent superposition is absent, a classical analogon of the
states for the $Mat_{n}(\mathbf{C})$ toy models would be a higher
dimensional simplex. The pure states are the vertices and every mixed state
is obtained by a unique convex combination of the vectors corresponding to
the vertices. This is quite different from the the ball-shaped region of
quantum physics where the representation of a point in the interior in terms
of pure states on the surface is highly nonunique. In fact this structure of
states is, more than anything else, the most characteristic property of
quantum physics. The presence of superselection rules tells us that at least
partially there exists a classical structure within quantum physics: the
central decomposition of a $\mathbf{C}$*algebra is unique (similar to the
classical case), and the unrestricted superposition principle holds only
within each component. In fact this classical aspect of the central
decomposition plays an important role in the measurement process and the
notion of ``commuting histories''.

The concepts needed in the infinite dimensional case are more subtle and
will be presented in a mathematical appendix.

The reader may ask the question of why superselection rules despite their
importance are rarely mentioned in quantum mechanics. The reason is the
validity of the Stone-von Neumann uniqueness-theorem on irreducible
(regular) representations of the Schr\"{o}dinger theory. In QM of \textit{%
finite degrees of freedom} it is only \textit{only through topological
nontriviality }of configuration- or phase-space that inequivalent
representations may enter the quantization procedure. Let us look at a
typical example, the quantum mechanics of a particle in a circle. The
geometric argument in favour of finding many inequivalent representations is
as follows. Diagonalizing first the algebra of the position operator we
represent the state vectors by (periodic) wave functions $\psi (\varphi )$
on $S^{1}$. In order to fulfill the Heisenberg-Weyl commutation relations,
the most general form for the momentum is $p=i\frac{\partial }{\partial
\varphi }+f(\varphi )$. Wheras on $\mathbf{R}$ the $f$ can be
(gauge-)transformed away, the best we can do on a circle is to transform to
a constant. In n dimension the n-component $f_{i}$ has to fulfill the
flatness condition $\partial _{i}f_{j}-\partial _{j}f_{i}=0.$ the can be
transformed The $p$ only commutes with itself iff the real function $f$ is a
constant (in higher dimensions the p would be like a gauge covariant
derivative, and the constancy of $f$ like the flatness of a connection or
the vanishing of the field strength associated with a vector potential). In
our one dimensional example on $S^{1},$ the $p$ which generates the circular
translation obeys the famous $\theta $-angle formula: 
\begin{equation}
p=i\frac{\partial }{\partial \varphi }+\theta ,\,\,\,\theta \,\,mod\,2\pi
\end{equation}
Hence there are many Schr\"{o}dinger theories corresponding to different
selfadjoint extensions of the nonselfadjoint hermitian differential operator
parametrized by a theta-angle which have different physical content (e.g.
the spectrum of $H_{0}^{(S_{1})}$). An equivalent form is obtained by
keeping the standard Schr\"{o}dinger-form of the operator $p$ but accepting
quasiperiodic wave functions. The ``$\theta $-obstruction'' is intrinsic, it
can be shifted from the algebra to the states, but only in a simply
connected space it can be removed alltogether by a nonsingular operator
transformation; this is the case in ordinary Schr\"{o}dinger theory.

This geometric viewpoint has one disadvantage: there is no natural way to
consider the different $\theta $-theories as just different manifestations
of one system, rather they are different geometrical objects (generally
inequivalent vectorial fibre bundles). Here the \textit{algebraic view} of
superselection rules is \textit{physically superior}: the different $\theta
- $theories are simply different representations of one abstract $C$*algebra
( in our case the ``rotational algebra'').

The mathematical example of a particle on a circle is closely related to the
Aharonov-Bohm effect. As long as the solenoid has not passed to the infinite
thin limit, the A-B system falls into the ordinary Schr\"{o}dinger
description. It is only through the \textit{limiting overidealization} that
the $\theta -$dependent circular mechanics with its nonsimple \textbf{C}%
*-algebra enters. There is a general message in this example: all
topologically nontrivial quantum mechanics result from an overidealization
of Schr\"{o}dinger theory. Only for infinite degrees of freedom in QFT it
becomes possible to encounter \textit{superselection rules which have a
natural fundamental origin} (e.g. a phase transition). We refer to section 5
of the last chapter where in particular the globalization through the
universal observable algebra (which mimics the geometrical compactification)
comes close to topologically nontrivial models of QM if one only retains the
global (nonlocalizable) degrees of freedom.

One lesson to be drawn for QFT from these illustrations is that one is not
limited by ``quantization'' methods. Rather one may use the superselection
idea and try to classify and construct QFT's by studying representations of
observable algebras instead of quantizing classical physics. It was realized
by Haag, Araki, Borchers and others \cite{Haag} already at the beginning of
the 60's that the principle of locality makes such a formulation very
consistent and structurally rich. But the path from those early studies to
the more recent advances in e.g. properties of low dimensional QFT with new
surprising nonperturbative insights was very thorny indeed. Our approach to
QFT and even these remarks on QM are strongly influenced by this ``algebraic
QFT''. In particular the problem of particle statistics will be presented as
part of the understanding of superselection charges.

\section{ Illustration of Important Quantum Concepts}

In this section some additional quantum-physical concepts will be introduced
in a finite dimensional setting (for simplicity). For the much richer
infinite dimensional concepts used to analyse $C^{*}$- and von Neumann
algebras, we refer to the mathematical appendix.

We start with the GNS-construction. It associates in a canonical way with a
given $\mathbf{C}$*-algebra \QTR{cal}{A}$\,$and a state $\omega $ on it a so
called GNS-tripel ($\mathcal{H}_{\omega }$,$\pi _{\omega }$($\mathcal{A}$),$%
\Omega _{\omega }$) which consists of a representation space \QTR{cal}{H}$%
_{\omega }$ with a distinguished vector $\Omega _{\omega }$ on which 
\QTR{cal}{A} through its representation $\pi _{\omega }(\mathcal{A})$ acts
cyclically. The construction is analogous to that of the regular
representation of $\mathbf{C}G$. Again one uses the state $\omega $ in order
to construct a positive semidefinite sesquilinear form on the linear space
of \QTR{cal}{A}:

\begin{equation}
\left( \psi _{A},\psi _{B}\right) :=\omega (A^{*}B)=\overline{\omega (B^{*}A)%
}
\end{equation}
Here we use a notation which distinguishes the vectors from the elements of
the algebra. The reason is that whereas the tracial state $\varphi $ on the $%
\mathbf{C}G$ was faithfull i.e. the sesquilinear form was strictly positive,
a general state $\omega $ on \QTR{cal}{A} leads to a nontrivial nullspace 
\QTR{cal}{N}$_{\omega }$%
\begin{equation}
\mathcal{N}_{\omega }=\left\{ A\in A\mid \omega (A^{*}A)=0\right\}
\end{equation}
Fortunately this nullspace \QTR{cal}{N}$_{\omega }$ is also a left ideal
(the Gelfand ideal) of \QTR{cal}{A} i.e. with \QTR{cal}{A}$\in \mathcal{N}%
_{\omega }$ also $BA\in \mathcal{N}_{\omega }$ for any $B\in \mathcal{A}.$
This follows from the Cauchy-Schwartz inequality for states (\ref{Schwa}): 
\begin{equation}
\left| \omega (A^{*}B)\right| ^{2}\leq \omega (A^{*}A)\omega (B^{*}B)
\end{equation}
if we write the latter in the adapted form: 
\begin{equation}
\left| \omega \left( (BA)^{*}BA\right) \right| ^{2}\leq \omega
(C^{*}C)\omega (A^{*}A),\quad C:=((BA)^{*}B)^{*}
\end{equation}
In this form it is obvious that the left hand side must vanish if $A\in 
\mathcal{N}_{\omega }$. The proof of the Cauchy-Schwartz inequality for
states is identical to that for scalar products of vectors in Hilbert space.

If one now defines the state vectors $\psi $ as elements of \QTR{cal}{A} mod 
\QTR{cal}{N}$_{\omega }$ and the action of \QTR{cal}{A} on these vectors as: 
\begin{equation}
\pi _{\omega }(B)\psi _{A}:=\psi _{BA},\quad B\in \mathcal{A}
\end{equation}
then one obtains the desired relation between the state and the
representation: 
\begin{equation}
\left( \Omega _{\omega },\pi _{\omega }(A)\Omega _{\omega }\right) =\omega
(A)
\end{equation}
Here $\Omega _{\omega }$ is the distinguished vector in $\mathcal{H}_{\omega
}$ which corresponds to $\mathbf{1}$ (vanishing Gelfand ideal on $\Omega
_{\omega })$. The only additional step for infinite dimensional algebras is
to form the closure of the linear space and to continue the definition of $%
\pi _{\omega }(A)$ to this Hilbert space closure of $\mathcal{A}$ mod $%
\mathcal{N}_{\omega }$. Since a dense set of vectors is obtained by applying 
\QTR{cal}{A} to $\Omega _{\omega },$ the proof is finished. It remains to be
added that every other cyclic representation $\pi _{\omega }(A)$ with the
same $\omega (A)$ turns out to be unitary equivalent to the canonical GNS
representation.

As a preparation for the next topic let us use a tracial state as a
reference state. On a factor $Mat_{n}(C)\otimes \underline{1}_{m}$ in the
central decomposition of a semisimple algebra a tracial state is unique and
has the standard form: 
\begin{equation}
\tau (A)=\frac{1}{n}TrA\equiv trA,\quad TrA=\sum_{i}A_{ii}
\end{equation}
On a semisimple algebra there is a family of tracial states parametrized by $%
\lambda _{i}\geq 0,$ $\sum_{i}\lambda _{i}=1:$%
\begin{equation}
\tau _{(\lambda _{1}\lambda _{2}....\lambda _{k})}\left( A\right)
=\sum_{i=1}^{k}\lambda _{i}\tau (A_{i})
\end{equation}
\begin{equation}
\tau _{(\frac{n_{i}}{N},\frac{n_{2}}{N}....\frac{n_{k}}{N})}\left( A\right)
=:\tau (A)=\frac{1}{N}TrA,\quad N=\Sigma _{i}n_{i}
\end{equation}
where $\mathcal{A}_{i}$ is the $i^{th}$ central component of $\mathcal{A}$
and $\tau $ is the standard faithful tracial state. We have seen before that
the most general state $\varphi $ may be described in terms of a density
matrix $\rho :$%
\begin{equation}
\varphi (A)=\tau (\rho A)  \label{repr}
\end{equation}
This formula is just a concrete realization of the GNS construction via the
trace formalism (in which case the nullspace vanishes and the correspondence
between matrices and vectors in the representation space is one to one): 
\begin{equation}
\begin{array}{c}
\tau (A^{*}B)=\left( \Omega _{\tau },\pi _{\tau }(A^{*})\pi _{\tau
}(B)\Omega _{\tau }\right) \\ 
\varphi (A)=\left( \Omega _{\varphi },\pi _{\varphi }(A)\Omega _{\varphi
}\right) =\tau (\rho ^{\frac{1}{2}}A\rho ^{\frac{1}{2}}) \\ 
\Omega _{\omega }=\rho ^{\frac{1}{2}}\Omega _{\tau },\quad \Omega _{\tau }=%
\underline{1}\quad \hbox{in space of matrices}\mathcal{H}_{\tau }
\end{array}
\end{equation}
Note the analogy of this construction with the regular representation of $%
\mathbf{C}G$: in both cases the algebra in its role as a space, together
with the trace, served as the the arena for GNS faithful representations.
Such a Hilbert space $\mathcal{H}_{\tau }$ may be written in a more
suggestive $L^{2}$-notation: 
\begin{equation}
\mathcal{H}_{\tau }=L^{2}(Mat_{N},\tau ),\quad \Omega _{\tau }=\underline{1}
\end{equation}
In addition to the ``left'' GNS representation $\pi _{l}=\pi _{\tau }=\pi ,$
we introduce an antilinear right representation: 
\begin{equation}
\pi _{r}(B)\psi _{A}=\psi _{AB^{*}}=\pi _{\tau }(A)\underline{1}\pi _{\tau
}(B^{*})
\end{equation}
In the group case, the right action unravels the multiplicity structure of
the irreducible representations (the irreducible representations occur
according to the Peter-Weyl theory with a multiplicity identical to the
dimension of the representation as a result of the complete symmetry between
left and right action on $\mathbf{C}G$ ). In our present more general
setting, the existence of the two commuting left-right actions furnish the
germ of a deep and general theory: the Tomita-Takesaki modular theory. One
first defines the antiunitary ``flip'' operator $J$: 
\begin{equation}
J\psi _{A}=\psi _{A^{*}}\curvearrowright j(\pi (A)):=J\pi (A)J=\pi _{r}(A)
\end{equation}
In this case the $J$ not only implements the flip, but it also transforms
the vector $A\Omega _{\tau }$ into $A^{*}\Omega _{\tau }.$ In the more
general $\varphi $-representation one has two different involutive operators
S and J: 
\begin{equation}
\begin{array}{c}
S\pi (A)\Omega _{\varphi }:=\pi (A^{*})\Omega _{\varphi }\quad \Omega
_{\varphi }=\rho ^{\frac{1}{2}}\underline{1} \\ 
J\pi (A)J=\pi _{r}(A)
\end{array}
\end{equation}
Fom this one reads off the new operator $S$ in terms of $J$ and $\rho $ (\ref
{repr}): 
\begin{equation}
S=J\Delta ^{\frac{1}{2}},\quad \Delta =\pi (\rho )\pi _{r}(\rho ^{-1})
\end{equation}
since the identity: $S\pi (A)\rho ^{\frac{1}{2}}\underline{1}=J\rho $$^{%
\frac{1}{2}}\pi (A)\rho ^{\frac{1}{2}}\underline{1}\rho ^{-\frac{1}{2}}=\pi
(A^{*})\rho ^{\frac{1}{2}}\underline{1}\,$ follows from the definitions. The
formula for $S$ agrees with the polar decomposition formula into an
``angular'' part $J$ and positive ``radial'' part $\Delta ^{\frac{1}{2}}$ .
The above formulas require $\rho $ to be invertible i.e. the $\varphi -$%
representation to be faithful (or equivalently $\Omega _{\varphi }$ to be a
separating vector in $\mathcal{H_{\tau }}$\QTR{cal}{)}. Remembering the
notation of the von Neumann commutant $\mathcal{A}^{\prime }$ and the
algebraic role of these operators, the following important properties are an
easy consequence: 
\begin{equation}
j(\mathcal{A})=\mathcal{A}^{\prime },\quad \sigma _{t}(A):=\Delta
^{it}A\Delta ^{-it}\in \mathcal{A}
\end{equation}
$j(\cdot )$ is the modular conjugation and $\sigma _{t}(\cdot )$ is called
the modular group. $\Delta ^{it}$ commutes with $J$ and hence the action of
the latter on the commutant $\mathcal{A}^{^{\prime }}$ is obtained by
replacing $\Delta ^{it}$ by its inverse. Now every $\rho $ may be written as
a Gibbs formula in terms of a (ad hoc) hamiltonian $H$: 
\begin{equation}
\rho =\frac{1}{Z}e^{-\beta H},\quad Z=Tre^{-\beta H}
\end{equation}
In this case $\Delta ^{it}=\pi _{l}(e^{-i\beta tH})\pi _{r}(e^{i\beta tH})$
and the modular automorphism on $\mathcal{A}$ is apart from a stretching
factor $-\beta $ equal to the hamiltonian automorphism. Note that the
infinitesimal generator $\mathbf{H}$ of the time translations is not simply $%
H$ but rather: $\mathbf{H}=\pi _{l}(H)\otimes \stackunder{-}{\mathbf{1}}%
\oplus \stackunder{-}{\mathbf{1}}\otimes \pi _{r}(-H).$ This fact becomes
important in the realistic (infinite dimensional) case, since the heat bath
fluctuations of the hamiltonian $H$ become infinitly large in the
thermodynamical limit. The generator of the modular automorphism acts on the
algebra as well as (in opposite fashion) on its commutant. In the context of
thermal physics: the generator of time translation is the sum of the
hamiltonian and its opposite. Only in this way the formalism controlls the
fluctuations of $\mathbf{H,}$ wheras those of the hamiltonian $H$ have the
well-known volume divergencies.

It is therefore not surprising that the modular theory was discovered
independently (and at the same time) by physicists in the study of
temperature states on bosonic or fermionic algebras.

In our toy case, instead of writing down the Gibbs formula, one may also
characterize the faithful state $\varphi $ by the so-called KMS condition: 
\begin{equation}
\varphi (A\sigma _{t}(B))=\varphi (\sigma _{t+i\beta }(B)A)
\end{equation}
In the finite dimensional case the analytic dependence of $\sigma $ on t is
automatic whereas for the general case the analyticity of a function $%
F_{z}(A,B):=\varphi (A,\sigma _{z}(B))$ (for $A,B$ from a dense subalgebra
of $\mathcal{A}$) and the formulas: 
\begin{equation}
\begin{array}{c}
F_{z}(A;B)\,\,\hbox{analytic in strip: }0\leq \hbox{ Im\thinspace }z\leq
\beta \\ 
\hbox{with }F_{z}\hbox{ =}\left\{ 
\begin{array}{c}
\varphi (A\sigma _{t}(B))\hbox{ for Im\thinspace \thinspace }z=0\,\,%
\hbox{i.e.}\ z=t+i0 \\ 
\varphi (\sigma _{t+i\beta }(B)A)\hbox{ for Im\thinspace }z=t+i\beta
\end{array}
\right.
\end{array}
\end{equation}
constitute part of the \textit{definition} of the KMS property of $\varphi .$
This characterization of thermal equilibrium states is more general than the
Gibbs formula, since the latter looses its meaning as a result of the volume
divergencies in the thermodynamic limit V$\rightarrow \infty $ with the
particle densities kept fix. In addition, even in case of finite volume, the
calculation of $\varphi $ by KMS boundary condition is much easier than by
calculating traces. This practical advantage was the reason why Kubo, Martin
and Schwinger introduced this condition, whereas the mathematical physics
connection was made much later by Haag, Hugenholtz and Winnink \cite{Haag}.

Besides the notion of states and representation, the concept of inclusions
of von Neumann algebras will play an important role in later sections. Here
we will only present a ``baby'' version. Suppose that $Mat_{2}(\mathbf{C})$
acts not on its natural irreducible space $\mathbf{C}^{2}$ but by left
action on the 4-dim Hilbert space $\mathcal{H}(Mat_{2}(\mathbf{C}),\frac{1}{2%
}Tr).$ In that space the commutant is of equal size and consists of $Mat_{2}(%
\mathbf{C})$ acting in the opposite order from the right which will be
shortly denoted as $Mat_{2}(\mathbf{C})^{opp}$. Explicitely the realization
of $\mathcal{H}$ as $\mathbf{C}^{4}$ may be defined as 
\begin{equation}
\left( 
\begin{array}{ll}
\xi _{11} & \xi _{12} \\ 
\xi _{21} & \xi _{22}
\end{array}
\right) \rightarrow \left( 
\begin{array}{l}
\xi _{11} \\ 
\xi _{21} \\ 
\xi _{12} \\ 
\xi _{22}
\end{array}
\right)
\end{equation}
and the action of $\mathcal{A}=Mat_{2}(\mathbf{C})$ takes the following
form: 
\begin{equation}
a=\left( 
\begin{array}{llll}
a_{11} & a_{12} & 0 & 0 \\ 
a_{21} & a_{22} & 0 & 0 \\ 
0 & 0 & a_{11} & a_{12} \\ 
0 & 0 & a_{21} & a_{22}
\end{array}
\right) \simeq \left( 
\begin{array}{ll}
a_{11} & a_{12} \\ 
a_{21} & a_{22}
\end{array}
\right) \otimes \underline{1}
\end{equation}
The most general matrix in the commutant $a^{\prime }\in \mathcal{A}^{\prime
}$ has evidently the form: 
\[
a^{\prime }=\left( 
\begin{array}{llll}
a_{11}^{\prime } & 0 & a_{12}^{\prime } & 0 \\ 
0 & a_{11}^{\prime } & 0 & a_{12}^{\prime } \\ 
a_{21}^{\prime } & 0 & a_{22}^{\prime } & 0 \\ 
0 & a_{21}^{\prime } & 0 & a_{22}^{\prime }
\end{array}
\right) \simeq \underline{1}\otimes \left( 
\begin{array}{ll}
a_{11}^{\prime } & a_{12}^{\prime } \\ 
a_{21}^{\prime } & a_{22}^{\prime }
\end{array}
\right) 
\]
The norm $\left\| \xi \right\| =\left( \frac{1}{2}Tr\xi ^{*}\xi \right) ^{%
\frac{1}{2}}$ is invariant under the involution $\xi \rightarrow \xi ^{*}$
which in the $\mathbf{C}^{4}$ representation is given by the isometry: 
\begin{equation}
J=\left( 
\begin{array}{llll}
K & 0 & 0 & 0 \\ 
0 & 0 & K & 0 \\ 
0 & K & 0 & 0 \\ 
0 & 0 & 0 & K
\end{array}
\right) ,\quad K:\hbox{natural conjugation in }\mathbf{C}
\end{equation}
We have: 
\begin{equation}
j(\mathcal{A}):=J\mathcal{A}J=\mathcal{A}^{\prime },\quad 
\hbox{antilin. map
}\mathcal{A}\rightarrow \mathcal{A}^{\prime }
\end{equation}
which may be rewritten in terms of a linear anti-isomorphism: 
\begin{equation}
a\rightarrow Ja^{*}J,\quad \mathcal{A}\rightarrow \mathcal{A}^{\prime }
\end{equation}
Consider now the trivial algebra $\mathcal{B}=\mathbf{C\cdot 1}_{2}$ as a
subalgebra of $\mathcal{A}=Mat_{2}(\mathbf{C})$. In the $\mathbf{C}^{4}$
representation the B-algebra corresponds to the subspace: 
\begin{equation}
\mathcal{H}_{B}=\left\{ \left( 
\begin{array}{l}
\xi \\ 
0 \\ 
0 \\ 
\xi
\end{array}
\right) ,\xi \in \mathbf{C}\right\} ,\quad \mathcal{H}_{B}=e_{B}\mathcal{H}%
,\quad e_{B}=\left( 
\begin{array}{llll}
\frac{1}{2} & 0 & 0 & \frac{1}{2} \\ 
0 & 0 & 0 & 0 \\ 
0 & 0 & 0 & 0 \\ 
\frac{1}{2} & 0 & 0 & \frac{1}{2}
\end{array}
\right)
\end{equation}
The projector $e_{B}$ commutes clearly with $\mathcal{B}$ i.e.$e_{B}\in 
\mathcal{B}^{\prime }$ . We now define a measure for the relative size of $%
\mathcal{B}\subset \mathcal{A}$ the Jones index: 
\[
\left[ A:B\right] =\tau _{B^{\prime }}(e_{B})^{-1},\quad \tau :%
\hbox{
normalized trace in }\mathcal{B}^{\prime } 
\]
In our example $\tau (e_{B})=\frac{1}{4}(\frac{1}{2}+0+0+\frac{1}{2})=\frac{1%
}{4}$ i.e. the satisfying result that the Jones index is 4. The same method
applied to the inclusion: 
\begin{equation}
Mat_{4}(\mathbf{C})\supset Mat_{2}(\mathbf{C})\otimes \mathbf{1}_{2}=\left\{
\left( 
\begin{array}{ll}
X & 0 \\ 
0 & X
\end{array}
\right) ,X\in Mat_{2}(\mathbf{C})\right\}
\end{equation}
also gives the expected result: 
\begin{equation}
\left[ A:B\right] =\frac{\dim Mat_{4}(\mathbf{C})}{\dim Mat_{2}(\mathbf{C})}%
=4
\end{equation}
If, as in the previous cases $B$ is a finite dimensional subfactor (i.e.a
full matrix algebra) of $A,$ the Jones index is the square of a natural
number. For inclusions of finite dimensional semisimple algebras the index
takes on more general values. For example: 
\begin{eqnarray}
Mat_{2}(\mathbf{C})\oplus \mathbf{C} &=&\left( 
\begin{array}{lll}
X &  &  \\ 
& X &  \\ 
&  & x
\end{array}
\right) \subset Mat_{2}(\mathbf{C})\oplus Mat_{3}(\mathbf{C}) \\
X &\in &Mat_{2}(\mathbf{C}),\quad x\in \mathbf{C}1  \nonumber
\end{eqnarray}
Here the index is 3. It is easy to see that instead of the projector formula
one may also use the incidence matrix formula: 
\begin{equation}
\left[ \mathcal{A}:\mathcal{B}\right] =\left\| \Lambda _{n}^{m}\right\| ^{2}
\end{equation}
The incidence matrix $\Lambda $ is describable in terms if a bipartite
graph. The number of say white vertices correspond to the number of full
matrix component algebras for the smaller algebra and the black vertices
labelled by the size of the components to the analogously labelled
irreducible components of the bigger algebra. A connecting line between the
two sets of vertices indicates that one irreducible component of the smaller
is included into one of the bigger algebra. In our case: 
\begin{equation}
\Lambda =\left( 
\begin{array}{ll}
1 & 1 \\ 
1 & 0
\end{array}
\right) ,\quad \left| \left| \Lambda \right| \right| ^{2}=3
\end{equation}
From a sequence of ascending graphs one obtains important infinite graphs
(Bratteli diagrams) which are very useful in the ``subfactor theory'' \cite
{Jones} which will appear in the mathematical appendix. In the infinite
dimensional case the inclusion of full matrix algebras corresponds to the
inclusion of von Neumann factors i.e. the ``subfactor problem''. In that
case the spectrum of inclusions shows a fascinating and unexpected
quantization phenomenon, the Vaughn Jones quantization formula for index $%
\leq 4.$ AFD (almost finite dimensional) C*-algebras obtained by sequences
of ascending Bratteli diagrams equipped with tracial states enter LQP via
the intertwiner algebra of charge transporters. A special case are the
combinatorial theories which result from Markov-traces on selfintertwining
transporters which contain the braid group and mapping class group
generators for arbitrary genus. They are better known under the name
``topological field theories'' (their differential geometric name). Their
physical role in 3-dimensional plektonic LQP will be briefly mentioned in
chapter 7.

\section{ Measurement and Superselection Rules}

The interpretation of quantum theory requires an observer which also may be
a registration apparatus outside the observed system. Therefore notions like
the ``state or wave function of the universe'' have to be handled with great
care and are in some cases meaningless, at least within the standard
physical interpretation of QFT. Somewhere between the (generally
microscopic) observed system and the observer a ``cut'' is needed. As
already pointed out by Heisenberg, this cut may be somewhat shifted, but it
must be there somewhere.

According to von Neumann the observed system is described by a selfadjoint
operator and the measured values are the eigenvalues $\alpha $ of the
``observable'' $A$ (with suitable mathematical adaptation in case of
continuous spectrum). The state immediately after the measurement is
obtained by a ``quantum jump'' i.e. cannot be computed via the
Schr\"{o}dinger time evolution of the observed system. Taking for $A$ a
projector $P$ whose eigenvalues are just 0 and 1, the new state created by
the measurement according to von Neumann can be described as follows: 
\begin{equation}
\begin{array}{c}
\omega _{after}(A\mathcal{)=}\omega _{before}(PAP)+\omega _{before}(\left(
1-P\right) A\left( 1-P\right) )\,,\,\, \\ 
\,\,\,A\mathcal{\in A},\,\,\hbox{algebra of observables }
\end{array}
\end{equation}
If $\omega _{before}$was a pure state described by a state vector $\psi
\,,\omega _{after}$ corresponds to the density matrix: 
\begin{eqnarray}
\rho _{after} &=&\left| P\psi \right\rangle \left\langle P\psi \right|
+\left| \left( 1-P\right) \psi \right\rangle \left\langle \left( 1-P\right)
\psi \right| = \\
&=&p_{1}\left| \psi _{1}\right\rangle \left\langle \psi _{1}\right|
+p_{2}\left| \psi _{2}\right\rangle \left\langle \psi _{2}\right|  \nonumber
\end{eqnarray}
\begin{eqnarray}
\psi _{1} &=&\frac{P\psi }{\left| \left| P\psi \right| \right| }%
,\,\,p_{1}=\left| \left| P\psi \right| \right| ^{2},\,\,\, \\
\,\psi _{2} &=&\frac{\left( 1-P\right) \psi }{\left| \left| \left(
1-P\right) \psi \right| \right| },\,\,p_{2}=\left| \left| \left( 1-P\right)
\psi \right| \right| ^{2}  \nonumber
\end{eqnarray}
The last formula represents the mixed state associated with $\rho _{after}$
as a sum of two orthogonal minimal projectors. However, as stressed before,
an impure state permits myriads of decompositions into minimal projectors.
If we could find a physical argument in favor of an orthogonal decomposition
(as for the one above in terms of $\psi _{1},\psi _{2})$ then uniqueness
follows. But no such principle is known. In addition, to have a change of
states as the above reduction of wave packet (or ``quantum jump'') which is
outside the unitary time development of the Schr\"{o}dinger equation is
somewhat mysterious and in order to sharpen this seemingly paradoxical
situation Schr\"{o}dinger replaced the pointer of a measuring apparatus by a
cat with the more dramatic alive/dead state replacing the two pointer
positions whose coherent superposition defies common sense. Any hamiltonian
dynamics leading to unitary propagation in time, necessarily preserves the
purity of states.

The important role of superselection rules for a resolution of these
mysterious aspects has been first pointed out and illustrated by a
mathematical model in the work of Hepp. The essential idea is that the
macroscopic measuring apparatus has superselection sectors\footnote{%
The physically realizable observable algebra which is restricted by locality
may be genuinly smaller than the (more global) mathematical algebra.
Therefore the superselection rules in the measurement process should be
understood in the ``effective'' sense.} which, as we have learned, is a
generic phenomenon for systems with infinitely many degrees of freedom.
Hepp's idea is that although the complete system including the apparatus is
governed by a unitary time development : 
\begin{equation}
\alpha _{t}(A)=e^{iHt}Ae^{-iHt},\,\,\,A\in \mathcal{A}
\end{equation}
the limit for t$\rightarrow \infty $ may well be only \textit{a positive map}
instead of an automorphism of the observable algebra. In terms of Hilbert
space concepts the limit of $e^{iHt}$ may only be an isometric mapping of
the total Hilbert space on a subspace. Therefore the initial state which may
be a pure state implemented by a vector in one superselection sector leaves
the coherent subspace in the limit and acquires components in other
superselection sectors. Although it remains formally a vector in the total
space, it describes physically a mixture since it has projections to several
coherent subspaces. 
\begin{equation}
\omega _{before}\rightarrow \omega _{after}=\sum_{i}\lambda _{i}\omega
_{after}^{i}
\end{equation}
Note that this central decomposition is completely intrinsic. It is a
special case of a partial orthogonal decomposition. Whereas the latter is
only unique within one superselection sector, the former is unique in
general and in this sense behaves like a classical decomposition.

Although the more modern treatment of the measuring process for good reasons
has discarded Hepp's idea about the importance of the superselection rules
in the Hilbert space of the apparatus and the $t\rightarrow \infty $ limit 
\cite{La1}, the crucial role of superselection rules and the uniquely
defined central decompositions has remained. In the modern treatment these
properties are generated by the ``environment'' of the apparatus i.e. the
apparatus is considered as an open system which remains in contact with its
infinite degree of freedoms surroundings (which also may consist of the
unobserved internal degrees of freedom). Therefore the total Hilbert space
consists of three tensor factors: the system to be observed and the
apparatus, both localized, and the infinite degree of freedom environment
which is best described by methods of QFT. As an infinite degree of freedom
system, the environment has natural superselection sectors (in distinction
to topological nontrivial QM's, which owes its superselection structure to
unnatural physical overidealizations viz. the Aharonov-Bohm system) and the
mathematical physics method is similar to that of determining the various
possible phases in statistical mechanics models of phase transitions. The
measurement interaction correlates first the system $S$ with the apparatus $%
A $ and then with the environment $E$, in case of a two-valued spin system 
\begin{eqnarray}
(\alpha \left| \uparrow \right\rangle +\beta \left| \downarrow \right\rangle
)\left| A_{I}\right\rangle \otimes \left| E_{I}\right\rangle &\rightarrow
&\left( (\alpha \left| \uparrow \right\rangle \left| A_{\uparrow
}\right\rangle +\beta \left| \downarrow \right\rangle )\left| A_{\downarrow
}\right\rangle \right) \otimes \left| E_{I}\right\rangle \\
&\rightarrow &\alpha \left| \uparrow \right\rangle \left| A_{\uparrow
}\right\rangle \otimes \left| E_{\uparrow }\right\rangle +\beta \left|
\downarrow \right\rangle \left| A_{\downarrow }\right\rangle \otimes \left|
E_{\downarrow }\right\rangle \equiv \left| \Psi \right\rangle  \nonumber
\end{eqnarray}
Interactions which accomplish such a Schr\"{o}dinger transition can be
constructed by suitable idealizations of the apparatus $A$ and the
environment $E.$ The last step which destroys the purity of the system $S+A$
is the averaging over the degrees of freedom of the environment $E$ (to
observe $S$ is to not observe $E)$: 
\begin{equation}
\rho _{S,A}=Tr_{E}\left| \Psi \right\rangle \left\langle \Psi \right|
\end{equation}

The trace stands symbolically for an averaging which is better described
(the trace is only well defined for type II von Neumann algebras) in terms
of a conditional expectation. It is important to note that the collapse of
the wave packet is not instantaneous \cite{Zu}. In order to find the
nonunitary time development of $S$ we assume that the total initial state is
a product state $\omega =\tilde{\omega}\otimes \sigma $

Here the von Neumann factualization of the measurement has been accomplished
by the nonobservation of $E.$ Since the crucial role is played by the
superselection sectors of the apparatus + environment (the observed system $%
S $ through its coupling with $A+E$ only activates the superselection
sectors of the latter), the question about the origin of the superselection
sectors arises. The standard sectors of infinite degrees of freedom models
of statistical mechanics are identical to the various phases in the theory
of phase transition. As mentioned before, in the environmental case a
detailed consideration reveals that the relevant observable algebra of $E$
(resp.$E+A$) is not the algebra of a full tensor factor $H_{E}$, but rather
a certain $C^{*}$-subalgebra of $B(H_{E}),$ since the operators which are
nonlocal with respect to the observer (those which could monitor phase
relations with a state of which one component is spatially far separated
from the rest) are missing \cite{La1}. So the restriction of the $E$%
-observables due to locality turns out to be crucial for the understanding
of the von Neumann collapse of the wave packet.

In the environmental approach the ``reduction of the wave packet'' is
achieved in the limit t$\rightarrow \infty $, i.e. it is a time-dependent
process. Assume that the initial state is a product state $\omega =\tilde{%
\omega}\otimes \sigma $ on the total observable algebra $\mathcal{A}=%
\mathcal{A}_{SA}\otimes \mathcal{A}_{E}$. Restricting $\omega $ to the
subalgebra $\mathcal{A}_{SA}\equiv \mathcal{A}_{SA}\otimes 1_{E}$ we can
compute the time development of $\tilde{\omega}$ from the original unitary
automorphism $\alpha _{t}$ as: 
\begin{equation}
\tilde{\omega}_{t}(A)=\omega (\alpha _{t}\left[ A\otimes 1_{E}\right]
),\quad A\in \mathcal{A}_{SA}
\end{equation}
One can define a reduced time development on the restricted state as: 
\begin{equation}
\tilde{\omega}_{t}(A)=\tilde{\alpha}_{t}^{*}(\tilde{\omega})
\end{equation}
This can be recast by ``dualization'' into a nonautomorphic time development
in terms of a positive map $\tilde{\alpha}_{t}$ on $\mathcal{A}_{SA}$%
\begin{eqnarray}
\tilde{\alpha}_{t}\left[ A\right] &=&E_{\sigma }(\alpha _{t}\left[ A\otimes
1_{E}\right] ) \\
E_{\sigma }(A\otimes B) &=&\sigma (B)A,\quad E_{\sigma }^{2}=E_{\sigma
},\quad \omega \circ E_{\sigma }=\omega  \nonumber
\end{eqnarray}
Here $E_{\sigma }$ is a $\omega $-preserving conditional expectation which
is defined on all $\mathcal{A}\rightarrow A_{SA}\subset \mathcal{A}$ by
linear extension of the above definition on tensor products. Typical
properties, as the above projection property, the formula $E_{\sigma
}(ABC)=AE_{\sigma }(B)C$ if $A,C\in \mathcal{A}_{SA}$ as well as the
positivity: $E_{\sigma }(A^{*}A)\geq E_{\sigma }(A^{*})E_{\sigma }(A)\geq 0$
follow from this definition. In order to calculate decoherence times $\tau
_{D}$ which measure the duration of the collapse, we need a model. An
interesting phenomenological class of models which has been used for the
first very recent experiments \cite{Brune} on Schr\"{o}dinger cat states
(rather on Schr\"{o}dinger ``kittens'', because the apparatus contained only
a few number of photons) is that proposed by Zurek and others\cite{Zu}. The
effective equation of motion projected onto the one particle density matrix
is the master equation: 
\begin{equation}
\frac{d\rho }{dt}=-\frac{i}{\hslash }\left[ H,\rho \right] -\gamma
(x-x^{\prime })\left( \frac{\partial \rho }{\partial x}-\frac{\partial \rho 
}{\partial x^{\prime }}\right) -\frac{2m\gamma k_{B}T}{\hslash ^{2}}%
(x-x^{\prime })^{2}\rho
\end{equation}
The second term is a dissipation term which results from an interaction of a
particle with a field which represents the infinite degrees of freedom of
the environment and $\gamma $ is the relaxation rate resulting from such an
interaction. T is the temperature of this field. One finds a relaxation time 
$\tau _{D}\simeq \tau _{R}\frac{\hslash ^{2}}{2mk_{B}T(\Delta x)^{2}}$ with $%
\tau _{R}=\gamma ^{-1}$ where $\Delta x$ is the nondiagonal separation for
the initial pure state which has been described by coherent superposition of
two gaussian wave functions $\psi (x)=\psi ^{+}(x)+\psi ^{-}(x)$ whose peaks
have a distance $\Delta x.$ Therefore the initial density matrix is $\rho
(x,x^{\prime })=\psi (x)\psi ^{*}(x^{\prime })$ and the decoherence time is
the time in which the loss of the coherent off diagonal terms i.e. the
collapse has taken place.

Closely related to the measurement process and the role of superselection
sectors is the problem of why molecules have a ``nuclear frame'', i.e. why
the nuclei do not have fuzzy positions corresponding to coherent
superpositions of position eigenstates. This ``Gestaltproblem'' is
mathematically connected to the validity and interpretation of the
Born-Oppenheimer approximation. For its conceptual understanding the
environment and the superselection rules play a similar role as in the
measurement process \cite{Am}.

We have seen that the important concepts in the measurement process are open
systems (the environment) and locality (responsible for the superselected $%
\mathcal{A}_{E}$ subalgebra of $B(H_{E})$). In QM these concepts have to be
enforced from the outside (by forming tensor products etc.) whereas in QFT
there is an intrinsic concept of openness. For example the split property
discussed in the last chapter and the affiliated statistical independence as
well as the relativistic localization concept and superselaction sectors are
intrinsic natural properties of QFT which allow to view the outside
environment as a split tensor factor of one QFT . So the proper arena for
the analysis of the measurement process seems to be QFT. But apart from a
beautiful structural discussion of the significance of the Bell inequalities 
\cite{Su}, there has been no series attempt to base the maesurement
investigation on the characteristic modular properties of QFT which recently
have given consderable insight into thermal and entropical aspects of
modular localization (see chapter 3).

\textbf{Literature for chapter 1}

R. Haag, ``Local Quantum Physics'' Springer Verlag, Berlin Heidelberg 1992

O. Bratteli and D. W. Robinson, ``Operator Algebras and

F. Goodman, P. de la Harpe and V. F. R. Jones, ``Coxeter graphs and towers
of algebras, MSRI Publications (Springer) \textbf{14 (}1989)

V. F. R. Jones, ``Subfactors and Knots'', CMBS Number 80, American

Mathematical Society.

V.F. Jones and V.S. Sunder, ``Introduction to Subfactors'', London
Mathematical Society, Lecture Note Series 234.

\chapter{The Construction of Fock Space.}

\section{The Bosonic Fock Space}

There are several reasons for combining $N$-particle spaces together with a
one-dimensional ``no-particle'' space (vacuum) into a big ``Master'' space,
the so called Fock-space: 
\begin{equation}
\mathcal{H}=\mathcal{H}_{0}\oplus \mathcal{H}_{1}\oplus \mathcal{H}%
_{2}\oplus .....
\end{equation}
One obvious reason is that relativistic local interactions do not conserve
the particle number but only total charges (i.e. particle-antiparticle
creation is allowed as long as it obeys the energy-momentum conservation).
This is also valid in the infinite volume limit (the so called thermodynamic
limit) in nonrelativistic systems for which the ground state (which has
generally a finite density of particles) becomes the reference state for
generally number-nonconserved (but charge-conserved) ``quasiparticle''
excitations. Another less obvious reason is that the Fock space together
with locality is also the natural setting for the formulation of
cluster-properties which are the most important imprints of local quantum
physics on QM. The textbook treatment of e.g. scattering theory in QM
strictly speaking applies to short range interactions because in that case
the cluster requirements on the S-matrix are automatically taken care of by
the plane wave boundary conditions, but not so for long range interactions.
For this reason the use of e.g. the Aharonov-Bohm interaction between
``dyons'' in order to produce exotic (anyonic) statistics is not treatable
as a standard quantum mechanical problem for fixed particle number n. Rather
the main unsolved problem is the elaboration of stationary scattering
boundary conditions which makes the n-particle problem for the S-matrix
physically consistent with the clustering of the n+1 particle scattering for
all n. It is doubtful that such nonrelativistic problems without real (mass
shell) particle creation, but with the effective particle number change
caused by shifting particles to infinity, can be solved without the use of a
field theoretic framework.

The bosonic Fock-space $\mathcal{H^{B}}$ is obtained by projecting the full $%
N$-particle spaces onto their symmetrized subspaces $\mathcal{H}_{N}^{B}$.
In the following we will introduce the creation and annihilation operator
formalism in x-space, having in mind wave functions in Schr\"{o}dinger
theory. If we interpret the formulas in momentum space however, there will
be no difference between the relativistic and nonrelativistic formalism
apart from possible changes in the normalization. A general vector in
Fock-space is given by a finite norm sequence of symmetric wave functions: 
\begin{equation}
\begin{array}{c}
\mathcal{H}_{F}^{B}\ni \Psi =\left( \psi _{0},\psi _{1}(\vec{x}),\psi _{2}(%
\vec{x}_{1},\vec{x}_{2}),.....\right) \\ 
\left( \Psi ,\Psi \right) :=\left| \psi _{0}\right| ^{2}+\sum_{n=1}^{\infty
}\left( \psi _{n},\psi _{n}\right) <\infty
\end{array}
\end{equation}
The creation operator depends linearly on the wave function f and the
n-particle component after its application on $\Psi $ is defined as: 
\begin{equation}
\begin{array}{c}
\left( a^{*}(f)\Psi \right) _{n}(\vec{x}_{1},.....,\vec{x}_{n})=\frac{1}{%
\sqrt{n}}\sum_{i=1}^{n}f(\vec{x}_{i})\psi _{n-1}(\vec{x}_{1}...\widehat{\vec{%
x}}_{i}...\vec{x}_{n}),\;\;n\geq 1,\;\;\; \\ 
\;\left( a^{*}(f)\Psi \right) _{0}=0
\end{array}
\end{equation}
Here the ``roof'' $\symbol{94}$ indicates deletion of the i-th coordinate.
The formula for the hermitian adjoint annihilation operator $a(f)$ follows
from the defining property: 
\[
\begin{array}{c}
\left( \Psi ,a^{*}(f)\Psi ^{^{\prime }}\right) =\left( a(f)\Psi ,\Psi
^{^{\prime }}\right) ,\quad \hbox{namely}
\end{array}
\]
\begin{equation}
\left( a(f)\Psi \right) _{n}(\vec{x}_{1}....\vec{x}_{n})=\sqrt{n+1}\int
d^{3}x\bar{f}(\vec{x})\psi _{n+1}(\vec{x},\vec{x}_{1}....\vec{x}_{n})
\end{equation}
The annihilation operator depends antilinear on f. In particular for the
vacuum $\Omega =\left( 1,0,0...\right) $%
\begin{equation}
\left( a(f)\Omega \right) _{n}=0\;\;\;\forall n
\end{equation}
i.e. $a(f)$ annihilates $\Omega .$

The multiple application of these operators leads to lengthy formulas,
however the commutators are very simple: 
\begin{equation}
\begin{array}{c}
\left[ a(g),a^{*}(f)\right] =\left( g,f\right) 1,\;\;\;\hbox{with }\left(
g,f\right) =\int \bar{g}(\vec{x})f(\vec{x})d^{3}x \\ 
\left[ a(f),a(g)\right] =0=\left[ a^{*}(g),a^{*}(f)\right]
\end{array}
\end{equation}
This simplicity was the reason for the choice of normalization in the
definition of $a^{*}.$

The number operator $\mathbf{N}$\QTR{cal}{\ }is defined to be that positive
semidefinite operator which multiplies each $N$-particle vector with $N$.
Its commutation rules with $a^{\#}$ (this notation is used if we do not want
to distinguish between $a$ and $a^{*}$) is: 
\begin{equation}
\left[ \mathbf{N},a(f)\right] =-a(f),\;\;\;\;\left[ \;\mathbf{N}%
,a^{*}(f)\right] =a^{*}(f)
\end{equation}
In terms of an orthonormal basis it looks as $\mathbf{N}%
=\sum_{i}a^{*}(f_{i})a(f_{i}).\,$

It is convenient to liberate the formalism from the wave functions by
introducing operator-valued distributions $a^{\#}(\vec{x}):$ 
\begin{equation}
\begin{array}{c}
a^{*}(f)=\int a^{*}(\vec{x})f(\vec{x})d^{3}x,\;\;a(f)=\int a(\vec{x})\bar{f}(%
\vec{x})d^{3}x \\ 
\hbox{with\ }\;\left[ a(\vec{x}),a^{*}(\vec{y})\right] =\delta (\vec{x}-\vec{%
y})\;\;\;\hbox{etc.}
\end{array}
\end{equation}
One can then introduce the improper basis (vector-valued distributions) in
Fock-space: 
\begin{equation}
\left| \vec{x}_{1},....\vec{x}_{N}\right\rangle =\frac{1}{\sqrt{N!}}a^{*}(%
\vec{x}_{1})....a^{*}(\vec{x}_{N})\left| 0\right\rangle ,\,\,\,\,\,\,\left|
0\right\rangle :=\Omega
\end{equation}
\begin{equation}
\left| \Psi \right\rangle =\psi _{0}\left| 0\right\rangle +\sum_{N}\int \psi
_{N}(\vec{x}_{1}....\vec{x}_{N})\left| \vec{x}_{1}....\vec{x}%
_{N}\right\rangle d^{3}x_{1}....d^{3}x_{N}
\end{equation}
The action of the $a^{\#}$ on the basis vectors is (as always)
contragredient to that on the wave functions. It is more common to use the
former.

Of frequent use (especially in the application of Fock-space in statistical
mechanics) is the so called occupation number representation. One chooses an
orthonormal set of wave functions$\;f_{i},$ $i=1,2,...\infty $ and defines a
basis in Fock-space by: 
\begin{equation}
\left| n_{i_{1}},n_{i_{2}},....n_{i_{r}}\right\rangle =\frac{1}{\sqrt{%
n_{i_{1}}!}}....\frac{1}{\sqrt{n_{i_{r}}!}}%
a^{*}(f_{i_{1}})^{n_{i_{1}}}....a^{*}(f_{i_{r}})\left| 0\right\rangle
,\,\,\,\,\,\sum_{k=1}^{r}n_{i_{k}}=N
\end{equation}

Often (in particular in Stat. Mech.) one encloses the system in a box and
uses the discrete set of plane waves as the orthonormal system for the
occupation number representation.

The creation and annihilation operators in x-space are useful for rewriting
the Schr\"{o}dinger theory into the Fock space formalism. One easily
verifies the validity of the following formulas: 
\begin{equation}
\begin{array}{c}
\mathbf{H=\int H}(x)d^{3}x\,\,\,\,,\,\,\,\mathbf{H(}x\mathbf{)=H_{0}(}x%
\mathbf{)\ +\ V(}x\mathbf{)} \\ 
\mathbf{H}_{0}(x)=\frac{1}{2m}\vec{\partial}a^{*}(x)\cdot \vec{\partial}%
a(x)\,\,\,\, \\ 
\mathbf{V}(x)=\frac{1}{2}a^{*}(x)\int d^{3}yV(x-y)a^{*}(y)a(y)a(x)
\end{array}
\end{equation}
applied to the previously introduced $N$-particle state $\left| \Psi
\right\rangle $ give the $N$-particle Schr\"{o}dinger-operator: 
\begin{equation}
\begin{array}{c}
\mathbf{H}\left| \Psi (t)\right\rangle =i\frac{\partial }{\partial t}\left|
\Psi (t)\right\rangle ,\,\,\,\,\,\Leftrightarrow \\ 
\,\left( \sum_{i=1}^{N}\frac{1}{2m}\Delta
_{i}+\sum_{i<j}V(x_{i}-x_{j})\right) \psi _{N}(\vec{x}_{1}....\vec{x}%
_{N};t)=i\frac{\partial }{\partial t}\Psi _{N}(\vec{x}_{1}....\vec{x}_{N}:t)
\end{array}
\end{equation}
The verification only uses the commutation relations of the$\;a^{\#}(x)$ and
the annihilation property of $a(x)$ applied to the vacuum. The various terms
in the $N$-body Schr\"{o}dinger operator result from the following
commutators which arise in the process of moving $\mathbf{H}$ through the $N$
\thinspace $a^{*}(x)^{\prime }s$ onto the no-particle state: 
\begin{equation}
\left[ \mathbf{H},a^{*}(\vec{x})\right] =-\frac{1}{2m}\Delta
_{x}a^{*}(x),\quad \left[ \mathbf{V},a^{*}(\vec{x})a^{*}(\vec{y})\right] =V(%
\vec{x}-\vec{y})a^{*}(\vec{x})a^{*}(\vec{y})
\end{equation}
The hamiltonian $\mathbf{H}$\textbf{\ }in\textbf{\ }Fock space is used to
define time-dependent operators: 
\begin{equation}
a(\vec{x},t)=e^{i\mathbf{H}t}a(\vec{x})e^{-i\mathbf{H}t}
\end{equation}
Only in the case $\mathbf{H}=\mathbf{H}_{0}$ and for ``external
interactions'' 
\begin{equation}
\mathbf{H}=\mathbf{H}_{0}+\int V(x)a^{*}(x)a(x)d^{3}x
\end{equation}
is the time dependent $a(x,t)$ linear in $a(x)$.

Before we consider an application some remarks on the mathematical status of
the $a^{\#}$'s and related operators are helpful. Since the N-particle
states for arbitrary N form a dense set of states, the $a^{\#}$ are densely
defined. Using the number operator it is easy to compute: 
\begin{equation}
\left| \left| a(f)\mathbf{N}^{-\frac{1}{2}}\right| \right| _{H_{F}^{\bot
}}=\left( f,f\right) ,\quad H_{F}^{\bot }=\hbox{subspace\ of}\ H_{F}\bot
\Omega
\end{equation}
We remind the reader that the norm of an operator $A$ is related to the
vector norm in the Hilbert space: 
\begin{equation}
\left| \left| A\right| \right| =\sup_{\psi }\frac{\left| \left| A\psi
\right| \right| }{\left| \left| \psi \right| \right| }
\end{equation}
The technique for computing such norms is always the same: first one uses
the defining formula for $a(f)$ in order to prove the inequality and then
one exhibits a particular vector for which the equality sign holds. In our
case $\int f(\vec{x})a^{*}(\vec{x})\Omega $ is such a vector. The norm of
the adjoint is the same $\left| \left| A\right| \right| =\left| \left|
A^{*}\right| \right| .$

The relative boundedness with respect to $\mathbf{N}$\textbf{\ }e.g\textbf{. 
}$\left| \left| a^{*}(f)\psi \right| \right| \leq \left| \left| \mathbf{N}%
\psi \right| \right| ^{\frac{1}{2}}$ may be used to show that these
unbounded operators are closable and hence admit e.g. a polar decomposition. 
\newline
Most physicist's calculations do not touch these fine points. They only
check equations for densely defined bilinear forms . In case of the above
formulas for $\mathbf{H}$\textbf{\ }this means that one checks this formula
for $\left\langle \Psi ,\mathbf{H}\Phi \right\rangle $ with the vectors
running through the dense set of smooth states of finite particle number.
The extension to a relation between densely defined closable or selfadjoint
operators is in most physically relevant cases possible and follows a
standard scheme (Reed-Simon). In those cases we will be satisfied with the
check for matrix elements which is easily done with the commutation
relations for $a^{\#}$and the above ``pulling through onto the vacuum''
rule. All perturbative calculation in Fock space are done with these rules
and this applies also to the derivation of Feynman rules in the real-time
operator setting of relativistic QFT \footnote{%
An important exception to the applicability of this cavalier attitude of
physicists (concerning dense set of vectors in a Hilbert space) is the
modular localization theory of section 5, chapter 3 and chapter 6. In that
case the relevant dense sets of vectors can be parametrized in terms of real
closed subspaces which describe the physical localization and carry the main
physical content of the theory in their net properties of real Hilbert
spaces.}.

In order to illustrate the application of bosonic (symmetrized) Fock space
techniques to coherent states, we first convince ourself that the $a^{\#}$-
formalism for an oscillator is a special case of the present formalism
(specialization to one degree of freedom). For a one-dimensional space $%
\mathcal{H}_{1}=\mathbf{C},$ all the tensor product spaces are also one
dimensional and the ``single degree of freedom'' operator $a^{\#}$ does not
require any additional label. An arbitrary vector may be written as: 
\begin{equation}
\left| \Psi \right\rangle =\psi _{0}\left| 0\right\rangle +\psi _{1}\left|
1\right\rangle +\psi _{2}\left| 2\right\rangle +....,\ \quad \left|
n\right\rangle =\frac{a^{*n}}{\sqrt{n!}}\left| 0\right\rangle ,\quad a\left|
0\right\rangle =0
\end{equation}
Writing instead of the $a^{\#^{\prime }}$s standard dynamical variables of
QM $p$ and $q$ (with natural units $\hbar =1):$%
\begin{equation}
a^{*}=\sqrt{\frac{\omega }{2}}x+i\sqrt{\frac{1}{2\omega }}p
\end{equation}
the $a^{\#}$- commutation relations go over into the Heisenberg c.r. and the
standard oscillator hamiltonian takes the form: 
\begin{equation}
H_{osc}=\frac{1}{2}p^{2}+\frac{\omega ^{2}}{2}x^{2}=\omega (a^{*}a-\frac{1}{2%
})
\end{equation}
The x-space wave functions $\left\langle x\mid n\right\rangle $ of the
eigenstates $\left| n\right\rangle $ turn out to be the well-known Hermite
functions. Coherent states are obtained by asking for the eigenstates of the
perturbed hamiltonian: 
\begin{equation}
H=H_{osc}+\lambda (a+a^{*})=H_{osc}+\lambda \sqrt{2\omega }x
\end{equation}
Apart from an uninteresting c-number term, the linear perturbation can be
obtained by applying a spatial translation by $a=\frac{\lambda }{\sqrt{%
2\omega }}$ to $H_{osc}.$ In terms of a$^{\#}$ this translation $U(a)$ is: 
\begin{equation}
U(\frac{\lambda }{\sqrt{2\omega }})=e^{-\frac{\lambda }{\omega }(a^{*}-a)}
\end{equation}
Different from the previous use of Fock space formalism, the number operator
for the oscillator quanta $\mathbf{N}=a^{*}a$ does not commute with the
perturbation. Hence the eigenstates of $H$ do not have a well defined number
of such quanta. In order to obtain explicit formulas for $U\left|
n\right\rangle $ we use the Campbell-Baker-Hausdorff formulas: 
\begin{equation}
e^{A}e^{B}=e^{A+B+\frac{1}{2}\left[ A,B\right] +....}
\end{equation}
where the.... stands for higher commutator terms. This is easily established
for matrices and (modulo domain problems) by perturbative arguments in the
general case. Due to the vanishing of higher commutators we get: 
\begin{equation}
e^{-\frac{\lambda }{\omega }a^{*}}e^{\frac{\lambda }{\omega }a}=e^{-\frac{%
\lambda }{\omega }(a^{*}-a)+\frac{1}{2}\frac{\lambda ^{2}}{\omega ^{2}}}
\end{equation}
Therefore the ground state of $H$ is an ``eigenstate'' of the annihilation
operator: 
\begin{equation}
\left| \Psi _{0}\right\rangle =U\left| 0\right\rangle =e^{-\frac{1}{2}\frac{%
\lambda ^{2}}{\omega ^{2}}}e^{-\frac{\lambda }{\omega }a^{*}}\left|
0\right\rangle ,\quad a\left| \Psi _{0}\right\rangle =-\frac{\lambda }{%
\omega }\left| \Psi _{0}\right\rangle
\end{equation}
Here in the first step we used the BCH formula to separate the annihilation
part of $U$ to the right (where an $e^{\alpha a}$ factor on $\left|
0\right\rangle $ becomes the identity). For the eigenvalue equation use the
translation property. On the higher eigenstates $U\left| n\right\rangle ,$
the application of $a$ leads to an additive modification of the eigenvalue
relation by $U\left| n-1\right\rangle .$ The probability distribution of the
oscillator quanta follows the Poisson distribution: 
\begin{equation}
\left| \left\langle n\mid \Psi _{0}\right\rangle \right| ^{2}=\frac{e^{-%
\frac{\lambda ^{2}}{\omega ^{2}}}}{n!}\left( \frac{\lambda }{\omega }\right)
^{2n}
\end{equation}
Physically the perturbed oscillator may be thought of as resulting from a
constant electric field: 
\begin{equation}
H=H_{osc}-eEx
\end{equation}
This field causes the expectation values of the $a^{\#}$'s to be
nonvanishing: 
\begin{equation}
\left\langle \Psi _{n}\mid a^{\#}\mid \Psi _{n}\right\rangle \sim E
\end{equation}
The free time development on the state vectors leads to the classical
oscillatory behaviour of expectation values: 
\begin{equation}
\begin{array}{c}
\left\langle \Psi _{n}(t)\left| x\right| \Psi _{n}(t)\right\rangle \sim
E\cos (\omega t-\varphi ) \\ 
\left\langle \Psi _{n}(t)\left| p\right| \Psi _{n}(t)\right\rangle \sim
E\sin (\omega t-\varphi )
\end{array}
\end{equation}
This oscillatory behaviour would be the result of a sudden switching on of
the field: 
\begin{equation}
H(t)=\left\{ 
\begin{array}{c}
H\quad for\;t<0 \\ 
H_{osc}\quad for\,\;t\geq 0
\end{array}
\right.
\end{equation}
by which the coherent states are created. The following classical behaviour
of expectation values of functions in $a^{\#}$ is characteristic for
coherent states: 
\begin{equation}
\left\langle \Psi _{0}\left| f(a^{\#})\right| \Psi _{0}\right\rangle
=f\left( \left\langle \Psi _{0}\left| a^{\#}\right| \Psi _{0}\right\rangle
\right)
\end{equation}

In case of a time dependent source: 
\begin{equation}
H(t)=H_{osc}+H_{int}(t),\quad H_{int}(t)=-eE(t)x
\end{equation}
we are dealing with time dependent unitary transformations which implement
the time dependent canonical transformations:

\begin{equation}
U(t)a^{\#}U^{*}(t)=a^{\#}-\sqrt{2\omega }E
\end{equation}
which lead from $H_{osc}$ to $H(t)$. In this simple case the $U(t)$ has the
same form as in the stationary case except that the constant in front of the 
$a-a^{*}$ term in the exponential is now time dependent. It is useful to
have a more systematic method which also works for cases for which the $U(t)$
is less simple. Such a method go back to Dirac (and flourished in QFT thanks
to Dyson). It treats the time-dependent problems in the ``interaction
picture'' which is between the Heisenberg picture and the Schr\"{o}dinger
picture. All these pictures agree on the level of physical states i.e. in
their expectation values, but they differ in how the total time development
is distributed between operators and state vectors. Whereas in the
Heisenberg - and Schr\"{o}dinger-picture the full time development is on the
operators respectively on the state vectors, the interaction picture is
characterized by the property that the operators only suffer the free time
development and the rest (the interaction picture time development V(t)) is
dumped on the vectors. According to this definition the interaction operator 
$H_{int}(t)$ becomes: 
\begin{equation}
H_{W}(t)=e^{iH_{0}t}H_{int}(t)e^{-iH_{0}t}
\end{equation}
The time development operator $V(t_{2},t_{1})$ which propagates the vector
state (or wave function) from one time to a later time is: 
\begin{equation}
V(t_{2},t_{1})=Te^{-i\int_{t_{1}}^{t_{2}}H_{W}(t)dt}
\end{equation}
It is a solution of the Schr\"{o}dinger equation in the interaction picture: 
\begin{equation}
i\frac{d}{dt}V(t,t^{\prime })=H_{W}(t)V(t,t^{\prime })
\end{equation}
The time- (or path-) ordering is defined as: 
\begin{equation}
\begin{array}{c}
TA_{1}(t_{1})A_{2}(t_{2})....A(t_{n})=A_{i_{1}}(t_{i_{1}})A_{i_{2}}(t_{i_{2}})....A_{i_{n}}(t_{i_{n}})
\\ 
for\quad t_{i_{1}}\geq t_{i_{2}}\geq ....\geq t_{i_{n}}
\end{array}
\end{equation}
and the above time ordered exponential is defined by the power series with
time ordered integrands or as the limit of subsequent products with
decreasing length of the time intervalls as: 
\begin{equation}
\begin{array}{c}
Te^{-i\int_{t_{1}}^{t_{2}}H_{W}(t)dt}= \\ 
\lim_{\Delta t\rightarrow
0}Te^{-i\int_{t_{n-1}}^{t_{n}}H_{W}(t)dt}....Te^{-i%
\int_{t_{2}}^{t_{3}}H_{W}(t)dt}Te^{-i\int_{t_{1}}^{t_{2}}H_{W}(t)dt}
\end{array}
\end{equation}

The proof consists in rewriting the Heisenberg time development operator $%
U(t,s).$ Just as in the case of time independent interactions, this operator
can be factorized into e$^{-iH_{0}(t-s)}$ and a remaining interaction $%
V(t,s) $: 
\begin{equation}
U(t,s)=e^{-iH_{0}(t-s)}V(t,s)\quad i.e.\,V(t,s)=e^{iH_{0}(t-s)}U(t,s)
\end{equation}
The Schr\"{o}dinger equation for $U(t,s)$ is then equivalent to the
following differential equation for $V(t,s)$: 
\begin{eqnarray}
i\frac{d}{dt}V(t,0) &=&e^{iH_{0}t}\left( H-H_{0}\right)
e^{iHt}=e^{iH_{0}t}\left( H_{int}\right) e^{-iH_{0}t}e^{iH_{0}t}e^{-iHt} 
\nonumber \\
&=&H_{W}(t)V(t,0)
\end{eqnarray}
The rest of the proof consists in deriving the time-ordered representation
from the formal integration of this differential equation. One first
converts this into an integral equation (using $V(0,0)=1$ as an initial
condition): 
\begin{equation}
V(t,s)=1-i\int_{s}^{t}dt^{\prime }H_{W}(t^{\prime })V(t^{\prime },s)
\end{equation}
Clearly the perturbative solution is the geometric series: 
\begin{equation}
V(t,s)=1+(-i)\int_{s}^{t}H_{W}(t^{\prime })dt^{\prime
}+(-i)^{2}\int_{s}^{t}dt_{2}%
\int_{s}^{t_{2}}dt_{1}H_{W}(t_{2})H_{W}(t_{2})+....
\end{equation}
where the $n^{th}$ term is integrated over the simplex $s\leq t_{1}\leq
t_{2}\leq ....\leq t_{n}\leq t.$ The use of the (nonlocal, by its very
definition) time-ordering prescription allows to convert the integration
over a simplex into one over the n-dim. hypercube $s\leq t_{i}\leq
t,\,i=1...n\,\;:$%
\begin{equation}
\begin{array}{c}
V(t,s)=1+(-i)\int_{s}^{t}H_{W}(t^{\prime })dt^{\prime }+\frac{1}{2!}%
\int_{s}^{t}\int_{s}^{t}TH_{W}(t_{2})H_{W}(t_{1})dt_{1}dt_{2}+.... \\ 
+\frac{1}{n!}\int ....\int ....+....
\end{array}
\end{equation}
which has the desired exponential time-ordered form. These somewhat formal
manipulations may be mathematically justified in two different ways. Either
one finds a bound for the $n^{th}$ term, or one shows the equivalence of the
time-ordered expression with an exact unitary transformation which, like the
one at the beginning of this section is a ``dressing transformation'' i.e.
applied to the free hamiltonian it generates the interaction. Let us briefly
explain this for the infinite degrees of freedom analog of the perturbed
oscillator: a bosonic field system under the influence of an external source
described by the hamiltonian: 
\begin{equation}
H(t)=H_{0}+a(j_{t})+a^{*}(j_{t}),\quad H_{0}=\int (\frac{\vec{p}^{2}}{2m}%
+\mu )a^{*}(\vec{p})a(\vec{p})d^{3}p,\quad
\end{equation}
where $j_{t}(\vec{x})=j(\vec{x},t)$ and we have added a chemical potential
term (in order to avoid infrared divergencies of the p-integrals in
subsequent calculations). The dressing transformation is: 
\begin{equation}
U(t)=e^{a(g_{t})-a^{*}(g_{t})},\quad g(\vec{x},t)=\frac{1}{\left( 2\pi
\right) ^{3}}\int \tilde{j}(\vec{p},t)(\frac{\vec{p}^{2}}{2m}+\mu )^{-1}e^{i%
\vec{p}x}d^{3}p
\end{equation}
The connection with the time development $U(t,s)$ is evidently (since it
dresses the free operator): 
\begin{equation}
U(t,s)=U(t)e^{-iH_{0}(t-s)}U^{*}(t)\quad or\quad
V(t,s)=e^{iH_{0}(t-s)}U(t)e^{-iH_{0}(t-s)}U^{*}(t)
\end{equation}
The direct calculation of the time-ordered representation for $V$ uses the
previously mentioned infinite product representation: 
\begin{equation}
V(t,s)=\lim_{\Delta t\rightarrow
0}\prod_{ord}e^{-i\int_{t_{i}}^{t_{i+1}}H_{w}(t^{\prime })dt^{\prime }}
\end{equation}
where the product is path-ordered (with ascending times going to the left),
but the time ordering within each factor is omitted. This formula is similar
to the famous Trotter product formula; in integrals over shrinking
intervalls the difference between the time ordered and the unordered
expression disappears. To this product form we may apply the BCH-formula in
order to collect all operators within one unordered exponential. The BCH
series ceases after the quadratic term in $H_{w}$: 
\begin{equation}
\begin{array}{c}
V(t,s)=\lim_{\epsilon \rightarrow 0}\exp
(-i\sum_{j=0}^{n-1}\int_{s+j\epsilon }^{s+(j+1)\epsilon }H_{w}(t^{\prime
})dt^{\prime } \\ 
-\frac{1}{2}\sum_{j\neq k}\int_{s+j\epsilon }^{s+(j+1)\epsilon }dt^{\prime
}\int_{s+k\epsilon }^{s+(k+1)\epsilon }dt^{\prime \prime }\left[
H_{w}(t^{\prime }),H_{w}(t^{\prime \prime })\right] )
\end{array}
\end{equation}
Since the commutators are c-numbers, the result is of the form: 
\begin{equation}
V(t,s)=\exp (-ia(j_{t,s})-ia^{*}(j_{t,s})-i\frac{\alpha }{2})
\end{equation}
where $\alpha $ is a numerical phase (resulting from the commutator) and $%
j_{t,s}(\vec{x})$ is the result of time propagation of the original source
function in $H_{int}.$ Hence we obtain agreement between the two methods.
Furthermore we learn that the time ordered exponential leads to a phase
factor which is not present in the dressing approach. Specializing now to
the limit $t\rightarrow \infty ,s\rightarrow -\infty $ (assuming that the
interaction only extends over a finite time or that the integrals over time
in V converge) we define the $S$-operator as the full interaction picture
transition operator $V$ which relates the free system before and after the
interaction: 
\begin{equation}
S=\lim_{t,s\rightarrow \pm \infty }V(t,s)
\end{equation}
Clearly the application of 
\begin{equation}
S=exp(-ia(g)-ia^{*}(g)-i\frac{\alpha }{2}),\;g=lim_{t,s\rightarrow \pm
\infty }(j_{t,s})
\end{equation}
onto the vacuum $\Omega $ gives a coherent state vector: 
\begin{equation}
S(g)\Omega =e^{-i\frac{\alpha }{2}}\Omega (ig),\quad
\end{equation}
The successive action of sources leads to the composition law: 
\begin{equation}
S(f)\Omega (ig)=c(f,g)\Omega (ig+if),\quad \left| c\right| =1
\end{equation}
Therefore even if we eliminate (by using the projective nature of QT) the
phase factor in the definition of $S\Omega $, it will reappear in form of a
so called 2-cocycle in the composition law. As in the case of the
oscillator, the source generates a coherent distribution of say photons with
a Poisson probability distribution. On a coherent state vector $\Omega (ig)$
the action of $S(f)$ will change the mean particle number to: 
\begin{equation}
\begin{array}{c}
\Delta N=\left( \Omega (if+ig),\mathbf{N}\Omega (if+ig)\right) -\left(
\Omega (ig),\mathbf{N}\Omega (ig)\right) \\ 
=\left| \left| f+g\right| \right| ^{2}-\left| \left| g\right| \right|
^{2}=\left| \left| f\right| \right| ^{2}+2Re(f,g)
\end{array}
\end{equation}
The interference term 2Re(f,g) describes induced absorption or emission
depending on the sign. Many important results of laser physics may be
developed in this formalism.

\section{The Fermion Fock Space}

The antisymmetric N-particle space was obtained by acting with the
antisymmetric projector P$_{a}$on the full N-fold tensor product \QTR{cal}{H}%
$_{N}$ of one-particle spaces: 
\begin{equation}
H_{N}^{\left( -\right) }=\pi (P_{a})H_{N},\qquad ~P_{a}=\frac{1}{N!}%
\sum_{P\in S_{N}}sign(P)P
\end{equation}
Here $\pi (P)$ $P\in S_{N}$ stands for the natural representation of S$_{N}$
on the full tensor space $H_{N}$. The fermionic Fock space is simply the
direct sum of all antisymmetrized $N$-particle spaces augmented by the one
dimensional no-particle state. 
\begin{equation}
H^{\left( a\right) }=H_{0}+H_{1}+\sum_{N=2}^{\infty }H_{N}^{\left( a\right) }
\end{equation}
The only difference to the bosonic case (besides the antisymmetry of the
wave functions) is the sign appearing in the formula for the creation
operator: 
\begin{equation}
\left( a^{*}(f)\psi \right) _{n}(\vec{x}_{1},....\vec{x}_{n})=\frac{1}{\sqrt{%
n}}\sum_{i}\left( -1\right) ^{i+1}f(\vec{x}_{i})\psi _{n-1}(\vec{x}_{1},..%
\widehat{\vec{x}}_{i,}..\vec{x}_{n})
\end{equation}
Here the roof on the $\vec{x}_{i}$ indicates omission of this variable. In a
completely analogous fashion we obtain the anticommutation relations: 
\begin{equation}
\left\{ a(f),a^{*}(g)\right\} =\left( f,g\right) ,\quad \left\{
a(f),a(g)\right\} =0,\quad \left\{ a^{*}(f),a^{*}(g)\right\} =0
\end{equation}
and its pointlike form by removing the wave packets: 
\begin{equation}
\left\{ a(\vec{x}),a^{*}(\vec{y})\right\} =\delta (\vec{x}-\vec{y}),\quad
\left\{ .,.\right\} =0\hbox{ in all other cases}
\end{equation}
There is no change in the formulas which carry the Schr\"{o}dinger theory on
antisymmetric N-particle wave functions to the Fock space (at least if one
writes $\mathbf{H}$ exactly in the same order in the $a^{\#}$'s ). A
significant difference to the bosonic case begins to show up, if one
realizes that as a consequence of: 
\begin{equation}
\left\{ a(f),a(f)\right\} =2a(f)^{2}=0
\end{equation}
and the hermitian adjoint relation, we obtain the Pauli exclusion principle:
in an occupation number representation any quantum level can maximally be
singly occupied: 
\begin{equation}
\left| n_{1},n_{2},...\right\rangle =\frac{a_{1}^{*n_{1}}}{\sqrt{n_{1}!}}%
\frac{a_{2}^{*n_{2}}}{\sqrt{n_{2}!}}....\left| 0\right\rangle ,\quad
n_{i}=0,1
\end{equation}
This principle holds only if all quantum numbers of a particle (including
spin and possible internal charges) have been taken into account. Closely
related is the ability of fermion-systems to form a new reference state by
simply \textit{occupying} a given set of levels ( orthonormal one-particle
vectors $f_{i}\;i=1...N$): 
\begin{equation}
\left| \Psi _{0}\right\rangle =a_{1}^{*}....a_{N}^{*}\left| 0\right\rangle
,\quad a_{i}^{*}\equiv a^{*}(f_{i})
\end{equation}
This vector is annihilated by the new annihilation operators: 
\begin{equation}
b(f)=\left\{ 
\begin{array}{c}
a^{*}(f)\quad if\ f\in H(f_{1},...f_{N}) \\ 
a(f)\quad if\ f\in H^{\perp }(f_{1},...f_{N})
\end{array}
\right.
\end{equation}
Here $H(f_{1}...f_{N})$ is the subspace of the one-particle space spanned by
the system of vectors $f_{i}$. Note that the $b^{\#}s$ obey the same
commutation relations as the $a^{\#}s$. The annihilation property $b\left|
\Psi _{0}\right\rangle =0$ is an easy consequence. Note that the hermitian
adjoint $b^{*}(f_{i})$ creates holes in $\left| \Psi _{0}\right\rangle .$
This ability to design states which are annihilated by transformed Fermion
variables $b^{\#}$ through occupying levels, is typical for CAR. On the
other hand for coherent states (relevant in e.g.laser physics) and
Poisson-distributions one needs Bosons. Mathematically the CAR- structure
(canonical anticommutation relations) is well behaved since the $a^{\#}(f)$
are bounded operators: 
\begin{equation}
\begin{array}{c}
\left( \Phi ,\left\{ a(f),a^{*}(f)\right\} \Phi \right) =\left( f,f\right)
\left( \Phi ,\Phi \right) \\ 
i.e.\quad \left| \left| a(f)\Phi \right| \right| ^{2}\leq \left( f,f\right)
\left( \Phi ,\Phi \right)
\end{array}
\end{equation}
By taking $\Phi =a^{*}(f)\Omega $ we establish saturation (=) and therefore $%
\left| \left| a(f)\right| \right| =\left( f,f\right) $ for the operator
norm. The counterpart of the one-dimensional oscillator is: 
\begin{equation}
\sigma _{x}=a+a^{*}\quad \sigma _{y}=i\left( -a+a^{*}\right) \quad \sigma
_{z}=aa^{*}-a^{*}a
\end{equation}
with $\sigma _{i}$ being the Pauli-matrices i.e. the smallest irreducible
representation of the Clifford algebra structure defined by $a^{\#}$ is in
terms of Pauli matrices. This observation can be generalized: \textbf{\
\quad \quad }

\begin{theorem}
$Alg($a$_{i}^{\#},i=1....N)\equiv Cliff(C^{N})=\otimes
^{N}Mat_{2}(C)=Mat_{2^{N}}(c)$
\end{theorem}

The proof consists in starting with a generating system of matrix units for
the $N$-fold tensor product of $Mat_{2}(\mathbf{C})$: 
\begin{equation}
e_{ij}^{\left( k\right) }=\underline{1}\otimes ...\underline{1}\otimes
e_{ij}\otimes \underline{1}...\underline{1}
\end{equation}
Clearly the four $2\times 2$ matrix units $e_{ij}$ are linear combinations
of the four Pauli-matrices and the system commutes for different k. The step
towards anticommutation requires the introduction of the famous Jordan-Pauli
transformormation. With the help of: 
\begin{equation}
\mu
_{k}=\prod_{i=1}^{k}(e_{11}^{(k)}-e_{22}^{(k)})=\prod_{i=1}^{k}(1-2a^{*}a)
\end{equation}
we define $a_{i}=\mu _{i-1}e_{12}^{(i)}.$ and its hermitean adjoint.
Commuting $e_{12}^{(i)}$factor in $a_{i}$ through the ($%
e_{11}^{(i)}-e_{22}^{(i)})$-factor in $a_{j}^{\#}$ for $j>i$ leads to the
-sign. The relation between the matrix units and the $a^{\#}s$ can be
inverted and the generated algebras are identical.

In the ``Paulion'' formalism, the filling operation is described by the
unitary: 
\begin{equation}
U=\sigma _{1}\otimes \sigma _{1}\otimes ....\otimes \sigma _{1},\quad \sigma
_{1}=\left( 
\begin{array}{cc}
0 & 1 \\ 
1 & 0
\end{array}
\right)
\end{equation}
From this one reads off the filling operator in the $a^{\#}$ representation.

Clearly the filling mechanism is as typical for Fermions as the coherent
state properties are for Bosons. Without the former there would be no
periodic table (nor we) and without the latter no laser.

The commutation relation of the charge-carrying fields in local quantum
physics are inexorably linked with the physical and mathematical properties
of the algebras which they generate, a fact which will be presented in
detail in the last section on algebraic QFT. There is only one exception: in
massive d=1+1 theories the statistics (either in the sense of particles or
field commutation relations) loses its intrinsic physical meaning, e.g. a
periodic table of elements in a d=1+1 world may also be described in terms
of bosons with long range interactions. In that case the inexorable link
between the charge properties (superselection sectors) and the particle
statistics is lost and particles become statistical ``schizons'' with well
defined (solitonic) charge properties \cite{S-S} but ill-defined commutation
relations (bosons, fermions or ``plektons'') of their interpolating fields.
We will take up the investigation of such 2-dimensional specialities in the
next chapter.

\section{The CCR and CAR Algebras.}

Whereas in the case of the Fermion Fock space an abstraction to a $C^{*}$%
-algebra is straightforward (just take the C$^{*}$-algebra generated by the a%
$^{\#}s$ subject to the anticommutation relations, uniqueness will be shown
later), a construction of a $C^{*}$-algebra from unbounded operators
generally meets serious obstacles. Following Weyl, one formally converts the 
$a^{\#}$ into unitary operators: 
\begin{equation}
W(f)=e^{i\Phi (f)},\quad \Phi (f)=\frac{1}{\sqrt{2}}(a(f)+a^{*}(f))
\end{equation}
The application of the BCH-formula leads to: 
\begin{equation}
W(f)W(g)=e^{-i\frac{1}{2}\sigma (f,g)}W(f+g)=e^{-i\sigma (f,g)}W(g)W(f)
\end{equation}
where $\sigma (f,g)=Im(f,g)$ is a non degenerate symplectic form. Formally
the unbounded operators $a^{\#}(f)$ may be reobtained by multiplying f with
a parameter $t$ and differentiating the modified W with respect to t at t=0.
We now take this Weyl relation or rather $Alg(W(f),f\in H)$ as our basic
definition of the Boson algebra. This algebra is clearly an infinite degree
of freedom generalization of the well-known Heisenberg-Weyl algebra which
underlies standard QM: 
\begin{eqnarray}
U(\vec{\alpha}) &:&=e^{i\vec{\alpha}\vec{p}},\quad V(\vec{\beta}):=e^{i\vec{%
\beta}\vec{q}}\quad \\
W(\vec{\gamma}) &:&=e^{i\frac{1}{2}\vec{\alpha}\vec{\beta}}U(\vec{\alpha})V(%
\vec{\beta}),\quad \vec{\gamma}=\vec{\alpha}+i\vec{\beta}  \nonumber
\end{eqnarray}
One easily checks with BCH, that the $W$ fullfills the above Weyl relation
with\ $f=\gamma \in C^{N}$ and the symplectic form being the form of
standard type known from $2N$ dimensional phase space of classical
mechanics. The following theorem collects the important structural
properties of the CAR and CCR(Weyl) algebras.

\begin{theorem}
\textbf{\ }The CAR and CCR algebras are simple (no ideals) $\mathbf{C}^{*}$%
-algebras generated by the CAR resp. CCR commutation relations.
\end{theorem}

We only indicate the proof and refer to Bratteli-Robinson Vol 2 for details.

In the CAR case we know from the previous consideration, that for finite
degrees of freedom Fermions $a_{i}^{\#}$ $i=1...N$ can be replaced by
``Paulions''. For infinite degrees of freedom we take a basis $f_{i},$ $%
i=1...\infty $ in the one particle space $H$. The uniqueness of the limiting
algebra follows from the continuity resulting from $\left| \left|
a(f)\right| \right| =\left| \left| f\right| \right| $. The full algebra is
in fact an inductive limit of finite degree of freedom algebras and its
separability is inherited from the $Mat_{N}$-algebras.

The proof in the CCR case is somewhat different. In this case one reduces
the problem to a projective unitary representation of an (infinite) abelian
group $H$ (associated with the linear space $H$) with the multiplier $%
exp-i\sigma (f,g)$ being a character. In this way the problem is reduced to
that of uniqueness of $C^{*}$-group algebras. The triviality of the ideal is
established by showing that the kernel of every representation is trivial.
However separability and the inductive uniform limit property do not hold.

As in the CAR case one may ask about the uniqueness of irreducible
representations (up to unitary equivalence). This indeed holds in the
important class of ``regular'' representations i.e. representations $\pi $
for which the unitaries $\pi (W(tf))$ are strongly continuous in t.

\begin{theorem}
(Stone-von Neumann uniqueness theorem). Every regular irreducible
representation of the Heisenberg-Weyl algebra for a finite number of degrees
of freedom is unitarily equivalent to the Schr\"{o}dinger representation, or
in other words: the algebraic structure of standard Q.M. has no nontrivial
super-selection rules.
\end{theorem}

The proof uses the infinitesimal generators $\Phi ,$ which thanks to the
regularity property turn out to have a densely defined domain which allows
to construct the $a(f)^{\#}$ and the number operator $\mathbf{N}=\sum
a_{i}^{*}a_{i}.$ The positivity of the latter requires the existence of of a
``lowest'' vector which is the required reference state for the annihilation
operators. If on the other hand we are dealing with infinite degrees of
freedom (i.e.dimH=$\infty $)$,\;$the sum in $\mathbf{N}$ need not to
converge. In such representations the number operator does not exist.
Examples are easily given.

\textbf{Bosonic illustration }If the shift function $c(x)$ in $%
a(x)\rightarrow b(x)=a(x)+c(x)$ is not square integrable (physically because
of short distance $\left[ \text{ultraviolet}\right] $ or long distance $%
\left[ \inf \text{rared}\right] $ divergencies), then $\mathbf{N}$\textbf{\ }%
does not exist in the corresponding representation and the formal expression
for the unitary implementer $U(c)$ (see (\ref{U}) below) cannot be given a
meaning.

\textbf{Fermionic illustration }If the ``occupied'' Hilbertspace is infinite
dimensional, no unitary implementer can exist. The reason is that a vector $%
\Phi =\prod_{i}a_{i}^{*}\Omega $ is orthogonal on each basis vector of the
particle number representation: 
\begin{equation}
\begin{array}{c}
\left( \Phi ,\Psi (n_{1},n_{2}....n_{N})\right) =0,\quad \\ 
\Psi (n_{1},n_{2}....n_{N})=a^{*n_{1}}a^{*n_{2}}....a^{*n_{N}}\Omega ,\quad
n_{i}=0\ or\ 1
\end{array}
\end{equation}
This is because for any arbitrary large $N$ the $\Phi $ contains infinitely
many creation operators which remain uncompensated. The formal expression
for $\Phi $ cannot be given a meaning in Fock space.

There is another way of looking at this illustration. The infinite sequence
of $0$ and $1$ in $\left( n_{1},n_{2}......\right) $ may be considered as a
binary fraction. Whereas the Fock-basis consists of binary fractions with $%
n_{i}=0$ for sufficiently large $i$ (which may become arbitrarily large),
the binary fraction for the above $\Phi $ is the constant sequence $\left(
1,1,1,..1....\right) .$ This sequence is not in the vacuum class associated
to the representative $\left( 0,0,..0....\right) ,$ where ``class'' here
means the set of sequences which deviate from each other only in an
arbitrary large but finite number of places. Each class belongs to a basis
in a seperate Hilbert space and the different basis elements $\Psi
(n_{1},n_{2},....)$ are obtainable from one reference element in the class
by the application of a finite (but arbitrarily large) number of $%
a^{\#\prime }s$ . The various irreducible representation spaces obtained
from the different classes are orthogonal subspaces of an inseparable
unwieldy (and unphysical) Hilbert space generated by all binary fractions
(which form a continuous set). The same idea of classes of sequences works
for bosons. In that case the $n_{i}$ run through all natural numbers
including zero. One obtains myriads of inequivalent irreducible
representations, and this construction is not even exhaustive. Most of them
are physically uninteresting, and one needs a physical selection principle.
Many of the physically interesting ones are in the subset of ``quasifree''
states explained in the next section.

We close this section by commenting on automorphisms of the CCR and CAR $C$%
*-algebras which are linear maps of the algebra onto itself which preserving
the algebraic structure. In physical terms they are symmetry transformations
which preserve the Weyl resp. CAR relation. Of particular interest are the
Bogoliubov automorphisms. They are induced by (anti-) linear invertible
transformations of the underlying linear wave function space $H$. In the CCR
case they are required to leave the symplectic form $\sigma $ on H invariant
and map the Weyl generators as follows: 
\begin{equation}
\sigma (Tf,Tg)=\sigma (f,g),\quad \alpha (W(f))=W(Tf)
\end{equation}

In the CAR case we must use (anti-)unitary operators in order to preserve
the anti-commutator: 
\begin{equation}
(Uf,Ug)=\left\{ 
\begin{array}{c}
(f,g)\hbox{ \quad unitary} \\ 
\overline{(f,g)}\quad \hbox{antiunitary}
\end{array}
\right.
\end{equation}
A slightly more general automorphism is obtained by combining these two
possibilities: 
\begin{equation}
\begin{array}{c}
\alpha (a(f))=a(Uf)+a^{*}(Vf) \\ 
UU^{*}+VV^{*}=1=U^{*}U+V^{*}V,\quad V^{*}U+U^{*}V=0=UV^{*}+VU^{*}
\end{array}
\end{equation}
where $U$ is linear and $V$ antilinear. The previous case is obtained by
specialization $V=0$ resp. alternatively $U=0$. Clearly the earlier
occupation transformation corresponds to the automorphism $\alpha
(a(f))=a^{*}(Vf).$ The crucial question is now whether the automorphism is
really a bona fide symmetry i.e. implementable by a unitary transformation.
Take as an example the shift $a^{*}(\vec{x})\rightarrow a^{*}(\vec{x})+c(%
\vec{x})$ which is formally implemented by the unitary: 
\begin{equation}
Ua^{*}(\vec{x})U^{*}=a^{*}(\vec{x})+c(\vec{x}),\quad U=e^{a(c)-a^{*}(c)}
\label{U}
\end{equation}
But without the condition $\int \left| c(\vec{x})\right| ^{2}d^{3}x<\infty $
the formal expression for $U$ would not define a bona fide unitary operator
in Fock space. Physical intuition would tell us to expect that Bogoliubov
tranformations in the one particle space $H$ need to be sufficiently close
to the identity in order to have an implementation in Fock space. This is
indeed the case; the deviation from \underline{1} should be in the
Hilbert-Schmidt class in order not to be thrown out of the representation
class (folium). Since Bogoliubov transformations leave the property of
``quasi-freeness'' invariant, the natural place for presenting the relevant
implementation formulas is the next section.

\section{Quasifree States}

The most convenient way to obtain representations of the CCR and CAR $%
\mathbf{C}^{*}$-algebras is through states $\omega $ on these algebras. We
have seen that the GNS-construction gives a canonical association of a
representation $\pi _{\omega }$ with a state $\omega $. Since there are too
many inequivalent states and associated representations on infinite algebras
which nobody has been able to classify, we need some limitation. It turns
out that the class of quasifree states and their representations can be
classified completely. They are defined by their two-point functions
together with a combinatorial formula which expresses their $n$-point
functions in terms of the given two-point functions. On the generators $%
a^{\#}$ we specify the state $\omega $ by giving first its two-point
functions: 
\begin{eqnarray}
&&\omega (a^{*}(f)a^{*}(g)),\quad \omega (a(f)a^{*}(g))\,,  \nonumber \\
&&\hbox{\quad or\quad }\omega (a^{*}(x)a^{*}(y)),\quad \omega (a(x)a^{*}(y))
\end{eqnarray}
The remaining two-point function $\omega (a(g)a(f))$ is (according to the
reality properties of states following from their positivity) just the
complex conjugate of $\omega (a^{*}(f)a^{*}(g))$ and $\omega (a^{*}(g)a(f))$
may be obtained by (anti-)commutation from $\omega (a(f)a^{*}(g)$. The
higher correlation functions of $\omega $ are given in terms of the
two-point function by the following combinatorical formula: 
\begin{equation}
\omega
(a^{\#}(f_{1})a^{\#}(f_{2})....a^{\#}(f_{2n}))=\sum_{pairings\,P}signP%
\prod_{i_{k}<i_{k+1}}\omega (a^{\#}(f_{i_{k}})a^{\#}(f_{i_{k+1}}))
\end{equation}
\[
\omega (a^{\#}(f_{1})a^{\#}(f_{2})....a^{\#}(f_{2n+1}))=0 
\]
We have to prove that $\omega $ is positive on the polynomial algebra
generated by the $a^{\#}$'s: 
\begin{equation}
\omega (A^{*}A)\geq 0\quad A=polyn(a^{\#})
\end{equation}
For the CAR-algebra the bound from the anticommutation relations: 
\begin{equation}
\omega (a(f)a^{*}(f))\leq \left| \left| f\right| \right| ^{2}
\end{equation}
gives immediately: 
\begin{equation}
\omega (a(f)a^{*}(g))=\left( f,Tg\right) \quad 0\leq T\leq 1
\end{equation}
The positivity on monomials $A=a^{\#}(f_{1})....a^{\#}(f_{n})$ is a result
of the basic two-point positivity: 
\begin{eqnarray}
\left| \left| \left( a(f)+a^{*}(g)\right) \Omega \right| \right| ^{2} &=& \\
\omega \left( a^{*}(f)a(f)+a(g)a^{*}(g)+a^{*}(f)a^{*}(g)+a(g)a(f)\right)
&\geq &0  \nonumber
\end{eqnarray}
(The latter holds as the result of the positivity of T and the
Cauchy-Schwarz inequality) and the combinatorial definition of the n-point
function.

A particular subclass of quasifree states are the gauge invariant quasifree
states. By definition only those correlation functions are nonvanishing
which contain the same number of $a$ and $a^{*}.$ Instead of working with
unbounded operators\label{before} one prefers to define the quasifree gauge
invariant states directly in the Weyl algebra: 
\begin{equation}
\omega (W(f))=\exp (-\frac{1}{4}\left| \left| f\right| \right| ^{2}-\frac{1}{%
2}\left| \left| T^{\frac{1}{2}}f\right| \right| ^{2})
\end{equation}
The standard Fock-representation reemerges as the special case T=0.

Quasifree states are regular (by construction) so that we can return to
unbounded $a^{\#}$ in a similar manner as we introduce Lie algebra
generators and their enveloping algebras in noncompact group representation
theory. The corresponding GNS representation is most conveniently written in
terms of an auxiliary ``doubled Fock space'', for CAR: 
\begin{equation}
\begin{array}{c}
a_{\omega }(f)=a(\sqrt{1-T}f)\otimes 1+\gamma \otimes a^{*}(K\sqrt{T}%
f),\quad T\leq 1 \\ 
a^{*}(f)=a^{*}(\sqrt{1-T}f)\otimes 1+\gamma \otimes a(K\sqrt{T}f)
\end{array}
\label{quasi}
\end{equation}
For CCR we obtain the analogous formula: 
\begin{equation}
\begin{array}{c}
a_{\omega }(f)=a(\sqrt{1+T}f)\otimes 1+1\otimes a^{*}(K\sqrt{T}f) \\ 
a_{\omega }^{*}(f)=a_{\omega }^{*}(\sqrt{1+T}f)\otimes 1+1\otimes a(K\sqrt{T}%
f)
\end{array}
\end{equation}
Here $K$ is the standard conjugation satisfying ($K$f,$K$g)=(g,f) and $%
\gamma $ (only defined in the CAR case) is the unitary operator which
implements the \textbf{Z}$_{2}$ gauge transformation (distinguishes even
from odd numbers of Fermions) in Fock space. The proof consists in a simple
calculation of the two-point function in the vector $\Omega _{double}=\Omega
\otimes \Omega .$

The irreducibility condition for gauge invariant quasifree representations
is that $T$ is a projector $T=P$. The equivalence criterion for two gauge
invariant quasifree representations is:

\begin{theorem}
Two irreducible representations given in terms of P and Q are equivalent iff 
$\left| \left| P-Q\right| \right| _{H.S.}^{2}<\infty .$ Here the H-S norm of
K is defined as 
\begin{equation}
\left\| K\right\| _{HS}^{2}\equiv TrK^{*}K<\infty
\end{equation}
\end{theorem}

\section{Temperature States and KMS condition}

For a finite quantization box (i.e. a discrete energy spectrum), finite
temperature states on the CCR or CAR-algebra are described in terms of the
Gibbs formula: 
\begin{equation}
\rho =\frac{1}{Z}e^{-\beta H},\quad Z=Tre^{-\beta H},\quad H=H_{0}(\mu
)+H_{int}
\end{equation}
since $e^{-\beta H}$ is then a trace class operator. Here $H_{0}$ includes
the chemical potential $\mu $: 
\begin{equation}
H_{0}(\mu )=\int \omega (\vec{p})a^{*}(\vec{p})a(\vec{p})d^{3}p,\quad \omega
(\vec{p})=\frac{\vec{p}^{2}}{2m}-\mu
\end{equation}
The box-enclosed version is of course a sum over discrete momenta which
results by extending the Laplace operator on smooth functions with support
in the volume V in a selfadjoint manner to square integrable functions in V.
The various ways of doing this correspond to the various boundary
conditions. The physical role of the chemical potential is that the ground
state energies for different particle numbers can be adjusted in such a way
that the averaged particle $n(x)$ and energy $h(x)$ density: 
\begin{equation}
\bar{n}=Tr\rho n(\vec{x})\quad \bar{\epsilon}=Tr\rho h(\vec{x})
\end{equation}
remain finite in the thermodynamical limit $V\rightarrow \infty $ and hence
can be expressed in terms of the two parameters $\beta $ and $\mu .$ For the
ideal Fermi or Bose gas ($H_{int}=0$) the approach of the (quasifree ) Gibbs
state to the limit KMS state is obvious by explicite calculation: 
\begin{equation}
\begin{array}{c}
\lim_{V\rightarrow \infty }\omega _{V}(a^{*}(f)a(g))=\omega (a^{*}(f)a(g))
\\ 
\omega _{V}(a^{*}(f)a(g))=\frac{1}{Z}Tre^{-\beta \mathbf{H_{0}}%
_{V}}a^{*}(f)a(g)=\left( g,T_{V}f\right) \\ 
T_{V}=(\exp -\beta H_{0V})(1+\exp -\beta H_{0V})^{-1}
\end{array}
\end{equation}
Here the (non-bold) $H_{0}$ are the one-particle operators acting on wave
functions whereas $\mathbf{H}_{0}$ acts in Fock-space. The $\omega \;$in the
thermodynamic limit is also of the quasifree form with $H_{0}$ replacing $%
H_{0V}.$ The simplest way of proving these relations is to use the KMS
property\footnote{%
The analytic continuation to imaginary times in the operator expressions can
be shown to exist on a dense set of operators in their operator algebra.}: 
\begin{eqnarray}
\omega _{V}(\sigma _{t}(a^{*}(f))a(g)) &=&\omega _{V}(a(g)\sigma _{t+i\beta
}(a^{*}(f))) \\
\sigma _{t}(\cdot ) &=&e^{it\mathbf{H}_{0}}\cdot e^{-it\mathbf{H}_{0}}
\end{eqnarray}
which for t=0 together with the (anti)commutation relation leads to: 
\begin{equation}
\omega _{V}(a^{*}(f)a(g)\pm a^{*}(e^{-\beta H_{0V}}f)a(g))=(g,e^{-\beta
H_{0V}}f)
\end{equation}
We used that the hamiltonian automorphism $\sigma _{t}$ is of the Bogoliubov
type. We rewrite this equation as 
\begin{equation}
\omega _{V}(a^{*}((1\pm e^{-\beta H_{0V}})f)a(g))=(g,e^{-\beta H_{0V}}f)
\end{equation}
Clearly this relation is solved by: 
\begin{equation}
\sigma _{t}(a^{*}(f))=a^{*}(e^{-\beta H_{0V}}(1\pm e^{-\beta H_{0V}})^{-1}f)
\end{equation}
After Fourier-transformation, $e^{-\beta H_{0V}}$becomes a multiplication
operator

$\exp -\beta (\frac{\vec{p}^{2}}{2m}-\mu )$ on the momentum space wave
functions and hence: 
\begin{equation}
\begin{array}{c}
n(\vec{p})=\exp -\beta \omega (\vec{p})(1\pm \exp -\beta \omega (\vec{p}%
))^{-1} \\ 
\omega _{V}(a^{*}(\vec{x})a(\vec{y}))=\frac{1}{\left( 2\pi \right) ^{3}}\int
e^{i\vec{p}(\vec{x}-\vec{y})}n(\vec{p}),\quad \;\omega (\vec{p})=\frac{\vec{p%
}^{2}}{2m}-\mu
\end{array}
\end{equation}
Since the + case belongs to the Fermions, one obtains for $\beta \rightarrow
\infty $ the expected occupation for the finite density ground state: 
\begin{equation}
\lim_{\beta \rightarrow \infty }n(\vec{p})=\left\{ 
\begin{array}{c}
1_{,}\quad \hbox{if }\vec{p}^{2}<\mu \\ 
0,\quad \hbox{if }\vec{p}^{2}>\mu
\end{array}
\right.
\end{equation}
The main difference between the finite volume expression and the
thermodynamic limit is that in the former case the p-values are discrete and
that in the latter case the trace class property of $\exp -\beta \mathbf{H}$
is lost and therefore the numerator and denominator in the Gibbs formula
(and hence the Gibbs formula itself) become meaningless. It is not difficult
to establish the thermodynamic limit for large classes of $\mathbf{H}_{int}.$

With the help of the KMS condition one may avoid the finite quantization box
and study statistical mechanics directly in the infinite system. It is
interesting to note that the KMS condition is equivalent to the stability of
the state under appropriately formulated local perturbations and to the
second law of thermodynamics (see Haag's book). The GNS construction with a
KMS state gives a GNS triple with a reference state $\Omega $ which, in
addition of being cyclic, also has the separating property i.e. an operator
from the algebra \QTR{cal}{A} which annihilates $\Omega $ must itself be
zero. In fact the hamiltonian, or more generally the KMS automorphism, is
the Tomita automorphism of the associated modular theory and vice versa, the
Tomita modular automorphism is characterized by its KMS property. Let us
illustrate this for quasilocal states on the CAR-algebra. Writing: 
\begin{equation}
\left\langle \Omega \left| a(f)a^{*}(g)\right| \Omega \right\rangle =\left(
f,Tg\right) ,\quad 0\leq T\leq 1
\end{equation}
the separability of $\Omega $ is warranted if $T$ has no eigenvalues 0 and 1
and the representation of the CAR algebra is even factorial if the
multiplicity of the eigenvalue $\frac{1}{2}$ is finite. The previous
considerations suggest that the modular operator is related to S by: 
\begin{equation}
T=\frac{\Delta _{T}}{1+\Delta _{T}}  \label{mod}
\end{equation}
and the GNS representation may be most naturally be described by
``doubling'' i.e. in a Fock space $\mathcal{H}_{double}=\mathcal{H}%
_{F}\otimes \mathcal{H}_{F}$ associated with the doubled one particle space $%
h_{double}=h\oplus h$ : 
\begin{equation}
\begin{array}{c}
\omega _{T}(A)=\omega _{P}^{double}(A) \\ 
P=\left( 
\begin{array}{cc}
T & T^{\frac{1}{2}}(1-T)^{\frac{1}{2}} \\ 
T^{\frac{1}{2}}(1-T)^{\frac{1}{2}} & 1-T
\end{array}
\right)
\end{array}
\end{equation}
The quasifree states in the doubled description are pure on the tensor
product algebra (and its representation is irreducible) since P is a
projector. But its restriction S to the first factor (which is the image of
the original CAR algebra under the doubling ) is impure and reducible. For
this reason the doubling is also called ``purification'' \footnote{%
The inverse mechanism, namely the incoherent mixture through
``nonobservation'' of degrees of freedom, is important in the environmental
approach to the measurement problem (section 4, chapter 1)}. In the
application to KMS states of statistical mechanics, the second factor in the
doubling is a ``shadow world'' i.e. a copy of the original one (
corresponding to the algebra of the previously discussed right action )
which has no spatial localization. Later we will also meet examples of the
modular theory for which the commutant algebra has a complementary
space-time localization. In those cases the modular theory has a deep
relation to TCP symmetry (the particle-antiparticle issue) and the Hawking
temperature.

In the previous section we have described the von Neumann algebras
associated with quasifree states on CCR or CAR algebras in terms of (\ref
{quasi}) of generators in tensor product form. In this form the commutant
and the Tomita involution $J$ may be easily read off. The modular operator $%
\Delta ^{it}$ is determined in terms of the KMS property and can be easily
written in terms of $T.$ E.g.in the CAR case it is the inversion of (\ref
{mod}): $\Delta ^{it}=T^{it}(1-T)^{-it}.$ The information about the type of
von Neumann algebra is contained in the structure of $T$ or $\Delta .$
Generically, i.e. without further conditions we have the hyperfinite type $%
III_{1}$(see mathematical appendix)$.$ In the CAR case all types of
hyperfinite factors (factor condition: $\dim \ker (T-\frac{1}{2})=0,\infty
,even)$ occur. If the restriction $T_{\left[ 0,1\right] }$ of $T$ to the
spectrum $\left[ 0,1\right] $ is of trace-class (Gibbs like behavior) the
factor is $I_{\infty },$ whereas for a Hilbert-Schmidt operator $(T^{\frac{1%
}{2}}-(\frac{1}{2})^{\frac{1}{2}})$ the factor type is $II_{1}.$ The type $%
II_{\infty }$ arises from a combination of the H-S property for ($T_{\left[
0,c\right] }^{\frac{1}{2}}-$ $(\frac{1}{2})^{\frac{1}{2}})$ and the trace
property for $T_{\left[ 0,c\right] }$ with $0<c<\frac{1}{2}.$ The remaining
cases are type $III.$ Generically one obtains hyperfinite $III_{1};$ in
order to get also hyperfinite $III_{0}$ and $III_{\lambda }$ $T$ must fulfil
more subtle spectral conditions. In the CCR case the hyperfinite type $%
III_{1}$ factors are also generic, but regular states on Weyl algebras do
not yield type $II$ factors. The local relativistic algebras (see next
chapter) belonging to spacetime regions with a nontrivial causal disjoint
are always hyperfinite type $III_{1}.$

\section{The CCR- and CAR-Functors\label{st, fac}}

In section 3 we introduced the CCR and CAR $C^{*}$-algebras as maps of
Hilbert spaces of functions into $C^{*}$-algebras. In particular the
Fock-representations of these $C^{*}$-algebras define functors from the
category of Hilbert spaces into von Neumann algebras.

Let us first look at the CCR-functor. Starting from a Hilbert space (always
complex unless stated otherwise) with a scalar product $f,g\rightarrow (f,g)$%
, we first describe the associated bosonic Fock space in the following way.
Let e$^{f}$ be the suggestive notation for the vector in the the Fock space $%
\mathcal{H}_{F}^{sym}\equiv e^{H}$ associated to $H$ with the following
n-particle components and inner product: 
\begin{equation}
e^{f}=1\cdot \Omega +\sum_{n}\frac{1}{\sqrt{n!}}\stackunder{n}{\underbrace{%
f\otimes ....\otimes f}},\quad (e^{f},e^{g})=e^{(f,g)}
\end{equation}
In this notation the vacuum is $\Omega =e^{0}$. These special vectors are
linear independent as well as ``total'' (i.e. they form a dense set) in $%
e^{H}.$ The Weyl operator $W(f)$ is defined on this dense set as: 
\begin{equation}
W(f)e^{g}=e^{-\frac{(f,f)}{4}}e^{-(f,g)}e^{f+g}
\end{equation}
The unitarity of $W$ and hence the extension to the whole space $H$ follows
from this formula. The isomorphic map $H\rightarrow e^{H}$ carries subspaces
of $H$ into subspaces of $e^{H}$ as well as the direct sum decompositions
into tensor products decompositions. Furthermore linear densely defined maps 
$A$ between one particle spaces $H\stackrel{A}{\rightarrow }K$ go over into $%
e^{H}\stackrel{e^{A}}{\rightarrow }e^{K}$ with the computational rules: 
\begin{eqnarray}
e^{A}e^{h} &=&e^{Ah},\quad (e^{A})^{*}=e^{A^{*}} \\
e^{A} &=&e^{U}e^{\left| A\right| },\quad A=U\left| A\right| \quad  \nonumber
\end{eqnarray}
the latter describing the fate of the polar decomposition under the map.

In order to use the Weyl-operators $W$ as a functor from the category of
linear spaces to von Neumann algebras, we need to understand a particular
family of real subspaces of $H$. Let M be a set of vectors in $H$. Define
the symplectic complement of $M$: 
\begin{equation}
M^{\prime }=\left\{ f\in H\mid \func{Im}(f,g)=0\,\,\forall g\in M\right\}
\end{equation}
Then $M^{\prime }$ is a closed real subspace ( the use of the symplectic
form $Im(f,g)$ requires the restriction to real linear combinations). The
following list of properties follows directly from the definition: 
\begin{eqnarray}
M &\subset &N\curvearrowright N^{\prime }\subset M^{\prime } \\
M\,\,dense\,\,in\,H &\curvearrowright &M^{\prime }=\left\{ 0\right\} 
\nonumber \\
(M+iM)^{\prime } &=&M^{\prime }\cap iM^{\prime }  \nonumber
\end{eqnarray}
As for von Neumann algebras, the two-fold application of the \'{}-operation
i.e. $M\rightarrow M^{\prime \prime }$ gives the (in this case symplectic)
completion i.e. the smallest closed real space generated by the set $M$. The
following definition strengthens the analogy with von Neumann algebras.

\begin{definition}
A real closed subspace M is called ``standard'' if $M+iM$ is dense and $%
M\cap iM=\left\{ 0\right\} .$ Every standard M defines a ``canonical
involution'' $\frak{s}$ via $\frak{s}(f+ig)=f-ig$ where $f,g\in M.$
\end{definition}

In other words, standard M's are +1 eigenspaces of an (unbounded) involution
antilinear \textsl{s}. We need its polar decomposition: 
\begin{eqnarray}
\frak{s} &=&j\delta ^{\frac{1}{2}},\quad j^{2}=1,\quad j\delta ^{\frac{1}{2}%
}=\delta ^{-\frac{1}{2}}j \\
\frak{s}h &=&h^{*}\quad \hbox{on dense set }\mathcal{D}_{M}=M+iM  \nonumber
\end{eqnarray}
with * referring to the reality concept defined by M. The important
relations: 
\begin{equation}
j(M)=M^{\prime },\quad \delta ^{it}M=M
\end{equation}
are rather direct consequences of the definitions.

We now define a von Neumann algebra R(M) associated with the real subspace
M: 
\begin{equation}
R(M)=\left\{ W(f)\mid f\in M\right\} ^{\prime \prime }=alg\left\{ W(f)\mid
f\in M\right\}
\end{equation}
The double commutant of a $^{*}$-symmetric family of operators in a Hilbert
space is identical to the von Neumann algebra generated by this family (see
appendix).

Note that although M is real, the von Neumann Algebras R are always complex.
The map: 
\begin{equation}
M\rightarrow R(M)
\end{equation}
turns out to be an ``orthocomplementary functor'' from the category of
Hilbert spaces $H$ and their standard real subspaces into the $B(H_{F})$
operator algebra on Fock space and von Neumann subalgebras in standard
position. Orthocomplementary means that the complement $M^{\prime }$
corresponds to the commutant $R(M)^{\prime }$ i.e. the validity of the
following ``duality'': 
\begin{equation}
R(M^{\prime })=R(M)^{\prime }
\end{equation}

The importance of this functor in QFT in quantum physics results from the
fact that the $R(M)^{\prime }$ describes all observables which are
compatible (simultaneously measurable) with an observable from $R(M),$ where
in important QFT cases M describes a space of real (classical) functions
localized in some region $\mathcal{O}$ in Minkowski space and $M(\mathcal{O}%
)^{\prime }=M(\mathcal{O}^{\prime })$ where $\mathcal{O}^{\prime }$ denotes
the causal disjoint region to $\mathcal{O}$. So the functor relates
classical localization regions with the quantum notion of simultaneous
measurability. The process of passing from classical functions with a
symplectic structure to operator algebras is often referred to as
``quantization''. Since this word creates the misleading impression that
quantum physics is founded on a parallelism to classical physics and in
particular that localization needs a classical function space, we prefer to
avoid it alltogether (Bohr's ``correspondence principle'' is the reverse,
namely to recover classical physics in some special limiting situations). In
some way algebraic QFT is the investigation of those structures which cannot
be obtained by ``quantization'' methods as Lagrangian canonical- and
pathintegral-methods.

The most interesting remaining problem is the connection between the
properties of $\frak{s},j$ and $\delta $ and their Fock space counterparts $%
S=e^{\frak{s}},J=e^{j}$ and $\Delta ^{it}=e^{\delta ^{it}}.$ As a result of: 
\begin{equation}
SW(f)\Omega =W(-f)\Omega =W(f)^{*}\Omega
\end{equation}
$S$ is Tomita's (unbounded) involution: 
\begin{eqnarray}
SA\Omega &=&A^{*}\Omega ,\quad A\in R(M) \\
S &=&J\Delta ^{\frac{1}{2}}
\end{eqnarray}
The Tomita-Takesaki theory states (see appendix) that any von Neumann $%
\mathcal{A}$ algebra acting in a Hilbert space $\mathcal{H}$ with a cyclic ($%
\mathcal{H}=\overline{\mathcal{A}\Omega })$ and seperating vector $\Omega $ (%
$A\Omega =0\curvearrowright A=0)$ has a closable operator $S$ defined as
above which admits a polar decomposition with $\Delta ^{\frac{1}{2}}$ being
the radial part and the antiunitary $J$ the angular part. The $\Delta ^{it}$
implements the modular automorphism: 
\begin{eqnarray}
\sigma _{t}(A) &\equiv &\Delta ^{it}A\Delta ^{-it}\in \mathcal{A}  \label{fo}
\\
JAJ &\in &\mathcal{A}^{\prime }  \nonumber
\end{eqnarray}
with $\mathcal{A}^{\prime }$ the commutant of $\mathcal{A}$ in $\mathcal{H}.$
In other words a pair $(\mathcal{A},\Omega )$ in ``general position'' leads
to modular objects $(\Delta ,J)\,$with the above properties.

In our case $\mathcal{A}=R(M)\subset B(H)$ and the $M$-dependence of $S$ is
solely encoded in its dense domain (whereas for $\frak{s}$ the star depends
on $M$). It is the simple part of the Tomita-Takesaki theory that $S$ and
the operators $J$ and $\Delta $ which result from polar decomposition
thereof always exist for general von Neumann algebras $\mathcal{R}$ in
standard position i.e. pairs $\left\{ \mathcal{R},\Omega \right\} $ with $%
\mathcal{R}$ $\in B(H)$ and $\Omega \in H$ cyclic and separating. Tomita's
deep theorem is the formula (\ref{fo})

It is not difficult to see that in our case $R(M)\cap R(M)^{\prime }=\mathbf{%
C}1$ (i.e. $R(M)$ is a factor) iff $M\cap M^{\prime }=\left\{ 0\right\} .$
This suggests the definition:

\begin{definition}
A real subspace $M$ is called factorial if $M\cap M^{\prime }=\left\{
0\right\} .$
\end{definition}

The family of standard von Neumann algebras which are in the range of this
functor are a subset of all standard von Neumann algebras in $B(H_{F}).$

There exists another functor which maps $H$ into $H_{F}^{antis}$ and the
standard real subspaces $M$ of $H$ into von Neumann algebras generated by
CAR operators: 
\begin{eqnarray}
\left\{ a(g),a^{*}(f)\right\} &=&(g,f)\mathbf{1} \\
\left\{ a^{\#}(g)a^{\#}(f)\right\} &=&0  \nonumber
\end{eqnarray}
\begin{equation}
CAR(M)=alg\left\{ A(f)=a^{*}(f)+a(f)\mid f\in M\right\}
\end{equation}
where $a(f)$ is the Fock space annihilation operator: $a(f)\Omega =0$.

The functorial constructions of the CAR appear somewhat simpler (and more
natural) if one follows Araki and interprets the complex Hilbert space $H$
as a ``doubled'' real Hilbert space. This is achieved by taking two copies $%
H_{\pm }$ and introducing an antiunitary involution $\Gamma $: 
\begin{eqnarray}
\Gamma \left( 
\begin{array}{l}
f_{+} \\ 
f_{-}
\end{array}
\right) &=&\left( 
\begin{array}{l}
\bar{f}_{-} \\ 
\bar{f}_{+}
\end{array}
\right) ,\quad f=\left( 
\begin{array}{l}
f_{+} \\ 
f_{-}
\end{array}
\right) \in K=H_{+}\oplus H_{-}, \\
\quad PK &=&H_{+}  \nonumber \\
f_{\pm } &\rightarrow &\bar{f}_{\pm }\hbox{ conjugation in }H_{\pm },\quad
(f,g)=(f_{+},g_{+})+(f_{-},g_{-})\quad  \nonumber
\end{eqnarray}
The selfconjugate subspace 
\begin{equation}
\func{Re}K=\left\{ f\in K\mid \Gamma f=f\right\}
\end{equation}
inherits on the one hand a real inner product and on the other hand this
real subspace is isomorphic with K$_{+}$ considered as a real space with the
isomorphism being: 
\[
f\rightarrow \sqrt{2}Pf,\quad f\in \func{Re}K 
\]
$\func{Re}K$ admits the following complex structure (P as above): 
\begin{equation}
if:=iPf-i(1-P)f,\quad f\in \func{Re}K
\end{equation}
This description of one particle spaces $K$ is the same for both functors.
The only difference is in the interpretation: instead of the symplectic
complement $M^{\prime }$ one uses the ``i-symplectic'' complement: $\tilde{M}%
^{\prime }=iM^{\prime }$. This could also be called the real orthogonal
complement. The relation with the vanishing anticommutator is: 
\begin{equation}
\left\{ A(f),A(g)\right\} =0,\forall g\in M,\curvearrowright f\in \tilde{M}%
^{\prime }
\end{equation}
An important distinction between the CCR and the CAR functor shows up if one
looks at the Tomita-Takesaki theory. In the CAR case one finds: 
\begin{equation}
S=J\Delta ^{\frac{1}{2}},\quad J=Te^{j},\Delta =e^{\delta }
\end{equation}
$T$ is the so-called Klein twist, a transformation which is only defined in $%
\mathcal{H}_{F}^{antis}$ and not in $H:$%
\begin{equation}
T=\frac{1+ie^{i\pi \mathbf{N}}}{1+i},\quad \mathbf{N}:\hbox{number op. in }%
\mathcal{H}_{F}^{antis}
\end{equation}
The general setting does not tell which of the two functors one must take in
concrete situations. In QFT this additional physical information is supplied
by localization properties.

\textbf{Literature to Chapter 1 and 2:}

Rudolf Haag : ``Local Quantum Physics'', Fields, Particles, Algebras.
Springer-Verlag 1992

Ola Bratteli and Derek W.Robinson : ``Operator Algebras and Quantum
Statistical Mechanics'' Vol.1 and 2, Springer-Verlag 1979

J.H.Roberts in ``The Algebraic Theory of Superselection Sectors,
Introduction and Recent Results'' Ed. D.Kastler, World Scientific 1990.

The detailed presentation of the Weyl functor is taken from some unpublished
notes of P. Leylands, J.Roberts and D.Testard \cite{Oster Leyland}.

\chapter{Poincar\'{e} Symmetry and Quantum Theory}

\section{ Symmetry in General Quantum Theory.}

In QM and elementary QFT the generator of a symmetry operation is described
by an hermitian operator (``charge'', in case of inner symmetry) which
commutes with the Hamiltonian.. Usually, in particular for continuous
symmetries, this operator has a geometric origin in terms of the
quantization of a Noether ``current''.

Since in relativistic QFT there is a characteristic causality and covariance
principle, a more intrinsic approach would suggest to avoid objects which
depend on the reference frame as the Hamiltonian $H$ and instead to use
concepts which are closer to LQP than those symmetry concepts obtained
through that parallelism to classical field theory referred to as
``quantization''.

Let $\psi $ a vector in a coherent subspace of a Hilbert space of a quantum
theory (example: an irreducible representation space of a CCR- or
CAR-algebra). The corresponding physical state, in the sense of expectation
values as defined previously, corresponds to the unit ray:

\begin{equation}
\underline{\psi }=\left\{ e^{i\alpha }\psi \mid \alpha \in \left[ 0,2\pi
\right] ,\left( \psi ,\psi \right) =1\right\}
\end{equation}

The probability for a ``source'' state \underline{$\psi $} containing a
``measured'' state \underline{$\varphi $} is: 
\begin{equation}
w(\underline{\varphi },\underline{\psi })=\left| (\varphi ,\psi )\right| ^{2}
\end{equation}
and does not depend on the representing vectors. A symmetry transformation $%
\underline{S}$ is defined to be a transformation of unit rays: 
\begin{equation}
\underline{\psi }\longrightarrow \underline{\psi }^{^{\prime }}\qquad %
\hbox{with}\qquad w(\underline{\varphi }^{\prime },\underline{\psi }^{\prime
})=w(\underline{\varphi },\underline{\psi })
\end{equation}

The physical full significance of such \underline{$S$} only becomes evident
through its action on local observables, an issue which we will take up in a
later section.

It is comforting to know, that this projective definition may be reduced to
the standard situation of (anti-)unitary operators in Hilbert space:

\begin{theorem}
(Wigner): Any ray representation \underline{S} may be rewritten in terms of
a (anti-)unitary vector representation $S$:
\end{theorem}

\begin{equation}
\psi =S\psi ^{\prime \quad }\hbox{with (}\varphi ^{\prime }\hbox{,}\psi
^{\prime }\hbox{)=}\left\{ 
\begin{array}{cc}
(\varphi ,\psi ) & \hbox{unitary} \\ 
(\psi ,\varphi ) & \hbox{antiunitary}
\end{array}
\right.
\end{equation}
In the antiunitary case, S may be written in terms of any conjugation $K$
(an antilinear operator which flips the bras and ket of a inner product) as $%
S=UK$ with $U$ unitary. A proof of this fundamental theorem of quantum
theory unfortunately only found its way into few QM textbooks. The reader
finds a fairly explicit proof in e.g. in \cite{Wei} Antiunitary operators
appear in quantum theory exclusively in symmetry transformations which
contain the operation of time reversal. For physical reasons one does not
want a symmetric spectrum, since the energy of systems at zero temperature
should be bounded below in order to avoid instabilities due to transitions
into arbitrarily negative energy eigenstates ( the same reason why Dirac
filled the negative energy Dirac sea).

The time reversal T flips the direction of time and therefore:

\begin{equation}
Te^{iHt}\psi =e^{-iHt}T\psi  \label{T}
\end{equation}

Taking $T$ unitary and $\psi \ $an energy eigenvector (the use of the
spectral representation for $H$ would be more rigorous), one would obtain a
symmetric energy spectrum which is in conflict with the existence of a
ground state (but not with the structure of finite temperature states).
Conversely, if $T$ is antiunitary, then by (\ref{T}) it commutes with $H$
and preserves its spectrum.

If the symmetry $S$ is part of a symmetry group whose group manifold is
connected (i.e. every element is continuously deformable into the identity),
evidently only unitary representor can occur.

The presence of superselection rules limits the previous consideration to
coherent subspaces. In the total space 
\begin{equation}
\mathcal{H=}\sum_{\oplus i}\mathcal{H}_{i}
\end{equation}
the phases between the $S^{\prime }$s in the subspaces are arbitrary and
without physical significance. Symmetries not related continuously with the
identity, as the various reflections: $P,T,$ $PT,$ as well as discrete
symmetries not related to space-time e.g. charge conjugation $C$, can in
principle transform one subspace into another. The various possibilities
require a careful discussion \cite{Wei}

If we apply the above consideration to symmetry groups, the two operators $%
U(g_{2})U(g_{1})$ and $U(g_{2}g_{1})$ need only to be identical up to a
phase factor: 
\begin{equation}
U(g_{2})U(g_{1})=e^{i\varphi (g_{2},g_{1})}U(g_{2}g_{1})
\end{equation}
This relation (``ray representation'') follows from the fact that $%
U(g_{2}g_{1})$ and $U(g_{2})U(g_{1})$ act the same way on observables (i.e.
like a classical symmetry) and here we assume that the latter generate an
irreducible operator algebra in each $\mathcal{H}_{i}$, i.e. one for which
Schur's Lemma holds. The associativity of the threefold composition yields a
consistency condition for the phase which depends on two group elements. It
is called a 2-cocycle condition. It is important to know under what
circumstances this phase may be absorbed into a redefinition of the $%
U^{\prime }s$, i.e. under what circumstances the ``cocycle is a
coboundary''. In passing from classical to quantum theory, physicists are
usually familiar with two ``obstructions'': phase factors coming from the
topology of groups, as e.g. the phase factor -1 in half-integer spin
representations (which becomes a projective representation if one considers
SO(3) and not SU(2) the represented group), and central extensions of Lie
algebras which, after exponentiation, lead to unremovable phase factors in
the associated groups. A famous physical illustration of the physical
relevance of central extensions is the Galilei group in Schr\"{o}dinger
theory. For semisimple Lie-groups the absence of Lie-algebraic central
extensions follows from the absence of a 2-cohomology $H^{2}(Lie$-$%
algebra)=0.$ In this case every ray representation is equivalent to a vector
representation of the universal covering group. The only known group in
physics which has nontrivial central extensions (and necessarily is not
semisimple) is the Galilei group. The physical interpretation is the
noncommutativity of the momentum operator $\vec{p}$ with the infinitesimal
generator of the velocity (Galilei) transformation $G=\vec{p}t-m\vec{x}.$
This group emerges from the Poincar\'{e} group by a process called
``contraction''. More surprisingly, it also appears in a 8-parametric
subgroup of the 10-parametric Poincar\'{e} group if one splits the
generators into longitudinal and transversal as suggested by the concept of
modular wedge localization which will be introduced in a later section.

\begin{theorem}
(Wigner, Bargmann) The projective unitary representations of the Poincare
group $\mathcal{P}$ are equivalently described by vector representations of
its universal covering $\widetilde{\mathcal{P}}$.
\end{theorem}

Physicists refer to the representations of the covering group sometimes as
multi-valued representations. From a topological point of view the two-fold
covering of $\mathcal{P}$ happens already inside the rotation subgroup SO(3)
whose covering is SU(2) i.e. the phenomenon of halfinteger spin.

As usual in Lie group theory, one describes representations in terms of
infinitesimal generators fulfilling Lie algebra relations. The best known
case in physics is the unitary representation theory of the $SU(2)=%
\widetilde{SO(3)}$ . If we characterize the rotation by an angle $\Theta $$%
\,\,$and axis $\overrightarrow{n},$ we have: 
\begin{equation}
U(\overrightarrow{n},\theta )=e^{i\theta \overrightarrow{n}\overrightarrow{%
\cdot J}}
\end{equation}
where$\overrightarrow{J}$ is the quantum mechanical rotation operator with
the Lie-algebra: 
\begin{equation}
\left[ J_{i},J_{j}\right] =i\epsilon _{ijk}J_{k}\hbox{ }
\end{equation}

The unitary irreducible representations are all finite dimensional and are
explicitly given by the following well-known matrices . For $\overrightarrow{%
J^{2}}=s(s+1)$: 
\begin{equation}
\begin{array}{l}
\left\langle s,m\left| J_{3}\right| s,m\right\rangle =m\qquad \\ 
\left\langle s,m+1\left| J_{+}\right| s,m\right\rangle =\left\langle
s,m\left| J_{-}\right| s,m+1\right\rangle =\left[ (s-m)(s+m+1)\right] ^{%
\frac{1}{2}}
\end{array}
\end{equation}
Here $-s\leq m\leq s,J_{\pm }=J_{1}+J_{2}$, and we have listed only the
nonvanishing matrix elements in the $2s+1$ dimensional representation space.

The distinction between a group and its covering does not show up in the Lie
algebra, but it can be seen in e.g. irreducible representations by looking
at the values of the Casimir (invariant) operators; in the present case the
distinction between halfinteger and integer spin in the eigenvalues of $\vec{%
J}^{2}$. The finite dimensional irreducible representations of the Lorentz
group are constructed in complete analogy to the rotation group. The
following construction shows that the knowledge of the above formalism
suffices. Choosing generators : 
\begin{equation}
\begin{array}{l}
\Lambda =e^{\frac{1}{2}iM_{\mu \nu }\theta ^{\mu \nu }}=e^{i\overrightarrow{M%
}\overrightarrow{n}\theta +i\overrightarrow{K}\overrightarrow{e}\chi } \\ 
\overrightarrow{M}\hbox{ and }\overrightarrow{K}\hbox{ related to }M^{\mu
\nu }\hbox{ as }\overrightarrow{B}\hbox{ and }\overrightarrow{E}\hbox{ to F}%
^{\mu \nu }
\end{array}
\end{equation}
the Lie algebra relations are : 
\begin{equation}
\left[ M_{i}^{\pm },M_{j}^{\pm }\right] =i\epsilon _{ijk}M_{k}^{\pm },\qquad
\left[ M_{i}^{+},M_{j}^{-}\right] =0\qquad \hbox{with }M_{j}^{\pm }=\frac{1}{%
2}(M_{j}\pm iK_{j})
\end{equation}

Our previous $SU(2)$ representation of the rotation extends to the $SL(2,C)$
(two-valued) representation space of the Lorentz group in terms of (undotted
or dotted \cite{Haag}) two-component spinors is: 
\begin{eqnarray}
\alpha (\Lambda ) &=&e^{i\frac{1}{2}\overrightarrow{\sigma }\overrightarrow{n%
}\theta +\frac{1}{2}\overrightarrow{\sigma }\overrightarrow{e}\chi }=\left\{ 
\begin{array}{l}
\cos \frac{\theta }{2}+isin\frac{\theta }{2}\cdot \vec{n}\vec{\sigma}%
\,\,,\,\chi =0 \\ 
\cosh \frac{\chi }{2}+sinh\frac{\chi }{2}\cdot \vec{e}\vec{\sigma}%
\,\,,\,\theta =0
\end{array}
\right. \\
\beta (\Lambda ) &\equiv &\alpha (\Lambda ^{-1})^{*}=i\sigma _{2}\overline{%
\alpha (\Lambda )}(i\sigma _{2})^{-1}
\end{eqnarray}
First we convince ourselves that $SL(2,C)$ matrices (i.e. complex matrices
with determinant =1) allow an exponential parametrization in terms of a
complex linear combination of the three traceless Pauli-matrices i.e. $%
\alpha =\exp i(\vec{c}_{1}-i\vec{c}_{2})\frac{\vec{\sigma}}{2}.$ The
interpretation of the $\vec{c}_{i}^{\prime }s$ in terms of the Lorentz group
parameters $(\vec{n},\theta ;\vec{e},\chi )$ follows from the relation of
the $L_{+}^{\uparrow }$ and $SL(2,C)$ groups via the conversion formula: 
\begin{eqnarray}
x^{\mu } &\rightarrow &\stackunder{\sim }{x}=x^{\mu }\sigma _{\mu } \\
\stackunder{\sim }{x} &\rightarrow &\alpha (\Lambda )\stackunder{\sim }{x}%
\alpha ^{*}(\Lambda )\,\,\,\curvearrowright x^{\mu }\rightarrow \Lambda
_{\nu }^{\mu }x^{\nu }  \nonumber
\end{eqnarray}
The matrix $\stackunder{\sim }{x}$ transforms therefore like a mixed spinor
with two spinor indices, one transforming with $\alpha $ and the other one
with $\bar{\alpha}.$

Using the notation $\overrightarrow{J}$ and $\overrightarrow{K}$ for the
representors of $\overrightarrow{M}$ and $\overrightarrow{K}$ in the
respective representations we find for this spinor representation : 
\begin{equation}
\overrightarrow{J}=\frac{\overrightarrow{\sigma }}{2}\hbox{ , }%
\overrightarrow{K}=-i\frac{\overrightarrow{\sigma }}{2}%
\hbox{ \quad or \quad
}\overrightarrow{J^{+}}=\frac{\overrightarrow{\sigma }}{2}\hbox{ , }%
\overrightarrow{J^{-}}=0\quad \hbox{with }\overrightarrow{J^{\pm }}=\frac{1}{%
2}\left( \overrightarrow{J}\pm i\overrightarrow{K}\right)
\end{equation}

The standard notation for this fundamental (undotted) $SL(2,C)$ spinor
representation is : 
\begin{equation}
D^{\left[ \frac{1}{2},0\right] }(\Lambda )=e^{i(\theta \overrightarrow{n}%
-i\chi \overrightarrow{e})\overrightarrow{J^{+}}}=\alpha (\Lambda )\qquad
J^{+}=\frac{\overrightarrow{\sigma }}{2}
\end{equation}

Similarly for the ``dotted'' spinors : 
\begin{equation}
D^{\left[ 0,\frac{1}{2}\right] }(\Lambda )=e^{i(\theta \overrightarrow{n}%
+i\chi \overrightarrow{e})\overrightarrow{J^{-}}}=i\sigma _{2}\bar{\alpha}%
(\Lambda )(-i\sigma _{2})=\beta (\Lambda )\qquad J^{-}=\frac{\overrightarrow{%
\sigma }}{2}
\end{equation}
The equivalence transformation with the Pauli-matrix $\varepsilon \equiv
i\sigma _{2}$ assures that both representations are identical if restricted
to $SU(2);$ this $\varepsilon $ plays a similar role in the spinor calculus
as the metric tensor $g_{\mu \nu }$ for the tensor calculus. It allows to
pass from lower (un)dotted indices to upper. $D^{\left[ 0,\frac{1}{2}\right]
}(\Lambda )$ refers to upper dotted indices and the $\stackunder{\sim }{x}$
matrix is a mixed spinor of the type $\stackunder{\sim }{x}_{\alpha ,\dot{%
\beta}}.$

Note that in these representations (as well as in all finite dimensional
representations of the Lorentz group) the generators $K$ are not hermitean
i.e. the associated $D$'s are not unitary. The general irreducible finite
dimensional representation are characterized in terms of two (half)integers $%
\frac{n_{\pm }}{2}$ which denote the formal $J_{\pm }$angular momenta$.$%
\begin{equation}
D^{\left[ \frac{n_{+}}{2},\frac{n_{-}}{2}\right] }(\Lambda )=e^{i(\theta 
\overrightarrow{n}-i\chi \overrightarrow{e})\overrightarrow{J^{+}}}\otimes
e^{i(\theta \overrightarrow{n}+i\chi \overrightarrow{e})\overrightarrow{J^{-}%
}}  \label{mat}
\end{equation}
Here $\overrightarrow{J^{\pm }}$ are the previously defined matrices of size 
$(2\frac{n_{\pm }}{2}+1)\times (2\frac{n_{\pm }}{2}+1)$ . In the spirit of
the spinor calculus, one should envisage these operators to act on tensor
products of (un-)dotted spinors. The $J_{\pm }$ act as Pauli matrices on $%
\otimes _{n_{\pm }}$Mat$_{2}(\mathbf{C})$ : 
\begin{equation}
\overrightarrow{J}_{i}=\left. \sum_{i=1}^{n}1\otimes ....\otimes \frac{%
\overrightarrow{\sigma }_{i}}{2}\otimes ....\otimes 1\right| _{symm.}
\end{equation}

The symmetrization in the $n_{\pm }$ spinorial indices assures the
irreducibility. The difference in undotted and dotted spinors is in the
action of L-boosts, whereas the rotations act the same way. The entries of
the matrices (\ref{mat}) can be expressed in terms of Jacobi functions.

\section{One Particle Representations}

We are now ready to study unitary representations of the covering of the
Poincar\'{e} $\widetilde{\mathcal{P}}$ and its subgroup $\widetilde{SO(3,1)}%
. $ The infinitesimal generators for noncompact groups are necessarily
unbounded operators. The domain problems for unbounded Lie-generators
(common dense domain etc.) have been studied, and we will ignore them unless
they are of direct physical significance (as e.g. in the relation between
the Tomita-Takesaki modular theory and symmetries described in a later
section).

The commutation relations of the Poincar\'{e} generators follow from the
composition property for the two-fold covering $\widetilde{\mathcal{P}}$%
\begin{equation}
\left( a_{2},\alpha _{2}\right) \cdot \left( a_{1},\alpha _{1}\right)
=\left( a_{2}+\alpha _{2}a_{1}\alpha _{2}^{-1},\alpha _{2}\alpha _{2}\right)
\end{equation}
The second translational term is simply the Lorentz-transformed vector $%
\Lambda _{2}a_{1}.$ From the special case 
\begin{equation}
\left( 0,\alpha ^{-1}\right) \left( a,0\right) \left( 0,\alpha \right)
=\left( \Lambda ^{-1}a,1\right)
\end{equation}
one abstracts for infinitesimal translations: 
\begin{equation}
U^{-1}(\alpha )P^{\mu }U(\alpha )=\Lambda _{\nu }^{\mu }P^{\nu
}\,\,\,\,\,\,\,\,\hbox{with \thinspace \thinspace }U(a)=e^{iaP}
\end{equation}
Analogously the transformation of the operator $U(\alpha )\footnote{%
Following habits of physicist, we will often write $U(\Lambda )$ instead of $%
U(\alpha (\Lambda )).$}=e^{iM^{\mu \nu }\theta _{\mu \nu }}$ by another
Lorentz transformation yields the tensor transformation property of $M^{\mu
\nu }$%
\begin{equation}
U^{-1}(\alpha )M^{\mu \nu }U(\alpha )=\Lambda _{\kappa }^{\mu }\Lambda
_{\lambda }^{\nu }M^{\kappa \lambda }
\end{equation}
In order to avoid clumsy notation, it is convenient to suppress the
unimodulars $\alpha $ inside unitaries and write simply $U$($\Lambda )$ with
the understanding that $\Lambda $ denotes an element of $\widetilde{\mathcal{%
P}}$ . Only for matrices (i.e. finite dimensional representations) the
notational distinction matters. The Lie-algebra relations are obtained from
the above transformation laws by expanding $U(\alpha )\equiv U(\Lambda )$
retaining only linear terms in $\theta _{\mu \nu }$: 
\begin{equation}
\begin{array}{c}
\left[ M^{\alpha \beta },P^{\mu }\right] =i(g^{\alpha \mu }P^{\beta
}-(\alpha \longleftrightarrow \beta )) \\ 
\left[ M^{\mu \nu },M^{\rho \sigma }\right] =i\left\{ g^{\nu \sigma }M^{\mu
\rho }-g^{\mu \rho }M^{\nu \sigma }-\left( \mu \longleftrightarrow \nu
\right) \right\}
\end{array}
\end{equation}
The last relation is the tensor form of the previous $J^{\pm }$ commutation
relations.

Approaching the Wigner theory via the infinitesimal generators $P^{\mu }$, $%
M^{\mu \nu }$, one first looks for the Casimir (invariant) operators which
take on characteristic values in irreducible representations : 
\begin{equation}
P^{\mu }P_{\mu }\,\,,\,\,\,\,\,W^{\mu }W_{\mu }\,\,\,\,\,\,\,
\end{equation}
The first invariant is the mass operator and the second one is usually
referred to as the Pauli-Lubanski invariant. It is formed with the
Pauli-Lubanski vector: 
\[
W_{\mu }=\frac{1}{2}\varepsilon _{\mu \alpha \beta \gamma }M^{\alpha \beta
}P^{\gamma } 
\]
Its commutation properties follow from those of the Poincare generators $%
M^{\mu \nu }$ and $P^{\mu }$: 
\begin{equation}
\left[ W_{\mu },P_{\nu }\right] =0\,\,\,\,\,\,\,\,\,\,\left[ W_{\mu },M_{\nu
\kappa }\right] =i(g_{\nu \mu }W_{\kappa }-g_{\kappa \mu }W_{\nu
})\,\,\,\,\,\,\,\,\,\left[ W_{\mu },W_{\nu }\right] =i\epsilon _{\mu \nu
\kappa \lambda }W^{\kappa }P^{\lambda }
\end{equation}

Since $P_{\mu }W^{\mu }=0$, there is no nontrivial third invariant. The
interpretation of $W_{\mu }$ and $W^{2}$ in terms of intrinsic angular
momentum becomes manifest, if we specialize to so called positive energy
representations.

Wigner classified the irreducible representations according to their
transitive p-space orbits (submanifolds of momentum space traced out by the
action of $\mathcal{L}$ to a given vector):\thinspace \thinspace $\,\,$

\begin{itemize}
\item  (i)$\,\,\,\,\,p^{\mu }p_{\mu }=m^{2}>0\,\,\,\,\,\,\,\,\,\,p_{0}>0$

\item  (ii)\thinspace \thinspace \thinspace \thinspace \thinspace $p^{\mu
}p_{\mu }=0$ \thinspace \thinspace \thinspace \thinspace \thinspace
\thinspace \thinspace $p_{0}>0$

\item  (iii)\thinspace \thinspace \thinspace \thinspace \thinspace
\thinspace $p_{\mu }=0$
\end{itemize}

\noindent and the corresponding orbits with negative energies $p_{0}<0$ , as
well as the spacelike orbits $p_{\mu }p^{\mu }<0.$ The first two exhaust the
positive energy representations. In order to construct them explicitly, we
look at the stability group (``little group'') of a point on the orbit.
Without loss of generality we may specialize to the stability group of a
selected reference momentum, since the stability group for other momenta are
equivalent (by Lorentz-boosts). In the case (i) we choose $p_{R}=(m,\vec{0})$
which yields the SO(3) respectively. its covering $SU(2)$ as the quantum
theoretically relevant little group. On the other hand the little group of
the lightlike reference vector which is chosen to be $p_{R}=(1,0,0,1)\ $
turns out to be the euclidean group $E(2)$ in two dimensions. Only the
rotation around the 3-axis is geometrically obvious , the interpretation of
the two euclidean ``translations'' is somewhat hidden and will be presented
later. Let us now look in detail at the massive case (i). We start with a
2s+1 dimensional representation of the little group. This irreducible
representation induces a unitary irreducible positive energy representation
of the Poincar\'{e} group $\widetilde{\mathcal{P}}$ as follows. We first
chose the momentum in rest $p_{R}=(m,\widetilde{0})$ as the reference vector
on the orbit$\;\;p^{2}>0,p_{0}>0.\;$The action on (improper, like plane
waves) reference basis vectors is: 
\begin{equation}
\begin{array}{l}
P^{\mu }\left| p_{R},s_{3};\gamma \right\rangle =p_{R}^{\mu }\left|
p_{R},s_{3};\gamma \right\rangle \;, \\ 
W_{0}\left| p_{R},s_{3};\gamma \right\rangle =0\;, \\ 
W_{k}\left| p_{R},s_{3};\gamma \right\rangle =\frac{m}{2}\epsilon _{k\mu \nu
0}M^{\mu \nu }\left| p_{R},s_{3};\gamma \right\rangle
\end{array}
\end{equation}
The last relation connects the spatial components of W with the Wigner spin
i.e. with the angular momentum in the rest frame: 
\begin{equation}
W_{k}\left| p_{R},s_{3};\gamma \right\rangle =\frac{m}{2}\epsilon
_{kij}M^{ij}\left| p_{R},s_{3};\gamma \right\rangle =mJ_{k}\left|
p_{R},s_{3};\gamma \right\rangle
\end{equation}
Since an invariant operator can be evaluated on any state vector, we have $%
W^{2}=-m^{2}\vec{J}^{2}$and therefore in an irreducible representation: $%
W^{2}=-m^{2}s(s+1)$. In this approach irreducibility just means the absence
of an additional degeneracy label, say $\gamma $ (such labels, which go
beyond spacetime characteristics as momentum and spin, are related to
internal symmetries and called charges.). One now uses a distinguished
family of Lorentz-transformations which link $p_{R}$ with a general point p
on the $p^{2}=m^{2}$ orbit. One chooses the family of rotational free
Lorentz-transformations (``boosts'') to relate the p-eigenstates: 
\begin{eqnarray}
\left| p,s_{3}\right\rangle &\equiv &U(L(p))\left| p_{R},s_{3}\right\rangle
,\,\,\,\left\langle p^{\prime },s_{3}^{\prime }\mid p,s_{3}\right\rangle
=2p^{0}\delta (\vec{p}^{\prime }-\vec{p})\,\,\,  \label{bas} \\
L(p) &=&\frac{1}{m}\left( 
\begin{array}{c}
p^{0},\,\,\,\,\,\,\,\,\,\,\,\vec{p}\,\,\,\,\,\,\,\,\,\,\,\, \\ 
\vec{p},m\delta _{ik}+\frac{p^{i}p^{k}}{p^{0}+m}
\end{array}
\right) =\Lambda (\vec{e},\chi ),\,\,\,\alpha (L(p))=\sqrt{\frac{p^{\mu
}\sigma _{\mu }}{m}}
\end{eqnarray}
\begin{equation}
\hbox{with: \thinspace \thinspace }\vec{e}=\frac{\vec{p}}{\left| \vec{p}%
\right| }\,\,\,\,\,\,\,\,\,\,\,\,\,ch\chi =\frac{p^{0}}{m}
\end{equation}

The invariant $\delta $-function which appears on the right hand side of the
inner product of relativistic (improper) momentum eigenstates corresponds to
the relativistic measure $d^{3}p/2\omega $ with $\omega =p^{0}=\sqrt{\vec{p}%
^{2}+m^{2}}.$ For later use, we also wrote the positive unimodular matrix $%
\alpha (L(p))$ which the spinor calculus affiliates with the boost.

We now are able to describe the $\left[ m_{+},s\right] $ Wigner
representation in global terms as follows. The one- particle Hilbert space
is: 
\begin{equation}
H_{\left[ m,s\right] }^{\left( 1\right) }=\left\{ \int \sum_{s_{3}}\tilde{%
\psi}(\vec{p},s_{3})\left| p,s_{3}\right\rangle \frac{d^{3}p}{2\omega }\mid
\int \sum \left| \tilde{\psi}\right| ^{2}\frac{d^{3}p}{2\omega }<\infty
\right\} \,\,\,\,\,\,\,\,\,
\end{equation}

and the transformation properties on the basis (\ref{bas}) are:

\begin{equation}
U(\Lambda )\left| p,s_{3}\right\rangle =U(L(\Lambda p))U(R(\Lambda
,p))\left| p_{R},s_{3}\right\rangle \,\,\,\,\,\,\,\,\,\,\,\hbox{with }%
R(\Lambda ,p)=L^{-1}(\Lambda p)\Lambda L(p)
\end{equation}
\begin{equation}
\hbox{and \thinspace \thinspace }U(R(\Lambda ,p))\left|
p_{R},s_{3}\right\rangle =\sum_{s_{3}^{^{\prime }}}\left|
p_{R},s_{3}^{^{\prime }}\right\rangle D_{s_{3}^{^{\prime
}}\,s_{3}}(R(\Lambda ,p))\,\,\,\,\,\,\,\hbox{we obtain:}
\end{equation}
\begin{equation}
\begin{array}{c}
U(\Lambda )\left| p,s_{3}\right\rangle =\sum_{s_{3}^{^{\prime }}}\left|
\Lambda p,s_{3}^{^{\prime }}\right\rangle D_{s_{3}^{^{\prime
}}\,s_{3}}(R(\Lambda ,p))\,\,\,\,\,\hbox{and for translations}\,: \\ 
\,U(a)\,\left| p,s_{3}\right\rangle =e^{ipa}\,\,\left| p,s_{3}\right\rangle
\,
\end{array}
\end{equation}
The successive transformations by a boost, followed by a Lorentz
transformation $\Lambda $ and the inverse boost yields a transformation $%
R(\Lambda ,p)$ which leaves $p_{R}$ $\,$invariant and therefore is called
the Wigner rotation. The appearance of $p$-dependent unitary matrices is
typical for relativistic quantum theory. It prevents a simple minded
transition to x-space covariant localizable functions via Fourier
transformation. As well known, one can rewrite the transformations from the
basis vectors in $H_{\left[ m,s\right] }^{\left( 1\right) }$ to the wave
functions on which one finds the contragredient action: 
\begin{equation}
\left( U(\Lambda )\tilde{\psi}\right) \left( \vec{p},s_{3}\right)
=\sum_{s_{3}^{^{\prime }}}D_{s_{3}s_{3}^{^{\prime }}}(R(\Lambda ,\Lambda
^{-1}p)\tilde{\psi}(\Lambda ^{-1}\vec{p},s_{3}^{^{\prime }})
\end{equation}
Besides this Wigner ``canonical'' representation, there exists the closely
linked ``helicity'' representation for which the spin quantization axis is
identified with the direction of the spatial momentum of the particle.
Calling the magnetic quantum number with respect to this direction $\lambda $
we define: 
\begin{equation}
\left| p,\lambda \right\rangle \equiv \sum_{s_{3}^{^{\prime }}}\left|
p,s_{3}^{^{\prime }}\right\rangle D_{s_{3}^{^{\prime }}\lambda }(R_{p,\check{%
p}})\,\,\,\hbox{with }R_{p\check{p}}=Rot(\varphi ,\theta ):
\end{equation}
being the ``minimal'' rotation which changes the z-direction into $\vec{n}=%
\frac{\vec{p}}{\left| \vec{p}\right| }$ i.e. a rotation around the y-axis
with latitude $\theta $ followed by a $\varphi $-rotation around the z-axis.
In the helicity basis the Wigner rotation is modified: 
\begin{equation}
\check{R}(\Lambda ,p)=R_{\Lambda p,\Lambda \check{p}}^{-1}R(\Lambda ,p)R_{p,%
\check{p}}\hbox{ \thinspace \thinspace leaves \thinspace \thinspace }\check{p%
}=(\sqrt{p^{2}+m^{2}},0,0\hbox{,}\left| \vec{p}\right| )\,\,%
\hbox{\thinspace
\thinspace invariant.}
\end{equation}

The evaluation of W$_{0}$ on the helicity reference state gives: 
\begin{equation}
W_{0}\left| \check{p},\lambda \right\rangle =\vec{J}\cdot \vec{p}\left| 
\check{p},\lambda \right\rangle =\left| \vec{p}\right| \lambda \left| \check{%
p},\lambda \right\rangle \,\,\,\,%
\hbox{or with \thinspace \thinspace
\thinspace }h\equiv \frac{\vec{J}\cdot \vec{p}}{\left| \vec{p}\right| }%
\,\,,\,\,\,\,\,\,\,h\left| \check{p},\lambda \right\rangle =\lambda \left| 
\check{p},\lambda \right\rangle
\end{equation}
The column vectors of $D^{\left( s\right) }(R_{p,\check{p}})$ furnish a
complete set of eigenstates of the helicity operator $h$, e.g. for $s=\frac{1%
}{2}$ we have: 
\begin{equation}
\begin{array}{c}
w^{\pm }=D^{(\frac{1}{2})}(R_{p,\check{p}})\chi ^{\pm }\,\,,\,\,\chi ^{\pm
}=\left( 
\begin{array}{c}
1 \\ 
0
\end{array}
\,\right) ,\left( 
\begin{array}{c}
0 \\ 
1
\end{array}
\right) \,\,i.e. \\ 
D^{\left( \frac{1}{2}\right) }(R_{p,\check{p}})=\left( 
\begin{array}{cc}
w_{1}^{+} & w_{2}^{\_} \\ 
w_{2}^{+} & w_{2}^{-}
\end{array}
\right) \,,\,\,\frac{1}{2}\vec{\sigma}\cdot \vec{p}\,w^{\pm }=\pm \frac{1}{2}%
w^{\pm }
\end{array}
\end{equation}
The advantage of the helicity basis is that one may take the limit $%
m\rightarrow 0.$ As expected, the helicity rotation matrix $D(\check{R})$
approaches a diagonal limit in terms of a ``Wigner phase'' $\varphi (\Lambda
,p)$ e.g. for $s=\frac{1}{2}$: 
\begin{equation}
\begin{array}{c}
\lim_{m\rightarrow 0}D^{\left( \frac{1}{2}\right) }(\check{R})=\left( 
\begin{array}{cc}
e^{i\frac{1}{2}\varphi (\Lambda ,p)} & 0 \\ 
0 & e^{-i\frac{1}{2}\varphi (\Lambda ,p)}
\end{array}
\right) ,\,\, \\ 
\,e^{i\frac{1}{2}\varphi (\Lambda ,p)}=\sqrt{\frac{p^{0}}{(\Lambda p)^{0}}}%
\left\langle w^{+}(\Lambda p)\left| \alpha (\Lambda )\right|
w^{+}(p)\right\rangle
\end{array}
\end{equation}
In the massless limit, the $(2s+1)$- component representation decomposes
into 2s+1 one-component representations.

A direct approach a la Wigner to the $m=0$ case would start with the
representation theory of the stability group of a light-like vector. In this
situation there is no such natural choice as before. Choosing a light-like
vector in the z-direction $p_{R}=(1,0,0,1),$ one \thinspace obtains the
following matrix realization of the 3-parametric euclidean group $E(2)$ in 2
dimensions: 
\begin{equation}
G(\alpha ,\beta )=\left( 
\begin{array}{cccc}
1+\frac{1}{2}\left| \rho \right| ^{2} & \alpha & \beta & -\frac{1}{2}\left|
\rho \right| ^{2} \\ 
\alpha & 1 & 0 & -\alpha \\ 
\beta & 0 & 1 & -\beta \\ 
\frac{1}{2}\left| \rho \right| ^{2} & \alpha & \beta & 1-\frac{1}{2}\left|
\rho \right| ^{2}
\end{array}
\right) ,\,\,\,\,R(\theta )=\left( 
\begin{array}{cccc}
1 & 0 & 0 & 0 \\ 
0 & \cos \theta & \sin \theta & 0 \\ 
0 & -\sin \theta & \cos \theta & 0 \\ 
0 & 0 & 0 & 1
\end{array}
\right)
\end{equation}
The first matrix is a Lorentz-transformation which leaves $p_{R}$ invariant
and transforms the time axis into $Gp=(1+\frac{1}{2}\left| \rho \right|
^{2},\alpha ,\beta ,\frac{1}{2}\left| \rho \right| ^{2}),\,\,\rho =\alpha
+i\beta .$ where the transformed vector, whose time component increases, has
been conveniently parametrized. Any other transformation having this
property can only deviate from $G(\rho )$ by a transformation which leaves
the two vectors $p_{R}$\thinspace and the time axis invariant i.e. a x-y
rotation $R(\theta )$. Obviously they \thinspace generate a stability group
which is easily checked to be isomorphic to $E(2)$, the euclidean
translations corresponding to $G(\rho )$. To be more precise, since the
euclidean group has to be considered as a subgroup of the covering of the
Poincar\'{e} group,\QTR{cal}{\ }only the two fold covering $\tilde{E}(2)$ is
relevant.

The description in terms of the corresponding subgroup of SL(2,C) and the
reference vector is somewhat simpler: 
\begin{equation}
\alpha (\rho ,\theta )=\left( 
\begin{array}{ll}
e^{i\frac{1}{2}\theta } & \rho \\ 
0 & e^{-i\frac{1}{2}\theta }
\end{array}
\right) ,\,\,p_{R}\sim \left( 
\begin{array}{ll}
2 & 0 \\ 
0 & 0
\end{array}
\right)
\end{equation}
The unitary representation theory of this noncompact group is somewhat more
complicated than that of $SU(2)$. But it is obvious that the representations
fall into two classes; the ``neutrino- photon'' class with $U(G(\rho ))=1$
i.e. trivial representation of the euclidean translations, and the remaining
``continuous spin'' representation with $U(G(\rho ))\neq 0$. The difference
also shows up in the spectrum of the operator $W^{2}.$ Whereas in the first
case $W^{2}=0$ (in fact $W^{\mu }=hP^{\mu }$), the value of $W^{2}$ in the
second case can be any negative number which is responsible for the name.
These representations of $\tilde{E}(2)$ are infinite dimensional. They are
usually discarded as a result of the apparent absence of such particles in
nature. We will later return to this representations, since a theoretician
should use theoretical arguments i.e. point to a property which makes this
positive energy representation appear less physical than the others.

As a curious side remark we mention that the two ``translations'' inside the
homogenous Lorentz group behave as transversal Galilei transformations if we
split Minkowski space into two longitudinal direction $e_{\pm }=(1,0,0,\pm
1) $ and the remaining transversal spacelike unit vectors $e_{x,y}.$ This
also simplifies the calculation of the Wigner phases $\theta (\Lambda ,k).$

It is comforting to know that the $\left[ m,s\right] $ representations admit
an extension of the Poincar\'{e} group which includes the reflections,
without enlarging the representation space. One obtains the well-known
formulas for the parity $\mathcal{P}$ and the time reversal $T$: 
\begin{equation}
\mathcal{P}\left| p,s_{3}\right\rangle =\xi _{P}\left| p_{0},-\vec{p}%
,s_{3}\right\rangle \,\,\,\,\,\,\,\,\,\,T\left| p,s_{3}\right\rangle =\xi
_{T}\sum_{s_{3}^{^{\prime }}}D_{s_{3},s_{3}^{^{\prime }}}^{\left( s\right)
}(i\sigma _{2})\left| p_{0,}-\vec{p},s_{3}^{^{\prime }}\right\rangle
\end{equation}
Here the $\xi ^{\prime }s$ are undetermined phase factors. This result
follows by first writing down the action of $\mathcal{P}$ and $T$ on the
reference vectors $\left| p_{R},s_{3}\right\rangle $ (the antiunitarity of $%
T $ brings in the ``spin-flip'' matrix $D(i\sigma _{2})$ which represents
the later appearing charge conjugation$)$. The rest follows from the
commutation relation of the reflections with the boost: 
\begin{equation}
\mathcal{R}_{\lambda }L(p)\mathcal{R}_{\lambda }^{-1}=L(p_{0},-\vec{p}%
)\,,\,\,\,\,\,\,\,R_{\lambda }=\mathcal{P}\,\,,\,\,T\,\,\,or\,\mathcal{P}T
\end{equation}
The corresponding operator relation may contain phase factors $D_{\lambda }$
i.e. 
\begin{equation}
\mathcal{R_{\lambda }}U(\Lambda )\mathcal{R}_{\lambda }^{-1}\mathcal{=}%
D_{\lambda }(\Lambda )U(R_{\lambda }\Lambda R_{\lambda }^{-1})
\end{equation}
These phase factors must form a representation of the Lorentz group. But
since there are no 1-dimensional representations, we have $D(\Lambda )=1.$
The above phases can be fixed. For unitary reflections we can achieve $%
\mathcal{R}_{\lambda }^{2}=1,$ whereas for antiunitaries $\mathcal{R}%
_{\lambda }^{2}=\pm 1.$ In the above special case we find: 
\begin{equation}
\mathcal{T}^{2}\mathcal{=}(-1)^{2s}
\end{equation}
The formulae for the $\left[ 0,s\right] $ representations are different as a
result of the different $p_{R}$ and its stability group which contain the
arbitrary z-direction : 
\begin{equation}
\mathcal{P}\left| p,s\right\rangle =\xi _{P}e^{\pm i\pi s}\left| p,-\vec{p}%
,-s\right\rangle \,\,\,\,\,\,\,\,\,\,\,\,T\left| p,s\right\rangle =\xi
_{T}e^{\pm i\pi s}\left| p,-\vec{p},s\right\rangle
\end{equation}
The $\pm $sign depends on the sign of $p_{y}$ (see Weinberg \cite{Wei}), and
this phase factor is only relevant if the states of opposite helicity are
not separated by a superselection rule.

The original motivation of Wigner was to classify relativistic wave
equations up to physical equivalence. Disregarding the continous spin class,
the classification of wave equations associated with finite positive energy
representations is as follows. We first present the three special cases s=0,$%
\frac{1}{2},1.$m$>0$

\begin{itemize}
\item  \textbf{s=0\thinspace \thinspace \thinspace \thinspace \thinspace
\thinspace \thinspace }
\end{itemize}

\textbf{\thinspace \thinspace \thinspace \thinspace \thinspace }The Fourier
transformation leads to covariant x-space wave function :

\begin{equation}
\psi \left( x\right) =\int e^{-ipx}\tilde \psi (p)\frac{d^3p}{2\omega }%
\,\,\,\,\,\,\,\hbox{\thinspace with}\,\,\left( U(\Lambda )\psi \right)
\left( x\right) =\psi \left( \Lambda ^{-1}x\right)
\end{equation}

The x-space function is a positive frequency solution of the Klein-Gordon
equation: 
\begin{equation}
\left( \partial ^{\mu }\partial _{\mu }+m^{2}\right) \psi \left( x\right) =0
\end{equation}

\begin{itemize}
\item  \textbf{s=$\frac{1}{2}\,\,\,\,\,\,\,\,\,\,$}
\end{itemize}

\textbf{$\,\,$} Here one has to convert the Wigner representation into a
covariant one involving (un)dotted spinors which transform covariantly even
under boost transformations. This is achieve by: 
\begin{equation}
\begin{array}{c}
\tilde{\Phi}_{a}(p):=\sum_{s_{3}}\alpha _{a,s_{3}}(L(p))\tilde{\psi}%
(p,s_{3})\,\,%
\hbox{\thinspace \thinspace with\thinspace \thinspace
\thinspace \thinspace }\alpha (L(p))=\sqrt{p^{\mu }\sigma _{\mu }}\,\,/\sqrt{%
m}\,\, \\ 
\hbox{a positive matrix, in short:}\,\,\tilde{\Phi}=\alpha (L(p))\tilde{\psi}
\end{array}
\end{equation}
\begin{equation}
\sigma _{\mu }\equiv (\underline{1},\overrightarrow{\sigma })
\end{equation}
As the notation suggests, $\tilde{\Phi}$ transforms like a (undotted)
spinor, a fact which follows by transforming the $\tilde{\psi}$ with the
Wigner rotation and then using its representation in terms of boosts : 
\begin{equation}
\begin{array}{c}
\alpha (L(p))\alpha \left( R(\Lambda ,\Lambda ^{-1}p)\right) =\alpha
(L(p))\alpha \left( L^{-1}(p\ \right) \alpha (\Lambda )\alpha \left(
L(\Lambda ^{-1}p)\right) \, \\ 
=\alpha (\Lambda )\alpha (L(\Lambda ^{-1}p)) \\ 
i.e.\hbox{ }\left( U(\Lambda )\tilde{\Phi}\right) (p)=\alpha (\Lambda )%
\tilde{\Phi}(\Lambda ^{-1}p)
\end{array}
\end{equation}
For later purpose it is helpful to rewrite the action of $\alpha (L(p))$ on $%
\widetilde{\psi }$ in terms of the column vectors of the boost matrix: 
\begin{equation}
\tilde{\Phi}(p)=\sum_{s_{3}}u(p,s_{3})\tilde{\psi}(p,s_{3})\,\,,\,\,\,\,\,%
\,u_{a}(p,s_{3}):=\alpha _{a,s_{3}}(L(p))
\end{equation}
$\alpha (L(p))$ and hence its column vectors $u$ has the intertwining
property between the Wigner and the covariant representation : 
\begin{equation}
\alpha (L(p))D^{\left( \frac{1}{2}\right) }(R(\Lambda ,\Lambda
^{-1}p))=D^{\left[ \frac{1}{2},0\right] }(\Lambda )\alpha (L(\Lambda ^{-1}p)
\end{equation}
$\,$

A similar intertwining relation is valid between the complex conjugate of
the Wigner representation $D^{*}$ and the same covariant $D^{\left[ \frac{1}{%
2},0\right] }$. In this case the intertwining matrix is $\alpha
(L(p))i\sigma _{2}$ and its columns are called $v$-spinors.

Fourier transformation gives the x-space wave function:

\begin{equation}
\begin{array}{c}
\Phi (x):=\int \tilde{\Phi}(p)e^{-ipx}\frac{d^{3}p}{2\omega }%
\,\,,\,\,\,\,\left( U(\Lambda )\Phi \right) (x)=\alpha \left( \Lambda
\right) \Phi (\Lambda ^{-1}x)\,\,,\, \\ 
\,\,\left( \partial ^{\mu }\partial _{\mu }+m^{2}\right) \Phi (x)=0
\end{array}
\end{equation}
In order to make contact with the Dirac theory, one defines a dotted spinor 
\begin{eqnarray}
\tilde{\chi}^{\cdot }(p) &:&=\frac{1}{m}p^{\mu }\tilde{\sigma}_{\mu }\tilde{%
\Phi}(p)=\alpha (L^{-1}(p))\tilde{\psi}(p) \\
\widetilde{\sigma }_{\mu } &\equiv &(\underline{1},-\overrightarrow{\sigma })
\nonumber
\end{eqnarray}
As indicated in the notation, $\tilde{\chi}^{\cdot }$ transforms as an upper
dotted spinor i.e. with a matrix $\beta (\Lambda )\equiv \alpha (\Lambda
^{-1})^{\dagger }.$ This is a result of the relation: 
\begin{equation}
(\Lambda p)^{\mu }\tilde{\sigma}_{\mu }\alpha (\Lambda )=\alpha (\Lambda
^{-1})^{\dagger }p^{\mu }\tilde{\sigma}_{\mu }
\end{equation}

This leads to the covariant transformation law for the corresponding x-space
wave function $\chi ^{.}(x):$%
\begin{equation}
(U(\Lambda )\chi )^{.}(x)=\beta (\Lambda )\chi ^{.}(\Lambda ^{-1})
\end{equation}

Defining a 4-component Dirac spinor, we immediately read off its properties: 
\begin{equation}
\tilde{\Psi}(p)=\sqrt{m}\left( 
\begin{array}{c}
\tilde{\Phi}(p) \\ 
\tilde{\chi}^{\cdot }(p)
\end{array}
\right) ,\,\hbox{\thinspace with\thinspace \thinspace \thinspace }\left(
p^{\mu }\gamma _{\mu }-m\right) \tilde{\Psi}=0\,\,\,%
\hbox{and\thinspace
\thinspace \thinspace }\gamma _{\mu }=\left( 
\begin{array}{cc}
0 & \sigma _{\mu } \\ 
\tilde{\sigma}_{\mu } & 0
\end{array}
\right)
\end{equation}
The first two components of the Dirac equation are identical to the
definition of $\tilde{\chi}^{\cdot }$ in terms of$\,\,\tilde{\Phi}$ and the
remaining equation is the Klein Gordon identity for$\,\tilde{\Phi}\,$: 
\begin{equation}
p^{\mu }\tilde{\sigma}_{\mu }\tilde{\Phi}=m\tilde{\chi}\,\,,\,\,\,p^{\mu
}\sigma _{\mu }\left( p^{\mu }\tilde{\sigma}_{\mu }\right) \tilde{\Phi}=m^{2}%
\tilde{\Phi}
\end{equation}
Rewriting the inner product in terms of $\Psi $ we obtain: 
\begin{eqnarray}
\left( \psi _{2},\psi _{1}\right) &=&\frac{1}{2m}\int \tilde{\Psi}_{2}^{*}%
\tilde{\Psi}_{1}\frac{d^{3}p}{2\omega }=\frac{1}{2m}\int \overline{%
\widetilde{\Psi }}\gamma _{0}\widetilde{\Psi }\frac{d^{3}p}{2\omega }\,\,\,\,
\label{Dir} \\
\,\hbox{with }\overline{\widetilde{\Psi }} &:&=\widetilde{\Psi }^{*}\gamma
_{0}\,\,\,\hbox{the Dirac adjoint}  \nonumber
\end{eqnarray}
where we have introduced the Dirac conjugate on the 4-spinors by a bar on
top of the symbol (sorry for the possible confusion with the complex
conjugate). Since the gamma matrices transform as a 4-vector, the Dirac
formalism permits to form tensors. In x-space we have: 
\begin{equation}
\Psi (x)=\int e^{-ipx}\widetilde{\Psi }(p)\frac{d^{3}p}{2\omega }%
\,\,,\,\,\,\left( i\gamma _{\mu }\partial ^{\mu }-m\right) \Psi
(x)=0\,,\,\,\,\,\,
\end{equation}
\begin{equation}
\overline{\Psi }(x)\Psi (x)=scalar,\,\,\overline{\Psi }(x)\gamma _{\mu }\Psi
(x)=vector\,\,etc.
\end{equation}

There are altogether 16 independent tensorial densities which one can form
in this way from products of $\gamma ^{\prime }s.$

Dirac's inner product is conveniently expressed in terms of the conserved
current:

\begin{equation}
j_\mu =\overline{\Psi }_2\gamma _\mu \Psi _1\,,\,\,\,\partial ^\mu j_\mu
=0,\,\,\,\,\,\left( \overline{\Psi }_2\Psi _1\right) :=\int j_0d^3x=2m(\psi
_2,\psi _1)\,
\end{equation}

The 4-component description allows a local matrix realization of the parity
symmetry: 
\begin{equation}
\left( \mathcal{P}\Psi \right) (x)=\gamma _{0}\Psi (x_{0},-\vec{x}%
),\,\,\,\,\,\,\,\gamma _{0}\gamma _{i}\gamma _{0}^{-1}=-\gamma _{i}
\end{equation}
It is helpful to define a fifth $\gamma $-matrix as the product of all four: 
$\gamma _{5}:=\gamma _{0}\gamma _{1}\gamma _{2}\gamma _{3}$. This matrix is
block- diagonal and behaves like a pseudoscalar under parity. Therefore
densities involving $\gamma _{5}$ are pseudo-scalars, -vectors etc. Finally
we mention the u- and v-intertwiners: 
\begin{equation}
\begin{array}{c}
u(p,s_{3})=S(L(p))u(p_{R},s_{3}),\,\,\,\,\,\,\,u(p_{R},\pm \frac{1}{2}%
)=\left( 
\begin{array}{c}
1 \\ 
0 \\ 
1 \\ 
0
\end{array}
\right) ,\left( 
\begin{array}{c}
0 \\ 
1 \\ 
0 \\ 
1
\end{array}
\right) , \\ 
\,\,S(L(p))=\left( 
\begin{array}{cc}
\sqrt{p^{\mu }\sigma _{\mu }} & 0 \\ 
0 & \sqrt{p^{\mu }\widetilde{\sigma }_{\mu }}
\end{array}
\right)
\end{array}
\end{equation}
\begin{equation}
v(p,s_{3})=Cu^{*}(p,s_{3})\,,\,\,\,\,\,\,\,\,C=i\gamma _{2}
\end{equation}
It is easy to check that $u$ and $v$ intertwine the $s=\frac{1}{2}$ Wigner
representations\thinspace $D^{\left( \frac{1}{2}\right) }(R)$ resp. $%
D^{\left( \frac{1}{2}\right) *}(R)$ with $D^{\left[ \frac{1}{2},0\right]
}\oplus D^{\left[ 0,\frac{1}{2}\right] }$ which is implemented by the
matrices $S$($\Lambda )$ . The so-defined $v$ fulfills: 
\begin{equation}
\left( -p^{\mu }\gamma _{\mu }-m\right) v=0;\,\,\,\,\overline{u}u=2m,\,\,\,%
\overline{v}v=-2m,\,\,\,\,\overline{u}v=0
\end{equation}

It is an interesting historical side remark that Dirac found his equation in
a more formalistic way. In order to overcome what he considered as a serious
shortcoming of the scalar Klein Gordon equation, Dirac searched for a first
order matrix differential operator which is a kind of square root of the
K-G. operator, i.e. $\left( i\partial ^{\mu }\gamma _{\mu }-m\right) \left(
-i\partial ^{\mu }\gamma _{\mu }-m\right) =\partial ^{\mu }\partial _{\mu
}+m^{2}$. The necessary and sufficient condition are the Clifford algebra
conditions: 
\begin{equation}
\left\{ \gamma _{\mu },\gamma _{\nu }\right\} =2g_{\mu \nu }
\end{equation}
There is only one finite-dimensional irreducible representation, it has
dimension equal to 4 (the Clifford algebra for a $2n$-dim. space has a $%
2^{n} $-dimensional irreducible representation). Our group-theoretical
approach has provided us with the so called chiral representation in which $%
\gamma _{5}$ is diagonal and which for m$\rightarrow 0$ decomposes naturally
into the two Weyl equations: 
\begin{equation}
p^{\mu }\widetilde{\sigma }_{\mu }\Phi =0,\,\,\,\,p^{\mu }\sigma _{\mu }\chi
^{\cdot }=0
\end{equation}
There are many equivalent representations which are useful for other
purposes. We will mention two of them. There is the representation used
first by Dirac: 
\begin{equation}
\gamma _{0}=\left( 
\begin{array}{cc}
1 & 0 \\ 
0 & -1
\end{array}
\right) ,\gamma _{i}=\left( 
\begin{array}{cc}
0 & \sigma _{i} \\ 
-\sigma _{i} & 0
\end{array}
\right)
\end{equation}

This representation is useful in calculations involving the nonrelativistic
limit as in the hydrogen-problem.\thinspace On the other hand for the field
theoretic application to selfconjugate $s=\frac{1}{2}$ particles and fields
the following Majorana representation is useful (with purely real $i\gamma
_{\mu }$ i.e. a real Dirac operator). 
\begin{equation}
\begin{array}{cc}
\gamma _{0}=\left( 
\begin{array}{cc}
0 & \sigma _{2} \\ 
\sigma _{2} & 0
\end{array}
\right) , & \gamma _{1}=\left( 
\begin{array}{cc}
i\sigma _{3} & 0 \\ 
0 & -i\sigma _{3}
\end{array}
\right) , \\ 
\gamma _{2}=\left( 
\begin{array}{cc}
0 & -\sigma _{2} \\ 
\sigma _{2} & 0
\end{array}
\right) , & \gamma _{3}=\left( 
\begin{array}{cc}
-i\sigma _{1} & 0 \\ 
0 & i\sigma _{1}
\end{array}
\right) ,
\end{array}
\end{equation}

\begin{itemize}
\item  \textbf{s=1 \thinspace \thinspace \thinspace }
\end{itemize}

In this case there are several low dimensional covariant intertwining
possibilities: 
\begin{equation}
D^{\left( 1\right) }(R)\rightarrow \left\{ 
\begin{array}{c}
D^{\left[ 1,0\right] } \\ 
D^{\left[ 0,1\right] } \\ 
D^{\left[ \frac{1}{2},\frac{1}{2}\right] }
\end{array}
\right.
\end{equation}
the first two have three components and the last is the 4-component vector
description which, if restricted to the rotation group decomposes as
follows: 
\begin{equation}
D^{\left[ \frac{1}{2},\frac{1}{2}\right] }(R)\cong D^{\left( 1\right)
}(R)\oplus D^{\left( 0\right) }(R)
\end{equation}
For an explicit description we apply the boost to the three spatial
coordinate vectors $e_{1},e_{2},e_{3}:$%
\begin{equation}
(e_{i}(p))_{\mu }\equiv e_{\mu }(p,i):=L_{\mu }^{\nu }(p)e_{\nu
}(0,i)\,\,\,\,\,\,\,\,\,e(0,i):=e_{i}
\end{equation}
Remembering the definition of the Wigner rotation, the transformation law is
(suppressing the vector indices): 
\begin{equation}
\Lambda e_{i}(p)=L(\Lambda p)R(\Lambda ,p)e_{i}=\sum_{i^{^{\prime
}}}L(\Lambda p)e_{i^{^{\prime }}}R_{i^{^{\prime }}i}(\Lambda
,p)=\sum_{i^{^{\prime }}}e_{i^{^{\prime }}}(\Lambda p)R_{i^{^{\prime
}}i}(\Lambda ,p)
\end{equation}
The covariant vector like wave functions are then: 
\begin{equation}
\tilde{V}_{\mu }(p)=\sum_{i}e_{\mu }(p,i)\tilde{\varphi}(p,i),\,\,\,\,\,V_{%
\mu }(x)=\frac{1}{\left( 2\pi \right) ^{\frac{3}{2}}}\int
e^{-ipx}\sum_{i}e_{\mu }(p,i)\tilde{\varphi}(p,i)\frac{d^{3}p}{2\omega }
\end{equation}
As always ,they fulfill the Klein-Gordon equation but, as a result of the
transversality $p^{\mu }e_{\mu }(p,i)=0$ which expresses the absence of the
scalar component $D^{\left( 0\right) },$ they also are divergenceless: 
\begin{equation}
\left( \partial ^{\kappa }\partial _{\kappa }+m^{2}\right) V_{\mu
}=0,\,\,\,\,\partial ^{\mu }V_{\mu }=0
\end{equation}
Both equations can be combined into a so called Proca-Wentzel equation: 
\begin{equation}
\left( \partial ^{\kappa }\partial _{\kappa }+m^{2}\right) V_{\mu }-\partial
_{\mu }\partial ^{\kappa }V_{\kappa }=0
\end{equation}
The covariance of this equation incorporates the transformation properties
of the field (just like for the Dirac equation) and is the Euler-Lagrange
equation of the Proca-Wentzel Lagrangian. Although Euler-Lagrange fields
exist for any spin (e.g. for s=$\frac{3}{2}$ the Rarita-Schwinger
equations), the Wigner approach, in contrast to the canonical or functional
integral approach, does not provide a preferential status to Lagrangian
fields.

From the definition one reads off the completeness relation: 
\begin{equation}
\sum_{i=1}^{3}e_{\mu }(p,i)e_{\nu }(p,i)=-g_{\mu \nu }+\frac{p_{\mu }p_{\nu }%
}{m^{2}}
\end{equation}
A limit $m\rightarrow 0$ does not exist i.e. there is no possibility to
intertwine the $\left[ m=0,s=1\right] $ Wigner representation with $%
D^{\left[ \frac{1}{2},\frac{1}{2}\right] }.$ The Maxwell description in
terms of field strength $F_{\mu \nu }$ corresponds to $D^{\left[ 1,0\right]
} $ or $D^{\left[ 0,1\right] }.$ This restriction together with the demand
that vector potentials are indispensable for describing the long-range
electromagnetic interaction in the context of quantum theory (in classical
physics vector potentials can be avoided) forces one to look for a
compromise slightly outside the Wigner scheme which will be presented in the
sequel.

\begin{remark}
The chosen covariant representations for s=0,$\frac{1}{2},1$ are
``Eulerian''. This means that they obey (multicomponent) spacetime
differential equations such that the Poincar\'{e} transformation properties
follow from the covariance of the (matrix-valued) differential operator. For
the Dirac operator this argument is well-known and can be found in any QFT
textbook. The matrices which appear in the infinitesimal
Lorentz-transformation are products of the matrices in the Euler equation..
In fact the above equations are even ``Euler-Lagrange'' because they are the
Euler equations of an action principle associated with a (free) quadratic
Lagrangian. In fact the Wigner theory always admits (precisely) one such
covariant description; for e.g. (m,s=$\frac{3}{2})$ it is the
Rarita-Schwinger equation. In this case the Wigner wave functions for (m,s)
particles and anti-particles correspond to the totality of all (positive and
negative frequency) solutions of this E-L equation. This also prevails to
the massless case, e.g. for the case below of (m=0,h=1) we obtain the
Maxwell equations. However since in the present approach we are not going to
use quantization methods, the existence of such classically preferred
covariant fields will be of no special interest to us. All covariant
representations of (m,s) are equally well suited to serve as local field
coordinates for the same local net of algebras; the intrinsic carriers of
the physical properties are not these field coordinates but the unique local
(free) (m,s)-net of algebras which each of them generates. This is not only
the correct underlying philosophy of our nonperturbative approach in chapter
6 but it also is behind our on shell (Bogoliubov-Shirkov-Epstein-Glaser)
causal perturbation theory which completely avoids off shell Lagrangians,
actions and functional integrals and only uses on-shell invariant
interaction operators $W$ (special elements of the free field Borchers
equivalence class).
\end{remark}

This point is the guiding thread through these notes and we will come back
to it and elaborate it on many occasions. It will become clear that whereas
the various quantization approaches start to enter QFT on the commutative
(quasi)classical side and develop their main strength in deforming around
free solutions, the more powerful nonperturbative method of chapter 6 starts
from modular theory which is extremely noncommutative in the sense that it
cannot even be formulated (and has no counterpart) in the context of
commutative algebraic structures, i.e. it is based on one of the rare
properties of LQP which are pure quantum and have no quasiclassical limit.

\begin{itemize}
\item  \textbf{s=1 m=0}
\end{itemize}

In order to obtain a formalism similar to the previous case of vector
mesons,one extends the two ``polarization vectors'' e$_{i}$ i=1,2 in x-and
y-direction by two orthogonal light-like vectors: 
\begin{equation}
e_{\pm }^{\mu }=e_{0}^{\mu }\pm e_{3}^{\mu }=\left( 1,0,0,\pm 1\right)
\end{equation}
We choose e$_{+}$ as the reference vector $k_{R}$ from which to start the
boost $L(k,k_{R})$. The latter consists of a rotation of the z-axis into the
momentum direction $\vec{n}=\frac{\vec{k}}{\omega }$ (fixed uniquely by the
standard prescription in terms of two Euler angles) and a subsequent L-boost
along this direction: 
\begin{equation}
\left( 
\begin{array}{c}
1 \\ 
\vec{n}
\end{array}
\right) \rightarrow k=\omega \left( 
\begin{array}{c}
1 \\ 
\vec{n}
\end{array}
\right)
\end{equation}
The decomposition of a general Lorentz transformation $\Lambda $ in terms of
its little group component $H(\Lambda ,k)$: 
\begin{equation}
\Lambda =L(\Lambda k,k_{R})H(\Lambda
,k)L^{-1}(k,k_{R}),\,\,\,\,\,\,\,\,\,H(\Lambda ,k)\in \tilde{E}(2)
\end{equation}
the twofold covering of the euclidean group in two dimensions which, as
explained before, is generated by two translations $\alpha ,\beta $ and one
rotation $\theta $ , where all euclidean parameters are functions of $%
\Lambda $ and k which can be computed from the previous formula. One defines
two transversal polarization vectors: 
\begin{equation}
\epsilon (k,\lambda )=L(k,k_{r})\left\{ 
\begin{array}{c}
\frac{1}{\sqrt{2}}\left( e_{1}+ie_{2}\right) ,\,\,\,\,\,\,\lambda =+ \\ 
\frac{1}{\sqrt{2}}\left( -e_{1}+ie_{2}\right) ,\,\,\,\,\lambda =-
\end{array}
\right.
\end{equation}
They are used as intertwiners in the attempt to define a vectorial wave
function: 
\begin{equation}
\tilde{A}_{\mu }(k)=k_{\mu }\tilde{\Phi}(k)+\sum_{\lambda =\pm }\epsilon
_{\mu }(k,\lambda )\tilde{\varphi}(k,\lambda )  \label{eich}
\end{equation}
Here the longitudinal first component is not determined by the Wigner
theory. We cannot consistently set it equal to zero, since the intertwiners
generate such an additive term under the action of the $\tilde{E}(2)$
``translations'': 
\begin{equation}
G(\rho )\epsilon (k_{R},\lambda )=\epsilon (k_{R},\lambda )+\left\{ 
\begin{array}{c}
-\frac{1}{2}\left( \bar{\rho},0,0,\bar{\rho}\right) ,\,\,\,\,\,\,\lambda =+
\\ 
+\frac{1}{2}\left( \rho ,0,0,\rho \right) ,\,\,\,\,\,\,\,\lambda =-
\end{array}
\right. \,\,\,\,\,\,\,\,\,\,\,\,\,\,\,\,\,\,\rho =\alpha +i\beta
\label{E(2)}
\end{equation}
whereas under $x$-$y$ rotations the $\epsilon $ picks up the standard Wigner
phase factor. The polarization vectors do not behave as 4-vectors since they
are not invariant under the euclidean translations in $\tilde{E}(2)$, as one
would have expected for a (nonexisting!) bona fide intertwiner from the $%
\left[ 0,h=1\right] $ Wigner representation to the $D^{\left[ \frac{1}{2},%
\frac{1}{2}\right] }$ covariant representation. Rather the intertwiner only
has L-covariance up to additive gauge transformations i.e. up to affine
longitudinal terms (for more details see section 6): 
\begin{equation}
(U(\Lambda )A)_{\mu }(k)=\Lambda _{\mu }^{\nu }A_{\nu }(\Lambda
^{-1}k)+k_{\mu }G(k)  \label{affine}
\end{equation}
This peculiar manifestation of the $\left( 0,h=1\right) $ little group $%
\tilde{E}(2)$ is the cause for the appearance of the local gauge issue in
local quantum physics. Unfortunately this quantum origin is somewhat hidden
in the quantization approach, where it remains invisible behind the
geometrical interpretation in terms of fibre bundles.

In the covariant quantization approach, contrary to the Wigner theory, the
gauge aspect becomes completely decoupled from the L-transformations. This
close relation to classical fibre bundles is only obtained at the expense of
leaving the realm of quantum physics and entering the world of ``ghosts'',
i.e. mathematical tricks which allow to maintain the benefits of the
standard formalism of pointlike fields in a situation which physically is
already outside that formalism. In fact it is my contention, that this is
one of the potential points of possible fruitful clashes between the
requirements of (classical) geometry and local quantum physics. In the
presence of electromagnetic interactions this problem will appear again in
our perturbative treatment of QED. Although we will try to sharpen this
apparent clash between the Wigner and the quantization framework in the
following section, the solution of the associated conceptual problem is not
yet known. But at least one has a recipe which works in perturbation theory.

Whereas the covariantization of the canonical Wigner $(m=0,h=1)$
representation can be done in terms of covariant field strength, the
requirement that the scalar product be expressible in terms of a local
tensorial formula necessitates the introduction of the above vector
potential $A_{\mu }.$ The Lorentz (gauge) invariant inner product for the $%
\tilde{A}_{\mu }$ is now only positive semidefinite on individual $A_{\mu }$
(but positive definite on gauge classes): 
\begin{eqnarray}
\left( A,A^{^{\prime }}\right) &=&-\int \tilde{A}_{\mu }^{*}\tilde{A}%
^{^{\prime }\mu }\frac{d^{3}p}{2\left| \vec{p}\right| }=\sum_{\pm }\int 
\tilde{\varphi}^{*}\tilde{\varphi}^{\prime }\frac{d^{3}p}{2\left| \vec{p}%
\right| }  \label{semidef} \\
p^{\mu }\tilde{A}_{\mu } &=&0  \nonumber
\end{eqnarray}
The connection of this space $H_{A}$ with the positive definite Wigner $%
H_{W} $ space of the $\tilde{\varphi}$-wave functions is give by factoring
out the null-space $H_{0}$: 
\begin{equation}
H_{W}=\frac{H_{A}}{H_{0}}
\end{equation}

This factor space representation of the Wigner space has however one
drawback: the subspaces have no natural behavior under multiparticle
tensoring; the well known impossibility to use a local condition $\partial
^{\mu }A_{\mu }\Phi =0$ for the characterization of physical equivalence of
vectors in the Gupta-Bleuler treatment (and the necessity to use matrix
elements) is a consequence. It turns out that a BRS-like cohomological
extension based on a nilpotent operator yields a more natural relation of
the Wigner space to the multiparticle tensor spaces. However the strongest
arguments in favor of such a cohomological approach comes from the
perturbative approach to renormalizable interactions of massive spin 1
particles. It turns out that without this cohomological trick there is no
renormalizable solution at all. Although by its use the problem becomes
soluble, the solution is extremely restrictive and essentially unique. This
is in contrast to the classical situation, where the increase of spin
(components) leads to an increasing number of interaction terms. The
classical restriction principle in the case of massless spin 1 fields is the
gauge principle. In our approach in the next chapter it is not imposed on
the QFT by quantization, but rather this (quasi)classical restriction is
obtained from the more fundamental (perturbative) QFT which obeys the
aforementioned severe consistency restrictions characteristic for higher
spin. This Bohr correspondence principle point of view is opposite to the
quantization point of view of gauge theory. It leads to significant
revisions about the Schwinger-Higgs mechanism and other concepts of
perturbative gauge theories.

As pointed out by Weinberg\cite{Wei}, this gauge aspect is common to all $%
\left[ 0,h=\frac{n}{2}\right] $ representations for $\frac{n}{2}\geq 1$.
There are simply no intertwiners from this Wigner representation to $%
D^{\left[ A,A\right] }$symmetric tensors, rather the possibility of
intertwining is restricted to $D^{\left[ A,B\right] }$ with $\left|
A-B\right| =h$ (h the Wigner helicity). The covariant vector potentials for $%
s=1$ and the covariant symmetric tensor $g_{\mu \nu }$ for $s=2$ of the
classical general relativity can at the quantum level only be introduced at
the prize of nonlocal (noncompactly localizable) $\psi ^{\left[ A,A\right] }$
fields or by the cohomological trick.

The case of general $\left[ m,s\right] $ intertwiners $u$ is a routine
exercise in Clebsch-Gordan gymnastics. One uses the intertwining relation
for $u$: 
\begin{equation}
u(p)D^{\left( s\right) }(R(\Lambda ,p))=D^{\left[ A,B\right] }(\Lambda
)u(\Lambda ^{-1}p)
\end{equation}
for the calculation of the $u$'s. Here we found it convenient to interpret
the intertwiner $u$ as a rectangular matrix with $2s+1$ columns and $\left(
2A+1\right) \left( 2B+1\right) $ rows. The first step consists in analyzing
this equation for $p=p_{R}$ (Weinberg) with the result that the $u(\vec{0})$
is proportional to the Clebsch-Gordan coefficients: 
\begin{equation}
u(\vec{0})\sim C_{AB}(s,s_{3};a,b)
\end{equation}
The second step consists in an application of a boost: 
\begin{equation}
u(p)=D^{\left[ A,B\right] }(L(p))u(0)
\end{equation}
For details we refer to Weinberg \cite{Wei}.

{\small The mathematical method behind Wigner's representation theory,
together with the Frobenius theory of induced representations of finite
groups, was extended by G. Mackey into the general theory of induced
representations which in turn is a special case of a theory of inductions
and reductions of (von Neumann) algebras by M. Rieffel and others (see
mathematical appendix).}

{\small For the case at hand, the starting point is the semi-direct product
of two locally compact groups of which one is abelian and denoted by T and
the other say K acts as automorphisms on T: } 
\[
G\equiv K\times _{\alpha }T:(k_{2},t_{2})\cdot
(k_{1},t_{1})=(k_{2}k_{1},t_{2}+\alpha _{k_{2}}(t_{1})) 
\]

{\small The action of K on the dual group \~{T}, which is defined as:} 
\[
(k\circ \tilde{t})(t)\equiv \tilde{t}(\alpha _{k^{-1}}(t)) 
\]
{\small foliates the \~{T} into transitive K-orbits }$\mathcal{O}${\small .
Each point \~{t} on one }$\mathcal{O}${\small \ defines a stability group:} 
\[
K_{\tilde{t}}\equiv \left\{ k\in K\mid k\circ \tilde{t}=\tilde{t}\right\} 
\]
{\small and apart from possibly singular points, these ``little groups'' for
different \~{t} are all inner equivalent (i.e. inside }$G).${\small \ Often
one finds a geometrically preferred reference point \~{t}}$_{0}${\small \
(e.g. the restframe momentum on the mass hyperboloid). Let }$\pi ${\small \
be an irreducible representation of }$K_{\tilde{t}}${\small \ and }$U_{\pi
}(k,\tilde{t})${\small \ the associated }$(K,O,U(H_{\pi }))-${\small cocycle
i.e.:} 
\begin{eqnarray*}
U_{\pi }(k_{2},k_{1}\circ \tilde{t})U_{\pi }(k_{1},\tilde{t}) &=&U_{\pi
}(k_{2}k_{1},\tilde{t}) \\
U_{\pi }(k,\tilde{t}_{0}) &=&\pi (k)
\end{eqnarray*}
{\small Introducing a Borel map }$L${\small \ : }$\mathcal{O}\rightarrow K$%
{\small \ (the family of boosts in Wigner's theory), one may take:} 
\[
U_{\pi }(k,\tilde{t})=\pi (L(k\circ \tilde{t})^{-1}kL(\tilde{t})) 
\]
{\small The Mackey induction ``machine'' associates to every pair (}$%
\mathcal{O},\pi )${\small \ an irreducible representation of }$G${\small \
on the Hilbert space }$H\equiv L^{2}(\mathcal{O},H_{\pi }):$%
\[
(V(k,t)\psi )(\tilde{t})=\tilde{t}(t)U_{\pi }(k,k^{-1}\circ \tilde{t})\psi
(k^{-1}\circ \tilde{t}) 
\]
{\small Here we assumed that there exists an }$K${\small -invariant measure
on }$\mathcal{O};${\small \ for a quasi-invariant measure the right hand
side has to be corrected by a Radon Nykodym derivative. For more details I
refer to \cite{Gri}}

\section{Wigner Theory and Free Fields}

We now use the Wigner representation theory in order to construct fields in
bosonic or fermionic Fock spaces. The creation operators in momentum space
should transform in the same way as the one particle states since their
application to the vacuum vector creates the latter. 
\begin{equation}
U(\Lambda )a^{*}(p,m)U^{*}(\Lambda )=\sum_{m^{^{\prime }}}a^{*}(\Lambda
p,m^{^{\prime }})D_{m^{^{\prime }}m}^{\left( s\right) }(R(\Lambda ,p))
\end{equation}

For computational convenience we identify the Wigner rotation with its
unimodular matrix representation: 
\begin{equation}
R(\Lambda ,p)\longrightarrow \left( \sqrt{\frac{1}{m}\left( \Lambda p\right)
^{\mu }\sigma _{\mu }}\right) ^{-1}\alpha (\Lambda )\sqrt{\frac{1}{m}p^{\mu
}\sigma _{\mu }}
\end{equation}
The corresponding relation for the annihilation operator contains the
complex conjugate matrix $D^{*}$ which is equivalent to $D$: 
\begin{equation}
D(i\sigma _{2})D^{*}(R)D(-i\sigma _{2})=D(R)
\end{equation}
If the particles are charged, there are also operators $b^{\#}(p,m)$ which
describe annihilation and creation of antiparticles with the same mass and
spin and hence the same transformation property as $a^{\#}(p,m)$. In order
to obtain covariant operators one uses the intertwiners u and v introduced
in the previous section. Interpreting these intertwiners as p-dependent
rectangular matrices of size $N\times (2s+1)$ with $N$= dimension of the
representation space on which the matrices$\;D^{\left[ A,B\right] }$($%
\Lambda $) act, we have:

\begin{equation}
\begin{array}{c}
D^{\left[ A,B\right] }(\Lambda )u(\Lambda ^{-1}p)=u(p)D^{\left( s\right)
}(R^{-1}(\Lambda ,p)) \\ 
D^{\left[ A,B\right] }(\Lambda )v(\Lambda ^{-1}p)=v(p)D^{\left( s\right)
*}(R^{-1}(\Lambda ,p)),\quad v(p)=u(p)D^{\left( s\right) *}(i\sigma _{2})
\end{array}
\,\,\,\,\,\,
\end{equation}

Therefore we find the following covariant creation and annihilation
operators: 
\begin{equation}
\begin{array}{c}
A^{\left( *\right) }(p)=\sum_{m}v(p,m)a^{*}(p,m) \\ 
B^{\left( *\right) }(p,)=\sum_{m}v(p,m)b^{*}(p,m) \\ 
A(p)=\sum_{m}u(p,m)a(p,m) \\ 
B(p)=\sum_{m}u(p,m)b(p,m)
\end{array}
\end{equation}
We have added a bracket to the * in order to indicate that the covariant
creation operator is not exactly the hermitian adjoint of the covariant
annihilator and we used $m$ (magnetic quantum number) instead of the
cumbersome $s_{3}$ notation.. The Fourier transform preserves covariance: 
\begin{equation}
\begin{array}{c}
\psi _{A}^{\left( -\right) }(x):=\frac{1}{(2\pi )^{\frac{3}{2}}}\int
e^{-ipx}A(p)\frac{d^{3}p}{2\omega }\,\,\,,\,\,\,\,\,\,\,\,\,\psi
_{A}^{\left( +\right) }(x):=\frac{1}{(2\pi )^{\frac{3}{2}}}\int
e^{ipx}A^{*}(p)\frac{d^{3}p}{2\omega } \\ 
\psi _{B}^{\left( -\right) }(x):=\frac{1}{(2\pi )^{\frac{3}{2}}}\int
e^{-ipx}B(p)\frac{d^{3}p}{2\omega },\,\,\,\;\;\;\psi _{B}^{\left( +\right)
}(x):=\frac{1}{(2\pi )^{\frac{3}{2}}}\int e^{ipx}B^{*}(p)\frac{d^{3}p}{%
2\omega }\quad
\end{array}
\end{equation}
obey the covariant transformation law: 
\begin{equation}
U(\Lambda )\psi (x)U^{*}(\Lambda )=D^{\left[ A,B\right] }(\Lambda )^{-1}\psi
(\Lambda x)
\end{equation}
We want to construct local covariant fields i.e. covariant fields which
(anti)commute for spacelike distances. The physical motivation is Einstein
causality for local observables. Prominent local observables associated with
charged fields are e.g. currents. Since they are typically second or higher
even degree polynomials in the fields, the (anti)commutation of the fields
is sufficient for the Einstein causality (spacelike commutativity) of the
local observables. Fields which are themselves observables as e.g. the
Maxwell field, must obey spacelike commutation relations.

It is well known that support properties in momentum space as the
restriction to the forward light cone prevent support properties of
(anti)commutators in x-space. The former give rise to analytic properties of
the latter. The standard example is the Fourier transform of a function with
support in the positive half-axis which is the boundary value of a function
analytic in the upper half-plane. According to the Schwarz reflection
principle, such function cannot vanish in a dense real subset without
vanishing identically. The above Fourier transforms are multidimensional
counterparts in which the halfline is replaced by the forward light cone and
the upper half plane by a tube $z_{\mu }=x_{\mu }+iy_{\mu }$ with y in the
dual cone i.e. the backward light cone. We therefore make the following
Ansatz for local fields: 
\begin{equation}
\begin{array}{c}
\psi _{A}(x)=\psi _{A}^{\left( -\right) }(x)+\psi _{A}^{\left( +\right)
}(x),\quad \psi _{B}(x)=\psi _{B}^{\left( -\right) }(x)+\psi _{B}^{\left(
+\right) }(x) \\ 
\psi (x)=\psi _{A}^{\left( -\right) }(x)+\psi _{B}^{\left( +\right) }(x)
\end{array}
\end{equation}
Complex coefficients in this linear combination bring no gain in generality,
since they can be absorbed into redefinitions. The following calculations
show that all these combinations between different frequency parts are local
covariant fields. The first two combinations are only physically useful if A
and B would be (accidentally equal mass and spin) selfdual particles. If on
the other hand, there is a charge superselection rule between A and B i.e. B
is the antiparticle of A, then we are forced to take the $\psi $ combination
because otherwise we would not be able to form local (Einstein-causal)
neutral observables. In this sense causality and the superselection
principle require the existence of anticharged particles of the same (m,s).

Returning to our notation for indices of irreducible finite dimensional
representations for the Lorentz-group, we find the following relation
between the spin and the spacelike (anti)commutativity:

\begin{theorem}
( Spin-statistics for free fields) : 
\begin{equation}
\left[ \psi _{a,b}^{\left[ A,B\right] }(x),\psi _{a^{^{\prime }},b^{^{\prime
}}}^{\left[ A,B\right] }(y)\right] _{\pm
}=0\,\,\,\,\,\,\,\,for\,\,\,\,\,(\,x-y)^{2}<0
\end{equation}
\end{theorem}

where the +sign i.e. the anticommutator is to be taken for A+B=halfinteger.

The proof consists in calculating the vacuum expectation value of the
product of two fields in the two different orders. Each two-point function
is the Fourier transform of a sesquilinear expression in the $u$ resp. $v$
intertwiners, e.g. 
\begin{equation}
\left\langle \psi _{a,b}^{\left[ A,B\right] }(x),\psi _{a^{\prime
},b^{\prime }}^{\left[ A,B\right] *}(y)\right\rangle _{0}=\frac{1}{(2\pi
)^{3}}\int e^{-ip(x-y)}\sum_{m}u_{ab}^{\left[ A,B\right] }(p,m)\overline{%
u_{a^{\prime }b^{\prime }}^{\left[ A,B\right] }}(p,m)\frac{d^{3}p}{2\omega }
\end{equation}
The computation of the $m$-sum in the integrand is (as the computation of
the intertwiners) a purely group theoretic problem and it yields a ($%
2A+1)\cdot (2B+1)\times (2A+1)(2B+1)$ matrix with $P_{a,b;\,a^{\prime
},b^{\prime }}(p)$ covariant $p$-polynomial entries. Therefore in x-space we
may write the correlation function in terms of the scalar two-point function 
$i\Delta ^{(+)}(\xi )$ (studied in detail below): 
\begin{equation}
\left\langle \psi _{a,b}^{\left[ A,B\right] }(x),\psi _{a^{\prime
},b^{\prime }}^{\left[ A,B\right] *}(y)\right\rangle _{0}=P_{a,b;\,a^{\prime
},b^{\prime }}(i\partial _{x})i\Delta ^{(+)}(x-y)\,  \label{twopoint}
\end{equation}
The polynomial matrix is even/odd under the transposition of matrix indices
together with $x\leftrightarrow y$ for $A+B$ =integer/halfinteger. Since the
scalar function $i\Delta ^{(+)}(\xi )$ is symmetric in $\xi $ for spacelike $%
\xi $, only the difference (commutator)/sum (anticommutator) vanishes for
spacelike distances, i.e integer/halfinteger $\leftrightarrow $
commutator/anticommutator. We mention that the commutation relation between
two $\psi ^{\prime }s$ must follow the same $\pm $ rule as the above $\psi
-\psi ^{*}.$ Since the two-point function vanishes for two fields with the
same charge, the proof requires the use of the 4-point function and will
only be mentioned in the case of interacting fields.

It is helpful to illustrate this spin-statistics connection with free fields
for $s=0$,$\frac{1}{2},1$ .

\begin{itemize}
\item  \textbf{s=0}\thinspace \thinspace \thinspace \thinspace \thinspace
\thinspace
\end{itemize}

\thinspace \thinspace \thinspace With\thinspace \thinspace \thinspace
\thinspace \thinspace \thinspace \thinspace \thinspace \thinspace $\psi (x)=%
\frac{1}{(2\pi )^{\frac{3}{2}}}\int \left(
e^{-ipx}A(p)+e^{ipx}B^{*}(p)\right) \frac{d^{3}p}{2\omega }$ \thinspace
\thinspace \thinspace \thinspace we obtain: 
\begin{equation}
\left[ \psi (x),\psi (y)\right] _{\pm }=\frac{1}{(2\pi )^{\frac{3}{2}}}\int
\left( e^{-ip(x-y)}\pm e^{ip(x-y)}\right) \frac{d^{3}p}{2\omega }=i\Delta
^{\left( +\right) }(x-y)\pm i\Delta ^{\left( -\right) }(x-y)
\end{equation}
Here $i\Delta ^{\left( -\right) }(\xi ):=i\Delta ^{\left( +\right) }(-\xi )$
\thinspace \thinspace and the momentum space integrals may be expressed in
terms of Hankel functions. One first uses the fact that $i\Delta ^{\left(
+\right) }(\xi )$ is analytic in the tube $\xi \rightarrow \zeta =\xi -i\eta 
$ with $\eta \in \bar{V}^{+}$, the closed forward light cone as a result of
the spectrum property $p\in V^{+}$. This means that the euclidean vector ($%
\xi _{4}=i\xi _{0}$,$\vec{\xi})$ is in the analyticity region at least if $%
\xi _{0}\geq 0.$ This analytic continuation is part of the so called
Wick-rotation. In this euclidean domain one now rewrites the integral for $%
i\Delta ^{\left( +\right) \,\,}\,$ in terms of an euclidean contour
integral: 
\begin{equation}
i\Delta ^{\left( +\right) }(x)=\frac{1}{(2\pi )^{4}}\int_{C}e^{ipx}\frac{%
d^{4}p}{p^{2}+m^{2}}
\end{equation}
The contour C in the complex p$_{0}-$plane is the imaginary $\xi _{0}$-axis
or the new (Wick-rotated) $\xi _{4}=i\xi _{0}+\epsilon $ axis. The proof of
this claim follows simply by closing the contour by an infinitely large
half-circle in the upper half plane on which the integrand vanishes
sufficiently and the subsequent application of the residuum theorem to the
pole at $p_{4}=$i$\sqrt{\vec{p}^{2}+m^{2}}$ . Since the Minkowski metric has
disappeared and there is no restriction on the Wick-rotated $\xi _{4}$(the
euclidean representation achieved an analytic continuation to all real $\xi
_{4}$), the remaining task is to perform a euclidean Fourier-integral with a
rotational invariant rational integrand. The d-dimensional integration in
polar coordinates requires the same amount of work as d=4. 
\begin{equation}
\frac{1}{\left( 2\pi \right) ^{d}}\int e^{ip\xi }\frac{d^{d}p}{p^{2}+m^{2}}=%
\frac{1}{\left( 2\pi \right) ^{d}}\frac{2\left( \sqrt{\pi }\right) ^{d-1}}{%
\Gamma (\frac{d-1}{2})}\int_{0}^{\infty }\frac{p^{d-1}}{p^{2}+m^{2}}%
dp\int_{0}^{\pi }\sin {}^{d-2}\theta d\theta e^{ipr\cos \theta }
\end{equation}
\begin{equation}
=\frac{2\left( \sqrt{\pi }\right) ^{d}}{\left( 2\pi \right) ^{d}}\int \frac{%
2^{\frac{d}{2}-1}}{\left( rp\right) ^{\frac{d}{2}-1}}\frac{1}{p^{2}+m^{2}}J_{%
\frac{d}{2}-1}(pr)p^{d-1}dp
\end{equation}
or the Bessel functions $J$ as well as a formula linking the Hankel function
of the first type to an integral over a Bessel function. The Hankel function 
$H_{\nu }(z)$ is analytic in the cut z-plane with a cut running from $%
-\infty $ to zero i.e. $K(z)$ has a cut for $z^{2}\leq 0.$ Specializing to d
=4, we obtain the following representation of the free field two-point
function as a boundary value of an analytic function: 
\begin{equation}
i\Delta ^{\left( +\right) }(\xi )=\lim_{\epsilon \rightarrow 0}\frac{1}{4\pi
^{2}}\frac{m}{\sqrt{-\left( \xi _{0}-i\epsilon \right) ^{2}+\vec{\xi}^{2}}}%
K_{1}(m\sqrt{-\left( \xi _{0}-i\epsilon \right) ^{2}+\vec{\xi}^{2}})
\end{equation}
As expected, the space-like points are (together with the euclidean points)
in the analytic domain. The distributional boundary value prescription
becomes important only in the time-and light-like region where the
transcription of the Hankel $H$ or Kelvin $K$ function in terms of $J$- and $%
N$-functions and the subsequent performance of the $\epsilon $-limit gives
the physical boundary values in terms of Schwartz distributions: 
\begin{equation}
i\Delta ^{\left( +\right) }(\xi )=\frac{1}{4\pi }\epsilon (\xi _{0})\delta
(\xi ^{2})+\frac{m}{\sqrt{-\xi ^{2}}}\frac{i}{8\pi }\left( J_{1}(m\sqrt{\xi
^{2}})\epsilon (\xi _{0})+iN_{1}(m\sqrt{\xi ^{2}})\right) ,\,\,\xi ^{2}\geq 0
\end{equation}
The strength of singularity on the light cone (determined by the
singularities of $K_{1}$ or $N_{1}$) is independent of the mass and given by
the zero mass two-point correlation function: 
\begin{equation}
iD^{\left( +\right) }(\xi )=\lim_{\epsilon \rightarrow 0}\frac{1}{4\pi ^{2}}%
\frac{1}{-\left( \xi _{o}-i\epsilon \right) ^{2}+\vec{\xi}^{2}}
\end{equation}
The particular $\varepsilon $-prescription is not only the consequence but
even equivalent to the positive energy property of its Fourier transform, a
fact which is often tends to be overlooked. The strength of the x-space
short distance behavior is independent of the state, e.g. the two-point
function in the ground state and that in any other vector or density matrix
state (e.g. in a temperature state, as considered later) on the same $^{*}$%
-algebra have the same leading light-cone behavior.. This also applies to
the generalization to curved space-time (chapter 4). The next-to leading
behavior (the log-term) in $K_{1}$ does however depend on the mass. The fact
that correlation functions have analyticity properties in space-like regions
is however very specific for the vacuum state; other states in the same
``folium'' of states have only spacelike smoothness but no analytic behavior
in their correlation functions. The dependence of the singularities on the
space-time dimension follows from the properties of the $K_{\nu }$
functions. It is conveniently encoded into the notion of ``operator
dimensions''of the fields e.g. one says e.g. that $dimA=1$ (in mass units)
for $d=3+1$ and $dimA=\frac{1}{2}$ for $d=2+1$ if the two-point function has
the leading singularity $\left( -\xi ^{2}\right) ^{-\dim A}$.

It turns out that the correlation functions of the higher spin free fields
can all be expressed in terms of $i\Delta ^{\left( +\right) }$ and its zero
mass limit $iD^{\left( +\right) }$ with matrix valued differential operators
in front. We again look at the important special cases $s=\frac{1}{2},1$
before we sketch properties of general free fields.

\begin{itemize}
\item  \textbf{s=$\frac{1}{2}$}
\end{itemize}

The ansatz for the positive and negative frequency parts for the local
spinor field (in analogy to the previous scalar field) is (using the
condensed notation from the beginning of this section for $A(p)$ and $B(p)$):

\begin{equation}
\psi (x)=\frac{1}{\left( 2\pi \right) ^{\frac{3}{2}}}\int \left(
e^{-ipx}A(p)+e^{ipx}B^{*}(p)\right) \frac{d^{3}p}{2\omega }
\end{equation}
For the (anti)commutators of the covariant creation and annihilation
operators one needs to know the completeness relations for the $u$-and $v$%
-spinors: 
\begin{eqnarray}
\sum_{s_{3}}u(p,s_{3})\bar{u}(p,s_{3}) &=&p^{\mu }\gamma _{\mu }+m=2m\Lambda
_{+}\,\,\,\,\,\, \\
\sum_{s_{3}}v(p,s_{3})\bar{v}(p,s_{3}) &=&p^{\mu }\gamma _{\mu
}-m=-2m\Lambda _{-}
\end{eqnarray}
where $\Lambda _{\pm }$ are projectors $\Lambda _{+}+\Lambda _{-}=1$ on the $%
\pm $ frequency subspaces in the 4-dim Dirac spinor space and the bar on the
4-spinors denotes the Dirac's conjugate (\ref{Dir}). With: 
\begin{eqnarray}
\left\langle \psi _{\alpha }(x)\overline{\psi _{\beta }}(y)\right\rangle &=&%
\frac{1}{\left( 2\pi \right) ^{3}}\int e^{-ip\left( x-y\right)
}\sum_{s_{3}}u_{\alpha }(p,s_{3})\bar{u}_{\beta }(p,s_{3})\frac{d^{3}p}{%
2\omega } \\
&=&\left( -i\partial _{x}^{\mu }\gamma _{\mu }+m\right) _{\alpha ,\beta
}i\Delta ^{\left( +\right) }(x-y)  \nonumber
\end{eqnarray}
\begin{eqnarray}
\left\langle \overline{\psi }_{\beta }(y)\psi _{\alpha }(x)\right\rangle &=&%
\frac{1}{\left( 2\pi \right) ^{3}}\int e^{ip\left( x-y\right) }\sum_{s_{3}}%
\overline{v}_{\beta }(p,s_{3})v_{\alpha }(p,s_{3})\frac{d^{3}p}{2\omega } \\
&=&-\left( i\partial _{y}^{\mu }\gamma _{\mu }+m\right) _{\alpha ,\beta
}i\Delta ^{\left( +\right) }(y-x)  \nonumber
\end{eqnarray}
we obtain with $\psi ^{\#}=\psi \,$or$\,\overline{\psi }$ 
\begin{eqnarray}
\left\{ \psi ^{\#}(x),\psi ^{\#}(y)\right\} &=&0\,\,\,if\,\,\,\,\left(
x-y\right) ^{2}<0\,\,\,\,\,\,\,\, \\
\left\{ \psi (x),\overline{\psi }(y)\right\} &=&\left( -i\partial _{x}^{\mu
}\gamma _{\mu }+m\right) i\Delta (x-y)
\end{eqnarray}
whereas the commutator is nonvanishing for spacelike distances. We get the
first glimpse at the spin-statistics connection.

The present construction of local $\psi ^{\prime }s$ also sheds some light
on the physical interpretation of the $v$-spinors in connection with the
charge conjugation symmetry. The latter transformation is defined in
Fock-space by : 
\begin{equation}
\mathcal{C}A(p,s_{3})\mathcal{C}^{*}=B(p,s_{3})
\end{equation}
Its action on the local fields is local and the transformation law involves
a matrix $C$ in Dirac space: 
\begin{equation}
\psi ^{C}:=\mathcal{C}\psi \mathcal{C}^{*}=C\psi ^{*}
\end{equation}

In the helicity representation used here, the matrix is $C=\gamma _{2}$ ,
whereas in the Majorana representation one finds $C=1$. This matrix
transforms the $u$-spinors into the $v$'s and vice versa and therefore is
the image of $D^{\left( s\right) }(i\sigma _{2})$ under the intertwining map
into the Dirac spinor space. It is an additional fringe benefit that via the
Dirac doubling all global Fock-space symmetries as $\mathcal{P}$, $T$ and $%
\mathcal{C}$ have local representations with constant matrices on Dirac
spinors. Furthermore the Dirac description goes over into the two decoupled
Weyl equation in the zero mass limit. Finally we notice that dim$\psi =\frac{%
3}{2}.$

\begin{itemize}
\item  \textbf{s=1,m$\neq 0$}
\end{itemize}

The massive (vector meson) case is straightforward. The local field is : 
\begin{equation}
V_{\mu }(x)=\frac{1}{\left( 2\pi \right) ^{\frac{3}{2}}}\int \sum_{i}e_{\mu
}(p,i)\left( e^{-ipx}a(p,i)+e^{ipx}b^{*}(p,i)\right) \frac{d^{3}p}{2\omega }
\end{equation}
Its 2-point function results from the completeness relation of the
polarization vectors : 
\begin{equation}
\left\langle V_{\mu }(x)V_{\nu }(y)\right\rangle =\left( -g_{\mu \nu }-\frac{%
\partial _{\mu }\partial _{\nu }}{m^{2}}\right) i\Delta ^{\left( +\right)
}(x-y)
\end{equation}
It is obvious that only the commutator can vanish for spacelike distances.
Different from the previous case,vector meson fields does not permit a zero
mass limit. Therefore we should not be surprised to meet some peculiarities
in the vectorial description of photons. Finally we have dim$V_{\mu }=2$ in
contrast to the expected classical dim$_{class}V_{\mu }=1$

\begin{itemize}
\item  \textbf{s=1,m=0}
\end{itemize}

Here we may use a physical description in terms of nonlocal (semiinfinite
string localized) vectorpotentials in the physical Fock space which under
the unitary Lorentz transformations suffer a affine transformation law (\ref
{affine}) which will be analyzed in more details in section 6 of this
chapter. In the following we briefly describe the traditional Gupta-Bleuler
photon formalism For a formally local\footnote{%
Local here means pointlike, i.e. fields which can be smeared with
unrestricted Schwartz test functions. Without the unphysical components, we
would have to restrict vector-valued test functions $f_{\mu }$ by demanding
transversality.} description in terms of vector fields, the longitudinal
part which the stability group transformation behavior of wave functions
demands (see previous section) is not enough; one also needs ``scalar
photons'': 
\begin{equation}
\tilde{A}_{\mu }(k)=e_{\mu }^{\left( +\right) }(k)c_{+}(k)+e_{\mu }^{\left(
-\right) }(k)c_{-}(k)+\sum_{\lambda =\pm }\epsilon _{\mu }(k,\lambda
)a(k,\lambda )
\end{equation}
\begin{equation}
A_{\mu }(x)=\frac{1}{\left( 2\pi \right) ^{\frac{3}{2}}}\int \left( e^{-ikx}%
\tilde{A}_{\mu }(k)+e^{ikx}\tilde{A}_{\mu }^{*}(k)\right) \frac{d^{3}k}{%
2\left| \vec{k}\right| }
\end{equation}
Here the $e^{\left( \pm \right) }$ are obtained by boosting the light-like
vectors $(1,0,0,\pm 1),$ i.e. $e^{(+)}$ is the old longitudinal part. We
obtain the covariant two point function : 
\begin{equation}
\left\langle A_{\mu }(x)A_{\nu }(y)\right\rangle =\frac{-g_{\mu \nu }}{%
\left( 2\pi \right) ^{3}}\int e^{-ik(x-y)}\frac{d^{3}k}{2\left| \vec{k}%
\right| }  \label{GB}
\end{equation}
from the completeness relation of the four vectors: 
\begin{equation}
\sum_{\lambda =\pm }\epsilon _{\mu }(k,\lambda )\epsilon _{\nu
}^{*}(k,\lambda )+\frac{1}{2}\left( e_{\mu }^{\left( +\right) }(k)e_{\nu
}^{\left( -\right) }(k)+e_{\mu }^{\left( -\right) }(k)e_{\nu }^{\left(
+\right) }(k)\right) =-g_{\mu \nu }
\end{equation}
This is the case iff the $a^{\#\prime }s$ behave in the standard way and the 
$c$'s have a nondiagonal inner product which, as a result of the presence of
the scalar photons $c_{-}$ corresponds to a genuine indefinite (not just
semidefinite) metric: 
\begin{equation}
\left\langle c_{+}(k)c_{-}^{*}(k^{\prime })\right\rangle =2\left| \vec{k}%
\right| \delta (\vec{k}-\vec{k}^{\prime })
\end{equation}
The $a$'s mix with the $c$'s under L-transformations viz. the comments on
gauge transformations in the wave function discussion of the last section.
But whereas in the wave function treatment of photons in terms of vectors
only an extension by $c_{+}$ ``photons'' was necessary, in our present Fock
space description of formally local point like vector potentials we need in
addition the negative metric causing $c_{-}$scalar contribution (or
alternatively a notion of pseudo-adjoint different from the bona fide adjoint%
\footnote{%
If one retains the positive metric, the Lorentz generators will be only
pseudo selfadjoint and the Poicar\'{e} symmetries do not operate as
automorphisms on $^{*}$-algebras, i.e. both descriptions lead away from $%
^{*} $-algebras. Whereas for the free case the metric can be changed at
will, the situation for interacting theories become mathematically
uncontrollable and physically questionable.}). Only in the weak sense of
matrix elements the condition of absence of scalar ``photons''can be
enforced in terms of a \textit{local condition}: 
\begin{equation}
\left\langle \psi \left| \partial ^{\mu }A_{\mu }\right| \varphi
\right\rangle =0\,\,\,\,\,
\end{equation}
Whereas in the case of the Wigner space it was possible to have the
transversality condition as a defining property of the Wigner representation
space, it is not possible to have transversal pointlike free vector
potentials in Fock space but rather only transversal matrix elements between
physical multiparticle states are compatible. This is equivalent to the use
of a nonlocal condition on vectors in terms of the annihilation part of $A$: 
$\,\,\partial ^{\mu }A_{\mu }^{\left( +\right) }\left| \varphi \right\rangle
=0.$ In the interacting case this Gupta-Bleuler formalism only works because
one can show that $\partial ^{\mu }A_{\mu }$ continues to fulfill the free
wave equation. In order to preserve the transversality condition under the
tensor product formation of multiparticle states, we will use in section 5
of the next chapter a cohomological BRS-like formalism. The main reason why
we postpone the introduction of that more stable (under interactions)
formalism is that the main argument in its favor is not so much the
incorporation of zero mass but rather the requirement of renormalizability
of massive and massless higher spin $\geq 1$ fields under the deformations
by interaction implementing local functions $W$ of free fields. We return to
the description of the Gupta-Bleuler formalism.

As we mentioned already in the previous section, all these problems are
absent (apart from the fact that the Wigner inner product does not permit a
local rewriting in terms of local field strength amplitudes), if we describe
the photons in terms of field strength instead of vector potentials. In that
case we only deal with physical photons: 
\begin{eqnarray}
F_{\mu \nu }(x) &=&\frac{1}{\left( 2\pi \right) ^{\frac{3}{2}}}\int
\sum_{\lambda =\pm }\left( e^{-ikx}u_{\mu \nu }(k,\lambda )a(k,\lambda
)+h.a.\right) ,\,\,\,\,  \label{Feld} \\
u_{\mu \nu }(k,\lambda ) &=&ik_{\mu }\epsilon _{\nu }-\left\{ \mu
\leftrightarrow \nu \right\}  \nonumber
\end{eqnarray}
But in order to formulate interacting QED with its specific long range
interaction\footnote{%
i.e. the quantum counterpart of the minimal external electromagnetic
coupling.} through the local renormalizable coupling of free fields, the
vector potential has been indispensable.. The fact that we do not know how
to employ string localized vectorpotentials $A_{\mu }(x,n)$ with $n$ being
the spacelike vector along the string direction, is the origin for certain
conceptual complications in spin $\geq 1$ LQP and the explanation for the
special role of the formalism of local gauge theories. In the standard
indefinite metric method, the descend from the unphysical vectorial
description defined by a free field with the two-point function \ref{GB} to
the physical photons in the sense of Wigner is done with the help of the
Gupta-Bleuler method. By the above transversality constraint one eliminates
the scalar $c_{+}$''photons''. This step leads from the indefinite metric
``Fock''-space to a \textit{positive semidefinite} subspace $\mathcal{H}%
_{ps} $ which still contains the zero norm longitudinal ``photons''. The
elimination of the latter can only be accomplished through descend to a
factor space (defined by equivalence classes): 
\begin{equation}
\mathcal{H}_{phys.}=\frac{\mathcal{H}_{ps}}{\mathcal{H}_{ps}^{\left(
0\right) }}\,,\,\,\,\,\,\mathcal{H}_{ps}^{(0)}=nullspace\,\,of\,\,zero\,%
\,norm\,\,vectors
\end{equation}
The Gupta-Bleuler method (as well as its BRS generalization) has a certain
formal geometrical elegance in renormalized perturbation theory, but its
conceptual physical aspects leave a lot to be desired. Of course the
conceptual and mathematical troubles start only with interactions.. I do not
know any controllable mathematics for indefinite metric algebras which could
be used for structural investigations i.e. spin\&statistics, localization
etc. The alternative to stay in physical and introduce nonlocal
vectorpotentials has not been seriously considered because the standard
perturbative framework requires pointlike fields. In the net approach this
problem seems to be related to finding a natural algebraic analogon of
semi-infinite ``axial gauges'' (\ref{nonloc}). More remarks on such ideas
will appear in a later section (\ref{modul}).

The higher spin cases are treated analogously. We only give a brief sketch,
the details may be found in Weinberg's book. Using the completeness
relations of the general $\left[ m,s\right] $ intertwiners one finds a two
point function of the form (\ref{twopoint}) with $P$ a covariant polynomial
in the derivatives. Again one observes the possibility of a matrix
realization of $\mathcal{P}$, $T$ and $\mathcal{C}$ if one uses the
``doubling'' $D^{\left[ A,B\right] }\oplus D^{\left[ B,A\right] }$. The
requirement of locality leads to the spin-statistics connection of the
previous theorem.

The zero mass case leads to a severe restriction between $A,B$ and the
helicity $h=s$ namely $\left| A-B\right| =h$. For $h=2$ the analogy with
classical general relativity and the long range nature of the graviton
interaction again demands to side step this rule by using a gauge theoretic
description in terms of a symmetric tensor $g_{\mu \nu }.$ in analogy (but
more complicated) with the vector potential for $h=1$\thinspace . The
massive $s=0,s=\frac{1}{2}$ and $s=1$ fields as well as their massless
helicity counterparts are ``Eulerian'' i.e. the transformation property is a
consequence of the matrix form of the differential operator which is the $%
4\times 4$ Dirac or the $4\times 4$ $s=1$ Proca-Wentzel operator. Also for
higher spins there are such Eulerian operators e.g. the Rarita-Schwinger
operator for $s=\frac{3}{2}$. But most of the covariantizations of the
Wigner representations are not ``Eulerian'' and can not be used for
Lagrangians and canonical quantization procedures (in particular all minimal
i.e. 2s+1 component descriptions for $s>0$). But this does not make them
less physical or useful. As already mentioned the helicity restriction of
zero mass fields only puts in evidence a problem which is also looming
behind the spin $\geq 1$ massive fields. As a result of their high operator
dimensions dim$\psi \geq 2$ (there is no covariant $\psi ^{\left[ A,B\right]
}$ with operator dimension below this value), a simple minded trilinear
coupling analysis for the interaction operator $W$ reveals that dim$W\geq 5,$
and hence $W$ is formally nonrenormalizable. The existence of the ``magic
cohomological trick'' (section 5) reveals that the idea that only local
functions of local fields with dim$W\leq 4$ (i.e. elements of the free field
Borchers class) can produce local deformations is not quite correct. As long
as one avoids the vectorpotentials and their higher spin generalizations,
the cohomological approach, which restricts the consistent expressions for $%
W $ severely, produces deformed local expressions whose operator dimensions
is only off by logarithmic terms from their free field dimensions. The best
way to deal with zero mass problems is to view them as zero mass limits of
the massive cohomological approach. As explained in section 5, one obtains
the classical gauge picture as the semiclassical manifestation of this more
fundamental LQP formalism in agreement with Bohr's correspondence principle
(which is the opposite of quantization).

Finally we make the following important observation. Despite the fact that
the Wigner theory gives a unique description for each mass and spin, we
completely loose this uniqueness on the level of local fields. We obtain a
countable covariant local family of fields which all share the same
Fock-space operators but differ in their $u$ and $v$ intertwiners. This is
true for any spin; even in case of $s=0$ we may use vectors or tensors which
of course turn out to be just derivatives of the standard scalar field. In
the next section we will show that these different fields generate the same
local algebras. With respect to those algebras they behave like different
coordinates in geometry. The intrinsic physical information is in the
``net'' of local algebras. As in geometry, it is of course not wrong to use
field coordinates in LQP.

\section{The Equivalence Class of a Free Field}

We have seen that the Wigner representation theory together with the
locality principle leads to a multitude of $(m,s)$ fields. Actually the set
of physically equivalent descriptions is even much larger. Let us understand
this first in the case of a neutral scalar field: 
\begin{equation}
A(x)=\frac{1}{\left( 2\pi \right) ^{\frac{3}{2}}}\int \left(
e^{-ipx}a(p)+e^{ipx}a^{*}(p)\right) \frac{d^{3}p}{2\omega }=A^{\left(
-\right) }(x)+A^{(+)}(x)
\end{equation}
Such operator-valued distributions cannot be pointwise multiplied as
classical functions can. In order to find a substitute for classical
pointwise multiplication, one studies first the matrix elements of products
of $A$ at different points e.g. 
\begin{equation}
\left\langle \Omega \left| A(x_{1})A(x_{2})....A(x_{n})\right| \Omega
\right\rangle
\end{equation}
Clearly the terms which become singular for coalescent points (more
generally if one of the difference vectors $x_{i}-x_{j}$ becomes light like)
results from ``Wick-contractions'': 
\begin{equation}
A^{\left( -\right) }(x_{i})A^{\left( +\right) }(x_{j})=i\Delta ^{\left(
+\right) }(x_{i}-x_{j})+A^{\left( +\right) }(x_{j})A^{\left( -\right)
}(x_{i})
\end{equation}
$i\Delta ^{\left( +\right) }$ which are generated by commuting the
annihilation components $A^{\left( -\right) }$ through the $A^{\left(
+\right) \prime }s$ to the right vacuum. The resulting terms in which the
annihilators are on the right of creators i.e. operator products of the
form: 
\begin{equation}
A^{\left( +\right) }(x_{i_{1}})....A^{\left( +\right) }(x_{i_{k}})A^{\left(
-\right) }(x_{i_{k+1}})....A^{\left( -\right) }(x_{i_{n}})
\end{equation}
have vanishing vacuum expectation values and finite matrix elements between
finite (but arbitrarily large) particle number vectors in Fock space. In
those ``Wick-ordered'' products the limit $x_{i}\rightarrow x$ of colliding
points can be taken without peril. Therefore one defines local functions of
the local field $A(x)$ in the sense of Wick-ordering as: 
\begin{equation}
:A^{n}(x):=\sum_{k-partitions}A^{\left( +\right) }(x)....A^{\left( +\right)
}(x)A^{\left( -\right) }(x)....A^{\left( -\right) }(x)
\end{equation}
i.e. the terms which result by simply ignoring the contractions. These are
the equal point limits of ``split-point'' Wick-products: 
\begin{equation}
:A(x_{_{1}})....A(x_{n}):=\sum_{k-partitions}A^{\left( +\right)
}(x_{i_{1}})....A^{\left( +\right) }(x_{i_{k}})A^{\left( -\right)
}(x_{i_{k+1}})....A^{\left( -\right) }(x_{i_{n}})
\end{equation}
where the order inside the k-partition is the original order. The usefulness
of the Wick-ordering results from the fact that despite their nonlocal
origin in terms of frequency separation, the resulting operators are local
or multilocal. This is because the above definition is equivalent to the
following obviously local inductive definition: 
\begin{equation}
A(x_{1})....A(x_{n})=:A(x_{_{1}})....A(x_{n}):+\sum_{m=1}^{\left[ \frac{n}{2}%
\right] }\sum_{m\;contr.}:A(x_{1})\underbrace{..\underbrace{...}.\underbrace{%
...}.}.A(x_{n}):
\end{equation}
where the lower brackets represent the Wick- ``contracted'' pairs and the
sum goes over all m-pairings and finally over all m. Clearly this formula
provides an inductive definition of Wick ordering (the right hand sum only
involves ordered products with a lower number of operators). The proof that
the previous frequency-ordering definition leads to this inductive formula
is elementary and left to the reader. The multi-localized (at $%
x_{1}....x_{n} $) product obviously approaches the one-fold localized Wick
power of the free field. Here the word ``local'' has a classical as well as
a quantum meaning. Classically it means that one only has to know the $A$'s
around the spacetime point x in order to compute $:A^{n}(x):$, whereas the
operational quantum meaning is that this pointlike composite commutes with
all the $A$'s whose localization is spacelike with respect to x (locality in
the sense of Einstein causality, which in local Quantum Theory means
simultaneous measurability). The best way to reconcile the classical with
the quantum aspects is the notion of Borchers class defined below. In order
to get a feeling for the properties of local composites, let us look at two
point functions of $n^{th}$ Wick powers: 
\begin{eqnarray}
\left\langle :A^{n}(x)::A^{n}(y):\right\rangle &=&n!\left( i\Delta
^{(+)}(x-y,m^{2})\right) ^{n} \\
&=&\int_{m^{2}}^{\infty }\rho ^{(n)}(\kappa ^{2})i\Delta ^{(+)}(x-y,\kappa
^{2})d\kappa ^{2}  \nonumber
\end{eqnarray}
here we indicated the dependence on mass parameters. According to a
well-known and easy to prove statement of Kall\'{e}n and Lehmann, any
two-point function of a Lorentz covariant scalar field in a theory with
positive energy conditions has a spectral representation in terms of a
positive $\rho $-function. In the present case the $\rho $ is the n-fold
convolution of the forward mass shell distribution $\theta (p_{0})\delta
(p^{2}-m^{2}).$ This only involves integrations over finite regions of $p$%
-space and may be carried out explicitly with the result that the $\rho
^{\prime }s$ are polynomial bounded (depending on n) functions.

The family of pointlike Wick-ordered composites is bigger than the above
illustrations; also derivatives as $:\partial _{\mu }A(x)\partial _{\nu
}A(x):$ etc. are included. It is very gratifying that also the inverse is
true:

\begin{theorem}
The set of fields in Fock space which commute for spacelike distances with
the free field A(x): 
\begin{equation}
\mathcal{B}(A):=\left\{ B\mid \left[ B(x),A(y)\right] =0\quad for\quad
\left( x-y\right) ^{2}<0\right\}
\end{equation}
is called the Borchers equivalence class $\mathcal{B}(A)$ and consists
precisely of the local composites which are generated by Wick powers.
\end{theorem}

The equivalence class aspects will be discussed in a later chapter in the
context of interacting fields. At the end of this section we will give a
proof of this theorem. It is important (e.g. for the derivation of the
Feynman rules) to be able to Wick-order products of local composites of free
fields. Let us look at examples: 
\begin{equation}
\begin{array}{c}
:A^{4}(x)::A^{4}(y):=:A^{4}(x)A^{4}(y):+4^{2}i\Delta ^{\left( +\right)
}(x-y):A^{3}(x)A^{3}(y):+ \\ 
+4^{2}3^{2}\left( i\Delta ^{\left( +\right) }(x-y)\right)
^{2}:A^{2}(x)A^{2}(y):+\left( 4!\right) ^{2}\left( i\Delta ^{\left( +\right)
}(x-y)\right) ^{3}:A^{3}(x)A^{3}(y):+ \\ 
+\left( 4!\right) ^{2}\left( i\Delta ^{\left( +\right) }(x-y)\right) ^{4}
\end{array}
\end{equation}
\begin{equation}
\begin{array}{c}
:\bar{\psi}(x)\gamma _{\mu }\psi (x)::\bar{\psi}(y)\gamma _{\nu }\psi (y):=:%
\bar{\psi}(x)\gamma _{\mu }\psi (x)\bar{\psi}(y)\gamma _{\nu }\psi (y): \\ 
+:\bar{\psi}(x)\gamma _{\mu }iS^{\left( +\right) }(x-y)\gamma _{\nu }\psi
(y): \\ 
+Tr\left\{ iS^{(-)}(y-x)\gamma _{\mu }:\psi (x)\bar{\psi}(y):\gamma _{\nu
}\right\} +Tr\left\{ iS^{(-)}(y-x)\gamma _{\mu }iS^{(+)}(x-y)\gamma _{\nu
}\right\}
\end{array}
\end{equation}
For a good understanding of the Wick-formalism of local functions a
knowledge of the following statements is indispensable..

\textbf{Statement 1:} \textit{Powers of the two-point functions are
well-defined distributions (singular functions), }

e.g. $F(x)=(i\Delta ^{\left( +\right) }(x))^{n}$ is again a distribution
with momentum space support properties. This is a multidimensional
generalization of the well-known statement that singular functions $F$ in
one variable, whose Fourier transform $\tilde{F}$ have support on the half
line, can be freely multiplied. The reason is that (as a result of the
support property) $i\Delta ^{(+)}(x)$ is the boundary value on the real axis
of an analytic function holomorphic in the upper half plane and therefore
this property is inherited by its $n^{th}$ power $F.$ Equivalently the
convolutions of $\tilde{F}$ only extend over a compact region. In the
multidimensional version the half lines are to be replaced by conic regions.
In standard QFT momentum space correlation functions are well behaved
functions, which at most have singularities at small momenta (infrared
problems). Their asymptotic increase is responsible for the x-space
singularities on the light cone.

\textbf{Statement 2: }\textit{The Noether conservation laws of classical
field theory also hold for the corresponding Wick-orderd objects in the free
field Borchers class.}

We provide two typical illustrations involving a Dirac field $\psi $ and a
scalar field $A$: 
\begin{equation}
\begin{array}{c}
\partial ^{\mu }j_{\mu }(x)=0,\quad j_{\mu }(x)=:\bar{\psi}(x)\gamma _{\mu
}\psi (x):\quad \psi =\hbox{ Dirac-field} \\ 
\,T_{\mu \nu }(x)=:\partial _{\mu }A(x)\partial _{\nu }A(x):-g_{\mu \nu }%
\frac{1}{2}:(\partial ^{\kappa }A(x)\partial _{\kappa }A(x)-m^{2}A^{2}(x)):
\\ 
\partial ^{\mu }T_{\mu \nu }(x)=0
\end{array}
\end{equation}
As in the classical case the covariant divergence hits both of the fields
and lead to operations on the u-and v-intertwiners which thanks to certain
identities (e.g. the vanishing of the momentum space Dirac operator on these
intertwiners) give the desired conservation law. In no stage of the argument
does one need the canonical formalism or the Euler-Lagrange form of equation
of motions, one only needs identities on intertwiners $u$ and $v$ which are
an immediate consequence of their definition.

\textbf{Statement 3: }\textit{In the relation between local ``currents'' and
global ``charges'':} 
\begin{equation}
Q=``\int d^{3}xj_{0}(x)",\quad P_{\mu }=``\int d^{3}xT_{\mu 0}(x)"
\end{equation}
\textit{the phenomenon of vacuum polarization enforces a nonclassic subtlety}
which is explained in the following.

A composite of a free field is more singular than the free field. In
particular for $d\geq 2+1$ it does not fit into the framework of canonical
equal time (anti)commutation relation, but rather has to be smeared with
test functions in d dimensions (in our case $d=3+1$). This can already be
seen by using the previously calculated two-point function of the composite
current operator j , e.g. 
\begin{equation}
\left\langle j_{\mu }(x)j_{\nu }(y)\right\rangle =\int (g_{\mu \nu }\partial
^{2}-\partial _{\mu }\partial _{\nu })i\Delta ^{\left( +\right) }(x-y,\kappa
^{2})\rho (\kappa ^{2})d\kappa ^{2}
\end{equation}
Since $\int \rho (\kappa ^{2})d\kappa ^{2}=\infty ,$ the smearing with test
functions supported on a spacelike hypersurface i.e. of the form $f(x)=\hat{f%
}(\vec{x})\delta (t)$ does not give a finite answer, one rather needs
smoothness in time as well. As in the classical case, one tries to obtain
the global charge $Q=\int (a^{*}(p)a(p)-b^{*}(p)b(p))\frac{d^{3}p}{2\omega }$
as a limit of ``partial'' charges referring to a finite region: 
\begin{equation}
\begin{array}{c}
Q(g.h):=\int j_{0}(x)g(\vec{x})h(t)d^{4}x,\quad \sup \hbox{p g }\subseteq
V+\delta V,\hbox{ supp h}\subseteq \left\{ \left| t\right| \leq \epsilon
\right\} \\ 
g\equiv 1\hbox{ in 3-volum V,\quad }\int hdt=1
\end{array}
\end{equation}
In words: g is a characteristic function of the 3-dim. volume region V which
has been smoothened outside, whereas h(t) is a smoothened $\delta $%
-function. It is easy to see that: 
\begin{equation}
\left[ Q(g,h),B\right] =\left[ Q,B\right] ,\quad 
\hbox{for locB in
completion of V}\quad
\end{equation}
i.e. for operators B localized in the causal completion of V (smearing
functions with support in completion of V) the commutator is already
independent of g,h (and identical to the global charge). However on the
vacuum vector $\Omega $ the partial charge has such strong vacuum
fluctuations (resulting from the presence of $a^{*}-b^{*}$ terms) that: 
\begin{equation}
\lim_{V\rightarrow \infty }\left| \left| Q(g,h)\Omega \right| \right|
^{2}=\infty ,\quad \hbox{but }\lim_{V\rightarrow \infty }(\psi ,Q(g,h)\Omega
)=0
\end{equation}
Here $\psi $ is from the dense domain on which the local functions of the
free field are defined i.e. the polynomial domain. The vacuum fluctuations
were discovered in the early days of QFT by Heisenberg and their physical
significance was studied by Weisskopf. One such manifestation is a
contribution to the Lamb-shift (see next chapter). This quantum phenomenon
has no counterpart in quantum mechanics and it has far going structural
consequences, e.g. it makes the local algebras of QFT essentially different
from the quantum mechanical Heisenberg-Weyl algebras (the former admit no
pure states or minimal projectors).

We now indicate the proof of the previous theorem on the structure of the
free Borchers class \cite{wightman}. We start with the general Wick
expansion of a translation covariant field $B(x)$ in terms of Wick products
of the free field $A.$ The first step consists in ``peeling off ''
iteratively the lower Wick monomials so that the new relatively local field
say $B_{m+n}(x)$ after $n+m$ steps starts with Wick monomials of degree $n+m$
and higher i.e. 
\begin{eqnarray}
&&\left\langle \Omega \left| :A(y_{1})...A(y_{m^{\prime
}}):B_{m+n}(0):A(x_{1})...A(x_{n^{\prime }}):\right| \Omega \right\rangle
\label{Ep} \\
&=&\left\langle \Omega \left| A(y_{1})...A(y_{m^{\prime
}})B_{m+n}(0)A(x_{1})...A(x_{n^{\prime }})\right| \Omega \right\rangle \,\,\,
\nonumber
\end{eqnarray}
for $m^{\prime }+n^{\prime }\geq m+n$ and all matrix elements
(``formfactors'') of $B_{n+m}$ with $m^{\prime }+n^{\prime }<m+n$ vanish. We
start the induction with $B_{0}(0)\equiv B(0)$ and first subtract the
constant vacuum expectation of $B$ in order to arrive at a $B_{1}(0)$ in Bo($%
A$)$.$ Since the various monomials do not mix if commuted with the free
field $A(x)$, the commutator of the first Wick degree contribution $%
B_{1}^{(1)}(0)$ in $B_{1}(0)$ with the free field $A(x)$ yields a c-number
which (as a result of causality) must be proportional to $i\Delta (x),$
possibly involving derivatives. But this means that $B_{1}^{(1)}(0)=(P(%
\partial )A)(0).$ We then subtract this local one-particle contribution $%
B_{1}^{(1)}$ and obtain $B_{2}$ and $B_{2}^{(2)}$ etc. The induction
consists in showing, that if $B_{i-1}^{(i-1)}$ is a local Wick monomial
(possibly including derivatives of $)$ of degree $i-1$ in $A$, then $%
B_{i}(0)\in $Bo($A).$ We then show that the lowest term $B_{i}^{(i)}(0)$ has
local formfactors (particle matrix elements in the $A$-particle basis
between $n,m$-particle vectors) identical to those of a $i=(n+m)$ degree
Wick polynomial. This follows from the structure of the multiple $i$-fold
commutator $K^{(i)}$ which fulfills a Klein-Gordon equation in every free
field coordinate. Therefore $K$ obeys a multi Cauchy initial value
representation on the $t_{j}=0$ space-like hypersurface (\ref{Ep}): 
\[
K(\underline{x},\underline{y}):=\left\langle \left| \left[ ...\left[ \left[
...\left[ B_{n+m}^{(i)}(0),A(x_{1})\right] ,...,A(x_{n})\right]
,A(y_{1})\right] ...,A(y_{m})\right] \right| \right\rangle 
\]
which is just a linear combination of ($n+m$)-degree formfactors. The Cauchy
initial value problem representation for the free fields $A(.)$ leads to the
following structure: 
\begin{equation}
K(\underline{x},\underline{y})=\prod_{i\leq n,j\leq m}\int ...\int \Delta
(x_{i}-x_{i}^{\prime })\Delta (y_{j}-y_{j}^{\prime })\overleftrightarrow{%
\partial }_{i}^{\prime }\overleftrightarrow{\partial }_{j}^{\prime }f(%
\underline{x}^{\prime },\underline{y}^{\prime })\mid _{\stackunder{%
y_{0}^{\prime }=y_{0}}{x_{0}^{\prime }=x_{0}}}d^{3}\underline{x}d^{3}%
\underline{y}
\end{equation}
Since the initial values $f$ on the equal time hypersurface is a product of $%
\delta $-functions and derivatives (here the causality of $K$ is used), $K$
is identical to the multiple commutator of a local Wick monomial $%
B_{i}^{loc}=mon.^{(i)}(A)$ i.e. we have the formal representation: 
\begin{equation}
B(x)=\sum_{i}B_{i}^{loc}(x)  \label{Wick}
\end{equation}
Since local fields are operator-valued Schwartz (tempered) distributions
with power singularities on the light cone for d$\geq 1+2$, the series must
be finite and $B$ is a Wick polynomial Q.E.D..

It is an interesting question whether this result still holds if one does
not know that $B(x)$ is relatively local to the free field $A(x),\,$but only
has an information about the absence of interaction in the sense of a
trivial scattering operator $S_{B}=1.$ In this case one would take for $A$
the free incoming field provided by scattering theory. Although there is no
direct information from locality, the information provided by analytic
properties of p-space formfactors of $B$ between incoming ket and outgoing
bra state vectors (see chapter 6) leads to the same result (\ref{Wick}). In
the present framework of equivalence classes this means that a ``weakly
local'' equivalence class consisting of all fields with the same $S$-matrix
in the special case $S=1$ contains only one local Borchers class, namely.
the standard free field class. This suggests that the $S$-matrix is a very
precise indicator of interactions. It also hints at the existence of a
general \textit{uniqu}e inverse scattering structure of Local Quantum
Physics.

\section{A First Look at Modular Localization \label{modul}}

Recently it turned out that the Wigner representation theory contains
information on localization which allows a direct access to the local
algebras, thus avoiding the use uf nonunique field coordinates \cite{S1}.
The starting point is the abelian subgroup of Lorentz boosts belonging to a
wedge, say the standard t-x wedge $x>\left| t\right| .$ Note that wedges can
be also characterized in terms of two light rays which for the standard
wedge are $e_{\pm }\sim (1,\pm 1,0,0).$ The Wigner theory also provides an
anti-unitary operator which reflects the standard wedge into its opposite
wedge. In the simplest case of irreducible representation for scalar neutral
particles, this reflection $j$ differs from the TCP operation by a $\pi $%
-rotation around the x-axis: 
\begin{equation}
(\Theta \varphi )(p)=\bar{\varphi}(p)\quad (j\varphi )(p)=\bar{\varphi}%
(p_{0},p_{1},-p_{2},-p_{3},)
\end{equation}
Define now an unbounded positive closed operator $\delta $ by functional
calculus from the selfadjoint standard ($x-t$ )boost generator K: 
\begin{equation}
\delta =e^{-K},\quad \delta j=j\delta ^{-1},\quad \hbox{since }e^{iK\chi
}j=je^{iK\chi }
\end{equation}
With the help of the Tomita-like unbounded involutive operator $\frak{s}%
:=j\delta ^{\frac{1}{2}}$ we define a closed ``real'' subspace $H_{R}$ of
the Wigner representation space $H$: 
\begin{equation}
H_{R}=\left\{ \varphi (p)\in H\mid \frak{s}\varphi =-\varphi \right\} ,\quad 
\frak{s}=j\delta ^{\frac{1}{2}}
\end{equation}
The $\pm $ eigenspaces (since $\frak{s}$ is antilinear, only real linear
combinations are possible) of the closed operator $\frak{s}$ can easily be
shown to form a dense set in $H$ and the above definition is also the unique
polar decomposition of $\frak{s}$. To be more specific, $\frak{s}$ acts as: 
\begin{equation}
\frak{s}:\quad h+ik\rightarrow -h+ik,\quad h,k\in H_{R}
\end{equation}

Here $H_{R}+iH_{R}$ is a dense subspace of the Wigner space (it is only
dense, even though $H_{R}$ is closed). It is the \textit{modular
localization space} for the standard wedge $W^{sta}.$ Using the standard
mathematical trick of introducing the graph norm affiliated with $\Delta ^{%
\frac{1}{2}},$ this dense space becomes a Hilbert space $H_{\func{mod}}$ in
its own right. As we will see in the next section, \textit{the Wigner inner
product restricted to the modular localization space can be rewritten as a
(Hawking) thermal inner product in this new Hilbert space }$H_{\func{mod}}$%
\textit{\ associated with the modular localization. }

A more explicit description of $H_{R}$ is obtained by introducing the
wedge-affiliated ``rapidity'' $\theta $ : 
\begin{equation}
p=m(qcosh\theta ,qsinh\theta ,n_{2},n_{3}),\quad q=\sqrt{%
1+n_{2}^{2}+n_{3}^{2}}
\end{equation}
The domain of the operator $\Delta ^{\frac{1}{2}}$ (and hence of $\frak{s)}$
in terms of rapidity-dependent wave functions consists of boundary values of
analytic functions which are holomorphic in the $\theta $-strip $0<\theta
<i\pi $ and $H_{R},$the real subspace of $\frak{s}$ with -1 eigenvalue, is
the closed real space of wave functions which are analytic in the strip and
fullfil the boundary condition: 
\begin{equation}
\overline{\varphi (\theta +i\pi )}=-\varphi (\theta )
\end{equation}
where we suppressed the dependence on $n_{i}$. Let us call this the ``$s$%
-reality property''. It is somewhat surprising that this concept did not
seem to have appeared in mathematical physics, e.g. it is absent from the
various books (including those by Reed and Simon).

For massive spin s representations the $s$- reality property reads: 
\begin{equation}
D^{(s)}(i\sigma _{2})lim_{\chi \rightarrow i\pi }D^{(s)}(R(p,\Lambda
^{sta}(\chi ))\overline{\varphi (\theta +i\pi )}=-\varphi (\theta )
\end{equation}
If particles are not selfconjugate, the 2s+1 component $\varphi $ must be
doubled and the action of $J$ on the direct sum involves a flip-operation on
the two Hilbert spaces. For zero mass, the rapidity parametrization for the
standard wedge is defined by $k=\Lambda ^{sta}(\theta )k_{0}$ with $k_{0}=$ $%
(1,\vec{n})$ and the Wigner rotation $R(k,\Lambda )$ is to be replaced by
the helicity representation in terms of the Wigner phase factor of the
euclidean group $E(2).$

Looking at the geometric interpretation of this construction, one
conjectures that the subspace $H_{R}$ of these momentum space wave functions
has something to do with localization in the standard wedge (or in the
opposite wedge in case of the -subspace). In fact the intertwiner formalism
of the previous sections allows to write $H_{R}$ (or equivalently $iH_{R})$
in terms of vector valued $W$-supported testfunctions $f$ i.e. the 2s+1
component wave function: 
\[
u^{*}(p)\tilde{f}(p)\mid _{p\in m.h.} 
\]
where $u^{*}$ is $u$ written as a row vector with complex conjugate entries
and the subscript indicates the restriction of $p$ to the forward mass
shell. Using the analyticity properties of $\tilde{f}$ which result from the
wedge support and the charge conjugation properties of $u,$ the eigenvalue
equation for $\frak{s}$ can be checked easily. The wedge localization can be
confirmed without the intertwiner formalism (the intertwiners are not
available in all positive energy representations, see later) by studying
coherence properties of the net of real wedge spaces generated via
Poincar\'{e} transformations g on the standard wedge : 
\begin{eqnarray}
H_{R}(W) &\equiv &U(g)H_{R},\quad W=gW^{sta}\quad \\
H_{R} &=&H_{R}(W^{sta})  \nonumber
\end{eqnarray}

For localization in the quantum sense, one needs a concept of ``outside''.
In Schr\"{o}dinger theory as well as in the relativistic work of Newton and
Wigner, one uses the orthogonality in wave function space: one calls $f$
localized in a 3-dim. region $R$ if a spatial translation which carries $R$
into its geometric complement transforms the wave function into the
orthogonal complement. This is of course the ``Born-localization'' based on
Born's probability interpretation of Schr\"{o}dinger wave functions at a
fixed time. It incorporates the fluctuation (uncertainty relation) aspect of
states in QT and leads to finite extension for bound states i.e. to a
distinction elementary versus bound.. For relativistic wave functions this
idea unfortunately (much to the dismay of Wigner) cannot be extended from
equal time localization to spacelike localization (apart from localization
in an ``effective'' sense i.e. modulo Compton tails). Fortunately for
relativistic local QFT there exists another more geometric notion of
localization\footnote{%
Even this more ``algebraic'' localization does not always agree with the
geometric one. The vectorpotential for the ($m=0,s=1$) representation and
the continuous spin representation are free field examples for such cases,
not to mention $d=2+1$ anyons.}. It is that one which underlies the
pointlike fields of standard QFT. Its transcription to wave functions is
related to the following notion of symplectic ``orthogonality'' (duality): 
\begin{equation}
H_{R}^{\prime }=\left\{ h^{\prime }\mid Im(h^{\prime },h)=0\right\} ,\quad
H_{R}^{\prime }(W)=U(g)H_{R}^{\prime }
\end{equation}
It then follows that: 
\begin{equation}
H_{R}^{\prime }(W)=H_{R}(W)^{\prime },\quad \hbox{and }H_{R}^{\prime
}(W)=H_{R}(W^{\prime }),\quad W^{\prime }\equiv W^{opp}
\end{equation}
where the last nontrivial equality is a consequence of: 
\begin{equation}
\frak{s}_{W}^{*}=\frak{s}_{W^{^{\prime }}}\quad \longleftrightarrow
\,\,\,\,\,\,\,j_{W}=j_{W^{^{\prime }\,}}\,,\,\,\,\,\delta _{W}^{-1}=\delta
_{W^{^{\prime }}}
\end{equation}
which in turn follows from the commutation relation of the standard (x-t)
reflection $j$ (which sends the wedge $W$ into $W^{opp}$) and the
Lorentz-boost $\delta ^{it}$ $.$ Again one ends up with real Hilbert spaces
which are standard and factorial in the sense of chapter 2, section 6.

Thus we arrived at a covariant net of wedge spaces and now we want to show
that this net is isotonous i.e. that if a wedge is contained in another one,
the same is true for the associated spaces. But in such a situation the
second wedge is obtained from the first by two lightlike translations which
carry it inside, so we have to show isotony for lightlike translations. For
such translations we have: 
\begin{equation}
\frak{s}_{\hat{W}}\subset \frak{s}_{W},\quad \hat{W}=g(\lambda l)W\subset
W,\quad \lambda >0
\end{equation}
where $g(\lambda l)$ is a translation along the lightlike vector $l$. In
order to show that $\frak{s}_{W}$ extends $\frak{s}_{\hat{W}}$ we rewrite
this relation as : 
\begin{equation}
U(\lambda l)j_{W}\delta _{W}^{\frac{1}{2}}U(\lambda l)^{*}\subset
j_{W}\delta _{W}^{\frac{1}{2}}
\end{equation}
For the bounded antilinear operator $j_{W}$ this gives the covariance law,
whereas for unbounded $\delta $ the required relation results from the
commutation relation of the lightlike translation with the standard
Lorentz-boost $U(\chi )$: 
\begin{equation}
U(\lambda l)U(\chi )=U(\chi )U(e^{\chi }\lambda l)
\end{equation}
One can show that the isotony is quite generally equivalent to the
positivity of the energy.

Wedge localization is too weak for a physical interpretation of the theory
(e.g. for the derivation of statistics and scattering theory). The
localization underlying standard (e.g. Lagrangian) theory is compact
localization which in our context means ($K$ stands for double cone): 
\begin{equation}
H_{R}(K)=\bigcap_{W\supset K}H_{R}(W)  \label{inters}
\end{equation}
\begin{equation}
H_{R}(K)+iH_{R}(K)\,\,\,\,\,\hbox{dense in }H,\quad \,H_{R}(K)\cap
iH_{R}(R)=\left\{ 0\right\}  \label{pro}
\end{equation}
\[
H_{R}(K)\cap H_{R}^{\prime }(K)=\left\{ 0\right\} 
\]
This property which previously (in section 2.6) was called standard and
factorial can be shown for all Wigner $(m,s)$ representations and even for $%
m=0$ (with the exception of the continuous spin representations, which do
not permit such a localization, since in this case the $H_{R}(K)$ spaces
turn out to be trivial). As a consequence the spaces fulfill the following
duality property: 
\begin{equation}
H_{R}^{\prime }(W)=H_{R}(W^{\prime }),\quad H_{R}^{\prime
}(K)=H_{R}(K^{\prime })
\end{equation}
\textit{In case of }$(m,s)$\textit{-representations one can prove this even
for disconnected and non simply connected regions in Minkowski-space. }

The abstract CCR functor introduced in the previous chapter converts the
real localization subspaces directly (i.e. without the interference of
pointlike fields) into a net of von Neumann algebras in Fock space. This and
the analogous statement for CAR is the only precise meaning of the word
second ``quantization''. Therefore the quantization approach to relativistic
QFT is limited to these cases and their formal Lagrangian perturbation which
also happen to be those to which differential geometric concepts as fibre
bundles are applicable. Beyond the word ``quantization'' has an intuitive
artistic connotation. But this territory beyond these functors and their
perturbative Lagrangian extension is precisely the region where the
algebraic approach takes over. There is however one important message in
these localization functors which will be studied in the sequel.

With these remarks we have entered the central issue of these notes: a
formulation of QFT which is independent of the choice of ``field
co-ordinates'' and refers directly to the map between localization regions
in Minkowski-space and observable algebras. Among the myriads of pointlike
fields there is of course no complete democracy. E.g. Noether currents are
physically distinguished. On the observational side it appears that all
prominent measured quantities can be represented in terms of matrix elements
of such currents. It is deeply gratifying that algebraic QFT attributes a
special role to such currents (via the ``split property'' see\cite{Haag}).
The rich physical harvest resulting from this new point of view outweighs by
far our present modest motivation in this section which was to reconcile the
multitudes of free fields with the uniquness of the positive energy Wigner
representations. We hope to be able to convince the reader about this in the
rest of these notes.

{\small It is interesting and in the spirit of this course to note that the
global spacetime symmetry can be encoded into the inclusions and
intersections of modular localization subspaces. Let us illustrate this by
two examples. }

\begin{itemize}
\item  {\small Suppose we shift the standard wedge W}$_{stan}$ {\small into
itself by applying a light like translation parallel to its longitudinal
light ray. This gives a ``halfsided modular inclusion'' with one of the
boundary surfaces of the shifted wedge W}$_{shif}${\small \ lying on the
corresponding boundary of the standard wedge. We note that the Lorentz
transformation associated with W}$_{stan}$ {\small compresses W}$_{shif}$ 
{\small into itself if we choose the sign of the Lorentz rapidity }$\chi $ 
{\small correctly. Since the modular group is equal to the appropriately
parametrized Lorentz boost this justifies the terminology halfsided modular
inclusion: }$\delta _{W}^{\pm it}H_{R}(W_{shif})\subset $ {\small H}$%
_{R}(W_{shif})$ {\small is }$\pm ${\small hs. In fact a simple computation
inside the Poincar\'{e} group shows that }$\frac{1}{2\pi }(ln\delta
_{W_{shif}}-ln\delta _{W_{stan}})$ {\small is the positive spectral
generator of the unitary light cone translation. The converse is also true.
If we start from a say a +hs modular inclusion of real Hilbert spaces H}$%
_{1}\subset H_{2}${\small , we note that }$\delta _{2}^{\frac{1}{2}}\leq
\delta _{1}^{\frac{1}{2}}$ {\small and assuming that is essentially
selfadjoint on D(ln}$\delta _{1}${\small )}$\cap D(\delta _{2})$ {\small we
may apply the Trotter product formula }$U(a)=expia(ln\delta _{1}-ln\delta
_{2})=s-lim_{n\rightarrow \infty }(\delta _{2}^{\frac{-ia}{n}}\delta _{1}^{%
\frac{+ia}{n}})^{n}${\small . The last step namely to prove that in this
abstract situation the so defined U(a) together with the }$\delta _{2}^{it}$ 
{\small and the modular reflection j define a 2-parametric group with }$%
\delta ^{it}${\small U(a)}$\delta ^{-it}=U(e^{-2\pi t}a)$ {\small and }$%
jU(a)j=U(-a)${\small , requires a bit more of the analytic techniques of
Borchers \cite{Bo}.}

\item  {\small \ By intersecting the standard wedge with another
L-transformed wedge W which has one light ray l in common with W}$_{stan}$ 
{\small (and not a hypersurface as above)}$,$ {\small one finds by
computations within the Poincar\'{e} group that the intersection W}$%
_{stan}\cap W$ {\small \ is halfsided modular included in W}$_{stan}$ 
{\small as well as in W}$.$ {\small This leads again to a positive
translation as above however this translation is not a geometric translation
inside the Poincare group; a reflection of the fact that the modular group
associated to the intersection W}$_{stan}\cap W${\small \ is not geometric.
But this time the W}$^{\prime }s${\small \ are not included and the
corresponding ``translation'' (the boost transformation transformations
remain inside the homogeneous Lorentz group) in the above formula (ln}$%
\delta _{W}^{\frac{1}{2}}-ln\delta _{W_{sta}}^{\frac{1}{2}}${\small ) has no
positive generator (it is a difference of the two previous nongeometric
translations in which the unknown nongeometric modular operator of W}$%
_{stan}\cap W${\small ) cancel out). In fact these transformations are
easily identified as one of the translations of the previously encountered
little group E}$(2).$ {\small The other one results from a modular
intersection in which the other light ray of W}$_{stan}$ {\small is shared
with a second wedge. The two ``translations'' are precisely the afore
mentioned transversal Galilei generators G}$^{\pm }.$

\item  {\small In fact the intimate relation between modular inclusions and
intersections with spacetime symmetries poses the question whether the full
Poincar\'{e} group may not be encoded ito a finite number of modular data.
This is indeed the case\cite{Bo Wie}, but the better context for explaining
such results is the later presented general framework of algebraic QFT.}
\end{itemize}

\section{Special Features of Zero Mass}

The above duality situation continues to be valid if one replaces $K$ by non
simply connected regions, \textit{but not so for the s}$\geq $\textit{1 zero
mass representations}. For example in the case of photons $(m=0,h=1),$ one
finds a duality violation for the toroidal ``corona'' region. Let \QTR{cal}{%
T\ }be the causal completion of a spatial torus which we call the
``corona''. A convenient mathematical description is to start from a unit
double cone in the longitudinal z-t plane centered at $\rho =\sqrt{%
x^{2}+y^{2}=}a>1,z=0,t=0$ and rotating it in the transversal x-y- plane once
around. The causal complement of $\mathcal{T}$ is causally multiply
connected. Then one obtains the following proper corona- inclusion: 
\begin{equation}
H^{R}\mathcal{(T)\subset }H^{R}\mathcal{(T^{\prime })^{\prime }}  \label{cor}
\end{equation}
\[
\curvearrowright \mathcal{A}(\mathcal{T})\subset \mathcal{A}(\mathcal{T}%
^{\prime })^{\prime } 
\]
where $H^{R}(\cdot )^{\prime }$ denotes the previously defined symplectic
complement and the \QTR{cal}{A}'s denote the corresponding von
Neumann-algebras as obtained from the $H^{R}(\cdot )$ by the Weyl
construction.

This obstruction (formally related to the appearance of $\delta ^{\prime }$
in the electric-magnetic canonical commutation relation) can be physically
understood in terms of a suitably regularized magnetic flux through a
surface which stretches from a circle inside the torus into the space-like
separated region inside. This flux does not change if one passes through
another spacelike surface subtended from the same circle. Hence such a flux,
although not being localizable within the 4-dim. toroidal region
nevertheless belongs to the symplectic complement of the spacelike
complement of the corona which is multiply spacelike connected. Via the CCR
functor this entails the above violation of Haag duality for the
corresponding algebras.

The issue of modular localization of (m=0,h=0) representations can also be
discussed in the $A_{\mu }$ description (\ref{semidef}). We can remove the
nullmodes explicitely in a fixed Lorentz frame by introducing a directional
dependent potential $A_{\mu }(x,n)$ by the following formulas: 
\begin{eqnarray}
\tilde{A}_{\mu }(k,n,\pm ) &=&\frac{n^{\nu }\tilde{F}_{\nu \mu }(k,\pm )}{%
n\cdot k+i\varepsilon },\quad \tilde{F}_{\nu \mu }(k,\pm )=k_{\{\nu
}\varepsilon _{\mu \}}(k,\pm )\psi (k,\pm )\quad  \nonumber \\
&\curvearrowright &\int \left| \psi (k,\pm )\right| ^{2}\frac{d^{3}k}{%
2\omega }=\int \overline{\tilde{A}}_{\mu }(k,n,\pm )\tilde{A}^{\mu }(k,n,\pm
)\frac{d^{3}k}{2\omega }\quad  \nonumber \\
&&\omega =\left| \vec{k}\right|
\end{eqnarray}
where wavy brackets denote the antisymmetrization in $\mu ,\nu $ and $%
\varepsilon _{\mu }(k,\pm )$ are the polarization vectors and the last
formula results from (\ref{Feld}). The singularity in k-space corresponds to
the semiinfinite line integral along $n$ in x-space. 
\begin{eqnarray}
A_{\mu }(x,n) &=&\int_{0}^{\infty }n^{\nu }\tilde{F}_{\nu \mu }(x-ns)ds
\label{nonloc} \\
&=&\int (e^{-ikx}\sum_{i=\pm }\tilde{A}_{\mu }(k,n,i)+h.c.)\frac{d^{3}k}{%
2\omega }  \nonumber
\end{eqnarray}
This vector potential has the following obvious properties under the Wigner
representation $U(\Lambda )$: 
\begin{eqnarray}
\partial _{\mu }A_{\nu }-\partial _{\nu }A_{\mu } &=&F_{\mu \nu }
\label{nonloc g.t.} \\
(U(\Lambda )A)_{\mu }(x,n) &=&\Lambda _{\mu }^{\nu }A_{\nu }(\Lambda
^{-1}x,n^{\prime })  \nonumber \\
&=&\Lambda _{\mu }^{\nu }A_{\nu }(\Lambda ^{-1}x,n)+\partial _{\mu }G(x) 
\nonumber \\
G(x) &=&\lim_{\varepsilon \searrow 0}\int e^{ikx}\frac{1}{(kn-i\varepsilon
)(kn^{\prime }-i\varepsilon )}n\cdot \tilde{F}(\Lambda ^{-1}k)\cdot
n^{\prime }\frac{d^{3}k}{2\omega }  \nonumber
\end{eqnarray}
i.e. the Lorentz transformation which acts on the Wigner wave function resp.
on the $F_{\mu \nu }$ tensor, transforms the potential covariantly except
for an additive nonlocal ``gauge'' term i.e. adds a nullmode contribution.
The nonlocality and the lack of covariance of this description in terms of
an affine transformation law is manifest. This peculiar ``gauge'' behavior
is a consequence of the nonfaithful helicity representation of the
noncompact little group $E(2).$ In particular as we already emphasized in
section 2 and 3 of this chapter,\textit{\ the quantum origin of gauge, and
gauge invariance in the sense of quantization of classical gauge theories
are somewhat different. }Unlike in classical theory\textit{\ }we either must
work with noncovariant vector potentials or (if we work with transverse
vector wave functions) or abandon the idea of pointlike transverse quantum
vectorpotentials.

In the presence of interactions one presently does not know a formulation
which uses the ghostfree semiinfinite stringlike localized Wigner potential $%
A_{\mu }(x,n).$ All the perturbative formulations of gauge theories use
covariant vectorpotentials which involve ghosts. We will return to this
issue in chapter 4. Here we only note that the nonlocal and noncovariant
behavior causes no problems if we could build up the net of localized
subspaces from wedges. The reason is that for spacelike vector $n$ in the
standard x-t wedge plane, the vector potentials corresponding to wedge
supported field strength $F$ are also wedge supported since the semiinfinite
spacelike line stays inside $W_{st}$: 
\begin{equation}
A_{\mu }(x,n)\in
H_{R}(W_{st})+iH_{R}(W_{st}),\,\,if\,\,\,support\,\,F\subset W_{st}
\end{equation}

The vectorpotentials are however not belonging to the simply connected
compact modular localization spaces (e.g. double cones) which result from
intersections except for closed spacelike contour integrals over $A$ which
loose their semiinfinite extension and can be expressed in terms of field
strength inside that localization region at least as long as the region
stays simply connected. The aforementioned peculiar behavior of the
non-simply connected corona region expressed in terms of vectorpotentials
means that the difference in the geometrical versus the quantum localization
i.e. the obstruction against Haag duality for nonsimply connected regions
can be accounted for in terms of a circular line integral over the
vectorpotential around the corona which looses its noncompact tail. Note
that in the covariant indefinite metric description the corona situation
behaves like for massive vector mesons i.e. there is no duality obstruction.
The duality obstruction only reappears after getting rid of ghosts. Hence
the mathematical locality and formal duality is another indication of the
unphysical nature of the ``ghostly'' description which only makes sense as
an auxilary intermediate bookkeeping.

In passing we remark that different from the massive case, there are many
positive energy nonvacuum representations of free photons. The appearance of
such non Fock ''infrared representations'' is characteristic for zero mass.
They show up in a richer representation theory of the associated Weyl
algebra. What we previously called Weyl functor should now be called the
Fock-Weyl functor. As before, the latter carries the net of real Hilbert
spaces into the vacuum observable net, but for zero mass there are other
functors from the same spaces to non Fock space algebras. Whereas free
massive nets have a unique positive energy representation, free zero mass
nets have inequivalent ``infravacua'' \cite{Kunhardt}. They also possess
electrically or magnetically charged ''infraparticle'' states.

Note that there is no problem in formulating the Weyl algebra based on the
vectorpotential description since only the imaginary part (symplectic form)
and not the full inner product is used in its definition. It just does not
possess a Fock representation in which the Lorentz group automorphisms of
the Weyl C$^{*}$-algebra possess a unitary representation. It is not clear
if in this description there are physically relevant infrared
representations which activate the zero ''modes'' i.e. the longitudinal
components of the vectorpotential.

Let us briefly return to a formal aspects of the corona obstruction. A
simple way to restore the harmony between the geometrical (classical) and
the quantum(modular) localization is to extend the corona localization space
by one type of object:

\begin{eqnarray}
&&\oint_{C\subset \mathcal{T}}A_{\mu }^{reg}(x,n)dx^{\mu } \\
A_{\mu }^{reg}(x,n) &=&\int \rho (\vec{x}-\vec{y})A_{\mu }(\vec{y}%
,x_{0},n)d^{3}y  \nonumber
\end{eqnarray}

This explains why the defect dimension of the two real Hilbert spaces in (%
\ref{cor}) is: 
\begin{equation}
dim\left[ H^{R}(\mathcal{T}^{\prime })^{\prime }:H^{R}(\mathcal{T})\right] =1
\end{equation}
The integration in the first formula is over a closed path $C$ inside $%
\mathcal{T}$ and we regularized the vector potential with a smooth function
of small support $supp\rho \in B_{\varepsilon }$ so that one maintains the
normalizability of the wave function and remains inside $\mathcal{T}.$ The
line integral represents the class of expressions of this kind, any two such
elements differ only by field strength localized in $\mathcal{T}.$ The line
integral is a L-invariant and may be expressed in terms of a magnetic flux
through any surface $S$ with the $C$ boundary. It is precisely this floating
surface stretching beyond $C,$ which in the quantum setting of commutativity
(or symplectic orthogonality) prevents the affiliation with $H^{R}(T)$ and
makes it a member of the nongeometric $H^{R}(T^{\prime })^{\prime }.$

It is a much more difficult question as to what becomes of this topological
obstruction in the presence of interactions. It is tempting to interpret
this obstruction as indicating the necessity of an interaction \footnote{%
This speculative remark is taken from \cite{Haag}, page 147.} i.e. of the
presence of non-vanishing electric or magnetic (or both) currents. 
\begin{equation}
\partial ^{\mu }F_{\mu \nu }(x)=j_{\nu }(x),\,\,\,\partial ^{\mu }\tilde{F}%
_{\mu \nu }(x)=\tilde{j}_{\nu }(x)  \label{Equ.50}
\end{equation}
One idea is that interactions are necessary to \textit{restore perfect Haag
duality} (i.e. for all multiple connected regions) which is violated in the
free theory. Such a point of view would attribute a very distinguished role
to electromagnetic duality and link it to Haag duality on a very fundamental
level. But lack of nonperturbative insight prevents a clear-cut resolution
of this duality connection. This is of course related to the non-understood
problematizing the notion of ``magnetic field'' on the same level of depth
as the notion of ``charge'' in the DHR superselection theory.

In low dimensional QFT the analogous issue of order-disorder duality and the
connection with Haag duality is much better understood. There, even in free
theories (see last section of this chapter), it is not possible to have 
\textit{\ no} charge sectors with both order and disorder. In the massive
case one charge comes out and the presence of both charges only occurs in
d=1+1 in the zero mass limit. The previous idea of maintaining corona
duality would enforce interactions and bring the interacting Maxwell-like
theories closer to the 2-dim. situation. A good understanding of modular
localization aspects of elctromagnetic interactions (in the vein of the
remarks about interactions in chapter 6 ) seems to be essential for future
progress.

As mentioned before, the corona inclusion may be constructed solely in terms
of the Wigner theory supplemented by the modular theory for wedges which
works with subspaces rather than with pointlike covariant amplitudes like $%
F_{\mu \nu }$. Helicities $h\geq 1$ present similar gauge problems and
corona obstructions as a result of the zero mass $E(2)$ little group. As in
the electromagnetic case, these duality obstructions are inexorably liked
with the gauge aspects of these massless theories. In the quantization
approach of the text books, this requires the introduction of ghosts and the
use of the BRS formalism including the cohomological control of the physical
factor space.

There remains in Wigner's list the positive energy zero mass representations
with so called ``continuous helicity'' which require infinite component
momentum space wave function. They are usually dismissed by saying that
``they are not realized in nature''. On the theoretical side there still
remains the question of whether their much weaker modular localization is
the theoretical reason why they may not behave as genuine particles.
Irreducibility (indecomposibility) and positive energy of the representation
of (space-time) symmetry is the only prerequisite for particles. Already
Wigner knew that this was insufficient and that one needs an appropriate
relativistic localization concept. In our present modular context we may
easily establish wedge localization. But how far beyond wedge localization
can one go? It turns out that the localization properties are rather similar
to those of d=2+1 anyons i.e. the best possible modular localization can
certainly not be better than (noncompact) spacelike cone localization.
Furthermore this analogy also suggests that the corresponding ``free field
theory'' does not have the Fock space structure. More comments will be
presented in the last chapter.

In such a situation one must first understand the physical consequences
before one rushes to the above dismissal. After all there are strange
particle-like objects as quarks which certainly cannot be identified with
the standard massive and massless Wigner particles.

Let us make some more remarks about the massive case. The principle of
locality requires to study intersection of wedges. Intersecting the
translated standard wedge $W_{a}^{st}$ with the opposite wedge $W_{opp}^{st}$
leads to an $x-t$ double cone which is cylindrically extended in the $y-z$
direction. Since the modular localization in $W_{opp}^{st}$ corresponds to a 
$s-$reality condition in the negative $\theta $-strip, the intersection of
both gives rise to a new ``edge of the wedge'' problem i.e. a Hilbert space $%
H_{R}(W_{a}^{st}\cap W_{opp}^{st})$ of analytic functions which are
meromorphic in both strips and fulfil a matching condition on the real $%
\theta -$axis in which the translation enters. Again $H_{R}(W_{a}^{st}\cap
W_{opp}^{st})$ is standard in the sense of the definition given in the
section on CCR and CAR functors.

The analytic situation for intersections of non coplanar wedges as one needs
them for double cones in(\ref{inters}) becomes very rich and is essentially
unexplored. In d=3+1 theories with halfinteger spin QFT of free fields
indirectly yield the information that the corresponding real subspaces are
standard and factorial.

If we apply this localization concepts to halfinteger spin, we find a very
interesting discrepancy by a factor $i$ between the action of $j$ and that
of the $\pi $-rotation of the wedge caused by the SU(2) transformation law
of the spin. Through this obstruction the Wigner theory already takes notice
of the spin-statistics.

We now explain the direct conversion of the net of Wigner subspaces into a
net of CCR- and CAR-algebras using the functorial formalism in section 6,
chapter 2.

Consider first the case of integral spin. The application of the
Weyl-functor to the subspace $H_{R}(W)$ gives the von Neumann-algebra: 
\begin{equation}
\mathcal{F}:H\rightarrow \mathcal{A}(H),\quad f\longmapsto W(f)
\end{equation}
\begin{eqnarray}
\mathcal{A}(W) &=&\hbox{v.Neumann Alg.}\left\{ W(f)|f\in H_{R}\right\} \\
&=&\mathcal{F}(H_{R}(W))  \nonumber
\end{eqnarray}
which inherits the following properties from the Hilbert spaces: 
\begin{eqnarray}
\hbox{isotony} &:&\hbox{\quad }\mathcal{A}(W)\subset \mathcal{A}(\tilde{W}%
),\quad \hbox{for }W\subset \tilde{W} \\
\hbox{Haag duality} &:&\hbox{\quad }\mathcal{A}(W^{\prime })=\mathcal{A}%
(W)^{\prime },\quad W^{\prime }=W^{opp.}  \nonumber \\
\hbox{covariance} &:&\hbox{\quad }U(g)\mathcal{A}(W)U^{*}(g)=\mathcal{A}%
(gW),\quad g\in \mathcal{P}\hbox{\quad }  \nonumber
\end{eqnarray}

In the halfinteger spin case we take the CAR functor $\psi ^{\#}(f)$: 
\begin{equation}
\mathcal{F}:H\rightarrow \mathcal{A}(H),\quad f\rightarrow \psi (f)\in B(%
\mathcal{H}_{F})=\mathcal{A}(H)
\end{equation}
\begin{eqnarray}
\mathcal{A}(W) &=&\hbox{v.Neumann Alg.}\left\{ \psi ^{\#}(f)\mid f\in
H_{R}(W)\right\} \\
&=&\mathcal{F}(H_{R}(W))  \nonumber
\end{eqnarray}
Different from the bosonic case, the operators $J$ and $S$ of this algebra
are not given by the application of the previous functor $\mathcal{F}$ but
the $J$ contains the famous Klein twist $K$ which changes geometrically
causally disjoint localized (hence anticommuting) operators into commuting
ones which one needs in the Tomita-Takesaki theory: 
\begin{equation}
J=K\mathcal{F}(j),\quad \Delta ^{it}=\mathcal{F}(\delta ^{it}),\quad
S=J\Delta ^{\frac{1}{2}}
\end{equation}
This is the T-T modular theory for wedge subalgebras of the CAR-algebra.

The same modular formalism can be used in order to construct relativistic
KMS states on free field algebras. In complete analogy to chapter 2.5, the
thermal two-point functions have the form (z=$e^{\beta \mu })$: 
\begin{eqnarray}
\left\langle \psi (x)\bar{\psi}(y)\right\rangle _{e,\beta ,\mu } &=& 
\nonumber \\
&&\frac{1}{(2\pi )^{\frac{3}{2}}}\int e^{-ip(x-y)}\sum_{s_{3}}u(p,s_{3})\bar{%
u}(p,s_{3})\frac{1}{1\mp ze^{-\beta p\cdot e}}\frac{d^{3}p}{2\omega } \\
&&+\frac{1}{(2\pi )^{\frac{3}{2}}}\int e^{ip(x-y)}\sum_{s_{3}}v(p,s_{3})\bar{%
v}(p,s_{3})\frac{ze^{-\beta p\cdot e}}{1\mp ze^{-\beta p\cdot e}}\frac{d^{3}p%
}{2\omega }  \nonumber
\end{eqnarray}
Here $e$ is a time-like vector which characterizes the rest frame of the
heat bath and the $\mp $sign corresponds to Boson/Fermion statistics.
Mistakes in the combinations of signs in front of the integrals can be
easily corrected by remembering that the thermal correlation functions must
have the same (anti)commutator functions as the standard free field
correlation functions (in addition to the KMS property), i.e. the thermal
aspect is an attribute of the state and not of the algebra. These
relativistic correlation functions have rather interesting analytic
properties \cite{BB}; they are analytic in $x-y=:\xi \rightarrow z,\,\,z\in 
\mathcal{T}_{\beta e}$ where $\mathcal{T}_{\beta e}$ is the tube $\mathcal{T}%
_{\beta e}=\left\{ z\in C^{d}:Imz\in V_{+}\cap (\beta e+V_{-})\right\} .$
The boundary values at the two edges fulfill as expected the KMS condition: 
\begin{eqnarray}
&&\exists F(z)\text{ analytic in }\mathcal{T}_{\beta e}\text{ s.t.} \\
\left\langle \psi (x)\bar{\psi}(y)\right\rangle &=&\lim_{z\rightarrow \xi
}F(z),\,\,\left\langle \bar{\psi}(y)\psi (x)\right\rangle
=\lim_{z\rightarrow \xi +i\beta e}F(z)  \nonumber
\end{eqnarray}
where the boundary values are taken from inside the analytic tube region.
All the statements are easily checked by explicit computations. Although the
boundary KMS condition is the standard one which relates the boundary values
on the two sides of the temperature strip, the relativistic aspect generates
a larger analytic tube in x-space which contains the strip in the $e$%
-direction. The temperature can be directly introduced as an extension of
the Wigner theory. It should be interesting to combine the modular
localization aspect with the heat bath temperature within the Wigner setting.

It is very interesting to note that the net of modular localization
subspaces and its modular automorphisms can be constructed from a finite
(with the size dependent on the space-time dimensions d) generating skeleton
of appropriately positioned (``half-sided modular'') real subspaces. This is
the analogue of a similar structure of nets of observables which developed
from its beginning in chiral conformal QFT in the form of the ``quarter
circle situation'' of the author \cite{S Ben}, via the more general
discussion of half-sided modular inclusions \cite{Wies}, to the
mathematically rigorous work of Araki and Zsido \cite{A Z}. In the present
modular subspace version which is adapted to the Wigner representation
theory, the mathematics is much simpler and very illustrative. We will
refrain here from a presentation. In particular the nonexistence of a
generalization of the CCR and CAR functors to d=2+1 anyonic Wigner spin and
to d=3+1 ``continuous spin'' Wigner representations is already visible in
the attempt to go beyond the net of modular wedge localized Wigner subspaces
(chapter 6).

It is very important to distinguish between the localized subspaces of the
Wigner representation space and localizes subspaces in Fock space. The
latter are not the image of the former under the CCR (CAR) functor $H(%
\mathcal{O})\rightarrow e^{H(\mathcal{O})}=\mathcal{H}(H(\mathcal{O}))$ but
one rather finds a genuine inclusion $\mathcal{H}(\mathcal{O})\subset 
\mathcal{H}(H(\mathcal{O}))$ where $\mathcal{H}(\mathcal{O})$ is the modular
localization subspace of Fock space which is identical to the domain of $S(%
\mathcal{O}).$

\section{Exotic Spin and Localization}

It is well known that in $d=1+2$ the Wigner spin is a priori not quantized.
This is the reason why in the generic case $s\neq ($half)integer, one uses
often the terminology ``any-on''. It is remarkable that these
representations reveal the nonexistence of a functor from a net of modular
localized subspaces to von Neumann subalgebras in the process of refining
the wedge localization.

First we note that the construction of $H_{R}(W_{stan})$ as an eigenspace of
the unbounded Tomita involution parallels that for (half)integer spin with
the only difference that instead of finding a phase factor $i$ (as for
halfinteger spin) which accounts for the difference between the quantum and
the geometric opposite we now find: 
\begin{equation}
H_{R}(W_{stan})^{\prime }=e^{is\pi }H_{R}(W_{stan}^{\prime })
\end{equation}
where the space $H_{R}(W_{stan}^{\prime })$ is defined by applying a
geometric rotation $W_{stan}\rightarrow W_{stan}^{\prime }$ to $%
H_{R}(W_{stan}).$ The net of wedges results from application of the
Poincar\'{e} transformations to $H_{R}(W_{stan})$ and as in the previous
cases, the positive energy is equivalent to the isotony properties of the
wedge net. Surprisingly these geometric properties become lost if we try to
refine the net by forming intersections. Whereas the triviality of the
compactly localized subspaces obtained by intersecting at least 3 wedges is
more or less expected on the physical grounds that anyonic spin should
require exotic statistics which needs noncompactly localized operators (see
our treatment of braid group statistics in a later chapter), the
nonexistence of a isotonic net structure for noncompact localizations as
e.g. spacelike cones (intersections of two wedges) is somewhat unexpected.
From the (doubled for antiparticle $\neq $ particle) Wigner theory for the
full (with reflections) Poincar\'{e} group we obtain $\delta ^{it\text{ }}$%
and the TCP related $j$: 
\begin{eqnarray}
(\delta ^{it}\phi )_{\pm }(p) &=&e^{is\varphi (\Lambda (-2\pi t),p)}\phi
_{\pm }(\Lambda (2\pi t)p) \\
j(\phi )_{\pm }(p) &=&e^{is\pi }\overline{\phi _{\mp }(-jp)}  \nonumber \\
\frak{s} &=&j\delta ^{\frac{1}{2}},\,\,\frak{s}%
H_{R}(W_{stan})=H_{R}(W_{stan})  \nonumber
\end{eqnarray}
where $\phi _{\pm }$ denotes the two component wave function, $\varphi
(\Lambda ,p)$ is the Wigner phase for the standard boost in x-direction and $%
e^{is\pi }$ the TCP phase. By covariance one obtains the net of wedge
spaces. By intersecting wedges we find the following obstruction for s$\neq
( $half)integer: 
\begin{eqnarray}
H_{R}(W_{stan}) &\frown &H_{R}(W_{rot})\varsubsetneq H_{R}(gW_{stan})
\label{durch} \\
\forall g &\in &\mathcal{P\,},\,\,\,W_{rot}=\vartheta \cdot W_{stan}\mathcal{%
\,}  \nonumber
\end{eqnarray}
where $\vartheta \in U(1)$ denotes a spatial rotation by the angle $%
\vartheta .$ So even if geometrically $W_{stan}\frown W_{rot}\subset
gW_{stan},$ the spaces will not be included! For the following it will be
convenient to have a representation of $H_{R}(W)$ in terms of wedge
supported test functions $f(x)$. We choose $f(x)$ with supp$f\subseteq W$
and construct the corresponding on-shell momentum space wave function
(leaving out the doubling in order to save on notation): 
\begin{eqnarray}
\phi (p) &\equiv &v(p)\tilde{f}(p)\mid _{h_{+}}  \label{phase} \\
v(p) &=&(p_{0}-p_{1})^{s}\frac{l(p)^{s}}{\overline{l(p)}^{s}},\,\,  \nonumber
\\
l(p) &=&p_{0}-p_{1}+m+ip_{2}  \nonumber
\end{eqnarray}
As expected the function $\phi $ is strip-analytic in rapidity and solves
the quantum modular localization equation $\frak{s}\phi =-\phi $ i.e. $\phi
\in H_{R}(W);$ a fact which is easily established. The only nontrivial
relation used is the intertwining relation: 
\begin{equation}
v(\Lambda (-2\pi t)\cdot p)e^{is\varphi (\Lambda (2\pi t,p)}=e^{2\pi st}v(p)
\end{equation}
which follows by straightforward calculation (from which one also obtains a
very compact formula for the Wigner phase). For $t\rightarrow i\pi $ the
exponential factor on the right hand side compensates the modular
conjugation phase (equal to the TCP phase). In fact all of $H_{R}(W)$ is
obtained in this way from the space of wedge supported test functions by
closure.

The proof now consists in exhibiting a family of wave functions, which by
construction belong to the left hand side of (\ref{durch}), but for no g are
contained in the right hand Hilbert space. Since $H_{R}(W_{stan}\frown
\vartheta \cdot W_{stan})\subseteq H_{R}(W_{stan})\frown H_{R}(\vartheta
\cdot W_{stan}),$ it follows that if we can prove that the left hand Hilbert
space contrary to the geometric inclusion is not contained in $%
H_{R}(g^{-1}\cdot (W_{stan}\frown \vartheta \cdot W_{stan}))$ then this is a
fortiori true for the right hand side. For this purpose we must get a better
understanding of the latter space of the g-transformed region. The relation (%
\ref{phase}) tells us how the standard boost acts on the rotated wave
function. A somewhat lengthy but simple calculation gives: 
\begin{eqnarray}
&&\exists \phi \,\,s.t.\,\,\,(u(\Lambda (2\pi t))u(g^{-1})\phi )(p)
\label{nonan} \\
&&is\,\,nonanalytic\,\,in\,\,t-strip  \nonumber
\end{eqnarray}
For the construction of these functions it is convenient to parametrize the $%
g$ with $g^{-1}(W_{stan}\frown \vartheta \cdot W_{stan})\subset W_{stan}$ in
the form $g^{-1}=\vartheta ^{\prime }\cdot \Lambda (-2\pi t^{\prime })\cdot 
\frac{\pi }{2}.$ By a rather lengthy calculation one shows that there are
wave functions $\phi $ on which the application uf the above transformations
(\ref{nonan}) yield a function whose analytic continuation of t into the
strip develops a cut for$.$ We will spare the reader the details.

This result means that within the Wigner theory for $2s\neq integer$ one
cannot refine the net of wedge localized subspaces by using intersections to
e.g. a net of spacelike cones. So the situation is very different from the
standard case $2s=integer$ where intersections even lead to nets which are
indexed by compact regions as double cones. The case of compact localization
is the only one which permits pointlike field generators. In the first case
the extended spin statistics theorem would demand braid group statistics
which is irreconcilable with a tensor product structure for multiparticle
state. This explains the nonexistence of a functor. Nevertheless it is
surprising that already the Wigner representation indicates a related
obstruction. In a later chapter we will use scattering theory in order to
determine the inner product structure of incoming free plektons and indicate
how we can construct an affiliated x-space ``free'' field theory.

We expect similar localization problems with the d=1+3 finite energy
``continuous spin'' representation. Our modular localization method makes it
possible for the first time to explore the physical consequences of these
Wigner representations.

\section{Localization and Hawking Temperature}

In this section we present two physical interpretations of the
Tomita-Takesaki formalism for the wedge localization of free field algebras
which are a special case of the more general Bisognano-Wichmann property of
interacting Wightman fields.

The first is a relation to Hawking-Unruh effect which was observed by Sewell 
\cite{Sew}. The modular group for the wedge is according to
Bisognano-Wichmann apart from a factor 2$\pi $ equal to the wedge associated
Lorentz-boost $\Delta ^{it}=U\left[ \Lambda (\chi =\frac{t}{2\pi })\right] $
. In the integer spin case we have seen that there is complete harmony
between the geometric and the quantum (in terms of von Neumann commutants)
notion of localization and in the non-bosonic cases one only needs
additional Klein factors. Furthermore thanks to the free field functors, the
explicit construction of the modular operators and the wedge localized
algebras can be delegated to the construction of an involutive unbounded
antilinear s-operator and its real closed subspaces $H_{R}$ of -1 or +1
eigenvalue. according to the previous section, the dense set of
wedge-localized wave function, is simply $H_{W}^{loc}=H_{R}+iH_{R}$ and $%
H_{R}$ consists of all momentum-space Wigner wave functions which are
analytic in a strip of the wedge rapidity $\chi $ and fulfill the s-reality
condition on the boundary. We know from the previous section:

\begin{equation}
H_{R}(W)+iH_{R}(W)=dom(\frak{s})\subset H_{Wigner}
\end{equation}

\begin{equation}
\mathcal{H}_{R}(W)+i\mathcal{H}_{R}(W)=dom(S)\subset \mathcal{H}_{Fock}
\end{equation}

These dense subspaces become Hilbert spaces in their own right if we use the
graph norm of the Tomita operators. For the $\frak{s}$-operators in Wigner
space we have: 
\begin{eqnarray}
\left( f,g\right) _{Wigner} &\rightarrow &\left( f,g\right) _{G}=\left(
f,g\right) _{Wig}+\overline{\left( \frak{s}f,\frak{s}g\right) }_{Wig}
\label{graph} \\
&=&\left( f,g\right) _{Wig}+\left( f,\delta g\right) _{Wig}  \nonumber
\end{eqnarray}
The graph topology insures that the wave functions are strip-analytic in the
wedge rapidity $\theta $: 
\begin{eqnarray}
p_{0} &=&m(p_{\perp })\cosh \theta ,\,\,\,\,p_{1}=m(p_{\perp })\sinh \theta
,\,\,\,m(p_{\perp })=\sqrt{m^{2}+p_{\perp }^{2}}  \label{rap} \\
&&strip:0<Imz<\pi ,\,\,\,\,\,\,z=\theta _{1}+i\theta _{2}  \nonumber
\end{eqnarray}

where this ''G-finiteness'' is precisely the analyticity prerequisite for
the validity of the KMS property for the two-point function. For scalar
Bosons we have for the Wigner inner product restricted to the wedge : 
\begin{eqnarray}
\left( f,g\right) _{Wig}^{W} &=&\left\langle A(\overline{\hat{f}})A^{*}(\hat{%
g})\right\rangle _{0}\stackrel{KMS}{=}\left\langle A^{*}(\hat{g})\Delta A(%
\overline{\hat{f}})\right\rangle _{0} \\
&&\stackrel{CCR}{=}\left[ A^{*}(\hat{g})A(\delta \overline{\hat{f}})\right]
+(f,\delta g)_{Wig}^{W},\,\,\,\,\delta =e^{2\pi K}  \label{thermal}
\end{eqnarray}

\begin{equation}
\curvearrowright \left( f,g\right) _{Wig}^{W}\equiv \left( f,g\right)
_{K,T=2\pi }=\left[ A^{*}(\hat{g})A(\frac{\delta }{1-\delta }\overline{\hat{f%
}})\right]  \label{int}
\end{equation}
Here we used a field theoretic notation ($A^{*}(\hat{g})$ is a smeared
scalar complex field of the type (\ref{int}) linear in $\hat{g}$ with $supp.%
\hat{g}\in W$ ) in order to emphasize that the temperature dependence on the
right hand side is explicit via the $\delta $ acting on the complex-valued
x-space smearing functions in the c-number commutator and not implicit, as
the restriction of the wave functions to the wedge region on the left hand
side. Of course the c-number commutator (without the state brackets) may be
rewritten in terms of p-space Wigner wave functions for particles and ($%
\delta ^{\frac{1}{2}}$-transformed) antiparticles in such a way that the
localization restriction is guarantied by the property that the resulting
expression is finite if the wave functions are G-finite. We mention for
experts that the difference between localization temperatures and heat bath
temperatures on the level of field algebras in Fock space corresponds to the
difference between hyperfinite type $III_{1}$ and type $I$ von Neumann
algebras.

In this way one obtains a thermal representation of the wedge restricted
Wigner inner product. The fact that the boost $K\,$appears instead of the
Hamiltonian $H$ for the heat bath temperature reveals one significant
difference between the two situations. For the heat bath temperature of a
Hamiltonian dynamics the modular operator $\delta =e^{-2\beta \mathbf{H}}$
is bounded on one particle wave functions whereas the unboundedness of $%
\delta =e^{2\pi K}$ in (\ref{thermal}) enforces the localization (strip
analyticity) of the Wigner wave functions.

This difference results from the two-sided spectrum of $K$ as compared to
the boundedness from below of $H.\,$In fact localization temperatures are
inexorably linked with unbounded symmetry operators.

The generalization to fermions as well as to particles of arbitrary spin is
easily carried out with the result that the localized thermal representation
involves the anticommutator. The differences between $K$ and $H$ also leads
to somewhat different energy distribution functions for small energies so
that Boson $K$-energy distributions may appear as those of $H$ heat bath
Fermions. In this context one is advised to discuss matters of statistics
not in Fourier space, but rather in spacetime where they have their
unequivocal physical interpretation.

One may of course consider KMS state on the same $C^{*}$-algebra with a
different $K$-temperature than $2\pi ,$ however such a situation cannot be
obtained by a localizing restriction. Mathematically $C^{*}$-algebras to
different $K$-temperatures are known to belong to different folii (in this
case after von Neumann closure to unitarily inequivalent $III_{1}$-algebras)
of the same $C^{*}$-algebra. Or equivalently: a scaled modular operator $%
\Delta ^{\alpha i\tau }$ cannot be the modular operator of the same theory
at a different temperature as it would be the case for type I algebras.

As in the Bisognano-Wichmann situation, the modular wedge localization in
the Wigner theory has led us to the Hawking-Unruh thermal situation. For
those readers who are familiar with Unruh's work we mention that the Unruh
Hamiltonian is different from $K$ by a factor $\frac{1}{a}$ where $a$ is the
acceleration (see below). Any modular localization (not necessarily the
wedge) leads to horizons and a thermal state. Only in very special cases one
has a geometrical picture in terms of Killing vectors in space-time.. In the
present setting it is the arena of Wigner space where one finds the
isometries. In order to remove any doubt that these thermal properties are
not typical for curved space-time QFT but constitute a general property of
QFT and as such are very relevant for ordinary QFT. In order to illustrate
this we look at the relation with crossing symmetry of formfactors.

More generally we may now consider matrix elements of wedge-localized
operators between wedge localized multiparticle states. Then the KMS
property allows to move the wedge localized particle state as an
antiparticle at the analytically continued rapidity $\theta +i\pi $ from the
ket to the bra. The simplest illustration is the two-particle matrix element
of a free current of a charged scalar field $j_{\mu }(x)=:\phi ^{*}\stackrel{%
\leftrightarrow }{\partial }_{\mu }\phi :$ smeared with the

wedge supported function $\hat{h}:$%
\begin{eqnarray}
&&\left\langle 0\left| \int j_{\mu }(x)\hat{h}(x)d^{4}x\right| f,\delta ^{%
\frac{1}{2}}g^{c}\right\rangle  \nonumber \\
\stackrel{KMS}{=} &&\left\langle 0\left| \left( \Delta \phi ^{*}(\delta ^{%
\frac{1}{2}}\overline{\hat{g}})\Delta ^{-1}\right) ^{*}\int j_{\mu }(x)\hat{h%
}(x)d^{4}x\right| f\right\rangle  \nonumber \\
&=&\left\langle \bar{g}\left| \int j_{\mu }(x)\hat{h}(x)d^{4}x\right|
f\right\rangle  \label{cro}
\end{eqnarray}
Here $\delta ^{\frac{1}{2}}g^{c}$ is the charged transformed antiparticle in
the Wigner wave function $g$ at the analytically continued rapidity $\theta
+i\pi $ whereas $\hat{g}$ denotes as before the wedge-localized space-time
smearing function whose mass shell restricted Fourier transform corresponds
to the boundary value of the analytically continuable Wigner wave function $%
g $. Moving the left hand operator to the left vacuum changes the
antiparticle charge to the particle charge. Since the $H_{R}(W)+iH_{R}(W)$
complex localization spaces are dense in the Wigner space, the momentum
space kernel for both sides of (\ref{cro}) takes the familiar form: 
\begin{equation}
\left\langle p^{\prime }\left| j_{\mu }(0)\right| p\right\rangle =%
\stackunder{z\rightarrow \theta +i\pi }{anal.cont.}\left\langle 0\left|
j_{\mu }(0)\right| p,p^{^{\prime }}(z)\right\rangle  \label{cross}
\end{equation}
where $p^{\prime }(z)$ is the rapidity parametrization of above (\ref{rap}).
This famous crossing symmetry, which is known to hold also in each
perturbative order of renormalizable interacting theories, has never been
derived in sufficient generality within any nonperturbative framework of
QFT. It is to be thought of as a kind of on-shell momentum space substitute
for Einstein causality and locality (and its strengthened form called Haag
duality).

There are two slightly different interpretations of the thermal wedge
situation.

The first physical interpretation is in terms of the Unruh effect i.e. the
temperature experienced in the vacuum by a uniformly accelerated observer 
\cite{Wald}. Such an observer moves on a world line: $t=\xi cosh\tau ,\,\,$ $%
x=\xi sinh\tau $ in an appropriately chosen Lorentz frame. In natural units $%
(c=1,$ $\hslash =1)$ his acceleration is $a=\frac{1}{\xi }$ and his proper
time $\tau _{a}$ is related to the wedge ``rapidity'' $\tau $ by $\tau _{a}=%
\frac{\tau }{a}$ and the corresponding Hamiltonian is $aK$ where $K$ is the
boost generator.$.$ The inside of the wedge (the Rindler world) as well as
its boundary (the horizon of the Rindler world) are invariant under the
boosts in the wedge direction $U(\Lambda _{W})$. On the positive part of the
wedge boundary $bdW_{+}$, the action of the positive light-like translation $%
T_{W}$ is also a transformation of that boundary into itself. The positive
spectral light-like translations together with the boosts form an
interesting two-parametric group which has a deep relation to so called
half-sided modular inclusions \cite{Wies}, but here we will confine
ourselves to more pedestrian methods.

Intuitively speaking, we expect that the global vacuum state appears similar
to a heat bath with respect to an uniformly accelerated observer. After all,
the wedge horizon generated by the acceleration signifies a loss of
information (the opposite wedge suffers a causal blackout for the
accelerated observer) which is also a characteristic feature of a heat bath
temperature state. The above formula shows indeed that the restriction of
the vacuum state to the wedge algebra (or the modular localization Wigner
subspace belonging to $W$) satisfies the KMS condition with respect to the $%
L_{W}$-Lorentz boost with the Hawking-Unruh temperature $\beta _{a}\equiv
kT_{a}=\frac{a}{2\pi }.$ Actually Unruh did all his calculations explicitly
(without recourse to modular localization theory) on a simple model of a
two-level detector coupled to a free field. The reader finds a beautiful
pedestrian review in\footnote{%
But beware of certain pitfalls, e.g. KMS states are not Gibbs states (the
latter need a quantization box which would wreck the wedge geometry) and
massless free fields in d=1+1 have infrared divergent two-point functions.
This requires an easy repair without change of conclusions.} \cite{BM}.

Sewell \cite{Sew} has shown that the Unruh effect may be generalized in such
a way, that it serves to understand the Hawking effect in a quite general
setting which includes black holes.

We now direct our attention to a second physical aspect of the wedge
situation which is the Hawking effect: of pair creation and the antiparticle
aspect, as well as its refined version namely crossing symmetry. For this
reason we study the complex $d=1+1$ Klein-Gordon field in a homogeneous
external electric field which points into the wedge direction. We pick a
gauge e.g. the axial gauge in time direction $A_{t}=Ex,A_{x}=0$. With the
Ansatz $\Phi =e^{-i\omega t}\varphi _{\omega }(x)$ the Klein-Gordon operator
takes the form: 
\begin{equation}
\left[ \partial _{x}^{2}+(\omega +Ex)^{2}\right] \varphi _{\omega
}(x)=m^{2}\varphi _{\omega }(x)
\end{equation}
By further canonical transformations $x\rightarrow \xi =\sqrt{E}(x+\omega
/E),\,-i\frac{\partial }{\partial x}\rightarrow i\frac{\partial }{\partial
\xi }$ and $u=\frac{1}{\sqrt{2}}(-i\frac{\partial }{\partial \xi }-\xi )$,
the equation looks like a eigenvalue equation of an auxiliary
Schr\"{o}dinger Hamiltonian which is isomorphic to the boost or scaling
generator: 
\begin{equation}
u\frac{\partial }{\partial u}\varphi (u)=(i\varepsilon -\frac{1}{2})\varphi
(u),\quad \varepsilon =\frac{m^{2}}{2E}
\end{equation}
That this boost had to eventually appear is clear from the classical
picture. The charged massive particles are uniformly accelerated and their
trajectory is identical to that of the Unruh accelerator. Classically there
is a perfect wedge horizon, but quantum field theoretically the vacuum state
does not factorize into product states on the wedge and its causal opposite.
Rather we expect particle-antiparticle pairs to appear near the horizon
which as a result of the action of the constant electric field will be split
into say right wedge particles and left (opposite) wedge antiparticles. A
pedestrian treatment of this external field problem would approximate the
constant field by a sequence of external fields which are different from
zero only during a time $T$ and inside a box of length $L$ around the tip of
the wedge with $T,L\rightarrow \infty $ at the end of the calculation. The
relation between the free field creation and annihilation operators before
(in) and after (out) is given in terms of a Bogoliubov automorphism: 
\begin{eqnarray}
a_{\omega }^{out} &=&\bar{\alpha}a_{\omega }^{in}+\beta b_{\omega }^{in*} \\
b_{\omega }^{out*} &=&\alpha b_{\omega }^{in*}+\bar{\beta}a_{\omega }^{in} 
\nonumber
\end{eqnarray}

However this global automorphism is only unitarily implemented as long as $%
T,L$ are kept finite (the spectrum of $\omega ^{\prime }s$ is discrete and
the density of $\omega $ states is proportional to $T,L$ and $E$)

The vacuum persistency probability follows the well known Schwinger formula: 
\begin{equation}
\left| \left\langle 0_{out}\right| \left. 0_{in}\right\rangle \right|
^{2}=exp\left( -\frac{ELT}{2\pi }ln(1+\left| \beta \right| ^{2}\right)
,\quad \left| \beta \right| ^{2}=e^{-m^{2}\pi /E}\quad
\end{equation}
which in turn is a consequence of the pair creation term in the formally
unitary implementer $U$ of the Bogoliubov transformation: 
\begin{equation}
\left| 0_{out}\right\rangle =U\left| 0_{in}\right\rangle
=e^{bil(a^{*},b^{*})}\left| 0_{in}\right\rangle
\end{equation}
where $bil.$ indicates the characteristic exponential bilinear dependence of
implementers of Bogoliubov transformations on the $a^{\#\prime }s$ and $%
b^{\#\prime }s.$ From the population ratio of: 
\begin{equation}
\left\langle n_{\omega }\right\rangle =\left\langle 0_{in}\left| a_{\omega
}^{out*}a_{\omega }^{out}\right| 0_{in}\right\rangle =\left| \beta \right|
^{2}
\end{equation}
for two charged free fields with slightly different masses $m$ and $m+\Delta
m$ one obtains: 
\begin{equation}
\frac{\left\langle n\right\rangle _{m+\Delta m}}{\left\langle n\right\rangle
_{m}}=e^{-2\pi \Delta m/a},\quad a=\frac{E}{m}
\end{equation}
i.e. the ratio of created pairs obeys a thermal distribution with
temperature $kT=2\pi /a$ as in the Unruh interpretation of the wedge
localization.

This last somewhat clumsy and conceptually controversial (the
box-quantization is too much of a brute force computational device in
scattering theory) derivation of the radiation aspects of the constant
external electric field model (the same which in many textbooks serves to
illustrate the Klein paradoxon) should be replaced by a more elegant method
using appropriate representation concepts which do lead to exponentials free
of $TL$-volume factors, in particular one expects a finite entropy density
ratios per unit horizon surface for say two models with different matter
contents. The afore-mentioned considerations suffer from the fact the type
III$_{1}$ nature of the localized algebras together with the fact that there
is no causally disjoint space-time between the wedge region and its
opposite, lead to uncontrollable fluctuations which strictly speaking wreck
the existence of Fock-space Bogoliubov operators and cause serious problems
with an intrinsic notion of entropy. Here the split property (with the split
distance approaching zero at the end of the calculation) is expected to have
beneficial consequences, but the calculation has not been carried out. This
split property which is related to the correct counting of degrees of
freedom in QFT ``phase space will be explained in a later chapter and in the
mathematical appendix.

As a curious side result we mention that the low energy distribution of a
system with a Hawking-Unruh temperature resulting from modular localization
can be significantly different from that of a standard heat bath
temperature. For free bosons and fermions, the energy distribution is given
in terms of the spectral decomposition of: 
\begin{equation}
\frac{1}{1\mp e^{\beta K}}  \label{dis}
\end{equation}
\newline
with $K$ being proportional to the infinitesimal generator of the modular
group. Only in the case of the heat bath situation is $K$ the positive
energy Hamiltonian in the rest frame of the heat bath. For the Rindler
localization $K$ is the wedge-affiliated Lorentz-boost. The most dramatic
difference between these two kinds of thermalizations show up for zero mass
in odd space-time dimensions d (where the ``reverberation'' leads to a
breakdown of Huygens principle). A convenient way to obtain the spectral
distribution of (\ref{dis}) is to Fourier transform the zero-mass two-point
function $W$ in the rapidity (boost) variable $\tau $: 
\begin{eqnarray}
\tilde{W}(\omega ) &=&\int_{-\infty }^{+\infty }e^{-i\omega \tau }W(\xi
(\tau )) \\
W(\xi ) &=&\frac{\Gamma (\frac{d}{2}-1)}{4\pi ^{\frac{d}{2}}}\frac{1}{\left( 
\sqrt{-(\xi _{0}+i\varepsilon )^{2}+\vec{\xi}^{2}}\right) ^{d-2}}  \nonumber
\\
\xi (\tau ) &=&\frac{1}{a}\left( sh\tau a,ch\tau a,0,..0\right)  \nonumber
\end{eqnarray}
For $d=2+1$ we obtain the well-known result\cite{Takagi} : 
\begin{equation}
\tilde{W}(\omega )=\frac{1}{2}\frac{1}{1+e^{\beta \omega }}
\end{equation}
which is spectacularly different from the heat bath result $\frac{1}{2}\frac{%
1}{1-e^{\beta \omega }}.$

This shows in addition, that one should avoid to analyze particle statistics
in momentum space.

The zero mass model is conformally invariant. This means in particular that
there are other situations with horizons which are conformally equivalent:
the forward light-cone (modular group = dilatations) and double cones
(modular group= radial conformal subgroup). They lead to identical energy
distributions but their physical interpretation is more involved. One very
surprising feature of modular wedge localization is that to compute thermal
expectation values with the KMS condition for given commutation structure
for low dimensional fields is often simpler than to handle ground state
problems with that algebraic structure.

There is another important lesson to be drawn from this section. The modular
localization aspects agree with the geometry of Killing isometries in the
Rindler situation (and its conformal transforms for zero mass matter) with
the Killing time being exponentially related to the (affine) geodesic time
on the horizon. However the modular automorphism group, which exists for any
region with a nontrivial causal complement, vastly generalizes these
concepts to situations where the quantum aspects of localization do not
allow for a geometric interpretation e.g. for general space-time regions as
double cones. In our free field case at hand, this generalization of the
Unruh picture is not describable in terms of finite dimensional geometric
data, but rather in terms of localized subspaces of the Wigner space and
their modular properties i.e. they depend more on the material content and
not only on geometric properties. \textit{This means that the notion of
``horizon'' becomes more ``quantum'' and inherits its properties not so much
from the geometry in space-time, but more from the position of modular
localized quantum subspaces inside Wigner resp. Fock-space.} Even those
algebras localized close to the horizon are expected to become increasingly
``fuzzy'' for increasing ``modular'' times in the modular group action.
Presently these general modular aspects are ill-understood. This remark also
applies to the closely related concept of entropy and the dependence on the
``split'' property. The latter leads to a factorizing Rindler state which
belongs to the same folium as the global vacuum. Of course paradoxes of
quantum theory in the presence of evaporating black holes cannot be
convincingly resolved without an entropy concept for nets.

Note that restriction of pure global states to local algebras always produce
impure (thermal) states on $\mathcal{A}(\mathcal{O})$. This is a
manifestation of the type III$_{1}$ nature of the local versus the type I
global algebras. Overlooking this effect would lead to fake causality
violations.

\section{Examples of Dis(order)-Fields}

In the following we will give an example for a field whose Borchers class is
associated with (but not equal to) that of a free field. This construction
is part of the d=1+1 duality (order- disorder) construction.

Let us start from a complex free Dirac field in d=1+1: 
\begin{eqnarray}
\psi (x) &=&\int \frac{dp}{2\omega }(e^{-ipx}u(p)a(p)+e^{ipx}v(p)b^{*}(p))
\label{Dirac} \\
u(p) &=&\sqrt{\frac{m}{2}}\left( 
\begin{array}{l}
e^{\frac{\theta }{2}} \\ 
-e^{-\frac{\theta }{2}}
\end{array}
\right) ,\quad v(p)=\sqrt{\frac{m}{2}}\left( 
\begin{array}{l}
-e^{\frac{\theta }{2}} \\ 
-e^{-\frac{\theta }{2}}
\end{array}
\right)  \nonumber \\
v(p) &=&u^{C}(p)=Ci\gamma _{0}u(p),\quad p=m(\cosh \theta ,\sinh \theta ) 
\nonumber
\end{eqnarray}
Here we took the following realization of the Dirac equation: 
\begin{eqnarray}
\left( i\gamma _{\mu }\partial ^{\mu }-m\right) \psi &=&0 \\
\,\,\gamma _{0}=\left( 
\begin{array}{ll}
0 & i \\ 
i & 0
\end{array}
\right) , &&\gamma _{1}=\left( 
\begin{array}{ll}
0 & -i \\ 
i & 0
\end{array}
\right)  \nonumber \\
C=i\gamma _{1} &=&\left( 
\begin{array}{ll}
0 & 1 \\ 
-1 & 0
\end{array}
\right)  \nonumber
\end{eqnarray}
This field is $U(1)$ covariant and the local generator is the conserved
current $j_{\mu }=:\bar{\psi}\gamma _{\mu }\psi :$. This $d=1+1$ current has
(relatively to $\psi )$ nonlocal pseudo-potential: 
\begin{eqnarray}
j_{\mu }(x) &=&\varepsilon _{\mu \nu }\partial ^{\nu }\phi (x),\quad \phi
(x)=:F_{x}(a^{\#},b^{\#}):=\int_{-\infty }^{x}\varepsilon ^{\mu \nu }j_{\nu
}d\xi _{\mu }= \\
&=&\int \left\{ 
\begin{array}{c}
\frac{-1}{\sinh \frac{1}{2}(\theta _{p}-\theta _{q}+i\varepsilon )}%
e^{i(p-q)x}\left[ a_{p}^{*}a_{q}-b_{p}^{*}b_{q}\right] + \\ 
+\frac{1}{\cosh \frac{1}{2}(\theta p-\theta _{q})}\left[
e^{i(p+q)}a_{p}^{*}b_{q}^{*}+e^{-i(p+q)}a_{q}b_{p}\right]
\end{array}
\right\} d\theta _{p}d\theta _{q}  \nonumber
\end{eqnarray}
As naively expected, the $\phi $ is a localized field which, although
relatively local with respect to the observables (generated by the current),
fails to be local relative to the field $\psi $ but instead fulfills: 
\begin{eqnarray}
\phi (x),\psi (y) &=&\theta (x_{0}-y_{0})\psi (y)\phi (x) \\
\lim_{x^{2}\rightarrow -\infty ,x^{1}\rightarrow \infty }\phi (x) &=&\sqrt{%
\pi }Q,\quad \quad  \nonumber
\end{eqnarray}
where $Q$ is the global charge. Formally we may write 
\begin{equation}
U_{{}}(2\pi \lambda )_{<}=\exp -2\pi i\int_{-\infty
}^{x_{1}}j_{0}(x_{0},y_{1})dy_{1}=\exp -2\sqrt{\pi }i\lambda \phi (x)
\end{equation}
is the representation for ``half-space'' rotation i.e. U$_{<}$ implements: 
\begin{equation}
\psi (x)\rightarrow \left\{ 
\begin{array}{l}
e^{-2\pi i\lambda }\psi (x)\quad x_{1}<0 \\ 
\psi (x)\quad x_{1}>0
\end{array}
\right.
\end{equation}
Such half-space transformations may be viewed as the point limit of
Bogoliubov transformations. The correct normal product which is necessary in
order to convert U$_{<}$ into a well defined expression, is the ``triple''
ordering. This is also recursively defined but different from the $\phi $%
-Wick product; one subtracts all connected correlation functions and not
just the two-point function. Formally we have the following simple
exponential formula: 
\begin{equation}
\vdots e^{ia\phi (x)}\vdots =\frac{e^{ia\phi (x)}}{\left\langle e^{ia\phi
(x)}\right\rangle }
\end{equation}
re-expressed in terms of the original $\psi $-Wick product we obtain a
nonlocal looking expression, which is best written in terms of rapidities: 
\begin{equation}
\mu (x)=\vdots \exp -2i\sqrt{\pi }\lambda \phi (x)\vdots =:\exp L_{\lambda
}(x):
\end{equation}
\begin{equation}
L_{\lambda }(x)=\frac{\sin \pi \lambda }{2\pi }\int \left\{ 
\begin{array}{c}
\frac{e^{-\lambda (\theta _{p}-\theta _{q})}}{\cosh \frac{1}{2}(\theta
_{p}-\theta _{q})}\left[ e^{i(p+q)x}a_{p}^{*}b_{q}^{*}+h.c.\right] + \\ 
\left[ -\frac{e^{-\lambda (\theta _{p}-\theta _{q})}}{\sinh \frac{1}{2}%
(\theta _{p}-\theta _{q}+i\varepsilon )}\right] e^{i(p-q)x-i\pi \lambda
}a_{p}^{*}a_{q}+ \\ 
\left[ \frac{e^{-\lambda (\theta _{p}-\theta _{q})}}{\sinh \frac{1}{2}%
(\theta _{p}-\theta _{q}-i\varepsilon )}\right] e^{-i(p-q)+i\pi \lambda
}b_{q}^{*}b_{p}.
\end{array}
\right\} d\theta _{p}d\theta _{q}
\end{equation}
Although $L_{\lambda }$ is represented in terms of nonlocal kernels in
rapidity space and is itself nonlocal, $\mu $ is a bosonic local field in
the quantum sense which is however nonlocal relative to $\psi $ i.e. outside
the $\psi $-Borchers-class.. It is easy to see that our special solution $%
\mu $ of the half-space commutation relation with $\psi $ belongs to a whole
family of solutions. We may modify the $\mu $ by any bosonic local function
of the $\psi ^{\#}$ without change in the relative commutation relations.
Within our construction method this is made manifest by the
``quasi-periodicity'' (up to local operators) in $\lambda $ mod 1. With the
help of $\mu $ one can now construct another field $\sigma (x)$ which
carries the same charge as $\psi ,$ but has quite different spacelike
commutation relations. Through the definition: 
\begin{eqnarray}
\sigma (x) &=&N\left[ \mu \psi \right] (x)=\lim_{y\rightarrow x}\mu (x)\psi
(y) \\
&=&\frac{1}{\sqrt{4\pi }}:\int (e^{-ipx}a_{p}+e^{ipx}b_{p}^{*})\mu
(x)d\theta _{p}:  \nonumber
\end{eqnarray}
one obtains the same soliton like relative commutation relations with $\mu $
as those between $\psi $ and $\mu :$%
\begin{equation}
\mu (x)\sigma (y)=\left\{ 
\begin{array}{l}
e^{i\alpha }\sigma (y)\mu (x)\quad x>y \\ 
\sigma (y)\mu (x)\quad y>x
\end{array}
\right.
\end{equation}
However the $\sigma $ carries a fractional spin and ``statistics''(see later
comments). Instead of dual commutation relations one finds symmetric
commutation relations associated with abelian representations of the braid
group i.e. 
\begin{equation}
\sigma (x)\sigma (y)=e^{-2\pi i\lambda sign(x-y)}\sigma (y)\sigma (x)
\end{equation}
These commutation relations appear as a interpolating continuous
generalization of Fermions and Bosons and are called ``anyonic''. Their
relation to particle statistics will be discussed later. The bosonic field
as well as the anyonic field are living in the same Fock space generated by
the free field $\psi $, but they are not members of the $\psi $-Borchers
class. As a local field $\mu $ generates its own Borchers-class (it is an
irreducible field in its own Hilbert space cyclically generated from the
vacuum). The question of whether the notion of equivalence classes of fields
can be generalized to anyonic fields will not be pursued here.

A physically more relevant illustration of duality and non-free Borchers
classes is obtained by starting from a Majorana (selfconjugate) spinor field
($a=b$). In this case the symmetry is the discrete $\mathbf{Z}_{2}$ and the
previous method based on a conserved Noether current is not applicable.
There are however several alternative methods which lead to the following
result ($c\equiv a=b$): 
\begin{eqnarray}
\mu (x) &=&:e^{\Lambda (x)}:,\quad \sigma (x)=s.d.l.:\hat{\psi}(x)\mu (x): \\
\Lambda (x) &=&\frac{i}{4\pi }\int d\theta _{p}d\theta _{q}\left\{ 
\begin{array}{c}
2\coth \frac{1}{2}(\theta _{p}-\theta _{q}+i\varepsilon )\cdot
e^{i(p-q)}c_{p}^{*}c_{q} \\ 
+\tanh \frac{1}{2}(\theta _{p}-\theta _{q})\cdot
(e^{i(p+q)x}c_{p}^{*}c_{q}^{*}-h.a.)
\end{array}
\right\}  \nonumber \\
\hat{\psi}(x) &=&\frac{1}{2\pi }\int d\theta
_{p}(e^{-ipx}c_{p}+e^{ipx}c_{p}^{*})\neq \psi (x)  \nonumber
\end{eqnarray}
Here s.d.l. stands for the operator contribution which carries the leading
short distance singularity). Whereas $\mu $ and $\sigma $ fulfill the
relative $Z_{2}$-duality relation, both fields are bosonic. Hence $\sigma $
generates a new Borchers class in $\mathcal{H}_{F}$ which is inequivalent to
the Fermion Borchers class. It is quite straightforward to show that the
Ising lattice model can be described in terms of lattice Fermions which in
the scaling limit (for fixed correlation length) become Majorana Fermions.
In addition the lattice (dis)order variables go over into ($\mu )\sigma $ if
one takes that limit from the disorder side ($T\rightarrow T_{c}+\varepsilon
)$. So we are justified to call our $\sigma $-fields the (order) ``Ising
fields''. Let us compare the free Majorana field with the Ising field
Borchers class. Consider the modular objects for the wedge algebras of the
two classes. The modular operators $\Delta ^{it}$ are identical and equal to
the wedge-based Lorentz transformations. However the modular reflections $J$
are different. For the free Fermion algebra the Wigner theory gave $J_{F}=K%
\mathcal{F}(j)$ with $j$ being the antiunitary wedge reflection, $\mathcal{F}
$ the CAR-functor and K the Klein transformation in Fock space. The Boson
algebra generated by $\sigma $ on the other hand has $J_{B}=KJ_{F}$ because
its commuting structure for space-like distances requires the absence of the
twist. This can also be read off directly from the TCP transformation
property of $\sigma $ under the TCP in Fock space. Note that the Klein
factor is a global operator whose half-space version is the disorder field $%
\mu $ (the Jordan-Wigner transformation in lattice theory). In our Ising
example $(\mathcal{N}_{F}$: fermion\#): 
\begin{equation}
K=\frac{1+iU}{1-iU},\quad U=e^{i\pi \mathbf{N}_{F}}
\end{equation}
Since the TCP symmetry $\theta $ of $\sigma $ is related to the free field $%
\theta _{0}$ TCP of the Majorana Fermion $\psi $ in the same way as the
above J's, namely by: 
\begin{equation}
\theta =K\theta _{0}
\end{equation}
and since (as will be shown in the section on scattering theory) the unitary
S-matrix is related to the antiunitary $\theta ^{\prime }s$ by $S=\theta
\theta _{0},$ we conclude $K=S$. this means that the S of $\sigma $ is
energy independent and $S^{(2)}=-1$ for two particles. On a somewhat formal
level we can understand this through: 
\begin{equation}
\lim_{t\rightarrow \infty }\mu (x)=\left\{ 
\begin{array}{l}
U \\ 
1
\end{array}
\right. ,\quad \lim_{t\rightarrow \pm \infty }\sigma (x)=\left\{ 
\begin{array}{l}
U\psi \\ 
\psi
\end{array}
\right.
\end{equation}
Writing $U\psi =K\psi K^{*}$, we read off: $S=K$ i.e. the Jordan-Wigner
transformation approaches the global symmetry whose square root is the Klein
transformation (which in this model coalesces with the S-matrix).

Returning for a moment to the $\lambda $-half-space rotation in the previous
complex free Dirac field, we find by the same method in case of rational $%
\lambda =\frac{1}{N}$ ($Z_{N}$ symmetry): 
\begin{eqnarray}
U &=&\sum_{n}e^{-2\pi i\frac{n^{2}}{N}}E_{n} \\
K &=&\sum e^{-i\pi \frac{n^{2}}{N}}E_{n}  \nonumber
\end{eqnarray}
The quadratic n-dependence of the spin-statistic phase on the charge
eigenvalues $\sim n^{2}$ is characteristic for anyonic commutation
relations. The ``exotic'' nature of the commutation relation of this anyonic 
$\sigma $ with itself shows up in the deviation of its modular reflection $J$
from its expected geometric action $J_{0}$. Such a twist factor $K=J\cdot
J_{0}$ should be distinguished from the S-matrix in $S=J\cdot J_{0}$ in
chapter 6. In the latter case $J_{0}$ has the interpretation of the incoming
modular involution and $J$ belongs to the interacting field algebra. Both
involutions are geometric and belong to fields with the same local
commutation relations. In this case there is no need for a twist which
repairs the lost connection between spin and statistics. It should however
be emphasized that in massive d=1+1 theories (different from chiral
conformal or massive d=1+2 theories), the distinction between a scattering
and a twist interpretation of $J\cdot J_{0}$ remains a delicate one. In some
sense statistics in $d=1+1$ massive QFT is an ambiguous notion. Consider the 
$a^{\#}(p)$ of a free Dirac field and write: 
\begin{eqnarray}
c(p) &=&e^{i\pi \lambda \int_{-\infty }^{p}n(p)dp}a(p)  \label{p-anyons} \\
n(p) &=&a^{*}(p)a(p)  \nonumber
\end{eqnarray}
The commutation relation of the c's is now ``anyonic'': 
\begin{eqnarray}
c(p)c^{*}(q) &=&-e^{i\pi \lambda }c^{*}(q)c(p)+2\omega \delta (p-q) \\
c(p)c(q) &=&-e^{i\pi \lambda }c(q)c(p),\quad etc.  \nonumber
\end{eqnarray}
The $c^{\prime }s$ live in the same Fock space, are covariant with respect
to the same representation of the Poincar\'{e}-group and (as it should be)
create the same one particle states, even though the c's have anyonic ( for $%
\lambda =1$ bosonic) commutation relations. This is a special instance of a
general phenomenon: particles in massive $d=1+1$ theories are really
statistical ``schizons'' \cite{S-S} i.e. the nature of their charges (fusion
laws etc.) does not determine their statistics (i.e. a Mendeleev periodic
table in a $d=1+1$ world allows for a bosonic description in terms of long
range interactions). This is very different from all other situations,
including chiral conformal QFT for which the field commutation relations
(the ``exchange algebra'') is uniquely determined in terms of the charges
carried by the fields. Warning: the statistical schizon phenomenon should
not be confused with ``bosonization'' in chiral conformal QFT (see next
section). In the latter case the commutation relations of the charge
carrying fields are uniquely determined by the characteristics of these
charges and cannot be changed. To the extend that the name ``bosonization''
indicates that a conformal fermion can live in the Fock space of bosonic
creation and annihilation operators, this is a misunderstanding.

An exponential bose field cannot be interpreted as just an exponential
function of an object in bosonic Fock space, rather it must be considered as
a short hand notation for an field which defines (through its correlation
functions) its own Hilbert space, which is a charged representation of the
neutral current algebra. In this sense ``bosonization'' is a generic
structure independent of space-time dimensions and conformal invariance: any
charged field (fermionic, for d\TEXTsymbol{<}1+3 also plektonic) can be
interpreted from arising from representing another sector on the
(necessarily) bosonic observable algebra, i.e. in this sense algebraic QFT
is ``bosonization'' par excellence. The only special feature is that the
process of creating a charge by dumping an anticharge ``behind the moon''
can be expressed in terms of a very simple formula e.g. 
\begin{equation}
\psi (x)=\lim_{y\rightarrow \infty }e^{i\pi \int_{x}^{y}jd\xi }
\end{equation}
in which the limit leaves Fock space and produces the new representation
space.

The schizon phenomenon is related to the fact that the natural framework for
massive $d=1+1$ theories is the ``soliton'' framework in which the concept
of braid-group statistics is replaced by the more general ``exotic'' (or
solitonic) commutation relations which can be changed at will (within
certain limitations) without effecting the superselection structure. The
most interesting new phenomenon in the algebraic QFT of solitons is that the
problem of multiple vacuum states, even in situations where this cannot be
blamed on spontaneous symmetry-breaking, becomes related in a profound way
with the superselection structure. We will return to these problems in the
sections on algebraic QFT (the net approach).

We conclude this section by some comments on the algebraic description \cite
{Mü} (independent of fields) of (dis)order. If one assumes that the theory
is given in terms of a field net $\mathcal{F}(\mathcal{O}),\mathcal{O}\in 
\mathcal{K}$ (family of double cones). As usual the observable algebra is
related to the field algebra by the invariance principle with respect to a
symmetry group $G$: 
\begin{equation}
\mathcal{F}(\mathcal{O})^{G}\mid _{H_{vac}}=\mathcal{A(O)}
\end{equation}

Whereas for $d\neq 1+1$ the so defined $\mathcal{A}(\mathcal{O})$
generically (apart from spontaneous symmetry breaking presented in a later
section) is Haag dual if $\mathcal{F}$ had this property (for fermionic $%
\mathcal{F}$ the duality must be twisted): 
\begin{equation}
\mathcal{F}^{tw}(\mathcal{O})^{\prime }=\mathcal{F}(\mathcal{O}^{\prime
})\curvearrowright \mathcal{A}^{\prime }(\mathcal{O})=\mathcal{A}(\mathcal{O}%
^{\prime })
\end{equation}
However in massive d=1+1 theories this conclusion is incorrect for double
cones but remains correct for wedges $\mathcal{O}=W$. The alternative
definition in terms of $\mathcal{A}(W):$%
\begin{equation}
\mathcal{A}^{d}(\mathcal{O})=\bigcap_{W\supset \mathcal{O}}\mathcal{A}%
(W),\quad \mathcal{O}\in \mathcal{K}
\end{equation}
gives a bigger algebra (equal to the dual net) which is Haag dual. It comes
as a bit of a pleasant surprise that the issue of $\mathcal{A}$ versus $%
\mathcal{A}^{d}$ is inexorably linked with the (dis)order structure. In any
massive two-dimensional QFT with an internal group symmetry G (i.e. not just
for free fields) there is a canonical way to introduce half-space
transformations $U_{l}^{\mathcal{O}}(g)$ which implement the full
g-transformation on the spacelike left of the double cone $\mathcal{O}$ and
is equal to the identity on its right. This construction uses the
``split-property'' (equivalent to the ``nuclearity'' i.e. a good phase-space
behavior of QFT), a concept which will be explained in the mathematical
appendix. Assume for the moment that G=$Z_{N}$ i.e. an abelian group which
leads to one half-space generator $U_{l}^{\mathcal{O}},$ $(U_{l}^{\mathcal{O}%
})^{N}\sim \mathbf{1.}$ We then extend the field algebra $\mathcal{F(O)}$ 
\footnote{%
Here the reader is asked to be content with an intuitive understanding of
local algebras and nets. The rigorous mathematics (see the appendix) is not
really necessary for a first glimpse without proofs.} by the disorder
operators $U_{l}^{\mathcal{O}}$%
\begin{equation}
\mathcal{F}_{ext}(\mathcal{O})=\mathcal{F}(\mathcal{O})\vee U_{l}^{\mathcal{O%
}}
\end{equation}
The map $\mathcal{O}\rightarrow \mathcal{F}\mathcal{(O)}$ is still an
isotonous net but it lost locality. The application of the invariance
principle yields: 
\[
\mathcal{F}_{ext}(\mathcal{O})^{Z_{n}}=\mathcal{A}^{d}(\mathcal{O})=\mathcal{%
A(O)}\vee U_{l}^{\mathcal{O}}=\mathcal{A}_{ext}\mathcal{(O)} 
\]
and we may arrange our result in form of the following ``commuting square''
of inclusions \cite{Mü}: 
\begin{eqnarray}
\stackunder{\cup }{\mathcal{A}^{d}\mathcal{(O)}} &\subset &\stackunder{\cup 
}{\mathcal{F}_{ext}\mathcal{(O)}} \\
\mathcal{A(O)\;\;} &\subset &\;\;\mathcal{F(O)}  \nonumber
\end{eqnarray}
Is there an invariance principle which describes the entire commuting
square, in particular of $\mathcal{A(O)}\subset \mathcal{F}_{ext}\mathcal{(O)%
}$? In the above abelian $Z_{N}$- illustration the half-space
transformations $U$ commute with $G=Z_{N}$ but they suffer a nontrivial
action of the dual group $\hat{Z}_{N}$ ($\sim Z_{N})$: 
\begin{eqnarray}
\alpha _{\chi }(U(g)) &=&\chi (g)U(g),\quad \mathcal{A}^{d}(\mathcal{O}%
)^{Z_{N}}=\mathcal{A(O)} \\
g &\in &Z_{N},\quad \chi \in \hat{Z}_{N}  \nonumber
\end{eqnarray}
Whereas on $\mathcal{F}$ only $G$ acts, $\mathcal{F}_{ext}$ is a natural
domain for the action of the ``double'' $G\times \hat{G}:$%
\begin{equation}
\mathcal{A(O)=(F}_{ext}\mathcal{(O)})^{G\times \hat{G}},\quad \mathcal{F(O)}=%
\mathcal{F}_{ext}(\mathcal{O})^{\hat{G}}
\end{equation}
It turns out that this has an interesting counterpart for nonabelian G's. In
that case the ``double'' has to be taken in the sense of Drinfeld which is
the cross product Hopf algebra which was introduced in the third section of
the first chapter: 
\begin{equation}
D(G)=C(G)\Join _{adj}G
\end{equation}
However in contradistinction to ordinary group symmetry, the double is
always spontaneously broken and maximally only G survives as an unbroken
symmetry. This mathematical manifestation of the (dis)order structure is
presently the only case for which Hopf algebras emerge naturally from
physical principles.

The expert reader recognizes the close relation with the global
Kramers-Wannier Duality of statistical mechanics for lattice systems. In
fact the formal scaling limit near a critical point towards a continuous QFT
maintains the local (the Kadanoff-)form of the K-W\ symmetry\footnote{%
The temperature becomes traded for the mass, but there is no ''dual mass''
which could substitute for the dual K-W temperature. Rather the dual
symmetry becomes a pure algebraic concept in the sense of Kadanoff. It is
always spontaneously broken, except in the conformal scale invariant zero
mass limit.}, although the Kramers-Wannier Duality and the notion of dual
temperature $T^{*}$ gets lost. The presentation of this section shows the
close relation of the statistical mechanics duality concepts to the Haag
duality of algebraic QFT. But whereas the former (in its relation to charge
sectors) is limited to $d=1+1$, Haag duality (and its controlled
breaking)\thinspace does not suffer such a limitation.

As already mentioned, all this rich structure (including the statistical
``schizon'' aspect of d=1+1 particles) may be subsumed into the algebraic
QFT framework for soliton sectors\cite{Schli} which will not be presented in
these notes.

Finally it is worthwhile to remark that all of the known $d=1+1$ (dis)order
models allow for a euclidean functional (Feynman-Kac)representation
involving an external Aharonov-Bohm potential or (in the case of additive
symmetries i.e. translations in field space) external ``electric and
magnetic'' sources. Some more details can be found in the section on
functional methods.

\section{Special Features of m=0, d=1+1 Fields}

It is well-known that the zero mass limit of massive free fields enhances
the space-time symmetry to the conformal group symmetry. In addition to the
general well understood peculiarities of such an extension (Einstein
causality ``paradox'' as the result of a continuous link through infinity of
the space- and time-like), there is a very surprising phenomenon which only
happens for $d=1+1$: there are continuously many local quantum theories in
the holomorphy region between the real and imaginary time boundary values.
Let us verify this for the massless Dirac field which results from the
formula (\ref{Dirac}): 
\begin{eqnarray}
\psi _{r}(v) &=&\frac{1}{\sqrt{2\pi }}\int_{0}^{\infty
}dp(e^{-ipv}a_{l}(p)+e^{ipv}b_{l}^{*}(p)),\quad v=t-x  \label{f} \\
\psi _{l}(u) &=&\frac{1}{\sqrt{2\pi }}\int_{0}^{\infty
}dp(e^{-ipu}a_{r}(p)+e^{ipu}b_{r}^{*}(p)),\quad u=t+x  \nonumber
\end{eqnarray}
where the right and left movers $a_{r}^{\#}(p),b_{r}^{\#}(p)$ and $%
a_{l}^{\#}(p),b_{l}^{\#}(p)$ anticommute with each other i.e. the chiral
fields $\psi _{r,l}$ define independent chiral theories. Therefore from now
on we will select one chirality and omit the r,l subscript.

As expected the two-point function can be rewritten in compact circular
coordinates $z=e^{i\varphi \text{ }}$by introducing an appropriately
normalized field $\psi _{c}$ via stereographic projection (\ref{ste}): 
\begin{equation}
\left\langle \psi (u)\psi ^{*}(u^{^{\prime }})\right\rangle =\frac{1}{%
u-u^{\prime }+i\varepsilon }\curvearrowright \left\langle \psi _{c}(z)\psi
_{c}^{*}(z^{\prime })\right\rangle =\frac{1}{z-z^{\prime }(1+\varepsilon )}
\end{equation}
where the $\varepsilon $-prescription in the last formula expresses the
radial ordering. From this one reads off the spatial invariance group. It is
the 3-parametric Moebius-group $SL(2R)$ in u or $SU(1,1)$ in $z$. The
analytic continuation of this two-point function has a positive definite
restriction onto any (Jordan) curve C circulating around zero i. e. 
\begin{equation}
\left( f,g\right) =\oint_{C}\oint_{C}\frac{1}{z(\gamma (t))-z^{\prime
}(\gamma (t^{\prime }))}\bar{f}(t)g(t^{\prime })dtdt^{\prime }
\end{equation}
is a (positive) scalar product, leading to a Hilbert space. Since the higher
point correlation functions are products of the two-point functions, we
obtain positivity. The algebra is still a CAR-algebra, but the quasifree
state defined through $C$ differs from the vacuum state. The unitary
equivalence of the representations is easily checked with the
Hilbert-Schmidt criterion of the first chapter. This means that the
transformation of the circle theory (the ``living space'' of the original
real time theory) to the parametrized curve $C$ theory is unitarily
implemented. This emergence of the spatial automorphism associated to the $%
Diff(S^{1})$ group and their unitary implementation from the existence of a
family of noncommutative Wightman theories between the original angular
circle and the euclidean (radial) cycle is the special property which fails
in any other dimension. The infinitesimal manifestation is the well-known
Lie-field structure of the energy-momentum tensor, in our case \footnote{%
The triple dot denotes Wick ordering according to the frequency
decomposition of $j$ whereas the double dot refers to Fermion Wick ordering.}%
: 
\begin{eqnarray}
T(z) &=&\vdots j^{2}(z)\vdots ,\quad j(z)=:\psi ^{*}(z)\psi (z): \\
\left[ j(z_{1}),j(z_{2})\right] &=&i\delta ^{\prime }(z_{12})  \label{cur} \\
\left[ T(z_{1}),T(z_{2})\right] &=&-i\frac{1}{12}\delta ^{\prime \prime
\prime }(z_{12})+i(T(z_{1})+T(z_{2}))\delta ^{\prime }(z_{12})  \nonumber
\end{eqnarray}
The generic $T-$algebra is obtained from this special case by replacing the
factor $\frac{1}{12}$ by $\frac{c}{12}$ with a positive $c$ which turns out
to take on any value above $c=1$ and is quantized below this free value. The
(nonlocal) Fourier components with respect to the rotation group lead to
Kac-Moody and Virasoro algebras. The latter have commutation relations: 
\begin{equation}
\left[ T_{n},T_{m}\right] =(n-m)T_{n+m}+\frac{c}{12}n(n^{2}-1)\delta _{n+m,0}
\end{equation}
In this form it has been discovered by Virasoro while studying the dual
S-matrix model of Veneziano. The local field theoretic version (\ref{cur})
on the other hand was found in 1973 (for the early history on
low-dimensional models see \cite{S2} within two-dimensional QFT while
persuing the construction of ``Lie Fields'', which were introduced in J. H.
Lowenstein's thesis in 1968. The latter should be considered as the early
version of what nowadays is called W-algebra.

Since all irreducible L-representations are one dimensional, the Lorentz
spin s (halfinteger) fields can all be represented in one bosonic (or
fermionic) Fock space generated by bosonic (or fermionic) $a^{\#},b^{\#}$
which are independent of s. It is also easy to see that it is not possible
to generalize this to arbitrary L-spin s i.e. to construct fields $\phi (u)$
with anyonic commutation relations: 
\begin{eqnarray}
\phi (u)\phi (u^{\prime }) &=&e^{2\pi i\lambda }\phi (u^{\prime })\phi
(u),\quad u>u^{\prime }  \label{any} \\
\phi (u)\phi ^{*}(u^{\prime }) &=&e^{-2\pi i\lambda }\phi ^{*}(u^{\prime
})\phi (u),\quad u>u^{\prime }  \nonumber
\end{eqnarray}
within the setting of Fourier transforms of creation/annihilation operators.
Only anyonic momentum space operators as in (\ref{p-anyons}) can be
constructed in this way.

The rich world of general chiral QFT begins to open if one realizes the
peculiar role of the scalar free field: 
\begin{equation}
\varphi (x)=\frac{1}{\sqrt{2\pi }}\int (e^{-ikx}a(k)+h.a.)\frac{dk}{2\omega }%
=\varphi (t+x)+\varphi (t-x)
\end{equation}
Due to the infrared divergence in this representation, the pointlike $%
\varphi (x)$ does not exist, only $\varphi (f)$ with $\tilde{f}(0)=0$
defines an operator in Fock space. In order to maintain well defined local
generators in Fock space, we consider the infrared finite first derivative $%
j(u)=\partial _{u}\varphi (u),$ u=t+x. A simple calculation shows that $j$
can also be obtained as the chiral current of a free fermion field (\ref{f}%
). Its commutation relations define the abelian current algebra (\ref{cur}).
We now use the Weyl functor (\ref{st, fac}): 
\begin{eqnarray}
W(f) &=&e^{i\int j(x)f(x)},\quad W(f)W(g)=e^{i\frac{1}{2}\sigma (f,g)}W(f+g)
\nonumber \\
\sigma (f,g) &=&\frac{1}{2}\int \left\{ f(x)\partial g(x)-g(x)\partial
f(x)\right\} dx  \label{Wey}
\end{eqnarray}
In order to make the M\"{o}bius-covariance of this algebra manifest, one
uses the angular parametrization for the compactified line: 
\begin{equation}
u\rightarrow z=\frac{i-u}{i+u}\,\,\quad z=e^{i\varphi }  \label{ste}
\end{equation}
In terms of this compact description, the above symplectic form $\sigma $
becomes: 
\begin{eqnarray}
\sigma (f,g) &=&\int \frac{dz}{2\pi i}f^{\prime }(z)g(z)=\sum nf_{n}g_{-n}
\label{sym} \\
f(z) &=&\sum_{n}f_{n}z^{n},\quad f_{n}^{*}=f_{-n}  \nonumber
\end{eqnarray}
Thanks to the aforementioned infrared property which forced us to define the
Weyl algebra in terms of currents instead of fields, the symplectic inner
product (\ref{sym}) is degenerate since it vanishes on constant functions
(''zero modes''). These are carried into the center of the abstract
C*-algebra which is generated freely from the $W^{\prime }s,$ subject to the
Weyl relation (\ref{Wey}). The center defines an abelian charge algebra and
there are continuously many superselected charge sectors obtained by
diagonalization of the center.

In order to come to a more interesting situation one must extend the Weyl
algebra by a lattice so that the extended algebra is not only characterized
by the linear space of functions, but in addition has an underlying lattice.
In mathematical terms the linear spaces are extended by ``noncommutative
tori''. In order to allow for sufficient generality, we start from a
multi-component abelian current algebra: 
\begin{equation}
\left[ J^{i}(z_{1}),J^{j}(z_{2})\right] =-g^{ij}\delta ^{\prime
}(z_{1}-z_{2}),\quad i,j=1...N  \label{current}
\end{equation}
where $\delta ^{\prime }(z_{1}-z_{2})$ is the appropriate circular $\delta -$%
function: 
\begin{equation}
\int \frac{dz^{\prime }}{2\pi i}f(z^{\prime })\delta ^{\prime }(z^{\prime
}-z)=-f^{\prime }(z)
\end{equation}
The symplectic form which now lives in $LV$ i.e. smooth loops in the $N$%
-dimensional vector space $V$ is given by: 
\begin{equation}
\sigma (f,g)=\int \frac{dz}{2\pi i}\left\langle f^{\prime
}(z),g(z)\right\rangle
\end{equation}
where $\left\langle \cdot ,\cdot \right\rangle $ denotes the inner product
in $V$ given in terms of the positive definite metric $g^{ij}.$ The
M\"{o}bius-group acts on $LV$ as: 
\begin{equation}
(u(g)f)(z)\equiv f_{g}(z):=f(g^{-1}z),\quad g\in PSU(1,1)
\end{equation}
and leaves $\sigma $ invariant.

We are interested to classify all positive energy representations.

\begin{theorem}
Every covariant positive energy representation $(\pi ,H_{\pi })$ of the $%
C^{*}$algebra $\mathcal{U}$ generated by the Weyl-operators $W(f)$
associated with $\sigma $ is a direct sum of irreducible ground state
representations i.e. $H=\sum_{i}H_{i}$, $H_{i}=\overline{U\Omega _{i}},$ $%
\Omega _{i}=$ground state in $H_{i}$
\end{theorem}

We may recover the current algebra fields (\ref{current}) if we restrict to
regular representations i.e. those which are related to states $\omega $
fulfilling continuity of the function $\lambda \rightarrow \omega (W(\lambda
f).$ This suggests the question whether the irreducible components can be
created by applying (smeared out) covariant fields to the vacuum i.e. if the
net point of view and the more standard field point of view are not only
based on the same physical pictures but are even mathematical equivalent.
Here we will only quote the result and leave the proof up to the last part
on algebraic QFT.

\begin{theorem}
Every regular ground state representation of the (abelian or nonabelian)
current- or the energy momentum tensor- algebra is generated by
charge-carrying localized fields. The currents and the energy momentum
tensor commute with the charge-carrying fields for noncoalescing points
whereas the latter can be chosen in such a way that they obey braid group
commutation relations (special case: permutation group) among themselves.
\end{theorem}

The representation theory of the above multicomponent Weyl-algebra is not
very interesting since it processes a continuous set of representations
labeled by additive (multicomponent) charges. They are generated by the
following localized automorphisms $\gamma _{\rho }$ \cite{BMT}: 
\begin{eqnarray}
\gamma _{\rho }(W(f)) &=&e^{i\rho (f)}W(f) \\
\rho (f) &=&\oint \frac{dz}{2\pi i}\rho (z)f(z)  \nonumber
\end{eqnarray}
Here the $N$-component function $\rho $ is local with support $\subset S^{1}$
so that $\gamma _{\rho }$ acts as the identity if $f$ and $\rho $ have
disjoint supports. The total charge $q=\rho (1)$ labels automorphism classes
which are ``inner equivalent'' i.e. for which there are unitaries $u(\sigma
,\rho )\in \mathcal{U}$ which intertwine between the two automorphisms: 
\begin{eqnarray}
\gamma _{\sigma }(W) &=&u(\sigma ,\rho )\gamma _{\rho }(W)u^{*}(\sigma ,\rho
)\quad W\in \mathcal{U}  \nonumber \\
u(\sigma ,\rho ) &=&W(f_{\sigma ,\rho })\quad if_{\sigma ,\rho }^{\prime
}=\sigma -\rho
\end{eqnarray}
These equivalence classes of automorphisms are also referred to as abelian
(superselection)sectors of $\mathcal{U}.$ They exhaust the locally generated
sectors of $\mathcal{U}$.

The properties of the automorphism immediately translate into properties of
the associated representations $\pi _{\rho }(W):=\pi _{0}\cdot \gamma _{\rho
}(W)$ where $\pi _{0}$ denotes the vacuum representation. So the charge
distribution $\rho (z)$ ``measures'' the local deviation from the vacuum.
The representation formalism is more close to the standard formulation of
QFT. Vice versa any representation of $\mathcal{U}$ which deviates only
locally from the vacuum (suitably defined) can be shown to allow a
representation of the above form in terms of a local automorphism. This is a
special case of a general theory, the so-called DHR (Doplicher, Haag,
Roberts) theory, which we will meet in the section on algebraic QFT. These
representations of the abelian current algebra was essentially known (though
in a more conventional field theoretic language which is less precise)
already by the author in collaboration with Swieca \cite{Swieca-S}\cite
{Swieca-S-V}.

The $C^{*}-$Weyl algebra $\mathcal{U}$ may be used as building blocks of
structurally richer and more interesting $C^{*}-$algebras. The first step in
this direction is the process of extensions by incorporating local sectors
into the algebra. The naturalness of the so-called lattice-(or
noncommutative torus-)extensions is best understood by looking first at
subalgebras of $\pi _{0}(\mathcal{U})$ belonging to disconnected
localization regions: 
\begin{eqnarray}
\mathcal{A}_{L}(I_{1}\cup I_{3}) &=&Alg\left\{ \pi _{0}(W(f))\mid f\in 
\mathcal{S}_{L}(I_{1}\cup I_{3})\right\} \\
\mathcal{S}_{L}(I_{1}\cup I_{3}) &=&\left\{ f\in \mathcal{S}\mid
f=const.\,in\;I_{2},I_{4};\;\;f(z_{2})-f(z_{4})\in 2\pi L\right\}  \nonumber
\end{eqnarray}
Here $Alg$ stands for the generated von Neumann algebra, $\mathcal{S}$ is
the Schwartz space of smooth test functions on the circle, $z_{2,4}$ are two
arbitrary points from $I_{2,4}$ and $L$ is an even lattice in $V$. The
commutant of this operator algebra acting on the vacuum Hilbert space is
not, as one could expect by a naive application of Haag duality equal to $%
A_{L}(I_{2}\cup I_{4}),$ but it rather equals the bigger algebra \cite{Sta}: 
\begin{eqnarray}
A_{L}(I_{1}\cup I_{3})^{\prime } &=&A_{L^{*}}(I_{2}\cup I_{4}) \\
L^{*} &=&dual\;of\;L  \nonumber
\end{eqnarray}
The reason for this state of affairs becomes clearer if one looks at the
physical interpretation of these algebras. The $I_{1}\cup I_{3}$ localized
algebra contains, in addition to the naively expected operators which are
separately neutral in $I_{1}$and $I_{3}$(zero values of $f$ in $I_{2,4}$),
also operators which are only globally neutral but locally charged with $%
I_{3}$ containing the compensating (anti-)charge to that in $I_{1}.$ The
dual charge (being described by the dual lattice $L^{*}=V/L$) consists
precisely of those values which lead to relative local commutativity: 
\begin{eqnarray}
W(f)W(g) &=&e^{i\sigma (f,g)}W(g)W(f)\quad f\in \mathcal{S}(I_{1}\cup
I_{3}),\,g\in \mathcal{S}(I_{2}\cup I_{4}) \\
\sigma (f,g) &=&2\pi \left\langle l\cdot .l^{*}\right\rangle =2\pi \cdot
integer\quad  \nonumber
\end{eqnarray}
The existence of these dual subalgebras of the vacuum representation of $%
\mathcal{U}$ suggests to look for extensions of $\mathcal{U}$ by lattices in 
$V.$ For this purpose it is convenient to introduce homogeneous
charge-transfer operators $\Gamma _{\alpha },\alpha \in V$ in a subspace $%
H_{L}\in H_{uni}$ defined in the following. $H_{uni}$ is the (nonseparable)
universal representation space which is simply the direct sum of all charged 
$q$ representations for all real values of $q.$ $H_{L}\in H_{uni}$ contains
only those charges lying on the $L-$lattice and $\Gamma _{\alpha }$ creates
a charge $\alpha \in L.$ As the charge $q$ representation we simply take for 
$H_{q}$ a copy of the vacuum representation Hilbert space $H_{0}$ but with $%
\mathcal{U}$ acting through $\pi _{0}(\gamma _{\rho }(W)).$ Hence the
universal representation is: 
\begin{eqnarray}
(\pi _{uni}(W)\phi )_{\alpha } &=&\pi _{0}(\gamma _{\rho }(W))\phi _{\alpha
}\quad \phi _{\alpha }\in H_{0} \\
(\Gamma _{\alpha }\phi )_{\beta } &=&\phi _{\beta -\alpha }  \nonumber
\end{eqnarray}
and the restriction to $L$ means that this formula is restricted to $%
H_{L}=\sum_{\alpha \in L}H_{\alpha }$ i.e. all charge indices $\alpha ,\beta
\in L.$ In particular the vacuum considered as a vector in $H_{\beta -\alpha
}$ is mapped into the vacuum but this time considered as a vector in $%
H_{\beta }.$ We will denote the universal representation restricted to $%
H_{L} $ as $\hat{\pi}.$ In order to speak about the ground state in each
charge sector we need a hamiltonian. In conformal field theory there are
two: the time translation and the rigid $S^{1}$ rotation generator $L_{0}$.
For the present discussion we only need the action of the rigid rotations: 
\begin{equation}
(R(\tau )\phi )_{\alpha }:=e^{\frac{i}{2}\left\langle \alpha ,\alpha
\right\rangle \tau }e^{iL_{0}\tau }\phi _{\alpha }
\end{equation}
Then ground states are mapped into ground states and $\Gamma _{\alpha }$
commutes with $R(\tau )$ (rotational homogeneity) and the ground state
energy in the sector $\alpha $ is $\frac{1}{2}\left\langle \alpha ,\alpha
\right\rangle .\Gamma _{\alpha }$ implements a nonlocalized automorphism: 
\begin{eqnarray}
\Gamma ^{*}\hat{\pi}(W)\Gamma &=&\hat{\pi}(\gamma _{\alpha }(W))\quad \gamma
_{a}(W(f))=e^{i\left\langle \alpha ,f_{0}\right\rangle }W(f) \\
f_{0} &=&\oint \frac{dz}{2\pi i}\frac{1}{z}f(z)\,\,i.e.\,\rho _{\alpha }(z)=%
\frac{1}{z}\alpha  \nonumber
\end{eqnarray}

Localized charge carrying operators in the same charge class with prescribed
support properties for $\rho _{\alpha }(z)\,$may be obtained by modifying $%
\Gamma _{\alpha }$ with a Weyl operator: 
\begin{equation}
\psi _{\rho _{\alpha }}^{\zeta }=\eta _{\xi }(\rho _{\alpha })\hat{\pi}(W(%
\bar{\rho}_{\alpha })\Gamma _{\alpha }  \label{psi}
\end{equation}
One easily checks that the necessary test function $\bar{\rho}_{\alpha }$
solves the first order differential equation: 
\begin{eqnarray}
\frac{d}{dz}\bar{\rho}_{\alpha }(z) &=&i(\rho _{\alpha }(z)-\frac{\alpha }{z}%
) \\
\bar{\rho}_{\alpha }(z) &=&i\sum_{n\neq 0}(\rho _{\alpha })_{-n}\frac{z^{n}}{%
n}-i\int \frac{dz}{2\pi i}\rho _{\alpha }(z)ln_{\zeta }(z)  \nonumber
\end{eqnarray}
Here $\zeta \in S^{1}$ denotes the direction of the cut along the line $%
\left\{ \lambda \zeta \mid \lambda \geq 0\right\} $ which is necessary in
order to define the branches of the logarithm. Remember that $\zeta =-1$
corresponds to infinity in the original $\mathbf{R-}$parametrization. Here
and in the following statement the lattice restriction is not yet needed,
everything holds within the universal representation for an arbitrary charge 
$\alpha .$ The numerical factor $\eta $ will be determined later. One finds:

\begin{theorem}
If $\zeta \notin $supp$\rho $ the charged operators $\psi _{\rho _{\alpha
}}^{\zeta }($\ref{psi}) are local with respect to the observables $U$ and
fulfill the following relations: 
\begin{eqnarray}
(i)\,\quad \psi _{\rho _{\alpha }}^{\zeta }\psi _{\rho _{\beta }}^{\zeta }
&=&e^{\pm i\pi \left\langle \alpha ,\beta \right\rangle }\psi _{\rho _{\beta
}}^{\zeta }\psi _{\rho _{\alpha }}^{\zeta },\quad if\,\,\,supp\rho _{\alpha
}\cap supp\rho _{\beta }=\emptyset \\
\quad (ii)\quad \psi _{\rho }^{\zeta _{1}}(\psi _{\rho }^{\zeta _{2}})^{*}
&=&e^{-\sigma i\pi \left\langle \alpha _{\rho },\alpha _{\rho }\right\rangle
}e^{2\pi i\left\langle Q,\alpha \right\rangle },\quad if\,\;supp\rho \subset
S^{1}\backslash \left\{ \zeta _{1},\zeta _{2}\right\}  \nonumber \\
(iii)\quad R(\tau )\psi _{\rho }^{\zeta }R^{*}(\tau ) &=&\psi _{r(\tau )\rho
}^{\zeta },\quad if\,\,\,\eta _{\zeta }=e^{\frac{i}{2}\int \frac{dz}{2\pi i}%
\left\langle \alpha _{\rho },\rho \right\rangle ln_{\zeta }(z)}  \nonumber
\end{eqnarray}
\end{theorem}

The sign in (i) is coupled to the orientation of the path going from $I_{1}$
to $I_{2}$ through $\zeta .$ The $\sigma $ in (ii) denotes $0,\pm 1$
according to whether the path which connects $\zeta _{1}$ with $\zeta _{2}$
and runs through supp$\rho $ in addition runs through -$1$ $(\sigma =0)$ or
not ($\sigma =\pm 1$, depending on the $\pm $ orientation). The charge
measuring operator Q is defined by: 
\begin{equation}
(\left\langle Q,\beta \right\rangle \Phi )_{\alpha }:=\left\langle \alpha
,\beta \right\rangle \Phi _{\alpha }
\end{equation}
The theorem is easily verified by explicit calculation.

Property (i) tells us that bosonic local fields correspond precisely to even
lattices: 
\begin{equation}
\left\langle \alpha ,\beta \right\rangle =2n,\quad n=0,\pm 1,\pm 2,...
\end{equation}
Restricting to such lattices $L$, the right hand side of (iii) applied to $%
H_{L}$ is equal to one and therefore independent of $\zeta $ i.e. those
fields live on $S^{1}$ (and not a covering thereof).

Now we change our standpoint by and consider the von Neumann algebra $%
\mathcal{A}_{L}$ generated by the extended operators together with its new
(neutral) vacuum and the representation space $H_{L}.$ It turns out that the
new net $\mathcal{A}_{L}$ has only finitely many additional positive energy
representations. They are labelled by points on the dual lattice $%
L^{*}\supset L$ modulo $L$ i.e. in $L^{*}/L.$ Lattices with $L^{*}=L$ are
called selfdual. They only have one sector (the vacuum sector) and they
fulfill the ``split'' Haag duality: 
\begin{equation}
\mathcal{A}(I_{1}\cup I_{3})=\mathcal{A}(I_{2}\cup I_{4})^{\prime }
\end{equation}
A famous illustration is the root lattice of $E_{8}$ as well as the Leech
lattice $\Lambda _{24}.$

The charge sectors of $\mathcal{A}_{L}$ corresponding to the abelian group $%
L^{*}/L$ can again be described in a manner similar to the previous formula: 
\begin{eqnarray}
\phi _{\rho _{\alpha }}^{\zeta } &:&=e^{i\pi (Q,\nu _{\alpha })}\psi _{\rho
_{\alpha }}^{\zeta }\mid _{H_{L^{*}}}\quad \nu _{\alpha }:=\lambda _{\alpha
}+\sum_{\beta }\left\langle \lambda _{\beta },\mu _{\alpha }\right\rangle
\lambda _{\beta }  \label{field} \\
\mu _{\alpha } &:&=\sum_{i=1}^{N-1}\left\langle \alpha ,\alpha
_{i}\right\rangle \lambda _{i}\quad i=2...N,\quad \mu _{1}=0  \nonumber
\end{eqnarray}
i.e. $\nu _{a}$ is a linear combination of the dual lattice basis vectors $%
\lambda _{j}:\left\langle \alpha _{i},\lambda _{j}\right\rangle =\delta
_{ij} $ $i,j=1...N$. The first factor in $\phi $ is a Klein factor which
plays a similar role as previously namely it adjusts certain commutation
relations to standard form, in this case relative commutation with the
observables $\mathcal{A}_{L}$ for disjoint localization. Again the unitary $%
\phi ^{\prime }s$ implement localized automorphisms. It is easy to see that
these sectors exhaust the possibilities of finite energy sectors. The
restriction to $L^{*}-$charges results from the requirement that the action
of $L^{*}$ exhaust the possibility of leaving the set of $L-$ charges
invariant. The phenomenon of charge quantization by charge extension is a
special case of the very general phenomenon of decrease in the number of
superselection sectors with increasing size of algebraic extension. Note
that the generators of $\mathcal{A}_{L}$ are in physical terms loops which
close modulo $2\pi L$ (and hence lead to univalued phase factors) instead of
the ``Weyl loops'' $f$ in $W(f)$ for which $f$ is strictly periodic.
Mathematically they consist of ``affine Hilbert spaces'' i.e. multicomponent
functions on the interval $\left[ 0,2\pi \right] $ which fulfill lattice
boundary conditions i.e. a combination of two well studied objects: Weyl
algebras over vector spaces with a (possibly degenerate) symplectic form and
Weyl-like algebras over (not necessarily even) lattices. The noncommutative
tori of the mathematicians as well as the external magnetic field problems
of Hofstedter are illustration of the latter. Whereas the von Neumann
uniqueness applies to regular representations of the Weyl algebras over
finite dimensional space with a nondegenerate $\sigma ,$ the tori algebras
are never simple and therefore have several representations.

Having constructed all the charge sectors of the extended observable algebra 
$\mathcal{A}_{L}$ one may consider the field algebra $\mathcal{F}_{L^{*}}$
generated by all the charge carrying fields (\ref{field}). It is easy to
establish the following theorem \cite{S Ben}.

\begin{theorem}
$\mathcal{F}_{L^{*}}(I)=\mathcal{F}_{L^{*}}(I^{\prime })^{tw}$
\end{theorem}

Here the twist $tw$ is a generalization of the fermionic twisted commutant.
As in that case one must ``twist'' the von Neumann commutant with a Klein
transformation which also in this case is a ``square root'' of the unitary
operator which represents the $2\pi $ rotation e$^{-i2\pi L_{o}}\cite{S Ben}$%
.This deviation of quantum physics and geometry increases with increasing
amount of non-commutativity (from Fermions to Plektons) and naturally also
makes the modular theory of e.g. anyonic field algebras for the wedge
regions (in chiral conformal theories just intervals) less geometric than
that of the observable algebras.

Besides the above extensions there is one other mechanism namely that of
factorizing the observable net $\mathcal{U}$ by a subgroup of its symmetry
group. In the case of one current there is just the charge conjugation: $%
j\rightarrow -j,$ whereas in the multicomponent case there are more
possibilities. One finds new representations for the fixed point algebras.
Some of these representations are not extendable to representations of the
original $\mathcal{U}(S^{1})$ but only to the noncompact $\mathcal{U}(R).$
These are called soliton representation because their charge distributions
behave differently for $x\rightarrow \pm \infty .$ If one prefers vague
analogies to differential geometry to concrete expressions from physics, one
may also call them ``orbifolds''.

The crucial remaining question is whether there exists a purely field
theoretical systematics which also leads to the more interesting
representations of algebras in which the charge sectors have branched fusion
laws (as current algebras associated to nonabelian groups and W-algebras).
The characteristic feature of those algebras is that they have
representations $\pi $ with nontrivial statistical (or quantum) dimensions $%
d_{\pi }>1$ and endomorphisms instead of automorphisms. Since both concepts
are far removed from standard QFT (Lagrangians etc.), their explanation will
be postponed to the last chapter. Here we will only sketch how by
``amplification'' and ``reduction'' one may get away from the
lattice-extended Weyl algebras.

By amplification we mean tensor products and in particular our interest is
to study nets formed by diagonal tensor products of extended observable
algebras: 
\begin{equation}
\Psi ^{(k)}(\rho )=\psi (\rho )\times \psi (\rho )\times ....\psi (\rho
),\quad \alpha _{\rho }\in L
\end{equation}
where the tensor factors are of the form (\ref{psi}) which we now write as $%
\psi (\rho )$. If we would follow the logic of loop-groups, we would chose $%
L=$ root lattice of e.g. $SU(n)$ and $exp\rho \in $loop-group. Technically
speaking one is dealing with a tensor product of k level one loop-group
representations. It is well known that by reduction one obtains the higher
level representations (with nontrivial branching laws) of the loop-group.
There are also arguments by which W-algebras are related with current
algebras through an invariance principle.

On the other hand a classification of admissible statistics by methods of
algebraic QFT (exchange algebras with braid-group commutation relations)
leads to 4-point functions which in simple cases exactly match the two
families of current- and W-algebras \cite{RS}. This strongly suggests that
the two families and possibly additional models with higher Virasoro
c-values modulo 8 (see chapter 7 section 4) exhaust the possible plektonic
(nonabelian braid group) commutation relations with finitely many sectors
(''rational''). A direct proof that the amplification and reduction
procedure leads to a family of irreducible nets, among which the nets with a
finite number of plektonic charge sectors (rational theories) are exhausted
by nonabelian current algebras and W-algebras, is still missing.

\section{Counting of Localized Degrees of Freedom}

In QM one finds that one degree of freedom occupies a phase space volume of (%
$2\pi )^{3}.$ This result simply follows from the discretization of momentum
space by enclosing the system in a box. Interactions modify this result
somewhat but the number of degrees in a finite phase space cell remains
always finite. As a consequence of the correct relativistic modular
localization, the number of degrees of freedom in a e.g. double cone region $%
\mathcal{O}$ with energy below $E$ turns out to be infinite but ''almost
finite''. Mathematically the following map $\Theta _{\Omega ,E}:$ $\mathcal{A%
}(\mathcal{O})\rightarrow \mathcal{H}$ is compact: 
\begin{equation}
\Theta _{\Omega ,E}(A)=P_{E}A\Omega ,\,\,\,A\in \mathcal{A}(\mathcal{O})
\end{equation}
Here $P_{E}$ denotes the projector onto the Fock space $\mathcal{H}$ below $%
E $ and compact means that the unit ball of the algebra $\mathcal{A}(%
\mathcal{O})$ is mapped into a compact set of vectors, i.e. one for which
each infinite sequence contains a convergent subsequence. Without loss of
physical insight we may take $\mathcal{A}(\mathcal{O})$ to be the previously
constructed free field algebra with $\mathcal{A}(\mathcal{O})\Omega $ the
dense modular localization space. This statement we owe to the far
sightedness of Haag and Swieca \cite{Haag}, who started the issue of
counting of degrees of freedom in phase space of local quantum physics way
back in the early 60$^{ies}.$ The reason why the number was not finite but
rather compact was precisely the mentioned denseness (the Reeh-Schlieder
property of local quantum physics) of localized states. The Newton-Wigner
localization (which is the localization of QM translated into QFT) would
give finiteness, but it is inappropriate because it is not Einstein-causal.
Actually the estimate of Haag and Swieca is unnecessarily conservative; by
using better estimates one finds that the above sets of vectors are not only
compact but also nuclear \cite{Haag}.

It turns out that there is an easier formulation of this physical property
if one does not use a sharp energy cutoff but takes a hint from thermal
aspects of either the heat bath-or the modular localization physics. The
following maps $\Theta :\mathcal{A}(\mathcal{O})\rightarrow \mathcal{H}$
turn out to be nuclear: 
\begin{equation}
\Theta _{\Omega ,\beta }(A)=e^{-\beta H}A\Omega ,\,\,\,A\in \mathcal{A}(%
\mathcal{O})
\end{equation}
\begin{equation}
\Theta _{\Omega ,\Delta }(A)\Omega =\Delta _{\hat{O}}^{\frac{1}{4}}A\Omega
,\,\,\,A\in \mathcal{A}(\mathcal{O})
\end{equation}
Here $\widehat{\mathcal{O}}$ is a region which contains $\mathcal{O}$ and a
collar around it; the damping modular factor $\Delta _{\hat{O}}$ refers to
the modular operator corresponding to ($\mathcal{A}(\widehat{\mathcal{O}}%
),\Omega ).$ Actually the two formulations are related to each other but for
their conversion one needs the more detailed concept of index of nuclearity.
Equivalently the following set of vectors is nuclear:

\begin{equation}
\mathcal{N}(O,\beta )=\left\{ e^{-\beta H}U\Omega :U\in \mathcal{A}(\mathcal{%
O}),U^{*}U=1\right\} \,\,is\,\,nuclear\,\,set  \label{nuc}
\end{equation}

A subset $\mathcal{N}$ of a Hilbert space $H$ is called a nuclear set if
there exists a sequence of unit vectors $\phi _{n}\in H$ which span $H$ and
linear functionals $l_{n}$ such that: 
\begin{eqnarray}
\sum_{n=1}^{\infty }\sup \left\{ \left| l_{n}(\psi )\right| :\psi \in 
\mathcal{N}\right\} &<&\infty \\
\sum_{n=1}^{\infty }l_{n}(\psi )\cdot \phi _{n}=\psi &&  \nonumber
\end{eqnarray}
The nuclear index is then defined as: 
\begin{equation}
\nu (\mathcal{N})=\inf \sum_{n=1}^{\infty }\sup \left\{ \left| l_{n}(\psi
)\right| :\psi \in \mathcal{N}\right\}
\end{equation}
On the basis of a naive picture which ignores the ''fuzziness'' generated by
the Gibbs factor in front of the localized vectors one would identify these
sets with those of a canonical ensemble occupying a box at temperature $%
\beta ^{-1}$ and hence expects the result: 
\begin{equation}
\nu (\mathcal{N}(\mathcal{O}_{r},\beta ))\leq e^{const.(r/\beta )^{3}}
\label{bound}
\end{equation}
This naive expectation is indeed what is borne out by the explicit
calculation below.. Remember that the zero temperature counting gave an
almost finite (nuclear) instead of the naive finite degrees per phase space
cell result in Schr\"{o}dinger theory.

Using the Weyl functor, the nuclearity may be obtained from a corresponding
property of suitably damped localization subspaces of the Wigner space. A
slight complication is caused by the fact that the map from Wigner space to
Fock space looses some of the functorial properties\footnote{%
e.g. the intersection of localization spaces is much bigger than the
localization space of the intersected region.}. The easy part is the nuclear
estimate for the $e^{-\beta h}$-damped localized wave functions: 
\begin{eqnarray}
e^{-\beta h}H_{R}^{(1)}(\mathcal{O}) &\,&\,is\,\,nuclear\,set \\
H_{R}^{(1)}(\mathcal{O}) &=&unit\,\,ball\,\,in\,\,H_{R}(\mathcal{O}) 
\nonumber
\end{eqnarray}
An equivalent charaterization in terms of a trace class operator in Wigner
space is: 
\begin{equation}
\left\| E(\mathcal{O})\varepsilon ^{-\beta h}\right\| _{1}<\infty
\end{equation}
Here $E(\mathcal{O})$ is the projector onto the real localization subspace $%
H_{R}(\mathcal{O})$ and the subscript denotes the trace norm. For the
nuclearity property we have to prove the following two theorems:

\begin{theorem}
The previously defined operator has the following properties $(r\geq 1)$: 
\begin{equation}
\left\| E(\mathcal{O})e^{-\beta h}\right\| <1
\end{equation}
\begin{equation}
\left\| E(\mathcal{O})e^{-\beta h}\right\| _{1}<c(r/\beta )^{3}  \label{est}
\end{equation}

This is the input for the following theorem:
\end{theorem}

\begin{theorem}
Let e$^{it\gamma }$ be a unitary group in Wigner space which commutes with
the TCP-operator and which in the standard way leads to a unitary operator e$%
^{itG}$ on $H_{Fock}$ which leaves the vacuum vector invariant. Then the
above boundedness and trace class properties with $\gamma =\beta h$ are
sufficient to establish the nuclearity of the set (\ref{nuc}) with the
nuclear index bounded by 
\[
v(\mathcal{O},\beta )\leq \det (1-E(\mathcal{O})e^{-\gamma })^{-2} 
\]
Inserting the estimate (\ref{est}) we obtain (\ref{bound}).
\end{theorem}

Similar estimates hold if one uses the modular damping with $\Delta _{\hat{O}%
}^{\frac{1}{4}}.$ In fact the nuclearity property and the nuclear index of
both versions are related. As in the case of the nonrelativistic counting of
degrees of freedom nuclearity is expected to be a stable property in the
presence of interactions with a possible change of the nuclear index. In
these notes we require nuclearity as a restriction on the interaction.
Grossly unphysical theories as those which posses exponentially increasing
energy level densities and lead to pathological thermodynamic behavior
(finite [Hagedorn] limiting temperature) are eliminated by the nuclearity
requirement. One also expects this property to play an important role in the
understanding of asymptotic completeness. We believe that the
nonperturbative interactions based on modular localization of chapter 6
fulfill this requirement.

\section{Split Property, Taming of Vacuum Fluctuations}

The physical problems which gave rise to the split property reach back to
the beginnings of QFT when Heisenberg observed that the QFT vacuum behaves
very differently to nonrelativistic ground states with respect to local
charges obtained by integrating conserved currents over a finite region. In
contemporary terminology such a partial charge leads to infinite
particle-antiparticle fluctuations at a sharp spatial boundary. Only if one
allows a smoothened decrease inside a a space-time collar around the
localization region of the charge, one is able to control these vacuum
fluctuations. In algebraic QFT this picture is the intuitive germ of a
powerful concept, the ''Split Property''. We will come back to it in the
later chapter on algebraic QFT. In its most practical version it states that
for two spacelike separated regions $\mathcal{O}_{i}$ such that they allow
spacelike collars, the von Neumann algebra generated by $\mathcal{A}(%
\mathcal{O}_{i}),i=1,2$ is unitarily equivalent to the tensor product: 
\begin{equation}
\func{alg}(\mathcal{A}(\mathcal{O}_{1})\vee \mathcal{A}(\mathcal{O}%
_{2})\simeq \mathcal{A}(\mathcal{O}_{1})\overline{\otimes }\mathcal{A}(%
\mathcal{O}_{2})  \label{indep}
\end{equation}
If the two regions touch each other (examples: a wedge and its geometric
opposite, a double cone and its outside spacelike complement), this property
is definitely violated. Physically this is blamed on the lack of control of
fluctuations near the common boundary. The statistical independence
expressed by the right hand side of (\ref{indep}) is interpreted as the
result of control of fluctuations thanks to the presence of a collar. It is
well known that a boundary in Schr\"{o}dinger QM (box-quantization) leads to
a split in an inside and outside Hilbert space with the interaction causing
cross contributions. The orthogonal sum of spaces in the conversion to the
multiparticle Fock space (2$^{nd}$ quantization) leads to
((anti)symmetrized) tensor products of two nonrelativistic Fock spaces
together with an associated tensor product factorization of the
nonrelativistic observable algebras $\mathcal{A}:$%
\begin{eqnarray}
H_{F} &=&H_{F}^{inside}\bar{\otimes}H_{F}^{outside} \\
\mathcal{L}(H_{F}) &=&\mathcal{A}^{inside}\bar{\otimes}\mathcal{A}^{outside}
\nonumber \\
\mathcal{A}^{outside} &=&(\mathcal{A}^{inside})^{\prime }  \nonumber
\end{eqnarray}
This is not what happens in the relativistic case of the causal (modular)
localization. Although in that case the algebra and its commutant still
generate the algebra of all operators $\mathcal{L}(H_{F}),$ they are of
hyperfinite type III$_{1}$ instead of the above quantum mechanical type I.
Whereas type I factors are similar to $\mathcal{L}(H_{F})$ in that they
admit pure states or maximal measurements, type III only have impure states,
a fact which is related to the thermal nature of modular localized states.
One cannot ignore these subtle properties of local quantum physics without
impunity. As an example of conceptual havoc, we cite some wrong statements
claiming that Fermi's conclusion that the limiting velocity in QFT is c (as
in its classical counterpart) is false \cite{Heger} and that instead QFT
allows for causality violations (and time machines). The mistake can be
traced back to the incorrect implicit assumption that the localized algebras
behave as type I factors \cite{Bu Yn}. Although experts have gotten tired to
refute the never ending stories about superluminal velocities (with ot
without tunelling), one may safely assume that they all suffer from the same
conceptual flaw.

An equivalent but mathematically more natural definition of the split
property is the following:

\begin{definition}
An inclusion $\mathcal{A}\subset \mathcal{B}$ is called split if there is a
type I factor $\mathcal{M}$ with: 
\begin{equation}
\mathcal{A}\subset \mathcal{M}\subset \mathcal{B}
\end{equation}
\end{definition}

In that case $\mathcal{A}\subset \mathcal{M}$ and $\mathcal{B}^{\prime
}\subset \mathcal{M}^{\prime }$ together with $\mathcal{L}(H_{F})=\mathcal{M}%
\bar{\otimes}\mathcal{M}^{\prime }$ results in the existence of an
isomorphism $\Phi $ of $\mathcal{A}\vee \mathcal{B}^{\prime }$ with $%
\mathcal{A}\bar{\otimes}\mathcal{B}^{\prime }:$%
\begin{equation}
\Phi (ab^{\prime })=a\bar{\otimes}b^{\prime }\in \mathcal{M}\bar{\otimes}%
\mathcal{M}^{\prime }
\end{equation}
In the case of interest $\mathcal{A}=\mathcal{A}(\mathcal{O}),\mathcal{B}=%
\mathcal{A}(\widehat{\mathcal{O}}),\mathcal{O}\subset \widehat{\mathcal{O}}$
we have more structure thanks to the fact that these two algebras as well as
their collar algebra $\mathcal{A}^{\prime }\cap \mathcal{B}\supset \mathcal{A%
}(\mathcal{O}^{\prime }\cap \widehat{\mathcal{O}})$ are ``standard'' with
respect to the vacuum vector $\Omega $ (cyclic+separating from the
Reeh-Schlieder property). In this case the inclusion $\Lambda \equiv (%
\mathcal{A}\subset \mathcal{B},\Omega )$ is called standard split. For such
standard split inclusions the product state: 
\begin{equation}
\phi _{\Lambda }(ab^{\prime })\equiv \left( \Omega ,a\Omega \right) \left(
\Omega ,b^{\prime }\Omega \right)
\end{equation}
has a normal extension to the v.N. algebra $\mathcal{A}\vee \mathcal{B}%
^{\prime }.$ Using the natural selfdual cone $\mathcal{P}_{\mathcal{A}%
^{\prime }\cap \mathcal{B}}$ one obtains a representation in terms of a
unique vector $\Omega _{\Lambda }\in \mathcal{P}_{\mathcal{A}^{\prime }\cap 
\mathcal{B}}$ $\subset H_{F}$ with: 
\begin{equation}
\phi _{\Lambda }(a)=\left( \Omega _{\Lambda },a\Omega _{\Lambda }\right)
,\,\,a\in \mathcal{A}\vee \mathcal{B}^{\prime }
\end{equation}
In this situation the above isomorphism has a unique unitary implementation $%
U_{\Lambda }:H_{F}\rightarrow H_{F}\bar{\otimes}H_{F}$%
\begin{equation}
U_{\Lambda }ab^{\prime }\Omega _{\Lambda }=a\Omega \bar{\otimes}b^{\prime
}\Omega
\end{equation}
This leads to a canonical interpolating type I factor $\mathcal{N}$%
\begin{equation}
\mathcal{N}\equiv U_{\Lambda }^{*}(L(H_{F})\bar{\otimes}\underline{1}%
)U_{\Lambda }  \label{N}
\end{equation}
This quantum mechanical like subalgebra has of course pure states. The
reader should be aware that the (im)purity of a state only has meaning with
respect to an algebra. The unitary map $U_{\Lambda }$ is called the \textit{%
universal localizing map} and it has some interesting physical consequences.
Among other things it leads to a purely intrinsic version of localized
symmetry transformations \cite{Haag} and represents a kind of local quantum
physical version of Noether's theorem without obstructions from vacuum
fluctuations. One simply takes the global symmetry operator $U(g),g\in G$
and defines: 
\begin{equation}
U_{\Lambda }(g)=U_{\Lambda }^{*}(U(g)\bar{\otimes}\underline{1})U_{\Lambda }
\end{equation}
It then follows that this operator is attenuated within the collar: 
\begin{eqnarray}
U_{\Lambda }(g) &\in &\mathcal{B}, \\
i.e.\,adU_{\Lambda }(g)b^{\prime } &=&b^{\prime }  \nonumber \\
adU_{\Lambda }(g)a &=&adU(g)a  \nonumber
\end{eqnarray}
This construction works for internal as well as for space-time symmetries.

In free field theories one can go beyond the mere existence and perform
explicit computations of the localizing map. In case of zero mass, the
geometrical nature of the modular theory of double cones simplifies the
calculation. We defer such a calculation to the appendix.

\section{Problems with Entropy}

Let us add some speculative remarks on entropy. As the thermal aspect of
modular localization has its characteristic properties which distinguish it
somewhat from its standard heat bath setting, one expects also some
peculiarities for the entropy of modular localization. By this we mean an
entropy concept which carries the (quasi)classical observation of Bekenstein
and its refinement by Hawking with its natural (God-given) black-hole
horizons deeply in to the noncommutative world of modular structures of
hyperfinite type III$_{1}$ localized factors (without Killing horizons).
Whereas for type I factors we have von Neumann's definition of entropy and
(in the thermodynamic limit) entropy-density (also relative entropy between
different states on the same algebra), there is no direct definition \cite
{Narn} in the hyperfinite type III$_{1}$ case. Following the previous logic,
we should start with e.g. two concentric double cones in order to create a
collar for the intermediate type I factor $\mathcal{N}$ (\ref{N}). The
global vacuum state restricted to $\mathcal{N}$ is a thermal state $\Omega $
relative to the factorizing ground state $\phi _{\Lambda }.$ One expects $%
\Omega $ to have a restriction on $\mathcal{N}$ which can be described by a
well behaved density operator $D=e^{-K},$ where $K$ is related to the
modular theory of the pair ($\mathcal{N},\Omega )$ and should be interpreted
as a kind of regularized version of the modular theory of the smaller double
cone ($\mathcal{A}(\mathcal{O}),\Omega )$ which in the massless case would
be geometric and represented by a one parametric subgroup of the
16-parametric conformal group.

One expects that the K-Hamiltonian is in fact well enough behaved in order
to allow for the existence of the von Neumann entropy: 
\begin{equation}
S=-trDlnD
\end{equation}
where $D=\frac{1}{tre^{-K}}e^{-K}$ is the density matrix defined in terms of
the modular Hamiltonian $K.$ This is a quantity which depends on the size of
the collar $\varepsilon $ and which diverges as $\varepsilon \rightarrow 0$
i.e. when the fuzzy type I factor becomes hyperfinite type III$_{1}.$ If the
result of the existing proposals \cite{Wil} is compatible with this idea, we
should expect a universal logarithmic divergence in the inverse size $%
\varepsilon ^{-1}$of the collar which controls the vacuum fluctuations: 
\begin{equation}
S=-trDlnD\sim Cln\varepsilon ^{-1}
\end{equation}
with $C$ related to the longitudinal 2-dim. conformal theory which according
to our previous discussion we expect to determine the geometric core of the
fuzzy modular group of the double cone algebra $\mathcal{A}(\mathcal{O}).$
Indeed the formula \cite{Wil} 
\begin{equation}
C\sim Area\sqrt{c}
\end{equation}
where $Area$ denotes the area of the double cone and $c$ the vacuum
fluctuation strength of the energy momentum tensor. Although the limiting
entropy is certainly infinite, we have not yet been able to confirm that
this infinity is universal and behaves exactly as argued by Larsen and
Wilszek \cite{Wil}.

In order to prove this one must do some new computations with modular
methods and in particular on the ``localizing map'' which is the most
convenient way to compute the distinguished type I factor \cite{Haag}. The
relevant degrees of freedom would ``live'', as we will argue later, inside
the collar and the ratios of this ``collar entropy'' stay finite for
vanishing collar size. This remains a fascinating program for the future.

Now we are able to formulate our two conjectures:

\begin{conjecture}
The modular group of the (nonconformal) massive double cone $\,$algebra $%
\mathcal{A}(\mathcal{O})$ with respect to the massive vacuum vector (i.e.
the physical vacuum state restricted to $\mathcal{A}(\mathcal{O})$ is
cocycle-related to the known geometric modular group of the associated
conformally invariant situation belonging to the pair ($A(\mathcal{O}%
),\omega _{m=0})$ where $\omega _{m=0}$ denotes the conformal invariant
vacuum state. For the equivalence of the massive with the massless algebra
one may either invoke the construction of the double cone algebra by
canonical quantization or the fact that local algebras are always
hyperfinite III$_{1}$factors and the latter is unique modulo unitary
equivalence. The cocycle accounts for the difference in the local
propagation of massless (Huygens principle) and massive theories and its
presence renders the action of the modular group ``fuzzy''. Only
asymptotically near the horizon i.e. the boundary of the double cone, the
fuzzyness decreases and the geometric conformal modular transformation
reappears. Although a single algebra $\mathcal{A}(\mathcal{O})$ of the
massive theory and its scale invariant limit may be identified, the two nets
inside $\mathcal{O}$ remain different. However the conjecture that the
difference is due to the different propagation suggests that the massive net
inside $\mathcal{O}$ may be obtained from the massless $\mathcal{A}(\mathcal{%
O})$ by adjoining the action of the Poincar\'{e} covariances inside $%
\mathcal{O}.$
\end{conjecture}

\begin{remark}
For a massive free Fermi field\footnote{%
We want to avoid the infrared problems of massless Bose fields} in d=1+1
this can be shown. One notes that the restriction of such a free massive
theory to the light rays which constitute the boundary of the d=1+1 double
cone is simply the restriction of the corresponding massless theory and that
by propagating the chiral conformal data on the one dimensional horizon
inside with the massive propagator, one regains the massive free field net
inside $\mathcal{O}.$ A general proof of this reduction of a d=1+1 situation
to its chiral conformal limit (+ possible covariance operators) would be
extremely desirable because it would explain the association of the degree
of freedoms of d=1+1 theories with the horizon, a phenomenon which has
attracted considerable interest and has been termed ``holographic behavior'' 
\cite{Bi Suss}.
\end{remark}

\begin{conjecture}
The double cone algebras $\mathcal{A}(\mathcal{O})$ are identical to any of
the two-dimensional double cone algebras $A(O^{(2)})$ obtained by cutting
the double cone by a two-dimensional plane which contains the t-axis and one
coordinate axis. The net inside $O^{(2)}$ may be obtained from the
associated chiral conformal net on the one-dimensional horizon and a local
representation of Poincar\'{e} covariances.
\end{conjecture}

\begin{remark}
The first part is actually a consequence of Haag duality and the fact that
the causal completion of $\mathcal{O}^{(2)}$ gives $\mathcal{O}:$ 
\begin{equation}
\mathcal{A}(\mathcal{O}^{(2)})\equiv \cap _{\delta }A(\mathcal{O}_{\delta })
\end{equation}
Where $\mathcal{O}_{\delta }$ denotes the middle slice of thickness $\delta $
by cutting the double cone parallel to the t-axis. Each $\mathcal{O}_{\delta
}$ has $\mathcal{O}$ as its causal completion and the property of ``Haag
Duality'' demands the equality of $A(O_{\delta })$ with the algebra of the
causal completion $\mathcal{A}(\mathcal{O}).$ The essential step in the
holographic reduction is the appearance of chiral conformal degrees of
freedom after removal of the angular degrees of freedom due to angular
symmetry (substituting the transversal symmetry in the case of the wedge).
The envisaged entropy is therefore not proportional to $area(\mathcal{O}%
^{(2)})\times angular\,volume$ but rather to horizon-length$(\mathcal{O}%
^{(2)})\times angular\,volume=volume\,of\,\,horizon(\mathcal{O}).$ Actually
such a situation would also suggest that there may be an infinite hidden
nongeometric (fuzzy) symmetry algebras in the nonperturbative structure of
any QFT\footnote{%
In principle every modular automorphism has the interpretation of a
(localized but fuzzy) physical symmetry. Some of these are
``semi-geometric'' i.e. they act geometric on subnets. The modular group of
the previous modular intersection situation (which led to the transversal
Galilei-transformation) is such a case of a ``semi-hidden'' symmetry \cite
{un}.}. Although they are local in the sense of keeping things inside say $%
\mathcal{O},$ their action within $\mathcal{O}$ is totally fuzzy. Such
symmetries of nonperturbative local quantum physics would escape
differential geometric methods. Note that the two conjectures cannot even be
formulated in terms of properties of expectation values of fields; the use
of the algebraic i.e. field coordinate independent concepts is indispensable
for the formulation. If algebraic QFT did not already exist, one would have
to invent it in order to understand the above thermal and entropic
properties.
\end{remark}

As we have seen, the thermal and entropic aspects which are erroneously
attributed \textit{exclusively} to black holes, are in fact a generic
nonperturbative feature of the modular localization structure of QFT. They
make their appearance e.g. in the formfactor bootstrap construction program
(vis.. the KMS origin of crossing symmetry as a generalization of TCP) and
also show up in CST QFT for the same (localization) reason. The latter case
is only distinguished by the fact that these concepts allow for a classical
(thermodynamic) interpretation which is of course the reason why they were
first noticed there. This begs the question of the algebraic point of view
about ``Quantum Gravity''. This time we put these words in quotation mark in
order to indicate their precarious physical status, especially in the
algebraic approach. It is agreed upon by most physicists that Quantum
Gravity, whatever it is, does not fit into the framework of local quantum
physics as say another spin=2 QFT. Therefore one may ask the more general
question of physically consistent theories outside the framework local
quantum physics. Surely there have been several attempts to imagine such
possibilities, e.g. the pure S-matrix approach, the modifications of
Lagrangians by formfactors or structure functions, SO(4) invariant cutoffs
in euclidean QFT as candidates of real time relativistic nonlocal theories
after analytic continuation and the peratization program (pairs of complex
conjugate poles in Feynman rules) of A. Pais and T. D. Lee. If the proposal
was not already defeated on the mathematical front, it turned out that the
physical interpretation was either inconsistent (existence of precursors
violating the indispensable macrocausality) or was not, as in local quantum
physics part of the theory, but (as a consequence of the missing
localization) had to be enforced from the outside. There is one recent
proposal \cite{DFR} which has survived recent years as a possible scenario
(but its future survival is by no means guarantied) which roughly speaking
consists in substituting the classical indexing of the algebras in a net by
space-time regions in Minkowski space by noncommutative versions in the
spirit of noncommutative geometry. Algebraic QFT would favor a situation in
which no a priori space-time indexing (neither commutative or
noncommutative) appears. Preferably one would like to have global algebra
with an intrinsic substructure such that it would contain our physical world
of localization and causality only in the germs of certain representations
(states). Such an idea would be much more radical\footnote{%
Note that this is the only radicalism in the algebraic approach since all
other deviations from the standard QFT approach, as uncommon and
revolutionary as they may appear at first sight, are all in accordance with
the physical principles of QFT.} than say string theory, because going from
pointlike to string extension does not mean that one abandons localization
altogether.

There is one intriguing property in chiral conformal QFT which has a certain
quantum gravity ``touch'' to it. This is the fact that the ``averaging over
a Fuchsian group'' of a chiral conformal QFT (assuming that the Poincar\'{e}
series converge in some sense) converts the Moebius invariant vacuum
expectation values into expectation values which loose this invariance but
gain deformation parameters (generalizations of ``compact temperatures'').
Formally the positivity property holds as in Wightman theory, but the old
localization region $S^{1}$ is now totally fuzzy. With other words there is
no causal complement in the quantum sense. In such a scenario there would be
no global concept (a priori knowledge of what is spacelike) of causality and
hence of localization, and the place to find the lost net properties would
be in certain states and even there they would only appear in their germs.
Our second conjecture which led via modular theory to the speculative
existence of a hidden ``fuzzy'' realization of the Moebius group would
suggest that such a scenario may also be possible in d=1+3 theories.

We have made these extremely speculative remarks about quantum gravitation
in connection with entropy, because we share with B. Kay \cite{Kay} the
believe that (holographic) entropy and quantum gravity belong to the same
circle of problems. We are attracted by his idea of quantum gravitational
degrees of freedom which have (apart from their classical long range tails)
no observable aspect (i.e. no localization) in line with the previous
conformal scenario, but we admit that we have not been able to reconcile the
remaining different viewpoints.. It is interesting that both ideas are very
removed from Weinberg's picture of quantum gravity as just a s=2 Lagrangian
QFT.

Summing up our excursion on nonperturbative QFT we would like to stress
again that the algebraic method allows for a completely intrinsic definition
and understanding of QFT independent of its Lagrangian or non Lagrangian
origin. Any quantum theory which fulfills the stability requirements of
positive energy and allows for a net interpretation and the associated
localization concepts is a QFT par excellence and enjoys all the general
structural properties which feature in this article as TCP, spin
\&statistics, crossing symmetry \& modular localization \& thermality,
wedge-localized fields without vacuum polarization, hidden modular
symmetries, Haag duality (an abstract form of the 2-d
Kramers-Wannier-Kadanoff Duality), nuclearity for the phase space degrees of
freedom \& the conjectured ``Holographic Entropy'' and all the other yet
unraveled properties of nonperturbative local quantum physics. The main
obstacle against progress is not so much the novel mathematics which these
new physical concepts require, but rather (as always in the past)
prejudices. One prejudice is that field theory has to be ``Lagrangian''. In
view of the many existing low-dimensional non-Lagrangian models and the fact
that they hardly rocked the Lagrangian boat, this appears to be the
mightiest prejudice.

\chapter{Perturbative Interactions}

\section{Kinematical Decompositions}

Before presenting an elementary approach to interactions and perturbation,
it is helpful to have a closer look at those observable quantities which one
wants to compute. Since among local ``field coordinates'' only currents have
a distinguished physical meaning, one is naturally interested in matrix
elements as: 
\begin{equation}
\Gamma _{\mu }(p^{\prime },p)=\left\langle p^{\prime }\left| j_{\mu
}(0)\right| p\right\rangle ,\quad W_{\mu \nu }(p,x)=\left\langle p\left|
j_{\mu }(\frac{x}{2})j_{\nu }(\frac{-x}{2})\right| p\right\rangle _{conn.}
\end{equation}
The first quantity (where possible spin quantum numbers have been
suppressed) is called the (electromagnetic) form factor of the p-particle
and its static limit $(p-p^{\prime })^{2}\rightarrow 0$ can be probed by
external magnetic fields and is related to the (anomalous) magnetic moment.

The second (diagonal) matrix element of two currents with the subscript
connected \footnote{%
connected part means removal of the ill-defined vacuum contribution $%
\left\langle p^{\prime }\mid p\right\rangle \left\langle 0\left| j_{\mu
}j_{\nu }\right| 0\right\rangle $ \textit{before} the limit p'$\rightarrow $%
p, which however does not influence the structure of the covariant
decomposition.} gives rise to the notion of ``structure function'' of the
p-particle and appears in the description of high-energy electron (more
general: lepton) inclusive scattering on nucleons (scattering in which one
does not observe the created outgoing hadrons).

Important energy shifts as the Lamb shift cannot be expressed in an elegant
form in terms of such matrix elements between particle states (only if one
defines ``off-shell'' extrapolations). These matrix elements between ``ket''
in-vectors and ``bra'' out vectors of localized (or multi-localized) fields $%
O(x)$ : 
\begin{equation}
^{out}\left\langle p_{1}^{\prime },..p_{m}^{\prime }\left| O(0)\right|
p_{1},..p_{n}\right\rangle ^{in}
\end{equation}
are referred to as ``generalized formfactors''. They are the most prominent
measurable quantities, but turn out to be also the most important building
blocks of the new constructive approach based on modular localization which
will be presented in chapter 6.

The most important experimental observables are the generalized formfactors
of the identity operator $O(x)=\stackunder{-}{1}$ which are identical to the
matrix elements of the $S$-matrix (or scattering operator): 
\begin{equation}
^{out}\left\langle p_{1}^{\prime },..p_{m}^{\prime }\left| \stackunder{-}{1}%
\right| p_{1},..p_{n}\right\rangle ^{in}=\,^{in}\left\langle p_{1}^{\prime
},p_{2}^{\prime }....p_{m}^{\prime }\left| S\right|
p_{1},p_{2}....p_{n}\right\rangle ^{in}
\end{equation}
from which via the momentum space kernel $T$ the cross sections can be
obtained: 
\begin{equation}
^{in}\left\langle p_{1}^{\prime },p_{2}^{\prime }....p_{m}^{\prime }\left|
S-1\right| p_{1},p_{2}....p_{n}\right\rangle ^{in}=\delta
(\sum_{i=1}^{m}p_{i}^{\prime }-\sum_{k=1}^{n}p_{k})T(p_{1}^{\prime
},...p_{m}^{\prime },p_{1},...p_{n})
\end{equation}
Here we used the fact that $S$ is a Poincar\.{e} invariant operator in the
Fock space of incoming particles (the energy-momentum conserving $\delta -$%
function results from translation invariance).

Kinematical properties means the decomposition of covariant into invariant
functions and the specification of the invariant variables on which the
latter depend. For the formfactor of $s=\frac{1}{2}$ particles one finds the
following decomposition (with $k=p\prime -p$) : 
\[
\Gamma _{\mu }(p^{\prime },p)= 
\]
\begin{equation}
\frac{1}{(2\pi )^{3}}\bar{u}(p^{\prime },s_{3}^{\prime })\left( \gamma _{\mu
}F(k^{2})-\frac{i}{2m}(p^{\prime }+p)G(k^{2})+\frac{1}{2m}k_{\mu
}H(k^{2})\right) u(p,s_{3})
\end{equation}
The fastest way to see this is to first use the free field formalism to
compute the matrix elements of the free current by ``Wick-gymnastics'': 
\begin{equation}
\left\langle 0\mid a(p^{\prime },s_{3}^{\prime }):\bar{\psi}(0)\gamma _{\mu
}\psi (0):a^{*}(p,s_{3})\mid 0\right\rangle =\frac{1}{(2\pi )^{3}}\bar{u}%
(p^{\prime },s_{3}^{\prime })\gamma _{\mu }u(p,s_{3})
\end{equation}
Then one has to construct the most general vector object from the $\gamma -$%
matrices and two mass shell momenta $p$ and $p^{\prime }$ subject to the
identity $\gamma _{\mu }p^{\mu }-m=0$ which is valid on the intertwiner $%
u(p) $. This leaves besides $\gamma ^{\mu }$ itself, which appears already
for a free current, only the above two momentum vectors (or linear
combinations thereof). Current conservation $k_{\mu }\Gamma ^{\mu }=0$ gives 
$H\equiv 0$ (because H is a nonsingular function) and the value $1$ of the
total charge $Q=\int j_{0}(x)d^{3}x$ between the one-particle states
requires $F(0)=1$.

Due to kinematical identities of the $u$ and $v$ intertwiners, there are
many different forms of covariant decompositions. For example the identity: 
\begin{equation}
\bar{u}(p^{\prime })i\sigma _{\mu \nu }q^{\nu }u(p)=\bar{u}(p^{\prime
})(2m\gamma _{\mu }+i(p^{\prime }+p)_{\mu })u(p)
\end{equation}
may be used to eliminate the $\gamma _{\mu }$ term in favor of the $%
(p^{\prime }+p)_{\mu }$ and the $\sigma _{\mu \nu }=\frac{i}{2}\left[ \gamma
_{\mu },\gamma _{\nu }\right] $ terms: 
\begin{equation}
\bar{u}(p^{\prime })\Gamma _{\mu }u(p)=\bar{u}(p^{\prime })\{\frac{i}{2m}%
(p^{\prime }+p)_{\mu }(F(q^{2})+G(q^{2}))+\sigma _{\mu \nu }q^{\nu
}F(q^{2})\}
\end{equation}
In this form the leading contribution for small spatial momenta p and p'
comes solely from the second term. The physical interpretation of $F$ (which
as $G$ can only depend on $k^{2}$ since this is the only invariant which one
can form two mass shell vectors) becomes clear if one rewrites the canonical
coupling of the current to an external (classical) vector-potential as
follows: 
\[
\left\langle p^{\prime }\left| -e\int d^{3}x\vec{j}(x)\vec{A}(x)\right|
p\right\rangle =-e\frac{1}{(2\pi )^{3}}\int e^{i(\vec{p}-\vec{p}^{\prime })%
\vec{x}}\bar{u}(p^{\prime },s_{3}^{\prime })\vec{\Gamma}(p^{\prime
},p)u(p,s_{3})\vec{A}(x)d^{3}x 
\]
\begin{equation}
\approx -eF(0)\frac{1}{(2\pi )^{3}}\frac{p_{0}}{m}\int d^{3}xe^{i(\vec{p}-%
\vec{p}^{\prime })\vec{x}}(\vec{A}(x)\cdot \left[ (\vec{p}-\vec{p}^{\prime
})\times \vec{J}\right] )
\end{equation}
Here the last line is the static approximation of $\bar{u}\vec{\Gamma}u$ in
first order of\thinspace $\vec{p}$-$\vec{p}^{\prime }$ which brings in the
angular momentum operator $\vec{J}=\frac{\vec{\sigma}}{2}$ (again just ``$%
\gamma $-gymnastics'' between $u$-intertwiner). The last step is to use $%
\vec{B}=\vec{\bigtriangledown}\times \vec{A}$ and to take $B$ constant
(static limit): 
\begin{equation}
\left\langle p^{\prime }\left| -e\int d^{3}x\vec{j}(x)\vec{A}(x)\right|
p\right\rangle \approx -\frac{e}{2m}F(0)2p_{0}\delta (\vec{p}-\vec{p}%
^{\prime })\vec{B}\vec{\sigma}_{s_{3}^{\prime },s_{3}}
\end{equation}
i.e. we obtain the magnetic moment interaction $-\vec{\mu}\cdot \vec{B}$
with $(\mu =\left| \vec{\mu}\right| )$%
\begin{equation}
\begin{array}{c}
\mu =\frac{e}{2m}F(0)=\frac{e}{2m}+\mu _{anom.} \\ 
\mu _{anom.}=\frac{e}{2m}(F(0)-1)=\frac{e}{2m}G(0)
\end{array}
\end{equation}
In a similar fashion one decomposes the structure function (the form factor
of two currents): 
\begin{equation}
\begin{array}{c}
\int W^{\mu \nu }(p,x)e^{iqx}d^{4}x\dot{=}:W^{\mu \nu }(p,q)= \\ 
-(\frac{q_{\mu }q_{\nu }}{q^{2}}-g^{\mu \nu })W_{1}(\nu ,q^{2})+\frac{1}{%
m^{2}}(p^{\mu }-\frac{p\cdot q}{q^{2}}q^{\mu })(p^{\nu }-\frac{p\cdot q}{%
q^{2}}q^{\nu })W_{2}(\nu ,q^{2})
\end{array}
\end{equation}
The invariant structure functions $W_{i}$ $i=1,2$ depend on two variables ($%
q $ is off-shell) $q^{2}$ and $\nu =\frac{p\cdot q}{m}$ (with $m$=target
mass). Again the number of invariants has been reduced by using current
conservation for the two currents. The $W^{\prime }s$ are measurable in
processes involving deep inelastic lepton scattering.

The $S$-matrix is not measured directly, but rather through the ensuing
scattering cross sections. The relevant formulas in most textbooks are
derived by inventing a ``box-quantization'' (in order to solve the problem
of ``squaring the energy-momentum $\delta -$function''). For scattering
theory as well as for statistical mechanics of open systems such a trick
goes against the basic concepts which require the infinitely extended
Minkowski space. Therefore it is comforting to know that that there are
suitable concepts and techniques which avoid such tricks in favor of working
directly in the infinite volume e.g. the KMS condition in the case of
statistical mechanics and techniques of wave packets in scattering theory.

\section{Perturbative Realization of Interaction}

In chapter 2.1 we solved the simple problem of a perturbation by an external
source on a free bosonic system and found that there are two methods, one
via unitary transformations (the so called ``dressing'' transformations),
and the other by the use of the interaction picture in the form of a \textit{%
time-ordered} exponential: 
\begin{equation}
S(j)=Te^{i\int A(x)j(x)d^{4}x},\quad A(x,j)=S^{*}(j)TA(x)e^{i\int
A(y)j(y)d^{4}y}  \label{S(j)}
\end{equation}
By Wick-ordering these expressions, we saw that they agree with the dressing
transformation method up to a phase (Feynman's famous ``vacuum phase'') and
that phases show up in the form of cocycle factors in the composition law of
the $S(j)$.

In translation-invariant interaction problems (i.e. without external
fields), there is a well-known obstruction against the existence of such a
unitary dressing operator, the \textit{Haag Theorem}. It says that in a
globally translation-invariant theory the ground state of an interacting
system cannot be described in the space of vector states of the free system 
\cite{Haag}, but it does not rule out the existence of a local dressing
transformation which depends on the localizing region. The traditional way
out is to overcome this No-Go theorem against Dirac's interaction picture in
QFT is to start with a system which fulfills only ``partial translational
invariance'' (similarly to the notion of ``partial charges'' in the free
field theory in chapter 3.4). We begin by defining ($A$ stands generically
for the ``would-be'' Heisenberg field which corresponds to the free field $%
A_{0}$) : 
\begin{eqnarray}
S(g) &=&Te^{i\int g(y)\mathcal{W}(x)d^{4}x},\,\,%
\;A(x,g)=S^{*}(g)TA_{0}(x)e^{i\int g(y)\mathcal{W}(y)d^{4}y} \\
&=&S^{*}(g)\frac{\delta }{\delta h(x)}S(g,h)\mid _{h=0},\;S(g,h)=Te^{i\int
\left\{ g(y)\mathcal{W}(x)+A_{0}(x)h(x)\right\} d^{4}x}
\end{eqnarray}
$\mathcal{W}$ is an invariant Wick-ordered polynomial in terms of free
fields which implements the notion of interaction via its use as a
deformation of free fields. Mathematically it is a Lorentz scalar polynomial
of at least trilinear degree in the free field Borchers class of the
interacting free fields For $g$ we choose a smooth function with compact
support in Minkowski space which can be thought of as a smooth version of
the characteristic function of a double cone with support in larger double
cone and the constant value $g(x)$ $\equiv g$ in a smaller cone placed
inside the bigger one. If $W$ contains several terms, there is one $g(x)$
for each monomial. Before we show some remarkable properties of these formal
operators in Fock space, some comments are in order.

\begin{itemize}
\item  (i)\quad Haag's theorem is not applicable to the $S(g)$ formalism (no
translation invariance), and we are allowed to do our calculations in Fock
space. One of the remarkable properties is that the local observables
localized within the smaller double cone fulfill \textit{partial}
invariances (including translations) in an algebraic sense explained later.

\item  (ii)\quad The standard derivation of the above formula for $A(x)$
(more precisely for the vacuum expectation values of time ordered products
of $A$) goes through the canonical formalism and is known under the name of 
\textit{Gell-Mann-Low formula}. Such derivations suffer from two conceptual
weaknesses. On the one hand they give (physically unmotivated) preference to
special field coordinates (only ``Eulerian'' free fields among the class of $%
(m,s)$ Wigner fields are fitting into the canonical formalism) and on the
other hand they rely on assumption that the fields $A$ to be constructed are
not more singular for short distances than the corresponding canonical free
fields. These assumptions are only valid in certain very special
low-dimensional models. The main culprit for the nonexistence of Fock space
operators as above (even not after renormalization) is the use of the
interaction picture with its time-ordered formalism\footnote{%
Even the spatial cutoff g is generally not sufficient for the existence of
the interaction picture or a dressing transformation.}. Physical concepts do
not require the existence of objects in addition to the Heisenberg fields $A$
and the asymptotic (in- or outgoing) free fields.

\item  (iii)\quad The interaction density $\mathcal{W}(x)$ is a local
function of free fields which (without the existence of a dressing
transformation) has no direct (outside an interpretation in terms of
infinitesimal deformations without global counterpart) physical
interpretation. This means that there is (apart from external perturbations
and some very special low dimensional models) \textit{no general physical
reason} to believe that after a certain necessary repair
(``renormalization''), one obtains a mathematically well-defined theory.
Perturbation theory generally has only a formal meaning as an \textit{%
infinitesimal deformation theory}. There is no reason whatsoever why the
``Bogoliubov axiomatics'' i.e. the above scheme of operators S(g) in Fock
space should have a solution (apart from the well-known superrenormalizable
models as e.g. P$\phi _{2})$. It is very unfortunate that we use such a
questionable framework for the baptization of interacting models. The
nonperturbative approach of chapter 6 which aims at the global existence is
based on quite different concepts.
\end{itemize}

Contrary to popular believes, it is not just the singular short distance
behavior as such which endangers the existence of the theory, but rather the
way interaction is introduced via the Ansatz in terms of a time ordered
operator $S(g)$ \footnote{%
The so called Bogoliubov axiomatics may have no solution in higher
dimensions and therefore the time-ordering method may be not appropiate for
introducing interactions. In the nonperturbative approach to low dimensional
QFT the time-ordering plays no role.} which creates an obstruction against
an intrinsic understanding of interactions. There is a very interesting
lesson in this respect from the d=1+1 ``bootstrap'' constructions which show
that short distance singularities can be worse than any given inverse power
of the Minkowski distance, but without the existence of the theory being
threatened. The simplest illustration of the fact that there is no obvious
relation between the deviation of the short distance behavior from that of
free fields is furnished by the Thirring model, for which the singular short
distance power can be made arbitrarily large by changing the coupling
parameter. If there is at all a general relation between the existence of a
QFT associated with $\mathcal{W}(x)$ and perturbative renormalization
aspects, then it must be linked to the vanishing of the $\beta $-function
i.e. the coupling parameter renormalization.

Let \underline{$h$ }be a collection of test functions which couple to
observable local fields (electro-magnetic. field strength, currents,..) and
let us introduce the following relative Bogoliubov-Shirkov operators (see
section 2 of this chapter): 
\begin{equation}
V(g,h)=S^{-1}(g)S(g+\underline{h})
\end{equation}
The causality of the local operators coupled to \underline{$h$} is then
expressed as: 
\begin{equation}
V(g,\underline{h_{1}}+\underline{h_{2}})=\left\{ 
\begin{array}{l}
V(g,\underline{h_{1}})V(g,\underline{h_{2}}),\,\,supph_{1}\gtrsim supph_{2}
\\ 
V(g,\underline{h_{2}})V(g,\underline{h_{1}}),\,\,supph_{2}\gtrsim supph_{1}
\end{array}
\right.
\end{equation}
i.e. the $V$ factorizes in the second argument if there is no future
directed causal curve from $supph_{1}$ to $supph_{2}$ or. vice versa. The
spacelike consistency of this formula implies in particular Einstein
causality. The double cone (=$\mathbf{C})$ algebra may than be defined as 
\begin{equation}
A_{g}(\mathbf{C})=\overline{\left\{ V(g,\underline{h}),\,supph\subset 
\mathbf{C}\right\} }
\end{equation}
The next step consists of showing that a change of $g(x)$ which maintains
the constant value g inside $W_{st}$ leads maximally (if the past ``collar''
is changed) to a h-independent unitary transformation. However such a common
transformation does not change the net of wedge algebras inside $W_{st}$ (by
the very definition of isomorphy of nets)$.$ By Lorentz transformations one
obtains the net of all wedge algebras and by intersecting wedges one finally
obtains the full net of observables (which includes the double cones).

Let us now show that on a formal level the Fock space operator $A(x,g)$
fulfills some remarkable formal properties. Suppose that we restrict the x
to the double cone K in which $g\equiv 1$ i.e. we consider $A(f,g)$ 
\begin{equation}
A(f,g)=\int d^{4}xf(x)S^{*}(g)TA_{0}(x)S(g)
\end{equation}
with $supp.f\subset K.$ Then as a generalization of the composition operator 
$S(j)$ in our old source model we find: 
\begin{equation}
\begin{array}{c}
S(g_{2}+g_{1})=S(g_{2})S(g_{1}),\quad \hbox{supp}g_{2}\geq \hbox{supp}%
g_{1}\quad \\ 
A(f,g)=A(f,g^{\prime }),\quad \hbox{if \thinspace supp}(g-g^{\prime
})\subset V_{-}(K)^{\bot }
\end{array}
\end{equation}
where the notation means that the points in supp$\ g_{2}$ are either
spacelike or timelike from those in supp $g_{1}$ and $V_{-}(K)^{\bot }$ is
the complement of the smallest backward light cone which contains the double
cone $K$. Furthermore any change of $g$ to $g^{\prime }$ localized in $%
V_{-}(K)^{\bot }\setminus K$ can be implemented by a unitary (''partial
dressing'') transformation $U(g)$ which is independent of f \cite{Bru}, i.e.
the same for all operators in the algebra \QTR{cal}{A}($K$): 
\begin{equation}
A(f,g^{\prime })=U(g^{\prime },g)A(f,g)U^{*}(g^{\prime },g)
\end{equation}
Formally this unitary has the same form as $S(h)$ where the smooth function $%
h$ is compactly supported in the intersection of $V_{-}(K)^{\bot }\setminus
K $ with a double cone $K$ which contains the support of both g's . For the
study of the net of double cone algebras localized in $\hat{K}$ inside $K$
the common unitary $U$ is irrelevant since nets which are related by one
common unitary are identical (isomorphic families define identical nets by
definition) i.e. it is only the relative positions of these algebras and not
the absolute position in the ambient space which counts. Hence even the
limit $K\rightarrow \infty ,$ the net of algebras may be described within
Fock space. Hence this Fock space is purely auxiliary. Physical states
strictly speaking are to be obtained as states on the net of operator
algebras with suitable localization properties. This would be the scenario
for the construction of interacting theories within the setting of time
ordered exponential of free field ``interaction densities'' $\mathcal{W}(x)$.

Before we look at the lowest nontrivial perturbative evaluation of these
formal operators, let us briefly notice that $A(x,g)$ fulfills Einstein
causality within $K$: 
\begin{equation}
\left[ A(x,g),A(y,g)\right] =0\quad (x-y)^{2}<0\quad and\quad x,y\in K
\end{equation}
The formal reason is that for spacelike separations the product can be
written in terms of one (cancellations between S's!) time-ordered free field
expression: 
\begin{eqnarray}
S^{*}(g)T(A_{0}(x)A_{0}(y)S(g)) &=&S^{*}(g)T(A_{0}(x)S(g))\cdot
S^{*}(g)T(A(y_{0})S(g))  \nonumber \\
\hbox{holds for (}x-y)^{2} &<&0
\end{eqnarray}
(remember: the $T$ only acts on all the $A_{0}$'s to the right). The
(Bogoliubov-Shirkov, Glaser-Epstein) renormalization approach allows to show
that these formal relations are valid at least in every order of
perturbation theory (expansion in $\mathcal{W}$). The idea is to reduce the
iterative determination of the operator T-products to a (Hahn-Banach type)
extension problem of time-ordered vacuum expectations. In the case of
renormalizable models this is possible with a finite number of parameters
(counter-terms). The issue of whether this latter requirement is more than
formal is still unsettled, although beyond the mere fact of computability of
higher orders without new parameters there seems to be considerable physical
success in this restriction.

Another remark, whose importance can only be fully appreciated later, is the
statement that the local algebras of a net are all unitarily equivalent and
there is (outside of perturbation theory) no relation between the particle
structure of the ambient Fock space and the physical content of local
algebras. The interaction generically speaking wrecks the one to one
correspondence between particles and fields which existed in the free theory%
\footnote{%
We are ignoring here the the use of interaction polynomials as a mere
placeholder for combining phenomenological correlation functions as in
``phenomenological'' Lagrangians. In those second derivative momentum space
Taylor coefficients the particle and field contents are identical by fiat.}.
For local observables described in terms of local nets of algebras, the
Hilbert space description allows great flexibility and the chosen massive
Fock space of the above formalism is not to be interpreted as a commitment
about physical parameters. This picture is unfortunately somewhat blurred by
perturbation theory which maintains an unrealistic rigid correspondence
between fields and particles (apart from the mentioned flexibility of
choosing the Fock space mass parameter different from the physical mass).
This (among other things) has created the misleading impression that QFT is
nothing more than a relativistic made form of quantum theory of particles.
Although it is a \textit{quantum theory} and it is \textit{relativistic }and
its principle physical aim is to describe\textit{\ particles}, it is
primarily a \textit{new physical realm} whose deep and unexpected concepts
(despite its 70 years of existence) still await exploration. This will
become much more evident in the later chapter on modular localization and
the bootstrap-formfactor approach than on the present level of perturbation
theory.

In these notes we only address some conceptual points of renormalized
perturbation theory. The $n^{th}$ order renormalization technology goes much
beyond and should be studied in appropriate review articles by Zimmermann,
Lowenstein, Piguet-Sorella \cite{Zi} and others.

\section{ Perturbation and Adiabatic Parametrization}

The naive expectation (i.e. by analogy to the external source problem in
chapter 1) about $S(g)$ would be that the limit of the theory for $%
K\rightarrow R^{4}$ exists and describes the physical S-matrix. Even in
perturbative evaluation as an infinitesimal deformation theory with suitably
adapted causality and positivity requirements, this picture needs two
corrections. One is related to the infrared divergence problem in certain
theories involving zero mass as QED, a somewhat special phenomenon whose
physical basis will be reserved for a later discussion. The other is of a
completely general nature related to the phenomenon of selfinteraction,
well-known already from classical field theory where it leads to the famous
problems of constructing consistent particle models within a classical field
theory, as studied by Poincar\'{e} and Lorentz at the beginning of this
century. As a result of selfinteraction, parameters with a physical name as
mass, charge etc. which entered the construction of $S(g)$ and $A(x,g)$, do
not represent the true measured value. Whereas for fields $A$ and their
correlation functions this does not matter\footnote{%
The reason why in the classical theory the divergencies are not so easily
disposed, is that unlike QFT the particle concept does not follow from the
Poincar\'{e} transformations of the fields but has to be imposed on the
classical field theory, a procedure which according to our best knowledge is
inconsistent. Hence although Kramers renormalization idea was historically
essential, it looses its importance as soon as one realizes, that unlike
classical fields, poinlike quantum fields are intrinsically singular objects
in the sense of Bohr and Rosenfeld. For those one needs Schwinger's
physically motivated point-split form of field equation. Knowing this one,
may safely deal with Feynman's simpler rules, as long as one confronts the
intermediate infinities created by his formalism without confusing them with
those of the classical theory studied by Lorentz and Poincar\'{e}. The
latter are a physical consequence of forcing classical particles upon a
classical field theory.} (the true physical values can be recovered from
asymptotic properties of correlation functions, see later), the large volume
limit of $S(g)$ for $\left| K\right| \rightarrow \infty $ represents the
physical S-matrix for the scattering of A-particles \textit{only if the true
physical mass is used}. The same applies to any quantity which is partially
``on-shell'' i.e. contains particles states as e.g. the electromagnetic
formfactor. The reason is that the adiabatic switching on and off by
multiplying $\mathcal{W}(x)$ with $e^{-\alpha \left| t\right| }$ and then
lim t$\rightarrow \infty $ is physically harmless only if the interaction
includes the effect of persistent selfinteraction ``counter-terms'' which
maintain the masses used in the Fock space in every order of $\mathcal{W}$
at their physical value. In case of a neutral scalar $\mathcal{W}%
=g:A_{0}(x)^{4}:$ model the modification is: 
\begin{eqnarray}
\mathcal{W}_{adiab}(x) &=&\mathcal{W}(x)+\frac{1}{2}\delta
m^{2}Z:A_{0}(x)A_{0}(x):  \nonumber \\
+\frac{1}{2}(Z-1) &:&(\partial _{\mu }A_{0}(x)\partial ^{\mu
}A_{0}(x)-m^{2}A_{0}(x)A_{0}(x)):
\end{eqnarray}
The ``selfmass'' $\delta m^{2}$ is chosen in every order to maintain $m$ as
the physical mass and $m_{0}^{2}=m^{2}-\delta m^{2}$ is an auxilary
unphysical mass (which loosely speaking corresponds to the mass without the
stabilizing counterterm which changes in every order of $\mathcal{W}$). The
second $Z$-counterterm has been added in order to obtain a nicer form of the
adiabatic principle which is the following requirement: 
\begin{equation}
\lim_{g\rightarrow 1}\left\langle 0\left| A(x,g)\right| p\right\rangle
=\left\langle 0\left| A_{0}(x)\right| p\right\rangle
\end{equation}
By adjusting $\delta m^{2}$ and $Z$ in every order such that this identity
holds we took all selfinteractions into account. A subsequent adiabatic
change of $\mathcal{W}_{adiab}$ i.e. 
\begin{equation}
\mathcal{W}_{adiab}\rightarrow e^{-\alpha \left| t\right| }\mathcal{W}%
_{adiab},\quad \alpha \rightarrow 0\,\,\,\hbox{at end of calculation}
\end{equation}
will not cause any harm i.e. does not change the one particle
characteristics. In theories without selfinteraction e.g. in Schr\"{o}dinger
theory, this is automatically fulfilled. Using our formal time-ordered
expressions we may rewrite the above requirement in second order: 
\begin{equation}
\frac{(\delta m^{2}Z)^{(2)}+(Z-1)^{(2)}(p^{2}-m^{2})}{p^{2}-m^{2}+i%
\varepsilon }=\frac{1}{2}\int \int \left\langle 0\left| TA_{0}(0)\mathcal{W}%
(x_{1})\mathcal{W}(x_{2})\right| p\right\rangle
\end{equation}
since the zero order terms agree and the second order term of the above
requirement: 
\begin{equation}
\lim_{g\rightarrow 1}\left\langle 0\left| A(0,g)\right| p\right\rangle
^{(2)}=0
\end{equation}
consists of a $W\cdot W$ contribution and the lowest counterterm
contribution (which we wrote on the left hand side). The evaluation of the
right hand side (omitting combinatorial factors) gives: 
\begin{equation}
g^{2}\int \int \Delta _{F}(0-x_{1})\Delta _{F}^{3}(x_{1}-x_{2})\frac{%
e^{ipx_{2}}}{(2\pi )^{\frac{3}{2}}}\simeq \frac{1}{p^{2}-m^{2}+i\varepsilon }%
\int e^{ip\xi }\Delta _{F}^{3}(\xi )d^{4}\xi
\end{equation}
Therefore $\delta m^{2(2)}\symbol{126}g^{2}\int e^{ip\xi }\Delta
_{F}^{3}(\xi )d^{4}\xi $ and $Z^{(2)}$ is the second Taylor coefficient in
the expansion of the integral around $p^{2}=m^{2\text{ }}$(note that $%
Z=1+Z^{(2)}+.$...).

Later we will (via time dependent scattering theory) meet a formalism which
relates off- shell quantities (correlation functions of interacting fields)
in a natural way with distinguished free fields which have the correct
physical mass ( the ``incoming'' /outgoing fields). This relation is
independent of the mass of the Fock space which we use for their
perturbative construction. Scattering theory may be viewed as an extension
of the adiabatic principle to multiparticle states. It is applied to
off-shell correlation functions and requires the introduction of the
physical mass i.e. a reparametrization which relates the mass of the
auxiliary Fock space to that of the in and out Fock space of scattering
theory. Since (Haag-Ruelle or LSZ-) scattering theory and (a fortiori) the
above adiabatic principle are consequences of the general framework of QFT,
a more systematic and conceptually clearer approach would have consisted in
presenting the perturbative treatment of interaction after an introduction
to the general framework. However a physical minded reader prefers to see
some important results and understand them at least partially with a modest
amount of mathematics and concepts before making a large formal investment.

Even though there are \textit{no physical reasons} to introduce counterterms 
\textit{for off-shell }quantities, the fact that the time-ordered products
of $\mathcal{W}^{\prime }s$ via Wick's theorem yield ill-defined (formally
infinite) expressions as e.g. $i\Delta _{F}^{3}(x_{1}-x_{2})$ forces
``renormalization'' for \textit{mathematical reasons}. With other words our
starting formula in Fock space although mathematically incorrect, was not
beyond physical redemption. The integrand in $S(g)$ i.e. $T\mathcal{W}%
(x_{1})....\mathcal{W}(x_{n}),$ though not defined on all (Schwartz)
testfunctions, is well-defined on the big class of testfunctions $%
f(x_{1,....}x_{n})$ which vanish of sufficient high order for coalescent
points. If $\mathcal{W}$ is a polynomial with $dim\mathcal{W}\leq 4$, the
order does not increase in $n$. This means that the (Hahn-Banach) extension
to all testfunctions will lead to time-ordered distributions which, although
lacking uniqueness, have well-controlled ambiguities (possibly additional
parameters) whose space-time dependence is given in terms of $\delta -$%
functions (which spread in the next order where new $\delta $-functions
arise) and derivatives up to a maximal finite order. With other words,
different extensions differ by finite local counterterms. These counterterms
may be used in order to achieve certain normalization conditions (as in the
case of the adiabatic principle), but there is no mathematical necessity to
take the ambient Fock space with a mass equal to the physical mass. The mass
of a particle does not belong to the set of observables which can be
measured locally. Globally however e.g. two free fields with different
masses cannot be unitarily equivalent.

In the following we will use the more pedestrian regularization methods
rather than the mentioned extension method (the latter will be used in the
later \textit{Curved Space Time} problems where it seems to be the \textit{%
only renormalization method}). Whereas for structural arguments we mostly
use the $A^{4}$model, the explicit second order calculations will be done in
Quantum Electrodynamics.. We now specialize to the $\mathcal{W}$ describing
QED\footnote{%
The reader should follow the more detailed calculations of Weinberg (\cite
{Wei}) or Itzykson-Zuber (\cite{I-Z}). We only sketch those computations
which we need in order to make some additional comments in line with our
aims.} (first without the counterterms): 
\begin{equation}
\mathcal{W}(x)=-ej_{0\mu }(x)A^{\mu }(x),\quad j_{0\mu }(x)=:\bar{\psi}%
_{0}\gamma _{\mu }\psi _{0}(x):\quad \psi _{0},A_{0\mu }\,\,\,%
\hbox{free
fields}
\end{equation}
and consider $S(g)$ in second order: 
\begin{equation}
S^{\left( 2\right) }(g)=\frac{e^{2}}{2!}\int \int g(x)g(y)Tj_{0\mu
}(x)j_{0\nu }(y)A_{0}^{\mu }(x)A_{0}^{\nu }(y)d^{4}xd^{4}y
\end{equation}
with the Wick-reordering from the previous chapter, we obtain for the
formfactor: 
\begin{eqnarray}
\left\langle p^{\prime }\left| j_{\mu }(0)\right| p\right\rangle ^{(2)}
&=&\left\langle 0\left| a(p^{\prime },s_{3}^{\prime })S^{*}T(j_{0\mu
}(0)S)a^{*}(p,s_{3})\right| 0\right\rangle ^{(2)}  \nonumber \\
&=&\left\langle 0\left| a(p^{\prime },s_{3}^{\prime })T(j_{0\mu
}(0)S)a^{*}(p,s_{3})\right| 0\right\rangle _{v.c.}^{(2)}
\end{eqnarray}
''vacuum-connected'' ($v.c.$) has the same meaning as before: leave out the $%
S^{*\hbox{ }}$in front of the $T$ and ignore the vacuum
``bubble''contributions. The evaluation of the right hand side amounts to
look for the $\bar{\psi}-\psi $ contribution in the Wick reordering of: 
\begin{equation}
Tj_{0\mu }(0)j_{0\nu }(x_{1})j_{0\kappa }(x_{2})A_{0}^{\nu
}(x_{1})A_{0}^{\kappa }(x_{2})
\end{equation}
In order to keep track of the combinatorial possibilities, it is customary
to draw graphs with vertices and edges (connecting lines). In our case their
are three interaction points 0, $x_{1}$and $x_{2}$, one connecting photon
line (one photon contraction ) and two e-lines so that one uncontracted $%
\psi _{0}$ and $\bar{\psi}_{0}$ remains. One easily sees that there are
three combinatorial distinct contributions according to whether the
remaining pair may come from $j_{0\mu }(0),$ from the $j(x)$'s or if it is
of mixed origin i.e. one from $j(0)$ and the other from a $j(x)$. The first
case only contributes in zero order since : 
\begin{equation}
\left\langle p^{\prime }\left| j_{0\mu }(0)\right| p\right\rangle
\left\langle 0\left| S^{*}S\right| 0\right\rangle ^{(2)}=0
\end{equation}
This cancellation is a general feature of all ``vacuum bubble''
contributions (which only contribute a phase factor to $S$ and the opposite
to $S^{*}).$ More interesting are the terms in which both of the $\psi $-$%
\bar{\psi}$ ``legs'' (the graphical representation of the fields which
remain after contractions) are contracted with legs in $S^{(2)}.$ This
constitutes the famous ``vacuum polarization'' contribution $\Gamma _{\mu
.pol}$ to the form factor and the so called ``one particle irreducible''
form factor $\Gamma _{\mu ,loop}$. The vacuum polarization contribution
contains the fluctuation of the zero order current: 
\begin{equation}
i\Pi _{\mu \nu }(x)=e^{2}\left\langle 0\left| Tj_{0\mu }(x)j_{0\mu
}(0)\right| 0\right\rangle
\end{equation}
The $\Gamma _{\mu ,loop}$-contribution originates from a contraction in
which one leg goes to one W-vertex and the other to the second. The
remaining ``electron selfenergy contribution'' arises from the mixed
contraction. It contain the electron selfenergy $\Sigma $ and therefore is
called $\Gamma _{\mu ,e.s.}$. The three types of terms are conveniently
pictured in terms of Feynman diagrams.(Fig.)

Inserting now the Fourier representation of the time ordered electron and
photon propagators we obtain: 
\begin{equation}
\left\langle p^{\prime }\left| j_{\mu }(0)\right| p\right\rangle =\bar{u}%
(p^{\prime },s_{3}^{\prime })(\Gamma _{\mu ,pol}+\Gamma _{\mu ,loop}+\Gamma
_{\mu ,e.s.})u(p,s_{3})
\end{equation}
where the vacuum polarization, vertex-loop and electron-selfenergy
contributions to the matrix-valued $\Gamma _{\mu }$ is 
\begin{equation}
\Gamma _{\mu ,pol}(p^{\prime },p)=\frac{-1}{(p^{\prime }-p)^{2}+i\varepsilon 
}\Pi _{\mu \nu }(k)\gamma ^{\nu }
\end{equation}
\begin{equation}
\Pi _{\mu \nu }(k)=\frac{-ie^{2}}{(2\pi )^{4}}\int d^{4}q\frac{Tr(\left[ -i%
\not{p}+m\right] \gamma _{\mu }\left[ -i(\not{q}-\not{k})+m\right] \gamma
^{\nu })}{(q^{2}-m^{2}+i\varepsilon )((q-k)^{2}-m^{2}+i\varepsilon )}
\end{equation}
\begin{equation}
\Gamma _{\mu ,loop}(p^{\prime },p)=\frac{ie^{2}}{(2\pi )^{4}}\int
d^{4}q\gamma ^{\rho }\frac{-i(\not{p}^{\prime }-\not{q})+m}{(p^{\prime
}-q)^{2}-m^{2}+i\varepsilon }\gamma _{\mu }\frac{-i(\not{p}-\not{q})+m}{%
(p-q)^{2}-m^{2}+i\varepsilon }\gamma _{\rho }  \label{ver}
\end{equation}
\begin{equation}
\Gamma _{\mu ,e.s.}=\left\{ i\Sigma (p^{\prime })S_{F}(p^{\prime })\gamma
_{\mu }+\gamma _{\mu }S_{F}(p)i\Sigma (p)\right\}  \label{sel}
\end{equation}
\begin{equation}
i\Sigma (p)=\frac{e^{2}}{(2\pi )^{4}}\int d^{4}q\frac{1}{q^{2}+i\varepsilon }%
\frac{\gamma ^{\kappa }(i\not{p}-i\not{k}+m)\gamma _{\kappa }}{%
(p-q)^{2}-m^{2}+i\varepsilon }
\end{equation}
The electron selfenergy $i\Sigma $ is not an observable quantity and
fortunately its contribution drops out by the adiabatic principle. This
happens because the two momentum variables $p^{\prime }$ and $p$ are on the
physical (mass m) mass shell and therefore the adiabatic principle forces us
to work with $\mathcal{W}_{adia}$ instead of $\mathcal{W}$ and we have to
fix the counterterm in such a way that the one particle matrix element of $%
\psi $ equals that of $\psi _{0}$. This is easily seen to be identical (in
second order) to: $i\Sigma (p)_{adia}u(p,s_{3})\mid _{\not{p}=m}=0$ with $%
i\Sigma _{adia}=i\Sigma +$ counter terms. The on-shell vanishing of the
selfenergy is just the mathematical expression that the persistent
selfenergy contribution to the large time asymptotics (equivalent to the
momentum space mass shell limit) has been correctly taken into account and
not ``switched off''. On the other hand as already mentioned, if we were to
compute the off-shell 3-point function $\left\langle T\psi \bar{\psi}A_{\mu
}\right\rangle ,$ we have the option to either use free fields with the
``bare'' mass $m_{0}$ and $\mathcal{W},$ or we can use free fields with the
physical mass and $\mathcal{W}_{adia}$. This observation is well known from
the Gell-Mann- Low representation or the Feynman-Kac representation of
correlation functions (naturally off-shell).

In passing we mention that the Gell-Mann- Low representation for the
correlation function of Heisenberg fields (in case of a scalar neutral
selfinteracting field) has the form: 
\begin{equation}
\left\langle TA(x_{1})....A(x_{n})\right\rangle =Z^{-1}\left\langle
TA_{0}(x_{1})....A_{0}(x_{n})e^{i\int W(A_{0}(x))d^{4}x}\right\rangle _{Fock}
\end{equation}
\[
\hbox{with }Z=\left\langle e^{i\int \mathcal{W}(A_{0}(x))d^{4}x}\right%
\rangle 
\]
We note that this is a special case of the previous formalism if we extend
it to products of fields and consider formally the adiabatic limit $%
g\rightarrow $1. The denominator $Z$ is a phase factor and represents the
(volume-dependent) ``vacuum-bubbles'' which cancel against similar
contributions from the numerator. The volume structure is completely
analogous to the thermodynamical limit of the Gibbs representation of
thermal correlation functions (at the end of chapter 1), with the only
structural difference that there is no equivalent to the KMS condition. For
physical reasons the correctly normalized ground state expectations should
result from the KMS theory for $\beta \rightarrow \infty .$

Looking formally at the momentum space representation for the second order
vacuum representation one would expect (by power counting in p) a quadratic
divergence. Invoking current conservation (or gauge invariance), only a
logarithmic divergence remains in $\pi _{\mu \nu }.$ A closer look at the
electron selfenergy term $i\Sigma $ reveals that the divergence is also
logarithmic and the same is obviously true (by power counting) for $\Gamma
_{\mu ,loop.}$ In the remainder of this section we present and explain the
result of the renormalization on the second order formfactor. The
presentation of the techniques and the actual calculation will be deferred
to the next section. We collect the results (omitting tildes in Fourier
transforms): 
\begin{eqnarray}
\Pi _{\mu \nu }(k) &=&i(k_{\mu }k_{\nu }-g_{\mu \nu }k^{2})\pi (k)%
\hbox{
\thinspace \thinspace \thinspace \thinspace with:}  \label{reg pol} \\
\pi ^{(2)}(k) &=&-\frac{\alpha }{3\pi }\left\{ +\frac{1}{3}+2(1+\frac{2m^{2}%
}{k^{2}})\left[ x{arccot}x-1\right] \right\} -Z_{3}^{(2)}  \nonumber \\
\hbox{where }x &=&(\frac{4m^{2}}{k^{2}}-1)^{\frac{1}{2}}\hbox{ for }%
k^{2}<4m^{2}\hbox{ and anal.cont.},  \nonumber \\
Z_{3}^{(2)} &=&\frac{\alpha }{3\pi }\ln \frac{\Lambda ^{2}}{m^{2}}  \nonumber
\end{eqnarray}
\begin{eqnarray}
\Sigma ^{(2)}(p) &=&\frac{\alpha }{2\pi }\left\{ 
\begin{array}{c}
(\not{p}-m)(1+\ln \frac{\mu ^{2}}{m^{2}})+2m\frac{m^{2}-p^{2}}{p^{2}}\ln (1-%
\frac{p^{2}}{m^{2}}) \\ 
-\not{p}\left[ \frac{3}{2}\frac{m^{4}-\left( p^{2}\right) ^{2}}{\left(
p^{2}\right) ^{2}}\ln (1-\frac{p^{2}}{m^{2}})+\frac{m^{2}-p^{2}}{p^{2}}%
\right]
\end{array}
\right\}  \nonumber \\
&&+\delta m+Z_{2}^{(2)}(\not{p}-m)  \label{reg self} \\
\hbox{with \thinspace \thinspace }\delta m &=&\frac{3\alpha }{4\pi }m(\ln 
\frac{\Lambda ^{2}}{m^{2}}+\frac{1}{2}),\;Z_{2}^{(2)}=\frac{\alpha }{2\pi }(%
\frac{1}{2}\ln \frac{\Lambda ^{2}}{m^{2}}+\ln \frac{\mu ^{2}}{m^{2}}+\frac{9%
}{4}+O(\frac{\mu }{m}))  \nonumber
\end{eqnarray}
\begin{eqnarray}
\Gamma _{\mu ,loop}^{(2)}(q^{2}) &=&\gamma _{\mu }F_{loop}^{(2)}(\theta )+%
\frac{i}{2m}\sigma _{\mu \nu }q^{\nu }G_{loop}^{(2)}(\theta )+\gamma _{\mu }B
\\
F_{loop}^{(2)}(\theta ) &=&\frac{\alpha }{\pi }\left\{ (\ln \frac{\mu }{m}%
+1)(\theta \coth \theta -1)-2\coth \theta \int_{0}^{\frac{\theta }{2}}\chi
\tanh \chi d\chi -\frac{\theta }{4}\tanh \frac{\theta }{2}\right\}  \nonumber
\\
G_{loop}^{(2)}(\theta ) &=&\frac{\alpha }{2\pi }\frac{\theta }{\sinh \theta }%
\quad \hbox{with }q^{2}=-4m^{2}\sinh ^{2}\frac{\theta }{2},\quad
\,\,\,\theta :"rapidity"  \nonumber
\end{eqnarray}
Here $\Lambda $ is a cutoff i.e. a formal device which cuts off certain
divergent momentum space integrals in a Lorentz-invariant manner. Although $%
\Lambda \,$carries no direct physical significance (and will be removed
shortly), it is important that the $\Lambda -$ dependent terms have at most
a polynomial p-dependence i.e. they are of the form of $\delta -$functions
and derivatives thereof. The infrared cutoff $\mu $ on the other hand has a
physical origin. The interaction of charged particles with ``soft'' ($\omega
\rightarrow 0)$ photons is very strong and changes the character of the one
particle states. Its projection on Wigner particles is zero and strictly
speaking we must abandon our formulation of the adiabatic principle (and the
standard forms of scattering theory) and think about ``infraparticles''. In
order to not drift too far away from elementary treatments we followed
standard practice and introduced a photon-mass $\mu $ into the $A_{\mu }$
propagator retaining at the end only the leading contribution for small $\mu
.$ Note that the $\Lambda -$dependent (unrenormalized) $\Sigma $ is
infrared-finite ($B$ contains a compensating contribution). In $\Gamma $
there is no such compensation. We have separated the $\Lambda $-cutoff
dependent $\delta m^{2}$ and $B$ terms in $\Sigma $ because the adiabatic
principle fixes the counter terms in $\mathcal{W}_{adiab}$ to be: 
\begin{equation}
\mathcal{W}_{c.t.}=\delta m^{(2)}\bar{\psi}\psi -B\bar{\psi}\not%
{\partial}\psi
\end{equation}
This leads to a modification (renormalization) of $\Sigma .$ We already
noted that the resulting $\Sigma _{adia}$ is the $\Lambda $-independent
content of the curly bracket. Insertion into the formula (\ref{sel}) gives $%
\bar{u}(p^{\prime })\Gamma _{\mu ,e.s}^{r}u(p)$ $=0$ i.e. the renormalized
contribution vanishes on the mass shell \/ \ \ \ ${p=m}^{2}$. Enforcing the
charge normalization for the diagonal matrix element: 
\begin{equation}
\bar{u}(p,s_{3})\Gamma _{\mu }^{r}u(p,s_{3})=\bar{u}(p,s_{3})\gamma _{\mu
}u(p,s_{3})
\end{equation}
we also eliminate the $\Lambda -$dependence in $\Gamma _{\mu ,pol}$ and $%
\Gamma _{\mu ,loop}.$ The result is: 
\begin{equation}
\Gamma _{\mu }^{r}=\gamma _{\mu }F^{(2)}(\theta )+\frac{i}{2m}\sigma _{\mu
\nu }q^{\nu }G^{(2)}(\theta )
\end{equation}
with $F$ and $G$ given by the previous formulae in the regime $q^{2}<0$ and
everywhere by analytic continuation (as $\pi (q^{2})$ and $\Sigma (q)$ they
can be represented by analytic functions with a cut on the real axis).
According to the previous section we obtain for the anomalous contribution
to the magnetic moment: 
\begin{equation}
\mu _{an}=\frac{e}{2m}(F(0)+G(0))=\frac{e}{2m}\frac{\alpha }{2\pi }
\end{equation}
Note that only the infrared-finite $G$ contributes to the zero Taylor term.

The calculation of the Lamb-shift is more complicated, computationally
(since atomic physics enters) as well as conceptually. Here one is
interested in the $2s_{\frac{1}{2}}-2p_{\frac{1}{2}}$ energy split which the
Dirac theory with external fields cannot explain. The idea is to study the
electron selfenergy term $\Sigma $ in the presence of an external field.
Whereas it is not difficult to represent the resulting energy shift in a
stationary $A_{\mu }^{ex}$as: 
\begin{equation}
\delta E_{N}=-\int \int d^{3}p^{\prime }d^{3}p\bar{u}_{N}(p^{\prime })\Sigma
^{r}(p^{\prime },p,E_{N};A_{\mu }^{ext})u_{N}(p),
\end{equation}
the evaluation of the low energy part of the renormalized external field
dependent selfenergy part $\Sigma ^{r}$ requires significant skills and
intuition. Since the atomic wave functions $u_{N}$ prevent the electron to
become free, one expects $\delta E_{N}$ to be a ``good'' variable i.e. $\mu
- $independent. In distinction to the previous case, this independence is
not manifest and can be used as a check on the approximation methods needed.
The only manageable approximation method involves a different treatment of
high and low energy contributions \cite{Wei}. For the low energy part one
uses the above formula. It turns out that the main modification consists in
replacing the $S_{F}(x-y)$ function by $S_{F}(x,y,A_{\mu }^{ext})$ as well
as a so-called ``tadpole'' term involving: 
\[
\left\langle 0\left| j_{0\mu }(x,A_{\mu }^{ext})\right| 0\right\rangle
=Tr\gamma _{\mu }S_{F}(x,x,A_{\mu }^{ext}) 
\]
The latter occurs because the charge conjugation invariance, which prohibits
any vacuum expectation with an odd number of $j^{\prime }$s, is broken by
the external field. This new term requires a ``tadpole'' counter term. The $%
S_{F}(x,y,A_{\mu }^{ext})\,$only involves Dirac's external field theory. The
high energy part is calculated in first order of $A_{\mu }^{ext}$ and has
the form: 
\[
\delta E_{N}\mid _{h.e.}=ie\int \int d^{3}pd^{3}p^{\prime }\bar{u}(\mathbf{p}%
^{\prime })\Gamma _{\mu }(\mathbf{p},E_{N};\mathbf{p}^{\prime }E_{N})u(%
\mathbf{p})A_{\mu }^{ext}(\mathbf{p}-\mathbf{p}^{\prime }) 
\]
The $\Gamma _{\mu }$ is almost the previous form factor $\Gamma ,$ except
for the fact that the hydrogen wave functions pushes it slightly off-shell.
One replaces this by the on-shell $\Gamma ,$ assuming a smooth continuation.
The infrared dependence of the latter causes a problem. One solution \cite
{I-Z} consists in converting the unphysical photon mass $\mu $ into an
infrared photon energy cutoff $\kappa $ via use of the soft bremsstrahlung. $%
K$ is then used to define an upper integration limit in $\delta E\mid
_{l.e.},$whereas the high frequency contribution $\delta E\mid _{h.e.}$is
calculated from the previous formula with $\mu $ in the formfactor $\Gamma
_{\mu }$ being replaced in favor of $\kappa $. The result for the s-p
splitting in hydrogen is: 
\begin{equation}
\delta E_{2s}-\delta E_{2p_{\frac{1}{2}}}=\frac{\alpha ^{5}m}{6\pi }\left\{
\ln \frac{\Delta E_{2p}}{\alpha ^{2}\Delta E_{2s}}+\frac{19}{30}+\frac{1}{8}%
\right\}
\end{equation}
Here the $\Delta E^{\prime }s$ are suitably averaged energies of the
hydrogen atom (only numerically accessible). This result corresponds to the
famous value 1052,19 MHZ. instead of a photon energy cutoff one may also
base the division on the decomposition for the photon propagator: 
\[
\frac{1}{k^{2}+i\varepsilon }=\frac{1}{k^{2}-\mu ^{2}+i\varepsilon }+(\frac{1%
}{k^{2}+i\varepsilon }-\frac{1}{k^{2}-\mu ^{2}+i\varepsilon }) 
\]
The first part leads to the $\Gamma $ contribution whereas the second faster
decreasing part enters the atomic physics calculation. Again the infrared
singular terms cancel. This somewhat more attractive calculation (invariant
cutoff) can be found in \cite{Wei}. Other physical problems related to the
formfactor are the radiative corrections to the Coulomb scattering i.e. the
second order correction to the Mott-formula and the bremsstrahlung
correction to the Mott formula. Both are separately infrared divergent for $%
\mu \rightarrow 0,$ but their joint cross section (for fixed photon infrared
resolution) approaches a finite limit. This is a special case of good
infrared behavior of photon inclusive cross sections which are the principle
observables of QFT's involving photons.

In this section we met two slightly different reasons for renormalization.
One is entirely physical: if we describe matrix elements between particle
states (i.e. on-shell quantities) we must use the $\mathcal{W}_{adiab}$ as
our interaction, independent of whether the counter terms in the formal
treatment have infinite coefficients or not. There is no other operator
description for such quantities than the one in a Fock space with the
correct mass. In the next section we will study off-shell quantities which
do not require $\mathcal{W}_{adiab}$. Any auxiliary mass Fock space may be
used for their perturbative evaluation. It will be shown later, that
scattering theory reconciles the description of on- and off-shell
quantities. The Hilbert space for scattering theory requires a
reparametrization from the off-shell auxiliary mass to the physical mass as
well as a multiplicative adjustment. In momentum space the difference
between these two type of quantities looks deceivingly simple: one just
sends certain p-variables to the physical mass shell. In x-space however
this distinction looks more dramatic: it is the difference between global
(particles involve asymptotic limits) and local quantities. In algebraic QFT
the fact that the local spacetime description does not ``perceive'' the
mass, corresponds to the local equivalence of algebras which belong globally
to inequivalent (different charges) representations (see later sections).

We end this section with some formal remarks on how to use the above
time-ordered formalism to obtain perturbative correlation functions. As a
prototype theory which is free of infrared problems, tensor-and
spinor-indices etc., we take the model with $\mathcal{W}(x)$=:$A_{0}(x)^{4}$:

Previously we have seen that the Gell-Mann Low representation for time
ordered $n$-point functions has the following form: 
\begin{equation}
\left\langle TA(x_{1})....A(x_{n})\right\rangle =\lim_{g(x)\rightarrow
g}\left\langle TA_{0}(x_{1})....A_{0}(x_{n)}e^{i\int g(x)\mathcal{W}%
(x)d^{4}x}\right\rangle _{0}^{v.c.}
\end{equation}
The subscript $0$ on the right hand side is a reminder that the free field
expressions are to be evaluated in the $A_{0}$-Fock space and the
superscript $v.c.$ (vacuum connected) indicates that ``vacuum bubbles'' in
the Wick-ordering must be omitted. We also mentioned the extension method of
distributions which succeeds to give an iterative definition of the expanded
right hand side. More popular with physicists (but not necessarily more
physical!) are the various regularization methods which we will discuss in
the next section. Let us consider the purely formal aspects of the $A^{4}$
model. This time we introduce counterterms $\mathcal{W}_{c}$ solely for the
elimination of the divergencies which arise from the removal of the
unphysical regularizations: 
\begin{eqnarray}
\mathcal{\hat{W}} &=&g:A_{0}^{4}:+\mathcal{W}_{c} \\
\mathcal{W}_{c} &=&\delta m^{2}Z:A_{0}^{2}:+Z(:\partial _{\mu }A_{0}\partial
^{\mu }A_{0}-m^{2}A_{0}^{2}:)+g(Z_{g}-1):A^{4}:  \nonumber
\end{eqnarray}
The claim (proven partially later) of renormalization theory is that $\delta
m^{2},Z$ and $Z_{g}$ can be chosen such that the correlation function stay
finite in the limit of removed regularization. This time the mass appearing
in Fock space does not have to be the physical one and the normalization of
A is not required to be standard. However in order to have a simple form for
scattering formulas it is convenient to implement the physical
parametrization and the standard field normalization already in every order
of perturbation theory.

\section{Invariant Parametrizations, Regularization}

The $x$-or $p$-space integrations of perturbation theory extend over
noncompact regions and are difficult to perform in their original form. An
efficient formalism which also allows to maintain the manifest Lorentz
covariance in the presence of regularizations and cutoffs is based on
Schwinger's $\alpha $-parametrization (another one is due to Feynman ): 
\begin{equation}
\frac{1}{p^{2}-m^{2}+i\varepsilon }=\int_{0}^{\infty
}e^{ia(p^{2}-m^{2}+i\varepsilon )}da
\end{equation}
\[
\frac{1}{\not{p}-m+i\varepsilon }=\frac{\not{p}+m}{p^{2}-m^{2}+i\varepsilon }%
=(\not{p}+m)\int_{0}^{\infty }e^{ia(p^{2}-m^{2}+i\varepsilon )}da 
\]
where the $i\varepsilon $ provides a damping factor for the upper
integration limit. Applying this representation to the second order vacuum
polarization we obtain: 
\begin{eqnarray}
\pi _{\mu \nu }(k) &=&-e^{2}\int \frac{d^{4}p}{(2\pi )^{4}}\left\{ \frac{%
Tr\gamma _{\mu }(\not{p}+m)\gamma _{\nu }(\not{p}-\not{k}+m)}{%
(p^{2}-m^{2}+i\varepsilon )((p-k)^{2}-m^{2}+i\varepsilon )}\right\} \\
&=&-4e^{2}\int \frac{d^{4}p}{(2\pi )^{4}}\left\{ \frac{p_{\mu }(p-k)_{\nu
}+\left\{ \mu \Leftrightarrow \nu \right\} -g_{\mu \nu }(p^{2}-p\cdot
k-m^{2})}{(p^{2}-m^{2}+i\varepsilon )((p-k)^{2}-m^{2}+i\varepsilon )}\right\}
\nonumber \\
&=&-4e^{2}\int \frac{d^{4}p}{(2\pi )^{4}}\left\{ \frac{\partial }{\partial
x^{\mu }}\frac{\partial }{\partial y^{\nu }}+\left\{ \mu \Leftrightarrow \nu
\right\} -g_{\mu \nu }(\frac{\partial }{\partial x}\cdot \frac{\partial }{%
\partial y}+m^{2})\right\}  \nonumber \\
&&\times \int \int\limits_{0}^{\infty }d\alpha _{1}d\alpha _{2}\exp i\left\{ 
\begin{array}{c}
\alpha _{1}(p^{2}-m^{2}+i\varepsilon )+\alpha
_{2}((p-k)^{2}-m^{2}+i\varepsilon ) \\ 
+x\cdot p+y\cdot (p-k)
\end{array}
\right\} _{x=0=y}  \nonumber
\end{eqnarray}
where in the last step we used the $\alpha $-parametrization and eliminated
the polynomial in the numerator by differentiation and setting the auxiliary
variables zero at the end. In this form the p-integration involves easy to
do oscillatory Gaussian integrals and the original divergence has been
shifted into the $\alpha $-integrals as divergencies at $\alpha =0$: 
\begin{eqnarray}
\pi _{\mu \nu }(k) &=&\frac{i\alpha }{\pi }\int \int\limits_{0}^{\infty }%
\frac{d\alpha _{1}d\alpha _{2}\alpha _{1}\alpha _{2}}{(\alpha _{1}+\alpha
_{2})^{4}}\left\{ 
\begin{array}{c}
2(k_{\mu }k_{\nu }-g_{\mu \nu }k^{2}) \\ 
-g_{\mu \nu }\left( k^{2}+\left[ m^{2}-\frac{i}{\alpha _{1}+\alpha _{2}}%
\right] \frac{(\alpha _{1}+\alpha _{2})^{2}}{\alpha _{1}\alpha _{2}}\right)
\end{array}
\right\} \times  \nonumber \\
&&\times \exp i\left\{ \frac{\alpha _{1}\alpha _{2}}{\alpha _{1}+\alpha _{2}}%
k^{2}-(\alpha _{1}+\alpha _{2})m^{2}\right\}
\end{eqnarray}
Here we split the polarization into a transversal and longitudinal part.
Note that the transversality property: 
\begin{eqnarray}
k^{\mu }\pi _{\mu \nu }(k) &=&-e^{2}\int \frac{d^{4}p}{(2\pi )^{4}}Tr(\not{k}%
\frac{1}{\not{p}-m+i\varepsilon }\gamma _{\nu }\frac{1}{\not{p}-\not%
{k}-m+i\varepsilon }) \\
&=&-e^{2}\int \frac{d^{4}p}{(2\pi )^{4}}Tr\gamma _{\nu }(\frac{1}{\not{p}-%
\not{k}-m+i\varepsilon }-\frac{1}{\not{p}-m+i\varepsilon })\stackrel{?}{=}0 
\nonumber
\end{eqnarray}
does not follow because the translation of integration variables is not
allowed. Instead of enforcing the transversality condition by ``brute
force'' (vanishing of the longitudinal term), we may also use
regularizations which maintain transversality (gauge invariance). There are
two gauge invariant methods: the Pauli-Villars method of auxiliary fields
and the more recent dimensional regularization method. The P-V method adds
fictitious spinor fields with masses $m_{i}=\lambda _{i}m$ and strength $%
C_{i}:$%
\begin{equation}
\pi _{\mu \nu }^{PV}(k,m_{1},m_{2}...)=\pi _{\mu \nu
}(k,m)+\sum_{i=1}^{n}C_{i}\pi _{\mu \nu }(k,m_{i})
\end{equation}
The power counting of the integrand indicates convergence for: 
\begin{equation}
1+\sum_{i=1}^{n}C_{i}=0
\end{equation}
For sufficiently convergent integrals one can shift integration variables
and obtain the transversality of $\pi _{\mu \nu }^{PV}.$ The transversal $%
\pi _{\mu \nu }^{PV}$ has the following $\alpha $-representation: 
\begin{eqnarray}
\pi _{\mu \nu }^{PV}(k) &=&-i(g_{\mu \nu }k^{2}-k_{\mu }k_{\nu })\pi (k) \\
\pi (k) &=&\frac{2\alpha }{\pi }\int \int\limits_{0}^{\infty }\frac{d\alpha
_{1}d\alpha _{2}\alpha _{1}\alpha _{2}}{(\alpha _{1}+\alpha _{2})^{4}}%
\sum_{i=0}^{n}C_{i}\exp i\left\{ k^{2}\frac{\alpha _{1}\alpha _{2}}{\alpha
_{1}+\alpha _{2}}-(m_{i}^{2}-i\varepsilon )(\alpha _{1}+\alpha _{2})\right\}
\nonumber \\
&=&\frac{2\alpha }{\pi }\int \int\limits_{0}^{\infty }d\alpha _{1}d\alpha
_{2}\alpha _{1}\alpha _{2}\delta (1-\alpha _{1}-\alpha _{2})\times  \nonumber
\\
&&\times \int_{0}^{\infty }\frac{d\rho }{\rho }\sum_{i=0}^{n}C_{i}\exp i\rho
\left\{ k^{2}\alpha _{1}\alpha _{2}-m_{i}^{2}+i\varepsilon \right\} 
\nonumber
\end{eqnarray}
where in the last line the identity $\int d\rho \delta (\rho -\alpha
_{1}-\alpha _{2})=1$ was used to introduce a radial variable $\rho .$ By
appropriate choice of the $C_{i}$ one improves the small $\rho $ behavior..
For our purpose it is sufficient that the above relation $\sum_{0}^{n}C_{i}$ 
$=1+\sum_{1}^{n}C_{i}=0$ leads to the vanishing of the sum in the integrand.
For this $n=1$ suffices. Higher order zeros could be obtained by requiring
higher moments to vanish as well i.e. $\sum_{i}m_{i}^{s}C_{i}=0.$ The $\rho $
integration gives (considering only leading contributions for large $\lambda
_{i}$): 
\begin{eqnarray}
\lim_{r\rightarrow 0}\int_{r}^{\infty }\frac{d\rho }{\rho }%
\sum_{i=0}^{n}C_{i}.... &=&\lim_{r\rightarrow
0}\sum_{0}^{n}C_{i}(-e^{i\sigma }\ln \sigma \mid _{\sigma =r(k^{2}\alpha
_{1}\alpha _{2}-m_{i}^{2})} \\
&&+\int_{0}^{\infty }d\sigma e^{i\sigma (1+i\varepsilon )}\ln \sigma ) 
\nonumber \\
&=&-\left\{ \ln (1-\frac{\alpha _{1}\alpha _{2}k^{2}}{m^{2}})-\ln \frac{%
\Lambda ^{2}}{m^{2}}\right\} ,\quad  \nonumber \\
\hbox{with }\sum_{1}^{n}C_{i}\ln \lambda _{i}^{2} &=&-\ln \frac{\Lambda ^{2}%
}{m^{2}}  \nonumber
\end{eqnarray}
This yields the Pauli-Villars regularized vacuum polarization: 
\begin{eqnarray}
\pi ^{PV}(k) &=&-\frac{2\alpha }{\pi }\int_{0}^{1}dxx(1-x)\left\{ \ln
(1-x(1-x)\frac{k^{2}}{m^{2}})-\ln \frac{\Lambda ^{2}}{m^{2}}\right\} 
\nonumber \\
&=&-\frac{\alpha }{3\pi }\left\{ 2(1+\frac{2m^{2}}{k^{2}})\left[ y{arccot}%
y-1\right] +\frac{1}{3}-\ln \frac{\Lambda ^{2}}{m^{2}}\right\} \\
\hbox{with }y &=&\hbox{(}\frac{4m^{2}}{k^{2}}-1)^{\frac{1}{2}}  \nonumber
\end{eqnarray}
Another more recent regularization scheme which also maintains gauge
invariance of $\pi _{\mu \nu }$ is the dimensional regularization. This
method only works in the euclidean formulation of perturbation theory which
will be discussed in a later section. The regularized expressions for $%
i\Sigma $ and $\Gamma _{\mu }$ are also conveniently derived in the $\alpha $%
-parametrization. since the integration over the loop momentum is always a
simple (oscillatory) Gaussian, we write directly: 
\begin{eqnarray}
\Sigma (p) &=&\frac{\alpha }{2\pi }\int \int \frac{d\alpha _{1}d\alpha _{2}}{%
(\alpha _{1}+\alpha _{2})^{2}}(2m-\frac{\alpha _{1}}{\alpha _{1}+\alpha _{2}}%
\not{p})\times  \nonumber \\
&&\times \exp \left\{ i(\frac{\alpha _{1}}{\alpha _{1}+\alpha _{2}}%
p^{2}-\alpha _{1}\mu ^{2}-\alpha _{2}m^{2})\right\} \\
&=&\frac{\alpha }{2\pi }\int_{0}^{\infty }\frac{d\rho }{\rho }\int \int
d\alpha _{1}d\alpha _{2}\delta (1-\alpha _{1}-\alpha _{2})(2m-\alpha _{1}\not%
{p})e^{i\rho (\alpha _{1}\alpha _{2}p^{2}-\alpha _{1}\mu ^{2}-\alpha
_{2}m^{2})}  \nonumber
\end{eqnarray}
As in the previous case we add a PV regularization term but this time
through an auxiliary ``heavy photon field'' of mass $\Lambda ^{2}.$ Again
only retaining the leading term, we obtain the answer by the substitution($%
C_{0}=1,C_{1}=-1$): 
\begin{equation}
\int \frac{d\rho }{\rho }e^{i\rho (\;above\;)}\rightarrow \int \frac{d\rho }{%
\rho }(e^{i\rho (\;above\;)}-e^{-i\rho \alpha _{1}\Lambda ^{2}})
\end{equation}
The $\rho -$integration and one of the $\alpha -$integrations say $\alpha
_{2}$ can be done and we are left with the following integral
representation: 
\begin{equation}
\Sigma (p,\Lambda )=\frac{\alpha }{2\pi }\int_{0}^{1}dx(2m-x\not{p})\ln 
\frac{x\Lambda ^{2}}{(1-x)m^{2}-x(1-x)p^{2}+x\mu ^{2}-i\varepsilon }
\end{equation}
If we stay within $p^{2}<m^{2}$ we may set $\mu =0$ and obtain the explicit
results 
\begin{equation}
\Sigma =\frac{\alpha }{2\pi }\left\{ 
\begin{array}{c}
\ln \frac{\Lambda ^{2}}{m^{2}}(2m-\frac{1}{2}\not{p})+2m(1+\frac{m^{2}-p^{2}%
}{p^{2}}\ln (1-\frac{p^{2}}{m^{2}}) \\ 
-\frac{1}{2}\not{p}[\frac{m^{4}-(p^{2})^{2}}{(p^{2})^{2}}\ln (1-\frac{p^{2}}{%
m^{2}})+\frac{m^{2}}{p^{2}}+\frac{3}{2}]
\end{array}
\right\}
\end{equation}
This is a matrix-valued analytic function in the cut p$^{2}-$plane which has
a diverging derivative on the mass shell \/ \ \ \ ${p}=m^{2}$ as a reminder
of the infrared problem. By keeping $\mu $ finite, the mass shell limit has
finite derivatives.

\begin{itemize}
\item  The regularization of the one particle irreducible second order
contribution to the vertex function (\ref{ver}) is more involved, since as a
result of the presence of three propagators one has to introduce 3 $\alpha
^{\prime }s.$ The $\alpha $-representation for the three denominators reads
as: 
\begin{eqnarray}
&&\int \frac{d^{4}q}{(2\pi )^{4}}\frac{e^{iqx}}{(q^{2}-\mu ^{2}+i\varepsilon
)(q^{2}-2p^{\prime }\cdot q+i\varepsilon )(q^{2}-2p\cdot q+i\varepsilon )} \\
&=&\frac{1}{(4\pi )^{2}}\int \int \int \frac{d\alpha _{1}d\alpha _{2}d\alpha
_{3}}{(\alpha _{1}+\alpha _{2}+\alpha _{3})^{2}}\exp -i\left\{ \alpha
_{1}\mu ^{2}+\frac{(\frac{x}{2}-\alpha _{2}p^{\prime }-\alpha _{3}p)^{2}}{%
\alpha _{1}+\alpha _{2}+\alpha _{3}}\right\}  \nonumber
\end{eqnarray}
where the exponential was added for the same reasons as in the previous case
of $\pi _{\mu \nu }$ namely to convert the polynomial q-dependent numerator
into a differentiation acting on the variable x. The (PV type)
regularization can be again implemented through the photon propagator. Since
there is no essential new idea involved but (even if one passes from the
off-shell vertex to the on shell formfactor) only some lengthy calculations,
we skip the details in the derivation of formula (\ref{reg pol})The
divergent $\Lambda $-dependent parts are evidently local (polynomial in the
external momentum variables) ant therefore can be compensated by counter
terms of the following kind: 
\begin{equation}
\mathcal{W}_{c.t.}=-\frac{1}{4}(Z_{3}-1)F^{2}+(Z_{2}-1)(\frac{i}{2}\bar{\psi}%
\overleftrightarrow{\not{\partial}}\psi -m\bar{\psi}\psi )+Z_{2}\delta m\bar{%
\psi}\psi -e(Z_{1}-1)\bar{\psi}\not{A}\psi
\end{equation}
Finite parts in counterterms would remain unspecified unless one imposes
normalization conditions. Natural normalization conditions are the
conditions which result from the adiabatic requirement (a must for on shell
quantities) augmented by the physical charge parametrization of the
formfactor. The regularized formulas of the previous section (\ref{reg pol})
have been written in such a way that the natural normalization means
omission of the $\delta m$ and $Z$-terms. It is easily seen that the
remaining second order $\pi ,\Sigma $ and $\Gamma $ terms have the correct
zeros required by the adiabatic principle and the physical charge (coupling
constant) parametrization. The proof of $n^{th}$ order renormalizability,
i.e. the statement that the old local counterterms iterated together with
the original $\mathcal{W}$ in the $S(g)$ expansion lead to higher order
correlation functions which in turn may be liberated from their infinities
by new higher order counterterms of the same local structure, requires a
significant extension of the regularization formalism. These notes are not
intended as a substitute for the rigorous treatment of $n^{th}$ order
renormalized pertubation theory. We only want to explain the physical
concepts behind as well as to present some of the famous second order QED
radiative corrections.
\end{itemize}

We already stressed the fact that the close connection between particles and
fields is an artifact of perturbation theory and not a result of the use of
the choice of a particular Fock space for the definition of local operator
algebras. In some sense the infrared singularities of Maxwell like (gauge)
theories can be interpreted as a perturbative indication that the theory is
not compatible with the imposed zero order particle content.

There exists a widespread misconception that a Lagrangian quantization
viewpoint is important for the intrinsic physical understanding of
interactions in QFT. From such a point of view the ultraviolet divergencies
appear as serious flaw of perturbation theory. The history of
renormalization \cite{Brown}. After all, renormalization arose from some
remarks of Kramers who suggested to use similar distinctions between bare
and physical masses (and other parameters) in QED as they were used by
Lorentz and Poincar\'{e} in in their attempt to understand the selfenergy
problem of particles grafted upon classical field theory. Additional support
came from the observation that for a few models (e.g. $A^{4}$ in d=1+1)
which were ``well-behaving'' in the perturbative treatment, it was possible
to prove their mathematical existence by extending the perturbative method.
In this connection also the functional integral method (based on the
euclidean Feynman-Kac representation) which furnishes a rather direct
relation between QFT and classical physics (more elegant than canonical
quantization) ought to be mentioned. However the recent progress in a
nonperturbative understanding of interactions from a different starting
point (sometimes called the ``bootstrap approach'') has cast grave doubts on
the universal correctness (apart from those few exceptions) of such
quantization approaches to interactions (beyond an infinitesimal deformation
picture). We will return to this important point at a more appropriate place.

It also should be stressed that renormalized perturbation theory does not
lend credibility to the idea of a physical cutoff i.e. a $\Lambda $ which
cannot be absorbed into the renormalization constants but rather enters the
physics. A physically interpretable nonlocal theory with an elementary
length in form of a cutoff $\Lambda \,$does presently not exist\footnote{%
The path of QFT is littered with proposals of theories which claimed to be
nonlocal and physical: cut-off in Feynman integrals, formfactors in
Lagrangians, ``peratization'' (pairs of complex poles in Feynman rules) etc.
The only attempt which is still not disproved (although the causality issue
has not yet been completely settled) is one in which the Einstein causality
in Minkowski spacetime is replaced by another structure which originates
from a model of noncommutative spacetime \cite{DFR}.}. If it ever comes into
existence, it probably will be quantum gravity par excellence. For large
families of lattice models many concepts of QFT as particle and scattering
notions exist, even though they are much harder to prove \cite{Barata} (as a
result of the missing knowledge about commutativity properties at different
times i.e. the substitute for spacelike commutativity). However even though
the lattice cutoff does not wreck the interpretation in terms of particle
excitations, the scarce rigorous results on scaling limits do not justify to
attribute a physical meaning in the QFT use of lattice model cutoffs.

\section{Specialities of Perturbative Gauge Theories}

The calculations of QED radiative corrections in the previous section made a
very naive use of covariant vectorpotentials in the spirit of classical
gauge theory and relied on the hope that classically gauge invariant
quantities will only involve the quantum physical degrees of freedom of
photons. The incorporation of this classical gauge principle into the
Heisenberg-Pauli-Fermi canonical field quantization then leads to the
quantized abelian gauge theory. In order to achieve something similar for
the nonabelian case, it is helpful to change from canonical to functional
integral quantization, in which case one encounters the presence of the more
complicated nonabelian Faddeev-Poppov ghosts which require the somewhat
involved BRS formalism for the extraction of physical quantities.

In the spirit of a more intrinsic approach, we should however avoid
quantization arguments altogether and start directly from the Wigner theory
applied to free fields in Fock space.

The gauge concept originated in the form of ``minimal coupling'' in QM with
external electro-magnetic interactions. As with many rules and recipes in QM
(e.g. spin-statistics), one expects that this gauge principle is a
consequence of the more fundamental spin=1 LQP together with perturbative
renormalizability. Indeed as we show below, the consistency requirements on
spin=1 perturbative interactions leave much less freedom than those between
lower spins, in stark distinction to classical physics, where it is the
other way around (and where one needs the gauge principle as a selection
criteria. Faithful with Bohr's correspondence principle, it is the more
basic LQP which explains the gauge principle of classical or external field
theory of vectorfields rather than the other way around, as the quantization
approach of the textbooks make believe. This leads not only to an
alternative viewpoint e.g. on the Schwinger Higgs mechanism, but relates
such disconnected issues as the decoupling of the alias Higgs field (in the
terminology of the standard approach) with the appearacce of semiinfinite
string-like localized charged fields in the zero mass limit for the mass of
the vectormesons.

In this section we first we remind ourselves that the general relation
between the (m,s) Wigner canonical creation and annihilation operators of
(anti)particles and the covariant pointlike free fields is (see previous
chapter): 
\begin{equation}
\psi ^{\left[ n_{+},n_{-}\right] }(x)=\int \sum_{s_{3}=-s}^{s}\left\{
e^{-ipx}u(p,s_{3})a(p,s_{3})+e^{ipx}v(p,s_{3})b^{*}(p,s_{3})\right\} \frac{%
d^{3}p}{2\omega }
\end{equation}
where $u(p,s_{3})$ are the columns of the rectangular ($2n_{+}+1)(2n_{-}+1)%
\times (2s+1)$ intertwiner matrix $U(p)$ which intertwines the canonical
Wigner representation with the covariant representation of the Poincar\'{e}
group built on the calculus of (un)dotted spinors: 
\begin{equation}
D^{\left[ n_{+},n_{-}\right] }(\Lambda )U(\Lambda ^{-1}p)=U(p)D^{\left[
s\right] }(R(\Lambda ,p)
\end{equation}
Independent of the choice of covariant field coordinates, for $s\geq 1$ one
always encounters a polynomial of degree $m\geq 2$ in p-space for the
two-point function: 
\begin{equation}
\left\langle \psi _{\cdot }(x)\psi _{\cdot }^{*}(y)\right\rangle =P_{\cdot
\,\,\,\cdot }^{(m)}(i\partial )i\Delta ^{(+)}(x-y)
\end{equation}
In the massless case the $\Delta $-functions have to be replaced by $D$%
-functions and in addition is has to be kept in mind that the family of
possible intertwiners is more restrictive as a result of the helicity rule $%
n_{+}-n_{-}=h.$ This increasing operator dimension is the cause of a quantum
obstruction against renormalizability: all spin (or helicity) $\geq 1$ field
operator have scale dimension $\geq 2$ and hence in the free field Borchers
class of the fields which enter in the invariant interaction $W,$ we must
have dim$W\geq 5$ (a $W$ describing interactions must be at least
trilinear!), whereas renormalizability demands dim$W\leq 4.$ This means that
even if one takes an algebraic point of view locality does not interacting
generators $W$ whose dimension is below 5. Note that their are certain
derivatives and composite fields which retain their classical dimensions as
e.g. $F_{\mu \nu },F^{\mu \nu }F_{\mu \nu },\psi $ etc. In fact the only
fields associated with an (m,s) Wigner representation in case of s$\leq 1$
which (even for an optimal choice of field coordinates) must have an
operator dimension beyond the classical one, is s=1 i.e. the vectorpotential.

Let us first look at the massive case which turns out to be conceptually
simpler. The minimal description is in terms of a transverse vectorpotential
with dim$A_{\mu }=2$ (this high operator dimension is resulting from the
transversality inherent in the Wigner theory. There is no known way to use
the milder nonlocal string-localized vectorpotentials in a perturbative
renormalization scheme, and the dream of a renormalizable higher spin
deformation theory could have ended right here. Fortunately one can outwit
the above No Go argument by a magic trick which produces
interaction-deformed observables (a power series *-algebra) generated by the
subset of physical (composite) operators in Fock space. The physical
generating fields in this process retained their classical dimensions modulo
logarithms, and the unphysical fields involving vectorpotentials attain
their classical value dim$A$=1 (mod logarithmic corrections) and become
formally local and covariant. In this approach the deformation theory only
depends on a finite number of parameters as expected for a renormalizable
theory. This magic is achieved by a cohomological representation of the
Wigner theory for which the application of the Weyl (pseudo)functor commutes
with the cohomological descend: 
\begin{eqnarray}
\stackunder{\downarrow }{H}^{ghost} &\longrightarrow &\stackunder{\downarrow 
}{\mathcal{F}}^{ghost} \\
H_{Wig}^{phys}\,\,\,\,\, &\longrightarrow &\mathcal{A}^{phys}  \nonumber
\end{eqnarray}
The horizontal arrows represent the (pseudo-)Weyl functors from wave
function spaces to (pseudo-)von Neumann algebras and the vertical arrows
denote the BRS descend for wave functions or algebras respectively. There is
a corresponding cohomological descend from a pseudo Fock space to a physical
(factor) Fock space. Here ``pseudo'' refers to the fact that the star
operation with respect to which the Lorentz group fulfils a unitarity-like
relation $U^{-1}(\Lambda )=U(\Lambda )^{*}$ in $H^{ghost}$ is not the
Hilbert space star related to the positive inner product. As in the
interaction free case of (m=0,h=1) in chapter3, one can (re)introduce
physical vectorpotentials in $H_{Wig}^{phys}$, but they will be necessarily
nonlocal. The magic trick spares one the conceptual pain to think about how
a theory manages to be strictly local, even if some of the objects as
vectorpotentials are in a physical sense mildly (semiinfinite stringlike)
nonlocal.

The simplest cohomological extension of the Wigner wave function space which
allows a nilpotent operation $s$ with $s^{2}=0,$ (such that the physical
transversality condition $p^{\mu }A(p)=0$ follows from the application of $%
s) $ needs besides two scalar ghosts wave functions $\omega $ and $\bar{%
\omega}$ another scalar field $\varphi _{a}$ :

\begin{eqnarray}
(sA_{\mu }^{a})(p) &=&p_{\mu }\omega _{a}(p) \\
(s\omega _{a})(p) &=&0  \nonumber \\
(s\bar{\omega}_{a})(p) &=&p^{\mu }A_{\mu }^{a}(p)-im_{a}\varphi _{a}(p) 
\nonumber \\
(s\varphi _{a})(p) &=&-im_{a}\omega _{a}(p)  \nonumber
\end{eqnarray}
Here we have already added a multiplett index since we have a selfcoupling
involving several vector mesons in mind with $m_{a}=m$ a common mass (the
interesting question of whether the stability of the deformation via
interaction can be maintained in the presence of unequal masses has not been
systematically investigated). One immediately realizes that $s^{2}=0$ and
that $s(\cdot )=0$ enforces the vanishing of $\omega _{a}$ and relates $%
\varphi _{a}$ to $p^{\mu }A_{\mu }.$ At this point there is no grading in
the formalism, i.e. the $\omega $ and $\varphi $ are simply ungraded wave
functions. \textit{The guiding principle for the cohomological extension of
the Wigner theory is that the associated two point function (or propagation
kernel) of the vectorpotential has a milder (renormalizable in suitable
interactions) high momentum behavior with the physically required behavior
only arising after the cohomological descend}. This is close to the particle
theory spirit of Lewellyn-Smith \cite{Lew}. In the geometrically motivated
Faddeev-Poppov method on the other hand, it was the gauge fixing in the
Lagrangian and the (euclidean) measure theoretical repair which assured the
existence of a renormalizable deformation of free spin=1 fields. On the
level of the Wigner theory it makes no sense to to impose a grading. However
the functorial transition from Wigner theory to QFT requires the
introduction of a grading with $\deg \omega =1,\deg \bar{\omega}=-1,$ and $%
\deg A_{\mu }=0,$ with $s$ transferring degree 1. The reason is that only
with this grading assignment, the $s$ allows a natural tensor extension to
multiparticle spaces with stable nilpotency, thus insuring the commutativity
of the above diagram (which represents the cohomological ascend and descend)
through the relation: $s(a\otimes b)=sa\otimes b+(-1)^{\deg a}a\otimes sb.$

This suggests to view the Fock space version $\delta $ of $s$ as the image
of a (pseudo)Weyl functor $\Gamma $ as $\delta =\Gamma (s)$ and to write the 
$\delta $ in the spirit of a formal Noether symmetry charge $Q$ associated
with the free field (linear) version of the BRS charge \thinspace \cite{BRS}%
\thinspace \cite{O}: 
\begin{eqnarray}
Q &=&\int (\partial _{\mu }A^{\mu }(x)+m_{a}\phi (x))\overleftrightarrow{%
\partial }_{0}\omega (x)d^{3}x=Q^{\dagger } \\
&&\delta F\equiv i\left\{ QF-(-1)^{\deg (F)}FQ\right\} ,\,\,\,F\in \mathcal{F%
}  \nonumber
\end{eqnarray}
which acts formally like an abelian gauge transformation (i.e. adds a
longitudinal contribution) on the vector potential. Here we keep the same
symbols for the (pseudo)quantum fields as for the Wigner (pseudo)wave
functions. As a result of the anticommutation relations of $\omega ,$ the
operator $Q$ and the $Z_{2}$ graded $\delta $ are nilpotent i.e. $\delta
^{2}=0$ and therefore defines a cohomology theory in the formal polynomial
free field algebra $\mathcal{F}$. This is a formal $^{*}$- algebra with a $%
Z_{2}$-graded antilinear $^{*}$-derivation. It is not difficult to show
that: 
\begin{eqnarray}
\mathcal{A} &=&\frac{\ker \delta }{\delta (\mathcal{F})}  \label{phys} \\
\ker \delta &=&\left\{ B\in \mathcal{F}\mid \delta (B)=0\right\}  \nonumber
\end{eqnarray}
where $\ker \delta $ is the nullideal (coboundary) associated with $s$, is
the algebra of physical photon operators. The $^{*}$-operation and the
associated notion of pseudo-hermiticity has an associated non-positive
definite sesquilinear form on $H_{Fock}(A,\omega ,\bar{\omega})$ with: 
\begin{equation}
\left\langle \phi ,F\psi \right\rangle =\left\langle F^{*}\phi ,\psi
\right\rangle
\end{equation}
In fact the indefinite nature of this sesquilinear form is a structural
consequence of the nilpotency and the pseudo-hermiticity of $Q:$%
\begin{eqnarray}
\left\langle Q\phi ,Q\psi \right\rangle _{ps} &=&\left\langle \phi
,Q^{2}\psi \right\rangle _{ps}=0  \label{Q} \\
\curvearrowright Q\equiv 0\,\,\, &\,unless\,&\left\langle \phi ,\psi
\right\rangle _{ps}\text{ }is\,\,indef\text{.}  \nonumber
\end{eqnarray}
i.e. any cohomological construction based on a nilpotent charge requires
indefinite metric. As expected, it is problem of Lorentz-covariance of
vectorpotentials which is the origin of this structure. In the (necessarily
indefinite) metric in which $Q$ is (pseudo)hermitian, also the
representation of the L-group comes out (pseudo)unitary. So in Wigner space
as well as in the cohomological representation it is \textit{always the
Lorentz-group representation} in which the deviation from the standard
situation shows up; in the Wigner space there was an additive term in the
transformation law, and the vectorpotential was nonlocal. whereas now these
aspects have been traded for the ``pseudoness'' of the transformation laws.
This trading is formally advantageous, because the mathematics of the
deformation does not care about correct physical behavior but only requires
the formal locality and covariance properties.

It is worthwhile to note that in our dichotomic division of QFT into
algebraic aspects and properties of states (and representations), the
``pseudoness'' is solely in the states and the GNS representation and not in
the abstract algebra. As already for standard spin$\leq 1$ renormalizable
couplings, the algebras of the formal deformation theory are only *-algebras
and not C*-algebras.

Now we come to the deformation of free fields via $W$-interaction
polynomials. If one stays within the framework of completely massive spin=1
theories, the formal validity of (LSZ, Haag-Ruelle) scattering theory
defines a reference space in which one can do all calculations: the incoming
(pseudo)Fock space referring to the incoming scattering situation. This
brings significant simplifications as compared to massless vectormesons: in
this reference space the conserved $Q^{\prime }s$ have a bilinear
representation in terms of free fields (with significant simplifications as
compared to the standard nonlinear BRS formalism). The bilinear structure of
conserved charges (including the Poincar\'{e}-generators) in terms of the
free incoming fields is absolutely crucial for the following. It gives a
fixed position of the physical cohomological space inside the pseudo Fock
space and it makes all basic fields as $F_{\mu \nu },\psi $ etc.
``physical'' (except the vectorpotential and the ghosts). Note that in the
standard approach based on quantization of gauge theories, the present
massive spin1 situation would be termed ``a completely broken gauge
theory''. The mentioned elementary physical fields would then be identical
with the composite fields $F_{\mu \nu }\cdot \Phi ,\psi \cdot \Phi ,$ where $%
\Phi $ is the Higgs field. This immediately poses the question of where and
in what form the Higgs field enters the present formalism. This has been
partially answered in the work of Scharf and collaborators \cite{D A S}. In
the following we will interpret and comment their findings. Take the
simplest case of selfinteracting first order massive spin one fields and
ghosts: 
\begin{equation}
W^{(1)}=W_{A}^{(1)}+W_{\omega }^{(1)}+W_{\phi }^{(1)}
\end{equation}
where the first term is the most general trilinear coupling $\sim \tilde{f}%
_{abc}A_{a}^{\mu }A_{b}^{\nu }\partial ^{\nu }A_{c}^{\mu }$ and the two
other terms contain the most general coupling to the $w$ and $\phi $-ghosts.
Physical consistency now leads to \cite{D A S}

\begin{enumerate}
\item  In first order one obtains a list of constraints for the couplings.
The result may be written in the form: 
\begin{eqnarray}
W^{(1)} &=&igf_{abc}(:\frac{1}{2}A_{a}^{\mu }A_{b}^{\nu }F_{c}^{\nu \mu
}:-:A_{a}^{\mu }\omega _{b}\partial _{\mu }\bar{\omega}_{c}:) \\
+igf_{abc}^{\prime } &:&\partial _{\nu }A_{a}^{\nu }\omega _{b}\tilde{\omega}%
_{c}:  \nonumber
\end{eqnarray}
where $f$ and $f^{\prime }$ are totally antisymmetric.

\item  In second order physical consistency leads above all to the
Jacobi-identity for the $f^{\prime }s$ i.e. the famous gauge group structure
comes from just physical consistency of the cohomologically extended spin 1
theory. The consistency for the remaining term demands that it be a total
divergence (surface term). But in the presence of the spin 1 mass, there is
the necessity of yet another compensation; this time (if the compensatory
field is taken as scalar field as the simplest possibility), it must be a
physical field and not a ghost. The result looks like that of the
spontaneous broken Yang Mills theory after the subtraction of the vacuum
expectation of the Higgs field, except that in our case there was no gauge
theory and the original $W^{(1)}$ did not contain a Higgs field to start
with but it rather enters in order to maintain the consistency of the theory
in higher order.
\end{enumerate}

What does this apparently very powerful ``physical consistency''\footnote{%
In the work of \cite{D A S}, the surface term property in (\ref{surface}) is
called ``quantum gauge principle'', but since this could be misunderstood as
if an additional structure besides physical consistency of spin=1
interactions is needed, we avoid this terminology.} mean in mathematical
terms? I is nothing but the requirement that the adiabatic limit: 
\begin{equation}
S=\lim_{g\rightarrow const}S(g)
\end{equation}
defines the physical S-matrix in a smooth way from the off-shell $S(g)$. The
natural mechanism for providing an S-matrix which commutes with $Q$ (the
physical consistency), is that the noncommuting contributions with $S(g)$
for arbitrary g's which have a constant value in a large double cone and
fall off to zero in a collar outside is: 
\begin{equation}
\left[ S(g),Q\right] =terms\,\,in\text{ }\partial g  \label{surface}
\end{equation}

In the Bogoliubov-Shirkov deformation theory this leads to a restriction for
the $T$-products: 
\begin{eqnarray}
S(g) &=&Te^{i\int W(x)g(x)d^{4}x}=  \nonumber \\
&=&\sum_{n}\frac{1}{n!}\int T_{n}(x_{1},....,x_{n})g(x_{1})....g(x_{n}) 
\nonumber \\
with &&\,\,T_{1}(x)=W^{(1)}(x)  \nonumber
\end{eqnarray}
d. terms involving only $\partial g,$ or in terms of the $T_{n}$: 
\begin{equation}
\left[ Q,T_{n}(x_{1},....,x_{n})\right] =\sum \frac{\partial }{\partial
x_{i}^{\mu }}T_{n,i}^{\mu }(x_{1},....,x_{n})
\end{equation}

Where the terms $T_{1,1}^{\mu }$ as well as all $n>1$ terms should be
iteratively determined and the Hahn-Banach extension problem for the $%
T_{n}^{\prime }s$ solved (the latter step being the renormalization aspect
in the Epstein-Glaser approach). It comes as somewhat of a surprise that
this requirement has a retroactive action even on the first order trilinear
input. For the computations up to second order which bring out the above
results we refer to \cite{D A S}.

It is worthwhile to reflect about this result. The physical consistency
requirement converted the apparent freedom of spin 1 couplings (which
classically seemed to be much larger than that for interactions between spin%
\TEXTsymbol{<}1 fields) into a very tight situation! The perturbative
treatment does not work without introducing another (this time physical)
scalar field. One may call this the Higgs field, however it should be clear
that its role is strictly perturbative and nothing is said about the
existence of a Higgs particle in a nonperturbative massive selfinteracting
massive spin=1 theory. Note that in this LQP consistency approach, there is
no intrinsic meaning to think of the Higgs particle as a ``fattened''
Goldstone mode. This underlines an old observation made in connection with
the Schwinger-Higgs mechanism in the d=1+1 Schwinger model, since there is
no Goldstone mechanism in d=1+1). The result means that there is no other
massive selfinteracting spin 1 theory other than that computed by the alias
broken Yang-Mills gauge theory i.e. the result is unique but this
terminology has no intrinsic physical meaning; it is at best a
mnemotechnical concept. It is interesting to note that the present approach
is natural for the massive theories which in the original Yang-Mills work
was the desired (but with those concepts unattainable) dream.

A direct application of the cohomological method to zero mass vectormesons
is necessarily more complicated as a result of the absence of a global
reference space. In that case there is no possibility of working with a
bilinear expression for $Q,$ rather one has to face the presence of
interacting trilinear contributions. As a consequence the physical
cohomology space changes its position inside the big (pseudo)Fock space with
the order of perturbation. Let us briefly look at the stability problem of
this more complicated situation. Start from the cohomological representation:

\begin{equation}
H_{Fock}=\frac{\ker Q}{rangeQ}
\end{equation}
\begin{eqnarray}
\mathcal{A} &=&\frac{\ker \delta }{\delta (\mathcal{F})},\,\,\,\delta
(F)=QF-(-1)^{\deg F}FQ \\
&&\ker \delta =\left\{ B\in F\mid \delta (B)=0\right\}  \nonumber
\end{eqnarray}
Check that $\ker Q$ is a positive semidefinite subspace and that range$Q$
agrees with the space of nullvectors in $\ker Q.$ The descend from $\mathcal{%
F}$ and $H_{Fock}^{pseudo}$of $\mathcal{A}$ to $H_{Fock}$ uses also a fairly
standard argument: an element of the form $B+\delta (F)$ with $B\in \ker Q$
applied to a vector of the form $\phi +Q\psi $ with $\phi \in \ker Q\,$%
(representative of the $A$-class applied to representative of the $H_{Fock}$%
-class) leads to $A\phi \in \ker Q$ and to: 
\begin{equation}
(A+\delta (F))(\phi +Q\psi )-A\phi \in QH_{Fock}^{ps}
\end{equation}
As mentioned before, for all gauge theories which allow an adiabatic limit
(i.e. for which all perturbative contributions to $S(g)$ in the limit $%
g(x)\rightarrow 1$ on all Minkowski space exist), one may use the incoming
free $Q.$ This excludes the (on-shell) infrared divergent theories as QED or
gauge theories without a complete Schwinger Higgs (-symmetry breaking,
better: -charge screening) mechanism.

Now we indicate of how to proceed, if such a free incoming reference
situation is not available. In that case we have to expand Q into a power
series. The main problem of the interacting theory is then to show that the
cohomological descend is stable against interactions. Of course intuitively
one expects the extended structure (which brings theories with photons on
the same formal level of locality and covariance in a physical space as
nongauge theories) to be stable under deformations as theories of spin zero
and $spin=\frac{1}{2}$ are. To prove this is not entirely trivial and only
possible if one works with a notion of positivity which is adjusted to
formal power series. We call a formal power series $s=\sum g^{n}s_{n}$
positive if there exists another power series $t=\sum g^{n}t_{n}$ with: 
\begin{eqnarray}
&&t^{*}t=s,\,\,t_{n}\in IR \\
\exists k &\in &IN_{0}\smile \left\{ \infty \right\}
\,\,\,s.t.\,\,t_{n}=0,\,\,if\,\,n<2k  \nonumber \\
&&and\,\,\,b_{2k}>0\,\,\,if\,\,\,k<\infty  \nonumber
\end{eqnarray}
We need this weak sense of positivity because in general the interaction
will change the position of $\ker Q$ and range$Q$ inside the interaction
independent fixed pseudo Fock space $H_{Fock}^{sp}$. The perturbative
interaction deforms the free field operators into formal power series: 
\begin{eqnarray}
F &=&\sum g^{n}F_{n} \\
Q &=&\sum g^{n}Q_{n}  \nonumber \\
\delta &=&\sum g^{n}\delta _{n}  \nonumber
\end{eqnarray}
where the previous free operators appear in zero order: 
\begin{equation}
Q_{0}=Q_{B},\;\delta =\delta _{previous},\,F_{0}=F_{previous},\,etc.
\end{equation}
with: 
\begin{eqnarray}
Q^{2} &=&0,\,\,\delta ^{2}=0 \\
\delta &=&AdQ  \nonumber
\end{eqnarray}
The new $\mathcal{A}$ and $H=\frac{rangeQ}{\ker Q}$ are now formal power
series in g. We have to prove the following three properties:

\begin{enumerate}
\item  $\left\langle \phi ,\phi \right\rangle \geq 0,\,\,\,$for $\phi \in
\ker Q$

\item  $\left\langle \phi ,\phi \right\rangle =0,\,\phi \in \ker
Q\,\,\,\,\curvearrowright \phi \in $range$Q$

\item  $\mathcal{A}$ is faithfully represented on $H\,\,$
\end{enumerate}

Here the positivity is meant in the previously explained weak sense of power
series. So the position of the $\ker Q$ inside $H_{Fock}^{ps}$ keeps
changing with the perturbative order and fulfills is in general only the
above weak positive semidefiniteness. An exception is the abelian case for
which the ghost field remains decoupled from the physical fields i.e. it
preserves its freeness, similar to the Gupta-Bleuler approach, even in the
presence of interactions. Therefore perturbative QED has a better positivity
status than QCD.

The proof of the three properties above is by induction. We refer the reader
to the work of Duetsch and Fredenhagen \cite{Due Fre}.

The conceptual simplicity of the massive theory versus the complex zero mass
situation means that the physical picture in LQP is opposite to that of
gauge quantization. It suggests another more indirect but conceptually more
interesting way than the above deformation theory in which the physical
space has a position inside the big (pseudo)Fock space which keeps changing
with the perturbative order. Whereas in the latter approach the massive
theory is interpreted as a broken ``symmetry'' (we use the parenthesis
because the gauge symmetry is really not a symmetry in any physical sense),
the LQP picture (with no Lie group structure put in!) is that of ``charge
liberation'' in the massless limit of the massive theory.

Charge liberation via the appearance\cite{B V} of new superselection sector
in the scaling limit is a well studied physical (in contradistinction to
gauge breaking) phenomenon in LQP. There are many illustrations in case of
more standard models, e.g. the two sectors of the massive Ising field theory
mutate into 3 chiral sectors for both chiral components (with the
identification of the two vacuum sectors). Actually the LQP picture is much
closer to Schwinger's little known contribution on the charge screening
phase \cite{schwinger} which led him to propose the Schwinger model. The
physical picture of m$\rightarrow 0$ based on the study of the Schwinger
model \cite{B V} as well as on physical intuition reveals that there exist
spacelike semiinfinite string localized operators (whose introduction into
the massive theory would ply no important physical role, as a result of the
excellent localization properties of a massive spin=1 theory), which in the
massless limit carry a string localized Maxwell charge. In other words the
charge liberation is just the opposite of Schwinger's screening mechanism.
The Higgs ``condensates'' are nonintrinsic description dependent quantities
which do not appear in the present spin 1 consistency approach. The
corresponding structural theorem of Swieca on charge screening/charge
liberation in Maxwell-like theories becomes, with the present perturbative
hindsight, a rigorous statement on consistent spin=1 interactions. The main
point of the present approach, which really merits strong emphasis, is that
it places the relation classical-quantum physics there were it should be (in
the spirit of Bohr): \textit{the consistent (in this case perturbative)
local quantum theory in semiclassical approximation selects the classical
gauge theory and not the other way around as in the gauge quantization
approach}. The correspondence principle of Bohr analyses the classical
situation by semiclassical approximations of the more fundamental quantum
theory and leaves the quantization parallelism as a piece of artistry
outside mathematical physics\footnote{%
Apart from the case of Weyl and CAR algebras, which can be rigorously
generated by applying an appropriate functor to classical function spaces.}.

To strengthen the present LQP approach to spin one problems one needs more
studies. In addition to the systematic $n^{th}$ order investigation of the
massive situation and a more profound understanding of the problem of
uniqueness of the physical particle (alias Higgs), there are the problems of
zero mass limits of this massive perturbation theory. Here one wants to
understand which modifications (adjustments) in the massive theory must be
done in order to have a smooth off-shell limits for the expected
semiinfinite string-localized charge-carrying fields. Another problem is to
understand the expected decoupling of the two scalar field (the $\phi $- and
alias Higgs-field). The change from the gauge QFT formulation to the present
one therefore causes an interesting change of paradigm in that the emphasis
is changed from the formal Higgs mechanism point of view to the more
physical problem of how the charge generating string like fields arise from
the massive theory in the limit. Another open question is whether the method
of cohomological representation for the sake of maintaining
renormalizability for spin=1 is the tip of an iceberg, i.e. if there are
higher spin cases for which this magic works. In that case the present
reformulation of the problem would have achieved much more than a clearer
conceptual setting of ( broken) Gauge QFT.

Needless to mention that the underlying idea of the original BRS work was
precisely to incorporate the observables of the massive case into the class
of perturbative renormalizable theories. The spirit of this early work on
renormalizable spin1 problems \cite{BRS}was more in the pragmatic vein of
Lewellyn-Smith\cite{Lew} and did not yet have to carry the present burden of
the predominance of differential geometry over local quantum physics.

In this section we have traded one miracle (the gauge principle) with
another one (removing renormalizability obstruction via a cohomological
trick of pseudo-structures). In the physical philosophy of Bohr and
Heisenberg one should always try to remove nonobservable (pseudo-physical)
vestiges even in intermediate steps of the calculation. One may hope that
ideas of modular localization, which already led to nonperturbative progress
in factorizing d=1+1 models, will give new concepts by which one could be
able to use directly string localized operators in a deformation approach,
since a theory which explains the origin of weaker localization of free
vectorpotentials should also give hints as to how to incorporate this
property into interactions. Although this has not been achieved in the
present LQP cohomology approach, we hope that our ideas may help to prepare
the ground for such a step. The cohomological magic trick creates a lesser
danger to be confused with a permanent achievement in QFT than gauge theory;
it is manifestly of a transitional nature and asks for a future more
fundamental physical approach in terms of quantum localization concepts.

\section{Interactions with External Fields, CST-Problems}

Interactions of quantum fields with external (classical) fields played an
important role in the development of full QFT. The simplest situation of
this kind one meets in case the quantum fields are free. In fact free Dirac
or Schr\"{o}dinger fields interacting with external electromagnetic fields
preceded QED and led, with some hindsight concerning interpretations, (see
the introduction in \cite{Wei}) to many correct formulas. If we look at
these external field problems from the point of view of
Poincar\'{e}-invariant QFT, we notice a conceptual problem. Since the vacuum
and also the particle states are defined in terms of Poincar\'{e}-covariance
properties, it is not immediately clear how one should define such reference
states if Poincar\'{e}-covariance is broken by external fields\cite{Wald}.
In an elegant formalism like Schwinger's (referring to his treatment of the
astrophysically relevant $\mu \bar{\mu}-$pair creation in electro magnetic.
fields), the formalism itself takes care of this problem without the user
being aware. A closer look reveals that the reference state built into the
Schwinger formalism is the ``adiabatic vacuum'', which in a more mundane
formalism corresponds to the approximation of the actual external
interaction by a sequence of softly switched on and off external
interactions. Whereas this is eminently reasonable for external
electro-magnetic interactions, this is generically speaking unreasonable for
problems in curved space time (CST) i.e. with external gravitational fields.
Since the Hawking radiation effect belongs to this class of problems, these
structural questions are not without (astro)physical interest and relevance.

In the following we will briefly sketch some ideas \cite{Rad} \cite{Bru}
which not only led to an answer for the correct reference states, but gave a
framework for the renormalized perturbation theory of \textit{interacting}
quantum matter in CST. Even if, as in the case of the present author, one is
not an actively working specialist in this area, one should take notice of
these developments for the following reason. General QFT as it stands, is
not quite that perfect quantum counterpart of the classical Faraday-Maxwell
theory with its ``action at the neighborhood principle''. Whereas the
algebraic part (the net theory) is local, the energy positivity and the
vacuum homogeneity are very nonlocal stability requirements. This is the
cause of the above mentioned difficulty. Therefore if QFT in CST requires to
think about a substitute, this may be beneficial even for Minkowski space
QFT.

Since all of the renormalization schemes use either euclidean space or
momentum space and none of these methods is meaningful in generic curved
space time situations, one is forced to re-think the renormalization
formalism. Looking at the literature, one notes that almost all the papers
(before the above work) on the subject are about euclidean CST QFT and
nobody ever tried to define a Wick-polynomial (needed e.g. for the real time
energy-momentum tensor etc.)

Since the answer to both questions requires the use of somewhat unfamiliar
concepts, let me make some qualitative comments on the ``microlocal''- (or
Fourier integral operator-) analysis developed by the mathematicians
H\"{o}rmander and Duistermaat around 1971. It is this analysis which gives
rise to the formulation of a ``microlocal spectrum condition'' in QFT \cite
{Rad}\cite{Bru}.

The basic idea is to refine the local analysis of singularity structure
Schwartz distribution $u$ which deals with singular supports $supp_{sing}u$
(the $supp_{sing}u$ is the complement of the largest open smoothness region
for u) and more generally of distribution densities on a manifold, by
shifting it from the base space to the cotangent bundle. In brief, one first
zooms in on a $u$-singularity and then one uses a directional Fourier
``telescope''. If $\phi $ is a localizing test function, one studies: 
\begin{equation}
\widetilde{\phi u}(\xi )=\left\langle u,e^{-i\left\langle \cdot ,\xi
\right\rangle }\phi \right\rangle
\end{equation}
where $\left\langle \cdot ,\cdot \right\rangle $ denotes dual pairing. This
is a fast decreasing function in $\xi $ as long as supp$\phi $ does not
touch the singularity points. If supp$\phi $ on the other hand does extend
into the singular region of $u,$ the singularity may be directional
dependent and in certain $\xi $-directions one may still encounter a fast
decrease. Therefore one uses the following mathematical definition ($\ V$
denotes the euclidean base space):

\begin{definition}
The wave front set, $WF(u),$ of $u$ is the complement in $V\times \mathbf{R}%
^{n}\backslash \left\{ 0\right\} $ of the points $(x_{0},\xi _{0})$ in
cotangent space $V\times \mathbf{R}^{n}\backslash \left\{ 0\right\} $ s. t..
for each $\phi $ there exists a neighborhood. $U\times \Gamma ,$ with $%
\Gamma $ conic (directional) neighborhood. of $\xi _{0},$ and an $N\geq 0$
with: 
\begin{equation}
\left| \left\langle u,e^{-i\left\langle \cdot ,\xi \right\rangle }\phi
\right\rangle \right| \leq C_{\phi ,N}\left( 1+\left| \xi \right| \right)
^{-N},\quad \forall \xi \in \Gamma \quad
\end{equation}
\end{definition}

Returning to physics, we recall that as the result of the positive energy
property, (unordered) correlation functions belong to a class of
distributions which can be freely multiplied. For example the product of two
Minkowski space two-point functions $w_{i}(x,y)=W_{i}(x-y)\,\,i=1,2$ is
again a well-defined distribution in the same class, because the convolution
of their Fourier transforms with the $V^{\uparrow }$ forward light cone
spectral support amounts to an integration over a finite (phase space)
region. Using the Kall\'{e}n-Lehmann spectral representation: 
\[
W_{i}(\xi )=\int_{0}^{\infty }i\Delta ^{+}(\xi ,\kappa )\rho _{i}(\kappa
)d\kappa ,\quad i=1,2 
\]
the convolution of the $\rho _{i}^{\prime }s$ extends over a finite mass
region. This property does not hold for time-ordered or retarded
distributions since they do not posses a spectral support in momentum space.

The main property of the wave front sets of distributions is that they allow
a simple criterion for the existence of the product: the conic neighborhoods 
$\Gamma _{i}$ must add up to a resulting conic neighborhood. in $V\times 
\mathbf{R}^{n}\backslash \left\{ 0\right\} $ . The coordinate free
adaptation to densities (distribution valued differential forms) on
manifolds $M$ is easy. The wave front sets are now cones in $T^{*}M$ and
they behave additively under multiplication in the following sense: 
\begin{equation}
WF(u_{2}u_{1})\subseteq \left\{ WF(u_{2})\oplus WF(u_{1})\right\} \cup
WF(u_{2})\cup WF(u_{1})  \label{prod}
\end{equation}
The product exists, if the zero section in $T^{*}M$ does not intersect $%
WF(u_{2}u_{1}).$ This is the analogue of the product structure of Wightman
functions.

Let us now test this idea for free QFT in CST. We start with the structure
of the algebra generated by a Klein-Gordon field $\phi $ in a globally
hyperbolic space time\cite{Wald}: 
\begin{eqnarray}
\left( g^{\mu \nu }\partial _{\mu }\partial _{\nu }-m^{2}\right) \phi (x)
&=&0,\quad \left[ \phi (f),\phi (g)\right] =E(f\otimes g)  \label{al} \\
\forall f,g\in C_{0}^{\infty }(M),\quad &&E(x,y)=\Delta ^{av}(x,y)-\Delta
^{ret}(x,y)  \nonumber
\end{eqnarray}
It is defined uniquely in terms of the manifold data i.e. the retarded
(advanced) functions belong to the algebraic characterization and are
uniquely determined by the geometry, whereas the unordered and time ordered
correlation functions are determined by the states. One now defines a wave
front set for the (yet unknown) two-point functions $\omega _{2}$: 
\begin{equation}
WF(\omega _{2})=\left\{ (x,k;x^{\prime },-k^{\prime })\in
T^{*}M^{2}\backslash \left\{ 0\right\} \mid (x,k)\sim (x^{\prime },k^{\prime
}),\,\,k\in \bar{V}_{+}\right\}  \label{Rad}
\end{equation}
where the equivalence relation $\sim $ means that there exists a light-like
geodesic from $x$ to $x^{\prime }$ s. t.. $k$ is co-parallel to the tangent
vector to the geodesic, and $k^{\prime }$ is its parallel transport from $x$
to $x^{\prime }.$ This physically appealing local requirement for the
selection of physical states was shown by Radzikowski to be mathematically
equivalent to the more global Hadamard condition (an older recipe to obtain
physical states). Free field structure, i.e. the Wick combinatorics means
that the higher point functions are products of $\omega _{2}$ i.e. that the
states $\omega $ on the algebra (\ref{al}) are so called quasi-free states
on a CCR algebra. Using the previous product structure (\ref{prod}) of
distributions with a known wave front, one then proves the existence of the
n-point functions and a formula for their $WF$ set. In a similar vein one
shows the existence of Wick-products as e.g. $:\phi ^{n}:.$ As in the
Minkowski case, the time-ordered propagator: 
\begin{equation}
iE_{F}(x,y)=\omega _{2}(x,y)+E_{ret}(x,y)
\end{equation}
does not have the one-sided spectral structure in order to allow for
pointwise multiplication. Its wave front set is: 
\begin{eqnarray}
WF(E_{F}) &=&\left\{ \text{same, but }x\neq x^{\prime },k\in \bar{V}_{\pm }%
\text{ if }x\in \mathcal{T}_{\pm }(x^{\prime })\right\} \\
&&\cup \left\{ (x,k;x,-k),x\in M,k\in T_{x}^{*}M\backslash \left\{ 0\right\}
\right\}  \nonumber
\end{eqnarray}
where $\mathcal{T}_{\pm }(x^{\prime })$ are the future/past of $x^{\prime }.$

Actually the formula (\ref{Rad}) turns out to be not general enough in order
to incorporate theories beyond free fields. The more general formula which
does not contain the restriction to light like geodesics and its stable
(under multiplication of n-point functions) generalization to $\omega _{n}$
(which is most conveniently expressed in terms of graphs with vertices $%
x_{i} $ and directed geodetic edges between them), can be found in the work
of Brunetti et. al..\cite{Bru}.

We still have to understand why the microlocal Spectrum Condition ($\mu SC$)
is not capable of a unique selection of a state and what kind of family it
selects. A local spectral condition is not able to single out states with
global symmetry, in fact for generic CST they do not exist. With one
particular state in this family, all other states in the same folium (vector
or density matrix in the same GNS Hilbert space) turn out to share the same $%
WF$ set. In fact the states with coalescing wave front sets form exactly one
folium of the set of all states on the $C^{*}$-algebra $\mathcal{A}.$ That
folium contains of course states with different superselection charges, a
situation which is vaguely reminiscent of the ``no hair'' property of black
holes (in the sense that the standard global characterizations of states
with their detailed assignments of quantum numbers loose their meaning in
local folia). There are two physical questions which enter ones mind. One is
whether physical properties (corrections to electro-weak effects as
anomalous moments, Lambshift etc.) change significantly in one folium. For
this one has to understand renormalization theory and the implementation of
physical normalization conditions (the adiabatic separation of interaction
discussed at the beginning of this chapter. The other is whether the $\mu SC$
can perceive interactions, i.e. whether the wave front set of an interacting
theory can be distinguished from that of a free theory. An affirmative
answer to this question would be extremely interesting even for Minkowski
QFT (since the global vacuum condition is not capable of such a distinction.
Both questions are presently under investigation \cite{Bru}. The
perturbative calculations of wave front sets is not an easy matter.

The last issue in this section is how to do renormalization theory for
interacting QFT in CST (i.e. how to avoid euclidean- and momentum-space). A
framework which stays in x-space is that of Epstein and Glaser \cite{Bru}.
The main problem in its adaptation to the present case, is how to avoid
translational invariance on which the E-G approach relies heavily. Let us
look at their starting formula for the n$^{th}$ order coefficients of the
Bogoliubov Shirkov operator $S(g)$ after Wick-ordering: 
\begin{eqnarray}
T_{n}^{k_{1}....k_{n}}(x_{1,}....,x_{n}) &=&\sum
t^{l_{1}....l_{n}}(x_{1},....,x_{n})\times \\
&:&\phi ^{k_{1}-l_{1}}(x_{1})....\phi ^{k_{n}-l_{n}}(x_{n}):  \nonumber
\end{eqnarray}
Where $t^{l_{1}....l_{n}}(x_{1},....,x_{n})$ are numerical time ordered
distributions. In the E-G approach it is crucial that Wick-products (i.e.
operator-valued distributions) can be multiplied with translational
invariant numerical distributions. The CST substitute is: 
\begin{equation}
WF(t_{n})\in \Gamma _{n}^{to},\quad to:time\,\,ordered
\end{equation}
where $\Gamma _{n}^{to}$ is a subset of $T^{*}M^{n}$ with a certain
graphical characterization. The construction of the time ordered operators $%
T_{n}^{k_{1}....k_{n}}$ is achieved by induction in two steps. First one
shows that $T_{n}$ for ($x_{1}....x_{n})\notin \Delta _{n}$ (the total
diagonal) can be patched together from all lower $n$ $T$'s. Let us denote
this $T_{n}$ on $M^{n}\backslash \Delta _{n}$ as $T_{n}^{0}.$ The second
step (the more difficult one) is the extension to the diagonal. For a
detailed presentation we refer to the recent literature \cite{Bru}. The main
point to be gathered from these highly technical investigations is that the
E-G approach allows a uniform treatment of renormalization as a distributive
extension problem with all the operators living in Fock space. As in the
Minkowski case in chapter 4.2 the E-G formalism defines a formal $^{*}$%
-algebra net with a folium of states on all local algebras. No general
principles are known which select a distinguished global state as the
Poincar\'{e} invariant vacuum or one-particle states.

The rigorous CST renormalization theory\cite{Bru} contains of course the
proof for the renormalizability of the standard theory as a special case.

It would have been too nice if QFT in CST could furnish a gateway into
``Quantum Gravity''. After all, QFT in external electro-magnetic fields was
an essential step towards QED. But unfortunately presently this does not
seem to be the case and Quantum Gravity continues to exist as only an
enigmatic idea, which for the time-being lacks concrete physical content.
With these remarks we conclude our short excursion into QFT on CST.

\textbf{Literature to chapter 4:}

S.Weinberg, ``The Quantum Theory of Fields'', Vol. I, Cambridge University
Press 1995

C.Itzykson and J-B.Zuber, ``Quantum Field Theory'', McGraw-Hill 1980

R.Haag, ``Local Quantum Theory'' Springer Verlag 1992

R. M. Wald, ``Quantum Field Theory in Curved Spacetime and Black Hole
Thermodynamics'', University of Chicago Press 1994.

\chapter{The General Framework of QFT}

\section{Model-independent Properties of pointlike Fields}

The conceptual situation of QFT after the discovery of renormalized
perturbation theory was at first somewhat confused. Despite the impressive
agreement of low order radiative corrections, the precise relations between
particles and fields as well as the mathematical consistency of QFT beyond
perturbation theory were ill-understood. Most of the post renormalization
progress was in the area of structural understanding about the
particle-field dichotomy and on mathematical formulations of well posed
physical requirements. These developments are often linked with the names of
the principle protagonists of those problems: Lehmann, Symanzik and
Zimmermann (LSZ) in the first case and Wightman in the second case (the
Wightman framework).

The strong return of perturbative methods in the 70$^{ies}$ via the Standard
Model, only led to a temporary lull in the ongoing research on general
structural properties of QFT, especially in view of the fact that those
nonabelian gauge theories, after some initial success (notably in the area
of small distance behavior off mass shell) run into tough nonperturbative
problems which appeared unsolvable in the standard approach. Up to this date
fundamental insights, as an intrinsic understanding of gauge theory on the
basis of its gauge invariant correlation functions (i.e. without reference
to the quantization method by which it has been constructed) are still
missing (see also chapter 4.5).

We already have explained the relation between free fields and the net of
local algebras generated by them in terms of an analogy to differential
geometry: the fields are like coordinates and the net corresponds to the
coordinate-free (intrinsic) approach to QFT. Many of the properties of
fields appear in a clearer physical light, if one thinks about them in terms
of local generators of nets, in analogy to the enveloping algebras of
noncompact Lie algebras. Therefore let us list some properties (the main
properties of the Haag-Kastler net theory) of nets before we write down the
(model-independent) postulates for fields.

\begin{itemize}
\item  (i) \thinspace \thinspace \thinspace There is a map of double cones $%
\mathcal{O}$ in Minkowski space into von Neumann operator algebras $\mathcal{%
A(O)}$ which are subalgebras of all operators $\mathcal{B(H)}$ in some
Hilbert space $\mathcal{H}$: 
\[
\mathcal{A}:\mathcal{O}\rightarrow \mathcal{A(O)} 
\]
The C$^{*}$-completion of this family yields the global C$^{*}$-algebra $%
\mathcal{A}_{quasi}:$%
\[
\mathcal{A}_{quasi}=\bigcup_{\mathcal{O\in M}}\mathcal{A(O)} 
\]

\item  (ii)\thinspace \thinspace \thinspace \thinspace The family $\mathcal{A%
}$ forms a ``net'' i.e. a coherent (isotonic) family of local algebras: 
\[
\hbox{if }O_{1}\subset O_{2}\hbox{ then }\mathcal{A(O}_{1}\mathcal{)}\subset 
\mathcal{A(O}_{2}\mathcal{)} 
\]
\end{itemize}

In case the local algebras represent observables one requires another
physically motivated coherence property namely Einstein causality: 
\[
\mathcal{A(O)\subset A(O}^{\prime }\mathcal{)}^{\prime } 
\]

\begin{itemize}
\item  (iii)\thinspace \thinspace \thinspace \thinspace covariance with
respect to the Poincare group. For observable nets: 
\[
\alpha _{(a,\Lambda )}(\mathcal{A)=A}(\Lambda \mathcal{O}+a) 
\]
\end{itemize}

As already mentioned, the subsequent properties of fields and their physical
interpretation is facilitated by thinking about them as coordinatizations of
generators for local nets. The main difference of the field approach versus
the net approach is that properties of charge-carrying fields are put in,
and not derived from those of observable neutral fields. In the net approach
charges (and their field carriers) are constructed via the superselection
theory. The latter approach is more fundamental and more suitable in
situations which are far away from quantization prescriptions and
Lagrangians (e.g. low dim.QFT with braid group statistics). In the sequel we
explain the properties of fields in the setting of Wightman. Here and in the
following we use the symbol $A$ as a generic notation for collection of
generating fields but the standard situation underlying illustrations and
proofs is mostly that of one generating scalar field.

\begin{description}
\item  \emph{Properties of Fields:}
\end{description}

\begin{itemize}
\item  \underline{A}\thinspace \thinspace \thinspace $\,\,\mathcal{H}$-space
and $\mathcal{P}$-group
\end{itemize}

1.\thinspace \thinspace \thinspace \thinspace \thinspace Unitary
representation $U(a,\alpha )$ of $\widetilde{\mathcal{P}}$ in $\mathcal{H}$, 
$\widetilde{\mathcal{P}}:$covering of $\widetilde{\mathcal{P}}$

2.\thinspace \thinspace \thinspace \thinspace Uniqueness of the vacuum $%
\Omega ,$ $U(a,\Lambda )\Omega =\Omega $

3.\thinspace \thinspace \thinspace \thinspace Spectrum condition: $specU\in 
\bar{V}_{+},$the forward light cone.

\begin{itemize}
\item  \underline{B}\thinspace \thinspace \thinspace \thinspace Fields

1.\thinspace \thinspace \thinspace \thinspace Operator-valued distributions: 
$A(f)$=$\int A(x)f(x)d^{4}x,\,\,\,f\in \mathcal{S}$ (the Schwartz space of
``tempered'' testfunctions) is an unbounded operator with a \thinspace dense
domain $\mathcal{D}$ \thinspace such that the function $\left\langle \psi
_{2}\left| A(x)\right| \psi _{1}\right\rangle $ exists as a sesquilinear
form for $\psi _{i}\subset \mathcal{D}$

\item  2.\thinspace \thinspace \thinspace \thinspace \thinspace Hermiticity:
with $A$, also $A^{*}$ belongs to the family of fields and the affiliated
sesquilinear forms are as follows related: $\left\langle \psi _{2}\left|
A^{*}(x)\right| \psi _{1}\right\rangle =\left\langle \psi _{1}\left|
A(x)\right| \psi _{2}\right\rangle $

\item  \underline{C}.\thinspace \thinspace \thinspace \thinspace $\widetilde{%
\mathcal{P}}$-covariance of fields: $U(a,\alpha )A(x)U^{*}(a,\alpha
)=D(\alpha ^{-1})A(\Lambda (\alpha )x+a)$
\end{itemize}

\begin{enumerate}
\item  For observable fields only integer spin representations ( i.e.
representations of $\mathcal{P}$ ) occur.
\end{enumerate}

\begin{itemize}
\item  \underline{D}.\ \thinspace \thinspace Locality: $\left[
A^{\#}(f),A^{\#}(g)\right] _{\mp }=0$ \quad for suppf$\times $suppg
(supports are spacelike separated).

\item  \underline{E}.\thinspace \thinspace \thinspace \thinspace \thinspace
Stability of local algebras under causal completion: $Alg(A,\mathcal{O})$=$%
Alg\mathcal{(}A,\mathcal{O}^{\prime \prime }),$ where (for $\mathcal{O}$
convex) the causal completion is the smallest double cone which contains $%
\mathcal{O}.\,$A weaker form of this requirement is the so called
``time-slice''property.
\end{itemize}

\textit{Comments:}

The domain requirements on (unbounded) smeared-out fields $A(f)$ are
reminiscent of properties which are required of generators of noncompact
groups. The motivation for this Wightman postulates are entirely pragmatic;
they insure that the standard calculational methods of physicists are
applicable. These more technical domain requirements will be absent in the
net approach. The only domain properties of the algebraic approach are the
very fundamental (and physical) domain properties of the Tomita-Takesaki
modular theory. \textit{But it turns out that as a result of the new concept
of modular localization, the Wightman domain properties, far from being
merely technical postulates, become carriers of important physical
informations} about properties of quantum versus geometric localization. The
Wightman domain turns out to be simply the intersection of all the dense
modular localization spaces.

The existence of the sesquilinear forms for pointlike fields is the
substitute for the classical notion of field strength. The $\widetilde{%
\mathcal{P}}$-transformation property of the hermitian adjoint field is that
of the complex conjugate transformation which is isomorphic to the
antiparticle field: $A^{(\bar{\lambda})}(x)=CA^{(\lambda )*}(x),$ $C$=charge
conjugation matrix.

Strictly speaking the causality requirement applies to locally observable
fields only (example: electromagnetic field strength and currents, but not
to vector potentials and charged matter fields). The restriction to local
fields, which by definition obey the $\mp $ commutation relations, is too
strong in $d\leq 2+1$ (see last chapter). In d=3+1 all compactly localizable
charge carrying fields are equivalent (by Klein-transformations) to local
Fermi- or Bose-fields. We will use the term ``localizable'' instead of local
for fields with noncompact localization properties.

The strong causal completion property is the substitute for an hyperbolic
equation of motion (which, due to ill-defined nonlinear terms, is a priori
meaningless in QFT). Its formulation and exploration is more natural in the
algebraic setting where it simply means that $\mathcal{A(O})=\mathcal{A}(%
\mathcal{O}^{\prime \prime }).$

Another physically important property which has been omitted here, but makes
its appearance in the net theory later on, is the nuclearity or compactness
property. This is the QFT counterpart of the statement that a finite cell in
phase space can only accommodate a finite (in QFT a nuclear instead of
finite set of vectors) number of degrees of freedom.

The most useful objects which one can form in such a Wightman setting of
fields are the vacuum expectation values or (terminology of condensed matter
physics) correlation functions: 
\begin{eqnarray}
w_{n}^{(\lambda _{1}....\lambda _{n})}(x_{1}....x_{n}) &=&\left\langle
0\left| A^{(\lambda _{1})}(x_{1})....A^{(\lambda _{n})}(x_{n})\right|
0\right\rangle \\
&=&W_{n}^{(\lambda _{1}....\lambda _{n})}(\xi _{1},....\xi _{n-1})\,\quad
\xi _{i}=x_{i}-x_{i+1}  \nonumber \\
&=&\left\langle 0\left| A^{(\lambda _{1})}(0)e^{-i\mathbf{P}\xi
_{1}}....A^{(\lambda _{n-1})}(0)e^{-i\mathbf{P}\xi _{n-1}}A^{(\lambda
_{n})}(0)\right| 0\right\rangle  \nonumber \\
&=&\idotsint \tilde{W}_{n}^{(\lambda _{1}....\lambda
_{n})}(q_{1}....q_{n-1})e^{-i\sum_{k}q_{k}\xi _{k}}d^{4}q_{1}....d^{4}q_{n-1}
\nonumber
\end{eqnarray}
The spectrum property: $spec(\mathbf{P})\subset \bar{V}_{+}$ evidently
implies that 
\[
supp\tilde{W}^{(..)}(q_{1}..q_{n-1})\subset \otimes ^{n}\bar{V}_{+} 
\]
and, as a property of a Fourier-Laplace transform of a cone supported
distribution we encounter the ``tube analyticity'', namely $W$ is boundary
value of a function $W^{(..)}(z_{1}....z_{n-1})$ analytic in the tube $%
T^{(n-1)}$ $z_{i}=\xi _{i}-i\eta _{i}$ with $\eta _{i}\in \bar{V}_{+}$
fulfilling the ``tempered'' bound (assuring the temperedness of the singular
boundary values): 
\begin{equation}
\left| W_{n}^{(..)}(z_{1}....z_{n-1})\right| \leq C\frac{\left(
1+\sum_{i}^{n-1}\left| z_{i}\right| ^{2}\right) ^{k}}{\left( \min_{j}(\eta
_{j}^{2})\right) ^{l}},\quad \left| z\right| ^{2}:=\sum_{\mu }\left| z_{\mu
}\right| ^{2}
\end{equation}
This tube analyticity together with the Lorentz-invariance of the W's (a
consequence of the invariance of the vacuum and the covariant transformation
properties of the fields) yields the invariance under the complex
Lorentz-group $L^{c}$: 
\begin{eqnarray}
L^{c} &=&\left\{ A,B\right\} \quad A,B\in SL(2,C) \\
\underline{z} &=&\sigma _{\mu }z^{\mu }\rightarrow A\underline{z}B^{*},\quad
z^{\mu }z_{\mu }=inv.  \nonumber
\end{eqnarray}
This complex extension is a rather direct consequence of the previous
analyticity and the fact that the finite dimensional representations $D^{(A,%
\dot{B})}$ permit an extension to a transformation in which the undotted and
dotted spinors transform independently. The details can be looked up in the
literature. $L^{c}$ has different from $L$, only two instead of four
connected components: det=$\pm .$ It takes some additional calculations to
prove that the extended tube is of the form $T_{ext}^{(n-1)}=L^{c}T^{(n-1)}.$
This is a natural analyticity region and the W are univalued. It is
remarkable that $T_{ext}$ contains real points. It is easy to see that the
convex real set: $\quad $%
\begin{equation}
\xi _{k}^{2}<0,\quad (\sum_{k}\lambda _{k}\xi _{k})^{2}<0\quad \lambda
_{k}\geq 0\quad \sum_{k}\lambda _{k}>0
\end{equation}
$\quad $(the so called Jost points) is contained in $T_{ext}$. The locality
binds all the n! different $w^{(n)}(x_{i_{1}}....x_{i_{n}})$ together to one
(anti-)symmetric holomorphic ``master function'' with $T_{ext}^{perm}$ the
extended permuted tube being the enlarged analyticity region on which the
permutations act: 
\begin{equation}
w_{n}^{(\lambda _{i_{1}}....\lambda
_{i_{n}})}(z_{i_{1}}....z_{i_{n}})=\left\{ 
\begin{array}{l}
w_{n}^{(\lambda _{1}....\lambda _{n})}(z_{1}....z_{n}) \\ 
signPw_{n}^{(\lambda _{1}....\lambda _{n})}(z_{1}....z_{n})
\end{array}
\right.
\end{equation}
The mathematical structure behind this extension is the so called ``edge of
the wedge'' theorem which generalizes the well known Schwarz reflection
principle from one to several complex variables. The resulting ``permuted
extended tube'' is not a natural holomorphy domain but its holomorphic
completion is difficult to understand (and fortunately physically not as
relevant as it appeared during the 60$^{ies})$. For a discussion of this and
related matters we refer to the literature. One physically relevant fact is
the univaluedness of the master function in d=3+1 theories. In d=1+1 the
possibility of richer spacelike commutation relations (e.g. braid group
statistics) which have multivalued master functions. Naturally its
restriction to the real analytic (Jost) points is always univalued (the
branching happens in the euclidean region), since otherwise the Hilbert
space setting of quantum physics would get lost \cite{wightman} \cite{Jo}.

The crucial question is now wether the family of $w$'s with those properties
following from the operator postulates suffice in order to reconstruct
uniquely (up to isomorphism) the quantum field theory. From our experience
with the GNS construction we would expect a positive answer. However the
``field algebras'' are not $C^{*}-$algebras of bounded operators and
therefore a special construction which is more adapted to this problem is
necessary. One uses the polynomial algebra $\mathcal{P}(M)$ : 
\begin{equation}
\left\{ f_{0}\underline{1}+\sum_{n=1}^{N}\idotsint
f_{n}(x_{1}....x_{n})A(x_{1})....A(x_{n})\mid f_{n}\in \mathcal{S}(\mathbf{R}%
^{dn}),\forall N\right\}
\end{equation}
Here we have again supressed all Lorentz- and charge-indices i.e. we used
our standard neutral scalar illustration. As in the case of CCR and CAR we
can interpret the expectation values $w^{(n)}$ as affiliating a positive
linear functional on a *- algebra of test functions: 
\begin{equation}
f=\left\{ f_{0}^{.}....f_{N}\right\} \in \bigoplus_{n}\mathcal{S}%
(R^{dn})\equiv T\mathcal{S}
\end{equation}
\begin{equation}
(f\cdot
g)_{n}(x_{1}....x_{n})=\sum_{k}f_{k}(x_{1}..x_{k})g_{n-k}(x_{k+1}..x_{n})
\end{equation}
\begin{equation}
(f^{*})_{n}(x_{1}....x_{n})=\overline{f_{n}(x_{n}....x_{1})}
\end{equation}
Note that different from the CCR and CAR case this is not a Hilbert space of
``one particle'' functions but a tensor algebra $T(M)$ on sequences of
functions. Here $M$ indicates that the testfunction space consists of
functions on Minkowski-space. The localized polynomial algebra $\mathcal{P(O)%
}$ is a subalgebra of $\mathcal{P(M)}$. The vacuum expectation values $w_{n}$
just define a state (positive definite normalized functional) on $T:$%
\begin{equation}
w(f)=\sum_{n}W_{n}(f_{n}),\quad w(f^{*}f)\geq 0
\end{equation}
In the operator way of writing, this is just the positivity of the norm
squared: 
\begin{equation}
\left\| \left( f_{0}\underline{1}+\sum_{n=1}^{N}\idotsint f_{n}^{(\lambda
_{1}....\lambda _{n})}(x_{1}....x_{n})A^{(\lambda
_{1})}(x_{1})....A^{(\lambda _{n})}(x_{n})\right) \left| 0\right\rangle
\right\| ^{2}\geq 0
\end{equation}
With an appropriately defined action $\alpha $ of $\mathcal{P}$ on the
tensor algebra, W is covariant: 
\begin{equation}
w(\alpha _{a,\Lambda }(f))=w(f)
\end{equation}
The reconstruction is completely analogous to the GNS situation. One obtains
a triple $(H,\pi ,\Omega )$ i.e. a *-representation (of the so-called
Borchers-Uhlmann tensor-algebra) which is covariant with positive spectrum
and a unique vacuum vector $\Omega $. Certain properties as the time-slice
requirement and its local version have no known equivalent in terms of
correlation function; they need the reconstructed operator theory for their
formulation.

\section{Simple Structural Properties}

\textbf{1. The Cluster decomposition property.} Its weak form is defined: 
\begin{eqnarray}
\lim_{\lambda \rightarrow \infty }W(f\alpha _{\lambda x}(g))
&=&\lim_{\lambda \rightarrow \infty }\left\langle 0\left| A_{f}^{*}U(\lambda
x)B_{g}\right| 0\right\rangle =W(f)W(g) \\
A_{f},B_{g} &\in &\mathcal{P}(M)  \nonumber
\end{eqnarray}
and results from the fact that only the discrete part of the energy momentum
spectrum (i.e. the assumed unique vacuum contribution) survives, whereas the
continuum oscillates to zero (Riemann-Lesbegue Lemma). The strong form is
conveniently formulated in terms of the connected correlation functions: 
\begin{eqnarray}
&&\left\langle 0\left| A_{f_{1}}(x_{1})....A_{f_{n}}(x_{n})\right|
0\right\rangle _{con}\stackrel{clustering}{\rightarrow }0 \\
A_{f}(x) &=&U(x)A_{f}U^{*}(x),\quad A_{f}\in \mathcal{P}(M)  \nonumber
\end{eqnarray}
It uses locality (in order to disentangle overlapping clusters) and needs
more mathematical effort for its derivation from the postulates. Note that a
vacuum degeneracy would show up in form of a very specific violation
(containing information about the dimension of the vacuum projector) of the
cluster property.

\textbf{2.The Reeh-Schlieder Theorem.} The localized polynomial algebra $%
\mathcal{P(O)}$ is cyclic and separating on $\Omega ,$ i.e. 
\begin{equation}
\overline{\left[ \mathcal{P(O)}\Omega \right] }=H,\quad A\Omega
=0\curvearrowright A=0,\quad A\in \mathcal{P(O)}
\end{equation}
For the cyclicity assume that $\psi \in \left[ P(O)\Omega \right] ^{\bot
},\psi \neq 0.$ Then for $A_{i}\in \mathcal{P}(\tilde{O}),\,\tilde{O}\ll O$
(no boundary touching) define: 
\begin{eqnarray}
F(x_{1}....x_{n}) &=&\left\langle \psi \left| \alpha
_{x_{1}}(A_{1})....\alpha _{x_{n}}(A_{n})\right| 0\right\rangle \quad \psi
\in \mathcal{D} \\
&=&0\quad on\,\left\{ (x_{1}....x_{n})\mid x_{i}\in V,\,\tilde{O}+V\in
O\right\}  \nonumber
\end{eqnarray}
The Fourier transform $\tilde{F}(p_{1}....p_{n})$ vanishes outside the
support: $\cap _{l}\left\{ \sum_{l}^{n}p_{i}\in \bar{V}^{+}\right\} ,$ as
follows from the spectrum condition. Therefore also the matrix element $F$
enjoys tube analyticity in $z_{1}....z_{n}$ (instead of n-1 z's as the W's
). They agree with the (obviously holomorphic) zero function in the above
real neighborhood.. The already mentioned multi-dimensional generalization
of the Schwarz reflection principle termed ``edge of the wedge theorem''
will then lead to the identical vanishing. But this contradicts the
assumption of nontriviality of $\psi ,$ q.e.d.

The proof of separability of $\Omega $ with respect to $\mathcal{P(O)}$ can
be reduced to cyclicity by using locality. We have: 
\begin{equation}
AA^{\prime }\Omega =A^{\prime }A\Omega ,\quad A\in \mathcal{P(O)},\quad
A^{\prime }\in \mathcal{P(O}^{\prime }\mathcal{)}
\end{equation}
Since $\mathcal{O}^{\prime }$ is non-void, $\mathcal{P(O}^{\prime }\mathcal{)%
}$ acts cyclically on $\Omega $ and therefore $A^{\prime }\Omega
=0\curvearrowright A^{\prime }=0$ on the dense set $\mathcal{P(O)}\Omega $
and hence $A\equiv 0.$

With the Reeh-Schlieder theorem we have met the first characteristic
property of local quantum physics. It has no counterpart in Schr\"{o}dinger
theory and general quantum theory. Indeed the idea that one can emulate
vacuum excitations ``behind the moon'' by operating with hardware localized
on an earthly laboratory with increasing accuracy, sounds somewhat exotic.
It has led to many misunderstandings especially with respect to causality.
One of the more spectacular conceptional mistakes even casts doubt on
Fermi's conclusion that Einstein's causality statements about classical
relativistic field theory are also valid in QFT \cite{S rem}. On the
positive side this property led to deeper thoughts about long range
correlation and the proper operational formulation of causality as well as
phase space localization of degrees of freedom (nuclearity).

\textbf{3. Irreducibility of }$\mathcal{P}\mathbf{(M)}$\textbf{\ }Starting
from the time-development automorphism which (according to the positive
energy assumption) is implemented by a positive Hamiltonian: 
\begin{equation}
\alpha _{t}(A)=e^{iHt}Ae^{-iHt},\quad H\geq 0,\quad A\in \mathcal{P(M)},
\end{equation}
we study the analytic properties of matrix elements of time translated
operators from the commutant: 
\begin{eqnarray}
f(t) &:&=\left\langle A_{1}\Omega \left| e^{-iHt}A^{\prime }e^{iHt}\right|
A_{2}\Omega \right\rangle =\left\langle \Omega \left| A_{1}^{*}\alpha
_{-t}(A^{\prime })A_{2}\right| \Omega \right\rangle \\
A_{i} &\in &\mathcal{P(M)},\quad A^{\prime }\in \mathcal{P(M)}^{\prime
},\quad  \nonumber \\
\mathcal{P(M)}^{\prime } &\equiv &\left\{ C\mid \left\langle A^{*}\phi
,C\psi \right\rangle =\left\langle \phi ,CA\psi \right\rangle \;\forall A\in 
\mathcal{P(M)},\,\phi ,\psi \in \mathcal{D}\right\}  \nonumber
\end{eqnarray}
One computes: 
\begin{eqnarray}
f(t) &=&\left\langle A^{\prime *}\Omega ,\alpha _{t}(A_{1}^{*}A_{2})\Omega
\right\rangle =\left\langle A^{\prime *}\Omega ,e^{itH}A_{1}^{*}A_{2}\Omega
\right\rangle \\
&=&\left\langle \alpha _{t}(A_{2}^{*}A_{1})\Omega ,A^{\prime }\Omega
\right\rangle =\left\langle A_{2}^{*}A_{1}\Omega ,e^{-itH}A^{\prime }\Omega
\right\rangle  \nonumber
\end{eqnarray}
The first line represents $f(t)$ as a matrix element of $e^{itH},$ and the
second of $e^{-itH}$. Therefore $f$ is a bounded function which is
simultaneously analytic in the upper and lower halfplane. According to a
theorem of Liouville this forces f to be a constant i.e. 
\begin{eqnarray*}
f(t) &=&\left\langle A^{\prime *}\Omega ,E_{0}A_{1}^{*}A_{2}\Omega
\right\rangle =\left\langle \Omega ,A^{\prime }\Omega \right\rangle
\left\langle A_{1}\Omega ,A_{2}\Omega \right\rangle ,E_{0}=%
\hbox{proj on
vac.} \\
&\curvearrowright &A^{\prime }=\left\langle \Omega ,A^{\prime }\Omega
\right\rangle \cdot \mathbf{1}
\end{eqnarray*}

\textbf{4. TCP Symmetry. }We first remind ourselves of the
TCP-transformation property of free particles: 
\begin{eqnarray}
\Theta \left| p,\lambda ,i\right\rangle &=&\sum_{\lambda ^{\prime }}\left|
p,\lambda ^{\prime },\bar{i}\right\rangle D_{\lambda ^{\prime },\lambda
}(i\sigma _{2})\quad \hbox{in }H_{1},\hbox{ antilinear} \\
\bar{i} &:&%
\hbox{antiparticles of type i,\thinspace \thinspace \thinspace
\thinspace }\Theta ^{2}=(-1)^{2s}\mathbf{1}  \nonumber
\end{eqnarray}
\begin{eqnarray}
\Theta \Phi ^{\left[ A,\dot{B}\right] }(x)\Theta ^{-1} &=&\left( -i\right)
^{F}(-1)^{\left| \dot{B}\right| }\Phi ^{\left[ A,\dot{B}\right] *}(-x)\equiv
\Phi ^{\theta }(-x) \\
F\quad \hbox{\# of fermions, }\left| \dot{B}\right| \hbox{{}} &=&%
\hbox{\# of
dotted spinor indices}  \nonumber
\end{eqnarray}
We will show that this formula holds in general (for local interacting
fields). If $\Phi =A=$scalar field, the proof starts from first rewriting
the content of TCP symmetry in terms of correlation functions: 
\begin{eqnarray}
&&\left\langle \Theta A(x_{m})....A(x_{1})\Omega ,\Theta
A(x_{m+1})....A(x_{n})\Omega \right\rangle \\
&=&\left\langle A(x_{m+1})....A(x_{n})\Omega ,A(x_{m})....A(x_{1})\Omega
\right\rangle  \nonumber
\end{eqnarray}
\[
\Leftrightarrow w(-x_{1},....-x_{n})=w(x_{n}....x_{1}) 
\]
where in the last line we used the above $\Theta $-action. We now take
notice of the fact that by combining the symmetry relation from locality in $%
T_{ext}^{perm}$ with the $L_{+}(C)$ invariance (which included the total
reflection) we have: 
\begin{eqnarray}
w(x_{n}....x_{1}) &=&w(x_{1}....x_{n})\stackrel{L_{+}(C)}{=} \\
&&w(-x_{1}....-x_{n})
\end{eqnarray}
This means that we obtain the above relation in $T,$ and hence also on the
physical boundary (the boundary $i\varepsilon $-prescription in the above
relation is the same on both sides) which is the desired relation for the
operators.

\textbf{5. Spin \& Statistics. }If we require the wrong local commutation
relations for $\Phi ^{\left[ A,\dot{B}\right] }$: 
\begin{eqnarray}
\left\{ \Phi (x)\Phi (y)\right\} &=&0,\quad \left( x-y\right) ^{2}<0,\quad A+%
\dot{B}=\frac{n}{2},\,n\,\hbox{odd} \\
\left[ \Phi (x)\Phi (y)\right] &=&0\quad \left( x-y\right) ^{2}<0,\quad n\,%
\hbox{\thinspace even}  \nonumber
\end{eqnarray}
then $\curvearrowright \Phi \equiv 0$. With other words within the framework
of local fields, the standard relation between spin and statistics is a
consequence of the postulates. The proof (again for neutral scalar fields)
only employs the two-point function $W(z)$ which is analytic in $T_{ext}$
and fulfills (as a consequence of $L_{+}(\mathbf{C})$-invariance)
W(z)=W(-z): 
\begin{eqnarray}
\left\langle 0\left| \left\{ \Phi (x),\Phi (y)\right\} \right|
0\right\rangle &=&W(\xi )+W(-\xi )=0,\;\left( x-y\right) ^{2}<0 \\
&\curvearrowright &2W(z)=0,\quad \curvearrowright \Phi (x)\Omega =0\quad \, 
\nonumber
\end{eqnarray}
Finally the Reeh-Schlieder theorem gives $\Phi \equiv 0$. The general case
with dotted and undotted spinors is left to the reader.

The TCP and Spin\&Statistics theorem are considered to represent the deeper
parts of structural QFT. They even gave the title for the first book on the
subject \cite{wightman}. Later we will see that they continue that role in
algebraic QFT with an additional gain in profoundness (in a particular
evident form in low dimensional QFT)..

\textbf{6. Normal C.R. and Klein transformations.} The previous theorem left
open the commutation relations between different Lorentz-multiplets. One
defines as ``normal'' the spacelike commutativity of two local fields of
which at least one is bosonic, as well as the spacelike anticommutativity
between two fermionic ($A+\dot{B}$=halfinteger) fields $\psi ^{(A,B)}$. As a
preparatory step towards proving that normal commutation relation can always
be achieved, let us prove that commutation relations remain stable under
transition to the hermitian adjoint field: 
\begin{eqnarray}
\left[ \Phi _{1}(x),\Phi _{2}^{*}(y)\right] _{\pm } &=&0\quad (x-y)^{2}<0 \\
&\curvearrowright &\left[ \Phi _{1}(x),\Phi _{2}(y)\right] _{\pm }=0\quad
(x-y)^{2}<0  \nonumber
\end{eqnarray}
The proof uses the cluster decomposition property (i.e. the uniqueness of
the vacuum): 
\begin{eqnarray}
&&\left\langle \Omega ,\Phi _{1}^{*}(f)\Phi _{2}^{*}(g)\Phi _{2}(g)\Phi
_{1}(f)\Omega \right\rangle \\
&=&\left\| \Phi _{2}(g)\Phi _{1}(f)\Omega \right\| ^{2}\geq 0  \nonumber \\
&&\stackrel{loc.}{=}\sigma \left\langle \Omega ,\Phi _{1}^{*}(f)\Phi
_{1}(f)\Phi _{2}^{*}(g)\Phi _{2}(g)\Omega \right\rangle  \nonumber \\
&&\stackrel{cluster}{\Rightarrow }\sigma \left\| \Phi _{1}(f)\Omega \right\|
^{2}\left\| \Phi _{2}(g)\Omega \right\| ^{2}\quad  \nonumber
\end{eqnarray}
Here $\sigma =\pm $. Consistency requires that $\sigma $ agrees with the $%
\pm 1$ in the original C.R.. between $\Phi _{1}$ and $\Phi _{2}^{*}$. Now we
are prepared to construct the Klein transformation which carries anomalous
into normal commutation relations. For the typical anomalous situation
assume that: 
\begin{eqnarray}
\left[ \varphi (x),\psi (y)\right] &=&0,\quad \left( x-y\right) ^{2}<0 \\
\varphi &:&\hbox{bosonic,\quad }\psi :\hbox{fermionic}\quad  \nonumber
\end{eqnarray}
Define: 
\begin{eqnarray}
\varphi ^{^{\prime }}(x)H_{even} &=&\varphi (x)H_{even} \\
\varphi ^{\prime }(x)H_{odd} &=&-\varphi (x)H_{odd}  \nonumber \\
\psi ^{\prime }(x)H_{even} &=&\psi (x)H_{even}  \nonumber \\
\psi ^{\prime }(x)H_{odd} &=&\psi (x)H_{odd}  \nonumber
\end{eqnarray}
or briefly: $\varphi ^{\prime }(x)=\left( -1\right) ^{\mathbf{F}}\varphi
(x),\quad \psi ^{\prime }(x)=\psi (x).$ One obtains: 
\begin{equation}
\left\{ \varphi ^{\prime }(x),\psi ^{\prime }(y)\right\} =0,\quad \left(
x-y\right) ^{2}<0
\end{equation}
The general situation is analogous.

\textbf{7. Characterizations of free fields.} Assume first that the
two-point function agrees with that of a free field, i.e. 
\begin{eqnarray}
\left\langle \Omega ,\varphi (x)\varphi (y)\Omega \right\rangle &=&i\Delta
^{(+)}(x-y)\curvearrowright \\
\varphi (x) &=&\frac{1}{(2\pi )^{\frac{3}{2}}}\int (e^{-ipx}a(p)+h.c.)\frac{%
d^{3}p}{2\omega }  \nonumber
\end{eqnarray}
in case of a neutral scalar field. The first step in the proof consists in
deriving the Klein-Gordon equation ($\partial ^{\mu }\partial _{\mu
}+m^{2})\varphi (x)=0.$ From the two-point function one obtains 
\begin{eqnarray}
\left\langle \Omega ,j(x)j(y)\Omega \right\rangle &=&0,\quad \hbox{where }%
j(x):=(\partial ^{\mu }\partial _{\mu }+m^{2})\varphi (x) \\
&\curvearrowright &j(x)\Omega =0  \nonumber
\end{eqnarray}
The analytic properties in the tube $T$ of the mixed $j$-$\varphi $
correlation functions together with the relative spacelike commutativity
bring about an edge of the wedge situation with: 
\begin{eqnarray}
&&\left\langle \Omega ,\varphi (x_{1})....j(x_{i})\varphi
(x_{i+1})....\Omega \right\rangle \\
&=&\left\langle \Omega ,\varphi (x_{1})....\varphi
(x_{i+1})....j(x_{i})\Omega \right\rangle =0  \nonumber
\end{eqnarray}
on an open set of the boundary and hence the vanishing of all matrix
elements of j on the dense domain $\mathcal{D}$ i.e. $j(x)\equiv 0.$
Therefore $\varphi $ indeed fulfills the free field equation and hence
permits a frequency decomposition: 
\begin{equation}
\varphi (x)=\varphi ^{(-)}(x)+\varphi ^{(+)}(x),\quad \varphi
^{(-)}(x)\Omega =0
\end{equation}
A characterizing property of free fields is their c-number (anti)commutator,
in our case: 
\begin{equation}
\left[ \varphi (x),\varphi (y)\right] =i\Delta (x-y)\mathbf{1}
\end{equation}
But this follows by again using analyticity properties. First we use the
spectrum condition to obtain: 
\begin{equation}
\varphi ^{(-)}(x)\varphi ^{(+)}(y)\Omega =i\Delta ^{(+)}(x-y)\Omega
\end{equation}
since $\varphi ^{(\pm )}$ transfers momentum on the forward (backward) mass
shell and hence the spectral transfer of the product is spacelike including
zero i.e. the intersection with the physical spectrum consists of just $p=0$
corresponding to the vacuum vector $\Omega .$ This is a much stronger
statement than the assumed two-point function structure. For the commutator
applied to $\Omega $ we now have: 
\begin{equation}
\left[ \varphi (x),\varphi (y)\right] \Omega =i\Delta (x-y)\Omega +\left[
\varphi ^{(+)}(x),\varphi ^{(+)}(y)\right] \Omega
\end{equation}
Let $\psi \in \mathcal{D}$ and consider the analytic properties of: 
\begin{equation}
F(x,y):=\left\langle \psi ,\left[ \varphi ^{(+)}(x),\varphi ^{(+)}(y)\right]
\Omega \right\rangle
\end{equation}
The momentum transfer of each $\varphi ^{(+)}$ is on the forward mass shell,
and hence this distribution is the boundary value of a function $%
F(z_{1},z_{2})$ analytic in $z_{i}=x_{i}-iy_{i},\;y_{i}\in V^{+}.$ Since
this function vanishes in a neighborhood of the real boundary, $%
\curvearrowright F\equiv 0.$ But: 
\begin{equation}
B(x,y):=\left[ \varphi (x),\varphi (y)\right] -i\Delta (x-y)\mathbf{1}
\end{equation}
is a bilocal operator-valued distribution ( $B(f,g)$ $\in \mathcal{P(O)}$ )
for which $\Omega $ is a separating vector, i.e. 
\begin{equation}
B(x,y)\Omega =0\curvearrowright B(x,y)\equiv 0\quad q.e.d.
\end{equation}
This property has no analogue in quantum mechanics i.e. the pair interation
does not show up in the nonrelativistic two-point function (related to the
absent selfinteraction). This conclusion continues to hold for semiinfinite
string-localized operator-valued solutions of free field equations\cite{Mund}%
.

Remembering that a free field has vanishing connected n-point correlation
functions for n\TEXTsymbol{>}2, the question arises whether this property is
typical. The affirmative answer is: 
\begin{eqnarray}
\hbox{if }\exists n &>&2\quad w_{n}(x_{1},....x_{n})^{conn}=0 \\
&\curvearrowright &\varphi (x)\hbox{ is a generalized free field}  \nonumber
\end{eqnarray}
A generalized free field shares with the free field the property of having a
c-number commutator. But this commutator is a (continuous) superposition of
free field commutators: 
\begin{equation}
\left[ \varphi (x),\varphi (y)\right] =\int d\rho (\kappa ^{2})i\Delta
(x-y,\kappa ^{2})
\end{equation}
We will not prove the above statement here.

Another characterization of (generalized) free fields not presented here is
in terms of gaps in the spacelike momentum transfer of fields. Reductions to
generalized free fields are effectively reductions to free fields in view of
the time slice property and in particular of the phase space nuclearity
property presented in a later section.

\textbf{8. Shape of Energy Momentum Spectrum. }The asymptotic factorization
or clustering of correlation functions suggests that the energy-momentum
spectrum \textbf{specP }is an additive set i.e. with\textbf{\ }$p_{1},p_{2}%
\mathbf{\in specP}$ also $p_{1}+p_{2}\mathbf{\in specP.}$ To see this
consider the vector: 
\begin{equation}
\psi _{21}(a)=U(a)A_{2}U^{*}(a)A_{1}\Omega ,\quad A_{i}\in \mathcal{P(M)}
\end{equation}
Assume that the energy-momentum transfer is limited to regions $\Delta
_{i}\in \mathbf{specP.}$ Then the Fourier-transform of $\psi $ has its
support in supp$\tilde{\psi}\in \Delta _{1}+\Delta _{2}$. The clustering: 
\begin{eqnarray}
\lim_{a\rightarrow \infty }\left\| \psi _{21}(a)\right\| ^{2}
&=&\left\langle \Omega ,A_{2}^{*}(a)A_{2}(a)\Omega \right\rangle
\left\langle \Omega ,A_{1}^{*}A_{1}\Omega \right\rangle \\
&=&\left\| \psi _{2}\right\| ^{2}\left\| \psi _{1}\right\| ^{2},\quad \psi
_{i}=A_{i}\Omega  \nonumber
\end{eqnarray}
serves to show that $\left\| \psi _{21}(a)\right\| \neq 0$ i.e. does not
vanish identically thus assuring the nontriviality of the vector carrying
the sum of the momenta.

Classically the hyperbolic causal propagation in classical field theory is
inexorably linked with Lorentz-covariance. By analogy one would expect that
causality, even if it does not extend translational covariance to full
Poincar\'{e} covariance, at least forces the energy-momentum spectrum to
have a Lorentz-invariant shape. Indeed, the implementer of the translation
can always be chosen in such a way: 
\begin{eqnarray}
\exists U(a)s.t.\alpha _{a}(A) &=&U(a)AU^{*}(a) \\
U(a) &=&e^{i\mathbf{P}a},\quad spec\mathbf{P}\hbox{ inv. shape}  \nonumber
\end{eqnarray}
This theorem is easier to prove in an algebraic setting and hence will be
deferred.

\section{Euclidean Fields}

Analytic continuations through Wick-rotation have been useful in
perturbation theory, because certain regularization techniques (e.g. the
dimensional regularization) only work if noncompact L-invariance can be
replaced by compact euclidean invariance. Therefore it is interesting to
know if a euclidean formulation is also possible outside perturbation theory
and whether it is useful there. Schwinger and later Symanzik and Nelson were
the first to realize that a euclidean framework indeed opens a useful
connection with statistical mechanics and classical probability theory.

The starting point for a nonperturbative euclidean approach is the
analyticity and univaluedness of the analytic extension of correlation
functions into the extended permuted tube $T_{perm}^{ext}.$ It is obvious
that the non coinciding $(\hat{x}_{i}\neq \hat{x}_{j}\,\,\forall i,j)$
euclidean points are inside this domain. The Wick-rotation $(\vec{x}%
,x^{0})\rightarrow $ $(\overrightarrow{\hat{x}}=\vec{x},\hat{x}^{0}=ix_{4})$
relates the Minkowski inner product with the Euclidean one and the group $%
\mathbf{O}_{+}\mathbf{(4)}$ with a subgroup of $\mathbf{L}_{+}\mathbf{(C).}$
Here and in most of what follows we present the euclidean formulation for
integer spin fields, the adaptation to halfinteger spin will be commented on
later. The restriction of the analytically continued correlation functions
to the euclidean points $(\vec{x},x_{4})\in \mathbf{E=}R^{d}$ are called
Schwinger functions: 
\begin{eqnarray}
\QTR{sl}{s}(x_{1}....x_{n}) &=&w(\hat{x}_{1}....\hat{x}_{2}) \\
&=&S(\xi _{1}....\xi _{n-1})  \nonumber
\end{eqnarray}
where we used translation invariance in the last line. As for time-ordered
functions, there is no spectrum condition which assures that they are
distributions on the Schwartz-space $\mathcal{S}$, rather their natural
domain of definition are those test-functions which vanish at coinciding
points of sufficiently high order. If the dimension of the fields is
canonical i. e. for scalar fields dim$\varphi =\frac{d}{2}-1=\dim ($free
field) then $S$ is naturally (i.e. without Hahn-Banach extension) integrable
and hence a $\mathcal{S}(\mathbf{E}^{n})$ distribution. We now collect those
properties of the Schwinger functions which allow to reconstruct a local
Poincar\'{e}-invariant QFT. These properties are called the
Osterwalder-Schrader axioms. In the following we present these axioms for
the illustrative case of scalar neutral fields.

\begin{itemize}
\item  \textbf{S1 }The Schwinger functions are translation invariant real
analytic function for non coinciding euclidean variables. They are
distributions in $\mathcal{S}^{\prime }(\mathcal{E}_{-}^{n-1})$ with $%
\mathcal{E}_{-}^{n-1}=\left\{ \xi \in \mathbf{E}^{n-1}\mid \xi _{1}^{4}<\xi
_{2}^{4}....<\xi _{n}^{4}\right\} $ where $\mathcal{S}(\mathcal{E}%
_{-}^{n-1}) $ is given a weaker topology which is defined by the following
system of seminorms: 
\begin{eqnarray}
\left\| f_{<}\right\| _{l,m}=\left\| \tilde{f}\right\| _{l,m}\quad f_{<}\in 
\mathcal{S}(\mathcal{E}_{-}^{n-1}) \\
\tilde{f}(q_{1}....q_{n-1})=\int ....\int e^{\sum_{1}^{n-1}(\xi
_{j}^{4}q_{j}^{0}+i\mathbf{\xi }_{j}\mathbf{q}_{j})}f_{<}(\xi _{1}....\xi
_{n-1})d^{d}\xi _{1}....d^{d}\xi _{n-1}  \nonumber
\end{eqnarray}
Here we used the property of the Laplace-Fourier transforms of mapping
continuously $\mathcal{S}(\mathcal{E}_{-}^{n-1})$ onto a dense set in $%
\mathcal{S}$($\mathcal{M}_{+}^{n-1})$ which are Minkowski-space test
functions $\tilde{f}$ with 
\begin{equation}
supp\tilde{f}\in \left\{ q\in \mathbf{M}^{n-1}\mid q_{i}^{0}\geq
0\,\,\forall i\right\} \equiv \mathcal{M}_{+}^{n-1}
\end{equation}
The above topology is the one which $\mathcal{S}(\mathcal{E}_{-}^{n-1})$
inherits from $\mathcal{S}$($\mathcal{M}_{+}^{n-1})$ through this map. The
Schwinger distributions are just the continuous linear functionals on $%
\mathcal{S}(\mathcal{E}_{-}^{n-1})$ in this topology. It is the analog of
the growth condition on the holomorphically extended correlation functions W
which insured the temperedness of their distributional boundary values and
often called the Osterwalder-Schrader growth condition.

\item  \textbf{S2. }Hermiticity. For the Schwinger functions of a scalar
neutral field: 
\begin{eqnarray}
s(x_{1}....x_{n})=\overline{s(Tx_{1}....Tx_{n})} \\
Tx=(\vec{x},-x_{4})\quad \hbox{euclidean time reversal}  \nonumber
\end{eqnarray}

\item  \textbf{S3.} Reflection-Positivity: 
\begin{equation}
\sum_{n,m}\int s(Tx_{m},...,Tx_{1},y_{1},...y_{n})\overline{%
f_{m}(x_{1},...x_{m})}f_{n}(y_{1},...y_{n})d^{d}x_{1}...d^{d}y_{n}\geq 0
\end{equation}
where the sum only involves a finite sequence of test functions

$(f_{0},f_{1}...f_{n}...f_{N})$ with their support on the time simplex 
\begin{equation}
\mathbf{E}_{<}^{n}=\left\{ x\in E^{n}\mid 0<x_{1}^{4}<...<x_{n}^{4}\right\}
\end{equation}
Clearly this property is the analogue of the Wightman-positivity for the
W's. In fact it results from the positivity of ``euclidean states'': 
\begin{equation}
\psi (x_{1}....x_{2})=\varphi (\vec{x}_{1,}ix_{1}^{4})....\varphi (\vec{x}%
_{n},ix_{n}^{4})\Omega ,\quad x\in \mathbf{E}_{<}^{n}
\end{equation}
Note that the spectrum condition allows to interpret the analytic
continuation as a smearing with a an exponential damping factor (fast
decreasing test function in time).

\item  \textbf{S4.} Euclidean covariance: 
\begin{equation}
s(Rx_{1}....Rx_{n})=s(x_{1}....x_{n})
\end{equation}

\item  \textbf{S5.} Permutation symmetry: 
\begin{equation}
s(x_{P(1)},....x_{P(n)})=s(x_{1}....x_{n})
\end{equation}

\item  \textbf{S6. }Cluster property: 
\begin{eqnarray}
\lim_{a\rightarrow \infty }\int
s_{n}(x_{1},...x_{m},x_{m+1}+a,...,x_{n}+a)f(x_{1},...x_{m})g(x_{m+1},...x_{n})
\\
=\int s_{m}(x_{1},...x_{m})f(x_{1},...x_{m})\times \int
s_{n-m}(x_{m+1},...x_{n})g(x_{m+1},...x_{n})  \nonumber
\end{eqnarray}
The generalization to charged fields with arbitrary finite spin is obvious:
the covariance law involves the representations of the $SU(2)\times SU(2)$
covering of $O(4)$ and the permutation symmetry carries an additional $%
sign(P)$
\end{itemize}

It is fairly obvious that a theory in terms of correlation functions
fulfilling positivity, hermiticity, P-covariance and locality leads to
Schwinger functions fulfilling S1-S6. One just defines euclidean vectors $%
\psi (x_{1},....x_{n})$ as above. The reflection positivity allows to equip
the linear vector space: 
\begin{eqnarray}
&&\left\{ \sum_{n=1}^{N}\int ..\int f_{n}(x_{1},....x_{n})\psi
(x_{1},....x_{n})\mid f_{n}\in \mathcal{S}(\mathbf{E}_{<}^{n})\right\} \\
&&\mathcal{S}(\mathbf{E}_{<}^{n})\hbox{ with }\left\| f\right\| _{l,m}-%
\hbox{topology}  \nonumber
\end{eqnarray}
with a positive semidefinite inner product. Factoring out the null-space and
forming the closure one obtains a euclidean Hilbert space which thanks to
the Reeh-Schlieder theorem is equal to the GNS space of the real time
correlation functions. The short-distance growth condition of the W's in the
tube (controlling the temperedness of the distributional boundary values)
are equivalent to the $\left\| \cdot \right\| _{l,m}$ topology of the
Schwinger functions: 
\begin{equation}
s(x....x)=\left\langle \Omega ,\psi (x....x)\right\rangle
\end{equation}
The permutation symmetry of $s$ is a result of that symmetry for the
analytic $w$'s (from locality). Actually already the $\psi ^{\prime }s$ are
symmetric as real analytic functions in the euclidean domain for $x_{i}\neq
x_{j},i\neq j,x_{j}^{4}>0,$ as can be shown by he application of the edge of
the wedge theorem. Note that the Osterwalder-Schrader (euclidean) reflection
positivity S3 cannot be interpreted as a state on a $^{*}$-algebra (the
Borchers-Uhlmann tensor algebra of functions) but only serves to define a
scalar product onon a linear space (finite sequences of test functions $%
f_{n}\in S(E_{<}^{n})$. The reconstruction of the real time theory can then
be carried out in two different ways. Either one uses functional analysis
(contractive properties of space-time semigroups) or the analytic properties
of the previous Laplace-Fourier transforms in S1 which relate the Schwinger
distributions $\in S(\mathcal{E}_{-}^{n-1})$ to the spectral supported
correlation functions $\tilde{W}\in \mathcal{S(M}_{+}^{n-1})$ and carries
the reflection positivity into the $W$ positivity. The latter method is more
appropriate in the present context whereas the first method also works in
situations without space-time analyticity e.g. the derivation of the
transfer matrix formalism in classical statistical mechanics on a lattice
(see a later section). We collect the result:

\begin{theorem}
(Osterwalder-Schrader) Every set of Schwinger functions with S1-S6 comes
from a real time QFT by analytic continuation and restriction to the
euclidean points.
\end{theorem}

The euclidean framework described here is primarily a structural
reformulation, it does not really solve any problem of the real time theory
which the latter is unable to solve by itself. In fact even from a
mathematical viewpoint some of the axioms look somewhat mocked up, since the
topology we used on $S(E_{<}^{n})$ is not natural. Only under very special
circumstances it becomes a powerful constructive tool of QFT. This happens
e.g. if the Schwinger functions allow an interpretation in terms of a
continuous classical statistical mechanics. Mathematically this amounts to
the Feynman-Kac representability of the Schwinger functions in terms of a
(infinite dimensional) functional measure theory e.g. (the $\varphi ^{4}$%
-theory): 
\begin{eqnarray}
s(x_{1},....x_{n}) &=&\frac{1}{Z}\int \left[ d\varphi \right] e^{-A\left[
\varphi \right] }\varphi (x_{1})....\varphi (x_{n}) \\
A\left[ \varphi \right] &=&\frac{1}{2}(\partial \varphi \partial \varphi
+m^{2}\varphi ^{2})+g\varphi ^{4}  \nonumber
\end{eqnarray}
A physically fruitful formal interpretation is in terms of a continuous
version of a Gibbs formula for classical statistical mechanics on a lattice: 
\begin{eqnarray}
\left\langle \varphi (x_{1})....\varphi (x_{n})\right\rangle _{Gbbs}
&=&\lim_{\Lambda \rightarrow \infty }\frac{1}{Z_{\Lambda }}%
\sum_{conf,\Lambda }e^{-\beta H_{\Lambda }\left[ \varphi \right] }\varphi
(x_{1})....\varphi (x_{n)} \\
Z_{\Lambda } &=&\sum_{conf,\Lambda }e^{-\beta H_{\Lambda }\left[ \varphi
\right] }  \nonumber
\end{eqnarray}
Here the dynamical variables $\varphi $ over each lattice point take on
either values in a discrete (e.g. \textbf{Z}$_{n}$ ) or continuous manifold
(e.g. $\mathbf{C}$, $SU(2)$ etc.) in which case the sum over configurations
represents an integral over the field values at each lattice point within
the volume $\Lambda $. There are two questions to be asked:

\begin{itemize}
\item  (i) can one work out a measure theory for stochastic variables such
that the above functional integrals have mathematical meaning?

\item  (ii) can one control ``critical limits'' (second order phase
transitions) of classical statistical mechanics precisely enough in order to
obtain possibly existing local QFT?
\end{itemize}

Deferring the second problem to a later section, we comment here only on the
first one, namely the relation between a Nelson-Symanzik stochastic
euclidean theory and real time QFT. Euclidean fields are continuous linear
maps $\phi $ from test function spaces $\mathcal{S}(E^{d})$ into random
variables over a probability space $(Q,\Sigma ,\mu )$ with $\mu $ a
normalized measure on $Q$ and $\Sigma $ the $\mu $-measurable subsets. Let
us define a generating functional W for the euclidean correlation functions
of $\phi $ in a reference state (the euclidean ``vacuum'') which has the
following properties: 
\begin{equation}
S(f)=\int_{Q}e^{i\phi (f)}d\mu ,\quad i.e.S(0)=1,\,\,S(f)=\overline{S(-f)}, 
\nonumber
\end{equation}
$S(f)$ is of positive type (\ref{Min}) and invariant under euclidean time
reflections.

Here we may declare any axis to be the time axis. According to a famous
theorem of Minlos, this measure-theoretical setting is equivalent to the
following (Nelson-Symanzik) positivity and covariance properties of the
functional S(f) (the functional Fourier-transform of $\mu $): 
\begin{eqnarray}
\sum_{i,j=1}^{n}\bar{c}_{i}c_{j}S(f_{i}-f_{j}) &\geq &0,\quad S(0)=1
\label{Min} \\
S(f) &=&S(\vartheta f),\quad S(f)=S(\alpha _{a,R}f)  \nonumber
\end{eqnarray}
the last line expressing the time reflection $\vartheta $ (the choice of the
time axis is arbitrary) and euclidean invariance. In addition S(f) is
continuous on $\mathcal{S}$ in the Schwartz topology.

This setting of euclidean fields is obviously appropriate for the
Feynman-Kac representation which assumes that the measure $\mu $ on the
space of field configurations is given by an invariant statistical
mechanics-like local ``Hamiltonian'' which consists of a quadratic free and
a polynomial interacting part. We already know that the validity of the
reflection positivity is a prerequisite for obtaining real time local
quantum physics. It is not difficult to prove that such a stochastic
euclidean theory with reflection positivity is equivalent to a special class
of real time QFT namely the so called stochastic positive QFT.

\begin{definition}
A QFT is said to fulfill stochastic positivity if its associated von Neumann
algebra $\mathcal{A}$ contains an abelian subalgebra $\mathcal{B}$ (''fields
at one time'') and an automorphism $\alpha _{t}$ (''time translation'') such
that: $\overline{\bigcup_{t}\alpha _{t}(\mathcal{B})}=\mathcal{A}$
\end{definition}

\begin{theorem}
\cite{KL} A reflection positive stochastic euclidean theory is equivalent to
a stochastic positive real time QFT.
\end{theorem}

Hence the equivalence requires the stochastic theory to have an additional
QFT positivity property (reflection positivity) and the QFT to posses an
additional stochastic (Nelson-Symanzik) positivity. We will not prove this
theorem since our main motivation here is pedagogical namely to counteract
the erroneous but widespread belief that QFT can be always be defined in
terms of measure theory or Feynman-Kac Formulas. Only theories which ``stay
close'' to the d=1+1 $\phi ^{4}$-theory (the standard relativistic
illustrative example of the above theorem ) allow for a Feynman-Kac
representation. Whatever the intuitive appeal of Lagrangian quantization and
functional integrals may be worth, one of its conceptual and mathematical
limitation is set by the above theorem. In quantum mechanics involving
charges coupled to vectorpotentials, it is possible to go somewhat beyond
the above standard setting of euclidean functional integrals at the expense
of loosing the tight physical relation of the euclidean theory to
statistical mechanics. But no such framework is known for e.g. the
Chern-Simons Feynman-Kac representations.

Note that we are here not concerned with mathematically fine points caused
by renormalization (e.g. $\phi ^{4}$ in d=1+2 or d=1+3) wrecking the
canonical (equal time) structure. Rather we mean that certain theories are 
\textit{structurally incompatible} with the stochastic Feynman-Kac
representations; they simply do not even posses a formal Feynman-Kac like
representation as the $\phi _{4}^{4}$ theory. Examples are chiral conformal
theories and, as mentioned before, $d=1+2$ theories with braid group
statistics (Chern-Simons actions). They are easily shown to fail on the
stochastic positivity property. The reason is the nonexistence of an abelian
subalgebra with the required density property.

On the other hand, if there is any quantum theory at all associated with the
Chern-Simons Lagrangian, then the combinatorial theory \cite{RTV} defined by
the Markov trace on the ribbon braid group $RB_{\infty }$ i.e. the theory
behind the knot invariant and the associated 3-manifold invariants of V.
Jones \cite{Jones} is the only reasonable candidate. Witten introduced rules
for Wilson loops which indeed give this result \cite{Witten}. Additional
support comes from algebraic QFT which finds these invariants in the type II$%
_{1}$ intertwiner algebras within the DHR theory of superselection sectors
(see last chapter). The system of intertwiners between different sectors
together with the Markov trace forms a ``combinatorial QFT'' par excellence%
\cite{FRS}. In this form the algebras underlying the topological theories
was known already in the famous 1969 DHR work \cite{Haag}. Of course the DHR
analysis becomes richer and more interesting for the more recent braid group
statistics in low dimensional field theory in which case one obtains the
Jones knot invariants as well as the invariants of 3-mf. of Witten's
approach (see appendix of \cite{FRS}). Mathematically very close related to
the algebraic QFT approach to intertwiner algebras is the combinatorial
theory which does not use physical principles but rather quantum group
methods. This is due to Reshetikin, Turaev and Viro\cite{RTV} and was
extended by Karowski and Schrader\cite{KaS}. Neither this method nor the
Witten method furnishes a physical interpretation .Only the intertwiner
approach of algebraic QFT which places the type II combinatorical
intertwiner algebra into a localizable (and hence interpretable) QFT carries
such information. Witten's rules for extracting the knot and 3 mf.
invariants from the Chern-Simons action is however somewhat surprising and
poses the question, whether this observation can also be understood in the
spirit of algebraic QFT by putting suitable states on e.g. appropriately
extended Weyl-like algebras. It is has been my firm belief that the use of
singular state will remove the ``meat'' of these Weyl-like algebras and just
leave the combinatorial ``bones'' \cite{S rem}\cite{S2}. This, if true,
would bridge the gap between the approach of Witten, which looks like
continuous QFT and the combinatorial approach and also answer some questions
concerning the relation between real time and euclidean time. As it stands,
the situation presents a very interesting paradox..

\textit{Singular} states harmonizes very well with the formal idea of
integrating over infinitely many \textit{gauge copies} in euclidean path
integrals, except that singular states are more noncommutative and in this
way reconcile the difference between real time and imaginary time theories:
the time development automorphism of the Weyl like algebras is wiped out by
the singular nature of states. Whereas in the standard formulation of gauge
theories there is no mathematical veto against considering non
gauge-invariant formally space-time dependent correlation functions,
singular states create such a mathematical veto. This aspect of singular
states is very desirable, because something which is unphysical, should also
be unmathematical. The idea is that whereas in ``full'' gauge theories with
physical photons and matter the singular states become regular on a huge
translation covariant subalgebra \cite{Morch Stro} (the gauge invariant
algebra of the quantization picture), in Chern-Simons theories the regular
part is so small \cite{S rem} that it can only support combinatorial type II$%
_{1}$ data. In fact singular states are the only states which are capable of
distinguishing a subalgebra. In quantum mechanics these states are usually
excluded by the regularity assumption, i.e. one is only interested in those
representations of the Weyl algebra (the von Neumann uniqueness theorem) for
which the translations $U(\alpha )$ are continuous in $\alpha .$ The only
known situation in QM where singular states are apparently needed is the
Hofstedter model of particles in a constant magnetic field (mathematically:
the ``noncommutative torus''). The abelian $\mathbf{Z}_{n}$-Chern Simons
theory in the real time formulation is a maximally extended QFT of one-forms
in d=2+1 where the extension is done by admitting semiinfinite one forms
which are closed but not exact in the angular variable. Since I am not going
to use such an approach in these lectures, I refer the interested reader to 
\cite{S rem} where he finds references to important previous work on
singular states. Unfortunately there exists presently no solution of this
interesting problem of topological versus combinatorial theory.

Excepting the correctness of this picture, it is not reasonable to attempt
to extract a theory of anyons from the pure Chern-Simons theories. In fact
the only physical use of topological field theory should be the illustration
of the working of singular states. In the last chapter we will use an
extension of the d=2+1 Wigner particle classification method because we
expect more concrete results on ``free'' anyons and plektons than with the
Chern-Simons approach. In this way we also follow the historical route,
since free Bosons and Fermions where first obtained by operator-methods and
Wigner theory before this description was transcribed (in connection with
perturbation theory) into functional language.

A closely related, conceptually more robust constructive idea is to try to
define QFT as scaling limits of mathematically controllable lattice systems
instead of working with formal Feynman-Kac representations. The guiding
principle (going back to Kadanoff, Wilson and others) was to use the
possible existence of second order phase transitions (''criticality'') to
loose the memory of the lattice and recover $\mathcal{P}$-covariance and
locality. This approach always has a ``light'' start since the mathematical
control of lattice systems is rather simple. But in the last step, the
investigation of criticality and the execution of the scaling limit, one has
to pay heavily for the easy life at the beginning. The mere control of
existence via the various lattice inequalities is not enough, the last step
requires a deep structural understanding of the lattice theory. Whereas it
is true that most of the QFT concepts as conserved charges, particles,
multiparticle scattering, antiparticles etc. can be transferred to the
lattice (albeit with much more sweat, since the helpful causality structure
is absent), a sufficiently detailed structural control is only possible
under special circumstances as integrability (meaning the Yang-Baxter
structure for 2-dim. lattice systems, as in the transition from the discrete
Ising model to the continuous Ising QFT). This kind of temporary practical
restriction is quite different from the above restriction through lack of
Feynman-Kac representability. In particular there is no limitation on the
short-distance behavior: the operator short-distance dimensions of e.g. the
Ising-, Thirring-, RSOS- etc. models in the scaling limit is too far away
from canonicity as to permit a euclidean F-K representation. Real time short
distance singularities which go significantly beyond canonical behavior do
not endanger the existence of real time QFT, but only limit certain methods
as quantizations by functional integrals. We do not really pursue a lattice
approach and refer the interested reader can find details on this subject in
a later section. Our main constructive contribution (presented only after
the chapter on algebraic QFT) will be based on the net approach.

\section{Scattering Theory}

Whereas scattering theory in e.g. Schr\"{o}dinger QM is very important for
the comparison of theory with experiments but less so for the formulation
and construction of quantum mechanical models, the S-operator takes on a
more fundamental significance in local quantum physics. The reason is
threefold: in addition to its standard role of permitting experimental
verification of the theory, S is an invariant of the net (i.e. S is attached
to a Borchers class and should not be affiliated with individual fields) and
finally S is related to the modular reflection $J$ for the wedge algebra and
the TCP-operator $\theta $ by $S=JJ_{0}=\theta \theta _{0}$ where the
subscript zero refers to the incoming fields (considered as a free theory).
In this section we will present the scattering content and the class
invariance property of S. The S-matrix adds an important aspect to the
particle-field dichotomy. Whereas the renormalization in the sense of
physical parametrization required the understanding of the relation of
one-particle properties and fields (or local observables in the algebraic
approach) the S-matrix deepens this connection by resolving continuum states
into multiparticle scattering states. With other words, all the important
aspects of the interpretation of the QFT formalism are determined by the
basic causality and spectral concepts of the theory and nowhere does one
have to add prescriptions from the outside. The completeness property
seperates QFT from other theoretical attempts about fundamental physics as
e.g. string theory.

In the perturbative approach we already met the S-matrix as the adiabatic
limit of $S(g)$. But we also realized that from a conceptual point of view
such limits should be avoided since that formalism is good for the local net
properties, but becomes unnatural for the calculation of ``on-shell''
quantities, in particular for the scattering operator. The conceptually most
satisfying method is to first calculate (the approximations for) the
correlation function and then to use the scattering theory for on-shell
quantities. Similar to the nonrelativistic theory, the main objective is to
use the time dependent formulation because of its physical clarity, and to
convert its content into analytically simple stationary formulas.

This aim is accomplished in the Lehmann-Symanzik-Zimmermann (LSZ) approach.
As quantum mechanical time dependent scattering theory relates interacting
wave functions for $t\rightarrow \infty $ with those of a free system,
scattering theory in QFT should relate interacting (Heisenberg) fields with
free fields. By checking with stationary external source models as well as
with renormalized perturbation theory these authors proposed the following
asymptotic condition (for the standard scalar situation): 
\begin{eqnarray}
lim_{t\rightarrow \pm \infty }\left\langle \phi \left| A_{f}(t)\right| \psi
\right\rangle &=&\left\langle \phi \left| A_{f}^{ex}\right| \psi
\right\rangle \quad ex=out,\,in  \label{as} \\
A_{f}(t) &=&\int_{x^{0}=t}f(x)\stackrel{\leftrightarrow }{\partial }%
_{0}A(x)d^{3}x  \nonumber
\end{eqnarray}
Here $f(x)$ is a solution of the Klein-Gordon equation, $A_{f}^{ex}$ is
defined by the same formula with $A$ replaced by the free incoming or
outgoing field (and therefore time-independent) and the state vectors $\phi
,\psi $ are taken from a dense set of in states (with nonoverlapping wave
functions in velocity space, as we know nowadays). Later the Haag-Ruelle
formulation, which is based on strong convergence, was derived from the
locality and spectral principles of QFT and it was shown that (\ref{as})
follows. But before we discuss these refinements, we will derive the useful
LSZ reduction formulas from (\ref{as}).

Let us start with the reduction of an incoming particle in the following
matrix-element: 
\begin{eqnarray}
&&^{out}\left\langle f_{n+1}....f_{n+m}\left| A(x)\right|
f_{1}....f_{n}\right\rangle ^{in} \\
&=&lim_{t\rightarrow -\infty }^{out}\left\langle f_{n+1}....f_{n+m}\left|
A(x)A_{f}(t)\right| f_{2}....f_{n}\right\rangle ^{in}  \nonumber \\
&=&\left\langle f_{n+1}....f_{n+m}\left| \int
K_{y}TA(x)A(y)f(y)d^{4}y\right| f_{2}....f_{n}\right\rangle ^{in}+c.t. 
\nonumber
\end{eqnarray}
where $T$ denotes the time ordering, $K$ is the Klein-Gordon operator and $%
c.t.$ (contraction terms) is the generic notation for terms in which $%
f^{\prime }$s in the in or out states have been contracted with resulting $%
(f_{i},f_{j})\times $lower terms (example: the annihilation part of $%
A_{f}^{in}$ may contract with $f_{i}$ in the in state if the overlap is
nonvanishing). In the third term the time ordering occurs since we want the
outgoing boundary contribution in: $lim_{t\rightarrow -\infty
}\{A_{f}(t)-A_{f}(-t)\}=$ $c.t.$ to appear on the left hand side of the
local operators whose matrix elements we are reducing (then its contribution
just produces outgoing contraction terms). The same statements apply
verbatim to the reduction of outgoing states. The iterative application of
this procedure therefore leads to the following reduction formula: 
\begin{eqnarray}
&&^{out}\left\langle f_{n+1}....f_{n+m}\left| A(x)\right|
f_{1}....f_{n}\right\rangle ^{in} \\
&=&\int ....\int \bar{f}_{n+1}(y_{n+1})...\bar{f}%
_{n+m}(y_{n+m})f_{1}(x_{1})...f_{n}(y_{n})\times  \nonumber \\
&&K_{n+1}...K_{n+m}K_{1}...K_{n}\left\langle 0\left|
TA(x)A(y_{1})....A(y_{n+m})\right| 0\right\rangle  \nonumber
\end{eqnarray}
Instead of $A(x)$ we could have also started with any multilocal product of
local fields. In the special case of $A\rightarrow \mathbf{1}$ we obtain the
reduction formula for the S-matrix: 
\begin{eqnarray}
&&^{out}\left\langle f_{n+1}....f_{n+m}\mid f_{1}....f_{n}\right\rangle ^{in}
\\
&&\int ....\int \bar{f}_{n+1}(y_{n+1})...\bar{f}%
_{n+m}(y_{n+m})f_{1}(x_{1})...f_{n}(y_{n})\times  \nonumber \\
&&K_{n+1}...K_{n+m}K_{1}...K_{n}\left\langle 0\left|
TA(y_{1})....A(y_{n+m})\right| 0\right\rangle +c.t.  \nonumber
\end{eqnarray}
By going to the limit of plane waves one obtains for the connected part of
the momentum space kernel of the S-matrix: 
\begin{eqnarray}
S(p_{n+1}..p_{n+m};p_{1}..p_{n})^{conn.} &=&lim_{p_{i}^{2}\rightarrow
m^{2}}\prod_{i}(p_{i}^{2}-m^{2})\times \\
&&\tau (-p_{n+1}..-p_{n+m},p_{1}..p_{n})  \nonumber
\end{eqnarray}
i.e. we obtain the residua on mass shell of the Fourier transforms of the
time ordered function $\tau .$ These reduction formulas are very suggestive
of the so called crossing symmetry: 
\begin{equation}
\text{incoming particle }p\text{ }\rightarrow \text{ outgoing antiparticle }%
-p
\end{equation}
for the generalized formfactores of local fields: 
\begin{eqnarray*}
&&^{out}\left\langle p_{1}...p_{n}\left| \mathcal{O}(0)\right|
p_{n+1}...p_{n+m}\right\rangle ^{in} \\
&=&^{out}\left\langle p_{1}...p_{n},-p_{n+1}\left| \mathcal{O}(0)\right|
p_{n+2}...p_{n+m}\right\rangle ^{in}
\end{eqnarray*}
where $-p_{n+1}$ stands for the momentum of the antiparticle on the backward
mass shell which by on-shell analytic continuation is related to the process
with the antiparticle momentum on the forward mass shell. With other words
such a ``symmetry'', in order to be physically meaningful, must be
interpretable as a relation between different boundary values of an on-shell
meromorphic ``master'' function. Although in renormalized perturbation
theory this turned out to be true in each checked case, a proof of the
necessary analyticity derived from the principles of QFT is only known in
special cases i.e. the reduction formula is only suggestive but does not
establish the crossing symmetry. A related question is the existence of
time-ordered functions outside perturbation theory. According to the best of
my knowledge, this has not been demonstrated in the general setting of QFT%
\footnote{%
In view of the fact that in the bootstrap-formfactor construction the time
ordering plays no role, it would be unreasonable to postulate its existence.}%
. A closer look at the derivation of the reduction formula reveals that a
pointlike covariant time ordering is not needed; any asymptotic ordering
will lead to the same on-shell values i.e. the residua on mass shell are
independent on the precise ordering prescription for finite space time
separations. In a later section we will see that time ordered fields is not
a natural concept in nonperturbative QFT. The more natural objects turn out
to be certain sesquilinear forms of the fields, the so called ``generalized
formfactors''.

In the following we will derive the Haag-Ruelle scattering theory in the
general setting of QFT and then comment in the derivation of the LSZ theory.

In $n$-particle Schr\"{o}dinger theory, the physical input for the existence
of scattering state vectors as large time limits of suitably chosen time
dependent vectors is the strong fall-off property of the two body potential.
Although one can somewhat relax those properties, potentials as the Coulomb
potential fall-off too weakly in order to belong to the standard situation
(the large time wave functions oscillate with a logarithmic factor which
does not contribute to the probabilities). In QFT the corresponding property
is the strong cluster property of correlation functions in spacelike
directions. A sufficient condition for this property is the existence of a
spectral gap in the mass operator.

An operator from the polynomial $\mathcal{P}$ algebra (see section 2 of this
chapter): 
\begin{eqnarray}
Q &=&\sum_{n}\int
f_{n}(x_{1},....,x_{n})A(x_{n})....A(x_{n})d^{d}x_{1}....d^{d}x_{n} \\
f_{n} &\in &\mathcal{S}^{4d}  \nonumber
\end{eqnarray}
will be called ``almost local'' (if $suppf_{n}\in O,$ $O$ is local). We will
be interested in the behavior of correlation functions of $%
Q(x):=U(x)QU(x)^{-1}.$ The relevant theorem is

\begin{theorem}
(Ruelle,1962, \cite{Haag}) In a local QFT with a spectral mass gap (isolated
one-particle mass shells) the quasilocal operators fulfill the strong
cluster property: 
\begin{equation}
\forall N\in \mathbf{N,}\,\,\exists \,C\,\,s.t.\,\,\left\langle
Q_{1}(x_{1})....Q_{n}(x_{n})\right\rangle _{con}<C_{N}R^{-N}
\end{equation}
Here R denotes the maximal space like distance: 
\begin{equation}
R=\max_{i,k}-(x_{i}-x_{k})^{2}
\end{equation}
\end{theorem}

We will not prove the theorem, but rather try to understand how it can be
used in order to understand the convergence for large times and the
structure of the incoming and outgoing multi-particle states. We first pick $%
Q_{i}s$ which applied to the vacuum create one-particle states with given
wave function $\tilde{\varphi}(p).$ By choice of $f_{n}\in \mathcal{S}^{nd}$
with appropriate energy-momentum support this is always possible. Then we
form the operators: 
\begin{equation}
Q_{i}(h_{i};t):=i\int_{x_{0}=t}Q_{i}(x)\stackrel{\longleftrightarrow }{%
\partial _{x_{0}}}h_{i}(x)d^{3}x
\end{equation}
where $h_{i}$ is a positive energy solution of the Klein-Gordon equation and
the derivative act with a minus sign to the left. Clearly: 
\begin{equation}
Q_{i}(h_{i};t)\Omega =\left| \psi _{i}\right\rangle ,\quad \tilde{\psi}%
_{i}(p)=\tilde{\varphi}(p)\tilde{h}_{i}(p)
\end{equation}
i.e. one obtains time independent one-particle states. On the other hand the
multiple application (at least two) of these operators leads to time
dependent states whose large time behavior is controlled by the following
theorem:

\begin{theorem}
(Haag 1958, \cite{Haag})
\end{theorem}

(i) \thinspace \thinspace \thinspace \thinspace The sequence of state
vectors 
\begin{equation}
\Psi (t)=\prod_{i}^{n}Q_{i}(h_{i};t)\Omega
\end{equation}
converge strongly for $t\rightarrow \pm \infty .$ The limiting states have
the physical interpretation of incoming and outgoing multiparticle
scattering states: 
\begin{eqnarray}
\Psi ^{in} &=&\lim_{t\rightarrow -\infty }\Psi (t)=\left| \psi _{1},....\psi
_{n}\right\rangle ^{in} \\
\Psi ^{out} &=&\lim_{t\rightarrow +\infty }\Psi (t)=\left| \psi
_{1},....\psi _{n}\right\rangle ^{out}  \nonumber
\end{eqnarray}

(ii)\thinspace \thinspace \thinspace \thinspace \thinspace The scalar
product of these scattering states has the Fock space structure: 
\begin{equation}
_{out}^{in}\left\langle \psi _{1}^{\prime }....\psi _{n}^{\prime }\mid \psi
_{1}....\psi _{m}\right\rangle _{out}^{in}=\delta _{nm}\sum_{P\in
S_{n}}\left\{ 
\begin{array}{l}
+ \\ 
sign(P)
\end{array}
\right\} \prod_{k}\left\langle \psi _{P(k)}^{\prime }\mid \psi
_{k}\right\rangle
\end{equation}
according to bosonic or fermionic spacelike behavior of the Heisenberg
fields $A(x).$ One should add that the Poincar\'{e} transformation act
naturally on the asymptotic Fock space structure, i.e. the in and out states
do not remember the special frame in which the time direction was defined.

The idea of the proof consists in showing that $\Psi (t)$ is a Cauchy
sequence i.e. that $\left\| \frac{d}{dt}\Psi (t)\right\| <Ct^{-\frac{3}{2}}.$
What one needs in addition to the cluster properties of the $Q(x)-$%
correlations is a refined asymptotic estimate on the single particle wave
functions which goes beyond the result of the well-known stationary phase
method: 
\begin{eqnarray}
h(\mathbf{x},t) &=&\frac{1}{(2\pi )^{\frac{3}{2}}}\int \tilde{h}(\mathbf{p}%
)e^{-i(\omega (\mathbf{p})t-\mathbf{px})}\stackunder{t\rightarrow \infty }{%
\rightarrow }  \label{stat.} \\
&=&const.t^{-\frac{3}{2}}\exp (-im\gamma ^{-1}t)(\gamma ^{\frac{3}{2}}\tilde{%
h}(m\gamma \mathbf{v})+O(t^{-1}))\quad  \nonumber \\
\gamma &=&\frac{1}{\sqrt{1-\mathbf{v}^{2}}},\quad \mathbf{v}=\frac{\mathbf{x}%
}{t}  \nonumber
\end{eqnarray}
The refined version determines the ``essential'' x-space support of h in
terms of the velocity support in momentum space $\Sigma =\left\{ \mathbf{v}=%
\frac{\mathbf{p}}{\omega }\mid p\in supp\tilde{h}\right\} .$ one has:

\begin{theorem}
(Ruelle 62) Let h be a positive energy solution of the Klein-Gordon equation
and $\Sigma $ its velocity support. With $\mathcal{U}$ an open set
containing $\Sigma $ we have:

(i)\thinspace \thinspace \thinspace \thinspace for v$\in \mathcal{U}:\left|
h(\mathbf{v}t,t)\right| <C\left| t\right| ^{-\frac{3}{2}}\quad as\ref{stat.}$

(ii)\thinspace \thinspace \thinspace \thinspace for v$\notin U:\left| h(%
\mathbf{v}t,t)\right| <C_{N}(1+\left| \mathbf{v}\right| )^{-N}\left|
t\right| ^{-N}$
\end{theorem}

If we now choose one-particle wave functions $h^{\prime }$ with
nonoverlapping velocity supports relative to the unprimed $h$ then 
\begin{eqnarray*}
&&\left\langle \Omega \left| Q_{1}(h_{1}^{\prime
};t)^{*}....Q_{m}(h_{m}^{\prime };t)^{*}Q(h_{n};t)....Q(h_{1};t)\right|
\right\rangle \\
&&\stackunder{t\rightarrow \infty }{\rightarrow }\delta _{nm}\left\{
\sum_{P\in S_{n}}\left\{ 
\begin{array}{l}
+ \\ 
sign(P)
\end{array}
\right\} \prod_{k=1}^{n}\left\langle \psi _{P(k)}^{\prime }\mid \psi
_{k}\right\rangle \right\} ,\quad 
\begin{array}{l}
bosonic \\ 
fermionic
\end{array}
\end{eqnarray*}
The connected part, upon integration with the dissipating wave packets, does
not contribute at all to the limit, as follows from the elementary
geometrical (essential) support pictures in Minkowski space.. The same holds
for any connected cluster with more than two operators $Q_{i}.$ this fixes
the structure of the in/out scalar products. The fall-off of $\left\| \frac{d%
}{dt}\Psi (t)\right\| ^{2}$ is even simpler to understand, because each term
which contributes to this norm square for large t contains one two-point
function where one operator is a time derivative of $Q(h;,t)$ which vanishes
upon acting on the vacuum.

The restriction to nonoverlapping wave packets has a physical origin:
parallel flying particles lead to a weaker convergence. The best strategy is
to prove formulas for the nonoverlapping situation and only at the end take
the plane wave limit. The formalism does not only allow to derive the LSZ
theory and the reduction formulas, but also gives higher order t corrections
to LSZ (\cite{Haag}, chapter II section 4).

The above scattering formalism needs to be modified in an essential and
interesting way if the fields have a spacelike commutation structure which
leads to braid group statistics. In the physically interesting case of d=2+1
dimensions, these ``plektonic'' fields have really a string-like spatial
extension i.e. they are not fields in the sense of Lagrangian QFT. Their
construction falls into the realm of general or algebraic QFT. One still can
prove the asymptotic convergence, but the asymptotic state vectors loose
their tensor product structure and the cut between kinematics (in/out
structure) and dynamics (genuine interactions) has to be essentially
modified. The fact that such theories are outside the Lagrangian framework
and even outside quantization ideas, does not make them any less physical or
susceptible to explicit and perturbative constructions, but the perturbation
around free ``plektons'' is expected to have more in common with ideas on
perturbing around chiral conformal theories than with Feynman perturbation
theory around bosonic/fermionic free fields. the structure of $d=2+1$
plektons will be investigated in the last chapter.

\textbf{Literature to chapter 5:}.

R.F. Streater and A.S. Wightman, ``PCT, Spin and Statistics and All That''
New York, Benjamin 1964

R.Jost ``The General Theory of Quantized Fields'', American Math. Soc. 1965

N.N.Bogoliubov, A.A.Logunov, A.I.Oksak and I.T.Todorov ``General Principles
of Quantum Field Theory'' Kluwer 1990

J. Glimm and A. Jaffe, ``Quantum physics. A functional integral point of
view''. Springer 1987

R. Haag ``Local Quantum Physics, Fields, Particles, Algebras'' Springer 1992

\chapter{Nonperturbative Constructions}

\section{Introductory Comments}

Presently QFT appears as being formed of several parts which seem to drift
apart into different directions. On the one hand there is the standard
approach presented in the previous chapters which is centered around
renormalized perturbation theory and the various quantization methods
(canonical, functional, the causal Bogoliubov-Shirkov approach). Enriched
with geometrical ideas the standard formulation based on Lagrangians and
actions has prepared the ground for many mathematical advances including the
duality structure of Seiberg-Witten as well as to string theory. On the
other hand there is the more algebra-based low dimensional approach which
has led to the construction of rich families of chiral conformal and
factorizing QFT without Lagrangians. This constructive approach, although
being somewhat conservative in its use of physical principles, has
nevertheless given many startling nonperturbative results of particle
physics concerning e.g. fusion charges, in particular of antiparticles from
particles, confined objects and solitons as being two opposite sides of the
same subject, and other more general (and somewhat surprising)
manifestations of the principle of ``nuclear democracy''. The third approach
to QFT which forms the backbone of these notes, was carried out by a rather
small group of theoretical physicist with a strong mathematical background
on operator algebras without restrictions of space-time dimensionality and
to integrability, with the aim to arrive at a general constructive
nonperturbative approach. This will be of main concern in this chapter.

The most interesting message of the low dimensional constructive
bootstrap-formfactor program is that the emphasis on the scattering matrix
advocated way back by Heisenberg and Wheeler and later by Chew, Stapp and
others, was in a certain sense physically well founded. What went wrong in
those early ``bootstrap'' attempts is mostly related to the enforced and
artificial separation of $S$ from local QFT and the (unfortunately
cyclically recurrent) working hypothesis of a ``TOE'', a theory of
everything (in this case of everything minus gravity). Looking at the old
articles it is hard to understand the fervor with which the $S$-matrix
concepts were cleansed from all field theoretic ``impurities''.

The main theme of this chapter is the realization that the S-matrix in
algebraic QFT acquires a new, hitherto unknown pivotal role in the
construction of local nets (whose generators are local fields). In this way
it is regaining some of its importance before the gauge theory began to
shift the emphasis from on-shell to off-shell properties. It turns out that
the S-matrix belongs to the foundation of the local field theory (in its
role as \textit{the} net invariant which carries local modular information)
as well as to its roof (in its role as describing scattering observables); a
truly vexing ``bootstrap'' situation. The fact that in d=1+1 factorizable
theories Chew's bootstrap ideas for the S-matrix work without mentioning
fields (but with the help of ``fusion'' and ``Yang-Baxter'') is not due to
the correctness of the underlying philosophy but rather to undeserved luck:
the physical rapidity (scattering) variable is at the same time the
uniformization parameter of the analytic properties\footnote{%
Even in d=1+1 the situation is very far removed from the desired uniqueness
of Chew's S-matrix approach.}. In higher dimensions or without the
factorization, Chew's program would fail without the use of local fields
(and it did fail). In that case one could hope hat an iterative procedure
which corrects the trial input S-matrix together with a locality improvement
of states and fields, may have a constructive chance, a situation which
could be more vaguely reminiscent of the Hartree-Fock iteration in
Schr\"{o}dinger theory than of renormalized perturbation theory.

In this section we will apply the modular localization introduced in
chapter3 to interacting theories \cite{sch}. In this way we will reconquer
the lost unity in QFT. In particular, we will learn a new and very
interesting lessons from the d=1+1 formfactor program. Far from being a
special ``exotic'' construction, remote from any ``real'' QFT, this
approach, if analyzed with general and deep concepts related to the TCP
theorem and the S-matrix (interpreted as an invariant of a local net),
reveals a surprising new and powerful nonperturbative construction principle
which, so we hope, may turn out to be the basis of a future new iterative
constructive approach in d=1+3 theories.

Locality of observables and localization of states (always relative to the
vacuum or some other distinguished reference state) in QFT is a conditio
sine qua non for its physical interpretation. Global topology (as in the
combinatorial or so called ``topological field'' theories, or in the vacuum
degeneracy structure beyond spontaneous symmetry), remains part of
mathematics, as long as its connection with the local structure remains
unclear.

The fastest way to get a glimpse at the ``modular localization'' \cite{sch}
is to recall that concept in connection with the Wigner representation
theory for positive energy representations of the Poincar\'{e} group and
free fields, as it was explained in chapter 3. There we learned that there
are infinitely many free fields in Fock-space and they constitute the linear
part (in creation and annihilation operators) of a huge local equivalence
class of fields, the so called Borchers class $\mathcal{B}(m,s)$ \cite{Haag} 
\cite{wightman} Any cyclic (with respect to the vacuum) representative field
from this class generates the same net of local von Neumann algebras in $%
H_{F}$: 
\begin{equation}
\mathcal{O}\rightarrow \mathcal{A(O)}
\end{equation}
In fact the emerging picture of pointlike fields that behave similar to
coordinates in differential geometry, was the prime historical motivation
for formulating algebraic QFT in terms of nets of algebras \cite{Haag}. For
a detailed recent presentation of the physical motivations and aims of this
net approach in algebraic QFT we refer to \cite{S2}$.$

In the following it is important to understand the\textit{\ direct}
construction of this net in terms of the ``modular localization'' principle.
Let us briefly review what we learned in chapter 3. We use the d=3+1 Wigner
(m,s)-representations as an illustrative example. In case of charged
particles (particles$\neq $antiparticles) we double the Wigner
representation space: 
\begin{equation}
H=H_{Wig}^{p}\oplus H_{Wig}^{\bar{p}}
\end{equation}
in order to incorporate the charge conjugation operation as an (antilinear
in the Wigner theory) operator involving the p-\={p}-flip. On this extended
Wigner space one can act with the full Poincar\'{e} group (where those
reflections which change the direction of time are antiunitarily
represented). For the modular localization in a wedge we only need the
standard L-boost $\Lambda (\chi )$ and the standard reflection $r$ which (by
definition) are associated with the $t$-$x$ wedge: 
\begin{equation}
\delta ^{i\tau }\equiv \pi _{Wig}(\Lambda (\chi =2\pi \tau ))
\end{equation}
\begin{equation}
j\equiv \pi _{Wig}(r)
\end{equation}
These operators have a simple action on the p-space (possibly) doubled
Wigner wave functions, in particular: 
\begin{equation}
(j\psi )(p)\simeq \left( 
\begin{array}{ll}
0 & -1 \\ 
1 & 0
\end{array}
\right) \bar{\psi}(p_{0},p_{1},-p_{2},-p_{3})
\end{equation}
By functional calculus we form $\delta ^{\frac{1}{2}}$ and define: 
\begin{equation}
s\equiv j\delta ^{\frac{1}{2}}
\end{equation}
This unbounded antilinear densely defined operator $s$ is involutive on its
domain: $s^{2}=1.$ Its -1 eigenspace\footnote{%
This is one of the few places where a sign mistake has no grave consequences
since a multiplication by $i$ transforms the $+$ into the $-$ real
eigenspace.} is a real closed subspace $H_{R}$ of $H$ which allows the
following characterization of the domain of $s:$%
\begin{eqnarray}
dom(s) &=&H_{R}+iH_{R} \\
s(h_{1}+ih_{2}) &=&-h_{1}+ih_{2}  \nonumber
\end{eqnarray}
Defining: 
\begin{equation}
H_{R}(W)\equiv U(g)H_{R},\,\,\,W=gW_{stand}
\end{equation}
where g is an appropriate Poincar\'{e} transformation, we find the following
theorem:

\begin{theorem}
$H_{R}(W)$ is a net of real Hilbert spaces i.e. $H_{R}(W_{1})\subsetneq
H_{R}(W_{2})$ if $W_{1}\subsetneq W_{2}$.
\end{theorem}

The proper containment is an easy consequence of a theorem by Borchers \cite
{Bo} which relates this property to the positivity of the energy (in fact
the geometric property is equivalent to the energy positivity.

If we now define: 
\begin{equation}
H_{R}(\mathcal{O})\equiv \bigcap_{W\supset \mathcal{O}}H_{R}(W)
\end{equation}
then it is easily seen (even without the use of the u,v-intertwiners) that
the spaces $H_{R}(\mathcal{O})+iH_{R}(\mathcal{O})$ are still dense in $%
H_{Wig}$ and that the formula: 
\begin{equation}
s(\mathcal{O})(h_{1}+ih_{2})\equiv -h_{1}+ih_{2}
\end{equation}
defines a closed involutive operator with a polar decomposition: 
\begin{equation}
s(\mathcal{O})=j(\mathcal{O})\delta (\mathcal{O})^{\frac{1}{2}}
\end{equation}
Although now $j(O)$ and $\delta (\mathcal{O})^{i\tau }$ have no obvious
geometric interpretation, there is still a bit of geometry left, as the
following theorem shows:

\begin{theorem}
The $H_{R}(\mathcal{O})$ form an orthocomplemented net of closed real
Hilbert spaces, i.e. the following ''duality'' holds: 
\begin{equation}
H_{R}(\mathcal{O}^{\prime })=H_{R}(\mathcal{O})^{\prime }=iH_{R}^{\bot }(%
\mathcal{O}).
\end{equation}
\end{theorem}

Here $\mathcal{O}^{\prime }$ denotes the causal complement, $H_{R}^{\bot }$
the real orthogonal complement in the sense of the inner product $Re\left(
\psi ,\varphi \right) $ and $H_{R}^{\prime }$ is the symplectic complement
in the sense of $\func{Im}\left( \psi ,\varphi \right) .$

The direct construction of the interaction-free algebraic bosonic net for
(m,s=integer) is now achieved by converting the ''premodular'' theory of
real subspaces of the Wigner space into the Tomita-Takesaki modular theory
for nets of von Neumann algebras using the Weyl functor:

The application of the Weyl functor $\mathcal{\Gamma }$ to the net of real
spaces: 
\begin{equation}
\mathcal{\Gamma }:H_{R}(\mathcal{O})\longrightarrow \mathcal{A}(\mathcal{O}%
)\equiv alg\left\{ W(f)\left| f\in H_{R}(\mathcal{O})\right. \right\}
\end{equation}
leads to a net of von Neumann algebras in $\mathcal{H}_{Fock}\,$which are in
``standard position'' with respect to the vacuum state with a modular theory
which, if restricted to the Fock vacuum $\Omega ,$ is geometric: 
\begin{eqnarray}
\Gamma (s) &=&S,\,\,\,SA\Omega =A^{*}\Omega ,\,\,\,A\in \mathcal{A}(W) \\
S &=&J\Delta ^{\frac{1}{2}},\,\,\,J=\Gamma (j),\,\,\,\Delta ^{i\tau }=\Gamma
(\delta ^{i\tau })  \nonumber
\end{eqnarray}
The proof of this theorem uses the functorial formalism of section 2.6. It
should be evident from the derivation that the wedge localization concept in
Fock obtained in this functorial way from the Wigner theory only holds for
interaction free situations. The Fock space is also important for
interacting QFT, but in that case the wedge localization enters via
scattering theory as in section 4, and not through Wigner's representation
theory.

Clearly the $W$- or $\mathcal{O}$- indexing of the Hilbert spaces
corresponds to a localization concept via modular theory. Specifically $%
H_{R}(\mathcal{O})+iH_{R}(\mathcal{O})$ is a certain closure of the one
particle component of the Reeh-Schlieder domain belonging to the
localization region $\mathcal{O}.$ Although for general localization region
the modular operators are not geometric, there is one remaining geometric
statement which presents itself in the form of an algebraic duality property 
\cite{Oster Leyland}: 
\begin{equation}
\mathcal{A}(\mathcal{O}^{\prime })=\mathcal{A}(\mathcal{O})^{\prime
},\,\,\,\,Haag\,\,Duality
\end{equation}
Here the prime on the von Neumann algebra has the standard meaning of
commutant. In the following we make some schematic additions and completions
which highlight the modular localization concept for more general free cases 
\cite{hep}.

In the case of $s=$ halfinteger, the Wigner theory produces a mismatch
between the ``quantum'' ( in the sense of the commutant) and the
``geometric'' opposite of $H_{R}(W),$ which however is easily taken care of
by an additional factor $i$ (interchange of symplectic complement with real
orthogonal complement). This requires, via the physical localization
property, the application of the CAR-functor instead of the CCR-functor as
well as the introduction of the well-known Klein transformation $K$ which
takes care of the above mismatch in Fock space: 
\begin{eqnarray}
J &=&K\mathcal{F}_{CAR}(ij)K^{-1} \\
\mathcal{A}(\mathcal{O}^{\prime }) &=&K\mathcal{A}(\mathcal{O})^{\prime
}K^{-1}  \nonumber
\end{eqnarray}
where the $K$ is the twist operator of the ``twisted''\thinspace \thinspace
Haag\thinspace \thinspace \thinspace Duality.

In section 3.8 we exposed the thermal properties of this modular
localization by constructing via the KMS property a representation which is
only defined in the thermal Hilbert space belonging to the wedge
localization. We also commented on the relation between these thermal
properties and the crossing symmetry of particle physics. This famous
crossing symmetry, which is known to hold also in each perturbative order of
renormalizable interacting theories, has never been derived in sufficient
generality within any nonperturbative framework of QFT. It was thought of as
a kind of on-shell momentum space substitute for Einstein causality and
locality (and its strengthened form called Haag duality). As such it played
an important role in finding a candidate for a nonperturbative S-matrix, an
important contribution known under the name of the Veneziano dual model.
Although it stood (in this indirect way) on the cradle of string theory, the
recent string theoretic inventions of duality result from formal
generalizations which appearantly have little do with the original physical
concepts of nonperturbative relatiistic scattering theory.

If crossing symmetry is a general property of local QFT, a conjecture
(proven in every order of perturbation theory) which nobody seems to doubt,
then it should be the on-shell manifestation of the off-shell KMS property
for modular wedge localization. In the construction of wedge localized
thermal KMS states on the algebra of mass shell operators satisfying the
Zamolodchikov-Faddeev algebraic relations\footnote{%
As will become clear in the next section, although these operators are
nonlocal, they generate the wedge localized states and as a consequence the
modular KMS formalism is applicable to them.} on the momentum space rapidity
axis \cite{Zam}, the derivation of crossing symmetry is similar (albeit more
involved) to the previous free field derivation \cite{S1}\cite{hep} and the
argument can be found in section 4 of this chapter. Very recently arguments
were proposed \cite{Nieder} which were bases on the idea that Haag-Ruelle
scattering theory can be adapted to the wedge situation. I think that such
ideas are untainable. In our approach it turns out that the general crossing
symmetry (in any dimension) is indeed related to the wedge KMS condition but
that this statement cannot be derived just from scattering theory alone. It
rather follows from the existence and the modular intertwining property of
the ``modular M\o ller operator'' $U$ (next section) which, unlike the
S-matrix, is not an object of scattering theory but is only defined in terms
of modular wedge localization (for this reason we maintain the prefix
``modular''). We intend to use this object in order to prove \cite{SWi} the
uniqueness of \textit{the main inverse problem} of QFT: $S_{scat}\rightarrow
QFT.$

In fact the previous free field formalism of the first two sections may be
generalized into two directions (or into both):

\begin{itemize}
\item  interacting fields

\item  curved space-time (without or with interacting quantum matter)
\end{itemize}

As mentioned before, low-dimensional interacting theories will be discussed
in the next section. For the generalization to curved space time (e.g. the
Schwarzschild black hole solution) it turns out that only the existence of a
bifurcated horizon together with a certain behavior near that horizon
(``surface gravitation'') \cite{Sew}\cite{S V} is already sufficient in
order to obtain the thermal Hawking aspect. In the standard treatment one
needs isometries in space-time, i.e. Killing vectors. In chapter3 we have
already presented the thermal aspects with the wedge situation. Here we
address a more general situation. The idea of modular localization suggests
to consider also e.g. double cones for which there is no space-time isometry
but only an isometry in $H_{Wigner}$ or $\mathcal{H}_{Fock}.$ Of course such
enlargements of spaces for obtaining a better formulation (or even a
solution) of a problem are a commonplace in modern mathematics, particularly
in noncommutative geometry. The idea here is that one trades the ill-defined
Killing isometries by a geometrically ``fuzzy'' but well-defined symmetry
transformations in quantum space, which only near the horizon loose their
space-time fuzziness. The candidates for these nongeometric symmetries are
the modular automorphisms of von Neumann algebras of arbitrary space time
regions together with suitable faithful states from the local folium of
admissable states. Although the restriction of the global vacuum state is
always in that folium, it is often not the most convenient for the
construction of the modular automorphism.

In this context one obtains a good illustration by the (nongeometric)
modular theory of e.g. the double cone algebra of a massive free field. From
the folium of states one may want to select that vector, with respect to
which the algebra has a least fuzzy (most geometric) behavior under the
action of the modular group. Appealing to the net subtended by spheres at
time t=0 one realizes that algebras localized in these spheres are
independent of the mass. Since m=0 leads to a geometric modular situation%
\footnote{%
The modular group is a one-parametric subgroup of the conformal group.} for
the pair ($\mathcal{A}_{m=0}(S),\left| 0\right\rangle _{m=0},$ and since the
nonlocality of the massive theory in the subtended double cones is only the
result of the fuzzy propagation inside the light cone (the breakdown of
Huygens principle or the ``reverberation'' phenomenon), the fuzziness of the
modular group for the pair ($\mathcal{A}_{m\neq 0}(C(S)),$ $\Omega _{m\neq
0})$ is a pure propagation phenomenon i.e. can be understood in terms of the
deviation from Huygens principle. In view of the recent micro-local spectrum
condition, one expects such nonlocal cases to have modular groups whose
generators are pseudo-differential instead of (local) differential operators 
\cite{Bru}. In fact for free massive theories in d=1+1 one can give a
rigorous proof together with an explicit formula for the modular
automorphism. In order to avoid the pathology of the d=1+1 scalar zero mass
field, we start with a massive free spinor field whose massless limit gives
a two-component field with the first component only depending on the left
light cone and the second on the right hand light cone. The same zero mass
theory results from Sewell's restriction to the light cone horizon
(boundary) of the double cone. But in the latter case we know the modular
group for the massless double cone algebra together with the massless vacuum
vector. It is a one-parametric subgroup of the conformal group \cite{Haag}.
The massless theory on the horizon is then propagated inside with the
massive causal propagator and this last step is responsible for the
delocalization inside the double cone. This and similar subjects will be the
content of a separate paper together with Wiesbrock.\cite{SW}.

The Hilbert space setting of modular localization offers also a deeper
physical understanding of the universal field domain $\mathcal{D}$ which
plays a rather technical role in the Wightman framework \cite{wightman} In
the modular localization approach the necessity for such a domain appears if
one wants to come from the net of localization spaces which receive their
natural topology from the (graphs) net of Tomita operators $\bar{S}(\mathcal{%
O})$ to a net of (unbounded) polynomial algebras $\mathcal{P}(\mathcal{O})$
such that: 
\begin{equation}
dom\,\,\bar{S}(\mathcal{O})\cap \mathcal{D}=\mathcal{P}(\mathcal{O})\Omega
=dom\,\,\mathcal{P}(\mathcal{O})
\end{equation}
this domain is of course also expected to be equal to $\mathcal{A}(\mathcal{O%
})\Omega .$ Here we used a more precise notation which distinguishes between
the operator $S$ defined on the core $\mathcal{A}(0)\Omega $ and its closure 
$\bar{S}$ which is defined on $\mathcal{H}_{R}(\mathcal{O})+i\mathcal{H}_{R}(%
\mathcal{O}).$ One may round off these new interpretations of old domain
problems by Fredenhagen's speculative remark about the modular role of
pointlike fields. This idea is based on modular observations in the
algebraic approach to chiral conformal field theory \cite{Fre Joe}. There it
is possible to extract the pointlike covariant fields without additional
technical assumptions directly from the net. This, together with the known
modular structure of the algebras in the net gives a beautiful
characterization of these fields: the ``one field states'' obtained by
applying a field to the vacuum and smearing with test functions form a
representation space for the universal modular group. The latter is defined
as the group generated by all modular groups belonging to arbitrary double
cones.

\section{Modular Localization and Interaction}

In order to obtain a clue of how to incorporate an intrinsic notion of
interactions into this modular localization setting, we remind ourselves
that if we do use pointlike fields, the modular localization for free fields
agrees with what we get by applying the polynomial in the localization
region supported smeared fields. In contrast to the conventional
characterization of localization in terms of x-space pointlike fields, the
modular characterization works in the \textit{momentum-(Fock)space} of the
(incoming) \textit{free particles}. It attributes a physical significance to
the precise position of the Reeh-Schlieder \cite{Haag} dense set of
localized vectors and the change of this position resulting from the change
of localization region, i.e. it primarily deals with subspaces and
subalgebras and only in second place with individual vectors and operators.

In order to formulate the modular localization principle in the case of
interactions, one must take note of the fact that the scattering matrix S of
local QFT is the product of the interacting TCP-operator $\Theta $ with the
free (incoming) TCP operator $\Theta _{0}$ and (since the rotation by which
the Tomita reflection $J$ differs from $\Theta $ is interaction-independent
as all connected Poincar\'{e} transformations are interaction-independent)
we have: 
\begin{equation}
S=\Theta \cdot \Theta _{0},\quad S=J\cdot J_{0}  \label{mo}
\end{equation}
and as a result we obtain for the Tomita involution $\check{S}\,\,$: 
\begin{equation}
\check{S}=J\Delta ^{\frac{1}{2}}=SJ_{0}\Delta ^{\frac{1}{2}}=S\check{S}_{0}
\end{equation}
Again we may use covariance in order to obtain $\check{S}(W)$ and the
localization domain of $\check{S}(W)$ as $\mathcal{D}(\check{S}(W))=\mathcal{%
H}_{R}(W)+i\mathcal{H}_{R}(W)$ i.e. in terms of a net of closed real
subspaces $H_{R}(W)$ of the incoming Fock space. However now the
construction of an associated von Neumann algebra is not clear since an
``interacting'' functor from subspaces of the Fock space to von Neumann
algebras is not known. In fact whereas the existence of a functor from the
net of localized \textit{Wigner subspaces} $H_{R}^{Wig}$ to a net of von
Neumann algebras is equivalent to the equality: 
\begin{equation}
H_{R}^{Wig}(W_{1}\cap W_{2})=H_{R}^{Wig}(W_{1})\cap H_{R}^{Wig}(W_{2})
\end{equation}
The equality becomes an inequality $\subset $ for the above localized
subspaces of \textit{Fock space }$\mathcal{H}_{R}$. It also turns into an
inequality for Wigner subspaces \textit{as soon as the Wigner spin becomes
anyonic or Wigner's continuous spin} \cite{Mund} as in section \textbf{3}.7.
We will make some remarks (still short of a solution of this important
problem) in the concluding section of this chapter and continue here with
some more helpful comments on modular localization of interacting state
vectors.

As in the free case, the modular wedge localization does not use the full
Einstein causality but only the so-called ``weak locality'', which is just a
reformulation of the TCP invariance \cite{wightman} Weakly local fields form
an equivalence class which is much bigger than the local Borchers class but
they are still associated to the same S-matrix (or rather the same TCP
operator). Actually the $S$ in local quantum physics has two different
interpretations: S in its role to provide modular localization in
interacting theories, and S with the standard scattering interpretation in
terms of (nonlocal!) large time limits. There is \textit{no parallel} 
\textit{outside local quantum physics} to this state of affairs. Whereas all
concepts and properties which have been used hitherto in standard QFT
(perturbation theory, canonical formalism and path integrals) as e.g. time
ordering \footnote{%
There is a conspicuous absence of the time-ordering operation in the
bootstrap construction of factorizable field models. Instead the basic
objects are generalized formfactors i.e. sesquilinear forms on a dense set
of state vectors.} and interaction picture formalism, are shared by
nonrelativistic theories, modular localization is a new structural element
in local quantum physics \footnote{%
This characteristic modular structure lifts local quantum physics to a new
realm by its own which cannot be obtained by specialization from general
quantum theory.} and its characteristic property. No physically viable
alternative (i.e. physically interpretable) to Einstein causality has ever
been found in the long history of QFT.

The simplest kind of interacting theories are those in which the particle
number is at least asymptotically (''on-shell'') conserved i.e. $\left[ 
\check{S},\mathbf{N}_{in}\right] =0.$

In the next section we will briefly review the d=1+1 bootstrap-formfactor
program in a manner which facilitates the later application of modular
localization.

\section{The Bootstrap-Formfactor Program}

In this section we will meet a constructive approach for ``integrable''
d=1+1 QFT. Our first task is to obtain an intrinsic QFT understanding of
integrability in a way which avoids classically inherited notions such as a
complete sets of conservation laws etc. For this purpose we note an
important d=1+1 peculiarity. Our generic expectation is that large spatial
separation of the center of wave packet of two particles in the elastic
two-particle scattering matrix leads to the weakening of scattering, or in
momentum space: 
\begin{equation}
\left\langle p_{1}^{^{\prime }}p_{2}^{^{\prime }}\left| S\right|
p_{1}p_{2}\right\rangle =\left\langle p_{1}^{^{\prime }}p_{2}^{^{\prime }}%
\mathbf{\mid }p_{1}p_{2}\right\rangle +\delta (p_{1}+p_{2}-p_{1}^{^{\prime
}}-p_{2}^{^{\prime }})T(p_{1}p_{2}p_{1}^{^{\prime }}p_{2}^{^{\prime }})
\end{equation}
where the identity contribution is more singular (has more $\delta $%
-factors) than the interaction $T$-term and therefore the second term drops
out in x-space clustering. This argument fails precisely in d=1+1 and
therefore the cluster property of the S-matrix is not suitable in order to
obtain an intrinsic understanding of interaction. In the large distance
clustering process, the two-particle S-matrix looses its higher particle
threshold structure, but it remains nontrivial (in distinction to d=3+1).
However for all higher particle scattering processes the behavior for d=1+1
is qualitatively the same as in higher dimensions: the decreasing threshold
singularities (which decrease with increasing particle number) are
responsible for the spatial decrease. Therefore any d=1+1 QFT is expected to
have a limiting S$_{\lim }$-matrix which is purely elastic and solely
determined by the elastic two-particle $S^{(2)}$-matrix. The Yang-Baxter
relation results as a consistency relation for the elastic 3$\rightarrow $3
particle $S_{\lim }^{(3)}$-matrix. If this limiting S-matrix would again
correspond to a localizable QFT, we would have a new class division of QFT,
this time based on a long distance limit (which in some sense is opposite to
the scale invariant short distance limit). It is this \textit{(long
distance) class property \footnote{%
Although I know of no article in which this has been spelled out, its
pervasive presence behind the scene is is recognizable in some publications.}
which makes these factorizing models so fascinating}, as much as the
fascination of chiral conformal QFT results from their role of representing
short distance universality classes. In d=3+1 $S_{\lim }=1$ and therefore
the limiting theory is expected to maintain the same superselection rules
but in the ``interaction-freest'' possible way (literally free theories as
we will argue later on). Hence in d=1+1 we are invited to speculate on the
validity of the following commutative diagram: 
\begin{equation}
\mathcal{F}\,_{\searrow }^{\nearrow } 
\begin{array}{l}
\mathcal{F}_{ld} \\ 
\stackrel{\downarrow }{\mathcal{F}}_{sd}
\end{array}
\end{equation}
Here $ld(sd)$ labels the long (short) distance limits. There are also
arguments \cite{Zam} that with the help of a perturbative idea one may
ascend from $\mathcal{F}_{sd}$ to $\mathcal{F}_{ld}.$ It is however
presently not clear how one can use the known properties of the $ld$
theories (i.e. the integrable models) in order to formulate a constructive
program for the nonintegrable members of the $ld$ equivalence class. We hope
that our modular localization principle (which is not restricted to
factorizable models) may turn out to be helpful for this purpose.

The constructive approach based on the bootstrap idea proceeds in two steps.
One first classifies unitary and crossing symmetric solutions of the
Yang-Baxter equations which fulfill certain minimal (or maximal, depending
on the viewpoint) requirements. Afterwards we use these factorizing
S-matrices together with the Watson equations (a notion from scattering
theory relating formfactors with the S-matrix) and analytic properties for
formfactors in order to compute the latter. One obtains the complete set of
multi-particle matrix elements of ``would be'' local fields, i.e. one
constructs the fields as sesquilinear forms. It is characteristic of this
method, that one does not use the ``axiomatic'' properties of the beginning
of this section, but rather less rigorously known momentum-space analytic
properties which, although certainly related to causality and spectral
properties, are more part of the LSZ+dispersion theoretic induced folklore
(example: crossing symmetry) than of rigorous QFT. As long as one
demonstrates at the end that the so obtained fields fulfill local
commutativity, this is a legitimate procedure\cite{KW}\cite{Smir}. It leaves
open the question of whether there exists a more direct conceptual link
between the S-matrix and the local fields or rather the field independent
local nets. That this is indeed the case will be shown after the
presentation of the formfactor program.

\subsection{Properties of Factorizing S-Matrices}

Consider first the analytic structure of an elastic S-matrix for a scalar
neutral particle. In terms of the rapidity variable $\theta $: 
\begin{equation}
\left| p_{1,}p_{2}\right\rangle ^{out}=S\left| p_{1},p_{2}\right\rangle
^{in}=S_{el}(p_{1},p_{2})\left| p_{1},p_{2}\right\rangle ^{in}
\end{equation}
\begin{eqnarray}
S_{el}(p_{1},p_{2}) &=&:S(\theta ),\quad p_{i}=m(cosh\theta _{i},sinh\theta
_{i}),\quad \theta :=\left| \theta _{1}-\theta _{2}\right| \\
^{in}\left\langle p_{1}^{\prime },p_{2}^{\prime }\left| S\right|
p_{1},p_{2}\right\rangle ^{in} &=&S(\theta )^{in}\left\langle p_{1}^{\prime
},p_{2}^{\prime }\mid p_{1},p_{2}\right\rangle ^{in}  \nonumber
\end{eqnarray}
Usually the elastic S-matrix is written in terms of the invariant energy $%
s=(p_{1}+p_{2})^{2}=2m^{2}(1+cosh\theta )$ and the momentum transfer (not
independent in d=1+1) $t=(p_{1}-p_{2})^{2}=$.$2m^{2}(1-cosh\theta ).$ As a
result of undeserved fortune, the rapidity $\theta $ turns out to be a
uniformization variable for the real analytic S i.e. the complex s-plane
with the elastic cut in $s\geq 4m^{2}$ is dumped into the strip $0\leq
Im\theta \leq \pi $ and the S-matrix becomes a meromorphic function $%
S(\theta )$ with $S(-\theta )=S^{*}(\theta )=S^{-1}(\theta ).$ (unitarity).
Hence the strip $-\pi \leq \theta \leq \pi $ is the physical strip for $%
S(\theta )$. Crossing symmetry in our special (neutral) case means a
symmetry on the boundary of the strip: $\theta \rightarrow i\pi -\theta $.
Note that the presence of inelastic thresholds would destroy the
uniformization.

The factorization implies the operator relation: 
\begin{eqnarray}
&&S_{12}(p_{1},p_{2})S_{13}(p_{1},p_{3})S_{23}(p_{2},p_{3}) \\
&=&S_{23}(p_{2},p_{3})S_{13}(p_{1},p_{3})S_{12}(p_{1},p_{2})  \nonumber
\end{eqnarray}
According to Liouville's theorem, the only minimal solution (minimal number
of poles,smallest increase at $\infty $) for this scalar diagonal case is $%
S=\pm 1.$ More general solutions are obtained by placing bound-state poles
into the minimal solution. In order to maintain unitarity, the pole factor
must be of the form: 
\begin{equation}
P(\theta )=\frac{sinh\theta +isin\lambda }{sinh\theta -isin\lambda }
\end{equation}
Transforming back this pole at $\theta =i\lambda $ into the original
individual particle variables, we obtain the following parametrization in
terms of a center of mass and relative rapidity: 
\begin{eqnarray}
p_{1} &=&m\left( cosh(\chi +\frac{i\lambda }{2}),sinh(\chi +\frac{i\lambda }{%
2})\right) \\
p_{2} &=&m\left( cosh(\chi -\frac{i\lambda }{2}),sinh(\chi -\frac{i\lambda }{%
2})\right)  \nonumber
\end{eqnarray}
Clearly the two-particle bound state has the momentum: 
\begin{eqnarray}
p_{1,2} &=&\left( p_{1}+p_{2}\right) _{at\,bd.state}=2mcos\frac{\lambda }{2}%
\left( cosh\chi ,sinh\chi \right) \\
p_{1,2}^{2} &=&m_{2}^{2},\quad m_{2}=\frac{m}{2sin\frac{\lambda }{2}}%
sin\lambda  \nonumber
\end{eqnarray}
The ``fusion'' of particles may be extended. For a 3-particle bound state we
would look at the 3-particle S-matrix which, as a result of factorization
has the form: 
\begin{equation}
S^{(3)}(p_{1},p_{2},p_{3})=S(\theta _{12})S(\theta _{13})S(\theta _{23})
\end{equation}
We first fuse 1with 2 and simultaneously 2 with 3 as before. The center of
mass + relative rapidity parametrization yields: 
\begin{eqnarray}
p_{1} &=&m\left( cosh(\chi +i\lambda ),sinh(\chi +i\lambda )\right) \\
p_{2} &=&m\left( cosh\chi ,sinh\chi \right)  \nonumber \\
p_{3} &=&m\left( cosh(\chi -i\lambda ),sinh(\chi -i\lambda )\right) 
\nonumber
\end{eqnarray}
Again we get the mass of the 3-particle bound state by adding the zero
components in the $\chi =0$ frame: 
\begin{equation}
m_{3}=(p_{1}+p_{2}+p_{3})_{0}=m_{2}cos\frac{\lambda }{2}+mcos\lambda =2\frac{%
m}{2sin\frac{\lambda }{2}}sin\frac{3\lambda }{2}
\end{equation}
Induction then gives the general fusion mass formula: 
\begin{equation}
m_{n}=2\mu sin\frac{n\lambda }{2},\quad \mu =\frac{m}{2sin\frac{\lambda }{2}}
\end{equation}
We will meet such trigonometric fusion formulas later in algebraic QFT where
they are related to the statistical dimensions of fused charge sectors. They
became first known through the Dashen-Hasslacher-Neveu quasiclassical
approach. The above fusion calculation was done as far back as 1976 \cite
{STW} and consisted in a synthesis of the quasiclassical work of DHN with
some suggetive ideas of Sushko using the factorization principle, but still
without the ideas of Yang and Baxter (which are not needed for this scalar
case). The decisive step towards a general factorizable bootstrap program
was taken two years later \cite{Ka}\cite{Za}$.$

The consistency of these particles as incoming and outgoing objects leads to
additional structures. Consider the scattering of the mass $m_{2}$ bound
state with a third m-particle. This S-matrix for the scattering of these two
different particles is obtained from $S^{(3)}$ by: 
\begin{equation}
S_{b.e.}(p_{1}+p_{2},p_{3})\mid _{(p_{1}+p_{2})^{2}=m_{2}^{2}}=\frac{1}{R}%
\stackunder{(p_{1}+p_{2})^{2}\rightarrow m_{2}^{2}}{Res}S_{12}S_{13}S_{23}
\end{equation}
where the projector $P_{12}$ together with a numerical residue value $R$ is
defined by: 
\begin{equation}
\stackunder{(p_{1}+p_{2})^{2}\rightarrow m_{2}^{2}}{Res}%
S(p_{1},p_{2})=RP_{12}
\end{equation}
and we used the subscript elementary $e$.and bound $b.$ as labels on the new
two-particle scattering operator: $S_{b.c.}.$ The factorization insures
that: 
\begin{equation}
P_{12}S_{13}S_{23}=S_{23}S_{13}P_{12}
\end{equation}
A prominent family of scalar S-matrices with $N-1$ bound state fulfilling
all these requirements are the $Z_{N}$ models \cite{Swieca}. Consistency
requires that the bound state of $N-1$ mass $m$-particles is again a mass $m$%
-particle (the antiparticle). For N=2 this family contains the Ising field
theory with $S_{Ising}^{(2)}=-1$ which we already met in the section on
(dis)order variables.

Instead of elaborating this scalar factorization situation, we pass
immediatly to the matrix case where we meet a new and interesting
phenomenon. We assume that the particle from which we start has an internal
``charge'' which can take on a finite number of values i.e. 
\begin{equation}
\left| p,\alpha \right\rangle \in H_{1}\otimes V,\quad dimV<\infty
\end{equation}
The two-particle S-matrix is then written as a matrix acting on $V\otimes V$
whose entries are operator-valued (represented as in the previous case by
momentum-space kernels): 
\begin{equation}
S\left| p_{1},....,p_{n}\right\rangle _{\alpha _{1}....\alpha
_{n}}^{in}=\left| p_{1},....,p_{n}\right\rangle _{\alpha _{1}^{\prime
}....\alpha _{n}^{\prime }}^{in}S_{\alpha _{1}....\alpha _{n}}^{\alpha
_{1}^{\prime }....\alpha _{n}^{\prime }}(p_{1},....p_{n})
\end{equation}
\begin{equation}
S^{(n)}(p_{1},....,p_{n})=\prod_{i<j}S^{(2)}(p_{i},p_{j})
\end{equation}
The factorization requires a specific order of the product of matrices.
Consistency demands the validity of the Yang-Baxter relation: 
\begin{eqnarray}
&&S_{12}(p_{1},p_{2})S_{13}(p_{1},p_{3})S_{23}(p_{2},p_{3}) \\
&=&S_{23}(p_{2},p_{3})S_{13}(p_{1},p_{3})S_{12}(p_{1},p_{2})  \nonumber
\end{eqnarray}
a relation which resembles the Artin (braid-group) relation, but upon closer
inspection reveals conceptual and mathematical differences to the latter.
The notation should be obvious: the subscript on S indicates on which of the
tensor factors in the 3-fold tensor product of one-particle spaces the
object acts. The relation with the Artin relations would be clear, if one
could ignore the p-dependence and rewrites the Y-B relation in terms of $%
\tilde{S}.=PS$, where $P$ is the permutation of two tensor factors.

We arrived via our $S_{\lim }$ arguments at the famous Yang-Baxter relation,
since as a result of the (independent) discovery by Yang and Baxter it was
clear, that such a mathematical structure had appeared before outside QFT in
a quite different setting of statistical mechanics models and
nonrelativistic scattering theory of $\delta $-function interactions. In our
context this identity permits to change the temporal order of individual
re-scatterings so that the n-particle scattering $S^{(n)}$ is independent of
those (graphically: invariance under parallel shifts of 2-momenta in
graphical illustrations of scattering processes). The problem of finding the
natural parametrization (e.g. Baxter's elliptic parametrization) for these
Yang-Baxter relations does not arise in QFT; the \textit{uniformizing
physical rapidity }$\theta $\textit{\ is already the natural Yang-Baxter
variable: } 
\begin{equation}
S_{12}(\theta )S_{13}(\theta +\theta ^{\prime })S_{23}(\theta ^{\prime
})=S_{23}(\theta ^{\prime })S_{13}(\theta +\theta ^{\prime })S_{12}(\theta )
\end{equation}
If fermion-antifermion pairs can go into boson-antiboson pairs, the object
which fulfills the Yang-Baxter relation is not $S$ but $\sigma S$ where $%
\sigma =\pm 1$ with + for bosons.

As the braid group relation, this is an overdetermined system of equations.
For the former one found a powerful mathematical framework within V. Jones
subfactor theory \cite{Jones}. Although the attempts to get an equally
powerful mathematical framework for the latter was less than successful (the
``Baxterization'' of the subfactor representations of Artin braids), one was
able to find many interesting families of nontrivial solutions of which some
even allowed a comparison with Lagrangian perturbation theory.

The S-matrix bootstrap idea originated in the early 60$^{ies}$ from
dispersion theory. Its revival in connection with d=1+1 factorization in the
late 70$^{ies}$ showed that its premises were physically reasonable, except
the idea that it could be seen as a ``theory of everything'' (TOE) which was
wrong and even absurd (for the more recent TOE's one would be hard pressed
to say friendly words about their physical content).

The basic new message \cite{Smir}\cite{BFK} is that one should use these
factorizing S-matrices as computational tools for the construction of local
fields and local nets as explained in the following subsection

\subsection{Generalized Formfactors}

Now we will probe the idea that these S-matrices belong to localizable
fields. Let A be any local field which belongs to a Borchers equivalence
class of local fields. We write the generalized formfactor of $A(x)$ as: 
\begin{equation}
_{\alpha _{1}....\alpha _{m}}{}^{out}\left\langle p_{1},....,p_{m}\left|
A(0)\right| p_{m+1},....,p_{n}\right\rangle _{\alpha _{m+1}....\alpha
_{n}}^{in}  \label{mixed}
\end{equation}

We are interested in its analytic p-space properties. ``On shell'' p-space
analytic properties are more elusive than x-space analytic properties. For
the latter the spectral support properties play the important role, whereas
p-space analyticity relies heavily on causality. The above matrix element
still contains energy-momentum $\delta $-functions resulting from
contracting incoming p's with outgoing. These are removed by taking the
connected parts of the formfactors. Only for the distinguished formfactor: 
\begin{equation}
\left\langle 0\left| A(0)\right| p_{1},....p_{n}\right\rangle _{\alpha
_{1}....\alpha _{n}}^{in}=\left\langle 0\left| A(0)\right|
p_{1},....p_{n}\right\rangle _{\alpha _{1}....\alpha _{n}}^{in,\,con}
\end{equation}
we have coalescence with its connected part. Similar to x-space analyticity,
one expects the existence of one analytic master-function whose different
boundary values correspond to the different n-particle formfactors: 
\begin{eqnarray}
&&_{\alpha _{1}....\alpha _{m}}{}^{out}\left\langle p_{1},....,p_{m}\left|
A(0)\right| p_{m+1},....,p_{n}\right\rangle _{\alpha _{m+1}....\alpha
_{n}}^{in,con} \\
&=&F_{\underline{\alpha }}^{A}(s_{ij}+i\varepsilon ,t_{rs}-i\varepsilon
,s_{kl}+i\varepsilon ),\quad i<j\leq m<k<l\leq n  \nonumber \\
t_{rs} &=&\left( p_{r}-p_{s}\right) ^{2},\quad r\leq m<s\leq n  \nonumber
\end{eqnarray}
There are Watson relations between the S-matrix and the formfactors. In the
d=1+3 dispersion theory setting it is well known that the cuts below the
inelastic threshold of $\left\langle 0\left| A(0)\right|
p_{1},p_{2}\right\rangle $ is related to the partial wave phase shifts in
that elastic region. In a factorizing d=1+1 theory these Watson relations
can be written down in general: 
\begin{eqnarray}
F_{\alpha _{1}....\alpha _{n}}^{A}(s_{ij}+i\varepsilon ) &=&\left\langle
0\left| A(0)\right| p_{1},....p_{n}\right\rangle _{\alpha _{1}....\alpha
_{n}}^{in} \\
&=&\sum_{out}\left\langle 0\left| A(0)\right| out\right\rangle \left\langle
out\mid p_{1},....,p_{n}\right\rangle _{\alpha _{1}....\alpha _{n}}^{in} 
\nonumber
\end{eqnarray}
\begin{equation}
\curvearrowright F_{\alpha _{1}....\alpha _{n}}^{A}(s_{ij}+i\varepsilon
)=F_{\alpha _{1}^{\prime }....\alpha _{n}^{\prime }}^{A}(s_{ij}-i\varepsilon
)S_{\alpha _{1}....\alpha _{n}}^{\alpha _{1}^{\prime }....\alpha
_{n}^{\prime }}(s_{ij})
\end{equation}
and for the mixed formfactors(\ref{mixed}): 
\begin{eqnarray}
&&F_{\underline{\alpha }}^{A}(s_{ij}+i\varepsilon ,t_{rs}-i\varepsilon
,s_{kl}+i\varepsilon ) \\
&=&S_{\alpha _{1}^{\prime }....\alpha _{m}^{\prime }}^{\alpha _{m}....\alpha
_{1}}(s_{ij})F_{\underline{\alpha }^{\prime }}^{A}(s_{ij}-i\varepsilon
,t_{rs}+i\varepsilon ,s_{kl}-i\varepsilon )S_{\alpha _{m+1}....\alpha
_{n}}^{\alpha _{n}^{\prime }....\alpha _{m+1}^{\prime }}(s_{kl})  \nonumber
\end{eqnarray}
Using the uniformazing $\theta ^{\prime }s,$ this is like a generalized
quasi periodicity property on $\theta $-strips for the F's (instead of the
periodicity of S). The first who considered formfactors beyond two-particles 
\cite{KW} and presented a system of axioms for their calculation was Smirnov 
\cite{Smir} Following a recent presentation by Babujian, Fring and Karowski 
\cite{BFK} in a more standard field theoretic setting (LSZ+ dispersion
theory), the formfactor program for the construction of d=1+1 QFT is as
follows. Introduce the ordered formfactors: 
\begin{equation}
f_{\underline{\alpha }}^{A}(\theta _{1},....,\theta _{n}):=\left\langle
0\left| A(0)\right| p_{1},....,p_{n}\right\rangle _{\underline{\alpha }%
}^{in},\quad \theta _{1}>...>\theta _{n}  \label{def}
\end{equation}
and define the value for reordered $\theta ^{\prime }s$ by analytic
continuation (starting with this ordering in the physical region). Demand
that the f's fulfill the following properties (``axioms''):

\begin{itemize}
\item  (i)$\quad f_{...ij...}^{A}(...,\theta _{i},\theta
_{j},...)=f_{...ji...}^{A}(...,\theta _{j},\theta _{i},...)S_{ij}(\theta
_{i}-\theta _{j})\quad \forall \,\,\theta ^{\prime }s$

\item  (ii)\quad $f_{12...n}^{A}(\theta _{1}+i\pi ,\theta _{2},...,\theta
_{n})=f_{2...n1}^{A}(\theta _{2},...,\theta _{n},\theta _{1}-i\pi )$

\item  (iii)\quad $f_{1...n}^{A}(\theta _{1},...\theta _{n})\stackunder{%
\theta _{1}\rightarrow \theta _{2}+i\pi }{\approx }\frac{2i}{\theta
_{1}-\theta _{2}-i\pi }C_{12}f_{3...n}^{A}(\theta _{3},...,\theta
_{n})(1-S_{2n}...S_{23})$
\end{itemize}

where $C_{\alpha \beta }=\delta _{\bar{\alpha}\beta }$ is the charge
conjugation matrix.

Here we have not mentioned the poles from bound states (states which appear
by the previous fusion) since they are automatically entering the
formfactors via the S-matrix. The word ``axiom'' in the context of this
paper has the significance of working hypothesis i.e. an assumption which
receives its legitimation through its constructive success. Physical
principles on the other hand, as the spectral and causality properties of
general QFT, will not be called axioms. Our main aim is to show how one can
reduce the above axioms of the bootstrap-formfactor approach to the
principles of QFT, and thereby recuperate the unity of this nonperturbative
approach with the rest of QFT.

The conceptually somewhat unusual property is the ``symmetry'' property (i).
Here one should bear in mind that from the point of view of the LSZ
formulation f is an auxiliary object to which the statistics property under
particle exchange does not apply (it would apply to the original
matrix-element). The above exchange property for $f$ is a statement about
analytic continuation. The statistics of incoming particle is only used in
order to get the charges (i.e. the tensor factors) into the same $j-i$ order
as the analytically interchanged $\theta ^{\prime }s.$ Following BFK \cite
{BFK} let us first remind ourselves of the standard argument for (i) in
somewhat detail. For the special case $\left\langle 0\left| A(0)\right|
\theta _{1}\theta _{2}\right\rangle ^{ex}$ $ex=in,out$ it is evident that: 
\begin{equation}
lim_{\varepsilon \rightarrow 0}F(s_{12}\pm i\varepsilon )=\left\{ 
\begin{array}{l}
\left\langle 0\left| A(0)\right| \theta _{1}\theta _{2}\right\rangle ^{in}
\\ 
\left\langle 0\left| A(0)\right| \theta _{1}\theta _{2}\right\rangle ^{out}
\end{array}
\right.
\end{equation}
i.e. there is one analytic masterfunction $f(z)$ (assuming identical
particles) with different boundary values on the $s\geq 4m^{2}$ cut, having
the $in,out$ interpretation. Assuming Bose statistics, the physical matrix
elements on the right hand side are symmetric under the interchange of the $%
\theta ^{\prime }s.$ In terms of the uniformization variable $\theta _{12}$
in $F$ the transition from $in\rightarrow out$ means a change of sign via
analytic continuation i.e. without changing the charge quantum numbers $%
\alpha $ i.e. the position of the tensor factors. After accomplishing this
last step by the bose commutation relation, the negative $\theta _{12}$
formfactor $F(\theta _{21})$ can according to the definition (\ref{def}) be
identified with $f_{21}(\theta _{2},\theta _{1})$ and the relation 
\begin{equation}
\left\langle 0\left| A(0)\right| \theta _{2}\theta _{1}\right\rangle
^{out}S(\left| \theta _{1}-\theta _{2}\right| )=\left\langle 0\left|
A(0)\right| \theta _{1}\theta _{2}\right\rangle ^{in}
\end{equation}
agrees with (\ref{def}). The generalization to $^{out}\left\langle \theta
_{3}...\theta _{n}\left| A(0)\right| \theta _{1}\theta _{2}\right\rangle
_{con}^{ex}$ has a problem, because replacing in by out means passing from
time-ordering to anti-time-ordering but the LSZ scattering theory produces
boundary terms contributing to the connected part. Although they are absent
for theories in which the number of in-particles are conserved (on-shell
conservation), it is unclear what property of general QFT is bringing about
(i) through specialization to factorizing d=1+1 models.

On the other hand (ii) and (iii) are consequences of the following standard
crossing formula \cite{BFK} which relate the connected part of the
generalized formfactors to the analytic master function $f$: 
\begin{eqnarray}
&&_{\bar{1}}\left\langle 0\left| A(0)\right| p_{2},...,p_{n}\right\rangle
_{2...n}^{in} \\
&=&\sum_{j=2}^{n}\,_{\bar{1}}\left\langle p_{1}\mid p_{j}\right\rangle
_{j}f_{2..\hat{j}..n}^{A}S_{2j}...S_{j-1j}+f_{12...n}^{A}(\theta _{1}+i\pi
_{-}...,\theta _{n})  \nonumber \\
&=&\sum_{j=2}^{n}\,_{\bar{1}}\left\langle p_{1}\mid p_{j}\right\rangle
_{j}f_{2..\hat{j}..n}^{A}S_{jn}...S_{jj+1}+f_{2...n1}^{A}(...,\theta
_{n}\theta _{1}-i\pi _{-})  \nonumber
\end{eqnarray}
The fastest way to understand this is to draw the corresponding graphs and
remember that a positive energy particle crosses into a negative energy
antiparticle. Successive application leads to a formula which expresses the
formfactors in terms of the analytic auxiliary function $f$ .The analytic
part of this relation gives (ii) whereas the $\delta $-function part is
responsible for (iii). A argument that these properties do not only insure
TCP-invariance (weak locality) but also Einstein causality, can be given
formally\cite{Smir}. Apart from the Ising model, the formfactor program has
not been carried out to the end although all of the two particle formfactors
associated with the computed S-matrices are known.

It would be nice to have a direct derivation of all the bootstrap-formfactor
axioms from the principles of QFT but this is still an open problem. It is
part of the complicated and incomplete momentum space analyticity problem.
Even the derivation of forward dispersion relations in particle physics took
several years, not to mention the derivation of the analytic aspects 
\footnote{%
Only together with the (mass shell) analytic properties the crossing
symmetry aquires a physical content.} of crossing symmetry which still
remain quite incomplete. It is precisely at this point where our modular
localization approach shows its strength. To anticipate one result, it shows
that the crossing symmetry is a kind of strengthened TCP-property, and that
the cyclicity\cite{sch}\cite{Nieder} it leads to is identical to the
KMS-temperature ($\equiv $Hawking -Unruh temperature in this special case)
characterization of the (Rindler-)wedge based Hawking-Unruh effect. From our
point of view, the most valuable result is that it opens for the first time
the way to a new constructive iterative (but non perturbative) approach to 
\textit{non-quantization,} non-Lagrangian based QFT. My confidence that this
may amount to more than just another fashion rests on the observation that
the tool of modular localization comes from a refinement of TCP which, as
anybody will immediately admit, \textit{the central structure} of local QFT.

\subsection{ Modular Theory and the Formfactor Program}

There is one more important idea which is borrowed from scattering theory
namely the existence of a ``modular M\o ller operator'' \cite{hep} $U$ which
is related to the S-matrix as: 
\begin{equation}
S_{s}=UJ_{0}U^{*}J_{0}
\end{equation}
This correspond to the well-known standard formula $S_{s}=(\Omega
^{out})^{*}\Omega ^{in}.$ This analogy with a scattering interpretation has
to be taken with a grain of salt however. Note that the Haag-Ruelle
scattering theory (as well as its more formal but better known LSZ
predecessor) in local quantum physics does not provide a M\o ller isometry
between Heisenberg states and incoming states because the scattering state
space and the space for the interacting fields are identical (in local
quantum physics, different from the nonrelativistic rearrangement scattering
theory one does not introduce a separate space of fragments). The name M$\o $%
ller appears here only in a modular context.

The idea of introducing such an object into our modular approach comes from
the unitary equivalence of the interacting and the free hyperfinite type III$%
_{1}$ wedge algebras: 
\begin{equation}
\mathcal{A}(W)=U\mathcal{A}^{in}(W)U^{*}  \label{Mo}
\end{equation}
together with the $U$-invariance of the vacuum $U\Omega =\Omega .$ Note that
this situation is very different from the problem of equivalence of the
canonical equal time commutation relations in the free versus the
interacting case. A unitary equivalence in this case (of an algebra
belonging to a region with trivial spacelike complement) would be forbidden
by Haag's theorem on the nonexistence of the interaction picture in QFT. The
above characterization of $U$ may be replaced by a slightly more convenient
one in terms of an intertwining property between modular operators: 
\begin{equation}
US_{0}=SU  \label{Mol}
\end{equation}
Such a M\o ller operator cannot commute with the Poincar\'{e} group (apart
from the boost associated with the wedge). The latter serves to define a
family of $W$-affiliated $U^{\prime }s$ from the standard wedge. In terms of
localized spaces, the $U$ has the property: 
\begin{equation}
U\mathcal{H}_{R}^{in}(W)=\mathcal{H}_{R}(W)
\end{equation}
Let us first illustrate these concepts in an explicitly known (including
pointlike fields) model. Between the two possibilities Ising field theory
with $S_{s}^{(2)}=-1$ and the (non-parity invariant) Federbush model with $%
S_{I,II}^{(2)}=e^{i\pi g\varepsilon (\theta _{1}-\theta _{2})}$ we chose the
latter because it allows also a Lagrangian interpretation (i.e. more old
fashioned nostalgia than conceptual necessity). The model consists in
coupling two species of Dirac fermions via a (parity violating)
current-pseudocurrent coupling \cite{Wigh}\cite{SW}: 
\begin{equation}
\mathcal{L}_{int}=g:j_{\mu }^{I}j_{\nu }^{II}:\varepsilon ^{\mu \nu
},\,\,\,\,j_{\mu }=:\bar{\psi}\gamma _{\mu }\psi :
\end{equation}
One easily verifies that: 
\begin{eqnarray}
\psi _{I}(x) &=&\psi _{I}^{(0)}(x)\vdots e^{ig\Phi _{II}^{(l)}(x)}\vdots
\label{local} \\
\psi _{II}(x) &=&\psi _{II}^{(0)}(x)\vdots e^{ig\Phi _{I}^{(r)}(x)}\vdots 
\nonumber \\
\psi _{I,II}^{(0)}(x) &=&\frac{1}{\sqrt{2\pi }}\int \left( e^{-ipx}u(\theta
)a_{I,II}(\theta )+e^{ipx}v(\theta )b_{I,II}^{*}(\theta )\right) d\theta
\end{eqnarray}
where $\Phi ^{(l,r)}=\int_{x^{\prime }\lessgtr x}j_{0}dx^{\prime }$ is a
potential of $j_{\mu 5}$ i.e. $\partial _{\mu }\Phi \sim \varepsilon _{\mu
\nu }j^{\nu }=j_{\mu 5}$ and the superscript $l,r$ refers to whether we
choose the integration region for the line integral on the spacelike left or
right of $x$. For the form of the $u$ and $v$ spinors we refer to (\ref
{Dirac}). The triple ordering is needed in order to keep the closest
possible connection with classical geometry and localization and in
particular to maintain the validity of the field equation in the quantum
theory; for its meaning we refer to the above papers. This conceptually
simpler triple ordering can be recast into the form of the analytically
(computational) simpler standard Fermion Wick-ordering in terms of the
(anti)particle creation/annihilation operators $a_{I,II}^{\#}(\theta
),b_{I,II}^{\#}(\theta )$. 
\begin{eqnarray}
\psi _{I}(x) &=&\psi _{I}^{(0)}(x):e^{L_{g}(x)}: \\
L_{g}(x) &=&\frac{sin\pi g}{2\pi }\int \left\{ 
\begin{array}{c}
\frac{e^{-g(\theta _{p}-\theta _{q})}}{\cosh \frac{1}{2}(\theta _{p}-\theta
_{q})}\left[ e^{i(p+q)x}a_{II}^{*}(\theta _{p})b_{II}^{*}(\theta
_{q})+h.c.\right] + \\ 
\left[ -\frac{e^{-g(\theta _{p}-\theta _{q})}}{\sinh \frac{1}{2}(\theta
_{p}-\theta _{q}+i\varepsilon )}\right] e^{i(p-q)x-i\pi g}a_{II}^{*}(\theta
_{p})a_{II}(\theta _{q})+ \\ 
\left[ \frac{e^{-g(\theta _{p}-\theta _{q})}}{\sinh \frac{1}{2}(\theta
_{p}-\theta _{q}-i\varepsilon )}\right] e^{-i(p-q)+i\pi g}b_{II}^{*}(\theta
_{q})b_{II}(\theta _{p}).
\end{array}
\right\} d\theta _{p}d\theta _{q}  \label{charge}
\end{eqnarray}
and similar formula for $\psi _{II}.$ Although in this latter description
the classical (manifest) locality is lost, the quantum exponential do still
define local Fermi-fields\cite{SW}; in the case of relative commutation of $%
\psi _{I}$ with $\psi _{II}$ the contributions from the exponential
(disorder fields) compensate. This model belongs to the simplest class of
factorizing models (those with rapidity independent S-matrix) and its
explicite construction via the formfactor program is almost identical to
that of the massive Ising field theory \cite{Karowski}. The reason why it
does not appear under this approach in the literature is that the bootstrap
classification was limited to strictly parity conserving theories. For our
present purposes it serves as the simplest nontrivial illustration of new
concepts arising from modular localization.

Despite the involved looking local fields (\ref{local}), the wedge algebras
are easily shown of utmost simplicity: 
\begin{eqnarray}
\mathcal{A}(W) &=&alg\left\{ \psi _{I}^{(0)}(f)U_{II}(g),\psi
_{II}^{(0)}(h);suppf,h\in W\right\}  \label{alg} \\
\mathcal{A}(W^{\prime }) &=&\mathcal{A}(W)_{Klein}^{^{\prime }}=alg\left\{
\psi _{I}^{(0)}(f),\psi _{II}^{(0)}(h)U_{I}(g);suppf,h\in W^{\prime }\right\}
\nonumber
\end{eqnarray}
i.e. the two wedge-localized algebras (W denotes the right wedge) are
generated by free fields ``twisted'' by global $U(1)$-symmetry
transformation of angle $g$ (coupling constant)\footnote{%
The equality of the $\mathcal{A}(W)$ net (\ref{alg}) to the net obtained by
the subsequent modular method adapted to the Federbush model is not a very
easy matter.} and the subscript ``Klein'' denotes the well-known Klein
transformation associated with the $2\pi $ Fermion rotation. The right hand
side follows from the observation that with $x$ restricted to W, one may
replace the exponential in $\psi _{I}$ in (\ref{local}) (which represents a
left half space rotation) by the full rotation since the exponential of the
right halfspace charge is already contained in the right free fermion
algebra etc. The following unitarily equivent description of the pair $%
A(W),A(W^{\prime })$ has a more symmetric appearance under the \textit{%
extended parity} symmetry $\psi _{I}(t,x)\leftrightarrow \psi _{II}(t,-x)$:%
\vspace{0in} 
\begin{eqnarray}
\mathcal{A}(W) &=&alg\left\{ \psi _{I}^{(0)}(f)U_{II}(\frac{g}{2}),\psi
_{II}^{(0)}(h)U_{I}(-\frac{g}{2});suppf,h\in W\right\}  \label{rep} \\
\mathcal{A}(W^{\prime }) &=&alg\left\{ \psi _{I}^{(0)}(h)U_{II}(\frac{g}{2}%
),\psi _{II}^{(0)}(f)U_{I}(-\frac{g}{2});suppf,h\in W^{\prime }\right\} 
\nonumber
\end{eqnarray}
The computation \cite{SW} of the scattering matrix $S_{s}$ from (\ref{local}%
) is most conveniently done by Haag-Ruelle scattering theory \cite{Haag}: 
\begin{eqnarray}
S_{s}\left| \theta _{1}^{I},\theta _{2}^{II}\right\rangle
&=&S_{s}^{(2)}\left| \theta _{1}^{I},\theta _{2}^{II}\right\rangle =e^{i\pi
g\varepsilon (\theta _{1}-\theta _{2})}\left| \theta _{1}^{I},\theta
_{2}^{II}\right\rangle \\
S_{s}^{(n)} &=&\prod_{pairings}S_{s}^{(2)}  \nonumber
\end{eqnarray}
These formulae (including antiparticles) can be collected into an operator
expression \cite{SW} : 
\begin{equation}
S_{s}=\exp i\pi g\int \rho _{I}(\theta _{1})\rho _{II}(\theta
_{2})\varepsilon (\theta _{1}-\theta _{2})d\theta _{1}d\theta _{2}
\label{Fed}
\end{equation}
Where $\rho _{I,II}$ are the momentum space charge densities in the rapidity
parametrization.

The surprising simplicity of the wedge algebra as compared to say double
cone algebras consists in the fact that one can choose on-shell generators.
We will show that modular wedge localization for factorizing models always
leads to on-shell generators though for rapidity dependent S-matrices they
are less simple than (\ref{rep}).

It would now be easy to solve the n-particle modular localization equation%
\footnote{%
The Tomita operator $S$ for Fermions is different from that of Bosons by a
Klein transformation. For a special family of d=1+1 solitons the correct TCP
operator has been computed by Rehren\cite{Rehr 97}. Since all the known
families of factorizing models are described by Fermions and Bosons and
since it is not clear whether this generalization is compatible with the
factorization we will ignore this more general TCP-situation in the present
context.}: 
\begin{eqnarray}
S\mathcal{H}_{R}^{(n)}(W) &=&-\mathcal{H}_{R}^{(n)}(W)  \label{F} \\
\mathcal{H}_{R}(W) &=&\left\{ \int F(\theta _{1},\theta _{2},...,\theta
_{n})\left| \theta _{1},\theta _{2},...,\theta _{n}\right\rangle d\theta
_{1}d\theta _{2}...d\theta _{n}\mid F\in H_{strip}\right\}  \nonumber
\end{eqnarray}
Here $H_{strip}$ denotes the space of square integrable function which allow
an analytic continuation into simplices inside a multi-strip: 
\begin{eqnarray}
Imz_{i_{1}} &>&Imz_{i_{2}}>....>Imz_{i_{n}} \\
0 &<&Imz_{i}<\pi ,\,\,\,i=1...n  \nonumber
\end{eqnarray}
This is just the p-space on-shell analyticity which comes from the wedge
localization. As for Wightman functions in x-space, the $n!$ different
boundary prescriptions $Imz_{i_{1}}>Imz_{i_{2}}>....>Imz_{i_{n}}\rightarrow
0 $ yield the generally $n!$ different boundary values $F(\theta
_{i_{1}},\theta _{i_{2}},...\theta _{i_{n}}).$ Similar statements hold for
the boundary values on the upper rim. This boundary prescription (which for
Wightman functions is a consequence of the energy positivity) follows from
the analytic aspects of the KMS properties and the remark that the group
parameters of the modular automorphisms are the rapidities. In the free case
i.e. for $S_{s}=1$, there are no discontinuities (i.e. the F's just
incorporate the Fermi statistics) but with the Federbush S-matrix the space
consists of strip-analytic functions which are a solution of a
Riemann-Hilbert boundary problem, i.e. the Tomita eigenspace equation for $%
\mathcal{H}_{R}(W)$ relates the boundary values (products of the two
particle S-matrices) on the various boundaries obtained by placing each
single rapidity $z_{i}$ on the lower/upper edge of the strip in the various
ordered manners of the (real parts fo the) rapidities. The general solution
of this problem (i.e. the characterization of the subspace $\mathcal{H}%
_{R}(W)$ within the full multiparticle wave function space) may be presented
as a special solution of the Riemann-Hilbert problem convoluted with the
general solution of the interaction free problem in $\mathcal{H}%
_{R}(W)^{in}. $ A physically more enlightening way consists in writing the
localization subspace in a field theoretic manner as: 
\begin{eqnarray}
\int d^{2}x_{1}...d^{2}x_{n}f_{n}(x_{1},....x_{n})
&:&Z_{I,II}(x_{1})....Z_{I,II}(x_{n}):\Omega ,\,\, \\
\,\,suppf_{n} &\in &W^{\otimes n},\,\,f_{n}\,\,real  \nonumber
\end{eqnarray}
where the $Z^{\prime }s$ are on-shell operators whose frequency positive and
negative momentum space components have to fullfil commutation relations
which must be compatible with the boundary relations governed by products of
two particle S-matrices. On immediately realizes that this leads to the
Zamolodchikov-Faddeev algebra relations for the Federbush S-matrix: 
\begin{equation}
Z_{I,II}(x)=\frac{1}{\sqrt{2\pi }}\int \left( e^{-ipx}u(\theta
)c_{I,II}(\theta )+e^{ipx}v(\theta )d_{I,II}^{*}(\theta )\right) d\theta
\label{Z}
\end{equation}
where the $c$ and the corresponding anti $d$ can be formally expressed in
terms of the incoming (anti)particle creation and annihilation operators: 
\begin{eqnarray}
c_{I,II}(\theta ) &=&a_{I,II}(\theta )e^{-i\pi g\int_{-\infty }^{\theta
}\rho _{II,I}(\theta ^{\prime })d\theta ^{\prime }}  \label{au} \\
d_{I,II}(\theta ) &=&b_{I,II}(\theta )e^{i\pi g\int_{-\infty }^{\theta }\rho
_{II,I}(\theta ^{\prime })d\theta ^{\prime }}  \nonumber
\end{eqnarray}
with the Zamolodchikov-Faddeev relations\footnote{%
The natural appearance of the (rapidity-dependent) Yang-Baxter structure in
these on shell Z-F operator algebras contains the interesting mathematical
message that whereas the Artin braid group, which represents statistics of
plektons, can be naturally represented in combinatorical type $II_{1}$
algebras (``topological field theory''), the natural representation of the
Yang-Baxter structure requires ``bigger'' algebras related to scattering
theory with spacetime aspects. There seems to be no easy ``Baxterization''
od the V. Jones tracial representations of the infinite braid group.}: 
\begin{eqnarray}
c_{I,II}(\theta _{1})c_{II,I}(\theta _{2}) &=&-S^{(2)}(\theta _{1}-\theta
_{2})c_{II,I}(\theta _{2})c_{I,II}(\theta _{1})  \label{c} \\
c_{I,II}(\theta _{1})c_{II,I}^{*}(\theta _{2}) &=&-S^{(2)}(\theta
_{1}-\theta _{2})^{-1}c_{II,I}^{*}(\theta _{2})c_{I,II}(\theta _{1})+\delta
(\theta _{1}-\theta _{2})  \nonumber \\
d_{I,II}(\theta _{1})d_{II,I}(\theta _{2}) &=&-S^{(2)}(\theta _{1}-\theta
_{2})d_{II,I}(\theta _{2})d_{I,II}(\theta _{1})  \nonumber \\
d_{I,II}(\theta _{1})d_{II,I}^{*}(\theta _{2}) &=&-S^{(2)}(\theta
_{1}-\theta _{2})^{-1}d_{II,I}^{*}(\theta _{2})d_{I,II}(\theta _{1})+\delta
(\theta _{1}-\theta _{2})  \nonumber \\
d_{I,II}(\theta _{1})c_{II,I}(\theta _{2}) &=&-S^{(2)}(\theta _{1}-\theta
_{2})^{-1}c_{II,I}(\theta _{2})d_{I,II}(\theta _{1})  \nonumber \\
d_{I,II}(\theta _{1})c_{II,I}^{*}(\theta _{2}) &=&-S^{(2)}(\theta
_{1}-\theta _{2})c_{II,I}^{*}(\theta _{2})d_{I,II}(\theta _{1})  \nonumber \\
\left\{ \cdot ,\cdot \right\} &=&0,\,\,\forall \,\,others  \nonumber
\end{eqnarray}
The simplicity of the model is reflected in the fact that interactions only
take place between species I and II and the independence of $S^{(2)}$ on $%
\theta .$ The interaction does make a distinction between left and right and
parity is only conserved if one also interchanges the two species. The
operators (\ref{c}) inserted into (\ref{Z}) lead to the same commutation
relations as those of the generators of $\mathcal{A}(W)$ in (\ref{rep}). We
still have to check that the $J$-transformed opposite algebra commutes with $%
\mathcal{A}(W)$ and is equal to the geometric opposite. Since there is no
proper Lorentz transformation which transforms $\mathcal{A}(W)$ into $%
A(W^{^{\prime }})$ we must take the previously mentioned unitary parity
transformation $P$ which in addition changes $I\leftrightarrow II$ for the
definition of the geometric opposite. This amounts to the following relation 
\begin{equation}
\mathcal{A}(W)^{\prime }=J\mathcal{A}(W)J=\mathcal{A}(W^{\prime })\equiv
PA(W)P  \label{geom}
\end{equation}
The first equality is a consequence of the following relation for the
generators: 
\begin{eqnarray}
j(c_{I,II}(\theta )) &\equiv &Jc_{I,II}(\theta )J=SJ_{0}a_{I,II}(\theta
)e^{-i\pi g\int_{-\infty }^{\theta }\rho _{II,I}(\theta ^{\prime })d\theta
^{\prime }}J_{0}S^{*}  \label{J} \\
&=&Ka_{I,II}(\theta )K^{*}e^{i\pi g\int_{\theta }^{\infty }\rho
_{II,I}(\theta ^{\prime })d\theta ^{\prime }}
\end{eqnarray}
where the $adJ_{0}$ transformation only changed the sign in the exponential
and the adjoint transformation with the S-matrix generates an exponential
factor $:\exp -i\pi g\int_{-\infty }^{\infty }\rho _{II,I}(\theta ^{\prime
})d\theta ^{\prime }:$ of which part of it compensates against the
sign-changed previous exponential factor. $K$ is the Fermion twist of the
modular theory for free Fermi-fields (chapter 3). Similarly we have 
\begin{equation}
j(d_{I,II}(\theta ))\equiv Jd_{I,II}(\theta )J=b_{I,II}(\theta )e^{-i\pi
g\int_{\theta }^{\infty }\rho _{II,I}(\theta ^{\prime })d\theta ^{\prime }}
\end{equation}
Therefore the TCP ($\equiv $J in d=1+1) transformed Z-F fields are: 
\begin{equation}
j(Z_{I,II}(x))=\frac{1}{\sqrt{2\pi }}\int \left( e^{ipx}j(c_{I,II}(\theta
))+e^{-ipx}j(d_{I,II}^{*}(\theta ))\right) d\theta
\end{equation}
The relative commutation relations of $Z_{I}^{\#}(x)$ for $x\in W$ with $%
j(Z_{I}^{\#}(y))$ for $y\in W^{\prime }$ are precisely those of a free field
since e.g. in the $Z_{I}$-$j(Z_{I}^{*})$ commutator the exponentials add up
to the $\theta $- independent term $\exp -ig\pi \int_{-\infty }^{\infty
}\rho _{II}(\theta ^{\prime })d\theta ^{\prime }$ which multiplies the free
(anticommutator) function with support outside the spacelike region. The
argument for the $Z_{II}$-$j(Z_{II}^{*})$ commutator with $locZ\in W$ is
identical, whereas all other $Z^{\#}$-$j(Z^{\#})$ commutators vanish without
giving rise to contraction terms. There is one more property of the wedge
algebra which ought to be checked: the opposite wedge $\mathcal{A}%
(W)^{\prime }$ is geometric i.e. the justification for the Haag duality $%
\mathcal{A}(W)^{\prime }=\mathcal{A}(W^{\prime }).$ Since the only geometric
interaction free transformation which links the two wedges in a unitary way
is the generalized parity covariance: 
\begin{equation}
a_{I}^{\#}(\theta )\longleftrightarrow a_{II}^{\#}(-\theta
),\,\,\,b_{I}^{\#}(\theta )\longleftrightarrow b_{II}^{\#}(-\theta )
\end{equation}
One easily checks that this transformation indeed transforms the wedge
generators $Z^{\#}$ into those of $\mathcal{A}(W)^{\prime }.$ To see this
one uses 
\begin{eqnarray}
a_{I}(\theta )e^{-ig\pi \int_{-\infty }^{\theta }\rho _{II}(\theta ^{\prime
})d\theta ^{\prime }} &\rightarrow &a_{II}(-\theta )e^{-ig\pi \int_{-\infty
}^{\theta }\rho _{I}(-\theta ^{\prime })d\theta ^{\prime }} \\
&=&a_{II}(-\theta )e^{-ig\pi \int_{-\theta }^{\infty }\rho _{I}(\theta
^{\prime })d\theta ^{\prime }}  \nonumber
\end{eqnarray}
and the analogous transformation law for the $b^{\prime }s.$ With this
relation we have completed the checks on the generators of the wedge
algebras. The surprising fact is that the wedge algebra which fulfills the
cyclic and seperating conditions of the Reeh-Schlieder theorem admits
on-shell generators generators which applied to the vacuum create
one-particle states without the vacuum polarization clouds. For compact
localization regions such operators cannot exist. If one looks at the more
general factorizing models one finds that such generating on-shell operators
always exist. As a matter of fact it is not even necessary to look for
formulas which represent the Z-F algebras in the incoming Fock space. Rather
the only important point is to start from the Ansatz of modular localized
states in the form: 
\begin{eqnarray}
&&\int \hat{F}(x_{1},...,x_{n})Z^{*}(x_{1})...Z^{*}(x_{n})\Omega \\
&=&\int F(\theta _{1},...,\theta _{n})\tilde{Z}^{*}(\theta _{1})...\tilde{Z}%
^{*}(\theta _{n})\Omega  \nonumber
\end{eqnarray}
where we have surpressed all indices distinguishing the various Z-F fields.
The multivariable strip $\theta $-analyticity comes from the localization
and the relation between the various boundary values of the on-shell
momentum space functions in rapidity variables function $F(z_{1},...,z_{n})$
for the different orderings $\theta _{i_{1}}\geq \theta _{i_{2}}\geq ...\geq
\theta _{i_{n}}$ is dictated by the Z-F commutation relations among the $%
\tilde{Z}^{*\prime }s.$ It is very helpful to write the wave functions which
characterize the Hilbert space $H_{R}^{(n)}$ in terms of the free functions $%
H_{R}^{(n)in}$ by splitting off a reference wave function $F^{ref}$: 
\begin{eqnarray}
F(\theta _{1},...,\theta _{n}) &=&F_{n}^{ref}(\theta _{1},...,\theta
_{n})f(\theta _{1},...,\theta _{n}) \\
H_{R}^{(n)} &=&F^{ref}H_{R}^{(n)in}  \nonumber
\end{eqnarray}
The reference wave function is most conveniently obtained by defining a
state on the $Z^{\#}$-generated wedge algebra $\mathcal{A}(W)$ which
fulfills the KMS condition with respect to the modular group generated by
the boost operator: 
\begin{eqnarray}
&&\left\langle Z^{\#}(x_{1})...Z^{\#}(x_{n})Z^{\#}(y_{1})...Z^{\#}(\rho
_{n},\chi -i\pi )\right\rangle _{T=2\pi }  \label{K} \\
&=&\left\langle Z^{\#}(\rho _{n},\chi +i\pi
)Z^{\#}(x_{1})...Z^{\#}(x_{n})Z^{\#}(y_{1})...Z^{\#}(y_{n-1})\right\rangle
_{T=2\pi }
\end{eqnarray}
Here all the mass shell Z-F operators have to be placed inside the wedge
i.e. $y_{i},x_{k}\in W,$ and the $\chi $ denotes the x-space rapidity and $%
\rho _{n}$ the radial coordinate of $y_{n}.$ Besides this KMS condition we
have enough boundary conditions from the commutation relation between the
positive and negative mass shell components of the $Z^{\#\prime }s$ which
lead to a recursive system linking the 2n correlation to the 2n-2 etc. This
is not only a very elegant way for finding a special solution of the above
Riemann-Hilbert problem, but it also emphasizes the physical role of the
auxiliary operators as being attached to the wedge algebra $\mathcal{A}(W)$
which does not have any pure state (it is hyperfinite type $III_{1}).$ The
Lorentz-invariance in the momentum space rapidity together with the
indicated iterative pairing enforces the following diagonal $\theta $%
-structure: 
\begin{equation}
(\tilde{Z}^{\#}(\theta _{1}^{\prime })...\tilde{Z}^{\#}(\theta _{n}^{\prime
})\Omega _{2\pi },\tilde{Z}^{\#}(\theta _{1})...\tilde{Z}^{\#}(\theta
_{n})\Omega _{2\pi })=\prod_{i}\delta (\theta _{i}^{\prime }-\theta
_{i})F_{n}^{ref}(\theta _{1},...,\theta _{n})
\end{equation}
The KMS condition relates this function to another one in which a particle
on one side is missing and an antiparticle on the other created. This KMS
property of the auxiliary fields is the germ for the crossing symmetry of
the local fields. The reference $F^{ref}$ ((suitably normalized) defines one
concrete realization of the n-particle component of the modular M\o ller
operator. As a matter of fact, any wedge localized vector $\psi =B\Omega
_{2\pi }$ could have been used to define via $(\psi
,Z^{\#}(y_{1})...Z^{\#}(y_{n})\Omega _{2\pi })$ a modular M\o ller
isomorphism $U^{(n)}$ from $H_{R}^{(n)in}$ to $H_{R}^{(n)};$ i.e. there are
as many transformations as there are localized vectors with n-particle
components. On the other hand if we would have a modular M\o ller operator $%
U $ in Fock space, the on mass shell auxiliary fields are defined as in (\ref
{Mo}) $Z^{\#}(0)=U\psi
^{\#}(0)U^{*},\,\,Z^{\#}(x)=U(x)Z^{\#}(0)U^{*}(x),\,\,x\in W,$ where in the
last formula $U(x)$ denotes the translation.

The $F^{ref}$ for different particle number are in principle all
independent. What we are really interested in for the formfactor program is
of course the case where $B$ is either a pointlike field $B(0)$ or an
operator from a double cone algebra. In that case the function transformed
to rapidity space is directly related to the $B$-formfactor between
n-particles and the vacuum and the rapidity transformed KMS condition
becomes the crossing relation since the vectors generated by the application
of the $\tilde{Z}^{\#}$ are intimately related to the on mass shell matrix
elements of $B(0)$ between incoming particle states. The raison d'etre for
the existence of the auxiliary wedge-localized fields becomes now clear:
these operators are the mediators between on-shell analytic properties and
the space-time localization properties. Despite their wedge localization
properties (which are reponsible for avoiding the consequences of the Haag
theorem), the algebra generated by them can resolve one particle states
which would not be the case for smaller localization regions which are not
left invariant by a boost subgroup of the Poincar\'{e} group. Their
existence cannot be derived from scattering theory since the wave packets in
scattering theory cannot be localized in a wedge. Note also that the
crossing symmetry does not hold for the incoming on-shell matrix elements of
the auxiliary fields, but it is the above KMS property which leads to the
crossing symmetry for \textit{local} fields. What is truly amazing is that
these fields bring together three aspects which up to now had their separate
places: The pure quantum aspects of the Bekenstein-Hawking-Unruh issue
(whose semiclassical manifestation shows up in black hole physics, the
on-shell crossing properties and the related nuclear democracy or duality
(that one which was suggestive for the Veneziano dual model) and the
uniqueness of the field theory related to an admissable S-matrix (the
inverse problem of LQP) and the nonperturbative construction idea based on
modular localization. This of course begs the question of whether the
existence of the mass shell vacuum polarization free fields can also be
guarantied outside of factorizing models, to which we have some comments in
the next section.

The last and difficult question, for which we will again use the Federbush
model for clarification is: how can we understand double cone algebras and
local fields? From the formula (\ref{local}) as well as from the formfactor
rules (axiom III) in the previous section it is clear that for local $\psi
^{\prime }s$ i.e. for formfactors $F_{B}$ of pointlike fields $B,$ the
n-point components become related by a kind of cluster property which
identifies the residuum of poles in the analytic continuation with lower
formfactors. The mechanism behind this can be better understood by looking
at the form of (\ref{local}) and asking the question how does the local
field manage to fulfill a pointlike TCP transformation $J\psi (0)J\sim \psi
^{C}(0)$ whereas the $Z^{\#}$-fields only insure the correct transformation (%
\ref{J}) of the entire wedge algebra $\mathcal{A}(W).$ So what kind of magic
in the formfactors $\left\langle \Omega \left| \psi ^{\#}(0)\right| \theta
_{1},...,\theta _{n}\right\rangle ,\,\,\,\theta _{1}\geq \theta _{2}...\geq
\theta _{n}$ is responsible for the pointlike TCP-property? Obviously the
pointwise (geometric) TCP property is most manifest in the representation (%
\ref{local}) whereas the (more quantum) ordering which allows a clear-cut
separation of n-particle components is the Fermion Wick-ordering. The
ordering we use for our auxiliary operators (\ref{au}) is different from the
Fermion ordering by a very simple cumulant expression in which the two terms
are only different by a simple c-number factor whereas in the presence of
fluctuation (particle-antiparticle) terms the operator terms change and the
Fermion ordered bilinear exponentials have a nontrivial $i\varepsilon $ pole
structure in rapidity space. It is precisely this structure which, similar
to the cluster decomposition property, relates the formfactors for different
n. Therefore the formfactor rule axiom $III$ which relates the pole
structure with a lower formfactor (i.e. the only difficult part of
bootstrap-formfactor program) corresponds to the step from the wedge
localization to the compact or pointlike localization. This was to be
expected on the basis of LQP physical intuition. For the wedge particle
massive state can still be identified and the modular localized n-particle
states can be chosen independently whereas for compact localization regions
it is not possible to separate single particles from the ``clouds'' which
accompany them and which regulate the relation of the formfactors (the
components of local field vector states) for different particle number.

The computation of the above reference wave functions from the KMS structure
(\ref{K}) is particularly simple for the Federbush model. All thermal
two-point functions are equal to the wedge restricted free field vacuum
expectations which according to chapter 3.8 have a manifest thermal
representation. This is a property of all local fields. Only for the
nonlocal auxiliary fields this identity between restriction and thermal
representation is violated and, as we have seen above, we must take the
thermal formula for the calculation of the coefficient functions $F$ which
appear in the particle rapidity representation of the wedge localized
states. Obviously the lowest nontrivial function for which our KMS formalism
becomes relevant, is the 4-point function: 
\begin{eqnarray}
&&\left\langle Z_{II}(\theta _{2}^{\prime })Z_{I}(\theta _{1}^{\prime
})Z_{I}^{*}(\theta _{1})Z_{II}^{*}(\theta _{2})\right\rangle _{2\pi } \\
&\sim &e^{-\frac{1}{2}i\pi g}\left\langle Z_{I}(\theta _{1}^{\prime
})Z_{I}^{*}(\theta _{1})\right\rangle _{2\pi }\left\langle Z_{II}(\theta
_{2}^{\prime })Z_{II}^{*}(\theta _{2})\right\rangle _{2\pi }+  \nonumber \\
&&+e^{\frac{1}{2}i\pi g}\left\langle Z_{II}(\theta _{2}^{\prime
})Z_{II}^{*}(\theta _{1})\right\rangle _{2\pi }\left\langle Z_{I}(\theta
_{1}^{\prime })Z_{I}^{*}(\theta _{2})\right\rangle _{2\pi }
\end{eqnarray}
where up to numerical factors the rapidity representation coalesces with the
modular group (boost variable) labeling. The thermal two-point functions at $%
T=2\pi $ are equal to the wedge restricted free expressions. The first
nontrivial functions are the 6-point functions: 
\begin{eqnarray}
&&\left\langle Z_{II}^{*}(\theta _{3}^{\prime })Z_{II}(\theta _{2}^{\prime
})Z_{I}(\theta _{1}^{\prime })Z_{I}^{*}(\theta _{1})Z_{II}^{*}(\theta
_{2})Z_{II}(\theta _{3})\right\rangle _{2\pi } \\
&\sim &e^{-\frac{1}{2}i\pi g}e^{-\frac{1}{2}i\pi g}\frac{e^{-g\pi (\theta
_{2}-\theta _{3})}}{\cosh \frac{1}{2}(\theta _{2}-\theta _{3})}\cdot 
\nonumber \\
&&\cdot \left\langle Z_{I}(\theta _{1}^{\prime })Z_{I}^{*}(\theta
_{1})\right\rangle _{2\pi }\left\langle Z_{II}(\theta _{2}^{\prime
})Z_{II}^{*}(\theta _{2})\right\rangle _{2\pi }\left\langle
Z_{II}^{*}(\theta _{3}^{\prime })Z_{II}(\theta _{3})\right\rangle _{2\pi } \\
&&+....
\end{eqnarray}
where the dots stand for three other terms which result from permutations of 
$\theta _{1}$ with $\theta _{2}$ and $\theta _{1}$ with $\theta _{3}$
weighted with appropriate phase factors. These expressions should be
consistent with $\left\langle \Omega ,Z_{II}(\theta _{2}^{\prime
})Z_{I}(\theta _{1}^{\prime })N(\psi _{I}^{*}(x)\psi _{II}^{*}(x))\Omega
\right\rangle $ and $\left\langle \Omega ,Z_{II}^{*}(\theta _{3}^{\prime
})Z_{II}(\theta _{2}^{\prime })Z_{I}(\theta _{1}^{\prime })N(\psi
_{I}^{*}(x)\psi _{II}^{*}(x)\psi _{II}(x))\Omega \right\rangle $ where $N$
denotes the leading operator term in a short distance expansion. This is
easily checked if one pays attention to the charge terms in the $%
i\varepsilon $ prescription which are contained in the exponential of (\ref
{charge}). The reference functions for the description of the wedge
localization spaces of the more general factorizable models are also
constructed by these thermal method. The important new contributions are the
minimal formfactors which. We will present the systematics together with
nontrivial illustrations in a seperate paper.

Although Gibbs states also share the KMS property, it should be stressed
that the thermality which originates from restriction in QFT, is described
by KMS without a Gibbs representation. There seems to be a confusion in the
literature on this point.

For the Federbush model it is very easy to solve all these problems since
the S-matrix does not depend on the individual rapidities but only on their
order. For more general factorizing models, the special wave functions
obtained from the KMS formalism as well as the formfactors are (as the
S-matrix) meromorphic functions in the rapidity variables which in this way
is elevated to a uniformization variable of the problem. Although this
simple Federbush model has no bound states, one expects that they show up in
the thermal expectations as soon as they contain one local local field in
addition to the wedge localized $Z^{\prime }s.$ Note that they would occur
slightly outside the strip region; if one $\theta $ is in the strip, another
one must be at the reflected point below the strip; only in this
pair-formation one encounters particle poles. Expressed in terms of the
difference variables this of course agrees with the old findings of
Karowski-Weisz \cite{KW}. We mention that in a previous paper\footnote{%
Unfortunately there were some errors in the formulas which characterize the
wedge localization spaces in \cite{hep}.} \cite{hep} we have discussed the
relation between the eigenvalue equations for the vectors in the modular
localization space $\mathcal{H}_{R}(W)$ and the Riemann-Hilbert problems
resulting the formfactor ``axiomatics'' following \cite{KW}\cite{Smir}\cite
{BKZ} without the thermal use of the Zamolodchikov-Faddeev algebra. But it
is only this thermal aspect which allows to incorporate the latter into the
general framework of QFT. Whereas the representation (\ref{rep}) for $%
\mathcal{A}(W)$ is a peculiar property of models with constant S-matrices,
the auxiliary thermal $Z$-fields exist for all factorizable models. The
modular study of the Federbush model has supplied us with physically rich
new nonperturbative concepts which are collected in the following remark:

\begin{remark}
For a given physically admissable $S$-matrix there exists a unique
interacting wedge algebra which is generated by semilocal on-shell fields
which for d=1+1 factorizing theories fulfill the Zamolodchikov-Faddeev
algebra in momentum space. Such a field applied to the vacuum creates a one
particle state. Although the generated algebra is TCP-covariant, these
fields are not yet TCP-covariant. In order to achieve the latter property of
covariant transformation property under TCP, one must find the generators
for double cone algebras of arbitrary small size. These generators (more
generally all compactly localized operators) have a rich virtual particle
structure such that the vector they generate from the vacuum has its higher
particle components determined in terms of the lower ones. In fact the
coefficient functions of the local operators are nothing else then their
generalized formfactors. Whereas the general analytic structure of an
n-particle formfactor is already determined by the auxiliary fields (in
particular the KMS property of the latter are responsible for the crossing
symmetry of the local operators), the relation between the formfactors for
different n as well as their pole-structure is only understood after having
passed from wedge to compact (double cone) modular localization. The
existence of the intermediate semilocal wedge algebra generating fields is
related to the existence of a modular M\o ller operator; in fact the
auxiliary semilocal field is nothing but the field $T(x)$ (\ref{trans}).
There are as many T's as there are free Wick-polynomials of the free fields
related to the S-matrix (including those which describe bound states).
\end{remark}

This situation for factorizable models is only the tip of an iceberg. Every
local QFT has thermal variables $T(x)$ without vacuum fluctuations which
generate the wedge algebra and are related to the modular M\o ller operator $%
U$: 
\begin{equation}
T(x)\equiv U(x)UA_{in}(0)U^{*}U^{*}(x),\,\,\,x\in W  \label{trans}
\end{equation}
where the translation $U(x)$ is acting on $T(0)\equiv UA_{in}(0)U^{*}.$ In
spite of our simplified notation, the $U$ depends on the wedge and we should
rather talk about a family $U(W).$

We will defer all explicit calculations concerning the construction of the
local model fields to a separate paper. The reason for this postponement is
that we want to present the construction of the factorizable models fully in
the spirit of algebraic QFT (where special local field coordinates are
avoided), i.e. by constructing the net of double cone algebras from the
wedge algebras\footnote{%
The main step in the algebraic constructive program is really the
calculation of the wedge algebra with a \textit{geometric} commutant.
Barring the possibility that all intersections of wedge algebras are empty,
the existence of local double cone algebras (and thei generating local
fields) is secured by the above main step.} and the relevant concepts for
these calculations are still unfinished. In particular our knowledge of the
interpretation of the modular objects ($J,\Delta ^{it})$ for double cones is
extremely scarce. Apart from the special case of conformal theories and a
conjecture about a geometrically ``fuzzy'' action of $\Delta ^{it}$ in
general, nothing is known. The wedge situation, for which all the
interaction resides in $J$ and $\Delta ^{it}$ is given in terms of Lorentz
boosts, is certainly no guide for the double cone localization. Even in the
case of the simple Federbush model one can show that there are no generators
which (if applied to the vacuum) create one particle states without clouds
of virtual particle-antiparticle pairs. This is true, although the model has
no real (on-shell) particle creation.

One can show that the modular construction of ``free'' anyons and plektons 
\cite{hep} in d=1+2 leads to similar mathematical problems. In this case the
whole construction takes place in a scattering space which, in
contradistinction to Fermions and Bosons, has no tensor product structure in
terms of Wigner spaces. The braid group commutation relation leads to a
Tomita $J$ which again involves a constant matrix S, however in this case it
does not carry scattering information but is identical to the braid group
representation R-matrix. Formally such a situation has some resemblance with
the Federbush model, apart from the fact that now the opposite of $\mathcal{A%
}(W)$ in the quantum sense of the von Neumann commutant is \textit{different
from the geometric opposite} $A(W)^{\prime }\neq A(W^{\prime }).$ This
difference is accounted for by a statistics ``twist'' $S_{twist}.$ Again the
model has a rich virtual particle structure even though no real particle is
created. But since this time the rich virtuality is a result of the
nontrivial statistics twist $S_{twist},$ there is no reason whatsoever to
expect that this fades away in the nonrelativistic limit. The fact that the
nonrelativistic limit of Fermions and Bosons leads to the Schr\"{o}dinger QM
is related to the existence of relativistic free fields in Fock space. But
since the Fock space structure in d=1+2 cannot support anyons and plektons,
there is all reason to expect a kind of nonrelativistic field theory which
can incorporate the virtuality which is \textit{necessary} to \textit{%
maintain the relation between spin and statistics in the nonrelativistic
limit}. Indeed all attempts to incorporate braid group statistics into QM
ever since the time of Leinaas and Myrheim have only led to a deformation of
(half)integer spin but not to nonrelativistic operators with the correct
spin-statistics commutation structure. A consistent multiparticle QM with
braid-group statistics i.e. a theory which leads to a multiparticle S-matrix
fulfilling the clusterproperties does presently not exist. The investigation
of two-particle Aharonov-Bohm scattering is not sufficient to settle the
issue of braid group statistics in QM. The above discussion casts doubt on
the quantum mechanical nature of braid group statistics i.e. on the
possibility to have a nonrelativistic description without virtual
particle-antiparticle creation. This could explain the negative results of
all pure geometric attempts in terms of Schr\"{o}dinger wave functions. The
method of using the Wigner representation and the correct multiparticle
structure from scattering theory \cite{FGR} together with the present
modular localization method looks well defined and promising, but still
needs to be carried out. A somewhat easier problem is the use of the modular
localization method in order to construct chiral conformal theories with
given plektonic statistics. Some remarks are contained in the next section.

In the remainder of this section we would like to make some pedagogical
remarks for readers with an insufficient knowledge of general properties of
nonperturbative QFT. In connection with rapidity dependent
Zamolodchikov-Faddeev algebra operators, as well as x-dependent chiral
conformal operators, one often finds the erroneous concept of ``analytic
field operators'' and ``holomorphic algebras.$\,$Because their use is so
widespread (the few articles where this misleading concept was not used are
an exception), it is interesting to ask where such incorrect ideas are
coming from. Since I am not an expert on string theory, I have limited my
search to QFT. The oldest paper which could be interpreted as alluding to
``analytic operators'' $A(z),z\in \mathbf{C}$ seems to be the famous BPZ 
\cite{BPZ} paper\footnote{%
In an older paper on conformal blocks e.g. \cite{Swieca-S}\cite{Swieca-S-V}
(called nonlocal components in a conformal decomposition theory with respect
to the center of the universal covering) such ``holomorphic'' terminology
was never used.} on minimal models. Although the authors do not use such
terminology in print, the notation used in that paper could be misunderstood
(and has been misunderstood by many physicist whose first experience with
QFT came through that famous work or was influenced by string theory). The
truth is that field algebras never have holomorphic properties. The analytic
properties of correlation functions and state vectors depend entirely on the
nature of states one puts on those algebras. Whereas vacuum ground states
lead to the famous BHW-domain (in chiral conformal QFT equal to a
uniformormization region with poles for coalescing coordinates), temperature
states will only yield strip analyticity. In the above thermal construction
the problem is to compute a L-boost KMS state at temperature $T=2\pi $ on
the Zamolodchikov-Faddeev algebra and not to construct a regularized Gibbs
state on an invented algebra of analytic operators $Z(\rho ,z_{boost})$ or $%
\tilde{Z}(z_{rapidity}).$

It is a fact that the associated analytic Bargman-Hall-Wightman domain for
conformal covariant QFT and factorizing models is larger than that of the
corresponding Poincar\'{e} covariant massive theories) and as a result of
braid group statistics one looses the unicity of these analytic
continuations (but never the univaluedness in the real time physical
localization points!) . But neither the old nor the new BHW domains have
anything to do with the ``living space'' of fields in the sense of quantum
localization of operators in this article. The analogy between x-space
analytic properties of conformal QFT and momentum rapidity space properties
of factorizing theories have been observed by many authors. They are related
to the similarity of p-space spectral support properties with x-space
localization properties. These analogies do however not justify the search
for a common formalism; the conceptual situations remain essentially
different.

In an attempt to attribute a constructive meaning to the above unfortunate
but existing terminology, one could point at a property which has \textit{no
analogue in higher dimension} (not even in the conformal invariant limit of
higher dimensional theories). Whereas generally with vacuum expectation
values one can relate at most two physical theories: a noncommutative real
time QFT and a commutative euclidean candidate for a statistical mechanics,
a chiral conformal theory on one light cone has infinitely many
noncommutative boundary values (and no commutative) each of which defines a
set of positive definite correlation functions and hence a theory. Namely
the restriction of the analytically continued correlation function defines a
positive QFT not only on the circle (the standard living space of chiral
theories) but also \textit{on each boundary encircling the origin} (with the
right $i\varepsilon $ Wightman boundary prescription). Although this does
not legitimize the notion of ``holomorphic operators'' in the literal sense,
the existence of a operator conformal QFT for each chosen boundary within
the BWH analyticity domain is a significant difference to the mentioned
higher dimensional situation on which one may base this terminology.. The
reader recognizes easily that this operator structure is equivalent to the
existence of the infinite dimensional diffeomorphism group which is related
to the Virasoro algebra structure in fact it is the LQP version of the
diffeomorphism covariance of the theory. The application of a symmetry,
which does not leave the vacuum reference state invariant, defines another
set of positive Wightman functions which at the case at hand belong to the
deformed boundary. Its deep relation to the modular theory will be commented
on in the next section.

\subsection{Modular Construction in General Case, open Problems}

One important general structural question is whether a physically admissable
scattering matrix $S_{s}$ belongs to only one QFT (or better LQP which is
the field-coordinate independent content of QFT). This is the famous inverse
problem of QFT. As we saw before, the modular M\o ller operators are highly
nonunique, so the question is if $\mathcal{A}(W)$ defined by (\ref{Mo}) is
uniquely determined. Suppose their would be two different $U^{\prime }s$ say 
$U_{1}$ and $U_{2}.$ Then $U=U_{2}^{*}U_{1}$ commutes with the free Tomita
involution $S_{0}$%
\begin{equation}
US_{0}=S_{0}U
\end{equation}
so the problem has been reduced to the question of whether $S_{s}=1$ implies
a free field theory. The proof of this old folklore statement of QFT is
surprisingly difficult. Here it helps to recall the physical picture that in
order to identify the vacuum and the one-particle state of a QFT it is not
necessary to know the theory globally, but it suffices to know the wedge
algebra and the geometric action of the modular L-boosts. Only if one also
in addition wants to identify scattering states, one of course needs the
whole Minkowski space.

The modular localization spaces for wedges are equal and agree with that of
a free field theory. From this we would like to conclude that the same is
true for the associated algebras i.e. $\mathcal{A}(W)\equiv U\mathcal{A}%
_{0}(W)U^{*}=\mathcal{A}_{0}(W)$. Since $J$ commutes with $U,$ the commutant
of $\mathcal{A}(W)$ is 
\begin{equation}
\mathcal{A}(W)^{\prime }=U\mathcal{A}_{0}(W)^{\prime }U^{*}=U\mathcal{A}%
_{0}(W^{\prime })U^{*}
\end{equation}
In fact from the equality of the modular operators and real spaces one could
try to prove a slightly stronger statement that the so called natural cones $%
\Delta ^{\frac{1}{4}}\mathcal{A}_{+}(W)\Omega $ and $\Delta ^{\frac{1}{4}}%
\mathcal{A}_{0,+}(W)\Omega $ are equal. According to a theory of Connes this
implies the equality of the algebras. A more direct physicist's argument
would consist in using the linear map $L$ between the algebras defined by 
\begin{eqnarray}
L &:&\mathcal{A}(W)\rightarrow \mathcal{A}_{0}(W) \\
A &\rightarrow &A_{0},\,\,A\Omega =A_{0}\Omega  \nonumber
\end{eqnarray}
This is a $^{*}$-linear map with the following identity for the special
matrix elements: 
\begin{equation}
\left\langle \Omega ,A\chi \right\rangle =\left\langle \Omega ,A_{0}\chi
\right\rangle ,\,\,\,\chi \in \mathcal{A}_{0}(W)\Omega =\mathcal{A}(W)\Omega
\end{equation}
Taking for $\chi $ wedge localized n-particle states and using free field
crossing symmetry, the equality generalizes to general bra-states $\chi
^{\prime }.$

As far as the strategy of constructing interacting models in d\TEXTsymbol{>}%
1+1 is concerned, we do not have the possibility to extract a bootstrap
program for the S-matrix. Hence the idea to introduce interactions via known
S-matrices is not a good one. A better object for the general case seems to
be the operator $U,$ which, as the formfactors, is partially off-shell (more
``local'' than $S)$ and has a quadratic relation to the S-matrix. In that
case the first step of the program namely to find an ``interacting pair'' $%
\mathcal{A}(W)$ and $\mathcal{A}(W)^{\prime }$ which is $U$-related to
corresponding free algebras such that $SU=US_{0},\,\,\left[ U,\Delta \right]
=0$ is easily carried out with the result that the candidate for the modular
scattering matrix would be $S_{s}=JJ_{0}=UJ_{0}U^{*}J_{0}.$ Certainly one
would not expect that an interaction introduced this way and used first for
the construction of a net of modular localized spaces and then for a net of
algebras, will lead all the way to nontrivial local nets of algebras of a
LQP. It would however be very interesting to find the obstructions and
understand them.

The analogy with the miraculous ``corner transfer matrix'' in 2-dim. lattice
problems is very tempting. In a relativistic theory the standard Hamiltonian
picture of implementing interactions (which distinguishes a time direction)
is not very natural, although one has accumulated a lot of experience and
intuition. The modular aspects of the formfactor program for factorizing
theories suggests to take the wedge proposal serious and to develop the
necessary intuition for ``good $U^{\prime }s"$. A guiding idea should be the
existence of mass shell auxiliary fields which create vacuum polarization
free states. Formally one would imagine to write for $x\in W$: 
\begin{eqnarray}
\psi _{aux}(x) &=&U(x)U\psi _{0}(0)U^{*}U^{*}(x)= \\
&=&\int \sum \left\{ e^{-ipx}u(\theta ,..)a_{aux}(\theta
,..)+e^{ipx}v(\theta ,..)b_{aux}^{*}(\theta ,..)\right\} d\theta .. 
\nonumber
\end{eqnarray}
where $\theta $ is the wedge affiliated rapidity and the dots stand for the
transversal variables. For the auxiliary fields one again expects a form 
\begin{equation}
a_{aux}(\theta ,..)=a(\theta ,..)e^{i\eta _{-}(\theta
)},\,\,b_{aux}^{*}(\theta ,..)=...
\end{equation}
with the operator $\eta _{-}(\theta )$ of the form: 
\begin{equation}
\eta _{-}(\theta )=\sum^{\prime }\int_{\theta }^{\infty }\eta (\theta
_{1},..,....\theta _{n},..):\tilde{\psi}(\theta _{1},..)....\tilde{\psi}%
(\theta _{n},..):
\end{equation}
Here $\tilde{\psi}$ denotes the Fourier transform of the free field, i.e.
the creation/annihilation operators of particles/antiparticles, and the
double dots represent the free field Wick-ordering. The integration extends
over all transverse variables and over the interval $\left[ \theta ,\infty
\right] $ in the longitudinal rapidities $\theta _{i}.$ The $\eta $%
-operators do not contain transitions between vacuum as well as one-particle
states to other states. This is expressed by the dash on the sum and is
nothing but the formulation of absence of vacuum polarization in the $\psi
_{aux}$ operators. In analogy with the integrable case, this operator should
be thought of as the $\left[ \theta ,\infty \right] $ part of the operator $%
\eta $ in a representation: 
\begin{equation}
Sa(\theta ,..)S^{*}=a(\theta ,..)e^{i\eta }
\end{equation}
An approach via interacting models of modular M\o ller operators is of
course not expected to lead to explicit 4-dimensional nets of operator
algebras. Rather one expects to obtain a controllable iterative framework
which is in analogy to QM more like the Hartree Fock theory than standard
perturbation theory. There is of course always a chance that there exist
simple and controllable LQP's which are not accessible from the perturbative
point of view (e.g. because they do not contain a deformation parameter).

The close relationship of the modular localization approach with its thermal
aspects leads to the question whether the wedge generating auxiliary fields
allow for a lattice version which could bring in a quantum mechanical
aspect. Lattice theories in infinite lattices with short range interactions
are conceptually somewhat similar to QFT in that they have a sharp particle
concept an fulfill cluster properties which (in the absence of causality)
are strong enough to allow for the derivation of scattering theory. The
(vague) analogy of $U$ with the corner transfer matrix also points into this
direction. The modular localization approach would however not support an ad
hoc momentum space cutoff since the modular structures are inexorably linked
with Einstein causality and furthermore since such a cutoff wrecks the basis
of time dependent scattering theory and more generally the physical
interpretability altogether..

Note that we have used the word ``nonperturbative'' in an intrinsic sense,
i.e. we do not expect that there are special field coordinates in terms of
which the theory becomes long- or short-distance perturbative. The modular
ideas are consistent with the construction of scale invariant limiting
theories but do not support the wild scale-sliding which seems to be a
hallmark of the quantization approach.

\section{Old Ideas on Conformal Theories in new Setting}

The desire of constructing nontrivial and dynamically interesting
nonperturbative models of QFT is as old as the Wightman framework. The
attempts in this direction can be divided into two groups. The main thrust
was to follow the formalism of perturbation theory enriched with ideas from
rigorous statistical mechanics. The constructive approach was subdivided
into two steps of which the first one was to control the system in a finite
space-time volume and the second step consisted in performing the
thermodynamic limit \cite{construc}. Since the first step in most cases used
the canonical formalism, it was severely limited to superrenormalizable
couplings of the $\phi _{2}^{4}$-type. Even with all the subsequent
improvements (partially due to better operator regularizations, in
particular the lattice regularization), it was not possible to control a
strictly renormalizable (i.e. non-superrenormalizable) model by these
methods. The later success with low dimensional models which started in the
late 60$^{ies}$ with the massless Thirring-model was based on different
construction methods.

Our nonperturbative methods in this chapter would suggest that any method,
which follows the conceptual framework of the perturbative approach too
closely, is liable to suffer from spurious short distance problems caused by
perturbative methods. These problems are known to every field theorist, and
they led to the questionable belief that there is (even outside of
perturbation theory) a direct connection between the existence of a QFT and
the short distance behavior of its fundamental fields. The algebraic
framework of LQP, with its emphasis on local algebras as the carriers of the
physical content, would cast doubt on this idea and the concrete
construction of factorizing theories based on modular wedge localization
with its emphasis on polarization free on-shell auxiliary fields has
substantiated this doubts. It seems that the traditional constructive
approach (``constructive QFT'') is very limited indeed.

On the other hand there were always attempts to look for nontrivial models
away from concepts related with perturbation theory. The first attempt to
get away from free fields was to study so called generalized free fields;
but unfortunately they turned out to be dynamically quite useless, although
their pathological aspect had the beneficial effect of directing attention
towards the important problem of ``degree of freedom counting'' i.e. the
issue of phase space behavior in QFT.

A second idea was to look for ``Lie-fields'', i.e. a set of fields whose
space-time commutation relation close on themselves (the predecessors of the 
$W$-algebras). This proposal apparently was made by W. Greenberg in the 60$%
^{ies}$. The special case of equal time current relations (current algebra)
was known to restrict, but not to completely determine the models. More
interesting was the idea of space-time Lie-fields, but in the first attempt
to analyse this situation \cite{Lowen}, the examples given were too trivial
and somewhat discouraging.

The obvious place to look for was 2-dimensional conformal QFT. Already in
1970 I presented some results on the conformal invariant aspects of the
massless Thirring model and their local generation by the conformal energy
momentum tensor at a ``Conference on Special Topics in Quantum Field'' July
27-31, St. Louis. This material entered a joint paper with John Lowenstein 
\cite{Lowen-Schroer}. It appears somewhat curious in retrospect that we did
not use the Lie-field idea immediately in order to liberate ourselves from
special models. Only in 1973 it became clear to me that a general structure
analysis on conformal energy-momentum tensors can be performed. The result I
presented at the January 1974 Rio de Janeiro V Brazilian Symposium and it
was published in the proceedings which the reader finds in their original
form attached as an appendix to these notes. The reason why the result was
not submitted to an international journal was that the topic was quite
marginal to most of the theoreticians at that time. In that article there
also appear for the first time differential identities (equ. 6.22) in
addition to the equations of motion which are characteristic of chiral
conformal theories and which many years later became known (together with
the equation of motion and in a somewhat more general context) as the
Knishnik-Zamolodchikov equations.The paper also shows that charge carrying
fields for nonabelian conformal current algebras were an object of
investigation already in the early 70$^{ies}$. In fact the main motivation
for my contribution was the observation that the interchange of equations of
motions with differential identities led to the erronous claim that an
appearantly new nontrivial conformal fixed point of the generalized Thirring
model had been found. In more recent times these models were called Wess
-Zumino models by Witten; the more precise terminology would have been W-Z-W
``field coordinates''\footnote{%
Group-valued fields in d=2 euclidean theories similar to those introduced by
Witten also made their natural appearance in the older literature on QCD$%
_{2} $ Fermion determinants \cite{Sc}.} in the old current algebra models.
These field coordinates are helpful if one wants to separate these conformal
invariant theories from the generic family of generalized (massive) Thirring
models within a Lagrangian framework; they are of lesser help in the actual
calculation of the correlation functions of the charge-carrying fields of
the conformal current algebras.

Three years before the $L_{n}$-algebra of Fourier coefficients had appeared
in a totally different context of the dual model S-matrix in a paper of
Virasoro \cite{Virasoro}. Apparantly Virasoro was not aware about the
ongoing work on d=2 conformal QFT, in particular the massless Thirring
model, since otherwise he could hardly have missed the central term c which
inevitably shows up in every conformal model. I certainly did not know
Virasoro's findings since, apart from sociological reason for not being a
member of any group close to one of the big laboratories, that global $L_{n}$
on-shell (S-matrix) formalism of the dual model was physically too far away
from my local (off-shell) mode of thinking which, as far as physical content
is concerned, was closer to the conceptual framework of the present notes.
The first physicists who also studied the conformal aspects of the Thirring
model and observed the formal relation to Virasoros contribution on the
algebraic structure of Veneziano's dual S-matrix model was an group of
Italian physicist \cite{Gatto}. Looking back at its content, one observes
the rare sociological curiosity that an author, whose prior printed work on
the same subject is not referered to, is nevertheless mentioned in the
acknowledgments. That very same paper introduced for the first time (I was
not able to find an earlier reference) the terminology ``Virasoro-algebra''
for the algebra generated by the Fourier coefficients of energy-momentum
tensor.

The above mentioned unfashionable aspect of algebraic structures of
two-dimensional QFT was related to their off-shell physics aspect, whereas
the Zeitgeist in those times was definitely on mass shell before it went off
shell with the renewed interest in conformal field theory of the 80$^{ies}$.
This is another of the curiosities of those times.

These findings in 1973 were sufficiently encouraging for Jorge Andre Swieca
and myself to collaborate on conformal QFT. In 1974/75 the results were
published in two papers \cite{Swieca-S}\cite{Swieca-S-V}. The conformal
blocks (called ``nonlocal component fields'') resulted from the
decomposition of the local fields with respect to the center of the
conformal covering. Physically this amounted to a resolution of the
``Einstein causality paradox'' which consisted in the observation that any
violation of Huygen's principle in massless theories means that the would be
local conformal fields are, contrary to popular opinion reducible and
decompose into irreducible components under the central decomposition of the
conformal covering, i.e. Einstein causality of the original local fields
requires the existence of nontrivial irreducible components under the
conformal covering; the idea that the local field is conformally irreducible
was too naive and responsible for the paradox.

The discovery of the nontrivial family of minimal models by BPZ \cite{BPZ}
10 years later was making use of the increased knowledge of special algebras
and their Verma module representation theory. In that case one is working
outside known quantum algebras and the problem of passing from a Verma
module to a Hilbert space with a representation of a *-algebra belongs to
the most subtle points. Here one uses special methods like factorizing out
certain zero states and appealing to special results of V. Kac which
unfortunately play no role in higher dimensional QFT. This always endangers
the main physical motivation for doing these low-dimensional calculations,
namely to develop a theoretical laboratory for getting some intuition about
the conceptual framework of nonperturbative QFT (and not for the purpose of
enriching mathematics). This was the motivation of Swieca and myself and
also that of Mack and Luescher, who two years later understood the origin of
the quantization of the cocycle constant c, the strength of the
energy-momentum tensor two-point function. In \cite{Mack} one finds an
interesting historical account. In order to complete the historical review,
I have attached a scanned version of my own paper which was published in a
(hard to access) Brazilian journal as a second appendix.

The reason for this historical excursion in this section is not just
historical piety or curiosity, but to point out that only now, with the
arrival of the modular techniques, we have for the first time the means to
approach the classification of chiral conformal QFT's with the help of
generally valid concepts of QFT and in this way relate it more tightly with
the rest of QFT. In this way one may hope to complete the old dream to use
conformal QFT as a theoretical laboratory for the study of the principles
which underlie nonperturbative QFT. In the case of conformal fields with
braid group statistics, the idea was to compute the plektonic correlation
functions in a similar spirit as in case of free Bosons/Fermions the
algebraic (anti) commutation structure together with the causality and
positivity of the energy spectrum determines the correlation functions (the
Jost-Schroer theorem \cite{wightman}\cite{Mund}). There one constructs
charge-carrying free fields directly without the intervention of the
representation theory of observable algebras. The guiding principle is that
chiral conformal QFT's are like free fields in that the (plektonic)
commutation relations and the internal symmetry determine the theory
uniquely, i.e. chiral theories have no possibility of continuously varying
coupling parameters. This turns out to be correct. In order to see this we
must first adapt our modular localization approach to zero mass problems.

In the presence of zero mass fields, the previous wedge localization method
is not suitable, because there is generally no in- Fock space from
scattering theory which may serve as a reference space. Only double cones,
whose modular theory is presently ill understood, would define compact
localization regions where the interacting algebras would be unitarily
equivalent to a free reference algebra of zero mass Fermions/Bosons. This
observation has a natural relation to the vanishing of the LSZ limits (from
the softening of the mass shell singularity through the infraparticle
structure). If the theory involving zero mass is however conformal
invariant, then the wedge region (via compactification) can be considered as
a ``double cone at infinity'' or in the case of a half line or circle
(chiral theories) as a special finite interval.

In plektonic chiral QFT one starts with the exchange algebra commutation
relations in momentum space. The multiparticle spaces on which these
operators act are represented by path spaces on a fusion graph for which the
vertices are labeled by the possible nonabelian quantum charges. The
n-particle Hilbert space is the direct sum over those charge sectors which
can be reached by the nonabelian addition of n nonabelian fundamental
charges which are represented by a path-space spanned by the iterated
fusions: 
\begin{eqnarray}
H_{n} &=&\sum_{\rho (n)}H_{\rho (n)} \\
H_{\rho (n)} &=&\sum_{path(\rho (n),0)}\int f(\chi _{n},...,\chi _{1})\left|
(\chi _{n},\rho (n)),...,(\chi _{1},\rho (1)),0\right\rangle d\chi
_{n}...d\chi _{1}  \nonumber
\end{eqnarray}
The $\chi ^{\prime }s$ denote the exponential parametrization of the momenta
on the light line of say the right line $p=e^{\chi }>0,$ whereas the $\rho
(i)$ label the charge vertices on a fusion graph which one can reach by
applying the fundamental charge i-times onto the vacuum. The plektonic
statistics i.e. the R-matrices (see chapter 8.7) appears, if one compares
two path which differ at one vertex. The path structure exactly agrees with
that obtained by the iterated application of exchange algebra operators in
momentum $\chi $-space. The natural order is again $\chi _{1}\geq ...\geq
\chi _{n}$ and the other orders are obtained by the following commutation
structure of annihilation and creation operators: 
\begin{eqnarray}
a_{\bar{e}_{1}}(\chi _{1})a_{e_{2}}^{*}(\chi _{2}) &=&\sum_{e_{1}^{^{\prime
}},e_{2}^{^{\prime }}}R_{\bar{e}_{1}e_{2}}^{e_{2}^{^{\prime }}\bar{e}%
_{1}^{^{\prime }}}a_{e_{2}^{^{\prime }}}^{*}(\chi _{2})a_{\bar{e}%
_{1}^{^{\prime }}}(\chi _{1})+\delta (\chi _{1}-\chi _{2})\delta _{\bar{e}%
_{1}e_{2}} \\
a_{e_{1}}^{*}(\chi _{1})a_{e_{2}}^{*}(\chi _{2}) &=&\sum_{e_{1}^{^{\prime
}}e_{2}^{\prime }}R_{e_{1}e_{2}}^{e_{2}^{^{\prime }}e_{1}^{^{\prime
}}}a_{e_{2}^{^{\prime }}}^{*}(\chi _{2})a_{e_{1}^{^{\prime }}}^{*}(\chi _{1})
\nonumber
\end{eqnarray}
Here the $e^{\prime }s$ label charge edges which consists of a source
charge, the charge $c$ transferred by the creation operator and the range
charge $e=(r(e),c,s(e))$ and $\bar{e}$ is the cojugate charge edge which
labels the annihilation operator. The new (non-Fock) feature is the source
and range dependence. Related to this is the fact that the Fourier
transforms are nonlocal fields in x-space. In order to obtain localized
fields which fulfil the exchange algebra relations in x-space, we have to
process these creation and annihilation operators through the modular
localization machine for the half-line (the one-dimensional wedge). Despite
the change of the n-particle spaces as compared to Fermions/Bosons, the
discussion of modular localization is similar. The deviation of $J$ from the
geometrically defined $J_{0}$ is determined by the statistics operators
which are piecewise constant in $\chi .$ For the Tomita involution we have
the same structure as previously:

\begin{equation}
\check{S}=SJ_{0}\Delta ^{\frac{1}{2}}
\end{equation}

The mathematics of modular localization does not care whether $S$ is (as in
the Federbush model) a rapidity independent scattering matrix or a Klein
factor or an R-matrix due to a change of statistics. The difference would
only become visible if one compares the quantum opposite $\mathcal{A}(%
\mathbf{R}_{+})^{\prime }$ with the geometric opposite $\mathcal{A}(\mathbf{R%
}_{-})$ where the latter must be constructed with a unita$\mathbf{R}_{+}$ry
geometric parity reflection. The first step constructs the $\mathbf{R}_{+}$
localized eigenspaces of $\check{S}.$ Here one again uses the thermal KMS
method. The generating semilocal operators are again non polynomial
expressions in the $a^{\#\prime }s$ but yet without vacuum polarization. The
last step from the semilocal polarization free auxilary operators to the
local fields is as involved as in the integrable case\footnote{%
A more direct way of using the x-space exchange algebra is not possible,
because in contradistinction to the momentum space situation the
distributional character at coalescing points is not directly known from the
algebraic structure.}.

Although at the time of writing I had not yet finished the construction of
the pointlike fields and their correlation functions, it is already evident
that the two simplifications render the conformal problem much simpler than
the construction of the massive factorizing models. The aim is to determine
the theory in terms of its quantized statistics (R-matrix) structure and to
obtain the FQS-quantization of the c-value as a consequence of the
statistics rather than the other way around. We hope to present these
conformal constructions in a separate paper.

Finally there is the question if the constant $c_{\pm }$ in the
energy-momentum commutation relation (or the central extension cocycle of
the Virasoro algebra) have a more fundamental modular significance within
the LQP setting based on observable nets. The Moebius symmetry group of the
vacuum which allows a modular construction (in terms of modular half-sided
inclusions) does not give any such information. The classical invariance
group of an interval on the circle is a transformed dilation group. It acts
fixed point free on the interior of the interval and has the two end points
as its only fixed points. All these groups are modular groups of algebras
localized in that interval. The generalization of this idea is a
diffeomorphism of the circle with more than two fixed points. Of particular
interest are those simple situations which are covering-equivalent to the
old Moebius situation. In terms of circular coordinates $z=e^{i\varphi }$
these diffeomorphisms are of the form: 
\begin{equation}
z\rightarrow \sqrt{\frac{a+bz^{2}}{c+dz^{2}}},\,\,\left( 
\begin{array}{ll}
a & b \\ 
c & d
\end{array}
\right) \in SU(1,1)
\end{equation}
The simplest illustration is in a somewhat symbolic short hand: 
\begin{equation}
Dil(t)^{(2)}\equiv \sqrt{z}\circ Dil(t)^{(1)}\circ z^{2}
\end{equation}
Our notation for the half-circle is $z=e^{i\varphi },\,\,0\leq \varphi \leq
\pi ,$ and $Dil(t)^{(1)}$ denotes the usual dilation $x\rightarrow e^{t}x$
rewritten into circular coordinates by stereographic projection. The complex
poles and zeros of such transformations lie outside (poles) or. inside
(zeros) the circle and may be joined by two cuts which do not cross the
circle so that the map is well defined as a circular diffeomorphism. The
point now is that the covering dilations have their own inner product in
which they act unitarily: 
\begin{equation}
\left( f,g\right) ^{(2)}=\frac{1}{2}\int dxdy\frac{(1+x^{2})(1+y^{2})}{%
\left( x-y+i0\right) ^{2}(1+xy)^{2}}f(x)g(y)
\end{equation}
The new inner product is only different from the old one by a multiplicative
change of the measure. It is easy to verify that both inner products have
the same symplectic form $\omega ^{(1)}(f,g)=\omega ^{(2)}(f,g)=\frac{1}{2}%
\int g(x)df(x)=\omega (f,g)$ hence belong to the same Weyl algebra 
\begin{equation}
W(f)W(g)=e^{i\omega (f,g)}W(f+g)
\end{equation}
The SL(2,R) group acts unitarily in both spaces, but the action on the
second space is different; it is the two-fold covering action 
\begin{eqnarray}
U^{(2)}(\alpha )f(x) &=&\Gamma f\left( \frac{a\frac{2x}{1-x^{2}}+b}{c\frac{2x%
}{1-x^{2}}+d}\right) ,\,\,\,(\Gamma f)(x)\equiv f\left( -\frac{1}{x}+\sqrt{%
1+(\frac{1}{x})^{2}}\right) \\
\alpha &=&\left( 
\begin{array}{ll}
a & b \\ 
c & d
\end{array}
\right) \in SL(2,C)  \nonumber
\end{eqnarray}
In particular the above $Dil(t)^{(2)}$ is a special one parametric subgroup
with 4 fixed points ($\mp )\infty ,-1,0,1$ which acts as: 
\begin{equation}
Dil(t)^{(2)}(f)(x)=f\left( -\frac{1-x^{2}}{2t^{2}x}+sign(x)\sqrt{1+\frac{%
1-x^{2}}{2t^{2}x}}\right)
\end{equation}

The suspicion that this group is the Tomita-Takesaki modular group of a
state on the two-fold algebra $\mathcal{A}\left[ -\infty ,-1\right] \vee 
\mathcal{A}\left[ 0,1\right] .\subset Weyl(\mathbf{R})$, i.e. a von Neumann
subalgebra of the the Weyl algebra on the line. The higher coverings lead to
a similar situation. This can be confirmed by checking the KMS property for $%
Dil(t)^{(2)}$. We will defer all proofs to a forthcoming publication in
collaboration of the author with H. W. Wiesbrock and limit ourselves to
comment on some results.

\begin{enumerate}
\item  The construction of the n-fold covering diffeomorphisms from the
modular groups of algebras of n disjoint intervals in different positions
follows the same pattern as the modular construction of the Moebius group.

\item  The underlying states (associated with the above inner products)
which lead to geometric modular situations are, with the exception of the
vacuum, not cyclic with respect to the algebra of a single interval. This
``defect'' is related to the fact that a geometric situation for Haag
duality for disjoint intervals overlooks the presence of higher
superselection sectors as charge-anticharge operators in the commutant of
disjoint interval algebras.
\end{enumerate}

These are fascinating developments since the modular point of view liberates
conformal QFT from its special dependence on structures (Virasoro-,
affine-algebras) which have no place in the conceptual framework of general
local quantum physics. By emphasizing multi-local algebras and their modular
structure we have the chance to discover a whole world of ``hidden''
symmetries, i.e. symmetry transformations which unlike diffeomorphisms have
no classical geometric aspect.

\section{Comparison with Ideas from String Theory}

One look at the content of theoretical publications under the heading of
``high energy physics'' reveals that what most physicist in this area are
doing is very different from the content of these notes. A more detailed
comparison shows that there are a few but perhaps significant concepts with
an interesting overlap.. One idea which immediately comes to ones mind are
certain structural properties which one affiliates with ``light cone
canonical quantization. This idea had a recent come back in relation with
the so called M-theory, both in the sense of Matrix theory and M-theory
proper \cite{Bi Suss}. The observation was that this quantization formally
leads to simpler fields which resemble the nonrelativistic situation in that
their application to the vacuum is free of particle-antiparticle
polarization clouds. This of course has a resemblance with the structural
properties of auxiliary fields which generate the wedge localized algebras.
The relation becomes even more striking if on realizes that the transversal
Galilei-group generators fit naturally into the subgroup G(8) of the
Poincar\'{e}-group which arises in connection with the split into
longitudinal, transverse and mixed generators. As with almost all standard
quantization methods, the assumptions (canonical structure, functional
Feynman-Kac representation) are not true properties of the results drawn
from this assumptions, but rather play the role of a working hypothesis or
better mental mark which suggests other structures which are true
properties, despite the fact that the original assumptions become violated
in the process of construction (e.g. through necessary renormalization
repairs. There are some more interesting connections, and the best procedure
seems to be to be to put these observations into a more systematic context.

The fastest way to understand the ideas underlying the present situation is
to look at history of post Feynman QFT. The common cradle of all present
frameworks of local quantum physics in general, and relativistic particle
physics in particular, is certainly Feynman's renormalized perturbation
theory and the ensuing understanding of QED. In the aftermath of this work,
three schools of thought have emerged. Although their aim was the same,
namely to abstract a nonperturbative framework (in order to incorporate
strong interactions), the paths taken towards this goal were quite
different. The first school of ideas which we will briefly refer to as
Lagrangian field theory, took the covariant perturbative formalism of
Lagrangian quantization as a starting point and interpreted it as a
realization of an extension of the very successful quantization idea from
quantum mechanics to the realm of infinite degrees of freedom. In this
parallelism to classical physics, renormalization was understood (following
a suggestion of Kramers) in analogy to the selfenergy problems of classical
field theory at the beginning of this century. At that time Lorentz,
Poincar\'{e} and others showed that these problems arise inevitably if one
adds to classical field theory models of particles. Hence the occurrence of
certain infinities was natural in a quantization approach and, apart from
problems of mathematical consistency, their ``dumping'' into physical
parameters was already partially anticipated in the classical theory. The
Lagrangian quantization approach led to functional integral representations
which, if it would not be for the necessity to perform infinite
renormalization, could serve as nonperturbative definitions of physically
relevant objects\footnote{%
In the lattice approach the discrete analogon to euclidean functional
integrals is indeed taken as a definition and the renormalization is merged
with the herculean task of controlling the scaling limit via second order
phase transitions. This task was only accomplished in cases of existence of
dynamical variables which are stable under scaling (e.g. the
Lieb-Matthis-Schulz Ising model fermions).}. But since in local QFT, if
analyzed from Wigner's particle picture, the one-particle aspects are
already a consequence of the Poincar\'{e} covariance via the covariant
fields which define the QFT, the infinities in a proper intrinsic
formulation should never have appeared. With this remark we are already
entering the territory of the second school : general QFT or in more recent
terminology LQP. This school of thought, after some conceptual
reformulations incorporated renormalized perturbation theory (see chapter 4)
as a deformation on free fields but believed that a nonperturbative
understanding would not come from generalizing Feynman's formalism but
rather require a different conceptual framework where the physical intuition
is mainly entering through local observables. The third school of thought,
the S-matrix school, insisted that the nonperturbative understanding will
come from an analysis of the scattering operator which is an important
global observable. Although this approach in form of the bootstrap program
came after 10 years to an end, certain of its notions survived up to this
date in the form of effective actions and (as a result of its emphasis on
crossing symmetry=duality) via Veneziano's S-matrix model in the present
string theory. Through the constructive approach based on modular wedge
localization as explained in this chapter, we also noticed that even in LQP
the S-matrix has interesting modular aspects. But if one only considers the
present sociological situation, the areas appear worlds apart. A comparison
therefore is difficult and endangered by misunderstandings.

Let us first establish some rules for comparison.. As we remarked above the
quantization approach works with structures as canonical commutation
relations, actions, functional integrals etc. which, although for themselves
are almost never true properties of the would be theory, nevertheless lead
to new properties which sometimes are at least perturbatively obeyed by the
resulting theory. In fact the initial canonical assumptions tie the
existence of the theory to a certain short distance behavior which even in
renormalized perturbation theory is violated. To think about a cutoff does
not help because a physically interpretable non Einstein-causal theory does
not exist\footnote{%
Such a cutoff would go against the modular localization ideas on which the
only explicitely known nonperturbative results are based. One can regularize
concrete Feynman integrals and even euclidean functional integrals, but
those objects cannot be related to a real time physically interpretable
(with time dependent scattering) QFT.}. So in some way any QFT quantization
approach has an artistic start. On the other hand all the properties on
which LQP and in particular modular localization is based are true
properties up to the very end. Hence the only reasonable way to compare LQP
with Lagrangian QFT or string theory is to look at the results at the latter
and not to pay attention to the words which enter the derivation. We first
collect schematically those properties, results and conjectures of LQP which
we want to compare. For a QFT with a mass gap with a complete scattering
interpretation, one can derive the following facts (chapter 3 and 6).

\begin{itemize}
\item  The modular theory of wedge algebras is geometric with the modular
group $\Delta ^{it}$ being the wedge associated Lorentz boost $\Lambda
_{w}\, $of the incoming particles and the modular reflection $J$ being
related to that of the incoming free field theory $J_{in}$ through the
scattering operator (S-matrix) $S_{scat}$: 
\begin{equation}
\Delta ^{it}=U(\Lambda _{w}(2\pi t)),\,\,\,J=S_{scat}J_{in}
\end{equation}
The dense set of wedge localized state vectors can be represented in the
form: 
\begin{eqnarray}
H_{loc}(W) &=&H_{R}(W)+iH_{R}(W) \\
H_{R}(W) &\equiv &\overline{\left\{ \psi ;S\psi =\psi \right\} },\;S\equiv
J\Delta ^{\frac{1}{2}}  \nonumber
\end{eqnarray}
where $H_{R}(W)$ is the real closed subspace generated by the +1\thinspace
eigenvectors of the antilinear unbounded Tomita operator which is involutive
on its domain $S^{2}=1.$ This brings the thermal Hawking Unruh aspects,
which one usually relates with black holes, into ordinary QFT (chapter 3).

\item  As a standard reference wedge W$_{stan}$ we may take the z-t wedge in
which case we call z,t the longitudinal and x,y the transversal coordinates.
This situation suggests to decompose the Poincar\'{e} group generators into
longitudinal, transversal and mixed generators 
\begin{equation}
\,P_{\pm }=\frac{1}{\sqrt{2}}(P_{0}\pm
P_{z}),\,\,M_{0z};\,\,M_{12},\,\,P_{i};\,\,G_{i}^{(\pm )}=\frac{1}{\sqrt{2}}%
(M_{i0}\pm M_{iz}),\,i=1,2
\end{equation}
The generators $G_{i}^{(\pm )}$ are precisely the ``translational'' pieces
of the euclidean stability groups $E^{(\pm )}(2)$ of the two light vectors $%
e^{(\pm )}=(1,0,0,\pm 1)$ which appeared for the first time in Wigner's
representation theory for zero mass particles. More recently these
``translations'' inside the homogenous Lorentz group appeared in the
structural analysis of ``Modular Intersections'' of two wedges \cite{Bo Wie}%
. Its role is analogous to that of the true translations $P_{\pm }$ with
respect to halfsided ``Modular Inclusions'' \cite{Bo Wie}
\end{itemize}

\begin{enumerate}
\item  As one reads off from the C.R., $P_{i},G_{i}^{(+)},P_{\pm }$ have the
interpretation of a central extension of a transversal ``Galilei group''%
\footnote{%
This G's are only Galileian in the transverse sense; they tilt the wedge so
that one of the light like directions is maintained but the longitudinal
plane changes.} with the two ``translations'' $G_{i}^{(+)}$ representing the
Galilei generators, $P_{+}$ the central ``mass'' and $P_{-}$ the
``nonrelativistic Hamiltonian''. The longitudinal boost $M_{0z}$ scales the
Galilei generators $G_{i}^{(+)}$ and the ``mass'' $P_{+}.$ Geometrically the 
$G_{i}^{(+)}$ change the standard wedge (it tilts the logitudinal plane) and
the corresponding finite transformations generate a family of wedges whose
envelope is the halfspace $x_{-}\geq 0.$ The Galilei group together with the
boost $M_{0z}$ generate an 8-parametric subgroup $G^{(+)}(8)$ of the
10-parametric Poincar\'{e} group: 
\begin{equation}
\,G^{(+)}(8):\,\,P_{\pm },\,\,M_{0z};\,\,M_{12},\,\,P_{i};\,\,G_{i}^{(+)}
\end{equation}
The modular reflection $J$ transforms this group into an isomorphic $%
G^{(-)}(8).$ All observation have interesting generalizations to the
conformal group in massless theories in which case the associated natural
space-time region is the double cone.
\end{enumerate}

\begin{itemize}
\item  The position of the subspace $H_{R}(W)$ within the incoming Fock
space allows to define a modular M\o ller operator $U(W)$ which intertwines
the wedge affiliated Tomita involution $S$ with that of the corresponding
incoming involution: 
\begin{equation}
SU(W)=U(W)S_{in}
\end{equation}
and leaves the vacuum unchanged. The interacting wedge algebra $\mathcal{A}%
(W),$ which together with the vacuum vector has $\Delta ^{it}$ and $J$ as
its modular data, is defined in terms of $\mathcal{A}_{in}(W)$ as: 
\begin{equation}
\mathcal{A}(W)\equiv U(W)\mathcal{A}_{in}(W)U^{*}(W)
\end{equation}
The Haag's theorem prevents the existence of analogous intertwining
unitaries for the type I equal time canonical algebras: 
\begin{equation}
A_{can}\equiv \cap _{\varepsilon }A(\varepsilon )
\end{equation}
which in the algebraic approach are represented as the intersection of time
slice algebras of thickness $\varepsilon .$ This leads to the nonexistence
of the interaction picture in local quantum physics and the necessity of the
artificial infinite volume limiting procedure involving a quantization box
(which is unfortunately not related to modular localization). The above
intertwining relation between the unique hyperfinite type III$_{1}$algebras
(all localized subalgebras in local quantum physics which have a nontrivial
causal complement are of this kind) is protected against such No Go
theorems. The existence of the \textit{modular} M\o ller operator \cite{Sch
W}$U(W)$ (in QFT it does not seem to be possible to define a ``scattering''
M\o ller operator) leads to the existence of generators of $\mathcal{A}(W)$
which are localized in $\mathcal{A}(W)$ but allow no smaller localization
inside $W$ i.e. they are nonlocal inside $W.$ They are ``on shell'' i.e.
contain a negative frequency part which annihilates the vacuum. Formally
they are given as $A_{W}(x)=U(x)U(W)A_{0}(0)U^{*}(W)U^{*}(x).$ It is
essentially the absence of vacuum polarization pairs i.e. the mass shell
support of their Fourier transform which makes these generators\footnote{%
Strictly speaking these polarization free operators are only well-defined
objects in the thermal Hilbert space associated with the modular wedge
localization. However they are very valuable as a basis for the pointlike
local fields which allow for a nonsingular extension outside the wedge \cite
{schroer 3}.} of $\mathcal{A}(W)$ extremely helpful. In factorizable d=1+1
theories, the positive and negative frequency components of these semiglobal
operators fulfill the Zamolodchikov Faddeev algebra \cite{schroer 3}. Be
aware that the $x$ in the $U$ transformed fields has nothing to do with
localization around $x$ inside $W$. Rather localization has to be
constructed via smaller algebras defined by intersecting wedge algebras: 
\begin{eqnarray}
\mathcal{A}(\mathcal{O}) &=&\bigcap_{W\supset \mathcal{O}}\mathcal{A}(W) \\
\mathcal{A}(W) &=&U(\Lambda )\mathcal{A}(W_{stand})U^{*}(\Lambda )  \nonumber
\\
W &=&\Lambda \cdot W_{stand}  \nonumber
\end{eqnarray}
where $W_{stand}$ is the standard x-t wedge and the net of W's is generated
from $W_{stand}$ by Poincar\'{e} transformations. It is very important for
the interacting case to realize that $U(W)$ depends on $W$ i.e. $%
U(W_{stand}) $ does not commute with the space-time transformations $U(L)$
except with the $W$-associated Lorentz boost. It should be clear that ideas
about how to construct such $U^{\prime }s$ should not be viewed in the
setting of the perturbative split $H=H_{0}+H_{int}$ (the free incoming
situation does not correspond to $H_{0}).$ Whereas the representations of
the Poincar\'{e} group of the interacting situation agrees with that of the
incoming fields, this is not so for the unperturbed theory belonging to $%
H_{0}.$ In fact the latter theory does not even live in the same Hilbert
space (only its local folium of states agrees with that of the interacting
theory).

\item  The M\o ller operator $U(W)$ can be explicitly computed for d=1+1
factorizable models and it is intimately related to the Riemann-Hilbert
properties of the modular localized real subspace $H_{R}(W)$ (chapter 6).
The mass shell components of the nonlocal generators $A_{W}(x)$ turn out to
satisfy \cite{schroer 3} the Zamolodchikov-Faddeev algebra and the modular
localization property defines a thermal KMS state on this algebra.

\item  The rich physical structure emerging from inclusions and
intersections of local algebras in the net. Algebraic QFT interprets the
external (space-time) and internal global symmetries from how one algebra is
positioned with respect to another one (see mathematical appendix):

\begin{enumerate}
\item  {}\textbf{``Shallow'' inclusions}. These are inclusions $\mathcal{N}%
\subset \mathcal{M}$ which posses a (noncommutative) conditional expectation
i.e. $\mathcal{N}=E(\mathcal{M})$ of the kind studied by Vaughn Jones. The
local endomorphism of algebraic QFT of the DHR theory and its extension to
low dimensional QFT are typical examples. The relation between the algebraic
QFT endomorphism and the Jones setting is well understood. Shallow
inclusions are related to inner symmetries\footnote{%
Note however that in low-dimensional QFT's there is no sharp distinction
between inner and spacetime symmetries because the charge structure relates
to the spatial covering aspects.}. For d=1+3 dimensional theories this leads
to Fermi-Bose Statistics and compact internal group symmetry whereas for low
dimensional theories this yields the class of (physically) admissable
unitary braid group representations through Markov traces on the braid group
B$_{\infty }.$

\item  \textbf{``Deep'' inclusions}. By this one means inclusions which have
no conditional expectation but obey a modular restriction which lead to
space-time symmetries. There are two ``modular inclusions'' whose geometric
consequences have been studied: halfsided modular inclusions and modular
intersections. The first case is illustrated by two touching wedges, a
situation resulting from a light like shift of a wedge into itself. A
halfsided modular inclusion leads to noncompact group isomorphic to the
longitudinal (d=1+1) Poincar\'{e} group. The second case of modular
intersections is illustrated by two wedges which have one light ray in
common. In that case the intersection data lead among other things to the
above Galilei generators. The full Poincar\'{e} group is obtained by the
relative ``modular position'' of a finite number of algebras (the minimal
number depends on the space-time dimensionality). In this way one may
generate the whole net from a finite ``modular skeleton net''.
\end{enumerate}

\thinspace \thinspace \thinspace \thinspace \thinspace \thinspace \thinspace
\thinspace \thinspace \thinspace \thinspace \thinspace Some comments are in
order.
\end{itemize}

Let us add to these rigorous results two conjectures which are important for
the later comparison. These conjectures are related to the Hawking-Unruh
issue of horizon physics of quantum matter in black hole solutions. Even
though our comments are only conjectures, we will try to stick to the
conceptual rigor of the rest of this article. For this reason we will not
use the word ``quantum gravity'' in this discussion and emphasize the fact
that the physical origin of the Hawking temperature is the \textit{modular
localization} in Minkowski- or curved- space-time and not primarily a black
hole horizon. The latter mainly plays the role of enforcing a natural
localization (by creating a bifurcated horizon via the e.g. black hole
metric) and constitutes a special case of the former. The notion of a
bifurcated horizon through modular localization is more abstract, since it
is not related to metric Killing vectors but rather to isometries in the
space of wave functions or the underlying Hilbert space of QFT \cite{schroer
3}. It nevertheless leads to the same physical consequences of thermal
behavior and Hawking radiation. For this reason the main concepts which are
usually attributed to gravitation theory can be perfectly understood in
terms of \textit{thermality through localization} (instead of the standard
heat bath thermal behavior). The main difference of the two thermalization
concepts can be traced back to that of the two sided spectrum of the modular
localization operators (e.g. the Lorentz boost) versus the one sided
spectrum of the heat bath Hamiltonian which leads to the boundedness of $%
e^{-\beta H}.$

This raises the question whether modular localization also leads to a
fundamental algebraic notion of entropy. Here it is helpful to mention the
``degrees of freedom'' counting in local quantum physics which deviates in
an interesting and significant fashion from that in e.g. Schr\"{o}dinger
quantum mechanics \cite{Haag}. In the latter case one learns, that the phase
space cells (standard notion of localization and momentum restriction) leads
to a finite number of degrees of freedom per $2\pi \hslash $ size phase
space cell. The first attempt in algebraic QFT by Haag and Swieca led to the
notion of compactness. Later this notion was sharpened to the ``nuclearity
criterion'' of Buchholz and Wichmann \cite{Haag} which does not use a sharp
cutoff in Hamiltonian- or. in the ``modular''-energy but rather an
exponential fall off. Contrary to the nonrelativistic case and to popular
opinion, the relativistic localization concept (as opposed to the standard
box quantization) together with the finite energy projection does not lead
to a finite number of quantum states (``bits'') but rather only to a compact
(Haag-Swieca) or nuclear (Buchholz-Wichmann) set. A computation for free
fields reveals that this behavior is optimal i.e. local quantum physics
cannot reproduce the finite degrees of freedom behavior of quantum
mechanics, but comes pretty close: 
\begin{eqnarray}
H-S\,\, &:&\,\,\,\left\{ P_{E}A\Omega \mid A\in \mathcal{A}(\mathcal{O}%
),\left\| A\right\| \leq 1\right\} =compact\,set \\
B-W &:&\,\,\,e^{-\beta H}\mathcal{A}(\mathcal{O})=nuclear\,set  \nonumber
\end{eqnarray}
The interrelation between these slightly different forms of relativistic
``local degree's of freedom counting'' has been discussed in \cite{Haag}.
This property forbids infinite towers of particles (as they occur e.g. in
genus $\leq 1$ string perturbation) and an associated limiting Hagedorn
temperatures. The most valuable consequence is a very profound
interpretation of the ancient issue of Heisenberg-Weisskopf vacuum
fluctuations: if a spatial volume is not interpreted as a quantization box,
but rather as a region for localization of a partial charge via a conserved
current (say inside e.g. an already defined Minkowski space free field
theory), then the vacuum fluctuation near the boundary are infinitely large.
In order to control them, it is necessary to allow a smooth transition to
zero charge density inside a ``collar'' around the localization region. The
``split property'' of algebraic QFT \cite{Haag}, which is a consequence of
the above ``nuclearity'' property of degree of freedom counting, just
provides the mathematical precision for this intuitive idea\footnote{%
The idea to define entropy with the help of the split property is due to
Heide Narnhofer (this was pointed out to me by H-W. Wiesbrock) \cite{Narn}.
It constitutes one important element in a joint unfinished project of
Wiesbrock with the present author.}: 
\begin{equation}
\exists \,\,type\,I\,\,factor\;\mathcal{N}\,\,s.t.\,\mathcal{\,A}(\mathcal{O}%
_{1})\subset \mathcal{N}\subset \mathcal{A}(\mathcal{O}_{2})
\end{equation}
Here one should imagine two concentric double cones $\mathcal{O}_{i}$ with
their associated hyperfinite III$_{1}$ factors. The type I factor $\mathcal{N%
}$ has a ``fuzzy'' localization inside the bigger double cone, and it is
just this fuzziness which allows the definition of partial charges without
infinite vacuum fluctuation and with a clear-cut split between the physics
``inside and outside'' \cite{Haag}. Needless to add that the algebras
underlying QM are always of type I, whereas the relativistic causality and
associated localization structure always lead to hyperfinite III$_{1}$
factors at least if the regions allow for a nontrivial causal complement. So
in order to find quantum mechanical structures inside local quantum physics,
one needs type I factors inside local hyperfinite III$_{1}$ factors. The
split property gives also a preferred candidate \cite{Do} for such an
interpolating type I factor $\mathcal{N}.$ The scenario for a definition of
entropy from first principles is in terms of the modular group of $\mathcal{N%
}$. As a consequence of type I this modular group is inner, i.e. there
exists a ``Hamiltonian'' described by a hermitian operator $K$ associated to 
$\mathcal{N}$ (this never happens for the $\mathcal{A}(\mathcal{O})$
factors).

The issue of entropy is then closely related to the problem of the modular
Hamiltonian $K$ of $\mathcal{N}$ which in turn is determined in terms of the
modular objects of the split data \cite{Haag}: 
\begin{equation}
\Delta _{\mathcal{O}_{1}}^{it},\Delta _{\mathcal{O}_{2}}^{it},\Delta _{%
\mathcal{O}_{1}^{\prime }\cap \mathcal{O}_{2}}^{it};J_{\mathcal{O}_{1}},J_{%
\mathcal{O}_{2}},J_{\mathcal{O}_{1}^{\prime }\cap \mathcal{O}_{2}}
\end{equation}
In zero mass conformal theories the double cone $J$ and $\Delta ^{it}$
relative to the massless vacuum are geometric transformations inside the
full (including reflections) conformal group \cite{Haag}, in particular $%
\Delta ^{it}$ transforms the r,t coordinates but not the two transversal
angular variables of the double cone. On the other hand the massive double
cone theory can be incorporated into the same Hilbert space or more
precisely, the massless and the massive. This suggests that the modular
object of the massive situation are nonlocal deformations of the conformal
massless split situation. One expects that the K-Hamiltonian is well enough
in order to allow for the existence of the von Neumann entropy: 
\begin{equation}
S=-trDlnD
\end{equation}
where $D=\frac{1}{tre^{-K}}e^{-K}$ is the density matrix defined in terms of
the modular Hamiltonian $K.$ This is a quantity which depends on the size of
the collar $\varepsilon $ and which diverges as $\varepsilon \rightarrow 0$
i.e. when the fuzzy type I factor becomes hyperfinite type III$_{1}.$ If the
result of the existing proposals \cite{Wil} is compatible with this idea, we
should expect a universal logarithmic divergence in the inverse size $%
\varepsilon ^{-1}$of the collar which controls the vacuum fluctuations: 
\begin{equation}
S=-trDlnD\sim Cln\varepsilon ^{-1}
\end{equation}
with $C$ related to the longitudinal 2-dim. conformal theory which according
to our previous discussion we expect to determine the geometric core of the
fuzzy modular group of the double cone algebra $\mathcal{A}(\mathcal{O}).$
Indeed the formula \cite{Wil} 
\begin{equation}
C\sim Area\sqrt{c}
\end{equation}
where $Area$ denotes the area of the double cone and $c$ the vacuum
fluctuation strength of the energy momentum tensor. Although the limiting
entropy is certainly infinite, we have not yet been able to confirm that
this infinity is universal and behaves exactly as argued by Larsen and
Wilszek.

In order to prove this one must do some new computations on the ``localizing
map'' (math. appendix) which is the most convenient way to compute the
distinguished type I factor \cite{Haag}. The relevant degrees of freedom
would ``live'', as we will argue later, inside the collar and the ratios of
this ``collar entropy'' stay finite for vanishing collar size. This remains
a fascinating program for the future.

Now we are able to formulate our two conjectures:

\begin{conjecture}
The modular group of the (nonconformal) massive double cone $\,$algebra $%
\mathcal{A}(\mathcal{O})$ with respect to the massive vacuum vector (i.e.
the physical vacuum state restricted to $\mathcal{A}(\mathcal{O})$ is
cocycle-related to the known geometric modular group of the associated
conformally invariant situation belonging to the pair ($A(O),\omega _{m=0})$
where $\omega _{m=0}$ denotes the conformal invariant vacuum state. For the
equivalence of the massive with the massless algebra one may either invoke
the construction of the double cone algebra by canonical quantization or the
fact that local algebras are always hyperfinite III$_{1}$factors and the
latter is unique modulo unitary equivalence. The cocycle accounts for the
difference in the local propagation of massless (Huygens principle) and
massive theories and its presence renders the action of the modular group
``fuzzy''. Only asymptotically near the horizon i.e. the boundary of the
double cone, the fuzzyness decreases and the geometric conformal modular
transformation reappears. Although a single algebra $\mathcal{A}(\mathcal{O}%
) $ of the massive theory and its scale invariant limit may be identified,
the two nets inside $\mathcal{O}$ remain different. However the conjecture
that the difference is due to the different propagation suggests that the
massive net inside $\mathcal{O}$ may be obtained from the massless $\mathcal{%
A}(\mathcal{O})$ by adjoining the action of the Poincar\'{e} covariances
inside $\mathcal{O}.$
\end{conjecture}

\begin{remark}
For a massive free Fermi field\footnote{%
We want to avoid the infrared problems of massless Bose fields} in d=1+1
this can be shown. One notes that the restriction of such a free massive
theory to the light rays which constitute the boundary of the d=1+1 double
cone is simply the restriction of the corresponding massless theory and that
by propagating the chiral conformal data on the one dimensional horizon
inside with the massive propagator, one regains the massive free field net
inside $\mathcal{O}.$ A general proof of this reduction of a d=1+1 situation
to its chiral conformal limit (+ possible covariance operators) would be
extremely desirable because it would explain the association of the degree
of freedoms of d=1+1 theories with the horizon.
\end{remark}

\begin{conjecture}
The double cone algebras $\mathcal{A}(\mathcal{O})$ are identical to any of
the two-dimensional double cone algebras $A(O^{(2)})$ obtained by cutting
the double cone by a two-dimensional plane which contains the t-axis and one
coordinate axis. The net inside $O^{(2)}$ may be obtained from the
associated chiral conformal net on the one-dimensional horizon and a local
representation of Poincar\'{e} covariances and this \textit{holographic}
modular picture is the local quantum physical property behind the
Bekenstein-Hawking quasiclassical entropical observations.
\end{conjecture}

\begin{remark}
The first part is actually a consequence of Haag duality and the fact that
the causal completion of $\mathcal{O}^{(2)}$ gives $\mathcal{O}:$ 
\begin{equation}
\mathcal{A}(\mathcal{O}^{(2)})\equiv \cap _{\delta }A(\mathcal{O}_{\delta })
\end{equation}
Where $\mathcal{O}_{\delta }$ denotes the middle slice of thickness $\delta $
by cutting the double cone parallel to the t-axis. Each $\mathcal{O}_{\delta
}$ has $\mathcal{O}$ as its causal completion and the property of ``Haag
Duality'' demands the equality of $A(O_{\delta })$ with the algebra of the
causal completion $\mathcal{A}(\mathcal{O}).$ The essential step in the
holographic reduction is the appearance of chiral conformal degrees of
freedom after removal of the angular degrees of freedom due to angular
symmetry (substituting the transversal symmetry in the case of the wedge).
The envisaged entropy is therefore not proportional to $area(\mathcal{O}%
^{(2)})\times angular\,volume$ but rather to horizon-length$(\mathcal{O}%
^{(2)})\times angular\,volume=volume\,of\,\,horizon(\mathcal{O}).$ Actually
such a situation would also suggest that there may be an infinite hidden
nongeometric (fuzzy) symmetry algebras in the nonperurbative structure of
any QFT\footnote{%
In principle every modular automorphism has the interpretation of a
(localized but fuzzy) physical symmetry. Some of these are
``semi-geometric'' i.e. they act geometric on subnets. The previous modular
intersection situation which led to the transversal Galilei-transformation
is such a case of a ``semi-hidden'' symmetry \cite{un}.}. Although they are
local in the sense of keeping things inside say $\mathcal{O},$ their action
within $\mathcal{O}$ is totally fuzzy. Such symmetries of nonperturbative
local quantum physics would escape differential geometric methods. Note that
the two conjectures cannot even be formulated in terms of properties of
expectation values of fields; the use of the algebraic i.e. field coordinate
independent concepts is indispensable for the formulation. If algebraic QFT
did not already exist, one would have to invent it in order to understand
the above thermal and entropic properties.
\end{remark}

Summing up our excursion on nonperturbative LQP we would like to stress
again that the algebraic method allows for a completely intrinsic definition
and understanding of QFT, independent of its Lagrangian or non Lagrangian
origin. Any quantum theory which fulfills the stability requirements of
positive energy and allows for a net interpretation and the associated
localization concepts is a QFT par excellence and enjoys all the general
structural properties of these notes as TCP, spin \&statistics, crossing
symmetry \& modular localization \& thermality, wedge-localized fields
without vacuum polarization, hidden modular symmetries, Haag duality (an
abstract form of the 2-d Kramers-Wannier-Kadanoff Duality), nuclearity for
the phase space degrees of freedom \& the conjectured ``Holographic
Entropy'' and all the other yet known properties of nonperturbative local
quantum physics. The main obstacle against progress is not so much the novel
mathematics which these new physical concepts require, but rather (as always
in the past) prejudices. One prejudice is that field theory has to be
``Lagrangian''. In view of the many existing low-dimensional non-Lagrangian
models and the fact that they hardly rocked the Lagrangian boat up to now
(they are simply ignored, because their construction does not fit a
quantization approach), this appears to be the mightiest prejudice.

Before we mention some recent results of string theory obtained with the
help of light cone quantization and its discretization as well as finite
truncation, we take a rapid look at its history since this allows a clearer
conceptual understanding than the somewhat confused picture obtained from
only looking at the recent hep-theory publications.

A good starting point is the dispersion theory which was the main
nonperturbative attempt of the 50 and 60 to go beyond the Feynman approach.
The main issue was to find sufficiently many ``on-shell'' properties of QFT
such that an S-matrix theory or at least a phenomenologically successful
scheme could emerge. Besides the obvious properties like unitarity and
certain analytic properties, the on-shell property which apparently was most
intimately and deeply related to the off-shell causality principle of QFT
was crossing symmetry. The complete nonperturbative understanding of
crossing symmetry including the requires on shell analytic behaviour and the
precise relation to TCP invariance and causality was never achieved.
Therefore it was considered a major achievement when Veneziano succeeded to
construct a S-matrix model which fulfilled crossing symmetry exactly and
allowed for a systematic unitarization which maintained crossing symmetry in
each step and in some sense was reminiscent of the perturbative systematics,
although it had little in common with ordinary perturbation theory. Later it
was realized that the quantum mechanics of strings can successfully describe
this model and its unitarization. The infinite tower of particles which, if
they would remain stable under unitarization, would violate the principles
of local quantum physics (the aforementioned degree of freedom behavior,
leading to a Hagedorn temperature) could, as in Feynman's perturbation
theory, become unstable particles i.e. poles in the second Riemann sheet of
the S-matrix and in this way the model could be perfectly consistent with
nonperturbative QFT. This was at least what I and many of my contemporary
QFT colleagues thought when we got used to those nice pictures involving
Regge trajectories.

But, as everybody knows, things happened differently. Instead we had to
witness the ``Bartholomew night massacre''\footnote{%
The name some of my QFT collegues attributed to this event expressed their
bewilderment after being told to forget the nice Regge trajectory
interpretation of Veneziano's dual crossing symmetric S-matrix model which
they already had gotten used to (as a proposal for structure of a strongly
interacting S-matrix which could originate from a nonperturbative QFT).} (a
bit poetic, but not completely unrealistic since part of the story really
happened in Paris \cite{Scherk}), also often referred to as the first string
revolution, in which the old string theory, which served as a laboratory of
certain aspects of nonperturbative QFT (notably strong interactions), was
killed and the mathematical formalism (first without any change) was pushed
upward in energy by more than 15 orders of magnitude and physically outed as
``quantum gravity''. Only later string theory obtained the modern
differential geometric wrapping, which partially expressed the increasing
mathematical sophistication of the string community. In fact in this modern
fashion it became an impressive source of mathematical innovations. It was
precisely the distance from any kind of laboratory physics, which protected
these developments from usual fate of theories whose relevant energy scales
stays close to the experimentally accessible region. Because of the involved
fantastic ``scale sliding'' and its innovative differential geometric
content, it is often referred to as the ``second string revolution''.

From the point of view of exhausting the scenarios offered by Lagrangian
quantization, the sliding up the energy scale was very logical indeed. It is
legitimate and even useful to stretch a framework (as the Lagrangian
quantization, canonical or functional), which was so successful as Feynman's
renormalized perturbation theory, to its physical limits set by the Planck
scale. A successful formalism in the history of physics was always pushed to
its limits. Such extensions only become somewhat counterproductive if, as a
result of apparent lack of alternatives, one identifies its consequences (as
the ``big desert'' region beyond the present laboratory energies up to the
Planck mass, or as the omnipotence of supersymmetry in its underlying
differential geometric mathematical formalism) with what should be expected
from nature. The very existence of alternative concepts as in LQP which do
not lead to such consequences shows that one is exploring the limits of
quantization approaches rather than of local quantum physics. Let us first
look at light cone quantization since most recent papers on string theory
use this structure. We briefly remind the reader about its meaning in the
simplest Lagrangian QFT context.

In analogy to canonical quantization but different from the discussion of
the bifurcated wedge situation one distinguishes one x-t plane as the
canonical quantization plane and the other for the definition of a
propagating time. In the Lagrangian setting we have for a selfinteraction of
a scalar field: 
\[
L=\int dx^{-}dx^{\perp }\left\{ \partial _{+}\varphi \partial _{-}\varphi -%
\frac{1}{2}(\partial _{\perp }\varphi )^{2}-\mathcal{L}_{int}\right\} 
\]
where $x^{\pm }=\frac{1}{\sqrt{2}}(t\pm x),\,\,x^{\perp }$= transversal
coordinates. Using longitudinal momenta $k_{-},$ the ``Hamiltonian''
becomes: 
\[
H=\int dk_{-}dx^{\perp }\left\{ \frac{\partial _{\perp }\varphi
^{*}(k_{-},x^{\perp })\partial _{\perp }\varphi (k_{-},x^{\perp })}{2k_{-}}+%
\mathcal{H}_{int}\right\} 
\]

The main difference to standard canonical quantization is the absence of
vacuum polarization in the fields $\varphi (k_{-},x^{\perp })$ \cite{Bi Suss}%
. The prize for this apparent simplification is a somewhat hidden nonlocal
interpretation of these auxiliary fields which went unnoticed by the authors
of (M)atrix theory\footnote{%
The LCQ field variables are very nonlocal with respect to the usual local
canonical fields. Only in this way these degrees of freedom manage to
surpress vacuum fluctuations.}. The light cone theory in the context of the
Unruh Hawking mechanism was studied by Sewell \cite{Sew}. His setting only
applies to free field theories and to those interacting fields which have
the same short distance behavior as e. g. $\varphi ^{4}$ in d=1+1. There is
one essential difference to the standard canonical commutation relations.
The canonical rules in light cone coordinates are \cite{Sew}: 
\begin{eqnarray}
&&\left[ \varphi (x_{+}=0,\partial _{-}f),\varphi (x_{+}=0,\partial
_{-}g)\right] =\int f\partial _{-}gdx_{-}dx_{\perp } \\
&&\varphi (x_{+},h)=\int \varphi (x_{+},x_{-},x_{\perp })h(x_{-},x_{\perp
})dx_{-}dx_{\perp },\,\,h\in \mathcal{S}(R^{3})_{0}\,\,
\end{eqnarray}
where the subscript 0 denotes Schwartz functions with vanishing integral
over all space $\int h=0$. Hence the infrared behavior of light cone fields
is analoge to that of zero mass fields in d=1+1. The characteristic initial
value problem is easily shown to be uniquely soluble if the initial data are
in $\mathcal{S}(R^{3})_{0},$ and the causality condition which leads to
localizability is easily reexpressed in terms of the light cone description.
There are two caveats in the interacting case. On the one hand the
restriction to a light front has the same limitations as the standard
canonical formalism: any theory with stronger than free field short distance
singularities does not permit such a restriction. This only leaves the
superrenormalizable $P\varphi _{2}$ interactions. Here one could be open
minded and argue that the light cone idea should not be taken literal, but
that the important message is that there are field coordinates without
vacuum polarizations \cite{Bi Suss}. However such degrees of freedom are
necessarily nonlocal, i.e. the interaction of an originally local theory
causes the appearance of nonlocal light cone degrees of freedom (even for
the $P\varphi _{2}$. theories). There is nothing wrong with the use of
nonlocal field coordinates in a local theory as long as one knows the
relation to the local fields. Unfortunately there is no mentioning of this
important problem in the literature on light cone quantization. Whereas
these short-comings may be viewed of a technical nature (as most of the
starting assumptions in the standard approach they must be repaired in the
course of calculations), the third one is more physical and hence more
serious. The light cone approach does not reveal how the interacting
nonlocal polarization-free light cone fields are related to the local
fields. The knowledge of this relation is crucially important for the
interpretation of the light cone approach.

The suggestion from our modular localization approach would be that the
light front situation is closely related with the wedge; the horizon of the
wedge consists of two light fronts. In this way the precise relation between
wedge localized fields without vacuum polarization and local pointlike
fields will be clear. In fact the division into wedge-affiliated
longitudinal and transversal degrees of freedom led us to the existence of
the 8-parametric ``Galileian extension'' $G^{(+)}(8)$ of the longitudinal
modular group. Interestingly enough, the Galilean subgroup of $G^{(+)}(8)$
constitutes the starting point of the BFSS \cite{BFSS} light cone
quantization framework. The standard obstacle against a quantum mechanical
description, namely the presence of vacuum polarization, has been taken care
of if one uses these semiloval fields and this is perhaps the main reason
why these variables and their QM discretization is relevant for the
exploration of ideas on M-theory. Here the terminology QM should be
understood in sufficient generality. Galilei-invariant field theories
without vacuum polarization, but with rich channel couplings between
different multiparticle sectors (as the T.D. Lee model , just to mention
one) can a priori not be excluded. In any case the analogy of the
polarization cloud free state vectors $\varphi _{light-cone}\Omega $ with
the wedge thermal space affiliated vectors $A_{W}\Omega $ of this chapter is
very startling.

By compactification of the light cone time $x^{-}$ one formally obtains a
discretization of LCQ called DLCQ. As with lattice discretization, DLCQ
allow a matrix approximation which, following BFSS, posses a natural
interpretation in the modern string setting. In our attempt to translate
these situations into the conceptual framework of algebraic QFT, we would
draw the analogy to the compact double cone situation $\mathcal{A}(\mathcal{O%
})$ which is the modular substitute for the box quantization in QM. This is
the spacetime localization by which one must substitute the wedge region if
there is no LSZ scattering theory as a result of the presence of infrared
photon clouds. In that case the electrically charged ``infraparticles'' have
vanishing LSZ limits and one only can work with a Fock reference space for
compact modular localization regions. The claimed proximity of the light
cone quantization idea and string- and M-theory begs the question whether
the latter theories are local in the presence of interactions. If they are,
than string theory is a special case of LQP and all the algebraic results
can be taken over. I expect that they are not local, i.e. that one is not
dealing with nonlocal variables in a local theory, but the string framework
itself is nonlocal. In QFT it is precisely the locality structure which
elevates it from a bunch of prescriptions to a complete theory in which e.g.
the particle- field dichotomy, the (time dependent) scattering theory and
other pivotal aspects of the interpretation follow from the theory itself
and need not to be imposed from the outside in form of recipes. The string
idea is not a theory in this sense.

There is another more speculative issue of potential agreement. It is the
issue of nonperturbative thermal and entropical behavior. LQP attributes a
thermal aspect to all local algebras in particular if the boundary of the
localization region defines a classical (Killing) horizon as in the case of
black holes. In fact this phenomenon was discovered in a pure classical
manner via the notion of a classical entropy and the use of thermodynamic
relations. Attempts to interprete this classical entropy in the quantum
sense led to the conjecture of holographic behavior of degrees of freedom
for such situations. Although there exists by now a fundamental general
understanding of the relation between horizon (with or without Killing
vectors), the direct understanding of entropy in LQP is still shaky, despite
our previous arguments concerning holographic behavior. The situation is not
any better in M-theory with light cone quantization. The only thing which
one can presently say is that if those speculations turn out to be correct,
than also the entropical properties have nothing to do with CST classical
Killing vectors but are general properties of LQP. Since in that case the
definition of entropy required a good phase space behavior (``nuclearity'')
of QFT, it is interesting to ask what string people say on this subject.
Their counting is that of a (nonrelativistic) spatial box quantization i.e.
the standard finite number per unit of phase space volume. LQP would
challenge the physical relevance of this concept since the correct
localization in LQP is the modular space-time localization and not the box
localization which in the relativistic context is the same as the unphysical
Newton-Wigner localization. As we have emphasized, the nuclearity property
(which even for free fields cannot be improved) assigns an infinite albeit
``small'' number (see mathematical appendix).

An even more basic discrepancy is the role of supersymmetry which in the
quantization approach is used to tame perturbative short distance behavior
and in this way to prevent getting too far into the quantum domain. It is an
essential symmetry for many recent results. On the other hand LQP does not
ask for such a symmetry and even questions its physical significance. One
reason is, as we already mentioned on several occasions that it is a
completely accidental symmetry which collapses under a contact with a heat
bath. Here accidental means that it plays no role in the understanding of
charge sectors, e.g. all known 2-dimensional soluble (tricritical Ising
etc.) models can be solved without using SUSY although it is not wrong to
use it. It also plays no role in the modular interpretation of space-time
symmetries. It is true that in low dimensions the charge sector structure
gets intermingled with the covering aspects of space-time, but SUSY has no
part in this nontrivial ``marriage'' which among other things is responsible
for the existence of 3-mf invariants. The mentioned thermal collapse is what
one expects of a accidental symmetry, but of course not of a bona fide inner
or space-time symmetry, e.g. the Lorentz symmetry only suffers the usual
spontaneous breakdown and not a collapse. If one is conservative in one's
judgement one may say that the physics of SUSY (if there is any physics) has
nothing to do with any known concept of symmetry and its possible meaning
may be outside of LQP. But such a mystification for maintaining a property
which entered physics in a completely formal way (and afterwards changed the
motivation for its original introduction), is probably going a bit too far.
Saying such things one is of course easily accused of ignorance about the
marvelous achievements by supersymmetric n=4 gauge theories which are d=1+3
conformal invariant field theories. This if true, is indeed somewhat
sensational and it would be nice to see a physical gauge invariant
correlation function (of necessarily composite fields) even in lowest order
perturbation theory, because then one could study the thermal stability
problems and understand to what degree SUSY was necessary to find such a
remarkable situation. Unfortunately despite the 15 year existence no such
result is available (certainly not because of lack of physical interest).

The biggest discrepancy between LQP and string theory however lies in the
physical philosophy of why one introduces a certain structure. In LQP this
is done because one wants to extend the range of a principle. For example
the introduction of extended operators which have a more general
localization than that of pointlike fields or compactly localized operators
would not be done for its own sake. The physical (Bohr-Heisenberg-Wigner)
esthetics is in the introduction of those structures which are necessary to
generalize the range of applications of (existing) principles, in this case
the causality or locality principle.. For theories with a mass gap in their
positive energy spectrum, the causality principle allows for charge carriers
with semiinfinite spacelike string localization and therefore they are
introduced and not because one wants to investigate string-like structures
in their own right. In fact in d=1+2 one finds that these objects must obey
braid group statistics; if one limits the carriers to be compactly localized
one falls back onto Fermions and Bosons and misses an important physical
realization of the same principle. The esthetics underlying string theory
and more generally any quantization approach is Dirac's (differential)
geometric esthetics which in the modern mathematical setting favors
geometrically natural generalizations. Gedankenexperiments which could
explain notions like stringiness versus QFT-ness play no role. What if the
infinite tower of particles in string theory converts in higher order in the
genus as in Feynman's perturbation into resonances in the second Riemann
sheet of the S-matrix; why is such a situation not compatible with LQP? Does
stringiness meant in the above sense of LQP charge carriers with noncompact
localization? What is the concept of localization, if any, in string theory?
What mathematical terms (correlation functions, algebras, states) should one
use for the characterization of string theory and what are the intrinsic
properties and related principles which allow an intrinsic characterization
independent of the prescriptions and recipes which went into its
construction? More questions than answers.

Let us add some concluding remarks. Although a \textit{direct} comparison of
nonperturbative QFT with string theory is presently not possible as a result
of the lack of an intrinsic physical definition, we tried an indirect
evaluation based on what may be called ``circumstantial evidence''. As a
geometrically based approach, string theory is strongest in the
quasiclassical domain. It need classical Killing horizons in order to
perceive thermal and entropical nonperturbative properties. Its similarity
to classical Kaluza-Klein ideas and its selection of high dimensions (26,
10, 11 etc.) probably has the same explanation: what in low dimensions
remains genuinely nonperturbative quantum, becomes more classical in higher
dimensions and hence more easily detectable by the string formalism. It is
only logical that it puts more confidence in differential geometric (and
since recently also noncommutative geometric) Diracian structure than in the
conceptual physical Bohr-Heisenberg mode of thinking.

\chapter{Introduction to Algebraic QFT}

\section{Some Useful Theorems}

Algebraic QFT requires more mathematical rigor and physical depth than QM or
even than standard perturbative QFT. In the latter case one has already
achieved a reasonable intuitive understanding whereas in the d=1+3
nonperturbative QFT we only have untested scenarios. So one tries to
safeguard the shaky intuition with mathematical rigor and conceptual
clarity. Once a good nonperturbative intuition has been achieved, such an
attitude may well appear as pedantic.

This means among other things, that the postulated physical requirements
should be rigorous properties of the resulting theory. This, as was already
mentioned before, is certainly not the case in the quantization approach.
Even the strongest defenders of Lagrangians in QFT are perfectly aware that
canonical commutation relations, functional integral representations etc.
are only mental marks for the inspiration, their purpose is to suggest
formal tricks and recipes which eventually lead to correct structures. They
are almost never properties of the constructed theory, and therefore
quantization is an art and not a mathematical theory. LQP on the other hand
only uses intrinsic properties which are valid throughout the process of
construction and maintains its validity for the resulting theory. The
original postulates may bring about a lot of other surprising properties,
but they never get lost in the results. Lagrangians and functional integral
representations are in general not reconstructible from e.g. the physical
correlation functions.

A representation ($\pi ,U$) of a $C^{*}$-algebra and an automorphism group ($%
\mathcal{A},\alpha _{t}$) is a representation $\pi $ of $\mathcal{A}$ in a
Hilbert space $H$ together with a strongly continuous unitary representation 
$U$ of \textbf{R} in $H$: 
\begin{equation}
U(t)\pi (A)U(t)^{-1}=\pi (\alpha _{t}(A)),\quad U(t)=e^{iHt},\,\,H\geq 0
\label{impl}
\end{equation}
For such representations the following theorems hold.

\begin{theorem}
(Reeh-Schlieder) Let $\left\{ \mathcal{A}(\mathcal{O})\right\} _{\mathcal{O}%
\in \mathcal{K}}$ be a local net with translation symmetry and $\mathcal{O}%
_{0}$ a space-time region such that there is $\mathcal{O}_{0}\subset 
\mathcal{O}$ and a neighborhood of zero $\mathcal{V}$ with: $\mathcal{O}%
\supset \mathcal{O}_{0}+\mathcal{V}$ and additivity: $\bigvee_{x}\mathcal{A}(%
\mathcal{O}_{0}+x)=\mathcal{A}^{\prime \prime }$. It follows that: $%
\overline{\mathcal{A}(\mathcal{O})\psi }=\overline{\mathcal{A}\psi }\,\,\,$%
for all $\psi \in \mathcal{D}(e^{aP}),\;a\in V^{\uparrow }$
\end{theorem}

\begin{remark}
The physical content is that there exist no annihilation operators as long
as the localization region allow a nontrivial causal complement. The
cyclicity is equivalent to the dual situation of absence of annihilation
operators in the causal disjoint region which is contained in the commutant
algebra.
\end{remark}

%TCIMACRO{\TeXButton{Proof}{\proof}}
%BeginExpansion
\proof%
%EndExpansion
One shows that for any vector $\phi \perp \mathcal{A}(\mathcal{O})$ implies $%
\phi =0$

For any such vector we have $(\phi ,\alpha _{x_{1}}(A_{1})....\alpha
_{x_{n}}(A_{n})\psi )=0,$ or in terms of boundary values of analytic
functions in the tube: 
\begin{equation}
\lim_{x_{i}\rightarrow z_{i}}(\phi
,e^{iz_{1}P}A_{1}e^{i(z_{2}-z_{1})P}A_{2}....A_{n}e^{-(iz_{n}+a)P}e^{aP}\psi
)=0
\end{equation}
where the limit is taken from the inside the tubes

\begin{equation}
\mathcal{T}^{(n,a)}:Im(z_{i+1}-z_{i})\in V^{\uparrow
},\,\,z_{0}=0,\,\,z_{n+1}=ia
\end{equation}
$\quad $in which the matrix element is an analytic function $F_{\phi ,\psi
}(z_{1}....z_{n}).$ Different boundary orders correspond to different orders
in the operator product. Since $F$ vanishes in an open set on the boundary
it is (thanks to the generalized Schwarz reflection principle) identically
zero. This proves the theorem, since $\phi $ is then orthogonal to a dense
set of vectors and hence on all vectors and therefore the cyclicity of $\psi 
$ follows.

Taking now locality into account we conclude that if $A\psi =0$ for a vector 
$\psi $ as above, and $A\in \mathcal{A}(\mathcal{O}^{\prime });$ from this
one concludes $BA\psi =AB\psi =0$ for all $B\in \mathcal{A}(\mathcal{O})$
and hence $A=0$ on a dense set and therefore $A\equiv 0$ or in words: $\psi $
is cyclic and separating. The separating property of $\mathcal{A}(\mathcal{O}%
)$ is equivalent to the cyclicity of $\mathcal{A}(\mathcal{O})^{^{\prime }}$
(with respect to the same vector $\Omega )$ and therefore guarantied by the
cyclicity for the spacelike complement $\mathcal{\ A}(\mathcal{O}^{\prime
})\subset \mathcal{A}(\mathcal{O})^{\prime }.$

This is a characteristic property in QFT of finite energy states (in
particular of the vacuum $\Omega $) with respect to local algebras $\mathcal{%
A}(\mathcal{O})$ such that $O^{\prime }$ is not empty i.e. it does not hold
for state vectors in standard QM.

If not handled with great care, one can easily get into pitfalls with
causality\footnote{%
One of the more ''spectacular'' recently published claims about apparant
causality violation has been critical reviewed in \cite{S rem}}. In fact
most of the more sophisticated apparent ``violations of causality'' in which
quantum mechanical properties are attributed to local algebras (including
those through tunneling) are due to some conceptual misunderstanding of QFT.

Literally speaking the R-S theorem says that by applying suitable operators
which have some time duration in a spatially limited laboratory, one can
approximate a state which describes the instantaneous creation of matter
here and antimatter ``behind the moon''. A closer look reveals that such a
sequence of approximations from the dense R-S set require more and more
exotic (increasing energy-momentum costs) pieces of hardware. This suggests
that the limited localization in phase space (i.e. a field theoretic
analogue of the finite number of degrees of freedom of standard QM) and not
just the localization in space-time becomes relevant for the cost balance.
Indeed, the precise formulation of this idea in the form of the ``nuclearity
property'' (of the degrees of freedom counting) has turned out to be
extraordinarily useful. Theories which do not obey this requirement, as e.g.
those with a e.g. ``Hagedorn temperature'', have physical pathologies. The
appropriate definition\footnote{%
Some of the properties already made their first appearance in the chapter 3
on free fields.} is the following:

\begin{definition}
\cite{Haag} A positive energy representation of an observable net is said to
fulfill the ``nuclearity property'' if the set of vectors of the form: 
\[
N_{\beta ,r}:=e^{-\beta H}\mathcal{A}^{(1)}(\mathcal{O}_{r})\Omega 
\]
is nuclear (intuitively: close to a finite set) and its nuclear index (see
mathematical appendix) is suitably bounded by $\beta $ and the size r of the
double cone $\mathcal{O}_{r}$.
\end{definition}

The notation is as follows: $\mathcal{A}^{(1)}(\mathcal{O})$ is the unit
ball of $\mathcal{A}(\mathcal{O}),$ i.e. observables with operator norms $%
\leq 1$. Nuclear means that the set of vectors in $H$ is contained in the
image of $TH^{(1)}$ where $T$ is a trace class operator, and $H^{(1)}$ the
unit ball in $H,$ i.e. that the infinite dimensional sphere in Hilbert space
becomes an ellipsoid with suitably decreasing higher semi-axis. There exists
a equivalent formulation of nuclearity, in which the global Hamiltonian is
replaced by modular operators which have a direct affiliation to local
regions, which is known under the name ``modular nuclearity'' \cite{Haag}.

All free theories and several interacting theories, for which it was
feasible to check such difficult (remote from pointlike fields)
requirements, fall into the range of validity of this property. On the other
hand for models with infinite towers of stable particles, one should be
prepared for violations. The nuclearity property not only secures the
nonperturbative existence of temperature states of infinite systems (whose
existence in the vacuum state has been assumed), but also give rise to a
number of interesting and surprising physical concepts. The most prominent
is the so called ``split property'':

\begin{definition}
For double cones $\mathcal{O}_{1}$ and $\mathcal{O}_{2}$ (or wedges in case
of massive theories) with strict inclusion $O_{1}\ll O_{2}$ (no touching of
boundaries), there exists a canonically constructed type I factor $\mathcal{N%
}$ with: 
\begin{equation}
\mathcal{A}(\mathcal{O}_{1})\subset \mathcal{N}\subset \mathcal{A}(\mathcal{O%
}_{2})
\end{equation}
\end{definition}

\smallskip Its existence can be used in order to split the apace and the
algebras in terms of tensor products as follows:

\begin{eqnarray}
\exists W\,\,H &\rightarrow &H\otimes H\,\,\,\,s.t.\,\,\,H=W^{*}(H\otimes
H)W\,\,\,s.t. \\
\mathcal{A}(\mathcal{O}_{1}) &=&W^{*}(\mathcal{A}(\mathcal{O})\otimes 
\mathbf{1)}W\mathbf{\subset }W^{*}(\mathcal{B}(H)\otimes \mathbf{1})W=%
\mathcal{N}  \nonumber \\
\mathcal{A}(\mathcal{O}_{2})^{\prime } &=&W^{*}(\mathbf{1}\otimes \mathcal{A}%
(\mathcal{O}_{2})^{\prime })W\subset W^{*}(\mathbf{1}\otimes \mathcal{B}%
(H))W=\mathcal{N}^{\prime }  \nonumber
\end{eqnarray}

This split property follows from the nuclearity property (Haag). It is
deeply related to one of the oldest concepts in QFT, the vacuum polarization
(first studied by Heisenberg and later elaborated by Weisskopf). One of the
observations in the old days was that a sharp spatial cutoff (e.g. by a
characteristic step function of a spatial or even a space-time region) leads
to infinite large vacuum fluctuations. This poses the question whether there
is a smoother causal way of splitting into $\mathcal{O}$ and the causal
complement, such that the Hilbert space factorizes in a manner well-known
from the QM box quantization (where such splits into a 3-dim. inside and
outside region are frequently used). If one leaves a ``collar'' around $%
\mathcal{O},$ then the above theorem yields such a factorization. In
ordinary QM, the carved out collar would prevent the tensor product space
from being equal to the full space (or rather one would have to work with
the tensor product of three factors including a factor for the collar).
Thanks to the Reeh-Schlieder theorem, the QFT situation is better. The fact
that one needs the collar in order to achieve the spatial split is very much
related to the hyperfinite type III$_{1}$ property of the $\mathcal{A}(%
\mathcal{O})$ local algebras. For the proof of hyperfiniteness, the
nuclearity enters in an essential way. Hyperfinite type III$_{1}$ factors
are unique (always meant modulo isomorphisms), and therefore the nuclearity
property leads to a universality of local algebras, thus convincingly
confirming the ideas of the founding fathers of algebraic QFT: a single
algebra is void of physical meaning (the ``no hair'' property of algebraic
QFT) and all physical properties reside in the net relations, i.e. in the
position of the algebras relative to each other.

Another technically important property is the following ``property B'' which
is due to Borchers \cite{Haag}.

\begin{theorem}
Let E be a local projector $E\in \mathcal{A}(\mathcal{O}).$ Then there
exists an isometry $V$ localizable in a possibly slightly bigger region $%
\widetilde{\mathcal{O}}\supset \mathcal{O}$ with $E=VV^{*}.$
\end{theorem}

Again this theorem points towards properly infinite (i.e. type $III)$
algebras; if it would not be for the possible enlargement of $\mathcal{O},$
the statement $E=VV^{*},\,\,V^{*}V=1$ (isometry) yields $E\sim 1,$ i.e. all
the projectors are ``infinite'' and therefore $\mathcal{A}(\mathcal{O})$ is
indeed of type $III$. All explicitly known local QFT algebras are actually
hyperfinite factors of type III$_{1}$ in the refined classification theory
of A. Connes. The subscript 1 refers to complete outerness of the action of
modular groups on the algebra, whereas ``hyperfinite'' is somewhat loosely
speaking a property of approximatability by finite degrees of freedoms
(prerequisite for lattice approximations) which can be shown to arise from
the QFT phase space structure which results from the nuclearity requirement.
It holds for the local algebras, but does not necessarily apply to
globalizations as GNS-representations of $\mathcal{A}_{quasi}$ and $\mathcal{%
A}_{univ}.$ Type $III$ are the ``biggest'' von Neumann factors in the sense
that they absorbs any other tensor factor. For the wedge region one can
actually prove that $\mathcal{A}(W)$ is a hyperfinite type III$_{1}$ factor.
Factor algebras are very natural also in physics since they generalize the
notion of irreducibility in those cases where their intrinsic impurity
prevents irreducibility. On the opposite end one finds type II$_{1}$ factors
which are absorbed into any other tensor factor and hence smallest in this
sense. The latter are big enough in order to incorporate all the
generalizations of group symmetry which recently emerged in V. Jones
inclusion theory of subfactors. In algebraic QFT only the so-called
intertwiner subalgebras (associated with the composition and reduction of
charge sectors of observable algebras) are of this kind. These subalgebras
give rise to combinatorial or so called topological QFT. Such small algebras
type II$_{1}$ algebras can only appear because the space-time symmetries
remain outside the intertwiner algebras. External symmetries, in particular
translations, require the presence of infinite projectors (typically type I)
as defined above, and finally restrictions to local algebras (with only
partial space-time symmetries) allow only infinite projectors (type III$_{1}$%
).

In connection with the limitation of energy-momentum and the nuclearity
formalism (mentioned below) it is convenient to have a mathematical
framework which makes precise the concept of energy-momentum transfer. This
was elaborated \cite{Bo book} by the mathematical physicist Borchers and the
mathematician Arveson. The idea is to first introduce a notion of spectrum
of the automorphism. The automorphism $\alpha _{t}$ of the $C^{*}$-algebra
may be extended via (\ref{impl}) to the enveloping von Neumann algebra $\pi (%
\mathcal{A})^{\prime \prime }.$ It should not lead to any confusion if we
stay sometimes with the same symbols for the extended objects. With the help
of L. Schwartz test functions $f\in \mathcal{S}(\mathbf{R})$ we form $\alpha
_{f}(A)=\int dtf(t)\alpha _{t}(A).$ It is easy to see that the extended
automorphism, and therefore $\alpha _{f},$ maps also the von Neumann
extension into itself (since it commutes with elements from the commutant $%
\pi (\mathcal{A})^{\prime }$ inside matrix elements). One now defines the
(Arveson-) spectrum of $A\in \pi (\mathcal{A})^{\prime \prime }$ as: 
\begin{equation}
spec_{\alpha }(A)=\left\{ \omega \in \mathbf{R}\mid \forall
nbhs\,N\;of\;\omega ,\exists \;f\in \mathcal{S}(\mathbf{R})\;s.t.\,supp%
\tilde{f}\subset N,\,\alpha _{f}(A)\neq 0\right\}
\end{equation}
The size of the individual $\alpha _{f}(A)$-contributions is evidently
limited by $supp\tilde{f}.$ We can manufacture operators $A$ with $%
spec_{\alpha }(A)\in I,I$ given, by smoothening a given $B$ with $f,supp%
\tilde{f}\subset I:$%
\[
A:=\alpha _{f}(B) 
\]
The (algebraic) subspaces with energy transfer $\geq E$ are defined as: 
\begin{equation}
\mathcal{A}_{E}=\left\{ A\in \mathcal{A}\mid sp_{\alpha }(A)\subset \left[
E,\infty \right] \right\}
\end{equation}
The usefulness of these concepts begins to show up if one relates this with
projection operators in the Hilbert space $H$ of the representation $(\pi
,U):$%
\begin{equation}
P_{E}:=proj\;on\,\,\bigcap_{E^{\prime }<E}\mathcal{A}_{E^{\prime }}H
\end{equation}
These projectors define a spectral family since they fulfill $P_{E}=1$ for $%
E\leq 0$ and $\lim_{E\rightarrow \infty }P_{E}=0,$ and are in addition upper
continuous. Hence one may associate a ``Hamiltonian'' $\mathbf{H}$ with $%
\alpha _{t}:$%
\begin{eqnarray}
\mathbf{H} &=&\int EdP_{E},\;V(t):=e^{i\mathbf{H}t}=\int e^{iEt}dP_{E} \\
&\curvearrowright &\alpha _{t}(A)=V(t)AV(t)^{-1}  \nonumber
\end{eqnarray}
Since: $\pi (\mathcal{A})^{\prime }\mathcal{A}_{E}H\subset \mathcal{A}_{E}H,$
we find: 
\begin{eqnarray*}
A^{^{\prime }}P_{E} &=&P_{E}A^{\prime }P_{E}=P_{E}A^{\prime }, \\
\curvearrowright \left[ P_{E},A^{\prime }\right] &=&0,\;\forall A^{\prime
}\in \pi (\mathcal{A})^{\prime }
\end{eqnarray*}
or using more physical terminology: the infinitesimal generator $\mathbf{H}$
of the symmetry $\alpha _{t}$ may always be chosen in such a way that $%
\mathbf{H}$ is associated to the algebra (the Arveson-Borchers theorem). The
algebraically determined $\mathbf{H}$ is called the ``minimal'' generator.
Although the general situation, unlike the well-known explicit Sugawara
expression for the chiral translation generator in chiral conformal QFT,
does not lead to a concrete functional expression, this should nevertheless
be interpreted as the generalized analogue of the situation known in
conformal QFT. The extension to abelian groups with several parameters
should be obvious.

In the same vein, but taking in addition locality into account and using
more powerful analyticity tools (''edge of the wedge techniques''), one
proves the following four interesting theorems \cite{Bo book}.

\begin{theorem}
(Locality and the shape of the spectrum) Let $\left\{ \mathcal{A}(\mathcal{O}%
),\mathcal{A},\mathbf{R}^{d},\alpha \right\} $ be a local net with
translation symmetry and positive energy. Let $V(a)$ denote the above
minimal positive energy representation. Then the lower bound of $specV$ is
Lorentz-invariant.
\end{theorem}

This is the counterpart of the classical fact that causal propagation can
only be satisfied with L-covariant equations. As a result of this inexorable
link in the classical theory, Einstein himself never separated the issue of
L-invariance from causality. The content of this theorem is that these
notions continue to stay inexorably linked in the local quantum theory
setting.

\begin{theorem}
(General cluster property) Let $\left\{ \mathcal{A}(\mathcal{O}),\mathcal{A},%
\mathbf{R}^{d},\alpha _{a}\right\} $ be a local net as before and $\omega $
a translation invariant state and $\left\{ \pi ,H,\Omega ,U(a)\right\} $ the
GNS-representation with $U(a)\Omega =\Omega .$ Denote by $P_{0}$ the
projector onto the subspace of pointwise invariant vectors i.e. $\Omega \in
P_{0}H.$ Assume furthermore that the center of $\pi (A)^{\prime \prime }$ is
pointwise invariant under $\alpha _{a}$. Then we have the following
relation: 
\begin{eqnarray}
&&\lim_{\lambda \rightarrow \infty }(\Omega ,\pi (A_{1})\pi (\alpha
_{\lambda b}B_{1})\pi (A_{2})...\pi (A_{n})\pi (\alpha _{\lambda
b}B_{n})\Omega ) \\
&=&(\Omega ,\pi (A_{1}A_{2}...A_{n})P_{0}\pi (B_{1}B_{2}...B_{n})\Omega ) 
\nonumber \\
&=&(\Omega ,\pi (B_{1}B_{2}...B_{n})P_{0}\pi (A_{1}A_{2}...A_{n})\Omega
),\quad b\,\,spacelike  \nonumber
\end{eqnarray}
\end{theorem}

Naturally some of the $A_{i},B_{j}$ may be omittable identity operators,
which allows to have $\#A\neq \#B$\textbf{.} In the case of the unique
vacuum and a spectral mass gap one may prove the strong (faster than any
inverse power in $\lambda $) approach of the right hand side which is the
standard form of the cluster property. This is then the starting point of
the derivation of scattering theory.

\begin{theorem}
(Additivity of spectrum) Let $\left\{ \pi ,H,U(a)\right\} $ be a factor
representation ( a von Neumann algebra with trivial center: $\mathcal{Z}=\pi
(\mathcal{A})^{\prime }\cap \pi (\mathcal{A})^{\prime \prime }$) of a theory
of local observables fulfilling the spectrum condition and assume that $%
U(a)\,$is the minimal representation. Then if $p_{1}$and $p_{2}$ are in $%
specP,$ so is $p_{1}+p_{2}.$ moreover if the mass spectrum consists of a
discrete part $m_{0}<m_{1}<...$and a continuum starting at $m_{c}>m_{i}$
then: 
\[
3m_{0}\geq m_{c} 
\]
\end{theorem}

The expected (from scattering theory) relation $m_{c}=2m_{0}$ remains still
unproven.

\begin{theorem}
(Absence of classical fields) There exist no classical field theories (i.e.
abelian algebras) which fulfill the spectrum condition.
\end{theorem}

\begin{theorem}
Let $\left\{ \mathcal{A}(\mathcal{O}),\mathcal{A},\mathbf{R}^{d},\alpha
\right\} $ be a theory of local observables and define: 
\begin{equation}
\mathcal{A}(a)=\bigcap_{a\in \mathcal{O}}\mathcal{A}(\mathcal{O})
\end{equation}
Then $\mathcal{A}(a)\subset \mathcal{Z}(\mathcal{A}).$
\end{theorem}

Clearly this may be interpreted as an algebraic generalization of the famous
Bohr-Rosenfeld argument on the nonexistence of finite electro-magnetic
quantum field strength at a point i.e. the necessity for smearing (or
avaraging) quantum field strength in order to avoid singularities. This
theorem which is due to Wightman has generalizations to subsets of Minkowski
space. For spacelike 3-d hypersurfaces and for time-like segments the
analogously defined algebras are nontrivial and equal to the algebras of
their causal completions i.e. $\mathcal{A}(\mathcal{O}^{\prime \prime })$

\section{Abstracting Superselection Principles}

If we use fields in standard QFT in order to define local nets of field
algebras $\left\{ \mathcal{F}(\mathcal{O})\right\} ,$ we find the following
properties\footnote{%
The reader interested in technical and conceptual details should follow the
historical path and look at Haag's book as well as the original articles.}
(see also chapter 5):

\begin{itemize}
\item  $\tilde{P}-$covariance, positive energy and uniqueness of the vacuum.

$\exists $ a strongly continuous representation U of the covering of the
Poincar\'{e} group $P_{+}^{\uparrow }:$%
\begin{equation}
U(L)\mathcal{F}(\mathcal{O})U(L)^{-1}=\mathcal{F}(L\mathcal{O})
\end{equation}
and the generators $P_{\mu }$ of the translations satisfy the spectrum
condition $spec\in V^{\uparrow }$ with $P\Omega =0,$ $\Omega $ being the
unique vacuum.

\item  $\exists $ a compact (global gauge) group\footnote{%
Llocal gauge groups are not directly related to symmetries in the same
theory.} $G$ and strongly continuous faithful representation $U$ of $G$
which commutes with the Poincar\'{e} group (factorization of internal and
external symmetries) s. t..: $U(g)\mathcal{F}(\mathcal{O})U(g)^{-1}=\mathcal{%
F}(\mathcal{O}),\quad U(g)\Omega =\Omega $

\item  $\exists \kappa \in G$ of order two i.e. $\kappa ^{2}=1$ s. t.. with $%
\mathcal{F}=\mathcal{F}_{+}+\mathcal{F}_{-}$ , $\alpha _{\kappa }(\mathcal{F)%
}_{\pm }=\pm \mathcal{F}$ and spacelike separated $\mathcal{O}_{1}$and $%
\mathcal{O}_{2};$ the following graded (or ``twisted'') locality relation
holds: 
\begin{eqnarray}
\left\{ \mathcal{F}_{-}(\mathcal{O}_{1}),\mathcal{F}_{-}(\mathcal{O}%
_{2})\right\} &=&0 \\
\left[ \mathcal{F}_{+}(\mathcal{O}_{1}),\mathcal{F}_{-}(\mathcal{O}%
_{2})\right] &=&0=\left[ \mathcal{F}_{+}(\mathcal{O}_{1}),\mathcal{F}_{+}(%
\mathcal{O}_{2})\right]  \nonumber
\end{eqnarray}
We write this in the condensed form: 
\begin{eqnarray}
\mathcal{F}(\mathcal{O}^{\prime }) &=&\mathcal{F}(\mathcal{O})^{tw},\quad 
\mathcal{F}(\mathcal{O})^{tw}:=K\mathcal{F}(\mathcal{O})^{\prime }K^{-1} \\
V\,\,s.t.\,\,V\mathcal{F}(\mathcal{O})V^{-1} &=&\kappa (\mathcal{F}(\mathcal{%
O}))\,\,\,and\,\,K=\frac{1+iV}{1+i}  \nonumber
\end{eqnarray}

\item  Additivity: $\mathcal{F}(\mathcal{O})=\bigvee_{i}\mathcal{F}(\mathcal{%
O}_{i}),\quad \mathcal{O}=\bigcup \mathcal{O}_{i}$

\item  Haag (twisted) Duality: 
\begin{eqnarray}
\mathcal{F}(\mathcal{O}) &=&\mathcal{F}(\mathcal{O}^{\prime
})^{tw}\,\,\curvearrowright \\
\mathcal{A}(\mathcal{O}^{\prime })^{\prime } &\mid &_{H_{0}}=\mathcal{A}(%
\mathcal{O})^{\prime \prime }\mid _{H_{0}},\,\,\,\mathcal{A}(\mathcal{O}):=%
\mathcal{F}(\mathcal{O})\cap U(G)^{\prime }  \nonumber
\end{eqnarray}
where the observable algebra is defined by this invariance principle and the
von Neumann algebra of a noncompact region $\mathcal{N}$, as e.g. the causal
complement of a double cone $\mathcal{O}^{\prime }$, are defined in terms of
an additive covering by double cones $\mathcal{O}_{i}$ together with von
Neumann closure: $\mathcal{A}(\mathcal{N})=\cup _{i}\mathcal{A}(\mathcal{O}%
_{i})$
\end{itemize}

Some comments are in order. Although some of these properties are evident or
at least plausible, I recommend to look up the proofs. The conclusion in the 
$\rightarrow $direction does not hold in the case of d=1+1 where the
order-disorder duality makes its appearance (see chapter 3, section 7). The
quantum intuition acquired from standard QT (as well as from Lagrangian
quantization) is treacherous in local QFT, an area for which a good
intuition still needs to be developed. This applies in particular to the
duality structure.

The notation $\mid _{H_{0}}$ denotes the restriction to the vacuum sector $%
H_{0}$ with: $U(G)H_{0}=H_{0}$ pointwise in $H_{0}$. What is referred to as
the observable algebra in these notes is not $\ \mathcal{A}$ in $H,$ but
rather the smaller (irreducible) algebra $\mathcal{A}\mid _{H_{0}}.$ The
gauge invariant part can also be obtained via the conditional expectation
(averaging with compact group): 
\begin{eqnarray}
m(F) &\equiv &\int_{G}dg\alpha _{g}(F),\quad \int_{G}dg=1 \\
with &:&\text{{}}(1)\text{ }m(\mathcal{F}(\mathcal{O}))=\mathcal{A}(\mathcal{%
O})  \nonumber \\
&&(2)\,\,m\,\,\text{is\thinspace \thinspace normal}  \nonumber \\
&&(3)\,\,m\text{ commutes with }\alpha _{g}\quad \quad \quad \quad  \nonumber
\end{eqnarray}
The continuity property $(2),$ defined in terms of the predual (appendix)
which is also equivalent to $m$ being ``$\sigma $-continuous'', allows the
continuation of $m$ to all operators $B(H)$ i.e. it is the natural kind of
continuity for operations on von Neumann algebras. We obtain $\mathcal{F}%
^{\prime \prime }=B(H),\mathcal{A}^{\prime \prime }=m(\mathcal{F}^{\prime
\prime })$ and hence $\mathcal{A}^{\prime \prime }=m(B(H))$ as well as $%
\mathcal{A}^{\prime }=U(G)^{\prime \prime }$

This gives us the desired tensor decomposition of the Hilbert space: 
\begin{equation}
H=\bigoplus_{\sigma }H_{\sigma }\otimes H_{\sigma }^{\prime }  \label{mul}
\end{equation}
where the first factor $H_{\sigma }$ is the irreducible representation space
for the irreducible representation $U_{\sigma }(G)$ of the internal symmetry
group, and $H_{\sigma }^{\prime }$ denotes its (infinite dimensional)
multiplicity space, which is an irreducible representation space of $%
\mathcal{A}$ corresponding to $\pi _{\sigma }$ . With other words we have: 
\begin{eqnarray}
A\left| _{H_{\sigma }\otimes H_{\sigma }^{\prime }}\right. &=&\mathbf{1}%
_{H_{\sigma }}\otimes \pi _{\sigma }(A)\quad A\in \mathcal{A} \\
U(g)\left| _{H_{\sigma }\otimes H_{\sigma }^{\prime }}\right. &=&U_{\sigma
}(g)\otimes \mathbf{1}_{H_{\sigma }^{\prime }}\quad g\in G  \nonumber
\end{eqnarray}

$\mathcal{A}$ in $H$ contains generally many other irreducible
representations $\pi _{\sigma }$ besides the vacuum representation, and the
primer into the theory of superselection sectors consists in classifying
these $\pi _{\sigma }$, in particular to understand what properties they
share. For this purpose we introduce minimal projectors in the algebra $%
U(G)^{\prime \prime }:$%
\begin{equation}
E=\int dgU(g)\left( \phi ,U_{\sigma }(g^{-1})\phi \right) \quad \phi \in
H_{\sigma }\text{ arbitrary, }\left\| \phi \right\| =1
\end{equation}
Since according to the Reeh-Schlieder theorem $\mathcal{F}(\mathcal{O})$
acts cyclically on $\Omega ,$ we always find elements $F\in \mathcal{F}(%
\mathcal{O})\,$with $EF\Omega \neq 0.$ The definition: 
\[
T\psi =EF\psi \quad \psi \in H_{0} 
\]
determines a partial intertwiner $T:H_{0}\rightarrow EH$ with the
intertwining property: 
\begin{equation}
T\pi _{0}(A)=\pi _{E}(A)T,\quad A\in \mathcal{A}(\mathcal{O}^{\prime })
\end{equation}
The reader easily checks that the vectors $T\Omega \in H_{E}$ and $\left|
T\right| \Omega \in H_{0}$ (since $T^{*}T:H_{0}\rightarrow H_{0})$ have the
same expectation values on $\mathcal{A}(\mathcal{O}^{\prime })$ i.e. induce
the same partial states. Using the Reeh-Schlieder cyclicity one shows that
there are sufficiently many partial intertwiners such that the set of states
over $\mathcal{A}(\mathcal{O}^{\prime })$ in all representation obtained
from the decomposition of $\pi (\mathcal{A})$ on $H$ agree i.e. the
restriction of the net $\mathcal{A}$ to $\mathcal{A}(\mathcal{O}^{\prime })$
gives the same folium (see mathematical appendix) of states independent of
the charge sector $\sigma .$

\begin{theorem}
All irreducible subrepresentations $\pi _{\sigma }$ satisfy the (DHR)
condition: 
\begin{equation}
\pi _{\sigma }\mid _{\mathcal{A}(\mathcal{O}^{\prime })}\simeq \pi _{0}\mid
_{\mathcal{A}(\mathcal{O}^{\prime })}\,\,\forall \,\mathcal{O}\in \mathcal{K}
\label{DHR}
\end{equation}
i.e. the representations of the observable algebra (obtained from an
invariance principle on the field algebra) are unitarily equivalent in the
causal complement of double cones (and more generally any space time region
which admits a nontrivial causal complement).
\end{theorem}

This is taken as a definition of (DHR) compactly localizable representations
for an arbitrary observable net.

\section{Starting the Reverse: the DHR Endomorphisms}

The previous DHR localization condition may be now be taken as the starting
point for the elaboration of the pivotal part of algebraic QFT: the DHR
superselection theory. Let us start with the classification of simple
(abelian) sectors because they are also simpler in the everyday use of the
word. In order to appreciate the following definitions, one should think of
one-dimensional representations of a group $G$ which form a subcategory of
representations closed under compositions. In terms of the projectors $%
E_{\sigma }$ on $H_{\sigma }\otimes H_{\sigma }^{\prime }$ from the previous
section which are elements of the center $\mathcal{Z}(U(G^{\prime \prime }))=%
\mathcal{Z}(\mathcal{A}^{\prime \prime }),$ we have $U(g)E_{\sigma
}=E_{\sigma }U(g)=\chi (g)E_{\sigma })$. In order to understand this
property in terms of observables $\mathcal{A}$ only (without $\mathcal{F})$,
we convince ourselves that the representation $\pi _{\sigma }$ satisfies the
Haag duality property, which up to now we only met in connection with the
vacuum representation: 
\begin{equation}
\pi (\mathcal{A}(\mathcal{O}^{\prime })^{\prime \prime }=\pi (\mathcal{A}(%
\mathcal{O}))^{\prime }\cap \pi (\mathcal{A})^{\prime \prime }  \label{Ha}
\end{equation}
Here the left hand side should be understood to as the von Neumann algebra
generated by $\mathcal{A}(\mathcal{O}_{1})$ for all $\mathcal{O}_{1}\subset 
\mathcal{O}^{\prime }.$ For $\pi $ irreducible the relation is often written
as $\pi (\mathcal{A}(\mathcal{O}^{\prime }))=\pi (\mathcal{A}(\mathcal{O}%
))^{\prime }.$ Replacing $=$ by $\subset ,$ we have the Einstein causality
relation, therefore (\ref{Ha}) represents a strengthening of causality
(maximal, as it turns out).

A $\pi _{\sigma },$ as obtained in the previous section by restriction from
a field algebra $\mathcal{F,}$ fulfills Haag duality since $\mathcal{A}%
(O)^{\prime }=(\mathcal{F}(\mathcal{O})\cap U(G)^{\prime })^{\prime }=%
\mathcal{F}(O)^{\prime }\vee U(G)^{\prime \prime }$ and acting with the
projection $E_{\sigma }$ as well as with $m$ on both sides (those actions
commute) the $U(G)^{\prime \prime }$ is killed and we finally obtain: $\pi
_{\sigma }(\mathcal{A}(\mathcal{O}))^{\prime }=E_{\sigma }(\mathcal{F}(%
\mathcal{O})^{\prime }E_{\sigma }=E_{\sigma }(\mathcal{F}(\mathcal{O}%
)^{\prime }\cap U(G)^{\prime })E_{\sigma }=\pi _{\sigma }(\mathcal{A}(%
\mathcal{O}^{\prime }))^{\prime \prime }.$ In the last step we used the
twisted duality of $\mathcal{F}.$ We will later see that representations of
the observable net $\mathcal{A}$ fulfill Haag duality iff they correspond to
simple sectors.

Let us now start to do the reverse, namely to construct a charge-carrying
field algebra $\mathcal{F}$ from the observable algebra $\mathcal{A}$ and
its DHR (\ref{DHR}) representations. We first must find some good
mathematical concepts to classify the DHR localized representations. The
unitary equivalence of $\pi (\mathcal{A}(\mathcal{O}^{\prime }))$ with $\pi
_{0}(\mathcal{A}(\mathcal{O}^{\prime }))$ in (\ref{DHR}) guaranties the
existence of partial intertwiners i.e. isometries $V$: $H_{0}\rightarrow
H_{\pi }$ with: 
\begin{equation}
V\pi _{0}(A)=\pi (A)V,\quad A\in \mathcal{A}(\mathcal{O}^{\prime })
\end{equation}
We define a representation $\hat{\pi}(\mathcal{A})$ in $H_{0}$ equivalent to 
$\pi (\mathcal{A})$ in $H_{\pi }$ as 
\begin{equation}
\hat{\pi}(A):=V^{-1}\pi (A)V,\quad A\in \mathcal{A}
\end{equation}
By construction this representation agrees with $\pi _{0}$ in $\mathcal{O}%
^{\prime }$. For sufficiently large regions namely $\mathcal{O}_{1}\supset 
\mathcal{O},$ the range of $\hat{\pi}$ is contained in that of $\pi _{0}$
i.e. $\hat{\pi}(\mathcal{A}(\mathcal{O}_{1}))\subset \pi _{0}(\mathcal{A}(%
\mathcal{O}_{1})),$ and hence a fortiori $\hat{\pi}(\mathcal{A})\subset \pi
_{0}(\mathcal{A}).$ This follows by using (vacuum) Haag duality, namely: $%
\left[ \pi _{0}(A^{\prime }),\hat{\pi}(A)\right] =\hat{\pi}\left[ A^{\prime
},A\right] =0$ for $A^{\prime }\in \mathcal{A}(\mathcal{O}_{1}^{\prime
}),\,A\in \mathcal{A}(\mathcal{O})$ and hence $\hat{\pi}(A)\in \pi _{0}(A(%
\mathcal{O}_{1}^{\prime })^{\prime }\subset \pi _{0}(\mathcal{A})$ by Haag
duality. Therefore $\rho $ defined by: 
\begin{equation}
\rho :=\pi _{0}^{-1}\circ \hat{\pi},\quad \rho :\mathcal{A}\rightarrow 
\mathcal{A}
\end{equation}

is an endomorphism of the $C^{*}$algebra $\mathcal{A}$ with the following
remarkable properties:

\begin{itemize}
\item  $\rho $ is localized in $\mathcal{O}$ (notation: loc$\rho \subset 
\mathcal{O}),$ i.e. $\rho (A)=A,$ $A\in \mathcal{A}(\mathcal{O}^{\prime })$

\item  transportable, i.e. $\forall \mathcal{O}_{1},\mathcal{O}_{2}$ with $%
\mathcal{O}_{2}\supset \mathcal{O}_{1}\cup \mathcal{O}\,\,\,\exists U\in 
\mathcal{A}(\mathcal{O}_{2})\;s.t.\,U\rho (A)U^{*}=A$ for $A\in \mathcal{A}(%
\mathcal{O}_{1}^{\prime })$

\item  $\rho (\mathcal{A}(\mathcal{O}_{1}))\subset \mathcal{A}(\mathcal{O}%
_{1}),\quad \forall \mathcal{O}_{1}\supset loc\rho $
\end{itemize}

The very simple proof of these properties is left to the reader. We follow
Haag and use the notation \thinspace $\Delta $ for the set of such $\rho
^{\prime }s$\thinspace \thinspace \thinspace $,$ and denote by $\Delta (%
\mathcal{O})$ the subset of $\rho ^{\prime }s$ with $loc\rho \in \mathcal{O}%
. $

In the constructive approach based on the observable net and its
endomorphisms with the above properties, one defines the \textit{sectors} as
the equivalence classes of $\rho ^{\prime }s$ modulo inner automorphisms.
The following structural investigation of localized transportable
endomorphisms is independent of the dimensionality of the QFT i.e. holds as
well for low dimensional theories. Let us first look at abelian sectors
which by definition are equivalence classes of automorphism i.e. $\rho
^{\prime }s$ with $\rho (\mathcal{A})=\mathcal{A}.$

\begin{theorem}
$\rho $ is automorphism$\Leftrightarrow \pi _{\rho }=\pi _{0}\circ \rho $ is
Haag dual $\Leftrightarrow \rho ^{2}$ is irreducible (no branched fusion)$%
\Leftrightarrow $Ind $\left[ A:\rho (A)\right] =1$ (trivial Jones index)
\end{theorem}

The reader should try to prove it for himself and consult Haag's book, if he
needs more than 5 lines.

In algebraic QFT the Jones index enters through the statistics operators $%
\varepsilon $ which we explain briefly in the sequel. They are special
intertwiners (``Verketter'' in the sense of Schur). An intertwining operator
is a $V\in B(H)$ which links a representation $\pi _{0}\rho $ with a
subrepresentation of $\pi _{0}\sigma $ i.e. $V\cdot \pi _{0}\rho (A)=\pi
_{0}\sigma (A)\cdot V$ \thinspace $\forall A\in \mathcal{A}$. In case that $%
\rho $ is irreducible, the two representations are equivalent and the
intertwining operator becomes a ``charge transporting'' operator. By Haag
duality\footnote{%
In the reverse approach which starts from the observable algebra $\mathcal{A}
$, the Haag duality is postulated for the vacuum representation. If it does
not hold for the original net, one passes to the dual net $\mathcal{A}^{d}$
which fulfills Haag duality by construction.} one obtains $V=\pi _{0}(T)$
with $T\in \mathcal{A}$ and the intertwining relations: 
\begin{equation}
T\rho (A)=\sigma (A)T\quad \forall \,\,\,A\in \mathcal{A}
\end{equation}
The space of self-intertwiners: ($\rho ,\rho )$ is the commutant $\rho (%
\mathcal{A})^{\prime }$ of $\rho (\mathcal{A})$ in $\mathcal{A}$ and by
Schur's lemma, equal to the scalars $\mathbf{C1}$ iff $\rho $ is
irreducible. Therefore, when $\rho $ is irreducible, the linear space of
intertwiners: $\rho \rightarrow \sigma $ is a \textit{Hilbert space within
the algebra of local observables} with the inner product $\left(
T_{1},T_{2}\right) :=T_{1}^{*}T_{2}.$ The notation for the space of
intertwiners $T\,$ from $\sigma $ to $\rho $ is $T\in (\rho ,\sigma ).$

For every pair of DHR endomorphisms there is a unitary local intertwiner $%
\varepsilon (\rho ,\sigma ):\rho \sigma \rightarrow \sigma \rho $ i.e.$%
\varepsilon \in (\sigma \rho ,\rho \sigma )\,.$ This flip operator is called
the \textit{statistic operator}. The collection of statistics operators is
uniquely determined by the coherence relations with local intertwiners and
among themselves: 
\begin{eqnarray}
\varepsilon (\sigma _{1},\sigma _{2})\sigma _{1}(T_{2})T_{1} &=&T_{2}\rho
_{2}(T_{1})\varepsilon (\rho _{1},\rho _{2})\quad \forall \,\,T_{i}:\rho
_{i}\rightarrow \sigma _{i} \\
\varepsilon (\rho _{1}\rho _{2},\sigma ) &=&\varepsilon (\rho _{1},\sigma
)\rho _{1}(\varepsilon (\rho _{2},\sigma ))  \nonumber \\
\varepsilon (\rho ,\sigma _{1}\sigma _{2}) &=&\sigma _{1}(\varepsilon (\rho
,\sigma _{2}))\varepsilon (\rho ,\sigma _{1})  \nonumber
\end{eqnarray}
together with the ``initial conditions'' 
\begin{eqnarray}
\varepsilon (\rho ,id) &=&\varepsilon (id,\rho )=1 \\
\varepsilon (\rho ,\sigma ) &=&1\quad \text{whenever }\sigma <\rho  \nonumber
\end{eqnarray}
where $\sigma <\rho $ means loc$\sigma $ is in the left spacelike complement
of loc$\rho .$ The Artin braid relation is a special consequence of the
above coherence relations 
\begin{eqnarray}
&&\rho _{3}(\varepsilon (\rho _{1},\rho _{2}))\varepsilon (\rho _{1},\rho
_{3})\rho _{1}(\varepsilon (\rho _{2},\rho _{3})) \\
&=&\varepsilon (\rho _{2},\rho _{3})\rho _{2}(\varepsilon (\rho _{1},\rho
_{3}))\varepsilon (\rho _{1},\rho _{2})  \nonumber
\end{eqnarray}
In particular, by assigning the local operators $\rho ^{i-1}(\varepsilon
(\rho ,\rho ))$ to the standard Artin generators $\sigma _{i}$ of the braid
group $B_{n}$ (see chapter1 ) we obtain a unitary representation of the
braid group $B_{\infty }$ in $\mathcal{A}$ which we call the statistics of
the endomorphism $\rho $ (for reasons which soon will become evident).

We introduce a conjugate endomorphism $\bar{\rho}$ to $\rho $ by demanding
that $\bar{\rho}\rho $ contains the vacuum sector, i.e. that there exists an
intertwiner $R\in (id,\bar{\rho}\rho )$ which induces a standard left
inverse $\phi $ of $\rho $%
\begin{equation}
\phi (A)=R^{*}\bar{\rho}(A)R\quad \forall \,\,A\in \mathcal{A}
\end{equation}
with finite statistics. Here we recall that the left inverse of an
endomorphism $\rho $ of $\mathcal{A}$ is a normalized positive linear map
satisfying the relation $\phi (\rho (A)B\rho (C))=A\phi (B)C.$ It is called
regular if it is of the above form, and standard, if in addition the
statistics parameter $\lambda _{\rho }:=\phi (\varepsilon (\rho ,\rho ))\in
\rho (\mathcal{A})^{\prime }$ is a nonvanishing multiple of a unitary (which
then depends only on the sector $\left[ \rho \right] )$. A sufficient
condition for the existence of a standard left-inverse and therefore of a
conjugate is that there is \textit{some} left-inverse with statistics
parameter $\lambda _{\rho }\neq 0$ (``finite statistics'') and that $\rho $
is translation covariant with positive energy condition. The uniqueness of
the standard left inverse is a consequence of its definition. Any theory
with a mass gap possesses a standard left inverse \cite{FRS1}. The standard
left inverse of $\rho $ turns out to be a trace on $\rho (\mathcal{A}%
)^{\prime }.$ The inverse modulus of $\lambda _{\rho }$ is called the 
\textit{statistical dimension} $d(\rho )\equiv d_{\rho }\geq 1.$ One easily
proves that $\lambda _{\rho }=\lambda _{\bar{\rho}}.$ For irreducible $\rho
^{\prime }s$ we have $\lambda _{\rho }=\frac{\kappa _{\rho }}{d_{\rho }}$
with $\kappa _{\rho }$ being the \textit{statistical phase. }If one computes
this numbers using the field formalism presented below, one finds $d_{\rho
}=dimH_{\rho }$ and $\kappa _{\rho }=\pm 1$ for Bosons/Fermions. In fact for
d=3+1 the statistics operator is easily shown on general grounds to fulfill $%
\varepsilon ^{2}=1$ (i.e. absence of monodromies) which leads to permutation
group statistics. The concepts are much richer in the case of braid group
statistics. Even in that case one succeeds to prove the identity of the spin
phase with the above statistics phase i.e. the spin-statistics theorem.

We will not enter a presentation of V. Jones inclusion (subfactor) theory,
but just mention that Ind$\left[ A:\rho (A)\right] =d_{\rho }^{2},$ i.e. the
inclusion index is the square of the statistical dimension.

In order to understand the reconstruction in the case of proper
endomorphisms i.e. for $\rho ^{\prime }s$ with Ind$\left[ A:\rho (A)\right]
>1,$ we need some more conceptual preparation. This is obtained by briefly
returning to the field algebra $\mathcal{F}$ in the case where $\mathcal{A}$
is the fixed point algebra under a nonabelian $G.$ In that situation an
irreducible endomorphism $\rho $ with $\pi _{0}\circ \rho \simeq \pi _{\rho
} $ and loc$\rho \subset \mathcal{O}$ gives via $\omega =\omega _{0}\circ
\rho $ a pure state localized in $\mathcal{O}$. The big Hilbert space $H$ in
which $\mathcal{F}$ acts has many vectors which induce $\omega :$%
\begin{equation}
H_{\omega }=\left\{ \phi \in H\mid (\phi ,A\phi )=\omega (A)\left\| \phi
\right\| ^{2}\right\}
\end{equation}
As the notation already anticipates, $H_{\omega }$ is a Hilbert space, a
fact which is easily verified using the purity of $\omega .$ Its dimension
is equal to the dimension $d_{\rho }$ of the $H_{\rho }$- tensor factor.
Physical intuition tells us that such vectors in $H_{\omega }$ can be
created from the vacuum by applying charge-carrying fields in $\mathcal{F}$.
In fact we have:

\begin{theorem}
Every $\phi \in H_{\omega }$ determines uniquely a field operator $\psi \in 
\mathcal{F}(\mathcal{O})$ with $\psi ^{*}\Omega =\phi $ and the intertwining
property $\psi A=\rho (A)\psi ,\,A\in \mathcal{A}.$
\end{theorem}

Encouraged by the intertwining relation in the previous theorem, we define
the following linear subspace of $\mathcal{F}:$%
\begin{equation}
H_{\rho }=\left\{ \psi \in \mathcal{F}\mid \psi A=\rho (A)\psi ,A\in 
\mathcal{A}\right\}
\end{equation}
The notation suggests the structure of a Hilbert space. Indeed for two
vectors $\psi _{i},i=1,2$ we have the following scalar product: 
\begin{equation}
\psi _{1}^{*}\psi _{2}\in \mathbf{C\cdot 1}
\end{equation}
The reason is that the inclusion of $\mathcal{A}\in \mathcal{F}$ is
irreducible i.e. $\mathcal{A}^{\prime }\cap \mathcal{F}=\mathbf{C\cdot 1.}$
This follows from $\mathcal{A}^{\prime }\cap \mathcal{F}=U(G)^{\prime \prime
}\cap \mathcal{F}$ and the statement that for any element $F_{0}$ with $%
F_{0}\mid _{H_{0}}=c\mathbf{1\mid }_{H_{0}}$ from the latter algebra the
conditional expectation of $F^{*}F$ with $F:=F_{0}-c\mathbf{1}$ vanishes i.e.%
$\pi _{0}(m(F^{*}F))=0$ hence $m(F^{*}F)=0,$ since $\pi _{0}$ is faithful.
But the expectation values of $m$ in any vector state $\phi \in H$ may be
written as an average with a positive integrand: 
\begin{eqnarray}
&&0=(\phi ,m(F^{*}F)\phi )=\int dg(\phi ,\alpha _{g}(F^{*}F)\phi ) \\
&\curvearrowright &\alpha _{g}(F^{*}F)=0\curvearrowright
F^{*}F=0\curvearrowright F_{0}=c\mathbf{1,}\,\,\,qed.  \nonumber
\end{eqnarray}
Algebraic Hilbert spaces of isometries inside von Neumann algebras can only
occur for von Neumann algebras of properly infinite type.

\begin{theorem}
For any set of field operators ($F_{i})_{i=1...d_{\rho }}\in \mathcal{F}(%
\mathcal{O})$ transforming like an irreducible tensor representation $%
U_{\rho }$ there exists a $\rho \in \Delta (\mathcal{O})$ and a $B\in A(%
\mathcal{O})$ s.t. $F_{i}=B\psi _{i}$ with ($\psi _{i})\in H_{\rho }$ a
orthonormal system of isometries spanning $H_{\rho }.$ The endomorphism $%
\rho $ is implemented by the $\psi _{i}^{\prime }s:$%
\begin{eqnarray}
\rho (F) &=&\sum_{i=1}^{d_{\rho }}\psi _{i}F\psi _{i}^{*} \\
\text{ }\phi (F) &\equiv &\frac{1}{d_{\rho }}\sum_{i=1}^{d_{\rho }}\psi
_{i}^{*}F\psi _{i},\,\,\,\phi (\rho (F))=F  \nonumber
\end{eqnarray}
The last property is the reason why $\phi $ is called the left-inverse of $%
\rho .$ The most interesting and useful emerging structure is the so called
Cuntz algebra $O_{d}$ i.e. the unique $C^{*}-$algebra generated by a family
of isometries $\left\{ \psi _{i}\right\} $ with a full range i.e. with $\sum
\psi _{i}\psi _{i}^{*}=\mathbf{1.}$ A detailed investigation (not done here)
reveals that this is a $\mathbf{Z}$-graded simple $C^{*}$-algebra i.e.
without two-sided ideals. Doplicher and Roberts observed that this algebra $%
O_{d}$ makes a perfect model for a characterization of the group dual which
is appropriate for the encoding of internal symmetries in QFT. The reason is
that since each compact group $G$ is a subgroup of some $U(d)$ for
sufficiently large N, there is a natural action $\alpha $ on $O_{d}$
(summation convention)$:$%
\begin{equation}
\alpha _{g}(\psi _{i})=\psi _{j}g_{ji},\quad \text{unitary in }H_{d}
\end{equation}
The tensor product structure is naturally contained in $O_{d}$ since $%
H^{k}\simeq \otimes ^{k}H.$ The fixed point algebra: 
\[
O_{G}=\left\{ A\in O_{d}\mid \alpha _{g}(A)=A\,\,\forall g\in G\right\} 
\]
gives rise to an inclusion $O_{G}\subset O_{d}$ which turns out to encode
the group structure, in analogy with Galois theory. It naturally contains
all the intertwiners of tensor representations $T:H^{\otimes r}\rightarrow
H^{\otimes s}.$ In terms of endomorphisms these may be characterized purely
algebraically i.e. without tensor products: 
\begin{equation}
T\rho ^{r}(A)=\rho ^{s}(A)T,\quad A,T\in O_{G}
\end{equation}
This should be compared with the ``classical'' Tanaka-Krein theory of group
duals in terms of representation spaces and intertwiners. In QFT based on
observables $\mathcal{A}$ only, one only knows the $\rho ^{\prime }s$ of $%
\mathcal{A}$ and neither the $H_{\rho }^{\prime }s$ nor $\mathcal{F}$. So
the question how to construct from an algebra of intertwiners a bigger
algebra with a group action is a ``baby version'' of the QFT symmetry
problem: how to reconstruct the symmetry from its shadow it leaves on the
observables (in analogy to the famous problem of Marc Kac: ``how to hear the
shape of a drum?''). For a successful treatment we must make sure that our
representation category i.e. the endomorphisms and their intertwiners are
big enough in order to contain the conjugates (antiparticle representations
in QFT). This is easily achieved by securing the existence of a faithful
selfconjugate representation because the tensor products of such a
representation contain every irreducible representation. Let us briefly look
at the special case $SU(d).$ The first tensor power which contains the
identity representation is $H^{d},$ explicitly the first invariant is: 
\begin{equation}
S=\frac{1}{\sqrt{d!}}\sum_{P\in S_{d}}sign(P)\psi _{P(1)}....\psi _{P(d)}
\end{equation}
Hence: 
\begin{equation}
\hat{\psi}_{i}=\frac{1}{\sqrt{(d-1)!}}\sum_{\stackunder{P(1)=i}{P\in S_{d}}%
}sign(P)\psi _{P(2)}....\psi _{P(d)}
\end{equation}
fulfill $\hat{\psi}_{i}=\sqrt{d}\psi _{i}^{*}S$ and therefore is a basis in $%
H_{d}^{conj}$ and $\bar{\rho}(A)=\sum_{i=1}^{d}\hat{\psi}_{i}A\hat{\psi}%
_{i}. $ Thus $O_{SU(d)}$ contains all irreducible representations of $SU(d)$
and every intertwiner. But how do we recognize that a $C^{*}$-algebra is
isomorphic to $O_{SU(d)}?$ The answer is surprisingly simple: in addition to
the operator $S$ we must find a copy of the (infinite) permutation group $%
S_{\infty }$. The model theory $O_{SU(d)}\subset O_{d}$ as such a
presentation: 
\begin{eqnarray}
\varepsilon (P) &=&\sum \psi _{(\alpha )}\psi _{(\alpha _{P})}^{*},\quad
P\in S_{n}\subset S_{\infty },\quad S_{n}\hookrightarrow S_{n+1} \\
(\alpha ) &=&(\alpha _{1},....,\alpha _{n}),\,\,(\alpha _{P})=(\alpha
_{P(1)}....\alpha _{P(n)})  \nonumber
\end{eqnarray}
where we used a multiindex notation. In particular the formula for the basic
transposition is: 
\begin{equation}
\varepsilon ((12))=\sum_{i,j}\psi _{i}\psi _{j}\psi _{i}^{*}\psi
_{j}^{*}=\pm \rho (U^{*})U,\quad \pm :\,F/B
\end{equation}
where in this Fermi/Bose alternative $U$ is an auxiliary charge transporter: 
$\psi _{i}^{\prime }=U\psi _{i}$ which shifts the charge into loc$\psi
_{i}^{\prime }\subset ($loc$\psi _{i})^{\prime }.$
\end{theorem}

One easily checks that the $\varepsilon ,S$ and $\rho $ are related by: 
\begin{eqnarray}
SS^{*} &=&E_{d}:=\frac{1}{d!}\sum_{P\in S_{d}}sign(P)\varepsilon (P) \\
S^{*}\rho (S) &=&(-1)^{d-1}d^{-1}\mathbf{1}  \nonumber \\
\rho (\varepsilon (P)) &=&\varepsilon (P^{\prime }),\quad P^{\prime }\in
S_{n+1},\quad P^{\prime }(1)=1,\,P^{\prime }(i+1)=P(i)\,\,i=1....n  \nonumber
\\
\phi (\varepsilon (P)) &=&\left\{ 
\begin{array}{l}
\varepsilon (P)\,\,\,\,\,\,\,P(1)=1 \\ 
\frac{1}{d}\varepsilon (P^{\prime })\;\;P(1)\neq 1,\quad P^{\prime
}(i)=((1P(1))P)(i+1)
\end{array}
\;\right.  \nonumber
\end{eqnarray}
Here $E_{d}$ is the antisymmetric representation projector in the $S_{d}$
group algebra and $\rho $ and its left inverse $\phi $ implement right and
(partial) left shifts on $S_{\infty }.$ The algebra $O_{SUd)}$ is generated
by these permutation group elements $\varepsilon (P)$ and the $S$%
-intertwiners (the so-called Brower elemements). If $G\subset SU(d)$ then $%
O_{G}\supset O_{SU(d)}$ and therefore there are more generators.

We will give the DR characterization of $G$ without proof:

\begin{theorem}
\cite{DR} Let $\hat{O}$ be a simple $C^{*}-$algebra with an endomorphism $%
\hat{\rho}$ and a unitary representation $\hat{\varepsilon}$ of $S_{\infty }$
with the following properties:

(i) \thinspace \thinspace $\hat{\varepsilon}(P)\in (\hat{\rho}^{n},\hat{\rho}%
^{n}),\,\,P\in S_{n}$

(ii)$\quad \hat{\varepsilon}((12...n+1))\hat{T}=\hat{\rho}(\hat{T})\hat{%
\varepsilon}((12...m+1)),\,\,\hat{T}\in (\hat{\rho}^{n},\hat{\rho}^{n})$

(iii)\thinspace \thinspace \thinspace $\exists \hat{S}\in (id,\hat{\rho}%
^{d}) $ with $\hat{S}^{*}\hat{S}=\mathbf{1},\,\,\hat{S}^{*}\hat{\rho}(\hat{S}%
)=(-1)^{d-1}\frac{1}{d}\mathbf{1}$ and $\hat{S}\hat{S}^{*}=\hat{E}_{d}$

(iv)\thinspace \thinspace \thinspace $\hat{O}$ is generated by the
intertwiners $\hat{T}\in $\thinspace $(\hat{\rho}^{n},\hat{\rho}%
^{m}),\,n,m\in \mathbf{N}$

Then there is a closed subgroup $G\in SU(d)$ (unique up to conjugation) and
an embedding of $\hat{O}$ in $O_{d}$ with $\hat{O}\in O_{G},$ s. t. $\rho
\mid _{O_{G}}=\hat{\rho},\hat{\varepsilon}=\varepsilon ,$ and $\hat{S}=S$
hold.
\end{theorem}

The basic idea on which the proof relies is actually reasonably simple: one
takes a kind of amalgamated product $\mathcal{B}$ of $\hat{O}$ with $O_{d}$
amalgamated over $O_{SU(d)}\subset \hat{O},$ i.e. we look for an algebra
with the following relations:

\begin{itemize}
\item  $\psi _{i}A=\hat{\rho}(A)\psi _{i},\quad A\in \hat{O},\quad \psi
_{i}\in O_{d},\,\,i=1....d$

\item  $\hat{\varepsilon}(P)=\varepsilon (P)$

\item  $\hat{S}=S$
\end{itemize}

The $SU(d)$ action on $\mathcal{B}$ is: 
\begin{equation}
\alpha _{g}(\psi _{i})=\sum_{i}\psi _{i}g_{ji},\quad \alpha _{g}(A)=A
\end{equation}
For the special case that $\hat{O}$ is generated by $\hat{\varepsilon}(P)$
and $\hat{S},$ and hence $\hat{O}=O_{SU(d)}\subset O_{d},$ $\mathcal{B}$ is
obviously $O_{d},$ as expected. If $\hat{O}$ is genuinely bigger, there
exist intertwiner $\hat{O}\ni T\notin O_{SU(d)}.$ The operators $\hat{T}%
_{(\alpha )(\beta )}:=\psi _{(\beta )}^{*}\hat{T}\psi _{(\alpha )}$ commute
with $A\in \hat{O}$ and hence $T_{(\alpha )(\beta )}\in \hat{O}\cap \mathcal{%
B}.$ Actually such operators are automatically in the center of $\mathcal{B}$%
, i.e. $\hat{O}^{\prime }\cap \mathcal{B}=\mathcal{Z}(\mathcal{B}).$ This
follows from the invariance of $\hat{O}^{\prime }\cap \mathcal{B}$ under the
action of $SU(d)$ which means that this subalgebra consists of invariant $%
SU(d)$ tensors $F_{i}$. With $F_{i},i=1....n$ being a tensor multiplet, we
determine an orthonormal basis $\tilde{\psi}_{i},i=1....n$ and obtain the
representation: 
\begin{equation}
F_{i}=B\tilde{\psi}_{i},\quad B=\sum_{i}F_{i}\tilde{\psi}_{i}^{*}
\end{equation}
If we could show that the $F_{i}$ commute with the generators of the Cuntz
algebra, we would be done. But this is an easy computational result of : 
\begin{equation}
B\in (\hat{\rho}^{n},id),\quad \tilde{\psi}_{i}\in (id,\rho ^{n})
\end{equation}
which according to the assumption (ii) of the previous theorem yields: 
\begin{eqnarray}
\hat{\rho}(B) &=&B\hat{\varepsilon}(n+1,....,1) \\
\rho (\tilde{\psi}_{i}) &=&\varepsilon (1,.....,n+1)\tilde{\psi}_{i} 
\nonumber
\end{eqnarray}
and hence $\psi _{j}F_{i}=\psi _{j}B\tilde{\psi}_{i}=\hat{\rho}(B)\rho (%
\tilde{\psi}_{i})\psi _{j}=......=\phi _{i}\psi _{j},$ Q.E.D..

Let us now return to the problem of construction of the field algebra. A
helpful and informative intermediate construction is the introduction of the
so-called ``reduced field bundle''. This is a bimodule over $\mathcal{A}$
which allows to use the $\rho ^{\prime }s$ in a direct manner.

\begin{definition}
As our Hilbert space for the reduced field bundle we take the direct sum of
vacuum Hilbert spaces $H_{\alpha }:=(\alpha ,H_{0})$ and define operators $F$%
($e,A)$ with $A\in \mathcal{A},$ $e=(\rho _{\beta },\rho ,\rho _{\alpha })$
where All the irreducible endomorphisms are taken from a pre-selected set
with one endomorphism $\rho _{\alpha }$ per sector $\left[ \rho _{\alpha
}\right] .$ We define the action of $F$ as: 
\begin{equation}
F(e,A)\cdot (\alpha ,\psi )=(\rho _{\beta },\pi _{0}(T_{e}^{*}\rho
_{a}(A))\cdot \psi )
\end{equation}
where $T_{e}\in \mathcal{A}$ are intertwiners from the space $T_{e}\in (\rho
_{\beta },\rho \rho _{\alpha }).$ Here $e$ may be pictured as an edge on a
fusion graph whose vertices are set of ``charge edges'' i. e. a triples of
source charge $s(e)=\rho _{\alpha }$ range charge $r(e)=\rho _{\beta }$ and
transferred charge $c(e)=\rho .$
\end{definition}

The $F^{\prime }s$ generate a $C^{*}$algebra $\mathcal{F}_{red}$ in $%
H=\oplus _{\alpha }H_{\alpha }.$ The commutation relations have the form of
an exchange algebra: 
\begin{eqnarray}
F(e_{2},A_{2})\cdot F(e_{1},A_{1}) &=&\sum_{f_{1}\circ f_{2}}R_{f_{1}\circ
f_{2}}^{e_{2}\circ e_{1}}(\pm )\cdot F(f_{1},A_{1})\cdot F(f_{2},A_{2}) \\
R_{f_{1}\circ f_{2}}^{e_{2}\circ e_{1}}(+) &=&T_{e_{2}}^{*}T_{e_{1}}^{*}\rho
_{\alpha }(\varepsilon (\rho _{2},\rho _{1})T_{f_{2}}T_{f_{1}},\quad
R(-):\rho _{1}\Leftrightarrow \rho _{2}  \nonumber
\end{eqnarray}
whenever $F_{2}$ is localized in the right /left spacelike complement of $%
F_{1}.$ In low dimensional theories where there is an invariant distinction
the $R^{\prime }s$ are related to the braid group whereas in d=3+1 one deals
with the special case, the permutation group.

This reduced field bundle only agrees with the field algebra of the standard
approach if $G$ is abelian. Its Hilbert space lacks the group theoretic
multiplicities incorporated in the formula (\ref{mul}) and the net inclusion 
$\mathcal{A}\subset \mathcal{F}_{red}$ is reducible and its index is the
square of the index of the irreducible inclusion $\mathcal{A}\subset 
\mathcal{F}$ (assuming finite $G).$ For d=3+1 this is the parastatistics%
\footnote{%
Beware that Kadanoff et al. use this terminology in conformal field theory
with a different meaning.} description which deals with higher Young
tableaux, but without an internal symmetry group.

The parastatistics fields are more noncommutative and do not allow an
interpretation in terms of ``quantization'' (e.g. they have no Lagrangians).
In case of the nonabelian braid group statistics of chiral conformal field
theory and d=2+1 plektons this is the only available description. In that
case the $F_{red}$-algebra was also called the ``exchange algebra'' \cite{RS}%
.

In case of d=3+1 theories and for the subcategory of permutation group
statistics sectors in d\TEXTsymbol{<}3+1 there exists the famous canonical
Doplicher-Roberts construction of a genuine field algebra in the sense of
the beginning of this section.

Let us first mention the special case of only automorphisms and assume
d=3+1, i.e. permutation group statistics. Since for automorphisms $d_{\rho
}=1,$ the permutation group representation is abelian, we are dealing with
Bosons/Fermions. In this case some very simple modifications of the reduced
field bundle will give the field algebra. The interested reader is referred
to \cite{Haag} page 185, 186. This case may also be obtained by
specialization of the construction of the reduced field bundle.

For the above mentioned nonabelian representations one applies the DR
theorem \cite{DR} about the construction of the group from a subalgebra of
the Cuntz algebra with a distinguished endomorphism. Without loss of
generality, we may assume that there exists a $\rho $ with statistical
dimension d s. t.. $id\subset \rho ^{d}$ which can always be achieved by
adding conjugates. A $C^{*}$-algebra $\hat{O},$ as needed in the theorem,
may be obtained via the inductive limit of intertwiner spaces: 
\begin{equation}
^{0}\hat{O}:=\bigcup_{n,m\geq 0}(\rho ^{n},\rho ^{m})
\end{equation}
where the induction uses the ``right'' embedding ($\rho ^{n},\rho
^{m})\rightarrow (\rho ^{n+1},\rho ^{m+1}):T\rightarrow T\times 1_{\rho }.$
This leads to a natural composition of two operators $S$ and $T$ by
embedding both in a space sufficiently shifted to the right. The algebra
contains the statistics operators $\varepsilon (P)$ (still in bosonized
form) as well as an isometry $S\in (id,\rho ^{d}),\,\,SS^{*}=E_{d}.$ The
endomorphism $\rho $ acts in a natural way on $^{0}\hat{O}$ and $\phi
(.)=S^{*}\rho (.)S$ defines a left inverse of $\rho .$ The properties of $%
\phi ,\rho ,\varepsilon (P)$ and $S$ are easily checked by computation. $^{0}%
\hat{O}$ has a unique $C^{*}$-norm and no ideal (i.e. $\hat{O}$ is simple).
Now the DR theorem leads to the identification with a subalgebra $%
O_{G}\subset O_{d}$ with a subgroup $G$ of $SU(d)$ which is determined up to
conjugation. The field algebra is now simply the free product of the
observable algebra $\mathcal{A}$ with the Cuntz algebra $O_{d}$ amalgamated
over the subalgebra $O_{G}$ with the following relations: 
\begin{equation}
\psi _{i}A=\rho (A)\psi _{i},\quad A\in \mathcal{A}
\end{equation}

\section{Remarks on Broken Symmetries}

The idea of spontaneously broken symmetries originated during the 60$^{ies}$
in Lagrangian QFT (Goldstone, Nambu). There were parallel developments in
condensed matter physics, in which case the understanding of the phase
transitions in the Heisenberg model was the main goal. Already by the end of
this decade there was a general model independent understanding \cite{Swie}
within the framework of QFT's possessing conserved quantum Noether currents
independent of their (Lagrangian or non-Lagrangian) origin. The main theorem
of this more general approach within the Wightman framework was the relation
of the nonexistence of the global charges (as a result of large distance
infrared divergencies in the spatial integrals over currents) with the long
distance property of the matrix elements of the current operator between
certain vector states as a result of the presence of ``Goldstone Bosons'' in
the energy-momentum spectrum. The nice feature of these rigorous methods is
that they apply to composite ``Goldstones'' ( i.e. they go beyond the family
of Goldstone Lagrangians for which a perturbative approach to the broken
phase is possible) as well. In this way the statement became a structural
theorem of general QFT.

Algebraic QFT offers an even more profound physical picture which we are
going to explain in the sequel. The starting point is the DR reconstruction
theory of the previous section. That theory deals with an unbroken symmetry $%
G$ because only those transformations are in a one to one correspondence
with the superselection sectors. Where to look for the bigger spontaneously
broken group $\Gamma $? The answer is contained in the breakdown of the
vacuum Haag duality of $\mathcal{A}$ \cite{Roberts}. The physical reason for
this is that certain operators which, if one only looks at their local
properties, carry charges and transform according to $\Gamma $-multipletts,
globally ``condense'' into the vacuum sector.

We have met a special case of this phenomenon in connection with the d=1+1
order/disorder discussion in the last section of chapter 3. The main point
there was that the original net violated the vacuum Haag Duality and the
order/disorder fields were required precisely in order to restore it. By
definition we called the field which did not belong to the original vacuum
representation, but has a nonvanishing vacuum expectation ``disorder''. It
is the adjunction of this field, which enlarges the observable net $\mathcal{%
A}$ to the Haag dual net $\mathcal{A}^{d}.$ Adjusting this to the situation
at hand, we assume that our original observable net $\mathcal{A}$ is smaller
than its unique dual extension $\mathcal{A}^{d}$ i.e.: 
\begin{equation}
\mathcal{A}\subset \mathcal{A}^{d}\subseteq \mathcal{F}
\end{equation}
where $\mathcal{F}$ is the unique DR field algebra determined by the
superselection theory of $\mathcal{A}^{d}$ . The DR group $G$ is the
unbroken gauge group $\Gamma \supset G,$ with $\Gamma $ defined to be the
group of automorphisms of $\mathcal{F}$ which leave $\mathcal{A}$ pointwise
fixed. $G$ is the unbroken part of $\Gamma .$ The following theorem
demonstrates the correctness of this interpretation.

\begin{theorem}
\cite{BDLR}

(i)\thinspace \thinspace \thinspace \thinspace Each $\gamma \in \Gamma $
leaves $\mathcal{F}(\mathcal{O})$ for each $\mathcal{O}\in \mathcal{K}$
globally stable and is locally normal.

(ii)\thinspace \thinspace \thinspace \thinspace $G$ is the $\mathcal{F}$%
-vacuum stabilizer in $\Gamma $

(iii)\thinspace \thinspace \thinspace \thinspace The normalizer of $G$ in $%
\Gamma $ is the invariance subgroup which act automorphically on $\mathcal{A}%
^{d}$
\end{theorem}

\textit{The Goldstone theorem (}or better the theorem on the Goldstone
mechanism), i.e. the prediction of a special kind of zero mass particle as a
result of spontaneous symmetry breaking, only \textit{follows} under more
stringent conditions and the \textit{standard situation of a conserved
Noether current} is certainly one such possibility. In order to understand
better the physics involved, let us look at the vacuum expectation of the
derivation defined by the generator $\delta $ of the automorphism, using
one-parametric subgroup: 
\begin{equation}
\delta (F)=\lim_{\lambda \rightarrow 0}\lambda ^{-1}(\gamma _{\lambda }(F)-F)
\end{equation}
The criteria for a spontaneous symmetry breaking in the general setting of
algebraic QFT without or with Goldstone particles are then formulated in
terms of behavior of the vacuum expectation of $\delta (F)$ for increasing
localization regions.

In low-dimensional QFT the charge sectors associated to an observable net
generally cannot be described in terms of group theoretical notions. Lacking
a group theoretic characterization of the position of the observable algebra
as the fixed points inside the field algebra under a compact group action,
one takes the breakdown of Haag duality for double cones in the vacuum
representation as the definition of ``broken quantum symmetry''. A
discussion in terms of Noether currents a la Goldstone and Nambu, if
possible at all in this case, is still missing.

From Lagrangian field theories one knows another mechanism of symmetry
breaking which was first conjectured and exemplified by Schwinger and then
brought into a perturbative setting by Higgs. Since it needs the formalism
of gauge theory, its intrinsic content has never been spelled out; up to
this date there is no rigorously known property of the observable part of
the theory which tells us that a massive particle received its mass in such
a way. Most of the folklore around this mechanism is not quite correct. For
example the idea that the mechanism could be thought of as a ``fattened''
Goldstone boson is contradicted by the Schwinger model, because in d=1+1
there aren't any Goldstone bosons but there is the Schwinger-Higgs
mechanism. In fact different from the previous Goldstone-Nambu mechanism of
spontaneously broken symmetries, the terminology ``spontaneously broken
gauge invariance'' is more a mnemotechnical calculational device than a
physical meaningful intrinsic property of the theory. However this does not
mean that there are no consistent nonperturbative conjectures which have
some chance to be proven in algebraic QFT. One appealing idea is to think of
a ``would be'' charged field with a Maxwellian (i.e. very nonlocal) charge 
\footnote{%
Formally a semiinfinite extended object formed from a Dirac field of QED
modified by a Mandelstam like $A_{\mu }$-flux to spatial $\infty $ serves as
a candidate for a local gauge invariant, but global U(1) charge carrying
field.} which is carried by a nonlocal (string-localized) matter field and
becomes screened as a consequence of the emergence of a massive vector
meson. To argue that such a ``phase'' of Maxwellian QFT exists was an
important contribution of Schwinger. The rigorous theorem about the charge
screening$\leftrightarrow $massive ``photon'' relation is again due to
Swieca \cite{Swie} In fact in order to show the existence of this mechanism,
Schwinger invented the ``Schwinger model''. In chapter 4.5 we demonstrated,
that the closely related opposite interpretation of ``charge liberation'' is
actually the more natural one and that there is no intrinsic physical
meaning to (Higgs) condensates. The understanding of the massless limit of
the selfinteracting massive vectormeson is actually the difficult part in
perturbation theory, since it requires the transition from local to nonlocal
string-like matter field coordinates (in order to get a matter field which
allows a massless limit). Only in the Schwinger model this process has been
carried out \cite{Buch}. As a result the semiinfinite string-like charge
carrier disappears from the physical stage and the resulting massive theory
will be more local (good infrared properties, in the sense of the LSZ mass
gap framework)) in the usual sense. There is still a lot of nonunderstood
physics hidden behind the deceiving terminology ``spontaneously broken gauge
symmetries''.

The concept of spontaneous symmetry breaking is also very important for
thermal QFT. The exploration of the physical consequences is however more
subtle than in the case of ground states as a result of the loss of Wigner's
particle concept..

All internal and external symmetries with one exception, suffer at most a
spontaneous symmetry breaking. The exception is supersymmetry which, as we
already mentioned in the introduction, collapses completely. The reason for
this radically different behavior is that supersymmetry is an ``accidental
symmetry''. By this we mean a symmetry of a field algebra, which is not
visible in the structure of charge sectors of the observable algebra and
plays no important role in the understanding of the model. An illustrative
example is the tricritical Ising model of chiral conformal field theory. Its
observable algebra completely determines all its (finite number of) charge
sectors. The fact that one can extend the energy-momentum tensor algebra by
fermionic sectors and obtain in this way an action of supersymmetry is only
an ornamental attribute and does not make the model any simpler than the
other models in the same family of minimal models. The physical content of
the model can be explored without taking notice of its supersymmetry. This
is a general observation: whenever supersymmetry leaves the twighlight of
its folklore and enters the clear air of controllable models (which
presently only happens in low-dimensional QFT), it reveals its accidental
structure. Too little is known about lowest order gauge invariant
supersymmetric gauge theories in order to check whether the claimed
properties (scale invariance, short distance compensations,...) are
characteristic for supersymmetry.

The modular Tomita-Takesaki theory which, as we have seen, unravels internal
(charge sectors) as well as external (Poincar\'{e}, conformal) symmetries of
the observable algebras, also never points in the direction of
supersymmetry. Whereas all such standard symmetries suffer at most a
spontaneous breaking in thermal (KMS) states, supersymmetry is unstable
under contact with a heat bath\footnote{%
The Lorentz boosts applied to the rest frame of the heat bath system are the
only spontaneously broken transformations of the Poincar\'{e} group.}; it
suffers a spontaneous ``collapse'' \cite{Bu O}, as one expects for an
accidental symmetry.. All symmetries with noncausal Noether currents (as the
fermionic currents of supersymmetry) are potential candidates for accidental
symmetries with unstable thermal behavior.

\section{Chiral Conformal Algebraic QFT}

Chiral conformal QFT has turned out to be an ideal theoretical laboratory
for algebraic (nonperturbative) QFT. Not only conformal QFT has profited
from this close relation, but the confidence in the algebraic method has
also significantly increased. To cite a recent example, within chiral
conformal QFT, one was able to rigorously prove the equivalence of the
standard approach using pointlike covariant fields with the net approach 
\cite{Joerss}. This is important because in formulating the net approach one
did not intend to widen the physical content, but rather only to put the
advanced theory of von Neumann algebras to the use for exploring the
physical principles of local QFT.

Since the literature on the subject, even if restricted by the above
guideline, is quite formidable, I will limit my attention to two points:

\begin{itemize}
\item  What is charge structure and quantum symmetry after conformal
compactification?

\item  How does one classify chiral conformal QFT?
\end{itemize}

The compactification of chiral conformal QFT is most efficiently done in
terms of a universal C*-algebra $\mathcal{A}_{uni}(S^{1})$ which is
different from the non-compact DHR quasilocal algebra $\mathcal{A}(\mathbf{R}%
).$ In order to understands its construction, we note that the net $\left\{ 
\mathcal{A}(I)\right\} _{I\subset S^{1}}$ is not directed (as the nets of
double cones in Minkowski space) towards infinity. Therefore we should think
of a globalization which is different from the inductive limit. For this we
use the following definition universal algebra $\mathcal{A}_{univ}:$

\begin{definition}
$\mathcal{A}_{univ}$ is the $C^{*}$algebra which is uniquely determined by
the system of local algebras $(A(I))_{I\in \mathcal{T}},$ $\mathcal{T}=$
family of proper intervals $I\subset S^{1}$ and the following universality
condition:

(i)\thinspace \thinspace \thinspace \thinspace there are unital embeddings $%
i^{I}:\mathcal{A}(I)\rightarrow \mathcal{A}_{univ}$ s. t.. 
\begin{equation}
i^{J}\mid _{\mathcal{A}(I)}=i^{I}\text{ \thinspace \thinspace }if\,\,\text{ }%
I\subset J,\,\,I,J\in \mathcal{T}
\end{equation}
and $\mathcal{A}_{univ}$ is generated by the algebras $i^{I}(\mathcal{A}%
(I)),\,\,I\in \mathcal{T};$

(ii)\thinspace \thinspace \thinspace \thinspace for every coherent family of
representations $\pi ^{I}:\mathcal{A}(I)\rightarrow \mathcal{B}(H_{\pi })$
there is a unique representation $\pi $ of $\mathcal{A}_{univ}$ in $H_{\pi }$
s. t.. 
\begin{equation}
\pi \circ i^{I}=\pi ^{I}
\end{equation}
\end{definition}

The universal algebra inherits the action of the M\"{o}bius group as well as
the notion of positive energy representation through the embedding.

The universal algebra has more global elements than the quasilocal algebra
of the DHR theory: $\mathcal{A}_{quasi}\equiv \mathcal{A}\subset A_{univ}$
with the consequence that the vacuum representation $\pi _{0}$ ceases to be
faithful and the global superselection charge operators which are outer for $%
\mathcal{A}$ become inner for $\mathcal{A}_{univ}$ . From this observation
emerges the algebra of Verlinde which originally was obtained by geometric
rather than local quantum physics arguments. The removal of a point $\xi $
from $S^{1}$ (this removal recreates the infinity of $\mathcal{A}_{quasi}$)
forces $\mathcal{A}_{univ}$ to shrink to $\mathcal{A}.$

Most of this new features can be seen by studying global intertwiners in $%
\mathcal{A}_{univ}.$ Let $I,J$ $\in T$ and $\xi ,\zeta \in I^{\prime }\cap
J^{\prime }$ (i.e. two points removed from the complements) and choose $\rho 
$ and $\sigma $ s. t.. loc$\rho ,$ loc$\sigma \subset I$ and $\hat{\rho}\in
\left[ \rho \right] $ with loc$\hat{\rho}\subset J.$ Then the statistics
operators $\varepsilon (\rho ,\sigma )$ and $\varepsilon (\sigma ,\rho )\in 
\mathcal{A}(I)\subset A_{\xi }\cap A_{\zeta }$ are the same (i.e. they don't
need a label $\xi $ or $\zeta )$ independently of whether we use the
quasilocal algebra $\mathcal{A}_{\xi }$ or $\mathcal{A}_{\zeta }$ for their
definition. By Haag duality a charge transporter $V:$ $\pi _{0}\rho
\rightarrow \pi _{0}\hat{\rho}$ lies both in $\pi _{0}(\mathcal{A}_{\xi })$
and $\pi _{0}(\mathcal{A}_{\zeta }).$ However its pre-images with respect to
the embedding are different. In fact: 
\begin{eqnarray}
V_{\rho } &\equiv &V_{+}^{*}V_{-}\text{ with}\,\,\,V_{+}\in \mathcal{A}_{\xi
},\quad V_{-}\in \mathcal{A}_{\zeta }  \label{V} \\
V_{\rho } &\in &\left( \rho ,\rho \right) _{glob}  \nonumber
\end{eqnarray}
is a global selfintertwiner, which is easily shown to be independent of the
choice of $V$ and $\hat{\rho}.$ The representation of the statistics
operators in terms of the charge transporters $\varepsilon (\rho ,\sigma
)=\sigma (V_{+})^{*}V_{+},\,\,\varepsilon (\sigma ,\rho )^{*}=\sigma
(V_{-})^{*}V_{-}$ leads to: 
\begin{equation}
\sigma (V_{\rho })=\varepsilon (\rho ,\sigma )V_{\rho }\varepsilon (\sigma
,\rho )\,\,\curvearrowright \pi _{0}\sigma (V_{\rho })=\pi _{0}\left[
\varepsilon (\rho ,\sigma )\varepsilon (\sigma ,\rho )\right]  \label{around}
\end{equation}
The first identity is very different from the relation between $\varepsilon
^{\prime }s$ due to local intertwiners. The global intertwiner $V_{\rho }$
is trivial in the vacuum representation, thus showing its lack of
faithfulness with respect to $\mathcal{A}_{univ}$. The global aspect of $%
V_{\rho }$ is only activated in charged representations where it coalesces
with monodromy operators. From its definition it is clear that it represents
a charge transport once around the circle $S^{1}\footnote{%
Note that in $A_{univ}$ which corresponds to a compact quantum world it is
not possible to ``dump'' unwanted charges to ``infinity''(as in the case for 
$A_{quasi})$, but instead one encounters ``polarization'' effects upon
charge transportation once around.}.$ As a result of its existence, the
monodromy which is defined as the above two-fold iteration of the braid
generator, takes on some of its geometric meaning which it has e.g. in the
theory of complex functions. The left hand side of the first equation in (%
\ref{around}) expresses a transport ``around'' in the presence of another
charge $\sigma ,$ i.e. a kind of charge polarization. Let us look at the
invariant version of $V_{\rho }$ namely the global ``Casimir'' operators $%
W_{\rho }=R_{\rho }^{*}V_{\rho }R_{\rho }:id\rightarrow id.$ This operator
lies in the center $\mathcal{A}_{univ}\cap \mathcal{A}_{univ}^{\prime }$ and
depend only on the class (=sector) $\left[ \rho \right] $ of $\rho .$ By
explicit computation\cite{FRS} one shows that after the numerical
renormalization $C_{\rho }:=d_{\rho }W_{\rho }$ one encounters the fusion
algebra: 
\begin{eqnarray}
(i)\,\,\,\,C_{\sigma \rho } &=&C_{\sigma }\cdot C_{\rho } \\
(ii)\,\,\,\,C_{\rho }^{*} &=&C_{\bar{\rho}}  \nonumber \\
(iii)\,\,\,\,C_{\rho } &=&\sum_{\alpha }N^{\alpha }C_{\alpha
}\,\,\,\,if\,\,\,\rho \simeq \oplus _{\alpha }N^{\alpha }\rho _{\alpha } 
\nonumber
\end{eqnarray}
Verlinde's modular algebra emerges upon forming matrices with row index
equal to the label of the central charge and the column index to that of the
sector in which it is measured: 
\begin{equation}
S_{\rho \sigma }:=\left| \sum_{\gamma }d_{\gamma }^{2}\right| ^{-\frac{1}{2}%
}d_{\rho }d_{\sigma }\cdot \pi _{0}\sigma (W_{\rho })
\end{equation}
In case of nondegeneracy of sectors, which expressed in terms of statistical
dimensions and phases means $\left| \sum_{\rho }\kappa _{\rho }d_{\rho
}^{2}\right| ^{2}=\sum_{\rho }d_{\rho }^{2},$ the above matrix $S$ is equal
to Verlinde's matrix $S$ \cite{Verlinde}which together with the diagonal
matrix $T=\kappa ^{-1}Diag(\kappa _{\rho }),$ with$\,\,\kappa
^{3}=(\sum_{\rho }\kappa _{\rho }d_{\rho }^{2})/\left| \sum_{\rho }\kappa
_{\rho }d_{\rho }^{2}\right| $ satisfies the modular equations of the genus
1 mapping class group 
\begin{eqnarray}
SS^{\dagger } &=&1=TT^{\dagger },\quad TSTST=S \\
S^{2} &=&C,\quad C_{\rho \sigma }\equiv \delta _{\bar{\rho}\sigma } 
\nonumber \\
TC &=&CT=T  \nonumber
\end{eqnarray}
It is remarkable that these properties are common to chiral conformal
theories and to d=2+1 plektonic models even though the localization
properties of the charge-carrying fields are quite different. In the former
case one has the additional phase relation: 
\begin{equation}
\frac{\kappa }{\left| \kappa \right| }=e^{-2\pi ic/8}
\end{equation}
where $c$ is the Virasoro constant which measures the strength of the
two-point function of the energy-momentum tensor. This relation may be
derived by studying the (modular) transformation properties of the Gibbs
partition functions for the compact Hamiltonian $L_{0}$ of the conformal
rotations under thermal duality transformations $\beta \rightarrow 1/\beta $%
. For d=2+1 plektons, no simple physical interpretation is known.

\begin{lemma}
The matrix $S$ is similar to the character matrix in section 2 of the first
chapter. However in distinction to nonabelian finite groups (which also
yield a \textit{finite} set of charge sectors of the fixed point observable
algebra) the present nonabelian sectors produce a symmetric ``character''
matrix $S$ which signals a perfect auto-duality between charge measurers $%
\left\{ Q\right\} $ and charge creators $\left\{ \rho \right\} .$
Furthermore the algebra $\mathcal{Q}$ generated by the central charges and
the action of the endomorphisms on those charges\footnote{%
This action leads out of the center and generates a global subalgebra of $%
\mathcal{A}_{univ}.$} do not contain the old ``group theoretical stuff''
since the phenomenon of charge ``polarization'' only perceives endomorphisms
with nontrivial monodromy.
\end{lemma}

This strongly suggests to try to understand the new ``quantum symmetry''
property in terms of the structural properties of $\mathcal{Q}.$ As a
generalization of $S$ one finds for the $Q^{\prime }s$ in the presence of
more than one polarization charges the entries of the higher genus mapping
class group matrices \cite{S rem}. The reason is that in addition to the the
process: 
\begin{equation}
vacuum\stackrel{split}{\longrightarrow }\rho \bar{\rho}\stackrel{%
global\,\,\rho }{\stackunder{selfintertw.}{\longrightarrow }}\rho \bar{\rho}%
\stackrel{fusion}{\longrightarrow }vacuum
\end{equation}
which led to the global intertwiner $W_{\rho }=R_{\rho }^{*}V_{\rho }R_{\rho
},$ there is the more involved global intertwiner associated with the
process in which the global selfintertwining occurs after a split of
nonvacuum charge $\sigma $ and a later fusion to $\mu $ which appear in a $%
\rho \bar{\rho}$ reduction: 
\begin{equation}
\sigma \stackrel{split}{\longrightarrow }\alpha \beta \stackrel{%
global\,\alpha }{\stackunder{intertw.}{\longrightarrow }}\alpha \beta 
\stackrel{fusion}{\longrightarrow }\mu ,\quad \sigma ,\mu \subset \rho \bar{%
\rho}\quad
\end{equation}
with the global intertwiner $V_{\alpha }\in \left( \alpha ,\alpha \right)
_{glob}$ being used in: $T_{e(\sigma )}^{*}V_{\alpha }T_{e(\mu )}$ where $%
T_{e(\mu )}$ is the $\alpha \beta \rightarrow \mu $ fusion intertwiner and
the hermitian adjoints represent the corresponding splitting intertwiner. As
in the vacuum case, the selfintertwiners $V$ become only activated after the
application of another endomorphism say $\eta ,$ i.e. in the presence of
another charge $\eta $ (hence the name ``polarization'' mechanism). It can
be shown that the following operators are the building blocks of the mapping
class group matrices $T_{e(\sigma )}^{*}V_{\alpha }T_{e(\mu )}$ which have
multicharge-measurer column and multicharge-creator row indices: 
\begin{equation}
\phi _{\lambda }((T_{g(\eta )}\eta (T_{e(\sigma )}^{*}V_{\alpha }T_{e(\mu
)})T_{f(\eta )}^{*})))
\end{equation}
$.$ here $T_{f(\eta )}$ and $T_{g(\eta )}$ are the intertwiners
corresponding to the charge edges $f(\eta ):\lambda \sigma \rightarrow \eta $
and $g(\eta ):\lambda \mu \rightarrow \eta ,$ whereas $\phi _{\lambda }$ is
the left inverse of the endomorphism $\lambda .$ Besides the global
intertwiners $V,$ we only used the local splitting intertwiners and their
hermitian adjoints which represent the fusion intertwiners. The main
question is: why do we organize the numerical data of the global
charge-measurer and charge-creator algebra $\mathcal{Q}$ as entries in a
multiindex matrix? What is the physical role of these matrices in a d=2+1
plektonic theory?

Closely related to these structures are the knot theoretical invariants of
3-manifolds. These objects also appear by analyzing certain formal
functional integrals with the hindsight of geometry and topology \cite
{Witten}. But in the context of algebraic QFT, the physical interpretation
is quite different because the new properties have nothing to do with the
``living space'' (in the sense of quantum theoretical localization) of
fields or algebras, but are rather manifestations of the inexorable link
between external (space-time) and internal symmetries which one encounters
in low dimensional real time (Minkowski space) QFT. It appears that they
generalize in some sense the angular momentum decompositions and one would
expect them to play a useful role in the understanding of e.g. the analysis
of scattering of d=2+1 plektons. Although these ideas of linking ``quantum
symmetry'' with a kind of universal mapping class group \cite{S rem}
(containing all genii) are highly seductive, I did not yet find an
convincing argument for \textit{why one should read the numerical aspects of
those polarization charges as entries of mapping class matrices acting on
``something''}. This open problem is closely related to the previously asked
questions.

It should be mentioned here that most attempts in the direction of quantum
symmetry have been aimed towards modified (``weak''...) Hopf algebras, thus
remaining near the spirit of the DR theory \cite{Re Hopf}. None of these
attempts was yet successful. One expects from a useful quantum symmetry
concept a clarification of the following two points:

\begin{itemize}
\item  A better understanding why in low dimensions the link between
external/internal symmetries is so strong, whereas in d=3+1 there was no
possibility to bring them together in any nontrivial way. This aims in
particular at a better physical understanding of the physical interpretation
of knot- and 3-manifold- invariants.

\item  A simplification of the problem of computing correlations of ``free''
plektons, i.e. the freest objects (in d=2+1 preferably with vanishing cross
sections) which fulfill the new braid group statistics. Since even the free
plektonic charges have an analytically more complicated spacetime structure
as a result of their semiinfinite localization, a symmetry concept which
does not split internal/external aspects is expected to have a better chance
to be useful for their understanding.
\end{itemize}

Concerning the classification of chiral conformal QFT's, it is reasonable to
approach this problem in two steps:

\begin{itemize}
\item  Classification of the physically admissable braid group
representations which go with the category of finitely many localizable
sectors (''rational'' representations).

\item  Construction of representative 4-point functions for the different
plektonic families.
\end{itemize}

The basic techniques for this two step approach is quite old \cite{RS} and
have been elaborated for the unitary braid group representations affiliated
with the special family of the Jones, Temperly-Lieb algebras. The more
general representations are those affiliated with the Hecke algebra and with
the Birman-Wenzl algebra explained in the next section. Even if one does not
know anything about these mathematical construction, the concepts of
algebraic field theory are so strong that they will lead us there.

Before we present the known (and presumably complete) set of families, it is
helpful to notice the changes in the structure of the exchange algebra which
result from the compactification through the $\mathcal{A}_{uni}$%
-globalization. The main difference to the standard exchange algebra
associated with $\mathcal{A}_{quasi}$ lies in the concept of localization.
The criterion of: $locF(e,A)\subset \mathcal{O}\Leftrightarrow F$ commutes
with all observables $B\in \mathcal{A}(\mathcal{O}^{\prime }),$ or in the
notation of the reduced field bundle formalism: 
\begin{equation}
\pi _{\beta }(B)F(e,A)=F(e,A)\pi _{\alpha }(B),\;c(e)=\rho ,\;s(e)=\alpha
,\;r(e)=\beta \;
\end{equation}
where we used the previously introduced notation concerning edges $e$ on a
fusion graph consisting of the source charge $\alpha ,$the range charge $%
\beta $ and the charge $\rho $ transported by $F.$ Written more explicitely,
this commutation relation is equivalent to the existence of a local unitary
charge transporter $U$ which simultaneously transports $\rho $ and $A$ into $%
\mathcal{O}:$ \thinspace \thinspace $locAd_{U}\circ \rho \subset \mathcal{O}%
,\,\,UA\in \mathcal{A}(\mathcal{O})$ . Hence the localization in $\mathcal{A}%
_{quasi}$ depends only on the pair $(\rho ,A).$ For the $\mathcal{A}_{uni}$
localization this characterization is too rough, because it ignores the
possibility of carrying charges several times around the circle $S^{1}.$ We
refine the definition as follows:

\begin{definition}
Let $J\subset \mathbf{R}$ be an intervall (of extension$<2\pi )$ of the
universal covering $\mathbf{R}$ of $S^{1}.$ For $\rho \in \Delta _{red}$ the
pair $(\rho ,A)$ is said to be localized in $J$ if there is a local operator 
$C\in \mathcal{A}(I)$ s. t.. all $F(e,A)$ with charge $\rho $ are obtained
from operators of the form $F(e,C)$ through Moebius-transformations. Here $I$
is an interval in the first sheet of $R$ (i.e. inside the $2\pi $ interval
which includes the zero).
\end{definition}

The following theorem shows how the new localization concept adapted to the
observable algebra $\mathcal{A}_{uni}$ changes the structure of the exchange
algebra in section 7.3.

\begin{theorem}
Let $(\rho _{1},A_{1})$ and $(\rho _{2},A_{2})$ be localized in intervals $%
J_{1}$ and $J_{2}$ on $\mathbf{R}$ which project onto disjoint intervals in $%
S^{1}.$ Define a relative winding number N s.t. $J_{1}+2\pi
N<J_{2}<J_{1}+2\pi (N+1)$. Then: 
\begin{eqnarray*}
F(e_{2},A_{2})\cdot F(e_{1},A_{1}) &=&\sum_{f_{1}\circ f_{2}}R_{f_{1}\circ
f_{2}}^{e_{2}\circ e_{1}}(N)\cdot F(f_{1},A_{1})\cdot F(f_{2},A_{2}) \\
R_{f_{1}\circ f_{2}}^{e_{2}\circ e_{1}}(N) &=&T_{e_{2}}^{*}T_{e_{1}}^{*}\rho
_{\alpha }(\varepsilon _{N}(\rho _{2},\rho _{1})T_{f_{2}}T_{f_{1}} \\
\varepsilon _{N}(\rho _{2},\rho _{2}) &=&\rho _{1}(Y_{\rho
_{2}}^{N})\varepsilon (\rho _{2},\rho _{1})Y_{\rho _{2}}^{-N},\quad Y_{\rho
}=e^{2\pi ih_{\rho }}V_{\rho }\quad
\end{eqnarray*}
Here $V_{\rho }$ is the selfintertwiner (\ref{V}). The $\varepsilon
_{N}^{\prime }s$ are associated with the \textit{cylinder ribbon braid group%
\cite{FRS}}.
\end{theorem}

Conformal QFT are in some way sophisticated free theories, where the
sophistication refers to their charge structure. Physically there can be no
genuine interaction on one light cone and mathematically one expects the
braid group statistics together with some gross features of the charge
structure to fix the observable algebras and their associated charge
carrying operators \footnote{%
This is analogous to the determination of d=3+1 free fields with a given
internal symmetry (a given charge structure).}. The simplest family of
plektonic chiral conformal theories are the so-called minimal theories; here
minimal is used in the sense of the smallest statistical dimensions.

The knowledge about the admissible braid group representations and in
particular about the statistical dimensions together with a bit of hindsight
on short distance behavior allows in many cases the determination of
plektonic 4-point functions \cite{RS}. There are other artistic methods to
fix the 4-point functions through there monodromy properties by an ad hoc
formal modification of the Coulomb-gas and contour representations. Since we
are not interested in conformal theory per se, but only in its aspects of
presenting a theoretical laboratory for LQP, we will not pursue this matter.
A potentially successful method in the spirit of these notes has been
sketched in chapter 6.8.

\section{Classification of Admissable Particle-Statistics}

The DHR method of classifying the admissable permutation group $S_{\infty }$
representations by defining sequences of projectors which are terminating
after finitely many terms (the consequence of positivity) was extended by
Ocneanu and Wenzl in such a way that it became a powerful tool for large
classes of families of subfactor models. The formalism is applicable to the
physically admissable braid-group representations in low-dimensional QFT and
leads to the famous Markov-trace formalism on the ribbon braid group $%
RB_{\infty }$. There are plausibility arguments (but no proof) that apart
from exceptional cases, the Markov traces on the Hecke algebras and the
Birman-Wenzl-Murakami algebras exhaust all possibilities. The physical
origin of the Hecke algebra is ``two-channel plektonic statistics'' and its
composites. This means that we are studying plektonic endomorphisms $\rho $
with $\rho ^{2}\simeq \rho _{1}\oplus \rho _{2}$ i.e. $d_{\rho }^{2}=d_{\rho
_{1}}+d_{\rho _{2}}$with $\rho ,\rho _{1},\rho _{2}$ irreducible as well as
all its higher composites (``fusion'') \cite{FRS1}. The BWM-algebras result
from plektonic 3-channel endomorphisms $\rho ^{2}\simeq \rho _{1}\oplus \rho
_{2}\oplus \alpha $ with $\alpha $ an automorphism i.e. $d_{\rho
}^{2}=d_{\rho _{1}}+d_{\rho _{2}}+1.$ Structural arguments based on the
4-point function suggest that each braid group representation family has two
field theoretic realizations, one in which the associated observable algebra
contains fields which transform under an internal symmetry group (the
current or Kac-Moody algebras) and the other without such group theoretic
multiplicities. In the following we present the arguments for this claims.

Our main tool is the application of the (iterated) left inverse $\phi $ on
the intertwiner algebras. In this way one obtains tracial states on those
algebras. Left inverses and some of its properties were already introduced
in section 3. Their use in this analysis is synonymous with the more
physical notion of ``conjugate'' or antiparticle. The latter is a particle
with the same Poincar\'{e}-group representation but ``opposite'' internal
symmetry behavior i.e. with possibility to annihilate into a state with the
quantum numbers of the vacuum. A sufficient condition for the existence of
antiparticles is the spectral gap (isolated one particle hyperboloids) which
is also the standard assumption of scattering theory. Let us briefly remind
ourselves that translated into the setting of endomorphisms of algebraic QFT
this means $\bar{\rho}\rho \supset id$ i.e. the existence of an localized
intertwiner $R$ with: 
\begin{equation}
R\cdot id(A)=\bar{\rho}\rho (A)\cdot R,\quad \quad A\in \mathcal{A}\quad
\end{equation}
where as a result of the localization of $R$ the global algebra $\mathcal{A}$
is either the quasilocal or the bigger universal algebra. Then: 
\begin{equation}
\phi (A):=R^{*}\bar{\rho}(A)R
\end{equation}
is a (unique for $\rho $ irreducible) left inverse of $\rho $ i.e. a
positive linear map with $\phi (\rho (A)B\rho (C))=A\phi (B)C$ for $A,B,C\in 
\mathcal{A}.$ The complex number $\lambda _{\rho }\underline{1}=\phi
(\varepsilon _{\rho })$ (for irreducible $\rho )$ is called the statistics
parameter and it is written as $\lambda _{\rho }=\frac{1}{d_{\rho }}\kappa
_{\rho }$ with $d_{\rho }$ the statistical dimension and $\kappa _{\rho
}=e^{2\pi ih_{\rho }}$ the statistical phase. We note in passing that the
spin-statistics theorem relates this to the spin-phase (in conformal QFT $%
h_{\rho }$ is related to the scale-dimension). Under this assumption of
irreducibility of $\rho $ (always assumed in the rest of this section) $\phi 
$ maps the commutant of $\rho ^{2}(\mathcal{A})$ in $\mathcal{A}$ into the
complex numbers: 
\begin{equation}
\phi (A)=\varphi (A)\underline{1},\quad A\in \rho ^{2}(\mathcal{A})^{\prime }
\end{equation}
and by iteration a faithful tracial state $\varphi $ on $\cup _{n}\rho ^{n}(%
\mathcal{A})^{\prime }$ with: 
\begin{eqnarray*}
\phi ^{n}(A) &=&\varphi (A)\underline{1},\quad A\in \rho ^{n+1}(\mathcal{A}%
)^{\prime }\quad \\
\varphi (AB) &=&\varphi (BA),\quad \varphi (\underline{1})=1
\end{eqnarray*}
Restricted to the $\mathbf{C}RB_{n}$ algebra generated by the ribbon
braid-group which is a subalgebra of $\rho ^{n}(\mathcal{A})^{\prime }$ the $%
\varphi $ becomes a tracial state, which can be naturally extended $%
(B_{n}\subset B_{n+1})$ to $\mathbf{C}RB_{\infty }$ in the above manner and
fulfills the ``Markov-property'': 
\begin{equation}
\varphi (a\sigma _{n+1})=\lambda _{\rho }\varphi (a),\quad a\in \mathbf{C}%
RB_{n}
\end{equation}
The terminology is that of V. Jones and refers to the famous russian
probabilist of the last century as well as to his son, who constructed knot
invariants from suitable functionals on the braid group.\textit{\ }The
``ribbon'' aspect refers to an additional generator $\tau _{i}$ which
represents the vertical $2\pi $ rotation of the cylinder braid group ($%
\simeq $ projective representation of $B_{n})\cite{FRS}.$

It is interesting to note that the Markov-property is the combinatorial
relict of the cluster property which relates the n-point correlation
function in local QFT to the n-1 point correlation or in QM the physics of n
particles to that of n-1 (rendering one particle a spectator by removing it
to infinity. The infinite permutation- and braid groups are the only groups
behaving like a russian ``matrushka'' i.e. the smaller ones are naturally
contained in the bigger. This picture is similar to that of cluster
properties which was already used in our attempts to understand statics in
the nonrelativistic setting of the first chapter. The existence of a Markov
trace on the ribbon braid group of (low dimensional) multi-particle
statistics is the imprint of the cluster property on particle statistics. As
such it is more basic than the notion of internal symmetry. It precedes the
latter and according to the DR theory it may be viewed as the other side of
the same coin on which one side is the old (compact group-) or new
(quantum-) symmetry. With these remarks the notion of internal symmetry
becomes significantly demystified.

Let us now return to the above 2-channel situation. Clearly $\varepsilon
_{\rho }$ has maximally two different eigenprojectors since otherwise there
would be more than two irreducible components of $\rho ^{2}.$ On the other
hand $\varepsilon _{\rho }$ cannot be a multiple of the identity because $%
\rho ^{2}$ is not irreducible. Therefore $\varepsilon _{\rho }$ has exactly
two different eigenvalues $\lambda _{1},\lambda _{2}$ i.e. 
\begin{equation}
(\varepsilon _{\rho }-\lambda _{1}\underline{1})(\varepsilon _{\rho
}-\lambda _{2}\underline{1})=0
\end{equation}
\begin{equation}
\leftrightarrow \varepsilon _{\rho }=\lambda _{1}E_{1}+\lambda
_{2}E_{2}\,\,,\quad E_{i}=\left( \lambda _{i}-\lambda _{j}\right)
^{-1}\left( \varepsilon _{\rho }-\lambda _{j}\right) ,\quad i\neq j
\end{equation}
which after the trivial re-normalization of the unitaries $g_{k}:=-\lambda
_{2}^{-1}\rho ^{k-1}(\varepsilon _{\rho })$ yields the generators of the
Hecke algebra: 
\begin{eqnarray}
g_{k}g_{k+1}g_{k} &=&g_{k+1}g_{k}g_{k+1} \\
g_{k}g_{l} &=&g_{l}g_{k}\,\,,\quad \left| j-k\right| \geq 2  \nonumber \\
g_{k}^{2} &=&(t-1)g_{k}+t\,\,,\quad t=-\frac{\lambda _{1}}{\lambda _{2}}\neq
-1  \nonumber
\end{eqnarray}
The physical cluster property in the algebraic form of the existence of a
tracial Markov state leads to a very interesting ``quantization'' \footnote{%
In these notes we use this concept always in the original meaning of Planck
as a discretization, and not in the modern form of a deformation.}. Consider
the sequence of projectors: 
\begin{equation}
E_{i}^{(n)}:=E_{i}\wedge \rho (E_{i})\wedge ...\wedge \rho
^{n-2}(E_{i})\,\,,\quad i=1,2
\end{equation}
and the symbol $\wedge $ denotes the projection on the intersection of the
corresponding subspaces. The notation is reminiscent of the totally
antisymmetric spaces in the case of Fermions. The above relation $%
g_{1}g_{2}g_{1}=g_{2}g_{1}g_{2}$ and $g_{1}g_{n}=g_{n}g_{1},$ $n\geq 2$ in
terms of the $E_{i}$ reads: 
\begin{eqnarray}
E_{i}\rho (E_{i})E_{i}-\tau E_{i} &=&\rho (E_{i})E_{i}\rho (E_{i})-\tau \rho
(E_{i})\,\,,\quad \tau =\frac{t}{\left( 1+t\right) ^{2}}  \label{TL} \\
E_{i}\rho ^{n}(E_{i}) &=&\rho ^{n}(E_{i})E_{i}\,\,,\quad n\geq 2  \nonumber
\end{eqnarray}
The derivation of these equations from the Hecke algebra structure is
straightforward. The following recursion relation is however tricky and will
be given in the sequel.

\begin{proposition}
(Wenzl, DHR) The projectors $E_{i}^{(n)}$ fulfill the following recursion
relation ($t=e^{2\pi i\alpha },\,-\frac{\pi }{2}<\alpha <\frac{\pi }{2}):$ 
\begin{eqnarray}
E_{i}^{(n+1)} &=&\rho (E_{i}^{(n)})-\frac{2\cos \alpha \sin n\alpha }{\sin
(n+1)\alpha }\rho (E_{i}^{(n)})E_{j}\rho (E_{i}^{(n)})\,\,,\quad i\neq
j,\quad n+1<q  \label{rec} \\
E_{i}^{(q)} &=&\rho (E_{i}^{(q-1)})\quad ,\quad q=\inf \left\{ n\in \mathbf{%
N,\,}n\left| \alpha \right| \geq \pi \right\} \,\,\,if\,\,\alpha \neq
0,\,\,q=\infty \,\,if\,\,\alpha =0  \nonumber
\end{eqnarray}
The DHR recursion for the permutation group $S_{\infty }$ is obtained for
the special case t=0 i.e. $\alpha =0.$ In this case the numerical factor in
front of product of three operators is $\frac{n}{n+1}.$
\end{proposition}

The proof is by induction. For n=1 the relation reduces to the completeness
relation between the two spectral projetors of $\varepsilon _{\rho
}:\,E_{i}=1-E_{j},\,i\neq j.$ For the induction we introduce the
abbreviation $F=E_{j}\rho (E_{i}^{(n)})=\rho (E_{i}^{(n)})E_{j}$ and compute 
$F^{2}.$ We replace the first factor $\rho (E_{i}^{(n)})$ according to the
induction hypothesis by: 
\begin{equation}
\rho (E_{i}^{(n)})=\rho ^{2}(E_{i}^{(n-1)})-\frac{2\cos \alpha \sin
(n-1)\alpha }{\sin n\alpha }\rho ^{2}(E_{i}^{(n-1)})\rho (E_{j})\rho
^{2}(E_{i}^{(n-1)})
\end{equation}
We use that the projector $\rho ^{2}(E_{i}^{(n-1)})$ commutes with the
algebra $\rho ^{2}(\mathcal{A})^{\prime }$ (and therefore with $E_{j}\in
\rho ^{(2)}(\mathcal{A})^{\prime }$), and that its range contains that of $%
\rho (E_{i}^{(n)})$ i.e. $\rho ^{2}(E_{i}^{(n-1)})\rho (E_{i}^{(n)})=\rho
(E_{i}^{(n)}).$ Hence we find: 
\begin{equation}
F^{2}=E_{j}\rho (E_{i}^{(n)})-\frac{2\cos \alpha \sin (n-1)\alpha }{\sin
n\alpha }\rho ^{2}(E_{i}^{(n-1)})E_{j}\rho (E_{j})E_{j}\rho (E_{i}^{(n)})
\end{equation}
Application of (\ref{TL}) with $\tau =\frac{1}{2\cos \alpha }$ to the
right-hand side yields: 
\begin{equation}
F^{2}=E_{j}\rho (E_{i}^{(n)})-\frac{\sin (n-1)\alpha }{2\cos \alpha \sin
\alpha }\rho ^{2}(E_{i}^{(n-1)})E_{j}\rho (E_{i}^{(n)})=\frac{\sin
(n+1)\alpha }{2\cos \alpha \sin n\alpha }F  \label{A}
\end{equation}
where we used again the above range property and a trigonometric identity.
For $n=q-1$ the positivity of the numerical factor fails and by $%
F^{2}E_{j}=(FF^{*})^{2}$ and $FE_{j}=FF^{*}$ the operator F must vanish and
hence $E_{j}$ is orthogonal to $\rho (E_{j}^{(q-1)})$ which is the second
relation in (\ref{rec}). For $n<q-1$ the right-hand side of the first
relation in (\ref{rec}) with the help of (\ref{A}) turns out to be a
projector which vanishes after multiplication from the right with $\rho
^{k}(E_{j}),k=1,...,n-2$ as well as with $E_{j}.$ The remaining argument
uses the fact that this projector is the largest with this orthogonality
property and therefore equal to $E_{i}^{(n+1)}$ by definition of $%
E_{i}^{(n+1)}$ Q.E.D..

The recursion relation (\ref{rec}) leads to the desired quantization after
application of the left inverse $\phi :$%
\begin{eqnarray}
\phi (E_{i}^{(n+1)} &=&E_{i}^{(n)}\left( 1-\frac{2\cos \alpha \sin n\alpha }{%
\sin (n+1)\alpha }\eta _{j}\right) ,\quad i\neq j  \label{pos} \\
\eta _{j} &=&\phi (E_{j}),\,\,0\leq \eta _{j}\leq 1,\,\,\eta _{1}+\eta
_{2}=1\quad  \nonumber
\end{eqnarray}
From this formula one immediately recovers the permutation group DHR
quantization for $\alpha =0.$ In that case positivity of the bracket
restricts $\eta _{j}$ to the values $\frac{1}{2}(1\pm \frac{1}{d}),\,\,d\in 
\mathbf{N}\cup 0.$ For $\alpha \neq 0$ one first notes that from the second
equation (\ref{rec}) one obtains (application of $\phi $): 
\begin{equation}
\eta _{j}E_{i}^{(q-1)}=\phi (E_{j}\rho (E_{i}^{(q-1)}))=\phi
(E_{j}E_{i}^{(q)})=0,\quad i\neq j
\end{equation}
where the vanishing results from the orthogonality of the projectors. Since $%
\eta _{1}+\eta _{2}=1$ we must have $E_{i}^{(q-1)}=0$ for i=1,2, q$\geq 4,$
because $E_{i}^{(q-1)}\neq 0$ would imply $\eta _{j}=0$ and $%
E_{j}^{(q-1)}=0. $ This in turn leads to $E_{j}\equiv E_{j}^{(2)}=0$ which
contradicts the assumption that $\varepsilon _{\rho }$ processes two
different eigenvalues. This is obvious for $q=3$ and follows for $q>3$ from
the positivity of $\phi $ (\ref{pos}) for n=2: 
\begin{equation}
\phi (E_{j}^{(3)})=-\frac{\sin \alpha }{\sin 3\alpha }E_{j}^{(2)}\quad
\curvearrowright E_{j}^{(2)}=0\quad \curvearrowright
E_{i}^{(q-1)}=0,\,\,i=1,2,\,q\geq 4
\end{equation}
Using (\ref{pos}) iteratively in order to descend in n starting from $n=q-2,$
positivity demands that there exists an $k_{i}\in \mathbf{N,\,}2\mathbf{\leq 
}k_{i}\leq q-2,$ with: 
\begin{equation}
\eta _{i}=\frac{\sin (k_{i}+1)\alpha }{2\cos \alpha \sin k_{i}\alpha }%
,\,\,i=1,2\,\,\,\,\,\curvearrowright \sin (k_{1}+k_{2})\alpha =0
\end{equation}
where the relation results from summation over $i.$ Since the only solutions
are $\alpha =\pm \frac{\pi }{q},\,k_{1}=d,\,k_{2}=q-d,\,d\in N,\,2\leq d\leq
q-2,$ one finds for the statics parameters of the plektonic 2-channel family
the value: 
\begin{equation}
\lambda _{\rho }=\sum_{i=1}^{2}\lambda _{i}\eta _{i}=-\lambda _{2}\left[
\left( t+1\right) \eta _{1}-1\right] =-\lambda _{2}e^{\pm \pi i(d+1)/q}\frac{%
\sin \pi /q}{\sin d\pi /q}
\end{equation}
a formula which allows for a nice graphical representation. We have
established the following theorem:

\begin{theorem}
. Let $\rho $ be an irreducible localized endmorphism such that $\rho ^{2}$
has exactly two irreducible subrepresentations. Then:
\end{theorem}

\begin{itemize}
\item  $\varepsilon _{\rho }$ has two different eigenvalues $\lambda
_{1},\,\lambda _{2}$ with ratio 
\begin{equation}
\frac{\lambda _{1}}{\lambda _{2}}=-e^{\pm 2\pi i/q},\quad q\in \mathbf{N}%
\cup \{\infty \},\,q\geq 4
\end{equation}

\item  The modulus of the statistics parameter $\lambda _{\rho }=\phi
(\varepsilon _{\rho })$ has the possible values 
\begin{equation}
\left| \lambda _{\rho }\right| =\left\{ 
\begin{array}{l}
\frac{\sin \pi /q}{\sin d\pi /q},\,\,q<\infty \\ 
\frac{1}{d},0\,\,\,\,\,\,\,\,\,\,\,\,\,q=\infty
\end{array}
\right. ,d\in N,\,\,\,2\leq d\leq q-2
\end{equation}

\item  The representation $\varepsilon _{\rho }^{(n)}$ of the braid group $%
B_{n}$ which is generated by $\rho ^{(k-1)}(\varepsilon _{\rho
}),\,k=1,...,n-1$ in the vacuum Hilbert space is an infinite multiple of the
Ocneanu-Wenzl representation tensored with a one dimensional (abelian)
representation. The projectors $E_{2}^{(m)}$ and $E_{1}^{(m)}$ are
``cutoff'' (vanish) for $d<m\leq n$ and $q-d<m\leq n$
\end{itemize}

respectively.

\begin{itemize}
\item  The iterated left inverse $\varphi =\phi ^{n}$ defines a Markov trace 
$tr$ on $B_{n}:$%
\begin{equation}
tr(b)=\varphi \circ \varepsilon _{\rho }(b)
\end{equation}
\end{itemize}

The ``elementary'' representation which is characterized by two numbers d
and q gives rise to a host of composite representation which appear if one
fuses the $\rho ,\rho _{1},\rho _{2}$ and reduces etc. We will not present
the associated composite braid formalism.

The problem of 3-channel braid group statistics has also been solved with
the projector method in case that one of the resulting channels is an
automorphism $\tau $: 
\begin{equation}
\rho ^{2}=\rho _{1}\oplus \rho _{2}\oplus \tau
\end{equation}
In that case $\varepsilon _{\rho }$ has 3 eigenvalues $\lambda _{i}$ which
we assume to be different: 
\begin{equation}
(\varepsilon _{\rho }-\lambda _{1})(\varepsilon _{\rho }-\lambda
_{2})(\varepsilon _{\rho }-\lambda _{3})=0
\end{equation}
The relation to the statistics phases $\omega _{\rho },$ $\omega _{i}$ is
the following: $\mu _{i}^{2}=\frac{\omega _{i}}{\omega ^{2}}.$ In addition
to the previous operators $G_{i}=\rho ^{i-1}(\varepsilon _{\rho
})=(G_{i}^{-1})^{*}$ we define projectors: 
\[
E_{i}=\rho ^{i-1}(TT^{*}) 
\]
where $T\in (\rho ^{2}\left| \tau )\right. $is an isometry and hence $E_{i}$
the projector onto the eigenvalue $\lambda _{3}=\lambda _{\tau }$ of $G_{i}.$
In fact one finds the following relations between the $G_{i}$ and $E_{i}:$%
\begin{eqnarray}
E_{i} &=&\frac{\mu _{3}}{(\mu _{3}-\mu _{1})(\mu _{3}-\mu _{2})}(G_{i}-(\mu
_{1}+\mu _{2})+\mu _{1}\mu _{2}G_{i}^{-1}) \\
E_{i}G_{i} &=&\mu _{3}E_{i}  \nonumber
\end{eqnarray}
This together with the trilinear relations between the $G_{i}^{\prime }s$
and $E_{i}^{\prime }s$ as well as the commutation of neighbors with distance 
$\geq 2$ gives upon renormalization the operators $g_{i}$ and $e_{i}$ which
fulfill the defining relation of the Birman-Wenzl algebra which again
depends on two parameters. The Markov tracial state classification again
leads to a quantization of these parameters except for a continuous
one-parameter solution with statistical dimension $d=2$ which is realized in
conformal QFT as sectors on the fixed point algebra of the $U(1)$ current
algebra (which has a continuous one-parameter solution) under the action of
the charge conjugation transformation (called ``orbifolds'' by people who
prefer geometrical pictures to physical concepts).

There exists another method which allows to construct Markov-traces on the
braid group $B_{\infty }$ which is a deformation theory of groups. It yields
a kind of Hopf-algebra called q-deformed group or ``Quantum Group'' (beware:
here the word ``quantum'' does not have the physical meaning given by Planck
and Heisenberg!). This deformation theory carries the tracial states on the
centralizer algebras of tensor representations of compact ie-groups into
Markov-traces on the braid group. If the value of the deformation parameter
approaches a unit root, these Markov-traces go smoothly over into the above
Markov tracial states on the braid group algebra whereas the ``quantum
group'' suffers a nonanalytic behavior.. This is a useful technique which
allows to obtain a rather rich family of physically admissable $B_{\infty }$
representations by more familiar Lie-group methods which avoids the
systematic DHR multi-channel classification. This is a very fine method as
long as one does not confuse the q-deformation technique with the new
quantum symmetry.

It is an interesting question, to what extend the combinatorial data of the
hyperfinite type II$_{1}$ intertwiner algebra fixes the chiral conformal
field theory. Whereas it is quite easy to see by the aforementioned methods
that a given admissable braid group representation of the previous type
always has a current algebra realization and a W-algebra (generalization of
minimal models) realization, the problem of finding all theories
corresponding to given fusion laws and braid group matrices is more
complicated. Since the charge-induction algebra introduced in the previous
section contains the c-value modulo eight of the energy-momentum algebra,
one expects a discrete family of theories. Indeed Fuchs \cite{Fuchs}
identified such a class of ``Ising-like'' models which have the same fusion
and braid-group structure as the Ising model. The following two statements
are presently only conjectures:

\begin{conjecture}
Conformal QFT are fixed by their associated type II$_{1}$ intertwiner
algebras (combinatorial, topological data) up to a possibly discrete
ambiguity related to the above $c\,\,\func{mod}\,8$ phenomenon.
\end{conjecture}

\begin{conjecture}
``Free'' d=2+1 plektons are in a one to one correspondence with chiral
conformal QFT.
\end{conjecture}

Here the restriction ``free'' is necessary, because braid group commutation
relations and associated combinatorial data do not fix general d=2+1
theories, inasmuch as the internal symmetry structure of d=3+1 fermions and
bosons does not determine interacting theories (which, depending on the
framework of description, require also the specification of couplings and
coupling strengths). ``Free'' plektons are not defined in terms of free
equation of motions (which, as will be shown in the next section, would
contradict the plektonic commutation relations) but rather in terms of
absence of ``on-shell'' processes. This will be explained in the next
section.

\section{Constructive Aspects of Plektons}

In view of the spin-statistics connection which, as we saw in the previous
section, extends for low dimensional theories beyond bosons and fermions to
plektons, it is reasonable to first investigate the properties of massive
d=2+1 Wigner representations with abelian $U(1)$-spin in the sense of the
little group. In passing we note that our discussion of covariant free
fields from Wigner's canonical momentum space representation in chapter 3
holds with small modifications for $U(1)$-spin, s=semiinteger. Here the $u$
and $v$ intertwiners intertwine between these abelian representations and
covariant representations of $\widetilde{SL(2,R)},$ where only two-fold
coverings have to be considered. The formulas for free fields in Fock-space
and the phenomenon of their nonuniqueness (leading again to Borchers
classes) are completely analogous to chapter 3. In particular the more
restrictive quantization method using Lagrangians and the canonical- or
functional-formalism for constructing free Bosons or Fermions works the same
way.

As in the case of d=3+1 theories there is however also the possibility to
avoid the $u,v$ intertwiners in favor of \textit{introducing localization
directly} via modular theory.

This modular localization approach is \textit{the only one for }$s\neq \frac{%
n}{2}$ i.e. for any(spin)ons. The definition of the net of real Hilbert
spaces which have their localization in wedges is the same as in chapter 3
(especially theorem..and will not be repeated here). But the descend to
compact space time regions fails. In fact it turns out that the problem
already starts with the noncompact spacelike cones, which may be obtained by
intersecting just two wedges. In principle two things could happen: such
intersections could be empty (or too small) or the spacelike cone based real
subspaces could fail to fulfill isotony. According to a recent result of
J.Mund \cite{Mund}, it is the latter possibility which actually happens. So
the prerequisite for having a functor from subspaces of the Wigner
representation space to von Neumann algebras fails. A different argument
with the same result on the impossibility of having a Fock space structure
can be obtained by generalizing the J-S theorem \cite{Mund}. From this one
learns that there are no on-shell x-space string-like (spacelike cone with
core-direction n and apex x) localized anyon fields, for which the
combination of $A(x,n_{1})A(y,n_{2})$ and $A(y,n_{2})A(x,n_{1})$
appropriately weighted by a relative phase vanishes for spacelike string
separation. Rather free anyons must have a virtual particle creation
(annihilation) structure. In this property free anyon fields are more
similar to the fields of d=1+1 dimensional fields of factorizable models,
than to d=3+1 free fields. Their two-point function must have a continuous
spectral mass contribution in addition to the on-shell Wigner component, a
fact well-known from from factorizable models which have real particle
conservation and (unlike free Bosons and Fermions) virtual particle
nonconservation.

An extremely useful additional information comes from the structure of
incoming and outgoing scattering states. In the following we will try to
give a systematic derivation starting from the structural analysis of
plektons with the algebraic method. We start with the charge sector theory
on d=2+1 observable algebras. The following theorem and its extension to
d=2+1 theories is crucial for plektonic sectors:

\begin{theorem}
\cite{BF} Covariant representations of an observable net $\mathcal{A}$ which
fulfill the spectral mass gap hypothesis admit localizations in spacelike
cones $S$ of arbitrary small width; in brief, the a-priori best possible
localization is semiinfinite string-like (the core of these cones): 
\begin{equation}
\pi \left| _{\mathcal{A}_{0}(S^{\prime })}\right. \cong \pi _{0}\left| _{%
\mathcal{A}_{0}(S^{\prime })}\right. \,\,\,\,S:=a+\bigcup_{\lambda
>0}\lambda \mathcal{O}
\end{equation}
\end{theorem}

Here are some notational explanations. $\mathcal{A}_{0}\equiv \mathcal{A}%
_{quasi}$ is the quasilocal algebra i.e. the $C^{*}$-algebra generated by
the norm closure of the union $\bigcup \mathcal{A}(\mathcal{O})$ of local
algebras. The $C^{*}$-algebra belonging to noncompact regions as $S$ and $%
S^{\prime }$ is defined analogously. The $\cong $ means unitary equivalence
on the indicated subalgebras where the apex a and the width of $\mathcal{O}$
is arbitrary as long as the cones are proper i.e. their causal completions
do not fill the whole space-like region. The construction of the $\pi $%
-associated vacuum representation $\pi _{0}$ is part of the proof \cite{Haag}%
. The positive effect which spectral gaps have on localization properties is
of course intuitivly plausible (and frequently used by condensed matter
physicists), however a convincing proof requires the application of a
sizable part of analytic techniques of algebraic QFT and will not be given
here. It was also shown there that the charge-carrying particles are
necessarily Bosons and Fermions with a internal compact symmetry structure
i.e. the standard DR-situation . However in d=2+1 this conclusion cannot be
drawn and braid-group statistics may occur . If the localization is DHR
(instead of $S$ a compact double cone $\mathcal{O})$ then the fields and the
associated particles are necessarily Bosons or Fermions with a possible
compact internal group symmetry.

Using now the vacuum representation on the vacuum GNS-space $H_{\pi
_{0}}\equiv H_{0}$ as the defining representation $\mathcal{A}_{0}$ and
making the standard assumption of Haag duality on the corresponding von
Neumann algebras: $\mathcal{A}(S^{\prime })=\mathcal{A}(S)^{\prime },$ one
may again trade $\pi $ for a homomorphism $\rho :\mathcal{A}%
_{0}(M)\rightarrow B(H_{0})$ which however as a result of the noncompact
nature in this case is not necessarily ``endo'' i.e. $\rho (\mathcal{A}%
_{0}(M))\subsetneq \mathcal{A}_{0}(M)$ and hence the composition of such
sectors is ill-defined. An elegant way out is to follow the construction of
a larger universal observable algebra $\mathcal{A}_{uni}$ of the previous
section. Since one only has to replace the proper intervals $I$ on the
circle by proper spacelike cones $S,$ we will not repeat that construction.
Now the net is generated by the $\mathcal{A}(S)$ and the universality
condition is formulated in terms of coherent families of representations $%
\pi ^{S}(S).$ The definition of the reduced field bundle as a $\rho (%
\mathcal{A})-\mathcal{A}$ bimodule with a $C^{*}$-algebra structure is as
before denoted by pairs $F(e,A),$with$\,\,e=(\rho _{\alpha },\rho ,\rho
_{\beta })$ which act on the reduced field bundle vectors $\{\rho _{\gamma
},\psi \}$ with $\rho _{\alpha }^{\prime }s$ being representative
endomorphisms (one chosen from each equivalence class): 
\begin{equation}
F(e,A)\left\{ \rho _{\gamma },\psi \right\} =\delta _{\alpha \gamma }\left\{
\rho _{\beta },\pi _{0}(T_{e}^{*}\rho _{\alpha }(A)\psi \right\} \quad
T_{e}\in (\rho _{\alpha }\rho ,\rho _{\beta })
\end{equation}
The exchange algebra relations follow in complete analogy to the circular
case. Again one has to pay attention to the localization concept for
operators in the exchange algebra. As in the circular case (which has the
same global topology), the localization of pairs $\left( \rho ,A\right) $
has a topological subtlety in that it depends on path classes:

\begin{definition}
$\left( \rho ,A\right) $ with $A\in A$ is said to be localized in the path
class of spacelike cones $\left[ S_{0},....,S_{n}\right] $ with either $%
S_{i}\subset S_{i-1}$ or $S_{i}\,\supset S_{i-1}$ if there exists a sequence
of unitary charge transporters $U_{i}\in \mathcal{A}(S_{i}\cup S_{i-1})$
such that $loc(AdU_{i}...U_{1}\circ \rho )\in S_{i}\,$and $U_{n}...U_{1}A\in 
\mathcal{A}(S_{n});$ shortly: $loc(\rho ,A)\subset $ $\left[
S_{0},....,S_{n}\right] .$
\end{definition}

Here we already anticipated the result that the localization depends only on
the homotopy class of paths with the beginning $S_{0}$ and the end $S_{n}$
fixed. In the vacuum representation $\pi _{0},$ the $U$-transport once
around is trivial (there are no nontrivial vacuum selfintertwiners), but in
nonvacuum sectors one expects a nontrivial representation of the first
homotopy-group.

All these consideration may look unnecessarily pedantic to readers familiar
with manipulations in supersymmetry or with differential geometric methods.
Why does one not write down the commutation relations of the exchange
algebra right away? Well, if anybody succeeds to guess the right answer, he
should go ahead. The point which I want to stress here is that one is
entering a new terrain in QFT, where the intuition based on known algebraic
structures or Lagrangians is of not much help. Therefore, as already
emphasized before, a careful presentation of arguments in algebraic QFT is
the result of a pragmatic and not a pedantic attitude.

Two operators $F_{1,2}$ with $S_{0}^{(1)}=S_{0}^{(2)}$are called mutually
spacelike localized if the ``endpoints'' of their paths also agree $%
S_{n_{1}}^{(1)}=S_{n_{2}}^{(2)}.$ The path topology is best pictured by
shifting the apex of the $S^{\prime }s$ into the origin and considering the
intersection of the shifted $S$ with the unit spacelike hyperboloid ( the
shadow which $S$ casts on the 2-dim. de Sitter space). The topology (in
particular the first homotopy group) of the resulting path model is the same
as that of the circle. Naturally the R-matrix structure of the exchange
algebra i.e. the commutation relations between the two $F_{1,2}$ is the same
as in the conformal case, namely the R-matrix structure constants represent
the generators of the cylinder braid group: 
\begin{equation}
F(e_{2},A_{2})\cdot F(e_{1},A_{1})=\sum_{f_{1}\circ f_{2}}R_{f_{1}\circ
f_{2}}^{e_{2}\circ e_{1}}(N)\cdot F(f_{1},A_{1})\cdot F(f_{2},A_{2})
\end{equation}
where the $R$-matrices, the statistics operators and the $T$- intertwiners
are defined as in chiral conformal QFT and the $N$ is the relative path-
winding on de Sitter hyperboloid. These d=2+1 exchange $C^{*}$-algebras are,
unlike chiral conformal QFT's but similar to d=3+1 bosonic or fermionic
fields with fixed internal symmetry, expected to have (for fixed $R^{\prime
}s$) continuously many representations corresponding to the idea of charge
structure preserving coupling constant deformations. We are interested in
the ``free'' plektons and therefore we now look at the inner product of
scattering states which is associated with the above exchange algebra. It is
well-known (see section on scattering theory) that the Haag-Ruelle
scattering theory determine the Fock space structure of the incoming
multiparticle states, in particular their inner products. The resulting ($%
\pm $symmetrized) tensor product structure of the multi-particle states in
terms of the one-particle Wigner space cannot be maintained in the presence
of the above exchange algebra. For the computation of the asymptotic limits
for plektons we choose a formulation of the Haag-Ruelle scattering theory
which makes the independence of the limits on the Lorentz-frame manifest .
We define: 
\begin{eqnarray}
f_{t}(x) &=&\frac{1}{(2\pi )^{\frac{3}{2}}}\int d^{3}pe^{-ipx+i(\frac{%
p^{2}-m^{2}}{2m})t}\,\tilde{f}(p),\,\,\,\,\,\,f(x)\in S(\mathbf{R}^{3}) \\
F(t) &=&\int d^{3}xf_{t}(x)\alpha _{x}(F(e,A))  \nonumber
\end{eqnarray}
and $locF(e,A)\subset \tilde{I},\,\,\,\tilde{I}=\left[ S_{0}....S_{n}\right]
,$ the path-class localization. The $F$ is chosen in the standard fashion of
H-R scattering theory namely the spectrum of $F\Omega $ contains an isolated
mass shell and $supp\tilde{f}\cap specF\Omega \subset H_{m},$ the mass shell
i.e. $F\Omega \equiv \psi $ is a one particle state. The quasilocal
operators $F(t)$ can be (as in the standard Haag-Ruelle theory) approximated
by operators $F_{\varepsilon }(t)$ localized in $\tilde{I}+tV_{\varepsilon
}(f)$ s. t.. $\left\| F(t)-F_{\varepsilon }(t)\right\| <c\left| t\right|
^{-N}$ where $V_{\varepsilon }(f)$ is the velocity support of f surrounded
by an $\varepsilon $-safety collar. We chose the $F_{i}^{\prime }s$ and $%
f_{i}^{\prime }s$ in such a way that the $F_{\varepsilon ,i}(t)$ are
spacelike for large t. Then the following generalized Haag-Ruelle theorem
holds.

\begin{theorem}
(\cite{FGR}) The sequence of vectors $F_{n}(t)....F_{1}(t)\Omega $, with $%
locF_{i}$ relatively spacelike, converges to a $L$-covariant limit with
braiding properties: 
\begin{eqnarray}
\lim F_{n}(t)....F_{1}(t)\Omega  &=&(\psi _{n},\tilde{I}_{n})\times
...\times (\psi _{1},\tilde{I}_{1})  \nonumber \\
U(L)(\psi _{n},\tilde{I}_{n})\times ...\times (\psi _{1},\tilde{I}_{1})
&=&(U(L)\psi _{n},L\tilde{I}_{n})\times ...\times (U(L)\psi _{1},\tilde{I}%
_{1}) \\
(\psi _{\sigma (n)},\tilde{I}_{\sigma (n)})\times ...\times (\psi _{\sigma
(1)},\tilde{I}_{\sigma (1)}) &=&\varepsilon (b)(\psi _{n},\tilde{I}%
_{n})\times ...\times (\psi _{1},\tilde{I}_{1})
\end{eqnarray}
\end{theorem}

Some comments are in order. In the standard theory there is no dependence on
the localization path, and we have a tensor product structure. In our
present case we have a similar situation, but only if we keep the
localizations $\tilde{I}$ fixed, as indicated by the above notation.
Applying the operators $F_{i}$ in a different order, the result can be
written as a unitary representer of a certain cylinder braid $b_{n}$ applied
to the original n-particle vector. The braid $b_{n}$ is from the groupoid of
colored (by the charge sectors) braids on the cylinder. It depends on the
permutation $\sigma $ and the spacelike paths $\tilde{I}_{1},...,\tilde{I}%
_{n}$ and can be obtained by a simple 3-dim. geometrical construction.
Asymptotically only the directions of the $\tilde{I}_{i}$ is relevant, i.e.
the hyperbolic angle of the translated $\tilde{I}_{i}$ with apex translated
to zero intersects with the unit spacelike hyperboloid. With other words the
path topology is that of the 2-dim. de Sitter space which is topologically
equivalent to $S^{1}.$ This de Sitter space is also the set of points at
spatial infinity by which one has to extend the Minkowski-space $%
M\rightarrow \tilde{M}$ in order to characterize the universal algebra in
geometrical terms as $\mathcal{A}_{uni}=\mathcal{A}(\tilde{M})$ in analogy
with the universal algebra of chiral conformal QFT $\mathcal{A}_{uni}=%
\mathcal{A}(S^{1}).$ These directions together with the non-coinciding
velocity space as well as the inner product in terms of the Markov trace on
the cylinder ribbon braid group may be elegantly encoded into an n-particle
momentum space structure of vector valued wave functions on the covering
space, which obey a covariance property\cite{FGR}. With the structure of the
in-space understood, the remaining problem is now the construction of the
n-particle modular localization subspaces and the associated plektonic nets.
We expect that similar ideas based on thermal methods for wedge localization
as in chapter 6.8 will be useful. Here our voyage ends for the time being.

\textbf{Literature to chapter 7:}

R.Haag ``Local Quantum Physics''. Springer 1992

S.Doplicher, J.E.Roberts: Why is there a Field Algebra with Compact Gauge
Group describing the Superselection Structure in Particle Physics?
Commun.Math.Phys \textbf{131,} 51(1990).

K.H.Rehren and B.Schroer ``Einstein Causality and Artin Braids'' Nucl.Phys. 
\textbf{B312}, 715 (1989)

K.Fredenhagen, K.H.Rehren and B.Schroer , Rev.Math.Phys., Special Issue
(1992) 113.

K.Fredenhagen, Proceedings of the 1991 Cargese Summer Institute.

\chapter{Tentative Resume and Outlook}

The most fruitful times in theoretical physics were those of clash with
principles. The resolution of a deep paradoxon can lead to an enormous
amount of progress, as exemplified by the rapid emergence of QM as a result
of Bohr's atomic model and the ensuing paradoxical situation with respect to
classical physics.

The post electro-weak period of stagnation in particle physics and QFT did
not lead yet to such a clear-cut clash. The present situation in QFT and
high energy physics seems to be somewhat similar to the years preceding the
progress of ``renormalization'' of QFT. Then there was also no clash with
fundamental principles, even though many physicist thought that the correct
handling of the ultraviolet problem requires radical new inventions.

In my opinion the present situation is the result of a significant
disequilibrium between two modes of thinking whose delicate balance is the
hall-mark of good progress in theoretical physics. I am thinking of Dirac's
method build on mathematics and esthetical appeal, versus the
Bohr-Einstein-Heisenberg-Wigner approach based on physical conceptual
analysis. The modern ``Diracian'' approach \cite{S2} is that of inventions,
based particularly on the formal \textit{geometric extension of formalism,}
leaving behind the physical principles which originally led to this
formalism. Recent illustrations of such ``Diracian inventions'' are the
introductions of supersymmetry, strings ore membranes. .

In such a situation it seemed to be helpful to revisit those old ideas in a
critical spirit, which were essential to the conceptual development of QFT.
Gauge theories also belong to this category. Even though they were intensely
investigated for almost three decades, there exists presently no intrinsic
characterization, i.e. if one would receive the physical gauge invariant
correlation on a divine silver plate, one would have no means to find out
whether they have a gauge theory behind without asking the person who
manufactured it. This is not to say that one lacks conjectures; one educated
guess is that if one needs charge carriers which have a semiinfinite
string-like extension, then the charges are Maxwellian charges of a gauge
theory. According to our discussion in chapter 4.5 every theory involving
spin=1 fields are gauge theories in which case the word (broken) ``gauge''
in LQP does not carry more information than renormalizable QFT of spin one.
Semiinfinite massive charge carriers play an important role in the
description of d=2+1 particles with braid group statistics (plektons). In
that case the asymptotic string direction is not frozen by infrared photon
clouds but rather fluctuating (topological charges in the sense of algebraic
QFT). The only way to picture this within a Lagrangian $``$straight jacket''
is through Mandelstam integrals over vectorpotentials which is a gauge
theoretical picture (but not yet a computational scheme).

We emphasized in these notes that despite its conservative way of dealing
with physical principles, algebraic QFT leads to a \textit{radical change of
paradigm}. Instead of the Newtonian view of a space-time filled with a
material content one enters the reality of Leibnitz created by relation (in
particular inclusions) between ``monades''($\sim $ the hyperfinite type III$%
_{1}$local von Neumann factors $\mathcal{A}(\mathcal{O})$ which as single
algebras are nearly void of physical meaning). Related to this is a very new
and surprising esthetics, namely the art of compressing relations between
very big objects as type III$_{1}$von Neumann factors\footnote{%
Here ``big'' is meant in the sense they absorbe any tensor factor with
another von Neumann algebra. Although V.Jones formulated his subfactor
theory in terms of the ``smallest'' infinite dimensional algebras (type II$%
_{1}$which gets absorbed by any other tensor factor), his theory applies
with only a few modifications to the present setting.}) into extremely
simple structures (which is very reminiscent to the esthetics of the V.
Jones subfactor theory). I expect that this new esthetics will be important
in understanding the connection of localization and entropy, i.e. the
generic incorporation of the Bekenstein-Hawking quasiclassical observations
into local quantum physics.

Closely related is the different mathematical way of thinking about a local
description of states. Whereas the standard approach uses the geometrical
idea of sections in fibre-bundles, algebraic QFT deals with sheafs. This is
because a partial state over $\mathcal{A}(\mathcal{O})$ is an equivalence
class of states on $\mathcal{A}$ which yield the same restriction on $%
\mathcal{A}(\mathcal{O})$ $\cite{Haag}.$ The algebraic net has the dual
description in terms of a co-sheafs. Whereas the fibre bundle approach and
Lagrangian QFT is limited to ``quantization algebras'' as CCR and CAR i.e.
algebras over classical function (or section) spaces and their infinitesimal
perturbative deformations, the algebraic approach is supposed to cover the
unknown territory beyond ( the generic nets are not indexed by classical
function spaces).

Another important distinction between the standard approach and algebraic
QFT is that the former deals already in its very formulation with global
concepts as e.g. functional integrals ( the restriction of integration to a
region does not describe the physics of that region), whereas the latter
starts with local concepts and makes contact with global aspects (as global
topology) only in a later stage. Those aspects of the vacuum structure,
which through the spontaneous breaking of localizable symmetries and
superselection rules are related to local properties of the theory, are
correctly accounted for. However vacuum degeneracies without any presently
visible local origin (as e.g. the vacuum structure in the Seiberg-Witten
duality construction), are presently out of reach by methods of algebraic
QFT.

It was our intention to apply concepts of algebraic QFT to those problems
which in our view are not appropriately taken care of by the standard
quantization formalism or which may even contain paradoxes and physically
fruitful contradictions. Examples are the local gauge concept, QFT's in
curved space-time, the structure of nonperturbative low dimensional QFT's,
and the role of various forms of ``Duality'' as well as ''Quantum
Symmetry''. Even in cases where definite answers are still missing,
algebraic QFT certainly casts a different and physically interesting light
on those problems. We reviewed the two notion of temperature, the \textit{%
standard} one being generated by a heath bath in a Lorentz frame and the
second one by a \textit{loss of information} through the creation of a
horizon (the Hawking-Unruh temperature or mathematically: the Tomita KMS
temperature originating from the vacuum representation of local observable
algebras), the latter having a surprising relation to the crossing symmetry
which represents the on shell aspect of causality and which played an
important role at the cradle of string theory.

The deep crisis in QFT and elementary particle physics at the end of this
century (with its ever increasing crave on sophisticated entertainment), is
plainly evident if one just looks into the titles of hep-theory preprints of
recent years. Never in its long and fascinating history has physics as a
human activity of fundamental endeavor been threatened in its existence as
presently. The crisis is to a large degree man made and its main cause is
the adoption of market ideology in science. One of the consequences in
particle physics is the emergence of (un)physical monocultures with the
danger of long term loss of deep knowledge and the demoralization of those
few young people who are willing to make profound intellectual investments
without the promise of instant return. The monocultures which are cultivated
within big electronically connected groups are easily identifyable as
supersymmetry\footnote{%
To avoid any misunderstanding, I am not talking about supersymmetry as a
theoretical structure whose investigation is completely covered by
intellectual curiosity and academic freedom, but I am pointing to the fact
that after 25 years of marginal scientific merits but great employment
capacity (since it can be easily manipulated in calculations) it become the
cropping ground for a second much more dangerous monoculture: string theory.}
with string theory on its back. The danger is not so much that many
physicist adhere to fashions which lead into physically fruitless
directions. This has happened before in this century and physics has
survived such situations. The most dangerous aspect is the quasi-religious
zeal and the signs of mass psychosis against which great intellectual
brilliance is no antidote, a phenomenon which hitherto was restricted to the
great political catastrophes of this century. The problem which threatens
the future of particle theory is not the work on those subjects itself, but
the monomaniacal zeal with which the protagonists make sure that the reign
of their monoculture will be total. The followers do not investigate
problems because they are led by intrinsic physical logic but rather because
one of their gurus drops a word e.g. the present fashion to apply a doses of
noncommutative geometry to strings. The picture the reader finds at the end
of this notes (in an attached ps file) is more serious than he thinks at
first glance: the theory divisions of almost all european and national
research institutions are headed by either protagonists or sponsors of those
monocultures. The future of particle physics as we have known it, is
extremely precarious indeed.

These notes are an (unfashionable and probably futile) attempt to go against
this tide and to recapture some of the earlier conceptual spirit which views
the laws of nature as the realization of physical principles. Whether such
attempts can lead to rational discourse, as they did in better times,
remains a matter of hope.

\textbf{Acknowledgment}: I thank Jos\'{e} Helayel-Neto for the invitation
and the hospitable atmosphere at the CNBPF, where some of the sections were
written in the southern summer of 1997 and the rest in 1998. My thanks for
reading parts of the manuscript and for interesting e-mail communications
goes to Karl-Henning Rehren. I am also indebted for some exchange of letters
with Raymond Stora concerning whose unflinching critical spirit has been a
source of encouragement.

\appendix

\chapter{Mathematical Appendix}

\section{Introduction}

Even at the risk of exaggeration, it is tempting to compare the role of
modular theory in von Neumann algebras with respect to local quantum physics
(LQP) with that of calculus in relation to Newtonian mechanics. I hope that
I succeed to convince the reader in this little mathematical appendix, that
this comparision is more than marketing. Whenever it helps to understand the
mathematical formalism and highlight the physical content, I will (as I
already did for the representation theory of groups in the first chapter)
use physical terminology in addition to the mathematical expressions. I try
to keep the notation of the main text and use large calligraphic letters for
nets of vNa$,$ large Latin letters\footnote{%
With all appologies for the intrusion into stringy and brany terminology.}
as $M,N...$ for vNa (and occasionally also calligraphic latin letters as $%
\mathcal{L}(H),\mathcal{A},\mathcal{C}...$ for individual vNa's and $C^{*}$%
-algebras), small Latin letters $x,y...$for individual operators and Greek
letters for vectors and states on $C^{*}$-algebras. Nets of vNa will always
be denoted by $\mathcal{A},\mathcal{B}..$. With a few exceptions I omitted
proofs, so the reader is asked to look up the literature given at the end
and to use the appendix mainly as an orientation aid (with a mathematical
physics bias) for the study of mathematical literature.

\section{States on C$^{*}$-algebras and representations}

Since (concrete) von Neumann algebras (abbreviated: vNa) are special
operator algebras in Hilbert\footnote{%
A Hilbert space will always have a positive definite inner product and in
the context of these notes it will be separable.} space $H$ (always a
complex Hilbert space unless stated otherwise), we first remind the reader
of some definitions and properties of single operators. If nothing is said,
an operator will always be a bounded linear map of $H$ into itself (bounded$%
\simeq $continuous) and $\mathcal{L}(H)$ or $B(H)$ is the notation for the
vNa (by definition) of all such operators. Operators have an adjoint and $%
x\rightarrow x^{*}$ is an involution which fulfills the so called $C^{*}$
identity: $\left\| xx^{*}\right\| =\left\| x\right\| ^{2}.$ The star is
inherited from the sesquilinear nature of the inner product (for physicist
antilinear in the left hand vector) by shuffling x from the ``ket to the
bra'' vector. There are three types of operators which every physicist knows
from QM: projections $P$ with $P^{2}=P$, selfadjoint operators $x=x^{*}$ and
unitary operators $u^{*}\cdot u=u\cdot u^{*}=1$. Projectors in operator
algebras are always selfadjoint and project onto closed (Hilbert)subspaces.
A positive operator $x\geq 0$ is a special kind of selfadjoint operator
which quantum physicist meet in the tracially normalized form of density
matrices (for the description of impure states in QM). In QFT more general
normal operators ($x\cdot x^{*}=$ $x^{*}\cdot x)$ as well as special
non-normal operators in the form of isometries ($v^{*}\cdot v=1,v\cdot
v^{*}=P)$ make their appearance. Another operator concept which is more
important in QFT than in QM, is the polar decomposition of a normal operator
into a unitary (``phase'') and positive (``radial'') part: $x=u\cdot h$ with 
$h=(x^{*}\cdot x)^{\frac{1}{2}}\equiv \left| x\right| .$ This exists also
for antilinear operators $x$ in which case $u$ will be antiunitary.

The spectrum of $x:$ spec$x$ is the set of $\lambda ^{\prime }s$ such that $%
(x-\lambda )^{-1}$ is not a bounded operator. For selfadjoint operatorsIf it
comes to unbounded operators $S$, only those which have a dense domain of
definition, and are closable feature in these notes (and in physics). Such
operators are most conveniently handled by the method of their graph: $%
G(S)=\left\{ \left( \xi ,S\xi \right) ;\xi \in domS\right\} $ with $G(S)^{-}$
denoting the closure$.$ For a closed operator the graph is a closed subspace
of $H\oplus H.$ Closable operators are defined as those whose graph presents
no ($0,\xi \neq 0)$ obstruction in the closure (i.e. the closure is again
the graph of an operator). The graph of the adjoint operator is $%
G(S^{*})=\left\{ (-S\xi ,\xi );\xi \in domS\right\} ^{\perp },$ where the
orthogonal complement refers to $H\oplus H.$ In this way one also sees that
the denseness of the domain and the property of having adjoints are ``dual''
properties under hermitian conjugation. At this point it is also helpful to
remind the reader on the relation between families of projectors $e(\lambda
) $ which are indexed by subsets $\left( -\infty ,\lambda \right) $ of the
spectrum of a selfadjoint operator and the selfadjoint operator itself,
which certainly everybody knows under the name of spectral resolution or
spectral decomposition of $x:$%
\begin{eqnarray}
x &=&\int \lambda de(\lambda )\longleftrightarrow \\
\left\langle \eta ,x\xi \right\rangle &=&\int \lambda d\mu _{\eta ,\xi
}(\lambda ),\,\,\mu _{\eta ,\xi }(\lambda )=\left\langle \eta ,e(\lambda
)\xi \right\rangle  \nonumber
\end{eqnarray}
This can be expanded to a functional calculus $f\in L^{\infty }(specx,\mu
)\rightarrow f(x)\in \mathcal{L}(H).$

The terminology selfadjoint is applied to unbounded operators only if they
are densely defined (hence closable). In that case the measure $\mu $
decides when a vector $\xi $ belongs to the domain of $x\ $by $\int \left|
\lambda \right| ^{2}d\mu _{\xi ,\xi }(\lambda )<\infty .$ The extension of
the polar decomposition of unbounded densely defined closed operators is
also straightforward. The physically most important application is the
decomposition of the Tomita (antilinear) involution $S.$ Quite generally,
unbounded operators in algebraic QFT are are mostly (in these notes
exclusively) entering via modular theory i.e. they are not individual
operators in the algebra but rather rather describe the modular
characteristics of the representation of the algebra or of the pair
(algebra,reference vector) where the most prominent example of a reference
vector is the vacuum of QFT.

Most concrete calculations in algebraic QFT take place in von Neumann
algebras. These are special operator algebras for which two prominent
examples already have been given: the algebra of all bounded operators $%
\mathcal{L}(H),$ and the commutative algebra of all essentially bounded
functions on a measure space $X:$ $L^{\infty }(X,\mu )$. QM leads to the
first kind of vNa, and the phase space observables of classical mechanics or
the functional calculus of the spectral representation theory on $%
L^{2}(specA,\mu )$ of a selfadjoint operator $A$ to the second. The reason
why vNa are more important for QT than any other type of algebra (and in
fact were introduced by von Neumann for this purpose) is that they implement
the physical idea of compatibility in the theory of quantum measurements
(``simultaneously measurable'') through their notion of commutant. Von
Neumann knew about other types of von Neumann algebras besides the two
mentioned example. Together with Murray he got the first rough
classification in terms of types I, II, and III in terms of projectors (see
next section), but he thought that only the type I $\mathcal{L}(H)$ algebra
is of physical relevance. The type II algebras received most of the
mathematical attention of Murray and von Neumann and their picture of type
III was even mathematically very vage. We know now that all three types
occur in physics, type III becomes important through the localization
concept in local quantum physics and type II enters in two ways: one is the
intertwiner calculus (of localized charges) together with its natural
tracial states (in the differential geometric Lagrangian approach called
``topological field theory'', but here the more appropriate name is
combinatorical QFT) and in magnetic QM for constant magnetic external fields
(the Hofsteadter theory). In form of the noncommutative torus it also has
more recent applications in which physicist were involved. Type II$_{1}$ is
too ``small'' to accomodate spacetime symmetries, but large enough (and
excellently suited) to incorporate inner symmetries described by compact
groups and ``quantum symmetry'' generalizations which are encountered in
low-dimensional QFT.

Von Neumann algebras are special C*-algebras. A C*-algebra is a normed
Banach algebra which fulfills the C* relation $\left\| x^{*}\cdot x\right\|
=\left\| x\right\| ^{2}.$ The conceptual relation of C*-algebras to vNa (for
a physicist) is similar to that of abstract groups\footnote{%
The relation of vNa to group theory is also very close from another point of
view: a vNa is generated by the group of its unitary elements. However in
mathematics the analogy of vNa with measure theory and that of C*algebras
with topology is more fruitful (noncommutative geometry).} to their concrete
representations. In the case of topologically nontrivial QT with quantum
ambiguities in the form of $\theta $-angles chapter 1.2), one wants to
consider the inequivalent theories for different $\theta ^{\prime }s$ as
different representations of one abstract object which is a C*-algebra $%
\mathcal{C}$, a viepoint which has no counterpart in the geometric
fibre-bundle approach where different $\theta $-angles correspond to
different fibre bundles. Therefore C*-algebras play an important conceptual
role, since they allow a unifying point of view. In particular in the
exploration of new symmetry concepts beyond group theory, the C*-algebras
which can be related to a bimodule have attracted a lot of attention. But
most of the calculations are done in the von Neumann closures of their
representations. Although one knows considerably less about C*-algebras than
von Neumann algebras, this did not hamper progress in algebraic QFT because
the von Neumann extensions of their representations make the rich body of
structural theorems and computational tools of vNa'a available to LQP.

The representation theory of C*-algebras is done via states. A state is a
normalized linear positive functional on a C*-algebra\footnote{%
States can be more generally defined on *-algebras e.g. the Wightman theory
and the closely related Borchers-Uhlmann tensor algebras of test functions.}
and the space of states is simply a normalized convex cone inside the dual
space $\mathcal{C}^{*}.$ Therefore all the notions of state decomposition
theory, as pure state, central decompostion etc. apply. The special feature
of a vNa $\mathcal{A}$, which distinguishes it within the C*-setting is the
fact that $\mathcal{A}$ is the dual of a Banach space $\mathcal{A}_{*}$ (the
``predual'') which contains the convex normalized cone $\mathcal{A}%
_{*}^{+(1)}.$ In the case of the vNa $\mathcal{L}(H),$ this is the space of
normalized positive trace operators i.e. density matrices in physical
terminology. For the standard commutative vNa $L^{\infty }(\mathbf{R},dx)$
one has $L^{\infty }(\mathbf{R},dx)_{*}=L^{1}(\mathbf{R},dx).$ Since
(concrete) vNa also have the standard definition of being weakly closed
*-subalgebras of of $\mathcal{L}(H),$ some states on $\mathcal{A}$ may be
simply obtained by restricting the density matrices to $\mathcal{A}\subset 
\mathcal{L}(H).$ At this place the reader should acquaint himself with the
notion (the ideal) of compact operators $K(H)\subset \mathcal{L}(H)$ and the
property $K(H)^{*}=\mathcal{L}(H)_{*}.$ He should also develop an awareness
for the existence of three different topologies: norm topology (operator
norm), strong topology and weak topology (only defined for Hilbertspace
operators). The convergence of sequences of operators on density matrices
gives a slightly stronger topology than the weak and is called ``$\sigma $%
-weak''. It is the cause for the existence of the predual. The reader will
find these definitions and related theorems in the first chapters of any
book on operator algebras; some of them are quoted at the end of the
appendix. More general than normal states on vNa's would behave in a very
pathological manner, whereas on C*-algebras which are not vNa's, the notion
of good and bad or rather regular and singular depends on the C*-algebra.
For example the Heisenberg-Weyl algebra which underlies standard QM
possesses singular states in the sense that there are no continuous
translations $e^{ipa},$ because the lack of continuity does not allow to
define an infinitesimal momentum operator p. In order to exclude such
possibilities which violate the Stone-von Neumann unicity theorem for the
Schr\"{o}dinger representation, one has to insist in regular representations
such that the translation $U(a)$ has continuity properties at $a=0.$ In this
way the so called topological (combinatorical) field theories (type $II_{1})$%
are thought to originate from algebras which formally (e.g. in attempting to
interprete them as the quantization of something) have spacetime aspects,
but loose them through singular representations.

The crucial role of the GNS representation as well as the outline of a proof
already appeared in the first chapter of the text. The difference in the
infinite dimensional case to the finite dimensional one is only in the
topological aspects and the adaption may be done by the reader. The simplest
situation is of course that of a faithful state for which the necessity to
form equivalence classes in order to represent vectors is absent.

The GNS construction is the most valuable construction method for vNa,
because the weak closure of the $C^{*}$-representation in the GNS Hilbert
space is a natural vNa extension of $\mathcal{C}.$ The state is then
automatically ``normal'' with respect to this generated vNa. (defined as
continuity in terms of its predual), however other states on $\mathcal{C}$
do not have this property unless they lie in the same ``folium'' of states
(i.e. states which correspond to density matrices on the vNa extension of
the representation of $\mathcal{C}$), in which case the vNa's are
quasi-equivalent. By this circumstantial definition it should already be
clear to the reader that normality is some continuity property which uses a
specific topology of vNa's (in fact it is the ``$\sigma $-weak
continuity''). The GNS construction permits to transfer properties from the
vNa closure of the canonical GNS representation to the states and
representations on the C*-algebra, e.g. one speaks of factorial
representations and factor states of $\mathcal{C}$. An important property in
QT is the dominance of one state by another:

\begin{theorem}
$\omega _{T}(A)\leq \omega (A)\,\forall A\in A\Longleftrightarrow \exists $
unique $T>0\in \pi _{\omega }(A)^{\prime },\left\| T\right\| \leq 1$ with $%
\omega _{T}(A)=(T\Omega _{\omega },\pi _{\omega }(A)\Omega )=(T^{\frac{1}{2}%
}\Omega _{\omega },\pi _{\omega }(A)T^{\frac{1}{2}}\Omega _{\omega })$
\end{theorem}

If a vNa is concretely given as an operator algebra in a Hilbert space,
there are many ways to represent states by spatial vectors i.e the ambiguity
is much larger than just a change of a representing vector by a phase. The
modular theory furnishes additional concepts which delimit these
possibilities. In particular the ``natural cone representation'' of states
leads to a unique correspondence of states and vectors (presented in a later
section).

As in elementary QM the notion of discrete eigenvectors was generalized by
Dirac to the ``improper'' vectors in the continuous spectrum ($\delta $%
-functions, or in more recent times also ``rigged'' Hilbert spacs), the
notion of state finds its generalization in the theory of weights on $C^{*}$%
-algebras. In order to give meaning to the value $\infty $ being taken by
``improper states'', one should start with a functional which is only
defined on the positive cone $\mathcal{C}_{+}.$ There is a very good
mathematical reason for introducing weights. There is also at least one
physica reason. But in these notes we did not take up those issues.

The C*-algebra setting and the formalism of vNa is more fundamental for
local quantum physics than the setting of functional integrals or other
quantization methods as well as the fibre bundle methods of differential
geometry. Besides to Schr\"{o}dinger QM, the latter essentially apply only
to those theories which allow a classical localization concept as the QFT
version of Weyl-algebras and CAR-algebras\footnote{%
This is why in QM and elementary QFT\ there is no need to leave the naive
Hilbert space formulation which identifies (pure) states with unit rays
created by vectors.}. These are noncommutative C*-algebras over classical
(test) function spaces. Apart from some extremely simple models (example
Schwinger's model), these methods just amount to a transcription of the
perturbation theory into the language of euclidean field theory. Its extreme
theoretical limitation (already the construction of the d=1+1 factorizing
models explained in these notes is outside these quantization schemes) is
only alleviated by the ease which with the greatest historical successes of
QFT, namely the measured effects of QED and its electro-weak unification,
were incorporated in that perturbative quantization framework. Again for
historical and sociological reasons, the mathematical sophistication of the
majority of quantum physicists during the last two decades developed more in
the direction of differential geomety than towards operator algebras, as
they were created with the inspiration from quantum physics by von Neumann.
This is the main reason why presently a text on local quantum physics based
on operator algebra methods needs a mathematical appendix as this.

\section{Von Neumann Algebras and their Classification}

Most of the following coarse grain classification of vNa's was already
accomplished in the three papers of the founding fathers Murray and von
Neumann . The strong connection with von Neumann's previous work on the
foundation of QT and the measurement process is plainly visible in the key
role of the projectors which are the simplest quantum observables.

We already mentioned the crucial notion of the commutant: 
\begin{equation}
M^{\prime }=\left\{ y\in \mathcal{L}(H)\mid \,\left[ x,y\right] =0,\forall
x\in M\right\}
\end{equation}
and its deep root in physics of quantum measurement as well as quantum
localization\footnote{%
In LQP observables are called localized in a spacetime region $\mathcal{O}$
if they commute with every spacelike disconnectedly localized observable,
i.e. the localization in $\mathcal{O}$ and its causal opposite (or dual) are
inexorably linked. For same observable algebras (CCR, CAR), this can be
implemented by the classical notion of support of functions.}. The physical
interpretation of the relation $M^{\prime \prime }=M$ is obvious. The
fundamental ``double commutant theorem'' of von Neumann $Q^{\prime \prime }=%
\overline{Q}^{w}$ ($\overline{Q}^{w}$ is the weak closure of a *subalgebra $%
Q $ of operators in $\mathcal{L}(H))$ converts aspects of topology into
algebraic properties. A physically important algebraic method for
manufacturing vNa's is to take commutants of representations of C*-algebras $%
M=\pi (C)^{\prime },$ the most prominent case being a group algebra: $%
\mathcal{C}=\mathbf{C}G$ in which case one obtains the trivial case $%
M=\left\{ \lambda \underline{1}\right\} $ only for irreducible
representations. Most interest is focused on factors (vNa's with trivial
center) 
\begin{equation}
M\cap M^{\prime }=\left\{ \lambda \underline{1}\right\}
\end{equation}
because general vNa's have a factorial decomposition. The von Neumann
extension of an irreducible representation of a C*-algebra is a special case
of a factor and the factorial property is transferred to the commutant. In
physics the notion of ``indecomposability'' of a system of observables is
both related to irreducibility or factoriality, depending of whether one
deals with global algebras which have pure states (e.g. vacuum, ground state
of lowest energy) or local algebras which in LQP are typically factors with
thermal properties. The only type of an vNa which admits pure states is $%
\mathcal{L}(H).$ They correspond to minimal one-dimensional projectors. The
system of all projectors $\mathcal{P}(M)$ in a vNa form a lattice. It is
clear that unitarily equivalent projectors should be considered as part of
an equivalence class and the first aim would be to understand the class
structure. In order to have coherence of this equivalence notion with
additivity of orthogonal projectors, one need to follow Murray and von
Neumann and enlarge the class of equivalent projectors in the following way:

\begin{definition}
Let $e,f\in \mathcal{P}(M),$ then

\begin{enumerate}
\item  the two projectors are equivalent $e\sim f$ if there exists an
partial isometry such that $e$ and $f$ are the source and range projectors: $%
u^{*}u=e,\,\,uu^{*}=f$

\item  $e$ is subequivalent to $f,$ denoted as $e\preceq f$ if $\exists g\in 
\mathcal{P}(M)$ such that $g$ is dominated by $f$ and equivalent to $e:e\sim
g\leq f$
\end{enumerate}
\end{definition}

One easily checks that this definition indeed gives a bona fida equivalence
relation in $\mathcal{P}(M).$ Via the relation between projectors and
subspaces, these definitions and the theorems of the Murray von Neumann
classification theory can be translated into relations between subspaces.
The main advantage to restrict to factors is the recognition that any two
projectors are then (sub)equivalent. One calls a vNa finite if a projector
is never equivalent to a proper subprojector. Example: in $\mathcal{L}(H)$
infinite dimensional spaces allow a partially isometric mapping on infinite
dimensional subspaces and therefore this factor is infinite. $Mat_{n}(C)$ is
of course a finite factor. It was a great dicovery of Murray and von
Neumann, that there are infinite dimensional finite factors. In fact they
defined:

\begin{definition}
A factor $M$ is said to be one of the following three types:
\end{definition}

\begin{enumerate}
\item  $I,$ if it pocesses pure normal states (or minimal projectors).

\item  $II$, if it not of type $I$ and has nontrivial finite projectors.

\item  $III,$ if there are no nontrivial finite projectors.
\end{enumerate}

Murray and von Neumann were able to refine their classification with the
help of the trace. In more recent terminology a trace without an additional
specification is a weight $Tr$ with $Trxx^{*}=Trx^{*}x$ $\forall x\in M.$ A
tracial state is a special case of a tracial weight.

The use of tracial weights gives the following refinement:

\begin{definition}
Using normal tracial weight one defines the following refinement for factors:
\end{definition}

\begin{enumerate}
\item  type$I_{n}$ if $ranTr\mathcal{P}(M)=\left\{ 0,1,...,n\right\} ,$ the
only infinite type$I$ factor is type$I_{\infty }$. Here the tracial weight
has been normalized in the minimal projectors (for finite n this weight is
in fact a tracial state).

\item  type$II_{1}$ if the $Tr$ is a tracial state with $ranTr\mathcal{P}%
(M)=\left[ 0,1\right] $; type$II_{\infty }$ if $ranTr\mathcal{P}(M)=\left[
0,\infty \right] .$

\item  type$III$ if no tracial weight exists i.e. if $ranTr\mathcal{P}%
(M)=\left\{ 0,\infty \right\} .$
\end{enumerate}

In particular all nontrivial projectors (including \textbf{1}) are
Murray-von Neumann equivalent.

The classification matter rested there, up to the pathbreaking work of
Connes which in particular led to an important gain in understanding and
partially resolving type$III$ factors. Since the modular theory is heavily
used, we will mention some results later on.

\begin{enumerate}
\item  Although nowhere explicitly stated, all indications are that von
Neumann believed on the basis of his analysis of physical observables, that
only the type$I$ with its pure states\footnote{%
In the general mathematical theory the notion of purity of states is not an
important one. For the decomposition theory of states the concept of factor
(central) or in physical terms into superselected charge sectors is more
important.} was of physical interest and that in particular type$II$ and the
associated idea of a ``continuous geometry'' (because of the continuous
range of the notion of ``dimension'') was of exclusive mathematical
importance. We know nowadays that the most important algebra of LQP is the
hyperfinite type$III_{1}$ factor. It is the carrier of the physical
localization properties (including the thermal aspects) and the spacetime
symmetries and is (from the modular point of view) the ``most
noncommutative'' factor in the sense that its folium contains only ``very
noncommutative states''. It also appears for heat bath thermal
representations in the thermodynamical limit. Type$I$ reappears through the
representation theory of globalizations of nets. One of the special very
comforting properties of type$III$ is that any isomorphism or endomorphism
(in particular automorphism) is unitarily implementable in a standard
representation. Type$II$ do not enter physics directly because they are too
small in order to incorporate spacetime symmetries and localization. However
they enter via the algebras generated from intertwiners in the decomposition
theory of localized charges by ``freezing'' the localization regions of the
latter, together with ``Markov traces'' on these intertwining algebras,
which results from the physical cluster decomposition property after
freezing. These combinatorical algebras are better known under the name of
topological field theories because this is the way they appear if one tries
to construct them by functional quantization of classical Chern-Simons
Lagrangians. Since localization is a condition sine qua non for physical
interpretation, it is always important (for physics) to know their position
inside a type$I$ or $III$ factor from which they originate. Type$II$ vNa's
also appear in Hofsteadters work on QM in a constant magnetic background
(noncomutative tori, well-known from the work of Connes and Rieffel). It is
believed that singular states on extended Weyl like $\mathcal{7}C^{*}$%
-algebras, associated with Chern-Simons Lagrangians via quantization, lead
directly to Type$II_{1}$ vNa's.and constitute the operator algebra version
behind Witten's differential geometric receipes on functional integrals
involving Chern-Simons actions which lead to topological field theories
outside the standard setting of euclidean Feynman-Kac representations of
real time QFT. The subalgebras on which singular states become regular
should then agree with the alias ``gauge invariant'' subalgebras.
\end{enumerate}

\section{Modular Theory}

Returning to the analogy with the commutative case $L^{\infty }(X,\mu )\sim
B(H)=\mathcal{L}(H)$, $L^{2}(X,\mu )\sim H,$ and adding the illustrative
example of the type II$_{1}$ tracable algebra, we realize that there are
different grades of noncommutativity. Whereas the first case is entirely
commutative and the second case noncommutative, the type II$_{1}$-algebra's
with a trace is somewhere in the middle, because although it is
noncommutative, the operators of course commute inside the tracial state.
For these noncommutative algebras the Tomita-Takesaki modular theory allows
to measure the degree of noncommutativity with respect to a GNS state. Since
there are no tracial states or even weights on type$III,$ they are in a way
the most noncommutative algebras. Among these the physically relevant (as a
carrier of spacetime localization) type$III_{1}$ sticks out since all states
are equally noncommutative i.e. there are no inner relations in the modular
group for different states.

Take as an example the $I_{\infty }$ factor (the modular concepts for the $%
I_{n}$ factor were already explained in chapter1 of the main text). In order
to describe it as the result of the GNS construction with a faithful state $%
\varphi $ on $\mathcal{L}(H)$, let $\rho $ be a trace class operator such
that in its spectral representation in terms of a orthonormal basis $\left|
i\right\rangle $ it has strictly positive (nonvanishing) components $\lambda
_{i}$ for all $i=1...\infty .$%
\begin{eqnarray}
\rho &=&\sum \lambda _{i}\left| i\right\rangle \left\langle i\right|
,\,\,\sum_{i}\lambda _{i}=1\, \\
&&\varphi (x)=tr\rho x  \nonumber
\end{eqnarray}
This garanties the faithfulness of the state and in complete analogy to the
matrix case in chapter1, the GNS construction leads to a Hilbert space of
Hilbert-Schmidt operators $\mathcal{H}$ which is generated by the algebra $%
\mathcal{L}(H)$ together with the inner product $(x;y)=tr\rho x^{*}y.$ The
reader immediatly realizes that in addition to the GNS left action of $%
\mathcal{L}(H)$ on $\mathcal{H}\equiv L^{2}(\mathcal{L}(H),\varphi )$ one
also has an opposite right action such that $\mathcal{L}(H)^{\prime }=%
\mathcal{L}(H)_{r}^{opp}.$ In fact at this point it is helpful to take some
hint from the standard Gibbs heat bath description by imagining a $\rho =%
\frac{e^{-\beta K}}{tre^{-\beta K}}$ with an hamiltonian $K$ with discrete
positive spectrum such that the trace exists. Then the same calculation as
done in chapter1 with matrices will lead to: 
\begin{eqnarray}
Sx\rho &=&x^{*}\rho \\
\curvearrowright S &=&J\Delta ^{\frac{1}{2}}  \nonumber
\end{eqnarray}
where in complete analogy to the finite dimensional case the $J$ represents
a antiunitary flip from the left action to its opposite and $\Delta ^{\frac{1%
}{2}}=\pi _{l}(\rho )\otimes \pi _{r}(\rho ^{-1}).$ The last formula for $%
\Delta ^{\frac{1}{2}}$ corresponds to the physical fact that $\Delta
^{it}=e^{-i\beta K_{th}}$ with the thermal hamiltonian (the generator of
time translations \textit{in the thermal state}) $K_{th}$ is the sum of the
left hand minus the right hand action of the original hamiltonian $K$ on: 
\begin{equation}
H\otimes H\simeq L^{2}(\mathcal{L}(H),\varphi )
\end{equation}
as in the finite dimensional case. Whereas in the adjoint action on the
observable algebra $\mathcal{L}(H)$ the difference cancels, in the
applications to vectors in $\mathcal{H}$, only $H_{th}$ gives the correct
time propagation without (in the thermodynamic limit $V\rightarrow \infty )$
thermal fluctuations in the reference state $\rho .$ This is the setting
which led Haag, Hugenholz and Winnik to there formulation of the KMS
condition and their emphasis of the fact that this condition survives the
thermodynamical limit (which is generally outside type$I)$ and is
responsible for the stability of these thermal states$.$ Another easily
accessible illustration is that of a type$II_{1}$ factor for which one can
directly work with the (unique) tracial state. The situation leads to
similar formulae i.e. the algebra acts left and right (the opposite) on
itself and the antiunitary $J$ act in the same way but now $\Delta =1,\,$%
i.e. the KMS formula is trivial$.$ In both cases the GNS construction led to
a GNS tripel ($M;H,\Omega )$ with $\Omega $ a cyclic and separating vector
in $H$ of $M.$ The separating property results from the faithfulness of the
state and means that $x\Omega =0\curvearrowright x=0.$ A vNa with these
extra trimmigs is called ``in standard form'' (strictly speaking this
terminology is only used if in addition one uses the natural cone described
below). Now we are well-prepared to state the result of the Tomita-Takesaki
theory in the most general setting:

\begin{theorem}
Let $M$ be a vNa in standard form (all vNa's with separable preduals permit
this standard form). Then the Tomita involution: 
\begin{equation}
Sx\Omega \equiv x^{*}\Omega
\end{equation}
is closable and hence permits a unique polar decomposition: 
\begin{equation}
S=J\Delta ^{\frac{1}{2}}
\end{equation}
The adjoint action of the antiunitary Tomita involution $J$ and the bounded
one parametric group generated by $\Delta ^{it}$ on the algebra is the
following: 
\begin{eqnarray}
adJM &=&M^{\prime } \\
ad\Delta ^{it}M &=&\sigma _{t}(M)=M  \nonumber
\end{eqnarray}
where $\sigma _{t}$ is called the modular automorphism group. The modular
construction can be generalized to the case of any separable vNa with a
faithful normal weight.
\end{theorem}

The proof of this theorem is simple for type$I$ and $II_{1}$but somewhat
demanding in the type$III$ cases, with the type$III_{1}$ being the most
demanding. In the most general case with unbounded $S$ one must use
analyticity techniques (which the older generation QFT-physicists is
familiar with because similar analytic methods have been used in Wightman
QFT) which entered through the relation with the KMS theory. We refer the
reader to the literature at the end of this appendix. The form of modular
operators for type$I$ factors is known whereas the general form for type$III$
factors is unknown. More profound than the above relation with heat bath
physics is the relation with local algebras $\mathcal{A}(\mathcal{O})$ of
the nets of LQP. These algebras are hyperfinite type$III_{1}$ and therefore
belong to the nontrivial application of the T-T modular theory (the
subscript refers to Connes modular refinement of the type classification.
The unboundedness of $\Delta $ and the fact that the modular automorphism is 
$\sigma _{t}$ outer (for all values of t) with respect to $M,$ makes this
algebra more noncommutative than the other types and it turns out, that it
is precisely this situation which is necessary in order to have modular
groups with relations to spacetime symmetry. The KMS\ condition allows a
direct check if an automorphism $\sigma _{t}$ is a a modular automorphism of
a given state $\varphi $ on $M$ by checking that it satisfies the KMS
boundary condition.

\begin{definition}
A state $\varphi \,$is said to fulfil the KMS boundary condition (at inverse
temperature $\beta =1)$ with respect to a an automorphism $\sigma _{t}$ of a 
$C^{*}$-algebra $\mathcal{C}$ if $\varphi (\sigma _{t}(x)y)$ is a continuous
(in t) boundary value of a strip analytic function $F(z),$ $-1<Imz<0$ with: 
\begin{eqnarray}
F(t) &=&\varphi (\sigma _{t}(x)y) \\
F(t+i) &=&\varphi (y\sigma _{t}(x))  \nonumber
\end{eqnarray}
\end{definition}

In extending the state to the vNa $M\equiv \pi _{\varphi }(\mathcal{C}%
)^{\prime \prime }$ there arises the question whether the automorphism
extends to this vNa in a natural way or in technical terms whether it has
has a $\sigma $-weak continuous extension. This turns out to have an
affirmative answer if the C*-algebra is unital and $\varphi $ is a $\sigma
_{t\text{ }}$invariant state. It furthermore turns out that this property is
characteristic i.e. such a $\sigma _{t}$ which fulfils the KMS condition for
a faithful state is necessarily the modular automorphism group a T-T theory
of any associated GNS representation. The importance of the KMS condition in
quantum statistical mechanics of open systems (i.e. in the thermodynamic
limit) is related to their stability against local perturbations.

Whereas the $\Delta ^{it}$ depends on the realization of the GNS
representation in the class of the unitarily equivalent realizations, the
modular automorphism only depends on the state and not on its representing
vector. In order to have the converse, one need to generalize from states to
weights. In that case Connes showed that each cocycle related modular group $%
u_{t}^{*}\sigma _{t}(\cdot )u_{t}$ is the modular group of a unique normal
weight 
\begin{equation}
\omega _{u}(\cdot )\equiv \lim_{t\rightarrow \frac{i}{2}}\omega (u_{t}\cdot
u_{t}^{*})
\end{equation}

A useful generalization is the relative modular theory which refers to two
cyclic and separating vectors say $\Omega ,\Omega ^{\prime }$
simultaneously. One defines the conjugate linear operator: 
\begin{equation}
S_{\Omega ^{\prime },\Omega }x\Omega =x^{*}\Omega ,\,\,x\in M
\end{equation}
Again the closability is easily established and the polar decomposition
leads to a positive operator $\Delta _{\Omega ^{\prime },\Omega }^{\frac{1}{2%
}}$ which is called the relative modular operator of the pair $\Omega
^{\prime },\Omega .$ If $\Omega ^{\prime }$ is in the natural cone of $%
\Omega $ (see below), $J_{\Omega ^{\prime },\Omega }=J_{\Omega ,\Omega }=J.$
Defining the unitaries $U_{\Omega ^{\prime },\Omega }(t)\equiv (\Delta
_{\Omega ^{\prime },\Omega })^{it}$ one finds the following symmetric
(between $M$ and $M^{\prime })$ relations: 
\begin{eqnarray}
U_{\Omega ^{\prime },\Omega }(t)xU_{\Omega ^{\prime },\Omega }^{*}(t)
&=&U_{\Omega ^{\prime }}(t)xU_{\Omega ^{\prime }}^{*}(t),\,\,\,x\in M \\
U_{\Omega ^{\prime },\Omega }(t)x^{\prime }U_{\Omega ^{\prime },\Omega
}^{*}(t) &=&U_{\Omega }(t)xU_{\Omega }^{*}(t),\,\,\,x^{\prime }\in M^{\prime
}  \nonumber
\end{eqnarray}
with $U_{\Omega ^{\prime }}(t)\equiv \Delta _{\omega ^{\prime
}}^{it},U_{\Omega }(t)\equiv \Delta _{\omega }^{it}.$ Using another
``spectator'' cyclic separating vector $\Omega ^{\prime \prime }$ the
definition 
\begin{equation}
u_{\Omega ^{\prime },\Omega }(t)\equiv U_{\Omega ^{\prime },\Omega ^{\prime
\prime }}(t)U_{\Omega ,\Omega ^{\prime \prime }}^{*}
\end{equation}
gives a unitary which commutes with $M^{\prime }$ and therefore lies in $M.$
The spectator $\Omega ^{\prime \prime }$ was chosen because this unitary
turns out to be independent of $\Omega ^{\prime \prime }.$ This operator is
the famous Connes cocycle which relates the modular operators of the two
states $U_{\Omega ^{\prime }}(t)=u_{\Omega ^{\prime },\Omega }(t)U_{\Omega
}(t)$: 
\begin{eqnarray}
u_{\Omega ^{\prime },\Omega }(t_{1}+t_{2}) &=&u_{\Omega ^{\prime },\Omega
}(t)\sigma _{t_{1}}^{\omega }(u_{\Omega ^{\prime },\Omega }(t_{2})) \\
\sigma _{t}^{\omega ^{\prime }}(\cdot ) &=&u_{\Omega ^{\prime },\Omega
}(t)\sigma _{t}^{\omega }(\cdot )u_{\Omega ^{\prime },\Omega }^{*}(t) 
\nonumber \\
(D\omega ^{\prime } &:&D\omega )(t)=u_{\Omega ^{\prime },\Omega }(t) 
\nonumber
\end{eqnarray}
The notation of the last equation suggest to view the Connes cocycle as the
noncommutative generalization of the classical Radon-Nikodym concepts. It
was already mentioned before that given the modular group $\sigma
_{t}^{\omega }$ and a Connes cocycle, one can construct a state $\omega
^{\prime }$ such that its modular group $\sigma _{t}^{\omega ^{\prime }}$ is
cocycle related to $\sigma _{t}^{\omega }$ with one caveat: the state may be
improper i.e. a weight.

The modular theory can be used for strengthening the relation between states 
$\omega $ and representing vectors $\xi (\omega ).$ To this end one starts
from the vNa in standard form ($M,H,\Omega )$ and defines in $H$ the
``natural cone'' $\mathcal{P}^{\#}$as the weak closure of a convex set: 
\begin{eqnarray}
\mathcal{P}^{\#} &=&\overline{\Delta ^{\frac{1}{4}}M_{+}\Omega } \\
&=&\overline{\left\{ JxJx\Omega ,x\in M\right\} }  \nonumber
\end{eqnarray}
The equality of the two expressions is derived from the relation $\Delta ^{%
\frac{1}{4}}xx^{*}\Omega =\sigma _{-\frac{i}{4}}(x)J\sigma _{-\frac{i}{4}%
}(x)\Omega .$ For entire analytic operators $x\in M_{0}$ i.e. such that $%
\sigma _{t}(x)$ has an analytic continuation to an entire function $\sigma
_{z}(x)$ this relation is easily derived and with $y\equiv \sigma _{-\frac{i%
}{4}}(x)\in M$ we obtain the desired identity which, as a result of the
denseness of $M_{0}$ in $M$ can be continued to all $M$. One then has the
following unicity theorem:

\begin{theorem}
$\xi (\omega )$ is the only vector in $P^{\#}$ such that 
\begin{eqnarray}
(\xi (\omega ),x\xi (\omega )) &=&\omega (x),\,\,\,x\in M\, \\
(\xi (\omega ),x^{\prime }\xi (\omega )) &=&\omega (AdJ(x^{\prime
*})),\,\,\,x^{\prime }\in M^{\prime }  \nonumber
\end{eqnarray}
the map $\xi \in \mathcal{P}^{\#}\rightarrow \omega _{\xi }\in M_{*+}$ is a
homeomorphism $\mathcal{P}^{\#}\rightarrow M_{*+}$ in the norm topology with 
$\left\| \xi -\eta \right\| ^{2}\leq \left\| \omega _{\xi }-\omega _{\eta
}\right\| \leq \left\| \xi -\eta \right\| \left\| \xi +\eta \right\| $ and
the inverse map $\omega \rightarrow \xi (\omega )$ is monotonously
increasing and concave with respect to the natural ordering of the two cones.
\end{theorem}

In the case of dominance $\omega \leq \omega _{\Omega },$ one obtains an
explicite formula for $\xi (\omega ):$%
\begin{equation}
\xi (\omega )=\left| A^{\prime }\Delta ^{-\frac{1}{2}}\right| \Omega
\end{equation}
Here $A^{\prime }\in M^{\prime }$ is the unique operator which corresponds
according to the dominance theorem to the state $\omega $ namely $\omega
(\cdot )=(A^{\prime }\Omega ,\cdot A^{\prime }\Omega ),$ whereas $\left|
B\right| $ denotes the positive part in the polar decomposition of $B.$ With
a little extra efford one can proof that $A^{\prime }=u_{\frac{i}{4}}$

Another physically useful aspect of the natural cone formalism is the fact
that automorphism (and endomorphisms) $\alpha \in Aut(\mathcal{A})$ can be
naturally unitarily implemented, i.e.$\exists U(\alpha )$ with: 
\begin{eqnarray}
U(\alpha )AU^{*}(\alpha ) &=&\alpha (A) \\
U(\alpha )\mathcal{P}^{\#} &\subseteq &\mathcal{P}^{\#}  \nonumber \\
\left[ U(\alpha ),J\right] &=&0  \nonumber
\end{eqnarray}
In fact $U(\alpha )\xi (\omega )=\xi (\alpha ^{-1*}(\omega ))$ where ($%
\alpha ^{*}\omega )(A)\equiv \omega (\alpha (A)).$ The natural cone theory
is also the starting point for Connes reconstruction of algebras from the
spatial data of the position of the natural cone inside the Hilbert space $%
\mathcal{P}^{\#}$ $\subset H_{\Omega }.$ Reconstructions of (nets of)
algebras from (nets of) positions of real subspaces $H_{R}=\mathcal{P}^{\#}+-%
\mathcal{P}^{\#}$ is an important issue in chapter 6. In the main text it
was not necessary to use this construction (because of the existence of a
reference algebra of the ``in net'' supplied by scattering theory permitted
the introduction of a modular M\o ller operator).

The application of modular theory to the problem of classification of von
vNa's led Connes to a complete understanding of all hyperfinite vNa's.
Hyperfinite means that they can be approximated by finite dimensional vNa's
or equivalently that the algebra is in the range of a conditional
expectation in $\mathcal{L}(H).$ The physical intuitive content is that the
associated physical systems allow an interpretation as a thermodynamic limit
of a sequence of finite systems or equivalently, in case of properly
infinite LQP, an approximation in terms of QM (type$I)$ systems. All local
vNa's $\mathcal{A}(\mathcal{O})$ are hyperfinite, but there globalizations
sometimes involve ``free product'' constructions which lead out of
hyperfiniteness.

\section{C$^{*}$-Algebras related with Bimodules}

In chapter 7 of the main text we have seen that the DHR theory leads to the
(reduced) field bundle which is really an algebraic bimodule on which the
observable algebra acts from the right through the identity representation
and to the left through its endomorphic image. This theory led to the theory
of which define a structure. The DR analysis in the case of d=1+3 QFT
finally led to a complete understanding of an associated field algebra with
a compact group action (symmetry). For this result to emerge algebraic
Hilbert spaces and associated Cuntz C$^{*}$-algebras played an important
role. For d$\leq 1+2$ QFT one neads a yet unknown generalization. It is a
natural idea to look for generalizations within the setting of bimodules
(containing Hilbert spaces as special cases) and their naturally associated
algebras. Assume that $\mathcal{A}\subset \mathcal{B}$ and denote by $%
\mathcal{C}$ the relative commutant $\mathcal{C}=\mathcal{A}^{\prime }\cap 
\mathcal{B}.$ Define a subset of $\mathcal{B}$: 
\begin{equation}
X_{\rho }=\left\{ \psi \in \mathcal{B}\mid \psi A=\rho (A)\psi \right\}
\end{equation}
where for the endomorphism of $\mathcal{A}$ we used the standard notation $%
\rho .$ Define now the $\mathcal{C}$-valued inner product: 
\begin{equation}
\left\langle \psi ,\psi ^{\prime }\right\rangle _{\mathcal{C}}=\psi ^{*}\psi
^{\prime }
\end{equation}
such that $\left\| \left\langle \psi ,\psi \right\rangle _{\mathcal{C}%
}\right\| =\left\| \psi \right\| _{\mathcal{B}}^{2}.$ If $X_{\rho }$ is
finite projective, we call $\rho $ inner in $\mathcal{B}.$ One immediatly
realizes that by setting $\mathcal{C}=$ $C\mathbf{1}$ but allowing $A\neq
B,A^{\prime }\cap B=C\mathbf{1\,}$i.e. irr. inclusion, $X_{\rho }$ is a
Hilbert space in $\mathcal{B}$ and $\rho $ the restriction to $\mathcal{A}$
of an inner endomorphism in $\mathcal{B}$ i.e. $\rho (B)=\sum_{1}^{n}\psi
_{i}B\psi _{i}^{*},$ with $\psi _{i},i=1,...,n$ an orthonormal basis in $%
X_{\rho }.$ In that case the DH-theory shows that $\mathcal{B}$ is a crossed
product of $\mathcal{A}$ by the action of a compact group. However the
universal algebra of the algebraic compactification of d$\leq 1+2$ QFT's
with braid group statistics in chapter 7 is a C*-algebra with nontrivial
center and there are indications that a would be field algebra $\mathcal{B}$
may be such that the relative commutant is nontrivial $A^{\prime }\cap B\neq
C\mathbf{1.}$ In that case it may be interesting to look for a symmetry
concept which is related to a more general crossed product $\mathcal{B}$
associated to the pair $\left\{ \mathcal{A},\rho \right\} $ and to find
reasonable conditions for $\mathcal{C}$ which insure uniqueness of $\mathcal{%
B}.$ A special situation of this type, which is mathematically interesting
in its own right is the situation where $X$ is given as a Hilbert
C*-bimodule with coefficients in $\mathcal{C}$, i.e.a right Hilbert $%
\mathcal{C}$-module with a monomorphism of C into $\mathcal{L}(X)$ defining
the left action. In that case the bimodule tensor powers $X^{r}$ may be
considered as the objects of a C*-category (with those adjointable right
module maps which commute with the left action of $\mathcal{C}$ being the
arrows of the category). There is a functorial construction (used in the DR
theory) which relates a C*-algebra $\mathcal{O}_{X}$ with such a situation.
Algebras as this and their subalgebras (example: Cuntz-Krieger algebras)
from such a bimodule viewpoint have been first studied by Exel and Pimsner.
For a presentation which is most close to the spirit of LQP I refer to the
paper of Doplicher et. al. below (where also furthergoing ivestigations
closer to the symmetry problem of LQP have been announced).

\section{Conditional Expectations, Canonical Endomorphisms}

The knowledge about conditional expectations on vNa's of QFT physicist is in
the majority of cases limited to the abelian case which is of course
probabilistic cradle of this concept. The renormalization group
manipulations in eucledian field theory where one often decimates degrees of
freedom as well as Nelson's Markov property uses either tacitly or
explicitely such concepts. As in the abelian case of measure spaes and the
associated $L^{\infty }(X,\mu )$ algebra, one defines a conditional
expectation $E:M\rightarrow N$ from $M$ to a subalgebra $N\subset M$ as a
projection of norm 1, i.e. a completely positive normalized (unit
preserving) map\footnote{%
All of our endomorphisms, operator valued weights and endomorphism are
always assumed to be injective so that they can be inverted on their image.} 
$E$ with $E(y^{*}xy)=y^{*}E(x)y,\,y\in N.$ The prototype noncommutative
example is obtained by having a group $G$ act on $M$ (example free field of
chapter 3 of a multicomponent field with a $SU(N)$ action) and using for $N$
the fixpoint algebra $N=M^{G}.$ A conditional expectation for this situation
is given by the Haar average as in chapter 7. If the group is not compact,
one obtains an improper version of $E$ which is called a operator-valued
weight and has the analog relation to a bona fide $E$ as a weight has to a
state.

There is one crucial difference to the commutative situation: a conditional
expectation (or its improper version) does not exist for each pair $N\subset
M.$ Instead one finds a fascinating relation between conditional
expectations and modular theory as described in the following theorem of
Takesaki:

\begin{theorem}
A modular group $\sigma _{t}$ of ($M,\Omega )$ leaves a subfactor $N$
invariant iff the inclusion $N\subset M$ possesses a normal ($\sigma $-weak
continuous) $\Omega $-invariant conditional expectation $E$ iff the modular
group $\sigma _{t}$ of $M$ restrict to the corresponding modular objects of $%
N$ on $\overline{N\Omega }$ and the equality $\overline{N\Omega }=\overline{%
M\Omega }$ holds only iff $M=N.$ In that case also the modular conjugation
of $\ M$ restricts to that of $N.$
\end{theorem}

Let us indicate the proof$.$ Assuming that the modular theory restricts, we
define $E:M\rightarrow N$ as $PxP=E(x)P$ where $P$ projects on $\overline{%
N\Omega }.$ This definition would be reasonable if we can show that $PxP=yP,$
$y\in N,$ because then by using the existence of a $U$ with $%
U^{*}U=P,UU^{*}=P^{\perp },$ one defines $E(x)=PxP+$ $P^{\perp
}UxU^{*}P^{\perp }$ which verifies the above relation and the properties of
a conditional expectation. In order to show the required property we start
from the evident inequality $NP\subseteq PMP$ and show the opposite
inequality $\supseteq :PMP\subseteq PJ_{M}N^{\prime }J_{M}P=J_{M}P^{\prime
}N^{\prime }PJ_{M}=J_{N}PN^{\prime }PJ_{N}=NP$ where in the first step we
used $M\subseteq J_{M}N^{\prime }J_{M}.$ This shows the $\rightarrow $ part.
Reversely, let us assume the existence of an $E:M\rightarrow N$ preserving
the vector state $\Omega .$ Define $Pa\Omega =E(a)\Omega \in H_{N}$ and
obtain ($a-E(a))\Omega \in H_{N}^{\perp }.$ From $a\in M,\,\;b,c\in N$ and $%
(\Omega ,c^{*}PaPb\Omega )=(\Omega ,c^{*}E(ab)\Omega )=(\Omega
,c^{*}E(a)b\Omega )$ obtain $PaP=E(a)P.$ Now study the Tomita operator on $%
H_{N}:S_{M}Pa\Omega =S_{M}E(a)\Omega =E(a^{*})\Omega =PS_{M}a\Omega .$ From
the denseness of states $M\Omega $ obtain $S_{M}P=PS_{M}$ which means that $%
S_{M}$ leaves $H_{N}$ invariant and restricts to $S_{N}.$ The rest follows
by polar decomposition.

As weights constitute a generalization of states, the natural generalization
of conditional expectations are the operator valued weights. In both cases
the generalizations are necessary in order to be able to invert or dualize
certain relations. The conversion back of a cocycle related modular group to
a state of which it is the modular group will have an obstruction which is
removed by allowing weights. Similarly another theorem of Connes and Hagerup
states that the dual inclusion $M^{\prime }\subset N^{\prime }$ for any
conditional expectation of the original inclusion $E\subset C(M,N)$
possesses a dual $E^{-1}:N^{\prime }\rightarrow M^{\prime }$ which generally
turns out to be an operator valued weight $E^{-1}\subset P(N^{\prime
},M^{\prime }).$ The latter is a singular (not defined on \textbf{1) }%
generalization of a conditional expectation in a similar sense as weights
are singular generalizations of states.

We will see in the next section that the group average example for a
conditional expectation is in a certain sense typical; conditional
expectations are always related to what a physicist calls an internal
symmetry which structure the relation between observable nets and field
nets. A different kind of inclusion which in the next section is shown to be
relates to space-time symmetries is given in terms of the so-called
canonical endomorphism which was introduced by Longo. The construction is as
follows. Start from an inclusion of properly infinite vNa's $N\subset M$ on
a separable Hilbert space $H.$ Then there exist vectors $\Omega \in H$ which
are cyclic and separating for both algebras (for finite algebras such
vectors do definitely not exist and therefore there is no canonical
endomorphism in the sense below). With the help of the corresponding $%
J_{M,N} $ and their adjoint actions $j_{M,N}$ we define the canonical
endomorphism $\gamma $: 
\begin{equation}
\gamma =j_{N}j_{M}|_{M}\in End(M)
\end{equation}
It maps $M$ into a subalgebra $N_{1}$ of $N$ and the $j^{\prime }s$ can also
be used in order to define a canonical extension: 
\begin{eqnarray}
restr. &:&\,\,N_{1}\equiv j_{N}j_{M}(M)\subset N  \label{inc} \\
ext. &:&\,\,\,\,M\subset M_{1}\equiv j_{M}j_{N}(N)=\gamma ^{-1}(N)  \nonumber
\end{eqnarray}
In fact without changing the Hilbert space one can keep on going into both
directions by defining $\rho \equiv j_{N_{1}}j_{N}\in End(N),\gamma
_{1}\equiv j_{M}j_{M_{1}}\in End(M_{1})$ with the standardness of the new
algebras with respect to $\Omega $ always being maintained. The equality
between modular conjugations of a vNa and its commutant leads to: 
\begin{eqnarray}
J_{N_{1}}J_{N} &=&J_{N}J_{N}J_{N_{1}}J_{N}=J_{N}j_{N}(J_{N_{1}})=J_{N}J_{M}
\\
&=&....=J_{M}J_{M_{1}}  \nonumber
\end{eqnarray}
and hence $\rho =\gamma |_{N}$ and $\gamma =\gamma _{1}|_{M}.$ Finally we
notice that the inclusions (\ref{inc}) are isomorphic thanks to $N=\gamma
_{1}(M_{1})$ and $N_{1}=\gamma (M)=\gamma _{1}(M).$ For plausible reasons
they are called ``dual'' relative to $N\subset M.$ They are anti-isomorphic
to $M^{\prime }\subset N^{\prime }$ by application of $j_{N}$ and in the
same vein $M_{1}^{\prime }\subset M^{\prime }$ is the anti-isomorphic image
of $N\subset M$ under $j_{M}.$ The ``tower'' and ``tunnel'' created by
positive and negative powers $\gamma ^{n}$ applied to $M$ resp. $N$ become
the even resp. odd parts of the Jones tower/tunnel associated with
subfactors if the inclusion $N\subset M$ permits a conditional expectation.

\section{Deep Inclusions and (spacetime) Geometry}

Geometric inclusions, as e.g. a double cone localized algebra in LQP
containing a localized algebra of a genuinly smaller double cone, do not
permit vacuum invariant conditional expectations because the modular groups
are very different. In the sense of the geometrical interpretation coming
from LQP they are ``deep''i.e they have a large relative commutant $%
N^{\prime }\cap M$. There is a particular deep family of inclusions which is
susceptible to profound investigations by modular methods. They are the
so-called ``half-sided modular'' inclusions as well as the ``modular
intersections''. Since many of their properties were already mentioned in
the main part of the notes, we will be content with referring to some
additional remarks in the small print at the end of chapter 3.5.

\section{Shallow Inclusions and (internal) Symmetry}

The existence of conditional expectations is the prerequisite for the Jones
theory of subfactors, as well as for its older LQP analogon, the DHR theory
of (localized) endomorphisms. In more recent times, there has been a merger
of the two into the sector theory of Longo, which is with increasing
frequency also used by mathematicians:. It is very closely related with
Connes theory of correspondences (bimodules). Bimodules also have been used
in the 1970 work of DHR on field bundles. In order to have unimpeded unitary
implementation of endomorphisms we will assume type$III$ factors. Let $%
L^{2}(M)$ be the standard Hilbert space with the modular conjugation $J_{M}.$
Let us denote $L^{2}(M)$ as $H_{\rho }$ if we use it as a representation
space of $M-M^{opp}$ acting as $m_{1}\cdot \xi \cdot m_{2}=\rho
(m_{1})J_{M}m_{2}^{*}J_{M}.$ In this way an endomorphism and $L^{2}(M)$ give
rise to a bimodule $H_{\rho }.$ Since according to our assumption every
isomorphism is spatial, one easily sees that each $M-M$ bimodule, whether
originating in the indicated way or not, can always be written in the form
of an $H_{\rho }$ and unitary equivalent bimodules have inner conjugate $%
\rho ^{\prime }s.$ We define: 
\begin{equation}
Sect(M)=End(M)/Int(M)
\end{equation}
i.e. sectors are equivalence classes of $End$ or $Bim$ according to taste.
Physicists prefer the $End$ because the composition rule follows more
standard procedures (ordinary composition of endomorphisms): 
\begin{equation}
H_{\rho _{1}}\otimes _{M}H_{\rho _{2}}=H_{\rho _{2}\rho _{1}}
\end{equation}
The natural setting for this bimodule calculus is the Connes spatial modular
theory. The sector notation is the standard equivalence class notation: $%
\left[ \rho \right] \in Sect,$ if $\rho \in End.$

Intertwiners are isometric operators $V$ in $M$ which relate endomorphisms $%
V:\rho _{1}\rightarrow \rho _{2}$. They form linear spaces: 
\begin{eqnarray}
(\rho _{1},\rho _{2}) &=&\left\{ V\in M;V\rho _{1}(x)=\rho _{2}(x)V\right\}
\\
&=&Hom(H_{\rho _{1}},H_{\rho _{2}})  \nonumber
\end{eqnarray}
In case $\rho _{1}$ is irreducible, i.e. $M\cap \rho _{1}=\mathbf{C\cdot 1}%
_{M}$ we can define an inner product: 
\begin{equation}
\left\langle V|W\right\rangle \mathbf{1}_{M}\equiv V^{*}W,\,\,\,V,W\in (\rho
_{1},\rho _{2})
\end{equation}
In particular the selfintertwiners correspond to relative commutants: 
\begin{equation}
(\rho ,\rho )=M\cap \rho (M)^{\prime }
\end{equation}
The index of an inclusion of factors is defined if $N\subset M$ possesses a
conditional expectation $E$. In terms of this conditional expectation, the
definition is: 
\begin{equation}
Ind(E):=E^{-1}(\mathbf{1})
\end{equation}
where $E^{-1}$ denotes the operator valued weight which corresponds to $E$
by applying the previously mentioned Connes bijection between operator
valued and their duals. The right hand side lies in the center of $M$ and
therefore is a multiple of unity:$Ind(E)\in \left[ 1,\infty \right] $ with $%
\infty $ if the identity is not in the range of the operator valued weight $%
E^{-1}.$ If it is finite, then in fact the operator valued weight $E^{-1}$
can be normalized by $E^{\prime }=Ind(E)^{-1}E^{-1}$ and defines the dual
conditional expectation which belongs to $M^{\prime }\in N^{\prime }$ and
fulfils $E^{\prime -1}=Ind(E)E$ and hence $Ind(E^{\prime })=Ind(E)$ with a
central valued $Ind(E).$ The set of normal conditional expectations consists
of more than one element if the inclusion is not irreducible. In that case
one uses the so called minimal index which corresponds to the previously
defined minimal conditional expectation $E_{0}$: 
\begin{equation}
\left[ M:N\right] :=\inf_{E}Ind(E)=Ind(E_{0})
\end{equation}
This index has a series of remarkable properties which are best formulated
and understood by introducing the concept of statistical dimension:

\begin{definition}
The statistical dimension of an inclusion is the square root of its minimal
index $d\equiv \left[ M:N\right] ^{\frac{1}{2}}$
\end{definition}

Using this concept in the above fusion and decoposition formalism of finite
index endomorphisms we find that $d_{\rho }$ follows precisely the rules of
a dimendional function in that it is additive and multiplicative: 
\begin{eqnarray}
d_{\rho } &=&\sum_{i}d_{\rho _{i}},\,\,\rho =\sum_{i}\rho _{i} \\
d_{\rho _{1}\rho _{2}} &=&d_{\rho _{1}}d_{\rho _{2}}  \nonumber
\end{eqnarray}
It is one of the advantages of type$III$, that one can also freely add
endomorphisms. For adding $n$ of them one selects $n$ isometries $%
v_{i},i=1...n$ with orthogonal ranges $v_{i}^{*}v_{j}=\delta _{ij}$ and
total range one $\sum_{i}v_{i}v_{i}^{*}=\mathbf{1}$ and defines the direct
sum endomorphism $\rho $: 
\begin{equation}
\rho =v_{1}\rho _{1}v_{1}^{*}\oplus ...\oplus v_{n}\rho _{n}v_{n}^{*}
\end{equation}

The presentation of sectors would be incomplete without introducing the
conjugate $\bar{\rho}$. For a LQP physicist, the conjugate is represented by
the ``anti-charge'' which is carried by the antiparticle. The latter if
brought together with the particle can annihilate into the vacuum sector $p+%
\bar{p}\rightarrow vac+....$ (In the case of abelian charges the resulting
charge is the vacuum charge without additional contributions).

\begin{theorem}
Let $\theta \in Sect(M)$ be an irreducible sector with $\theta \bar{\theta}=%
\bar{\theta}\theta .$ Then $\theta $ has finite index iff the sector $\theta 
\bar{\theta}$ contains the identity. In that case $\theta \bar{\theta}$ and $%
\bar{\theta}\theta $ contain the identity with multiplicity one and if $%
\theta ^{\prime }$ is any other sector such that $\theta ^{\prime }\theta
\supset id,$ then $\theta ^{\prime }=\bar{\theta}$
\end{theorem}

In the case of finite statistical dimensions there exists a canonical way to
implement conditional expectations by isometries. An important special case
is the following:

\begin{theorem}
Let $N\subset M$ be an irreducible inclusion with finite index. Then there
exists two isometries $v\in M,w\in N$ with $vm=\gamma (m)v$ $\forall m\in M$
and $wn=\gamma (n)w\,\,\forall n\in N$ such that 
\begin{equation}
w^{*}v=\left[ M:N\right] ^{-\frac{1}{2}}\mathbf{1=}w^{*}\gamma (v)
\end{equation}
\begin{equation}
E(m)=w^{*}\gamma (m)w:M\rightarrow N
\end{equation}
Furthermore $M$ can be represented in terms of $N$ by adjunction of a single
basis element $v$: $m=$ $\left[ M:N\right] \cdot E(mv^{*})v,$ i.e. $M=Nv.$
\end{theorem}

\begin{remark}
The last representation is the type$III$ adaption of the well known
representation in terms of a Popa-Pimsner basis which for finite factors
consists if a basis whose (minimal) dimension is related to the index of the
inclusion.
\end{remark}

Having explained the meaning of conjugate for finite index endomorphisms,
one may formulate the Frobenius reciprocity relations for intertwiner
spaces. They are completely analogous to their classic group theoretical
counterpart: 
\begin{equation}
(\rho _{\gamma },\rho _{\alpha }\rho _{\beta })\stackrel{\eta }{\rightarrow }%
(\rho _{\gamma }\bar{\rho}_{\beta },\rho _{\alpha })
\end{equation}
Here the Frobenius isomorphism $\eta $ is a linear map which can be
explicitely in terms of an intertwiner basis and allows (as all other
intertwiner relations) a nice graphical representation. In fact the
intertwiners and their properties form a so called C*-category.

From the point of view of subfactor theory the endomorphisms are are special
realizations of a given inclusion $\rho (M)=N\subset M.$ Any type$III$
inclusion allows for a endomorphic representation: just compose any
isomorphism of $M$ with $N$ with the injection map $N\hookrightarrow M$ .
Since type$III$ factors have myriads of isomorphisms, the endomorphic
representations are highly arbitrary. In terms of an $\rho $ the canonical
endomorphism can be written as $\gamma =\rho \bar{\rho}$ and the two
``charge-anticharge intertwiners $R\in $($id,\rho \bar{\rho})$ and $\bar{R}$
may be shown to fulfil: 
\begin{equation}
\rho (R^{*})\bar{R}=d_{\rho }^{-1}\mathbf{1=}\bar{\rho}(\bar{R}^{*})R
\end{equation}
This situation invites to define a left inverse $\phi $ of $\rho .$ This is
a completly positive normal linear map defined by: 
\begin{equation}
\phi (x)=R^{*}\bar{\rho}(x)R,\,\,x\in M
\end{equation}
It gets its name from the relation $\phi \rho =id.$which is a consequence of
the easily established relation: 
\begin{equation}
\phi (\rho (x_{1})x\rho (x_{2}))=x_{1}\rho (x)x_{2},\,\,x_{i},x\in M
\end{equation}
Whereas automorphisms have an inverse, injective endomorphisms only possess
a left inverse in the above sense unique under the assumed finite index and
irreducibility conditions). The left inverse is an important tool which the
physicist introduced in order to obtain natural trcial states on the ($\rho
^{n},\rho ^{n}),n=1,2,...\infty $ intertwiner algebras which contain the
multiparticle statistic operators. The inverse product $E=\rho \phi $ is
easily shown to define a conditional expectation $M\rightarrow \rho (M).$
for the reducible case there are several left inverses and conditional
expectations and one defines the minimal left inverse which is also called
the standard left inverse) such that it corresponds to the minimal
conditional expectation which was related to the minimal index $\left[
M:N\right] $ (Its square root is the statistical dimension d$_{\rho }).$

\section{Split Property, Localizing Map and Nuclearity}

As a generalization of the standardness of a representation of a single vNa
(always with separable predual), one calls a pair $(N,M)$ of vNa $M$ and sub
vNa $N\subset M$ standard, if there exists a faithful normal state $\omega $
such that the GNS representation ($\pi _{\omega },\Omega _{\omega
},H_{\omega })$ gives rise to an inclusion $\pi _{\omega }(N)\subset \pi
_{\omega }(M)$ which is standard to the vector $\Omega _{\omega }.$ This
last terminology means that in addition to $\pi _{\omega }(M),$ this vector
is also standard with respect to and the ``collar'' $\pi _{\omega
}(N)^{\prime }\cap \pi _{\omega }(M).$ We will identify the abstract vNa and
their GNS representations $\pi _{\omega }.$

An inclusion $(N,M)$ is called split if there exists an intermediate type$I$
factor $B:$%
\begin{equation}
N\subset B\subset M
\end{equation}
A type$I$ factor defines a tensor product split of the Hilbert space $H=H_{1}%
\bar{\otimes}H_{2}$ such that $B=B(H_{1}).$ In this case there exists an
isomorphism $\Phi $ of $N\vee M^{\prime }$ (the commutant of the collar) to
the tensor product $N\bar{\otimes}M^{\prime }$%
\begin{equation}
\Phi (n\cdot m^{\prime })=n\bar{\otimes}m^{\prime },\,\,n\in N,\,\,m^{\prime
}\in M^{\prime }
\end{equation}
The additional property resulting from standard split is that the middle
factor $B$ can be adjusted so that this isomorphism can be canonically
implemented. For this we use the standardness of $N\cap M^{\prime }$ with
respect to $\Omega $ and that of $N\bar{\otimes}M^{\prime }$ with respect to 
$\Omega \bar{\otimes}\Omega .$ This results in the unitary implementability
of $\Phi $ in terms of the following unitary$:$%
\begin{eqnarray}
U_{can}n\cdot m^{\prime }\Omega &=&n\Omega \bar{\otimes}m^{\prime }\Omega \\
U_{can}:H &\rightarrow &H_{1}\bar{\otimes}H_{2}  \nonumber \\
H_{1}=\overline{N\Omega } &,\,H_{2}=&\overline{M^{\prime }\Omega }  \nonumber
\end{eqnarray}
Using the modular involution $J$ of the collar $N^{\prime }\cap M,$ the
canonical type$I$ factor is defined as: 
\begin{eqnarray}
B_{can} &=&N\vee JNJ \\
&=&U_{can}^{*}B(H_{1})\bar{\otimes}1_{H_{2}}U_{can}
\end{eqnarray}
One of the most useful consequences of the split property is the existence
of a universal localizing map $U_{can}.$ It gets its name from the
remarkable property of ``localizing'' global automorphisms of $B(H)$ inside $%
M$ such that the action on the subalgebra $N$ is identical to the
restriction of the global action. In the context of the geometric action of
the Poincar\'{e} group on the net, where $M$ is the algebra localized in
e.g. a bigger double cone, the definition: 
\begin{equation}
U_{loc}(g)\equiv U_{can}^{*}U(g)\bar{\otimes}1U_{can}
\end{equation}
has the correct property and may be viewed as a kind of algebraic
pre-version of the Quantum Noether theorem.

In QFT applications this property is garantied for inclusions of localized
algebras with a causal collar between them if the local degrees of freedom
have a reasonable phase space behavior called ``Nuclearity Property''. Phase
space degree of freedom counting in LQP is based on the concept of $%
\varepsilon $-content $N(\varepsilon )$ of a map $\Theta $ between Banach
spaces. $N(\varepsilon )$ is defined as the maximal number of elements $%
E_{i} $ in the unit ball of the source space such that $\left\| \Theta
(E_{i}-E_{j})\right\| >\varepsilon $ for $i\neq j.$ The order q of the map
is then defined to be: 
\begin{equation}
q=lim_{\varepsilon \searrow 0}sup\frac{lnln\varepsilon }{ln\frac{1}{%
\varepsilon }}
\end{equation}
In heat bath LQP the map would be $\Theta :\mathcal{A}(\mathcal{O}%
)\longrightarrow H$, given by $\Theta _{\Omega ,\beta }(\mathcal{A}%
)=e^{-\beta \mathbf{H}}\Omega ,A\in A(\mathcal{O})$ with $H$ the
Hamiltonian, whereas in case of thermality through localization one takes $%
\Theta _{\Omega ,\Delta }(A)\Omega =\Delta _{\hat{O}}^{\frac{1}{4}}A\Omega
,\,\,\,A\in \mathcal{A}(\mathcal{O})$ with $\Delta $ the modular operator of
the localization region. For the derivation od the split property from the
nuclear properties of these maps with a certain value of q (which measures
the strength with which the number of relativistic degrees of freedom per
unit phase space volume approaches infinity) we refer to \cite{Haag}.

\section{Inverse Problem of Modular Theory}

Mathematically the inverse problem of modular theory is the construction of
a vNa in standard position i.e. pairs $(M,\Omega )$ such that a given would
be pair of modular operators ($\Delta _{0},J_{0})$ of ($M_{0},\Omega _{0}),$
with $M$ isomorphic to $M_{0},$ are also the modular operators of $(M,\Omega
_{0}).$ The appearantly weaker (more general) version in which one only
demands that the new algebra ($M,\Omega _{0})$ shares the same $\Delta _{0}$
turns out to be reducible to the previous problem and is the kind of inverse
problem close to the physical one described below. There are other inverse
problems which have been solved, the most prominent being Connes inverse
problem of reconstructing vNa's from the position of their natural cones in
Hilbert space which (as the below physical problem) has a unique solution.

The above isomorphy classes $IC(M_{0},\Omega _{0})$ have many elements. For
example $M=VM_{0}V^{*}$ with a unitary $V$ commuting with $\Delta _{0},J_{0}$
and obeying $V\Omega _{0}=\Omega _{0}$ gives a family of vNa's. This
construction does not exhaust the possibilities. Already for matrices i.e. $%
I_{n}$ factors one can construct yet another family of inverse problem
classes by choosing $V$ as $V=V_{1}U$ with $V_{1}$having the same
commutation properties as above but $V_{1}\Omega _{1}=\Omega _{0}.$ Here $%
\Omega _{1}$ is such that ($\Delta _{0}^{-1},J_{0})$ are the modular objects
of ($M_{0},\Omega _{1})$ and $U$ commutes with $J_{0}$ and obeys $\Delta
_{0}^{-1}=U^{*}\Delta _{0}U,U\Omega =\Omega ,U\Omega _{1}=\Omega _{1}.$ This
construction only works for $I_{n}$ factors because only there $\Delta
_{0}^{-1}$ can be the modular operator of a pair ($M_{0},\Omega _{1}).$ For
type$I$ factors one has a rather complete solution of the inverse problem.
For type$III$ on the other hand extremely little is known.

The physical aspects of the inverse problem which deal with a special
setting within hyperfinity type$III_{1}$ factors or in LQP terminology wedge
or double cone localized algebras. There one has always a given reference
algebra which in the case of massive particles would be the wedge algebra
associated to the incoming free fields say $\Delta _{0}$. The first step in
the construction of interacting algebras is the recognition that the would
be interacting factor algebra $M$ for the wedge shares the modular $\Delta
_{0}^{it}$ which is the Lorentz boost. This is a corrolar of the fact that
the interaction does not manifest itself in the representation of the
(connected) Poincar\'{e} group but only in the reflections as explained in
chapter5. On the other hand the $J$ of the interacting $M$ is related to $%
J_{0}$ by the physical S-matrix $J=S_{scat}J_{0}.$ In this case there are
sufficiently many additional requirements from LQP so that the interacting
wedge algebra is uniquely determined by the physical S-matrix. The unitary
equivalence transformation between $M_{0}$ and $M$ is analogous to the M\o
ller operator from scattering theory and called the modular M\o ller
operator. It is very interesting that in this physical case the solution of
the inverse problem in the sense of modular theory coalesces with that of 
\textit{the} \textit{inverse problem of LQP}: the unique relation of the
(physically admissable) S-matrix with the associated QFT.

The inverse problem of modular theory and its nonuniqueness, the field
theoretic setting and relation to the inverse problem of QFT (admissable
scattering matrix $S\rightarrow local\,\,QFT),$ the modular M\o ller
operator and the nonperturbative constructive approach.

\textbf{Suggested Literature to this Appendix}

O. Bratteli and D. W. Robinson, ``Operator Algebras and Quantum Statistical
Mechanics'' I, Springer, New York 1979

V. S. Sunder, ``An Invitation to von Neumann Algebras'', Springer, New York
1987

R. V. Kadison and J. R. Ringrose, ``\textit{Fundamentals of the Theory of
Operator Algebras}'' I ans II, Academic Press, New York 1986

M. Takesaki, ``Conditional expectations in von Neumann algebras'', J. Funct.
Anal. \textbf{9}, 306-321 (1972)

R. Longo and K.-H. Rehren, ``\textit{Nets of subfactors}'', Rev. Math.
Phys., \textbf{7}, (1995), 567

S. Doplicher, C. Pinzari and R. Zuccante, ``\textit{The C*-algebra of a
Hilbert Bimodule}'', funct-an/9707006 and previous work of Rieffel, Exel and
Pimsner cited therein.

S. Doplicher and R. Longo, ``Standard and split inclusions of von Neumann
algebras'', Invent. math. 75, 493 (1984)

To my Brazilian collegues: olha para a home page do Ruy Exel:
http://www.ime.usp.br/\symbol{126}exel/

A. Connes, ``Noncommutative Geometry'' Academic Press 1994

\end{document}